\documentclass{msuphddissertation}
\usepackage{amsmath, amssymb, array}
\allowdisplaybreaks
\usepackage{dsfont}
\usepackage[square]{natbib}
\usepackage{mathtools}
\usepackage{adjustbox}
\usepackage{graphicx,xcolor}
\usepackage[colorlinks, linkcolor=black, citecolor=black, urlcolor=black]{hyperref}
\usepackage{slashed, bm, physics}
\usepackage[normalem]{ulem}
\usepackage{xspace}
\usepackage{tocloft}
\usepackage[nottoc,notlof,notlot]{tocbibind}

\author{Zhite Yu}
\title{Exclusive QCD Factorization and Single Transverse Polarization Phenomena at High-Energy Colliders}
\unit{Physics --- Doctor of Philosophy}

\renewcommand{\bibname}{\vspace{-2\baselineskip}\centering\normalsize BIBLIOGRAPHY\vspace*{-2\baselineskip}}

\newcommand{\pp}[1]{\left(#1\right)}
\newcommand{\bb}[1]{\left[#1\right]}
\newcommand{\cc}[1]{\left\{#1\right\}}
\newcommand{\vv}[1]{\langle #1 \rangle}

\newcommand*{\Scale}[2][4]{\scalebox{#1}{\ensuremath{#2}}}

\newcommand{\ch}[1]{Ch.~\ref{#1}}
\renewcommand{\sec}[1]{Sec.~\ref{#1}}

\newcommand{\fig}[1]{Fig.~\ref{#1}}
\newcommand{\figs}[2]{Figs.~\ref{#1} and~\ref{#2}}

\newcommand{\eq}[1]{Eq.~\eqref{#1}}
\newcommand{\eqs}[2]{Eqs.~\eqref{#1} and~\eqref{#2}}
\newcommand{\eqss}[3]{Eqs.~\eqref{#1}\eqref{#2} and~\eqref{#3}}

\newcommand{\eqfig}[2][0.5]{\adjincludegraphics[valign=c, scale=#1]{#2}}

\newcommand{\beq}[1][]{\begin{equation}\label{#1}}
\newcommand{\eeq}{\end{equation}}
\newcommand{\bse}[1][]{\begin{subequations}\label{#1}}
\newcommand{\ese}{\end{subequations}}
\newcommand{\nn}{\nonumber}

\newcommand{\out}{{\rm out}}

\newcommand{\LQCD}{\Lambda_{\rm QCD}}
\newcommand{\TeV}{{\rm TeV}}
\newcommand{\GeV}{{\rm GeV}}
\newcommand{\MeV}{{\rm MeV}}
\newcommand{\ivfb}{{\rm fb}^{-1}}
\renewcommand{\pb}{{\rm pb}}

\newcommand{\ab}{{\rm ab}}\newcommand{\fm}{{\rm fm}}
\newcommand{\sgn}[1]{{\rm sgn}\left[#1\right]}
\newcommand{\wt}[1]{\widetilde{#1}}
\newcommand{\st}[1]{\tilde{#1}}
\newcommand{\la}{\langle}
\newcommand{\ra}{\rangle}
\renewcommand{\P}{\mathcal{P}}
\newcommand{\M}{\mathcal{M}}

\newcommand{\F}{\mathcal{F}}
\newcommand{\Ft}{\widetilde{\mathcal{F}}}
\newcommand{\G}{\mathcal{G}}
\newcommand{\Gt}{\widetilde{\mathcal{G}}}

\newcommand{\htt}{ht\bar{t}}

\renewcommand{\k}{\kappa}
\newcommand{\kt}{\widetilde{\kappa}}
\newcommand{\lum}{\mathcal{L}}
\newcommand{\Poincare}{Poincar\'e\xspace}

\def\lab{{\rm lab}}
\def\A{\mathcal{A}}
\def\At{\widetilde{\mathcal{A}}}

\def\M{\mathcal{M}}

\def\P{\mathcal{P}}
\def\Pb{\bar{\mathcal{P}}}
\def\F{\mathcal{F}}
\def\Ft{\widetilde{\mathcal{F}}}
\def\H{\mathcal{H}}
\def\C{\mathcal{C}}
\def\T{\mathcal{T}}
\def\Ht{\widetilde{\mathcal{H}}}
\def\E{\mathcal{E}}
\def\Et{\widetilde{\mathcal{E}}}
\def\Ct{\widetilde{C}}

\def\O{\mathcal{O}}
\def\UV{{\rm UV}}
\def\CO{{\rm CO}}

\renewcommand{\slash}[1]{#1 \hspace{-0.45em} / }
\newcommand{\Slash}[1]{#1 \hspace{-0.66em} / }

\newcommand{\lrD}{\raisebox{0.09em}{$\stackrel{\raisebox{-0.03em}{$\scriptstyle\leftrightarrow$}}{D}$}{}}
\newcommand{\lD}{\raisebox{0.09em}{$\stackrel{\raisebox{-0.03em}{$\scriptstyle\leftarrow$}}{D}$}{}}
\newcommand{\rD}{\raisebox{0.09em}{$\stackrel{\raisebox{-0.03em}{$\scriptstyle\rightarrow$}}{D}$}{}}

\newcommand{\lrp}{\raisebox{0.09em}{$\stackrel{\raisebox{-0.03em}{$\scriptstyle\leftrightarrow$}}{\partial}$}{}}
\newcommand{\lp}{\raisebox{0.09em}{$\stackrel{\raisebox{-0.03em}{$\scriptstyle\leftarrow$}}{\partial}$}{}}
\newcommand{\rp}{\raisebox{0.09em}{$\stackrel{\raisebox{-0.03em}{$\scriptstyle\rightarrow$}}{\partial}$}{}}

\begin{document}

\maketitlepage

\begin{abstract}
This Ph.D.~thesis is divided into two distinct parts. The first part focuses on hard exclusive scattering processes in Quantum Chromodynamics (QCD) at high energies, while the second part delves into spin phenomena at the Large Hadron Collider (LHC).
 
Hard exclusive scattering processes play a crucial role in QCD at high energies, providing unique insights into the confined partonic dynamics within hadrons, complementing inclusive processes. Studying these processes within the QCD factorization approach yields the generalized parton distribution (GPD), a nonperturbative parton correlation function that offers a three-dimensional tomographic parton image within a hadron. However, the experimental measurement of these processes poses significant challenges. This thesis will review the factorization formalism for related processes, examine the limitations of some widely used processes, and introduce two novel processes that enhance the sensitivity to GPD, particularly its dependence on the parton momentum fraction $x$.
 
The second part of the thesis centers on spin phenomena, specifically single spin production, at the LHC. Noting that a single transverse polarization can be generated even in an unpolarized collision, this research proposes two new jet substructure observables: one for boosted top quark jets and another for high-energy gluon jets. The observation of these phenomena paves the way for innovative tools in LHC phenomenology, enabling both precision measurements and the search for new physics.
\end{abstract}

\begin{dedication} \centering
Dedicated to Lijiang, C.-P., and Jianwei
\end{dedication} 

\begin{acknowledgment}
Perhaps every Ph.D. student has a long and struggling story. It is hardly conceivable that I could accomplish this thesis without the support from people around me. 

First and foremost, I owe my deepest gratitude to my Ph.D. advisor, Prof.~C.-P. Yuan. Throughout my graduate study, I navigated several research topics, and each time, C.-P. was fully supportive and helped me in all possible ways he could. His guiding principle---that I must teach him something before I could graduate---has shaped my approach and led to immense personal growth. I am particularly appreciative of his willingness to make time for me, despite his highly demanding schedule. Each interaction guided me in organizing my thoughts, challenging my preconceived understanding, and probing deeper into my research subjects. I owe him a profound debt of thanks for instructing me in the art of research, for ceaselessly encouraging me to explore new horizons, and for his invaluable assistance in expanding my network and connections. The warmth and hospitality extended by C.-P.'s wife, Ping Ma, during the difficult times of Covid-19, deserve my heartfelt gratitude. Her kindness, support, and the wonderful dinners she prepared provided a comforting sense of home, even while I was far away from my family.

I must also express my profound thanks to Prof.~Jianwei Qiu, with whom I began collaborating during my visit to Jefferson Lab in 2020. This association has been nothing short of enriching. Jianwei's generous sharing of his knowledge, life philosophy, work attitudes, research insights, and innovative ideas has had a transformative effect on my approach to my studies. His guidance and encouragement, coupled with our deep and stimulating discussions, enabled me to navigate the intricacies of QCD factorization with clarity and confidence. Additionally, I owe him immense gratitude for building numerous academic connections on my behalf and facilitating my attendance at several significant conferences. These opportunities have not only expanded my horizons but have also considerably benefited my career. The lessons I've learned from working closely with him continue to resonate with me, and I am eager to carry these forward in my future endeavors.

The professors in the MSU high-energy physics group have been instrumental in my growth, assisting me in various aspects of my research and personal development. I extend my heartfelt thanks to my Ph.D. committee members, Profs.~Andreas von Manteuffel, Wade Fisher, Andrea Shindler, Alexei Bazavov, and Dean Lee, for their continuous encouragement, unwavering support, thoughtful suggestions, and genuine care for both my academic progress and personal well-being.
Andreas deserves my special thanks for the stimulating discussions, considerate suggestions, and warm invitations to his home gatherings. I'm grateful to Prof.~Kirtimaan Mohan for enriching physics dialogues and his tireless assistance with enhancing my programming skills. Prof.~Carl Schmidt guided me through my first physics project and opened up a world of fascinating physics topics that I had the privilege to explore.
I am deeply indebted to Prof.~Wayne Repko, whose groundbreaking papers on polarization phenomena inspired my works on the top quark, $W$, and gluon polarization. His encouragement and enlightening historical insights have been a constant source of motivation. My thanks also go to Profs.~Huey-Wen Lin, Joey Huston, and Reinhard Schwienhorst for their multifaceted support, engaging discussions, and the sharing of crucial career insights.
I must also acknowledge Prof.~Vladimir Zelevinsky for his excellent teaching in quantum mechanics and the unique opportunity he provided me to teach a class under his guidance.
Lastly, I want to express my appreciation to Brenda Wenzlick and Kim Crosslan, who skillfully handled numerous non-academic tasks on my behalf. Their efficiency and dedication saved me much time, energy, and distraction, allowing me to focus on my academic pursuits.

I would also like to take this opportunity to extend my sincere gratitude to several professors back in China who have been guiding me and watching over my academic progress throughout my educational journey. My graduate study in the U.S. was made possible because of the generous help from my undergraduate advisor, Prof.~Qing-Hong Cao. His invaluable advice, constant encouragement, and unwavering support were instrumental during the crucial decisions and preparation stages. I am also profoundly grateful to Profs.~Jiang-Hao Yu and Bin Yan. Their guidance, mentorship, and assistance, particularly in the early stage of my Ph.D. study, not only helped me navigate complex challenges but also instilled in me a strong foundation and passion for my research field.

Despite the wealth of guidance and support I received, the pursuit of a Ph.D. was often fraught with frustration and moments of despair. The challenges were made bearable, however, by the friendships I forged during my time in graduate school. I am profoundly grateful for the companionship, encouragement, and joy brought into my life by the following friends: Bakul Agarwal, Shohini Bhattacharya, Lisong Chen, Zhouyou Fan, Yao Fu, Syuhei Iguro, Peter Kong, Lisa Kong, Zhen Li, Yang Ma, Matteo Marcoli, Xiaoyi Sun, Xudong Tian Tang, Keping Xie, 
Shuyue Xue, Tongzhi Yang, Kang Yu, Rui Zhang, Fanyi Zhao, and Yiyu Zhou. 
In particular, I must extend my most sincere thanks to my friend Boyao Zhu, who has been accompanying and helping me on various important occasions.

After all is said, words cannot fully express my gratitude to my family, especially my parents and brother. They have been standing by me unconditionally, supporting me in every way throughout my entire life. Their unwavering love has filled me with warmth and strength, constantly reminding me that I am cherished. I cannot imagine myself reaching this point and acquiring this Ph.D. degree without their understanding, companionship, encouragement, and steadfast support. I thank them from the bottom of my heart for backing me along this journey, sharing my burdens, and for always being my anchor and my inspiration.

Finally, and most specially, I want to express my heartfelt thanks to my wife, Lijiang Xu. She came into my life at my most frustrated time, bringing colors and sunshine into my dark sky. I thank her for always taking pride in me, supporting every decision of mine, and for imbuing me with more confidence, courage, and determination. Her embrace has filled my life with warmth, help, encouragement, and love. This thesis would not have been possible without her, yet it still does not seem sufficient as a gift in return, to which, I would like to devote the rest of my life.

To all who have walked this journey with me, I extend my deepest thanks and dedicate this work to you. Your faith, guidance, support, and love have shaped not just this thesis, but the scholar and person I have become.

\end{acknowledgment}

\TOC

\newpage
\pagenumbering{arabic}

\begin{doublespace}

\part{QCD Factorization for Exclusive Processes}

\chapter{Introduction}
\label{ch:intro}

Within the Standard Model of particle physics, Quantum Chromodynamics (QCD), the quantum theory describing 
the strong interaction, is the most special part. The Lagrangian governing the dynamics of the theory has colored 
fields, quarks and gluons, as the fundamental degrees of freedom, while the physical spectrum consists of color-neutral 
particles, hadrons, that are composite states of the former. 
The colored particles are never observed in isolation, a property called color confinement, which is the defining feature
of QCD but which is still established as an experimental fact instead of having been derived from the first principles of QCD.
Although it is believed that QCD has all the components for confinement to emerge, it has not been explicitly shown. 

As a result of confinement, the study of the strong interactions among quarks and gluons is always directly involved with hadrons.
The color interactions are so strong that the quarks and gluons are strongly bound inside the hadrons.
On the other hand, however, the color interactions get weaker as the interacting scales become higher. This phenomenon is 
called asymptotic freedom~\citep{Gross:1973id, Politzer:1973fx}, 
which was historically sought first as a necessary condition for describing the strong interaction. 
In a hard collision involving hadrons in the initial or final states, interaction happens at a short time and distance scale so that
the strength of the QCD interaction is very weak, and it is the quark and gluon degrees of freedom that are directly involved.
The latter now behave almost as free particles, so they are collectively called partons.
\footnote{In many contexts, the term ``parton'' loosely refers collectively to a quark or gluon, with no regard to the interaction scale.
Here we restrict its meaning to hard scattering regime because in the bound state of a hadron, the interaction among the colored 
degrees of freedom is so strong that the latter do not have clear particle properties. So when referring to them as partons, 
it is meant that we are working in the kinematic regime where the hadron is probed at a short distance scale, so that the color interaction 
becomes weak and the {\it particle} nature of quarks and gluons emerges. In this sense, the concept of ``particles'' is by itself a 
concept for perturbative interactions, but not for a theory with strong interactions.
}
As partons move away from the hard interaction, the distance between them becomes larger and larger, and the color interaction
becomes stronger and stronger, which eventually turns the partons into hadrons, a process called hadronization. 

Therefore, hard scattering processes of hadrons typically involve QCD in a full range of scales, from the hard scale characterized 
by the hard probe, where the color interactions are weak and perturbative, down to the low scale characterizing the hadronization,
where the color interactions become strong enough to confine the colored degrees of freedom into hadrons.
It is perhaps an ultimate goal to be able to describe such a full process at all scales within QCD, especially the low scales dominated
by the nonperturbative regime of the color interaction. This task, however, is unprecedentedly difficult. Even a semi-complete solution
has not been achieved, yet. The only nonperturbative method so far is given by Lattice QCD~\citep{Workman:2022ynf};
it, however, still suffers from limitations from computational resources and timing and from the intrinsic Euclidean nature instead of
the real Minkowski nature. Furthermore, Lattice QCD is more of simulating the physical results with the QCD Lagrangian 
rather than describing the physical mechanisms. A fully analytic solution is still desirable. With that said, though, the results from
Lattice QCD are still valuable inputs to our endeavor to understand the nonperturbative QCD interaction.

Another approach, given such a situation, to understanding how colors are confined in hadrons is by probing the hadronic structures
in a phenomenological way. This is done through hard scattering experimental processes in which the hadrons are hit by 
high-energy beams of elementary or hadronic particles or large-momentum hadrons are produced in the collision of elementary particles. 
First of all, such processes are probing the partonic degrees of freedom, so are indeed sensitive to the color interactions.
Second, the hard scale implies that there is one stage in the process where the QCD interaction becomes weak and one can utilize
perturbative calculation method, by virtue of the asymptotic freedom.

The QCD factorization theorem~\citep{Collins:1989gx, Collins:2011zzd} has been developed to make full use of the asymptotic freedom
by factorizing the dynamics at the hard perturbative momentum scale and that at the low nonperturbative scale.
Effectively, it gets around the nonperturbative region by identifying good cross sections (or good physical observables) 
whose {\it leading} nonperturbative dynamics can be organized into some distribution functions characterizing the full nonperturbative
partonic dynamics within hadrons, while whose other non-leading nonperturbative contributions are shown to be suppressed 
by inverse powers of the large momentum scale of the collision.
By neglecting the non-leading contributions, the remaining part involves purely the hard momentum scales so as one can reliably 
use perturbation theory on the weakly coupled on-shell partons.
It comes as a result of the factorization theorem that the nonperturbative dynamics associated with each explicitly observed hadron
is independent of each other and the resultant distribution function only depends on each hadron itself, 
but not on the specific processes.
Such property is termed universality of the distribution functions, which, albeit not perturbatively calculable, can be fitted from 
experimental data. In reality, these distribution functions can be represented as correlation functions of quark or gluon fields
between hadronic states. Such field-theoretic definitions also allow them to be calculated using nonperturbative approaches 
like Lattice QCD~\citep{Constantinou:2020hdm}. 
Once obtained, they can be used as inputs to make predictions for different hard scattering processes at different energies. 
In this way, although universality does not come as a prerequisite of establishing factorization in the first place, 
it is the universality that equips factorization with a predictive power.

On the other hand, the universality of the parton correlation functions, especially with the field-theoretic operator definitions, 
enables them to be studied on their own, and thereby uncover certain aspects of the confined partonic dynamics. 
This is how factorization serves as a phenomenological way to probe the hadron structures.
Since the colors are fully entangled inside hadrons, the hadron structures in terms of partons are far more complicated than
the atomic structures in terms of electrons. The best one can do phenomenologically to probe hadron structures is to study
various parton correlation functions. Those correlation functions are in turn embedded by factorization formalism 
into physical observables in hard scattering processes. Different processes give probe to different correlation functions.

The simplest process is the deeply inelastic scattering (DIS), in which a high-energy electron beam is scattered off
the hadronic target, be it a single hadron like a proton or neutron or a nucleus, with a large momentum transfer $q = l - l'$,
where $l$ and $l'$ are the momenta of the electron before and after the scattering, respectively, and $Q \equiv \sqrt{-q^2}$
is much greater than the hadronic scale, $\LQCD \simeq 200~\GeV$. In the final state, only the scattered electron is 
identified and measured, and everything else is {\it inclusively} summed over. At leading power of $\LQCD / Q$, the 
cross section is dominated by the scattering configuration where the target enters the hard interaction via one single
energetic parton 
(which can also be accompanied by arbitrarily many gluons of scalar polarization in a gauge theory with a covariant gauge).
The inclusiveness of the hadronic final states causes the soft gluon exchanges between the scattered parton 
and the beam remnants to be suppressed by $1/Q$, and thus makes the dynamics of the target evolution independent of the
rest of the scattering. In this way, the DIS cross section is factorized into a set of parton distribution functions (PDFs) $f_{i/h}(x)$, 
which, loosely speaking, count the parton number densities at a given longitudinal momentum fraction $x$ 
of a fast moving hadron $h$, for each parton flavor $i$ being a quark $q$, antiquark ($\bar{q}$), or gluon ($g$).

The factorization formula of the DIS cross section and the concept of PDF date back to 1969 before the invent of QCD
when Feynman first proposed the parton model~\citep{Feynman:1969ej}. 
Nevertheless, a carefully formulated factorization formalism based on the first principles of QCD 
leads to many fruitful results. Below are a few relevant ones to this thesis.
\begin{enumerate}
\item[(1)] It identifies the factorization formalism as being separating hard and low energy scales, and makes a consistent use of 
asymptotic freedom.
\item[(2)] It provides a clear operator definition for the PDF, allowing it to be studied by itself within field theory,
and an unambiguous procedure for perturbatively calculating 
the hard parton scattering cross sections $\hat{\sigma}(x)$, 
whose convolution with the PDFs gives the full cross section at leading power of $1/Q$,
allowing one to go to any perturbative orders in principle, which in turn allows the fitting of PDFs to data at any perturbative order.
\item[(3)] By carefully separating hard and low scales with renormalization effects taken into account, the factorization formalism
introduces a factorization scale $\mu$ to the PDF and hard scattering coefficients, so that they are now dependent on one more
variable and shall be written as $f_{i/h}(x, \mu)$ and $\hat{\sigma}(x, \mu)$. 
Then the full hadronic cross section $\sigma$ of the DIS is expressed as,
\beq[eq:intro-DIS-factorize]
	\sigma_h(x_B, Q) = \sum_{i = q, \bar{q}, g} \int_{x_B}^1 dx \, f_{i/h}(x, \mu) \, \hat{\sigma}(x_B / x, Q / \mu) 
		+ \order{\frac{\LQCD}{Q}},
\eeq
where $x_B = Q^2 / 2 P\cdot q$ is the Bjorken variable with $P$ being the hadron momentum.
\item[(4)] The physical requirement that the whole cross section $\sigma_h$ not depend on $\mu$ leads to a set of evolution equations,
called DGLAP equations~\citep{Dokshitzer:1977sg, Gribov:1972ri, Lipatov:1974qm, Altarelli:1977zs},
for the PDFs and hard coefficients, controling their dependence on $\mu$, 
whose solution gives an efficient way to resum all logarithms of $Q / \Lambda$, improving the precision of \eq{eq:intro-DIS-factorize}.
\item[(5)] The factorization procedure can be generalized to other processes besides DIS, 
including semi-inclusive DIS (SIDIS) in lepton-hadron collisions, and Drell-Yan processes in hadron-hadron collisions. 
With the clear operator definitions of the PDFs, one can show their exact universality, 
so as to maximize the predictive power of factorization.
\item[(6)] The full spin dependence of both the hadrons and partons can be consistently included, together with their evolution equations.
\item[(7)] The systematic formulation brings the power suppressed terms in \eq{eq:intro-DIS-factorize} under control, such that one can
also include higher-power terms, with new parton distribution functions, if desired.
\item[(8)] It can be easily generalized (although with practical complications) to different kinds of inclusive processes 
involving more than one scales, especially when they are widely separated.
This leads (consistently) to, among others, a new kind of factorization, called transverse-momentum-dependent (TMD) factorization,
giving rise to a new plethora of distribution functions for probing the hadron structure.
\item[(9)] The same factorization formalism can be applied to exclusive processes where one observes all final state particles. 
Such processes complement the inclusive ones by probing hadron structures in further different aspects, which will form the focus of
this thesis.
\end{enumerate}

QCD factorization formalism has been extremely successful in interpreting high energy experimental data from all facilities 
around the world, covering many orders in kinematic reach in 
both $x$ and $Q$ and as large as 15 orders of magnitude in difference in the size of observed scattering cross sections, 
which is a great success story of QCD and the Standard Model at high energy and has given us the confidence and the 
tools to discover the Higgs particle in proton-proton collisions~\citep{CMS:2012qbp, ATLAS:2012yve}, 
and to search for new physics~\citep{CidVidal:2018eel}.

However, the probe with a large momentum transfer $Q$ is so localized in space, 
$1 / Q \ll R$ with $R \sim 1/\LQCD$ being the typical hadron size,
that it is not very sensitive to the details of confined {\it three-dimensional} (3D) structure of the probed hadron, 
in which a confined parton should have a characteristic transverse momentum scale $\langle k_T\rangle \sim 1/R \ll Q$ 
and an uncertainty in transverse position $\delta b_T  \sim R \gg 1/Q$.  
This calls for the need to go beyond the longitudinal hadron structures described by PDFs, probed by the DIS.
Recently, new and more precise data are becoming available for the so-called {\it two-scale} observables, 
which have a hard scale $Q$ to localize the collision so as to probe the partonic nature of quarks and gluons,
but at the same time entail a {\it controllable} soft scale to give a handle for the dynamics taking place at ${\cal O}(1/R)$.  
Such two-scale observables can be well described by generalizing the factorization formalism for the fully inclusive DIS.
Distinguished by their inclusive or exclusive nature, the generalized factorization theorems enable 
quantitative matching between the measurements of such two-scale observables 
and the 3D internal partonic structure of a colliding hadron.

For inclusive two-scale observables, one well-studied example is the production of a massive boson that 
decays into a pair of measured leptons in hadron-hadron collisions, known as the Drell-Yan process, 
as a function of the pair's invariant mass $Q$ and transverse momentum $q_T$ in the lab frame~\citep{Collins:1984kg}.
When $Q \gg 1/R$, the production is dominated by the annihilation of one active parton from one colliding hadron 
with another active parton from the other colliding hadron, including quark-antiquark annihilation to a vector boson 
($\gamma$, $W/Z$) or gluon-gluon fusion to a Higgs particle.  When $Q\gg q_T \gtrsim 1/R$, 
the measured transverse momentum $q_T$ of the pair is sensitive to the transverse momenta of the two colliding partons 
before they annihilate into the massive boson, providing the opportunity to extract the information 
on the active parton's transverse motion inside the colliding hadron,
which is encoded in the TMD PDFs (or simply, TMDs), $f_{i/h}(x, k_T, \mu^2)$~\citep{Collins:2011zzd},
whose dependence on the factorization scale $\mu$ has been included. 
Like PDFs, TMDs are universal distribution functions to find a parton $i$
with a longitudinal momentum fraction $x$ {\it and} transverse momentum $k_T$ from a colliding hadron of momentum $p$ 
with $xp \sim \mu \sim Q \gg k_T$, and describe the 3D motion of this active parton, its flavor dependence and its correlation 
with the property of the colliding hadron, such as its spin~\citep{Bacchetta:2006tn,Diehl:2015uka}. 
For a spin-$1/2$ target, there are 8 different types of TMDs, for both quark and gluon partons,
categorized by the dependence on the parton and target spins, 
which greatly generalizes the 3 types of quark PDFs and 2 types of gluons PDFs.
Although this poses new challenges for a full extraction of TMDs from experimental data, 
it undoubtedly provides more opportunities to probe multifaceted aspects of the hadronic structures.

However, due to the inclusive nature,
the probed transverse momentum $k_T$ of the active parton in the hard collision is {\it not} the same as 
the intrinsic or confined transverse momentum of the same parton inside a bound hadron~\citep{Qiu:2020oqr}. 
As the colliding hadron is broken by the large momentum transfer $Q$, the fast-moving partons travel with bare colors,
whose strong interactions among each other trigger a complex series of parton emissions and evolutions.
Such collision-induced partonic radiation, called parton shower, generates an additional transverse momentum 
to the probed active parton, which is encoded in the evolution equation of the TMDs and could be non-perturbative, 
depending on the hard scale $Q$ and the phase space available for the shower. 
With more data from current and future experiments, including both lepton-hadron and hadron-hadron collisions, 
better understanding of the scale dependence of TMDs could provide us with valuable information on the 
confined motion of quarks and gluons inside a bound hadron
~\citep{Accardi:2012qut, AbdulKhalek:2021gbh, Liu:2021jfp, Liu:2020rvc}.

In contrast, in exclusive hadronic scattering processes, the colliding hadron(s) are not broken, 
so the corresponding observables could be directly related to intrinsic properties of hadrons, 
without being interfered by parton showers.
In order to employ asymptotic freedom for perturbative calculation, it is necessary to have a hard scale 
$Q \gg 1 / R$ for good exclusive observables. Then, as will be discussed in detail in this thesis, 
the amplitudes of such exclusive processes can also be factorized into nonperturbative parton correlation functions,
with coefficients capturing the hard scattering of the partons that can be calculated perturbatively.
The resultant correlation functions also have field-theoretic definitions, which can be studied on their own 
and encode information on the confined parton dynamics complementary to inclusive processes.

The simplest hard exclusive process is large-angle meson scattering~\citep{Lepage:1980fj, Brodsky:1989pv}, 
with the simplest example being the scattering of electron $e$ and neutral pion $\pi^0$, 
\beq[eq:intro-em2ea]
	e + \pi^0 \to e + \gamma, 
\eeq
which in the center-of-mass (c.m.) frame produces an electron and photon pair with a large scattering angle. 
At a high collision energy, this process contains a hard scale $Q$ that is characterized by the large 
transverse momentum of the final-state $e$ or $\gamma$. 
Then at the leading power of $1/Q$, the annihilation of the $\pi^0$ happens through {\it two} collinear parton lines,
which are constrained to be a quark-antiquark pair by isospin symmetry. 
By slightly generalizing the factorization of DIS, one can express the amplitude of \eq{eq:intro-em2ea} in terms of
the convolution of the {\it distribution amplitude} (DA), $D(z)$, of the pion and a hard coefficient $C(z)$. 
In contrast to inclusive processes, whose factorization is at cross section level, 
the factorization of exclusive processes works at amplitude level, 
and the resultant correlation functions correspond to amplitudes.
In this way, $D(z)$ is the probability amplitude of turning the pion into a pair of quark and antiquark,
carrying longitudinal momentum fractions $z$ and $1-z$, respectively, of the pion, 
in a way analogous to a hadron wavefunction. 
The parton distribution function, on the other hand, is analogous to square of the hadron wavefunction, 
(with certain degrees of freedom traced over).
The process in \eq{eq:intro-em2ea} can also be reversed, with the pion being produced in the final state.
The factorization equally applies and results in a DA $\bar{D}(z)$ that gives the transition amplitude from a pair of collinear
quark-antiquark pair into a pion. 
The operator definitions of $D(z)$ and $\bar{D}(z)$ simply related them by a complex conjugate.

The single-meson process in \eq{eq:intro-em2ea} can be generalized to involve more mesons and also to large-angle
scattering involving baryons. The factorization formalism can be similarly applied, with a DA associated with each hadron
in the initial or final state. A detailed knowledge of the $z$ dependence of DAs entails how the hadron momentum 
is distributed among the valence partons.

Slightly more complicated exclusive processes involve diffracted hadrons. 
The simplest example is given by replacing the pion in \eq{eq:intro-em2ea} by a $h$ of any flavor 
and also adding another hadron $h$ in the final state with a slightly diffracted momentum,
\beq[eq:intro-eh2eha]
	e(l) + h(p) \to e(l') + h(p') + \gamma(k),
\eeq
where the same label $p$ is used for both the protons and initial-state proton momentum. 
In the c.m. frame of the scattering, we require the transverse momenta of the final-state electron and photon to be much greater
than that of the diffracted proton,
\beq
	l'_T \sim k_T \sim q_T \gg p'_T \sim \sqrt{-t},
\eeq
with $t = (p-p')^2$. 
Consider the scattering channel when the photon $\gamma(k)$ is not radiated by the electron, the diffracted hadron is then
connected by {\it two} collinear parton lines to the hard scattering, which is characterized by a hard scale $Q \sim q_T$.
By generalizing the factorization for \eq{eq:intro-em2ea}, we can factorize the amplitude of \eq{eq:intro-eh2eha} into
a new type of parton correlation functions, generalized parton distributions (GPDs) $F^i_h(x, \xi, t)$, 
convoluted with hard coefficients $C_i(x, xi; Q)$ that can be perturbatively calculated. 
The GPD combines the PDF and DA into a coherent picture, 
with $x$ playing the role of the longitudinal parton momentum fraction, which is like the $x$ variable of the PDF,
and $\xi$ characterizing the longitudinal momentum transfer from the hadron $h$ to the hard interaction, which plays the role
of the pion momentum in the DA.

More importantly, the GPD contains one more soft scale $\bm{\Delta}_T \equiv \bm{p}_T - \bm{p'}_T$ that is controllable, 
similar to the $k_T$ dependence in TMDs. Thus the diffractive hard exclusive processes provide another type of good
two-scale observables. 
Here, by Fourier transforming the GPD with respect to $\bm{\Delta}_T$ to the position space $\bm{b}_T$ in the forward limit ($\xi\to 0$), 
the transformed GPD $f_{i/h}(x, \bm{b}_T)$ as a function of $b_T$ provides a transverse spatial distribution of the parton $i$ 
inside the hadron $h$ at a given value of the parton momentum fraction $x$~\citep{Burkardt:2000za, Burkardt:2002hr}. 
That is, measuring GPDs could provide an opportunity to study QCD tomography to obtain 3D parton images in the 
$x$ and $\bm{b}_T$ space, which complements the 3D images encoded in TMDs in the $x$ and $\bm{k}_T$ space.
The spatial $b_T$ dependence could allow us to define an effective hadron radius in terms of its quark (or gluon) spatial distributions, 
$r_q(x)$ (or $r_g(x)$), as a function of $x$, in contrast to its electric charge radius~\citep{Hofstadter:1955ae, Hofstadter:1956qs, 
Simon:1980hu, A1:2010nsl, A1:2013fsc, Zhan:2011ji, Mihovilovic:2016rkr, Mihovilovic:2019jiz, Xiong:2019umf}, 
allowing us to ask some interesting questions, such as should $r_q(x) > r_g(x)$ or vice versa, 
and could $r_g(x)$ saturate if $x\to 0$, which could reveal valuable information on how quarks and gluons are bounded inside a hadron.
By virtue of the exclusiveness, $\bm{\Delta}_T$ is measured in experiments and directly correspond to the GPD variable, 
without contamination from any parton showers. The 3D pictures entailed in GPDs can thus be unambiguously extracted.

However, there are obstacles in the way to use exclusive processes to probe hadron structures. 
First, the exclusiveness dictates each hard scattered hadron or each diffracted hadron to be connected to the hard scattering by
two collinear parton lines. Compared to inclusive processes, this causes a penalty of one power suppression of $1/Q$. 
As a result, the cross sections become lower as one goes to higher energies. 
The accessible data are therefore limited to a finite range of $Q$.
Second, the intriguing parton pictures encoded in DAs and GPDs require a precision knowledge of them as functions of 
$z$ or $x$. This is, however, hard to extract, for two following reasons.
(1) The exclusive factorization happens at the amplitude level, and the convolution variable $z$ or $x$ is the parton loop momentum, 
flowing through the active parton pair defining the DAs or GPDs, whose integration is always from $0$ to $1$ or $-1$ to $1$, 
and is never pinned down to a particular value. 
This is in sharp contrast to the factorization of inclusive processes like the DIS, which happens at the cross section level.
As shown in \eq{eq:intro-DIS-factorize}, the probed $x$ is constrained within the range $[x_B, 1]$.
At the leading perturbative order, $x$ is also equal to the $x_B$, which is a direct experimental observable. 
(2) For most of the known DA or GPD-related processes, the convolutions of the hard coefficients with DA or GPDs 
only give ``moment-type" information, like the integral $\int_0^1 dz \, D(z)/z$ for the DA or 
$\int_{-1}^1 dx \, F(x,\xi,t) / (x - \xi + i\varepsilon)$ for the GPD.
Extracting full details of DAs or GPDs from such moments does not yield a unique solution.
Third, for diffractive processes, there is one extra channel where the diffracted hadron is connected to the hard scattering by 
one virtual photon if its quantum state is allowed. As we will show, the single photon exchange channel has one power enhancement
compared to the GPD channel, so could dominate the contribution to the total amplitude and also interfere with the GPD-sensitive channels.
This causes a large background for extracting the GPDs. 
It will be the main focus of this thesis to try to improve the extraction of GPDs, especially their $x$ dependence.

The structure of this thesis is organized as the following.
In \ch{ch:factorization}, we will review the factorization formalism, in the context of simple processes like the 
representative Sudakov form factor and inclusive processes like the DIS. 
The key elements of factorization, including the Libby-Sterman analysis, power counting, 
subtraction formalism, and Ward identity will be explained in a fair amount of detail. 
The important use of unitarity in inclusive hadron production and Drell-Yan processes 
will be left out for lack of direct relevance to the subsequent content.
In \ch{ch:exclusive}, we will apply the factorization formalism to exclusive processes. 
First we will show the factorization for $2\to2$ large-angle meson scattering, 
starting with the one-meson process in \eq{eq:intro-em2ea},
gradually increasing the meson number, and 
ending at the meson-meson scattering into two mesons. 
The extension to baryonic and $2\to n$ ($n > 2$) cases should be straightforward and will not be discussed.
Then we will generalize the large-angle factorization to the single-diffractive case. 
This will introduce a complication due to a partial pinch in the Glauber region. 
Thanks to the single-diffractive constraint, this pinch will be avoided by deforming contours of the soft gluon momenta.
This will lead to a unified factorization for a general type of hard exclusive processes that only involves one single diffraction.
Our argument will also indicate that as one goes beyond single-diffractive cases, the contour deformation will not be allowed to 
get away with the Glauber pinch, and thus they will not be factorizable, to be discussed in \sec{ssec:double-diffractive}.
There is an intrinsic similarity between the exclusive process factorization at leading power 
and inclusive process factorization at subleading powers, which will be briefly discussed in \sec{ssec:high-twist}.

After setting up the factorization formalism for the single-diffractive processes, \ch{ch:GPD} will be devoted to the phenomenological
study of GPDs, especially to their $x$ dependence. we will review the definitions of GPDs and the parton pictures encoded therein,
and then briefly discuss the most popular processes for probing GPDs, especially their drawbacks in the $x$ dependence.
Then after a systematic discussion on the sensitivity to $x$ dependence, we will introduce two new processes that provide enhanced
$x$-sensitivity. We will first give a detailed description of their hard coefficient calculations, and then organize
them into observables, including both unpolarized differential cross sections and various polarization asymmetries. 
Finally, we will demonstrate how enhanced $x$ sensitivity can help us determine the GPDs.

In \ch{ch:summary-qcd}, we conclude this thesis and present the outlook for the future.

\chapter{Review of QCD Factorization: General principles}
\label{ch:factorization}


QCD is a renormalizable non-Abelian SU(3) gauge theory describing the color interaction among quarks and gluons~\citep{Workman:2022ynf}. 
The non-Abelian gauge interaction structure leads to the following renormalization group equation for the strong coupling 
$\alpha_s = g_s^2 / (4\pi)$,
\beq[eq:QCD-RGE]
	\mu^2 \frac{d \alpha_s}{d \mu^2} = \beta(\alpha_s) = - b_0 \alpha_s^2 + \order{\alpha_s^3},
\eeq
where $b_0 = (11 C_A - 4 n_f T_F) / (12\pi) > 0$, with the Casimirs $C_A = 3$, $T_F = 1/2$ being the SU(3) color factors, and
$n_f$ the active quark flavor number at the scale $\mu$.
Truncating \eq{eq:QCD-RGE} at leading order (LO) gives the simple solution,
\beq[eq:running-as]
	\alpha_s(\mu) = \frac{1}{b_0 \ln(\mu^2 / \LQCD^2)}
\eeq
with the Landau pole
\beq
	\LQCD = \mu_0 \exp\pp{ - \frac{1}{2 b_0 \alpha_s(\mu_0)} }
\eeq
being determined by an experimental measurement of $\alpha_s(\mu_0)$ at a certain scale $\mu_0$. 
Given the input $\alpha_s(m_Z) = 0.118$ at the $Z$ pole $m_Z = 91.18~\GeV$ (where $n_f = 5$), we get 
$\LQCD \simeq 87.8~\MeV$. Evaluating \eq{eq:QCD-RGE} at a higher oder can modify this value. A more
precise computation at $\order{\alpha_s^5}$ gives $\LQCD \simeq 208~\MeV$~\citep{Chetyrkin:2000yt}.

\eq{eq:QCD-RGE} establishes QCD with drastically distinct physics phenomena 
at the two opposite ends of the energy spectrum~\citep{Gross:1973id, Politzer:1973fx}. From \eq{eq:running-as} we can infer that 
at high-energy (or short-distance) scale, the coupling $\alpha_s$ decreases asymptotically to zero 
such that quarks and gluons are asymptotically free particles, with their interactions as perturbations on their free motions, 
whereas at the low-energy (or long-distance) side, the coupling strength becomes increasingly larger
and blows up at the Landau pole $\LQCD$, and hence quarks and gluons strongly interact with each other 
and the perturbative picture breaks down.
Although this picture is obtained within perturbation theory, the RGE solution has resummed a certain logarithmic order 
to all perturbative orders in $\alpha_s$ and reflects some aspects of the nonperturbative domain, 
which is also confirmed by the fact that $\LQCD$ is of the same order of the hadron size $R\sim 1~\fm \simeq 200~\MeV$.
So while it is not likely to be the case that the coupling $\alpha_s$ indeed becomes infinite at $\LQCD$, 
it certainly implies that the perturbation description cannot be extrapolated beyond that scale. 
Some nonperturbative mechanism must kick in at $\mu \gtrsim \LQCD$ to resolve the perturbative singularity, and confine 
the quarks and gluons within hadrons.
In this way, elementary fields in the QCD Lagrangian do not appear individually in nature, but it is their bound states, hadrons, 
that are directly observed in reality. 

This is the basic intuitive picture of QCD. 
Its speciality in this aspect is that one cannot separate the strong interaction from the particle constituents, 
like what one may consider for the electroweak sector of the SM. 
The QCD by itself is a nonperturbatively strongly interacting theory, as well as self-contained and self-explained.
Therefore, any experiments one may possibly conceive to probe the QCD dynamics are directly involved with hadrons, 
and the associated nonperturbativity. 
QCD factorization is a method or formalism that applies to certain kinds of processes involving one (or more) hard scale $Q$
by separating the short-distance dynamics, which can be perturbatively treated with the quark and gluon degrees of freedom, 
from the long-distance dynamics, which is proved to correspond to certain nonperturbative universal process-independent 
parton correlation functions in the hadrons.
Those functions can be obtained by fitting various different processes to the experiments and facilitate the predictive power of QCD.
On the other hand, a precise knowledge of those functions also uncovers valuable aspects of the hadron structures.

In this section, I will review important concepts and technicalities of QCD factorization, using the Sudakov form factor and
some familiar inclusive processes as examples.
This will serve as a useful background and comparison for the exclusive processes to be treated in the next section.


\section{Factorization as a power expansion}
\label{sec:fac-power}

\subsection{Feynman's intuition and parton model}
\label{ssec:parton-model}
Factorization is the rigorous mathematical formulation of the Feynman's parton model~\citep{Feynman:1969ej} from the first principles of QCD.
To motivate the starting point of factorization, I briefly review the Feynman's intuition leading to the parton model.

In the deep inelastic scattering (DIS) of electron $e$ and proton $p$, 
\beq
	e(k) + p(P) \to e(k') + X,
\eeq
the electron exchanges a virtual photon $\gamma^*$ of momentum $q = k - k'$ that has a high virtuality $Q = \sqrt{-q^2}$ to 
localize the interaction within the size $\delta r \sim 1/Q$.
When working in the center-of-mass (CM) frame, the fast-moving proton undergoes Lorentz contraction and dilation, such that 
\begin{enumerate}
\item [(1)]
	from the perspective of the hard interaction, the proton appears as a flat plate of transverse width $R \sim 1/\LQCD$,
\item [(2)]
	the parton constituents, called partons by Feynman and later identified as quarks and gluons, are more or less evenly distributed in this ``flat plate'', and
\item [(3)]
	the interaction among the partons happens in a time scale $\tau \sim 1/ \LQCD$ in the proton rest frame, and now becomes dilated to be 
	$\tau' \sim (Q / m) \tau \sim Q / \LQCD^2$, where we used the fact that the proton mass $m$ is of the same order as $\LQCD$.
\end{enumerate}
Now (2) implies that when the electron (or the virtual photon) hits one parton in the proton, which happens in the distance scale $\delta R \sim 1/Q$ 
with a duration $\delta t \sim 1/Q$, the probability for a second parton to participate in the hard interaction is of the order 
$\delta r / R \sim 1 / (QR) \sim \LQCD / Q$, which is suppressed as $Q \gg \LQCD$.
The partons confined in the hadrons are never freely on shell, but engage in strong interactions with other partons and have virtualities of order $\LQCD^2$. 
During the hard interaction, the interaction between partons is suppressed by $\delta t / \tau' \sim \LQCD^2 / Q^2$, by (3),
so the role played by the parton virtualities in the hard interaction is also suppressed.

As a result, the hard interaction between the electron and proton is actually between the electron and a single free on-shell parton. 
The whole DIS cross section can be approximated by the product of the probability of finding one parton out of the proton 
and the cross section of the electron scattering off a free parton. 
The factorization formula written down following this intuition constitutes the Feynman's parton model~\citep{Feynman:1973xc}.
The Feynman's parton model predates the establishment of QCD and will receive further corrections and developments from QCD.
But it makes clear the important principle of factorization that we are approximating the full hadron cross section by expanding 
in terms of the power of $\LQCD / Q$. Factorization is valid up to a power correction.

\subsection{Assumptions of factorization}
\label{ssec:fac-assumption}
In order to derive factorization from the first principles of QCD, we need to systematize the power expansion of the full hadronic cross sections. 
Without being able to solve QCD nonperturbatively, this cannot be constructed from zero, but has to rely on certain assumptions on the 
nonperturbative nature of QCD. These assumptions need not be made very precise, and should be moderate enough so as not to contradict our 
first intuitions and experiments. For our purpose, we take the following assumptions:
\begin{enumerate}
\item[(1)]
	A hadron entering the interaction is connected to a group of parton lines that make up a correct quantum number. The connection vertex can be thought of
	as some wave function which does not need to be made clear. All possible parton configurations should be included, with different probability amplitudes.
	This assumption makes concrete the discussion of the hadron scattering in terms of its parton constituents.
\item[(2)]
	Inclusive processes involve a sum over final states, which in reality are all kinds of hadronic states. In perturbative picture, those hadronic states emerge
	from the hadronization of partonic states. We take the assumption that the sum over hadronic states is equivalent to the sum over partonic states,
	\beq
		\sum_X | X_h; \out \rangle \langle X_h; \out |
		\Longleftrightarrow
		\sum_X | X_{q,g}; \out \rangle \langle X_{q,g}; \out |.
	\eeq
	Note that this is not an equal sign, because the partons can easily make up a color non-singlet state. 
	Since initial states are usually color singlets, those non-singlet states necessarily give zero results.
	This assumption avoids dealing with the details of parton to hadron transitions in the final states, as the latter is not clearly understood.
\item[(3)]
	The sum over all perturbative Feynman diagrams in terms of partons, whether it converges or not, represents the true nature. This assumption also
	underlies the perturbation theory for the electroweak interaction, but has more significance for the nonperturbative theory, QCD. It is also the foundation of 
	the previous two assumptions.
\item[(4)]
	The configuration in which all the partons connected to the hadrons are highly virtual is strongly suppressed. In the opposite limit, when the partons have low
	virtuality, we expect nonperturbative dynamics to kick in and slowly saturates the kinematic region with the parton virtuality $k^2 \lesssim \LQCD^2$. 
	This will be made more precise in Sec.~\ref{ssec:pinch} that if the low-virtuality region is not associated with a pinch singularity, one can deform the 
	contour of the parton momentum to stay away from a low virtuality. It is only at {\it pinched} low-virtuality regions that genuine nonperturbative dynamics dominates.
\end{enumerate}

\section{Libby-Sterman analysis}
\label{sec:libby-sterman}

\subsection{Two examples of pinch singularity}
\label{ssec:pinch-ex}
To understand the significance and get some feelings of pinch singularity, let us first study two simple toy examples.

The first example is given by the one-dimensional integral 
\beq[eq:I1-pinch]
	I_1(m) = \lim_{\epsilon \to 0^+} \int_{-\infty}^{\infty} \frac{dx}{2\pi}\, \frac{1}{x^2 - m^2 + i\epsilon} \,,
\eeq
where the limit $\epsilon \to 0^+$ is to remind that the same $i\epsilon$ prescription as in Feynman integrals applies here 
to shift the poles $\pm m$ on the integration contour to lower and upper half planes respectively. 
We will suppress the limit $\epsilon \to 0^+$ later if no confusion occurs.
Because of the $i\epsilon$ prescription, the integration variable $x$ should be considered as a complex variable whose integration 
contour is on the whole real axis.
By the normal trick of Wick rotation $x \to i x$, this integral can be easily evaluated,
\beq[eq:I1-exact-sol]
	I_1(m) = - \frac{i}{2m}.
\eeq

\begin{figure}[htbp]
	\centering
	\begin{tabular}{cc}
	\includegraphics[scale=0.75]{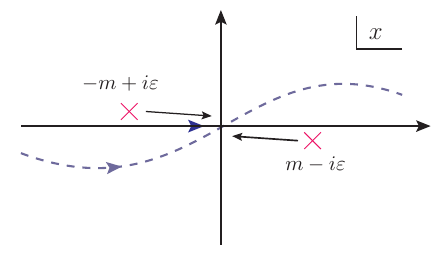} & \hspace{4em}
	\includegraphics[scale=0.75]{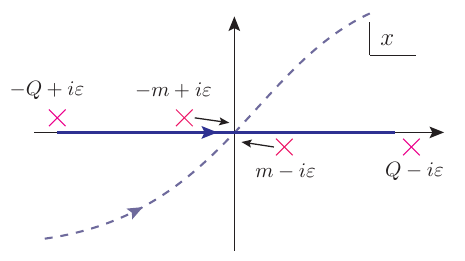} \\
	(a) & \hspace{4em} (b)
	\end{tabular}
	\caption{Integration contours of $x$ in (a) \eq{eq:I1-pinch} and (b) \eq{eq:I2-pinch} on the complex plane, respectively. 
		The poles $\pm m \mp i \epsilon$ approach the origin as $m = 0$ and pinch the the contour in both cases. 
		When $m \neq 0$, we can deform the contours up to $m$, as indicated by the dashed blue lines.
	}
	\label{fig:x-contour}
\end{figure}

Why is the integral $I_1(m)$ singular at $m = 0$? 
As $m \to 0$, the two poles $\pm m \mp i \epsilon$ approach each other and coalesce at $0$ as $m = 0$. These two coalescing poles lie on 
different sides of the integration contour and pinch it at the origin such that no matter how we deform the contour, it must pass through the poles
as $0$. Around the pole, we are dealing with an integral $\int_0 dx / x^2$, which diverges in a power form, and hence $I_1(m) \propto 1/m$ has
a power singularity of $m$ as $m \to 0$. Such singularity due to a pair of coalescing poles pinching the integration contour is called pinch singularity.

When $m \neq 0$, the pinch becomes inexact. And then we can deform the contour to stay away from the two poles $\pm m \mp i \epsilon$, but only up to the
extent of $\order{m}$, i.e., in the region near the poles, the distance between the contour and the poles is at most $m$, $|x - (\pm m)| \lesssim m$, as shown in 
\fig{fig:x-contour}(a) by the dashed blue line. In this region, the integral has a power counting
\beq
	\int_{\sim m} \frac{dx}{x^2 - m^2 + i\epsilon} \sim m \cdot \frac{1}{m^2} \sim \frac{1}{m},
\eeq
where we count $|dx|$ as $m$ and $|x^2 - m^2|$ as $m^2$.
This agrees with the exact solution in \eq{eq:I1-exact-sol}. The region with $|x| \gg m$ does not suffer from any pinched poles, so we can deform the contour 
to make $x$ arbitrarily large, which gives infinitesimal contribution to the integral, 
\beq
	\int_{\gg m} \frac{dx}{x^2 - m^2 + i\epsilon} \sim \int_{x \gg 1} \frac{dx}{x^2} \sim 0.
\eeq
In this way, the main contribution to the integral $I_1(m)$ comes from the region near the two poles $\pm m$ that would become pinched as $m \to 0$.

We can understand this further by taking the two poles $\pm m$ to be on the same half plane, which gives a modified integral
\beq[eq:I1']
	I'_1(m) = \int_{-\infty}^{\infty} \frac{dx}{(x - m + i\epsilon)(x + m + i\epsilon)} 
		= \int_{-\infty}^{\infty} \frac{dx}{x^2 - m^2 + i\epsilon \, \sgn{x}}.
\eeq
Since both poles $\pm m - i\epsilon$ lie on the lower half plane, we can uniformly deform the integration contour to the upper half plane all the way to infinity, which 
kills the whole integral,
\beq
	I'_1(m) = \lim_{K \to \infty} \int_{-\infty}^{\infty} \frac{dx}{(x + i K - m + i\epsilon)(x + i K + m + i\epsilon)} = 0,
\eeq
as can be easily verified from the direct evaluation of \eq{eq:I1'} by residue theorem.

Therefore, the pinch singularity at $m = 0$ becomes the region that gives important contribution to the integral when $m \neq 0$.

Now we consider the second toy example,
\beq[eq:I2-pinch]
	I_2(m, Q) = \int_{-\infty}^{\infty} \frac{d^2 \bm{x}}{(\bm{x}^2 - m^2 + i\epsilon)(\bm{x}^2 - Q^2 + i\epsilon)},
\eeq
which is a two-dimensional integral with an extra hard scale $Q \gg m$.
A direct evaluation gives
\beq[eq:I2-exact-sol]
	I_2(m, Q) = - \frac{1}{4\pi (Q^2 - m^2)} \ln \frac{Q^2}{m^2} \simeq - \frac{1}{4\pi Q^2} \ln \frac{Q^2}{m^2} \times \bb{ 1 + \order{\frac{m^2}{Q^2}} },
\eeq
where in the second step we also gave the approximation to the leading power of $m^2 / Q^2$.

Similar to \eq{eq:I1-exact-sol}, $I_2$ is also singular as $m\to 0$, but logarithmically. This is also due to the pinched poles $\pm m \mp i\epsilon$ as $m \to 0$. Around
the poles, the integral now counts as $I_2 \sim d^2 \bm{x} / (\bm{x}^2 Q^2) \sim dx / (x Q^2)$, which diverges logarithmically due to the two-dimensional integration measure.
When $m \neq 0$, we can deform the contour to avoid the poles $\pm m$ but only up to $m$, so that the contribution of the region near those poles is of the order
\beq[eq:I2-m]
	\int_{\sim m} \frac{d^2 \bm{x}}{(\bm{x}^2 - m^2 + i\epsilon)(\bm{x}^2 - Q^2 + i\epsilon)}
	\sim \frac{m^2}{m^2 \cdot Q^2}
	\sim \frac{1}{Q^2},
\eeq
where we counted $d^2\bm{x}$ as $m^2$, and neglected $m^2$ and $x^2$ with respect to $Q^2$. 

Apart from the poles $\pm m$, $I_2$ has two additional poles $\pm Q \mp i\epsilon$. We take $Q \gg m$ to be large, so these two poles do not give pinch singularity, but they still
keep the contour from being arbitrarily deformed. In the region with $|x| \sim Q$, the distance between the contour and the poles $\pm Q$ is at most of order $Q$, as shown in 
\fig{fig:x-contour}(b), so this region gives a power counting
\beq[eq:I2-Q]
	\int_{\sim Q} \frac{d^2 \bm{x}}{(\bm{x}^2 - m^2 + i\epsilon)(\bm{x}^2 - Q^2 + i\epsilon)}
	\sim \frac{Q^2}{Q^2 \cdot Q^2}
	\sim \frac{1}{Q^2},
\eeq
where we have counted both $d^2\bm{x}$ and $|\bm{x}^2 - Q^2|$ as $Q^2$.

In the scenario with $Q \gg m$, the region $|x| \sim m$ is usually called the soft region, and $|x| \sim Q$ the hard region. This is an example of two regions, and we notice from Eqs.~\eqref{eq:I2-m} and~\eqref{eq:I2-Q} that both regions have the power counting $1/Q^2$. The intermediate region should also have the same power counting and give the result
\beq[eq:I2-mQ]
	\frac{1}{Q^2} \int_{\sim m}^{\sim Q} \frac{dx}{x} \sim \frac{1}{Q^2} \ln \frac{Q}{m}.
\eeq
The region with $|x| \gg Q$ is referred to as the UV region in a Feynman integral context. Here because of the UV counting $d^2x / x^4 \sim 1/ x^2$ as $|x| \to \infty$, it does not contribute to $I_2$.

In this way, we can understand the result in \eq{eq:I2-exact-sol}. The integration of $\bm{x}$ runs over the whole domain from $-\infty$ to $\infty$, and the scales $m$ and $Q$ appear 
in the final integral result through pinch singularities. Here we are using the term ``pinch singularity" in a more general sense, 
not necessarily related to a genuine singularity, but referring to the fact that the contour is constrained by two poles on different sides so that it is forced to go through the region
set by those poles.
The pinch singularity is a necessary condition for the associated region to make an important contribution to the integral, true for both soft and hard regions. 
As we have seen from \eq{eq:I2-mQ}, if two regions have the same power counting, their intermediate region leads to a logarithmic contribution interpolating the two regions.
 
\subsection{Power expansion and pinch singularity}
\label{ssec:pinch}
Physical amplitudes are represented by Feynman diagrams and are given by Feynman integrals of the the parton loop momenta, which can be written as
\beq[eq:Feynman integral]
	I(p, m) = \int d^d k \frac{X(k; p, m)}{\prod_{j = 1}^N (D_j(k; p, m) + i \epsilon)^{n_j}},
\eeq
where $k$, $p$, and $m$ denotes the array of all the loop momenta, external momenta, and masses, respectively, all being multidimensional.
The denominator $D_j(k; p, m)$ is at most a quadratic form of its arguments, and the numerator $X(k; p, m)$ is a polynomial.
Typically, $n_j = 1$ for all $j$.

The external momenta $p = \{p_1, p_2, \cdots, p_n\}$ define $n + C_n^2 = n(n+1)/2$ scales, given by their virtualities $Q_i^2 = p_i^2$ and 
scalar products $Q^2_{ij} = p_i \cdot p_j$. We examine the simplest case when there is a single hard scale $Q$, provided by one or more of the $|Q^2_{ij}|$'s,
which is much larger than all the other scales, i.e., mass, virtuality, and $Q_{ij}^2$ scales\footnote{The case of a highly virtual photon as in DIS should be considered 
as being embedded into the full diagram, with the electron and proton being the external particles.}, which are taken to be of the same order, to be referred to as 
the soft scale and denoted as $M$. Such a two-scale integral can be examined through the power expansion, which can be schematically written as
\beq[eq:power expansion]
	I(Q, M) = Q^{\dim I} \cdot \sum_{n = -n_0}^{\infty} \pp{\frac{M}{Q}}^n \bb{ \sum_{i = 0}^{\infty} I_{n, i} \, \ln^i \pp{ \frac{Q}{M} } },
\eeq
where $n, n_0$ and $i$ are integers, $I_{n, i}$ are scaleless functions of $M/Q$, and the logarithmic dependences on $Q/M$ have been explicitly separated out. 
In the kinematic regime $Q \gg M$, it would be a good approximation to only keep the leading term (or first few terms) in the power series [\eq{eq:power expansion}].
The Feynman's parton model described in Sec.~\ref{sec:fac-power} motivates the conjecture that the leading-power approximation should give a factorization structure.

How do we systematically obtain such a power expansion? 
First, we rescale all the variables in \eq{eq:Feynman integral} by $Q$,
\beq[eq:rescale by Q]
	k \to Q \, \st{k}, 
	\quad
	p \to Q \, \st{p},
	\quad
	m \to Q \, \st{m}.
\eeq
This separates a factor $Q^{\dim I}$ and converts \eq{eq:Feynman integral} into a scaleless integral,
\beq[eq:Feynman integral rescale]
	\st{I}(p, m) = I(p, m) / Q^{\dim I} 
		= \int d^d \st{k} \frac{X(\st{k}; \st{p}, \st{m})}{\prod_{j = 1}^N (D_j(\st{k}; \st{p}, \st{m}) + i \epsilon)^{n_j}}.
\eeq
The scenario $Q \gg M$ can be approached by taking the limit $Q \to \infty$, under which all the external particles become massless and on shell 
and all the mass scales vanish,
\beq
	\st{p}_i^2 = p_i^2 / Q^2 \to 0, 
	\quad
	\st{m}_i^2 = m_i^2 / Q^2 \to 0.
\eeq
The scalar products $\st{p}_i \cdot \st{p}_j$ becomes a scaleless constant of order $1$ if $i$ and $j$ are separated by a constant angle, 
or vanishing if they become collinear to each other.
In this way, the the high-$Q$ limit is equivalent to taking all the mass scales to 0 and all the external particles to be massless and on shell. 
This massless limit implies singularities for the terms in \eq{eq:power expansion} with $n \leq 0$. More leading terms correspond to more severe mass divergences.
Hence, the problem of obtaining the power expansion in $M/Q$ is converted to the problem of finding mass divergences in the corresponding massless theory.

Where does the mass divergence come from? 
The loop momentum $\st{k}$ in \eq{eq:Feynman integral rescale} is integrated from $-\infty$ and $\infty$, and the $i\epsilon$ prescription implies that we should consider 
the integral as a multidimensional integration of a complex variable $\st{k}$ on the real axis. There are poles along the integration contour, given by the zeros of one or 
more $D_j(\st{k}; \st{p}, \st{m})$'s. If the contour is not trapped around the poles, we can deform the contour to stay away from them such that on the deformed contour,
$|D_j| = \order{1}$ instead of 0. This gives a contribution of order 1 to the integral. However, if the contour deformation is forbidden by a pair of or more pinched poles, the 
integration is forced to include the region where one or more $D_j$ is vanishing. Such regions give singular contribution to the integral, which may or may not be remedied by
the numerator $X$ and integration measure. 

This is the case for the massless theory with $\st{m}, \st{p}^2 \to 0$. For finite $\st{m}$ and $\st{p}^2$, the pinch is no longer exact, and we are allowed to deform the contour to
stay away from the poles that would become pinched as $\st{m}, \st{p}^2 \to 0$, but only up to the extent of order $M$, such that in the region near those poles, 
the previously vanishing propagators $D_j(\st{k}; \st{p}, \st{m})$ are now of order $M^2 / Q^2$ or $M/Q$, depending on their dimensions. 
Without considering the numerator and integration measure, this pinched region would give power enhanced contribution to the integral with respect to 
the hard region where all propagators are of order 1.
Therefore, we conclude that {\it the pinch singularities in the corresponding massless theory specify the important integration regions in the massive theory.}

The above discussion only concerns the pure perturbative Feynman integrals and assumes all particles are massive to regulate the mass divergences.
In the real case of QCD, there is indeed a massless gluon, which can cause exact pinch singularity to the Feynman integral. However, around the pinched poles,
there are some parton propagators with vanishing virtualities. By the fourth assumption in Sec.~\ref{ssec:fac-assumption}, 
when partons have virtualities that are less than or of the same order of $\LQCD^2$, 
we should expect nonperturbative dynamics to come in and rescue the perturbative singularities. 
In this way, the soft scale in the full theory is not given by the masses of partons, but by the intrinsic QCD scale $\LQCD$, or the hadron mass scale.
And we expect the nonperturbative effects not to change the power counting of the perturbative theory, 
but only to smoothly regulate the singular behavior, playing a role similar to a mass scale.
Following this, {\it the perturbative pinch singularities do not lead to genuine divergences in the amplitudes or cross sections, 
but imply the regions in the parton momentum integrations that are sensitive to nonperturbative QCD dynamics.} 

This idea is an underlying (usually implicit) foundation of the applications of perturbative QCD. 
Without the knowledge of the nonperturbative solution to QCD, it indicates two approaches for predicting high-energy scattering experiments, 
(1) to separate the part of a diagram containing the propagators that are pinched on shell from the propagators with high virtualities (with or without contour deformation),
and
(2) to design suitable observables for which the perturbative pinch singularities cancel, so that the infrared sensitivity is cancelled.
The first approach leads to factorization, in which the subdiagrams containing the pinched propagators will be organized into universal parton correlation functions, 
and the subdiagrams with highly virtual propagators have little sensitivity to the infrared QCD dynamics and constitute a hard scattering coefficient.
Following the second approach are defined the so-called infrared-safe observables, which can be reliably calculated using perturbative method.
We will see that to obtain the factorization for most processes, both approaches are needed; in particular, we need to show the cancellation of soft gluon connections.

\subsection{Landau criterion for pinch singularity}
\label{ssec:landau}
The first step to derive factorization is to identify the pinch singularities in the corresponding massless theory, obtained by rescaling all the momentum and mass scales 
by the hard scale $Q$, as in \eq{eq:rescale by Q}. This task is easy for simple low-dimensional integrals like Eqs.~\eqref{eq:I1-pinch} and \eqref{eq:I2-pinch}, but not 
for multidimensional Feynman integrals; even the simple one-loop Sudakov form factor [\fig{fig:sudakov-1-loop}(a)] becomes not trivial. A systematic criterion for pinch singularity
is therefore needed, which is given by Landau equations~\citep{Landau:1959fi, Collins:2020euz}.
For the Feynman integral in \eq{eq:Feynman integral} (or the corresponding massless one), a singular point $k_S$ makes a subset of the denominators $D_j$'s vanish. 
The sufficient and necessary condition for this singularity to be pinched is that the first derivatives of those $D_j$'s at $k_S$ are linearly dependent, with non-negative coefficients,
i.e., 
\bse\label{eq:Landau condition}
\begin{align}
	D_j(k_S) &= 0 , \mbox{ for } j \in S_N \subset \{1, 2, \cdots, N\} ,
	\quad
	\mbox{ and } \\
	\sum_{j \in S_N} \lambda_j \frac{\partial D_j(k_S)}{\partial k_S^{\mu}} &= 0,
	\mbox{ with } \lambda_j \geq 0 , \mbox{and at least one $\lambda_j$ is strictly positive}.
	\label{eq:Landau eq2}
\end{align}
\ese
Note that in the notation $k$, we have assembled all the loop momenta $\{k_1, k_2, \cdots, k_L\}$ by a direct sum, so 
the index $\mu$ in \eq{eq:Landau condition} actually runs over $4L$ components. Hence \eq{eq:Landau eq2} is true for each loop momentum.
\beq[eq:Landau eq loop]
	\sum_{j \in S_N, \, j \in \mbox{ loop } l} \lambda_j \frac{\partial D_j(k_{Sl})}{\partial k_{Sl}^{\mu_l}} = 0,
	\quad
	l = 1, 2, \cdots, L,
\eeq
where $k_l$ is the loop momentum for the loop $l$, and $L$ denotes the total loop number.

Finding all solutions to the Landau equation is made simple by the Coleman-Norton theorem~\citep{Coleman:1965xm}. A single diagram may be associated with different
solutions in which different sets of propagators $D_j$ go on shell. For those propagators that are not on shell, we contract the corresponding lines in the Feynman diagram 
to points and obtain a {\it reduced diagram}. 
In the reduced diagram, each loop gives an equation as \eq{eq:Landau eq loop}. Suppose a certain loop momentum $k$
flows through $n$ lines, we assign to each vertex a spacetime coordinate, $x_a^{\mu}$, ($a = 1, 2, \cdots, n$), 
and define the spacetime distance between two adjacent vertices $j_a$ and $j_b$ that are connected by the line $j$,
\beq
	\Delta x_{j_a j_b}^{\mu} = \Delta x_{j_a}^{\mu} - \Delta x_{j_b}^{\mu} = \lambda_j \frac{\partial D_j(k)}{\partial k_{\mu}}.
\eeq
Recall that we are dealing with massless theory, and the condition $D_j(k) = 0$ implies that each propagating line is on shell. 
In most cases, $D_j(k)$ takes a quadratic form like $(k + p)^2$, such that $\partial D_j(k) / \partial k^{\mu}$ gives the momentum of the internal line, whose direction
is oriented according to the direction of $k$. By interpreting $\lambda_j$ as the ratio of the travel time $\Delta x_{j_a j_b}^0$ to the propagating energy
$\partial D_j(k) / \partial k_0$,
\beq
	\lambda_j = \Delta x_{j_a j_b}^0 \bigg/ \frac{\partial D_j(k)}{\partial k_{0}},
\eeq
we have
\beq
	\Delta x_{j_a j_b}^{\mu} = \Delta x_{j_a j_b}^0 \cdot v_j^{\mu},
\eeq
where $v_j^{\mu} = \partial^{\mu} D_j(k) / \partial^{0} D_j(k) = (1, \bm{p}_j / E_j) = (1, \bm{v}_j)$ is the four-velocity of the particle on line $j$. 
In this way, $\Delta x_{j_a j_b}^{\mu}$ becomes the spacetime elapse of a physically propagating on-shell (massless) particle, whose velocity is set by its momentum
$\partial^{\mu} D_j(k) \equiv \partial D_j(k) / \partial k_{\mu}$. 
And then the condition in \eq{eq:Landau eq loop} is equivalent to 
\beq[eq:Coleman-Norton]
	\sum_{j \in S_N, \, j \in \mbox{ loop } l} \Delta x_{j_a j_b}^{\mu} = 0,
\eeq
where the direction $j_b \to j_a$ is the same as the loop momentum $k$.
\eq{eq:Coleman-Norton} gives a consistent condition for a physically realizable classical process of particle propagation: by orienting all lines as going forward with positive
energy, each line represents a on-shell particle propagating with a certain velocity determined by its on-shell momentum, 
and they can scatter, split, and merge at each vertex, subject to momentum conservation.

The above conclusions apply to both massive and massless theories. But for the concern of determining leading-power contributions, we are interested in the massless limit 
of the massive theory. Then the task of determining the reduced diagrams is very easy. All the internal lines in the reduced diagrams carry on-shell lightlike momenta and 
propagate in certain directions at the speed of light, or they have zero momenta and can be attached anywhere.

\subsection{Example: Sudakov form factor}
\label{ssec:sudakov}

\begin{figure}[htbp]
	\centering
	\begin{tabular}{ccccc}
	\includegraphics[scale=0.6]{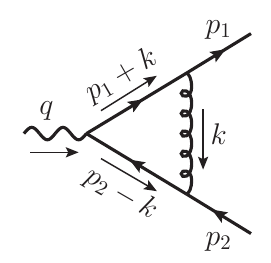} & 
	\includegraphics[scale=0.6, trim={0 -1em 0 0}, clip]{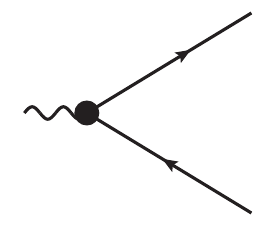} &
	\includegraphics[scale=0.6, trim={0 -1em 0 0}, clip]{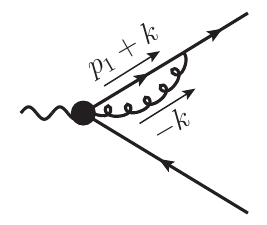} &
	\includegraphics[scale=0.6, trim={0 -1em 0 0}, clip]{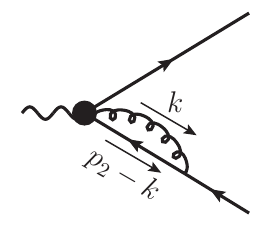} &
	\includegraphics[scale=0.6, trim={0 -1em 0 0}, clip]{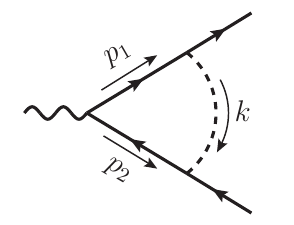} \\
	(a) & (b) & (c) & (d) & (e) 
	\end{tabular}
	\caption{The one-loop diagram of Sudakov form factor in QCD (a) and its reduced diagrams (b)-(e).
	}
	\label{fig:sudakov-1-loop}
\end{figure}

As a simple example, in \fig{fig:sudakov-1-loop}(a) is shown the one-loop diagram for Sudakov form factor, where a virtual photon with momentum $q = (Q, 0, 0, 0)$ decays into
a quark-antiquark ($q\bar{q}$) pair, which go to opposite directions along the $z$ axis, with momenta
\beq
	p_1 = \frac{Q}{2} (1, 0, 0, 1), 
	\quad
	p_2 = \frac{Q}{2} (1, 0, 0, -1).
\eeq
To obtain the reduced diagram, we contract internal lines and identify the resultant diagram as a physically realizable process. 
For the simplest example, we contract all the three propagators into the reduced vertex, which is called the hard vertex, and obtain the reduced diagram in \fig{fig:sudakov-1-loop}(b).
Since it reduces to a tree-level diagram, the meaning of physical process is evident. Now since there is no internal line, this diagram does not contain a pinch. Strictly speaking,
this is not an example of pinch singularity, nor a solution to Landau equation because it requires all $\lambda_j = 0$ in \eq{eq:Landau eq2}. But we will see that this diagram still
gives an important contribution to the integral.

For a less trivial example, we only contract the propagator $p_2 - k$ and obtain the reduced diagram in \fig{fig:sudakov-1-loop}(c). For it to correspond to a physical process, 
we need both $p_1 + k$ and $-k$ to be lightlike and propagating along the same direction, 
\beq[eq:col-q-pinch]
	\lambda_q (p_1 + k) = \lambda_g (-k),
\eeq
with $\lambda_{q,g} > 0$,
which is just \eq{eq:Landau eq2} with $\lambda_{\bar{q}} = 0$ for the coefficient of the antiquark propagator. 
In this case, the gluon is propagating in a direction collinear to the quark line.
Similarly, contracting the propagator $p_1 + k$ gives the reduced diagram in \fig{fig:sudakov-1-loop}(d), in which the gluon is propagating collinearly to the antiquark line.

If we do not contract any propagators, the diagram cannot correspond to a physical process unless the gluon has a zero (or {\it soft}) momentum, $k = 0$. 
A zero-momentum particle does not exist as a real particle, so has no meaning in the sense of ``propagating with a certain velocity". 
To embed it into the picture depicted by Coleman and Norton, 
we interpret the soft particle as having an infinite wavelength, which is not a local particle and can instantaneously connect any two vertices in the reduced diagram.
In terms of Landau condition, a soft propagator gives $\partial D_s(k) / \partial k^{\mu} = 0$ at the point $k = 0$, which automatically satisfies \eq{eq:Landau eq2} if
$ \lambda_s = 1$ and all the other $\lambda_j$'s are zero. This pinch singularity is given by the reduced diagram in \fig{fig:sudakov-1-loop}(e) in which the soft gluon is
represented by the dashed line. The soft pinch singularity is the endpoint of the collinear pinch singularity by taking $\lambda_q = 0, \lambda_g = 1$ and $k = 0$
in \eq{eq:col-q-pinch}.

\begin{figure}[htbp]
	\centering
	\begin{tabular}{cc}
	\includegraphics[scale=0.75, trim={0 0 -2em 0}, clip]{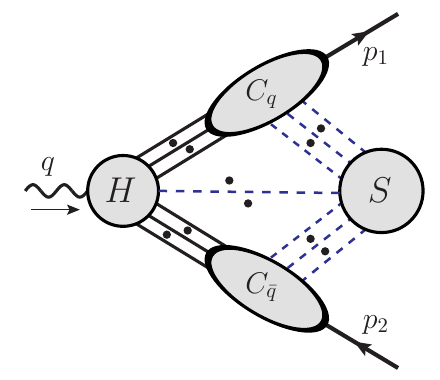} \quad&
	\includegraphics[scale=0.75, trim={0 0 -2em 0}, clip]{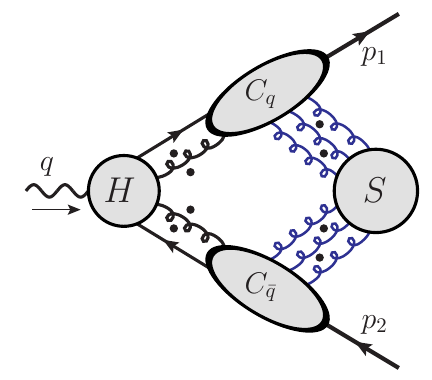}\\
	(a) & (b)
	\end{tabular}
	\caption{(a) General reduced diagram for Sudakov form factor. $H$ is the hard subgraph that contains arbitrarily many propagators which are not pinched and 
	whose virtualities are of order $Q^2$ after proper contour deformations. $C_q$ and $C_{\bar{q}}$ are collinear subgraphs, which are connected to $H$ by
	arbitrarily many collinear propagators. $S$ is the soft subgraph, which is connected to $C_q$, $C_{\bar{q}}$, and $H$ by arbitrarily many soft propagators.
	$S$ is not necessarily connected. The dots refer to any arbitrary collinear or soft lines.
	(b) The leading region for Sudakov form factor. The dots refer to any arbitrary collinear or soft longitudinally polarized gluon lines that can be added.
	}
	\label{fig:sudakov-reduced-diagram}
\end{figure}

Such procedure can be easily generalized to an arbitrary diagram. In any reduced diagram, the collinear lines coming out of the hard vertex move away from each other 
at the speed of light. They can only split and combine in their moving directions, and lines of different collinear directions never meet again. 
Therefore, the collinear sectors are defined by the external particles. Each external lightlike particle with a momentum of order $Q$ defines a collinear direction, 
along which there can be arbitrarily many collinear lines, as shown in \fig{fig:sudakov-reduced-diagram}, 
where there are two collinear sectors $C_q$ and $C_{\bar{q}}$, associated with the quark and antiquark, respectively.
The hard vertex $H$ contains arbitrarily many propagators whose virtualities are of order $Q^2$.
On top of these, there can be arbitrarily many soft lines connecting onto $C_q$, $C_{\bar{q}}$, and $H$, as indicated by the blue dashed lines.
They are collected by the soft subdiagram $S$ (which is not necessarily connected).

\section{Power counting of pinch singularity}
\label{sec:power-counting}

\subsection{Pinch surface: intrinsic and normal coordinates}
\label{ssec:pinch-surface}
For the example of one-loop Sudakov form factor discussed at the end of Sec.~\ref{ssec:landau}, 
the soft pinch singularity is a point $k^{\mu} = 0$ in the 4-dimensional Minkowski space of the loop momentum $k$, 
while the collinear pinch singularities are two straight lines, 
\beq[eq:q-col pinch]
	k = (\alpha \, p_1^+, 0^-, \bm{0}_T) \mbox{ with } \alpha \in (0, 1)
\eeq 
for the quark-collinear region and 
\beq
	k = (0^+, \beta \, p_2^-, \bm{0}_T) \mbox{ with } \beta \in (0, 1)
\eeq 
for the antiquark-collinear region.
In a general multi-loop diagram, the solutions to Landau equation form a multidimensional {\it pinch surface} in the (real-valued) loop momentum space.
Illustrated in \fig{fig:sudakov-reduced-diagram}(a) is a generic pinch surface for the Sudakov form factor. This pinch surface is characterized by a set of soft momenta, 
$k_s^i = 0$, a set of collinear momenta along the quark direction, 
$k_q^i = (\alpha_i \, p_1^+, 0^-, \bm{0}_T)$, and along the antiquark direction, 
$k_{\bar{q}}^j = (0^+, \beta_j \, p_2^-, \bm{0}_T)$, 
and a set of hard momenta in the hard subgraph $H$ whose virtualities are of order $Q^2$.

As discussed in Sec.~\ref{sec:libby-sterman}, the Libby-Sterman analysis relates the power counting of a diagram $I(Q, M)$ in the ratio $M/Q$ of the hard scale to soft scale
to the mass singularity of the same diagram at the limit of $M \to 0$. The mass singularity is in turn given by the pinch singularity of the massless theory, as determined by
the Landau criterion. 
In the case with finite masses, the pinch singularity of the massless theory becomes inexact and regulated by the masses, 
as visualized by the simple examples in Sec.~\ref{ssec:pinch-ex}. 
Even though the loop momentum contour is no longer pinched by poles to give zero-virtuality lines, it is indeed trapped by pairs of close poles 
separated by distances much smaller than $Q$, which forbids it from being arbitrarily deformed. The maximum virtualities of the pinched propagators in this region are still 
much smaller than $Q^2$, which lead to a large integrand. 
As a result, the region around the pinch surface is likely to give an important contribution to the integral, in the sense of having 
a leading power counting behavior in $M/Q$ in \eq{eq:power expansion}.
However, this is not necessarily true, because the numerator and integration measure of the Feynman integral may rescue the singular behavior and reduce the power counting.
Therefore, locating the pinch surfaces is only necessary but not sufficient to determine the mass singularities in the massless integral, or the leading power counting
contributions in the massive integral. We need to further formulate a power counting rule for the divergence degree around the pinch surface.

Around the pinch surface, we define a set of intrinsic coordinates to describe the points on that surface, and a set of normal coordinates to characterize the deviations from the
surface. As an example, for the quark-collinear pinch surface in \eq{eq:q-col pinch}, we can use $\alpha$ as the intrinsic coordinate, and $(k^-, \bm{k}_T)$ as the (three-dimensional) 
normal coordinates. In contrast, for the soft pinch ``surface" in \fig{fig:sudakov-1-loop}(e), there is no intrinsic degree of freedom; any nonzero component $k^{\mu}$ is a normal coordinate.
The integration of normal coordinates leads to the singularity. Since it is the virtualities of pinched propagators that cause a large integrand, we further redefine
the normal coordinates as a radial coordinate $\lambda$ and angular coordinates, such that for a fixed $\lambda$, the virtualities stays constant. 

Around the pinch surface in \eq{eq:q-col pinch}, $k^-$ and $k_T$ are much smaller than $k^+$, which is of order $Q$, 
the pinched quark and gluon propagators have virtualities
\begin{align}\label{eq:col-q-g propagators}
	(p_1 + k)^2 &= 2(1 + \alpha) p_1^+ k^- - k_T^2 , \\
	k^2 & = 2 \alpha p_1^+ k^- - k_T^2,
\end{align}
which are linear with $k^-$ but quadratic with $k_T$. So we choose the radial coordinate $\lambda$ such that
\beq[eq:col-normal-coor]
	k^- = \lambda^2 \bar{k}^- / p_1^+, 
	\quad
	\bm{k}_T = \lambda \bar{\bm{k}}_T,
\eeq
where $\lambda$ has the mass dimension, and $\bar{k}^-$ and $\bar{\bm{k}}_T$ are (two-dimensional) dimensionless angular variables subject to the condition
$| \bar{k}^- | + |\bar{\bm{k}}_T|^2 = 1$ if we choose the radial coordinate $\lambda$ as
\beq[eq:col-lambda]
	\lambda = \sqrt{|p_1^+ k^-| + |k_T|^2}.
\eeq
Note that Eqs.~\eqref{eq:col-lambda} and \eqref{eq:col-normal-coor} are 
just a change of variables from a flat coordinate system into an angular coordinate system. 
For a fixed nonzero $\alpha$ and $\lambda$, the integration of the angular variables 
$\bar{k}^-$ and $\bar{\bm{k}}_T$ do not touch any singularities, so gives a regular result.
The two propagators in \eq{eq:col-q-g propagators} together with the integration 
$d^4k \sim \lambda^3 d\lambda$ (modulo the $\alpha$ and angular integrations)
gives an integral 
\beq[eq:col-lam-int]
	\int_0^{\lesssim Q} \frac{\lambda^3 d\lambda}{(\lambda^2)^2} = \int_0^{\lesssim Q} \frac{d\lambda}{\lambda}.
\eeq
The collinear pinch means that the singularity at $\lambda = 0$ cannot be avoided. 
\eq{eq:col-lam-int} then has a divergence degree $p = 0$ for a logarithmic divergence. 

This analysis is for a massless theory, for the massive case with both the quark having a mass $m$, the pinch becomes inexact, and we can avoid the pole by 
order of $m$, so that $\lambda$ never reaches $0$. This smoothly cuts off the integral in \eq{eq:col-lam-int} at $\lambda \to 0$, and gives
\beq[eq:col-lam-int-m]
	\int_{\gtrsim m}^{\lesssim Q} \frac{d\lambda}{\lambda} \sim \ln\pp{ \frac{Q}{m} },
\eeq
which manifests the logarithmic divergence. 
This massive discussion can be used to analyze the form factor in QED with a massive electron, but obviously not to the real QCD. 
In QCD, partons never appear as external on-shell lines, which must be replaced by hadrons. The quarks are indeed massive, but the light quark masses are 
much less than $\LQCD$, so the mass regulation to the pinch singularity does not come into play before the nonperturbative dynamics kicks in. 
Hence we should regard the mass scale in \eq{eq:col-lam-int} as of the same order as $\LQCD$, i.e., the scale $\lambda \lesssim \LQCD$ 
should be controlled by nonperturbative QCD.

Similarly, around the soft pinch in \fig{fig:sudakov-1-loop}(e), we have $|k^{\mu}| \ll Q$. The gluon propagator $k^2 = 2k^+k^- - k_T^2$ is quadratic with 
both $(k^+, k^-)$ and $k_T$. So we choose the radial coordinate $\lambda_S$ as
\beq[eq:soft-lambda]
	k^{\mu} = \lambda_S \bar{k}^{\mu}, 
	\quad
	\mbox{with} 
	\quad
	\lambda_S = \sum_{\mu} |k^{\mu}|,
\eeq
where the $\bar{k}^{\mu}$ is a (three-dimensional) dimensionless angular coordinates subject to the constraint $\sum_{\mu} |\bar{k}^{\mu}| = 1$.
Then the three pinched propagators have the scaling
\bse\label{eq:soft-q-g propagators}\begin{align}
	k^2 &= \lambda_S^2 (2\bar{k}^+ \bar{k}^- - \bar{k}_T^2) 
		&&= \order{\lambda_S^2}, \label{eq:soft-g propagator}\\
	(p_1 + k)^2 &= 2p_1^+ k^- + k^2 = 2\lambda_S \, p_1^+ \bar{k}^- + \lambda_S^2 (2\bar{k}^+ \bar{k}^- - \bar{k}_T^2) 
		&&= \order{\lambda_S Q}, \label{eq:soft-q propagator}\\
	(p_2 - k)^2 &= -2p_2^- k^+ + k^2 = -2\lambda_S \, p_2^- \bar{k}^+ + \lambda_S^2 (2\bar{k}^+ \bar{k}^- - \bar{k}_T^2) 
		&&= \order{\lambda_S Q}. \label{eq:soft-qb propagator}
\end{align}\ese
We note that now the collinear lines have higher virtualities than the soft line, being much larger than $\lambda_S^2$ but still much smaller than $Q^2$. 
This is because it is the large momentum components of the collinear lines, $p_1^+$ or $p_2^-$, that interact with the soft gluon. 
Together with the integration measure $d^4k \sim \lambda_S^3 d\lambda_S$, \eq{eq:soft-q-g propagators} gives the scaling for the soft region,
\beq[eq:soft-lam-int]
	\int_0^{\lesssim Q} \frac{\lambda_S^3 d\lambda_S}{\lambda_S^2 \, (\lambda_S Q)^2}
	\propto \int_0^{\lesssim Q} \frac{d\lambda_S}{\lambda_S},
\eeq
which is logarithmically divergent, for a divergence degree $p = 0$, similarly to \eq{eq:col-lam-int}.

Now in the massive theory, suppose both the quark and gluon carry masses, $m_q$ and $m_g$, respectively. 
If the quark and antiquark are on shell, $p_1^2 = p_2^2 = m_q^2$, the same scalings in Eqs.~\eqref{eq:soft-q propagator}\eqref{eq:soft-qb propagator} hold, 
but the gluon propagator becomes 
$\order{\lambda_S^2} + m_g^2$, which smoothly cuts off the $\lambda_S \to 0$ region and brings \eq{eq:soft-lam-int} to a form like \eq{eq:col-lam-int-m} with
$m$ replaced by $m_g$. 

On the other hand, if the gluon is massless but the quark and antiquark are off shell by $\order{\LQCD^2}$, 
their virtualities would become $\order{\lambda_S^2}$ and $\order{\lambda_S Q} + \order{\LQCD^2}$ for the gluon and quark/antiquark, respectively. 
This situation resembles the real QCD more since the partons are never on shell. But this brings a more intricate power counting. 
Compared to the hard region in \fig{fig:sudakov-1-loop}(b), which has the same 
power counting as the leading-order (LO) diagram, the soft region has a power counting
\beq
	I_S = Q^2 \int_0^{\lesssim Q} \frac{\lambda_S^3 d\lambda_S}{\lambda_S^2 \, (\lambda_S Q + \LQCD^2)^2}
	= \int_0^{\lesssim Q} \frac{\lambda_S d\lambda_S}{(\lambda_S+ \LQCD^2 / Q)^2}.
\eeq
Now we examine three subregions in the soft region, 
\begin{itemize}
\item
	$\lambda_S \ll \LQCD^2 / Q$, where $I_S \ll \order{1}$ is power suppressed;
\item 
	$\lambda_S \sim \order{\LQCD^2 / Q}$, which gives $I_S \sim \order{1}$;
\item 
	$\order{\LQCD^2 / Q} \ll \lambda_S \lesssim \order{\LQCD}$, which gives $I_S \sim \order{1}$,
\end{itemize}
where we stop at $\lambda_S \lesssim \order{\LQCD}$ beyond which all the three propagators have virtualities much greater than $\LQCD^2$, 
and start entering the hard region.
We found that the whole region $\LQCD^2 / Q  \lesssim \lambda_S \lesssim \LQCD$ gives a leading-power contribution.
In the low end with $\lambda_S \sim \order{\LQCD^2 / Q}$, the quark propagators have virtualities of order $\LQCD^2$, but the gluon has $\LQCD^4 / Q^2 \ll \LQCD^2$.
In the high end with $\lambda_S \sim \order{\LQCD}$, the gluon propagator has a virtuality of order $\LQCD^2$, but the quarks have $Q\, \LQCD \gg \LQCD^2$.
Given the fourth assumption in Sec.~\ref{ssec:fac-assumption}, the whole soft region $\lambda_S \lesssim \LQCD$ is in the nonperturbative regime. 
But for the perturbative analysis, we usually make the second assumption in \sec{ssec:fac-assumption} to
convert the sum over final hadron states into a sum over parton states, so on-shell partons do appear in the final states, 
for which the region $\lambda_S \ll \LQCD^2 / Q$ also becomes important and contribute to soft divergences. 
In such cases, we need to show that the whole soft region is cancelled.
Therefore, since factorization is rooted in a perturbative analysis, we need to consider the whole soft region $\LQCD^2 / Q  \lesssim \lambda_S \lesssim \LQCD$.

Such complication arises because in the soft region, the collinear lines and soft lines have different virtualities, which causes an extra scale $\LQCD^2 / Q$.
In contrast, the power counting analysis of the collinear region is much simpler because all collinear lines have virtualities 
$\order{\lambda^2} + \order{\LQCD^2}$, and hence only the region $\lambda \sim \LQCD$ needs to be considered.
In a more complicated diagram, we can have (multiple) soft and collinear momenta at the same time. 
Each collinear momentum scales as
\beq[eq:col-scaling]
	k_c = (k_c^+, k_c^-, \bm{k}_{cT}) \sim (Q, \frac{\lambda^2}{Q}, \lambda),
	\quad
	\mbox{ with } 
	\quad
	\lambda \sim \order{\LQCD},
\eeq
and each soft momentum has the scaling
\beq[eq:soft-scaling]
	k_s = (k_s^+, k_s^-, \bm{k}_{sT}) \sim (\lambda_S, \lambda_S, \lambda_S),
	\quad
	\mbox{ with } 
	\quad
	\frac{\lambda^2}{Q} \lesssim \lambda_S \lesssim \lambda.
\eeq
Eqs.~\eqref{eq:col-scaling} and~\eqref{eq:soft-scaling} constitute the {\it canonical scaling} for the pinched regions, with 
$\lambda$ and $\lambda_S$ being the radial normal coordinates
parametrizing the distance from the pinch surfaces.

\subsection{Power counting}
\label{ssec:power counting}
Now we derive the power counting around the pinch surface, based on the canonical scaling in 
Eqs.~\eqref{eq:col-scaling} and \eqref{eq:soft-scaling}, with $\lambda_S = \order{\lambda^2 / Q}$.
We will take a simpler approach than the treatment in \citep{Collins:2011zzd} 
by examining the power counting with respect to the leading-order diagrams.
The derivation is for a generic quantum field theory (QFT), and we work in the Feynman gauge for a gauge theory involving a vector boson.

Each pinched momentum belongs either to a collinear sector or the soft sector, so a general pinch surface represented by a reduced diagram 
is decomposed into a hard subgraph $H$, a set of collinear subgraphs $C_i$, and a soft subgraph $S$. 
Normally we work in the CM frame of the hard subgraph $H$, a momentum $k_H$ in which has all its components of order $Q$.
Each collinear subgraph $C_i$ is defined by one (or more collinear) external hard particle $p_i$
and is connected to the hard subgraph $H$ via a set of collinear lines $\{ k_{iH} \}$. 
We include all the propagators of $\{ k_{iH} \}$ and their integrations $\prod_{k_{iH}} \int d^4 k_{iH}$ in $C_i$.
Within each $C_i$, the collinear lines can interact with each other in all arbitrary ways under the constraints of fixed $\{ k_{iH} \}$ and $p_i$.
The soft subgraph $S$, which may contain one or more connected parts, is connected to each collinear subgraph $C_i$ and/or hard subgraph $H$ 
by soft lines, $\{ k_{iS} \}$ and/or $\{ k_{HS} \}$, respectively. 
Similarly, all the propagators and integrations of $\{ k_{iS} \}$ and/or $\{ k_{HS} \}$ are included in $S$.
A concrete example is given by the Sudakov form factor in \fig{fig:sudakov-reduced-diagram}(a), but the discussion in this section applies more generally.

\subsubsection{Leading order and hard region} 
First, for a given process, the leading-order diagram can be easily worked out, 
whose power counting in the scaling limit $Q \to \infty$ is determined by its dimension.
For example, the Sudakov form factor $\Gamma^{\mu}$ has dimension one, so it simply scales as $Q^1$ at leading order. 
The purely hard region, as illustrated in \fig{fig:sudakov-1-loop}(b) where 
all internal propagator denominators are of order $Q^2$, has the same structure as the leading
order and gives the same power counting. 
This is the feature of a renormalizable quantum field theory, as is the case of QCD, for which the coupling is dimensionless; 
otherwise, we would have a suppression from a power of $g / Q^{\dim(g)}$. 

\subsubsection{Collinear subgraph} 
For ease of notation, now we look at a particular collinear subgraph and denote it as $C$,
In the simplest case, $C$ only comprises a single line of the external particle, which does not cause additional power counting analysis beyond the previous discussion. 
This situation can be trivially generalized to the case where $C$ is connected to $H$ by a single propagator but with an arbitrary two-point function included in $C$.
If an extra line of the field $\Phi(x)$ connects $C$ to $H$, their convolution will be modified to 
\begin{align}\label{eq:dim CH}
	&\int \frac{d^4 k_{CH}^{\Phi}}{(2\pi)^4} H^{\alpha}(k_{CH}^{\Phi}) \, C^{\alpha}(k_{CH}^{\Phi}) \nn\\
	&\hspace{2em}
	= \int \frac{d^4 k_{CH}^{\Phi}}{(2\pi)^4} H^{\alpha}(k_{CH}^{\Phi}) 
		\bb{ \int d^4 x \, e^{-i k_{CH}^{\Phi} \cdot x} \langle C | T\{ \cdots \Phi^{\alpha}(x) \cdots \} |0 \rangle } \,,
\end{align}
where we have only explicitly indicated the extra dependence on the new particle $\Phi$, $k_{CH}^{\Phi}$ is its momentum, 
and $\alpha$ describes its spin quantum number.
The extra dimensions of $C$ and $H$ due to the appearance of $\Phi$ are
\beq[eq:dim C-H]
	\Delta C^{\alpha} = -4 + \dim(\Phi), 
	\quad
	\Delta H^{\alpha} = -\dim(\Phi),
\eeq
where $\dim(\Phi)$ is the dimension of the field $\Phi(x)$, which is $1$ for a scalar or vector field, and $3/2$ for a fermion field.
The dependence of $C^{\alpha}$ on $\alpha$ can be easily worked out using a boost analysis. 
If we choose the direction of $C$ as the $z$ axis, then each collinear momentum in $C$ scales as in \eq{eq:col-scaling}. 
Now we consider boosting $C$ back to its rest frame, which causes the $C$-collinear momenta to scale as $(\lambda, \lambda, \lambda)$.
Hence each component of $\alpha$ should scale in the same way, and the power counting $C^{\alpha}$ is simply given by its dimension, 
\beq
	C^{\alpha} \simeq \lambda^{\dim(C)} 
	\mbox{ for each } \alpha \mbox{ in $C$ rest frame.}
\eeq
Then we boost $C^{\alpha}$ back to the lab frame, where it is highly boosted along the $z$ direction. 
This gives an enhancement $(Q/\lambda)^s$ to the largest component of $C^{\alpha}$, with $s$ being the spin of $\Phi$.
In contrast, each component of $H^{\alpha}$ scales the same, being $Q^{\dim(H)}$. 
Including the power counting $d^4 k_{CH}^{\Phi} \sim \lambda^4$, we then have the extra power counting due to $\Phi$:
\begin{itemize}
\item
$\Phi = \phi$ (scalar): $\dim(\phi) = 1$ and $s = 0$, leading to $Q^{-1} \cdot \lambda^1 = \lambda/Q$; (Note that this case also applies to ghost fields.)
\item 
$\Phi^{\alpha} = \psi^{\alpha}$ (fermion): $\dim(\psi) = 3/2$ and $s = 1/2$, leading to $Q^{-3/2} \cdot \lambda^{3/2} \cdot \sqrt{Q / \lambda} = \lambda/Q$;
\item 
$\Phi^{\alpha} = A^{\alpha}$ (vector boson): $\dim(A) = 1$ and $s = 1$, leading to $Q^{-1} \cdot \lambda^1 \cdot (Q / \lambda) = 1$. 
This only applies to the unphysical $A^+$ component which is proportional to its momentum. 
The physical transverse polarization $A^{\perp}$ receives no
enhancement from the boost, so gives a power counting $\lambda/Q$, the same as the scalar case. 
The remaining component $A^-$ undergoes a suppression $\lambda/Q$ by the boost, so gives the power counting $(\lambda/Q)^2$.
\end{itemize}
Therefore, attaching a collinear subgraph to the hard subgraph by a scalar, fermion, or transversely polarized vector boson brings a power suppression by $\lambda/Q$,
while by a longitudinally polarized vector boson brings no power suppression.

\subsubsection{Soft subgraph connection to a collinear subgraph} 
Now we consider the power counting due to soft lines. 
Adding an extra line of the field $\Phi^{\alpha}(x)$ between the soft subgraph $S$ and some collinear subgraph $C$ (taken to be along the $z$ direction) changes their convolution to
\beq[eq:dim CS]
	\int \frac{d^4 k_{CS}^{\Phi}}{(2\pi)^4} C^{\alpha}(k_{CS}^{\Phi}) \, S^{\alpha}(k_{CS}^{\Phi})
	= \int \frac{d^4 k_{CS}^{\Phi}}{(2\pi)^4} C^{\alpha}(k_{CS}^{\Phi}) 
		\bb{ \int d^4 x \, e^{-i k_{CS}^{\Phi} \cdot x} \langle 0 | T\{ \cdots \Phi^{\alpha}(x) \cdots \} |0 \rangle } \,,
\eeq
where $k_{CS}^{\Phi}$ is the soft momentum that scales as $(\lambda^2 / Q, \lambda^2 / Q, \lambda^2 / Q)$, and $\alpha$ is the spin index. 
This new attachment changes the dimensions of $S^{\alpha}$ and $C^{\alpha}$ by
\beq[eq:dim C-S]
	\Delta S^{\alpha} = -4 + \dim(\Phi), 
	\quad
	\Delta C^{\alpha} = -\dim(\Phi).
\eeq
This is similar to \eq{eq:dim C-H} with $H$ replaced by $C$ and $C$ by $S$, but now the 
collinear subgraph $C^{\alpha}$ has different scalings for different $\alpha$ components. So we count \eq{eq:dim CS} as
\begin{align}\label{eq:counting CS}
	\Delta\bb{ \int \frac{d^4 k_{CS}^{\Phi}}{(2\pi)^4} C^{\alpha}(k_{CS}^{\Phi}) \, S^{\alpha}(k_{CS}^{\Phi}) }
	&\sim
	\pp{\frac{\lambda^2}{Q}}^{\dim(\Phi)} \times \bb{ \lambda^{-\dim(\Phi)} \cdot \mbox{(spin enhancement)} } \nn\\
	&\sim
	\pp{\frac{\lambda}{Q}}^{\dim(\Phi)} \times \mbox{(spin enhancement)},
\end{align}
where in the first step, the first factor is from the soft subgraph and the integration measure, and the second factor is from the collinear subgraph.
By the same boost argument as in the previous situation, $C^{\alpha}$ receives a power enhancement for fermions and vector bosons.
\begin{itemize}
\item
$\Phi = \phi$ (scalar) or $A^{\perp}$ (transversely polarized vector boson), $\dim(\Phi) = 1$ without boost enhancement, leading to
a $\lambda / Q$ suppression; (Note that this case also applies to ghost fields.)
\item 
$\Phi = \psi$ (fermion), $\dim(\Phi) = 3/2$ with a boost enhancement $\sqrt{Q / \lambda}$, leading to
a $\lambda / Q$ suppression;
\item 
$\Phi = A^+$ (longitudinally polarized vector boson), $\dim(\Phi) = 1$ with a boost enhancement $Q / \lambda$, leading to a power counting of $1$.
\end{itemize}
Therefore, attaching the soft subgraph to a collinear subgraph by a scalar, fermion, or transversely polarized vector boson brings a power suppression by $\lambda/Q$,
while by a longitudinally polarized vector boson brings no power suppression.

\subsubsection{Soft subgraph connection to the hard subgraph} 
Adding an extra soft line to the hard subgraph $H$ works in a similar way and leads to the power counting formula
\beq[eq:counting HS]
	\Delta(S\otimes H) \sim \pp{\frac{\lambda^2}{Q}}^{\dim(\Phi)} \cdot Q^{-\dim(\Phi)} = \pp{\frac{\lambda}{Q}}^{2\dim(\Phi)}.
\eeq
Compared to \eq{eq:counting CS}, the power factor $Q^{-\dim(\Phi)}$ instead of $\lambda^{-\dim(\Phi)}$ generally suppresses the soft connections to $H$, 
and there is power enhancement from the boost.
Therefore, attaching the soft subgraph to the hard subgraph by a scalar or vector boson brings a power suppression by $(\lambda/Q)^2$,
while by a fermion brings a power suppression $(\lambda/Q)^3$.

To conclude, we list in Table~\ref{tab:power-counting} the power counting rules for adding an extra line in a certain reduced diagram. 
The rules work in a fashion of construction so give the power counting relative to a certain diagram, e.g., the leading-order diagram. 
As an example, for the Sudakov form factor in \fig{fig:sudakov-reduced-diagram}(a), the external $q\bar{q}$ lines dictate the 
two collinear subgraphs to be connected to $H$ by at least a fermion line separately. 
From our power counting rules, having additional line connections generally brings power suppressions except for 
longitudinally polarized gluons connecting $C_{q, \bar{q}}$ to $H$ or $S$. 
This leads to the reduced diagram in \fig{fig:sudakov-reduced-diagram}(b) that has the same power counting as the leading-order diagram or the pure hard region. 
Any other pinch surfaces give more suppressed power counting. So it is called the {\it leading region}.
\begin{table}
    \centering
	\caption{The counting of the $(\lambda/Q)$ power associated with each extra line attachment between a collinear subgraph $C$ and the hard subgraph $H$,
		the soft subgraph $S$ and $C$, or $S$ and $H$. 
		In the second and third columns, we take $S$ to refer to the soft region with the momentum scaling $(\lambda^2 / Q, \lambda^2 / Q, \lambda^2 / Q)$,
		while in the last two columns, $S^\prime$ refers to the soft region with the scaling $(\lambda, \lambda, \lambda)$. 
		For $S^\prime$ we denote $\Delta n_{cs} = n_{cs} - 1$ as the number of collinear propagators that the soft momentum flows through
		with respect to the minimal configuration ($n_{cs} = 1$).}
	\label{tab:power-counting}
	\begin{tabular}{>{\centering}m{1.2cm} >{\centering}m{1cm}  >{\centering}m{1cm} >{\centering}m{1cm} >{\centering}m{2.5cm} >{\centering\arraybackslash}m{1cm}}
	\hline
			  			& $C$-$H$  	&  $S$-$C$  	& $S$-$H$ 	& $S^{\prime}$-$C$	& $S^{\prime}$-$H$\\ 
	\hline
	$\phi, c, \bar{c}$ 			&		1			&		1			&		2			&			1	+ $\Delta n_{cs}$				&				1	\\ 
	\hline
	$\psi$			&		1			&		1			&		3			&			1/2 + $\Delta n_{cs}$				&				3/2	\\ 
	\hline
	$A^+$			&		0			&		0			&		2			&			0 + $\Delta n_{cs}$				&				1	\\ 
	\hline
	$A^{\perp}$ 	&		1			&		1			&		2			&			1 + $\Delta n_{cs}$				&				1	\\ 
	\hline
	$A^-$			&		2			&		2			&		2			&			1 + $\Delta n_{cs}	$				&				1	\\
	\hline
	\end{tabular}
\end{table}

\subsubsection{Power counting for alternative soft scalings}
\label{sssec:power counting-soft}
The previous subsection assumes the soft gluon momenta scale by a uniform scaling $\lambda_S \sim \lambda^2 / Q$.
Such soft momenta do not change the collinear propagator virtualities when flowing through the latter.
In general, we can have $\lambda_S$ to vary between $\lambda^2 / Q$ and $\lambda$, for $\LQCD \lesssim \lambda \ll Q$.
This generic soft scaling does not affect the power counting for the collinear-to-hard coupling, but alters that for the soft attachments
to the collinear and hard subgraphs.

When a soft momentum $k_s \sim (\lambda_S, \lambda_S, \lambda_S)$ flows along a collinear momentum $k_c \sim (Q, \lambda^2 / Q, \lambda)$,
it changes the virtuality to
\beq
	(k_c + k_s)^2 = k_c^2 + k_s^2 + 2k_c\cdot k_s \sim \lambda^2 + \lambda_S^2 + \lambda_S Q 
	\sim \max(\lambda^2, \lambda_S Q) = \lambda^2 \cdot \max(1, \lambda_S Q / \lambda^2).
\eeq
Thus if this soft momentum flows through $n_{cs}$ collinear propagators, the collinear subgraph gains an extra factor
$\bb{ 1 / \max(1, \lambda_S Q / \lambda^2) }^{n_{cs}}$ apart from the dimensional counting $\lambda^{-\dim(\Phi)}$ times a boost enhancement factor
in \eq{eq:counting CS}. Since we take $\lambda_S \gtrsim \lambda^2 / Q$, the term $\lambda_S Q / \lambda^2$ is at least of order 1, so we can simplify
$\max(1, \lambda_S Q / \lambda^2)$ to $\lambda_S Q / \lambda^2$.
The power counting of the soft subgraph should also be modified by $\lambda_S$. 
Therefore, an extra soft attachment between $S$ and a collinear subgraph leads to an additional power counting
\beq[eq:counting CS modify]
	\Delta(C\otimes S) \sim \pp{ \frac{\lambda_S}{\lambda} }^{\dim(\Phi)} 
	\cdot \pp{ \frac{Q}{\lambda} }^s \cdot 
	\pp{ \frac{\lambda^2}{ \lambda_S Q} }^{n_{cs}}
	= 
	\pp{ \frac{\lambda}{Q} }^{\dim(\Phi) - s}
	\pp{ \frac{\lambda_S}{\lambda^2 / Q} }^{\dim(\Phi) - n_{cs}}
\eeq
where $s$ is the spin of $\Phi$.
This introduces an extra factor $(\lambda_S Q / \lambda^2)^{\dim(\Phi) - n_{cs}}$ with respect to \eq{eq:counting CS}.
For $\lambda_S = \lambda^2 / Q$, it recovers the same counting. 
As $\lambda_S$ increases, we gain a power enhancement if $\dim(\Phi) > n_{cs}$,
which can only happen for fermion case with $n_{cs} = 1$, but otherwise a suppression.
The minimal configuration $n_{cs} = 1$ yields a power counting 
\beq
	\pp{ \frac{\lambda}{Q} }^{\dim(\Phi) - s}
	\pp{ \frac{\lambda_S}{\lambda^2 / Q} }^{\dim(\Phi) - 1}, 	
	\quad (n_{cs} = 1)
\eeq	
which does not affect the power counting for scalar and vector bosons, but changes the fermion case to
$\sqrt{\lambda_S / Q}$, enhancing the $\lambda / Q$ counting in Table~\ref{tab:power-counting} if $\lambda_S \sim \lambda$.
For $n_{cs} \geq 2$, a large $\lambda_S \gg \lambda^2 / Q$ leads to a suppression for all cases. 
Hence, for the leading region, one usually needs $n_{cs} = 1$ for the large scaling.

For the soft momentum $k_s$ flowing into $H$, we can still neglect it in $H$ since $\lambda_S \ll Q$. Then \eq{eq:counting HS} 
is modified to 
\beq[eq:counting HS-modify]
	\Delta(S\otimes H) \sim \lambda_S^{\dim(\Phi)} \cdot Q^{-\dim(\Phi)} = \pp{\frac{\lambda_S}{Q}}^{\dim(\Phi)},
\eeq
which enhances the counting in \eq{eq:counting HS} if $\lambda_S \gg \lambda^2 / Q$.

To summarize, we include in the last two columns of Table~\ref{tab:power-counting} the power counting for the high end of soft region
$S^\prime$ with $\lambda_S = \order{\lambda}$, and with $n_{cs} = 1$. 
Since the soft attachments still lead to power suppression except for longitudinally polarized gluons, the leading region graph for 
Sudakov form factor takes the same form as \fig{fig:sudakov-reduced-diagram}, now for the whole soft region with 
$\lambda^2 / Q \lesssim \lambda_S \lesssim \lambda$.

\section{Basic approximation for a single region}
\label{sec:approx-region}

It is the regions around the pinch surfaces that give important power contributions to the Feynman integrals. 
Having identified the pinch surfaces that are associated with leading power contributions, we may then make certain approximations 
to extract those contributions based on the power counting of the related momenta in Eqs.~\eqref{eq:col-scaling} and~\eqref{eq:soft-scaling}.

As shown in Table~\ref{tab:power-counting}, the major complication from the gauge theory is that there is no penalty from adding arbitrarily many 
vector boson lines attaching the collinear subgraphs to the hard or soft subgraphs. 
But the polarizations must be proportional to the collinear momenta. This is the key for the factorization of gauge theory as it will allow the use of 
Ward identity.

To present the approximators, it is helpful to confine ourselves to a particular process or amplitude. 
So in the following discussions, we will mainly be using the Sudakov form factor as an example,
but other processes like DIS will also be referred to for completeness.

\subsection{Approximation of collinear-to-hard connections}
\label{ssec:approx-c-h}
In the Sudakov form factor, each collinear subgraph $C_i$ is connected to the hard subgraph $H$ by one quark line and a series of gluon lines. 
They are convoluted by the integrations of the loop momenta, which can be written as
\begin{align}
	C_q \otimes H \otimes C_{\bar{q}} 
	=  &\, \int \frac{d^4k_q}{(2\pi)^4} \frac{d^4k_{\bar{q}}}{(2\pi)^4} \bb{ \prod_{i = 1}^n \frac{d^4k_i}{(2\pi)^4} } \bb{ \prod_{j = 1}^m \frac{d^4l_j}{(2\pi)^4} }
		\nn\\
		&\hspace{2em} \times
		C_{q, \alpha, \mu_1 \cdots \mu_n}(k_q, k_1, \cdots, k_n) \cdot g^{\mu_1 \nu_1} \cdots g^{\mu_n \nu_n} 	\nn\\
		&\hspace{2em} \times
		H_{\alpha, \nu_1 \cdots \nu_n; \beta, \sigma_1 \cdots \sigma_m}(k_q, k_1, \cdots, k_n; l_{\bar{q}}, l_1, \cdots, l_m)	\nn\\
		&\hspace{2em} \times
		g^{\sigma_1 \rho_1} \cdots g^{\sigma_m \rho_m}  \cdot C_{\bar{q}, \beta, \rho_1 \cdots \rho_m}(l_{\bar{q}}, l_1, \cdots, l_m) 
\label{eq:CHC-conv}
\end{align}
for a certain region of some diagram that has $n$ and $m$ collinear gluons connecting $C_q$ and $C_{\bar{q}}$ to $H$, respectively. 
In \eq{eq:CHC-conv}, $\alpha$ ($k_q$) and $\beta$ ($l_{\bar{q}}$) are the spinor indices (loop momenta) of the quark and antiquark respectively, 
and $\{\mu_i\}$ ($\{k_i\}$) and $\{\nu_j\}$ ($\{k_j\}$) are the Lorentz indices (loop momenta) of the $C_q$- and $C_{\bar{q}}$-collinear gluons, respectively.
All the collinear (pinched) propagators have been included in $C_q$ or $C_{\bar{q}}$, and we will eventually include the loop integrations into them as well.

We take the quark and antiquark to move along the $\pm z$ directions, respectively. Around the pinch surface, the collinear momenta scale as 
\beq
	k_i \sim (Q, \lambda^2 / Q, \lambda), \; (i = q, 1, \cdots, n), \quad
	\mbox{and} \quad
	l_j \sim (\lambda^2 / Q, Q, \lambda), \; (j = \bar{q}, 1, \cdots, m).
\eeq
These momenta circulate between the hard subgraph $H$ and collinear subgraphs. 
The propagators inside $H$ are not pinched, and proper deformations can be done
to make them have high virtualities of order $Q^2$. 
Then we may expand each of them with respect to the small parameter $\lambda$ 
without encountering any singularities. The leading-power contribution can be simply
obtained by neglecting $\lambda$ in $H$, so we approximate $H$ by
\beq
	H(\{ k_i \}; \{ l_j \}) \to H(\{ \hat{k}_i \}; \{ \hat{l}_j \}), 
\eeq
with
\beq[eq:sudakov-col-mom]
	\hat{k}_i^{\mu} = (k_i^+, 0^-, \bm{0}_T) = (k_i \cdot n) \, \bar{n}^{\mu},
	\quad
	\hat{l}_j^{\mu} = (0^+, l_j^-, \bm{0}_T) = (l_j \cdot \bar{n}) \, n^{\mu},
\eeq
where $i = q, 1, \cdots, n$, $j = \bar{q}, 1, \cdots, m$, and we have introduced two lightlike auxiliary vectors
\beq[eq:n-nbar]
	n^{\mu} = (0^+, 1^-, \bm{0}_T) = \frac{1}{\sqrt{2}}(1, -\vec{z}),
	\quad
	\bar{n}^{\mu} = (1^+, 0^-, \bm{0}_T) = \frac{1}{\sqrt{2}}(1, \vec{z}).
\eeq

For the spinor contraction between $C_q$ and $H$, the boosted factor $C_q$ has a large component in the spinor space, which can be projected out by~\citep{Collins:2011zzd}
\beq[eq:sudakov-projector-P]
	C_{q, \alpha} (k) 
	\to C_{q, \delta} (k) \, \P_{\delta \alpha}
	= C_{q, \delta} (k) \bb{ \frac{\gamma\cdot n \sum_{s} u_s(\hat{k}) \bar{u}_s(\hat{k}) }{2 \hat{k} \cdot n} }_{\delta \alpha}
	= C_{q, \delta} (k) \bb{ \frac{\gamma^+ \gamma^-}{2} }_{\delta \alpha} ,
\eeq
where $k$ stands for a generic collinear momentum along the $z$ direction and $u_s(\hat{k})$ is the massless spinor with momentum $\hat{k} = (k\cdot n)\bar{n}$ and spin $s$.
The projector $(\P)_{\delta \alpha}$ has the property
\beq
	\P^2 = \P, \; 
	\bar{u}_s(\hat{k}) \P = \bar{u}_s(\hat{k}), \;
	\mbox{and } \P (\gamma \cdot \hat{k}) = 0,
\eeq
such that it projects $C_{q} (k) $ onto a massless spinor along the $z$ direction.
Perturbatively, the fermion propagator numerator $\slash{k} + m$ contracted with $\P$ keeps its large component $k^+\gamma^-$ intact,
so $\P$ keeps the leading-power accuracy.
$\P$ inserting between $C_q$ and $H$ contracts the massless spinor $\bar{u}_s(\hat{k})$ 
with $H$ to give the hard factor a physical interpretation of massless quark interaction.

Similarly, for the spinor contraction between $H$ and $C_{\bar{q}}$, we insert a projector
\beq[eq:sudakov-projector-Pb]
	C_{\bar{q}, \beta} (l) 
	\to \bar{\P}_{\beta\kappa} C_{\bar{q}, \kappa} (l)
	= \bb{ \frac{ \sum_{s} u_s(\hat{l}) \bar{u}_s(\hat{l}) \gamma \cdot \bar{n} }{2 \hat{l} \cdot \bar{n}} }_{\beta\kappa} C_{\bar{q}, \kappa} (l)
	= \bb{ \frac{\gamma^+ \gamma^-}{2} } C_{\bar{q}, \kappa} (l),
\eeq
which is the same as $\P$. 
After these two approximations, \eq{eq:CHC-conv} becomes
\begin{align}
	C_q \otimes H \otimes C_{\bar{q}} 
	\simeq  
		&\, \int \frac{d^4k_q}{(2\pi)^4} \frac{d^4k_{\bar{q}}}{(2\pi)^4} \bb{ \prod_{i = 1}^n \frac{d^4k_i}{(2\pi)^4} } \bb{ \prod_{j = 1}^m \frac{d^4l_j}{(2\pi)^4} }
		\nn\\
		&\hspace{2em} \times
		C_{q, \alpha, \mu_1 \cdots \mu_n}(k_q, k_1, \cdots, k_n) \cdot g^{\mu_1 \nu_1} \cdots g^{\mu_n \nu_n} 	\nn\\
		&\hspace{2em} \times
		\P_{\alpha\delta} \, 
		H_{\delta, \nu_1 \cdots \nu_n; \kappa, \sigma_1 \cdots \sigma_m}
		(\hat{k}_q, \hat{k}_1, \cdots, \hat{k}_n; \hat{l}_{\bar{q}}, \hat{l}_1, \cdots, \hat{l}_m) \,
		\P_{\kappa\beta}	\nn\\
		&\hspace{2em} \times
		g^{\sigma_1 \rho_1} \cdots g^{\sigma_m \rho_m}  \cdot C_{\bar{q}, \beta, \rho_1 \cdots \rho_m}(l_{\bar{q}}, l_1, \cdots, l_m),
\label{eq:CHC-conv1}
\end{align}
where the hard factor $H$ is surrounded by two $\P$'s which amputate and put on-shell the two quark lines connected to $H$. 
This fact will be important when applying Ward identity to the collinear gluons.

Based on our power counting rules, only the longitudinal polarizations of the collinear gluons connecting $C_q$ or $C_{\bar{q}}$ to $H$ are of leading power.
This means that we shall have $\nu_i = +$ and $\sigma_j = -$ in \eq{eq:CHC-conv1}, which extracts the $g^{-+}$ components for all the metric tensors.
So we make the approximations
\begin{align}\label{eq:g-approx}
	g^{\mu_i \nu_i} \, \mapsto \frac{n^{\mu_i} \, \hat{k}_i^{\nu_i} }{ \hat{k}_i \cdot n} 
		= \frac{n^{\mu_i} \, \hat{k}_i^{\nu_i} }{ k_i \cdot n}, 
		\quad
	g^{\sigma_j \rho_j} \, \mapsto \frac{\hat{l}_j^{\sigma_i} \, \bar{n}^{\rho_j} }{ \hat{l}_j \cdot \bar{n}} 
		=  \frac{\hat{l}_j^{\sigma_i} \, \bar{n}^{\rho_j} }{ l_j \cdot \bar{n}}. 
\end{align}
These are equivalent to $g^{\mu\nu} \mapsto n^{\mu} \bar{n}^{\nu}$, but writing as \eq{eq:g-approx} has the advantage that 
the gluon connection to the hard factor $H$ will be replaced by its momentum contracted with $H$,
\begin{align}
	&C_q \otimes H \otimes C_{\bar{q}} 
	\simeq  
		\int \frac{d^4k_q}{(2\pi)^4} \frac{d^4k_{\bar{q}}}{(2\pi)^4} \bb{ \prod_{i = 1}^n \frac{d^4k_i}{(2\pi)^4} } \bb{ \prod_{j = 1}^m \frac{d^4l_j}{(2\pi)^4} }
		\nn\\
		&\hspace{3em} \times
		C_{q, \alpha, \mu_1 \cdots \mu_n}(k_q, k_1, \cdots, k_n) \cdot \bb{ \prod_i  \frac{ n^{\mu_i}}{ k_i \cdot n} } 	\nn\\
		&\hspace{3em} \times
		\pp{ \prod_i \hat{k}_i^{\nu_i} } 
		\bb{
			\P_{\alpha\delta} \, H_{\delta, \nu_1 \cdots \nu_n; \kappa, \sigma_1 \cdots \sigma_m}
			(\hat{k}_q, \hat{k}_1, \cdots, \hat{k}_n; \hat{l}_{\bar{q}}, \hat{l}_1, \cdots, \hat{l}_m) \,
			\P_{\kappa\beta} 
		}
		\pp{ \prod_j \hat{l}_j^{\sigma_j} } 	\nn\\
		&\hspace{3em} \times
		\bb{ \prod_j \frac{\bar{n}^{\rho_j} }{ l_j \cdot \bar{n}} }  \cdot C_{\bar{q}, \beta, \rho_1 \cdots \rho_m}(l_{\bar{q}}, l_1, \cdots, l_m),
\label{eq:CHC-conv1}
\end{align}
which will in turn allow the use of Ward identity.

\subsection{Approximation of soft-to-collinear connections}
\label{ssec:approx-s-c}
From the power counting rules in Table~\ref{tab:power-counting}, soft connections are generally power suppressed except for 
soft gluons that are attached to collinear subgraphs with polarizations proportional to the collinear momenta. 
For the leading region graph in \fig{fig:sudakov-reduced-diagram}(b), a soft momentum $k_s$ scales as $(\lambda_S, \lambda_S, \lambda_S)$
and can be taken to circulate from $S$ to $C_q$, to $H$, to $C_{\bar{q}}$, and then back to $S$. 
When $k_s$ flows through $H$, each of its component is much smaller than $Q$, so we can neglect it in $H$ to the leading-power accuracy.
When it flows through $C_q$ along a collinear line with momentum $k_i$, it modifies the momentum to $k_i + k_s$, which does not change
the leading component $(k_i + k_s)^+ = k_i^+ = \order{Q}$ thus does not change the collinear propagator numerator, 
but it modifies the propagator denominator by
\beq
	(k_i + k_s)^2 - m^2 = (k_i^2 - m^2) + 2k_i \cdot k_s + k_s^2.
\eeq
Since $k_s$ has a uniform scaling $\lambda_S$ for all components, it is the term $2k_i^+ k_s^- = \order{Q \lambda_S}$ that is the most important
among all $k_s$-related terms. Therefore, we may only keep $k_s^-$ when $k_s$ flows through $C_q$, which gives the approximation,
\beq[eq:soft-approx-k-in-A]
	k^{\mu}_{qs} \mapsto \hat{k}^{\mu}_{qs} = (k_{qs} \cdot \bar{n}) n^{\mu},
\eeq
where $k_{qs}$ denotes the soft momentum flowing through $C_q$.
This applies for the whole range of $\lambda_S \in (\lambda^2 / Q, \lambda)$. Even though for $\lambda_S \sim \lambda$, the whole quark propagator 
is dominated by $2k_i^+ k_s^- = \order{Q \lambda}$, we do not modify the term $(k_i^2 - m^2)$ in order for a unified approximation.
Similarly, for a soft momentum $k_{\bar{q}s}$ flowing in $C_{\bar{q}}$, we approximate it by
\beq[eq:soft-approx-k-in-B]
	k_{\bar{q}s}^{\mu} \mapsto \hat{k}_{\bar{q}s}^{\mu} = (k_{\bar{q}s} \cdot n) \bar{n}^{\mu}.
\eeq

Those soft momentum approximation decouples the soft momenta from the hard subgraph and simplifies the soft-collinear couplings to 
\begin{align}\label{eq:soft-c-c-coupling}
	C_q \otimes S \otimes C_{\bar{q}}
	\simeq &\, \int \bb{ \prod_i \frac{d^4 k_{qs, i}}{(2\pi)^4} } \bb{ \prod_j \frac{d^4 k_{\bar{q}s, j}}{(2\pi)^4} }
		C_{q; \mu_1 \cdots \mu_n}(\{ \hat{k}_{qs, i} \}) 
		g^{\mu_1 \nu_1} \cdots g^{\mu_n \nu_n}	\nn\\
		& \hspace{2em}\times 
		S_{\nu_1, \cdots, \nu_n; \sigma_1, \cdots, \sigma_m} (\{ k_{qs, i} \}, \{ k_{\bar{q}s, j} \})
		\nn\\
		& \hspace{2em} \times
		g^{\sigma_1 \rho_1} \cdots g^{\sigma_m \rho_m}
		C_{\bar{q}; \rho_1 \cdots \rho_m}(\{ \hat{k}_{\bar{q}s, j} \}),
\end{align}
where we have suppressed the collinear momentum dependence, and $\hat{k}_{qs, i}$ and $\hat{k}_{\bar{q}s, j}$ are defined as 
Eqs.~\eqref{eq:soft-approx-k-in-A} and \eqref{eq:soft-approx-k-in-B}, respectively. 
The soft factor $S$ includes all the soft gluon propagators. Here we are separately examining the soft momenta attaching to $C_q$
and $C_{\bar{q}}$, which are related by necessary delta functions included in $S$; eventually we will also include the soft integrations into $S$.
Similar to the collinear gluon coupling $C_i$ to $H$, here only the $g^{-+}$ components of all the metric tensors give leading-power contributions.
So we make the approximations,
\beq[eq:soft-col-coupling-approx]
	g^{\mu_i \nu_i} \mapsto \frac{\hat{k}_{qs,i}^{\mu_i} \, \bar{n}^{\nu_i} }{\hat{k}_{qs,i} \cdot \bar{n}} 
	= \frac{\hat{k}_{qs,i}^{\mu_i} \, \bar{n}^{\nu_i} }{k_{qs,i} \cdot \bar{n}},
	\quad
	g^{\sigma_j \rho_j} \mapsto \frac{n^{\sigma_j} \, \hat{k}_{\bar{q}s, j}^{\rho_j} }{n \cdot \hat{k}_{\bar{q}s, j}}
	= \frac{n^{\sigma_j} \, \hat{k}_{\bar{q}s, j}^{\rho_j} }{n \cdot k_{\bar{q}s, j}}.
\eeq
Similar to \eq{eq:g-approx}, this is equivalent to $g^{\mu\nu} \mapsto n^{\mu} \bar{n}^{\nu}$. It simplifies \eq{eq:soft-c-c-coupling} to
\begin{align}\label{eq:soft-c-c-coupling}
	C_q \otimes S \otimes C_{\bar{q}}
	\simeq &\, \int \bb{ \prod_i \frac{d^4 k_{qs, i}}{(2\pi)^4} } \bb{ \prod_j \frac{d^4 k_{\bar{q}s, j}}{(2\pi)^4} }
		C_{q; \mu_1 \cdots \mu_n}(\{ \hat{k}_{qs, i} \}) 
		\pp{ \prod_i \hat{k}_{qs,i}^{\mu_i} } 	\nn\\
		& \hspace{2em}\times 
		\bb{ \prod_i \frac{\bar{n}^{\nu_i} }{k_{qs,i} \cdot \bar{n}} } 
		S_{\nu_1 \cdots \nu_n; \sigma_1 \cdots \sigma_m} (\{ k_{qs, i} \}, \{ k_{\bar{q}s, j} \})
		\bb{ \prod_j \frac{n^{\sigma_j} }{n \cdot k_{\bar{q}s, j}} }
		\nn\\
		& \hspace{2em} \times
		\pp{ \prod_j \hat{k}_{\bar{q}s, j}^{\rho_j} } 
		C_{\bar{q}; \rho_1 \cdots \rho_m}(\{ \hat{k}_{\bar{q}s, j} \}).
\end{align}
The soft gluon couplings to the collinear factors are replaced by their approximated momenta in the collinear subgraphs, which will 
allow the use of Ward identity.

\section{Glauber region and modified approximations}
\label{sec:glauber-region}
The previous discussion on the soft momenta all relies on the uniform scaling in \eq{eq:soft-scaling},
assuming the integration of the angular variable $\bar{k}^{\mu}$ in \eq{eq:soft-lambda} has a uniform bound 
in its whole range. 
This includes the power counting rules, determination of leading regions, and the soft approximations.
The missing regions surrounding the soft pinch surface concern two types:
\begin{itemize}
\item
	$|k_s^+ k_s^-| \sim k_{sT}^2 \ll Q^2$ but $|k_s^+| \gg |k_s^-|$ or vice versa. 
	An example is $k_s \sim (\lambda, \lambda^3 / Q^2, \lambda^2 / Q)$. This is still a soft momentum but 
	with a large rapidity $y \sim \ln(Q / \lambda)$, so it is also collinear to the quark.
	We can call this scaling soft-collinear scaling.
	When it flows through $C_q$,
	it is no longer a good approximation to only keep $k_s^-$ as in \eq{eq:soft-approx-k-in-A}. 
	As we will see in Secs.~\ref{sec:subtraction} and \ref{sec:sudakov-factorization}, 
	a correct treatment needs to consider $k_s$ as a collinear momentum, and 
	its overlap with the soft region will be taken care of by the subtraction formalism.
\item 
	$Q^2 \gg |k_s^+ k_s^-| \gg k_{sT}^2$. This does not raise any new issue compared to the uniform scaling and soft-collinear scaling.
\item 
	$|k_s^+ k_s^-| \ll k_{sT}^2 \ll Q^2$. This transverse-component-dominated region is called {\it Glauber region}, to which we now turn our discussion.
\end{itemize}

\subsection{Glauber region}
\label{ssec:glauber-region}
The Glauber region $|k_s^+ k_s^-| \ll k_{sT}^2 \ll Q^2$ is a subset of the soft region. A typical Glauber momentum scaling is
\beq
	k_s^{\rm Glauber} \sim (\lambda^2 / Q, \lambda^2 / Q, \lambda),
\eeq
where the plus and minus components are taken of the same order. 
When it flows through the collinear subgraphs $C_{q,\bar{q}}$, it does not change the virtualities of collinear lines,
\begin{align}\label{eq:glauber-col-scaling}
	(k_q + k_s^{\rm Glauber})^2 
		&\, 
		\simeq k_q^2 - (\bm{k}_{sT}^{\rm Glauber})^2 + 2 k_q^+ \, (k_s^{\rm Glauber})^- 
			- \bm{k}_{qT} \cdot \bm{k}_{sT}^{\rm Glauber}
		\simeq \order{\lambda^2},	\nn\\
	(k_{\bar{q}} - k_s^{\rm Glauber})^2 
		&\, 
		\simeq k_{\bar{q}}^2 - (\bm{k}_{sT}^{\rm Glauber})^2 - 2 k_{\bar{q}}^- \, (k_s^{\rm Glauber})^+
			+ \bm{k}_{\bar{q}T} \cdot \bm{k}_{sT}^{\rm Glauber}
		\simeq \order{\lambda^2},
\end{align}
where we only retained the terms of the highest scaling.
The soft propagator also has the same scaling 
\beq
	(k_s^{\rm Glauber})^2 \sim (k_{sT}^{\rm Glauber})^2 \sim \order{\lambda^2}.
\eeq
In this way, the Glauber region also makes a leading-power contribution. 
But it is clear from \eq{eq:glauber-col-scaling} that the transverse component of the soft momentum becomes non-negligible
in the collinear subgraphs, so that the approximations in Eqs.~\eqref{eq:soft-approx-k-in-A} and \eqref{eq:soft-approx-k-in-B}
are no longer valid. Even though we may still take the approximations in \eq{eq:soft-col-coupling-approx}, the soft momentum
$\hat{k}_s$ that couples to collinear subgraphs is not the same soft momentum that flows in them. As a result, the Glauber 
region violates the soft approximations that allow the exact use of Ward identities.

As we will see, it is crucial for the use of Ward identities to factorize soft gluons from collinear subgraphs in a gauge theory. 
So the presence of the Glauber region endangers factorization, which we must deal with particularly.

\subsection{Contour deformation}
\label{ssec:glauber-deform}
The Glauber region (or any other soft region) is in the neighborhood of the soft pinch surface, but is not itself a pinch surface. 
When getting away from the pinch surface, the momentum contour is no longer exactly pinched at a singular point to give 
zero virtualities. But if we are close to the pinch surface, the contour deformation is normally still restricted to keep the virtualities
from getting too large. Therefore, we need to investigate whether the contour is pinched in the Glauber region. 
If not, we may still deform the contour to avoid the Glauber region.

Since the characteristics of the Glauber region is that the longitudinal components $k_s^{\pm}$ are much smaller than the 
transverse component $k_{sT}$, we would identify the poles of $k_s^{\pm}$ around 0 given $k_{sT}$. 
First, the denominators of $C_{q}$-collinear propagators
\begin{align}
	(k_{qi} + k_s)^2 + i\epsilon 
	&= 2 (k_{qi}^+ + k_s^+)(k_{qi}^- + k_s^-) - (\bm{k}_{qi,T} + \bm{k}_{sT})^2 + i \epsilon 	\nn\\
	&\simeq 2 k_{qi}^+ (k_{qi}^- + k_s^-) - (\bm{k}_{qi,T} + \bm{k}_{sT})^2 + i \epsilon
\end{align}
contribute to poles of $k_s^-$ on the lower half plane, of the order $\lambda^2 / Q$.
Since all the $C_{q}$-collinear lines propagate from $H$ to the future with large positive plus momenta, and we can always choose
$k_s$ to flow along $C_{q}$-collinear lines in the same direction without detouring back and forth inside $C_q$, 
all the $C_{q}$-collinear lines only contribute to small $k_s^-$ on the lower half plane.
Similarly, all the denominators of $C_{\bar{q}}$-collinear propagators only contribute to $k_s^+$ poles on the upper half plane,
\begin{align}
	(k_{\bar{q}j} - k_s)^2 + i\epsilon 
	&= 2 (k_{\bar{q}j}^+ - k_s^+)(k_{\bar{q}j}^- - k_s^-) - (\bm{k}_{\bar{q}j,T} + \bm{k}_{sT})^2 + i \epsilon \nn\\
	&\simeq 2 k_{\bar{q}j}^- (k_{\bar{q}j}^+ - k_s^+) - (\bm{k}_{\bar{q}j,T} + \bm{k}_{sT})^2 + i \epsilon,
\end{align}
of order $\lambda^2 / Q$.
Both these $k_s^+$ and $k_s^-$ poles are in the Glauber region, but only on the same half plane respectively, and not pinched.
Around those poles, the gluon propagator also contribute to poles for $k_s^+$ and $k_s^-$, but of order 
$k_{sT}^2 / k_s^{\pm} \sim \order{Q}$, which is far away.

Therefore, we can deform the contour of $k_s^{\pm}$ such that their magnitudes stay much greater than $\lambda^2 / Q$. 
Due to the $k_s^-$ poles from $C_q$ propagators, we deform the $k_s^-$ contour to the upper half plane,
\beq[eq:glauber-deform-k-]
	k_s^- \mapsto k_s^- + i \, v(k_s^-),
\eeq
where $v(k_s^-) > 0$ kicks in when $k_s^- \sim \order{\lambda}$ and keeps $k_s^-$ on the deformed contour to be at least 
of order $\lambda$. A simple choice can be, e.g., 
\beq[eq:glauber-deform-func]
	v(k) = \lambda \, e^{-k^2 / 2\lambda^2},
\eeq
which only modifies the Glauber region. 
This deformation does not change the $k_s^+$ poles from the $C_{\bar{q}}$ propagators, but changes the $k_s^+$ pole from the 
gluon propagator to the order $k_{sT}^2 / k_s^{-} \sim \order{\lambda}$. Hence, it is still compatible to deform the $k_s^+$ contour
to the lower half plane by $\order{\lambda}$,
\beq[eq:glauber-deform-k+]
	k_s^+ \mapsto k_s^+ - i \, v(k_s^+),
\eeq
where we chose the same deformation function in \eq{eq:glauber-deform-func}, which is not necessary.

Eqs.~\eqref{eq:glauber-deform-k-} and \eqref{eq:glauber-deform-k+} deform 
the contours of $k_s^+$ and $k_s^-$ by the same amount.
This is called symmetrical deformation. To avoid possible obstruction from 
the poles of the gluon propagator, the maximum extent
of the symmetrical deformation is $\order{\lambda}$.
Symmetrical deformation is not always necessary. 
We will also see that in certain cases, e.g., in \sec{sssec:dvmp}, 
symmetrical deformation is not allowed by a partial pinch in the Glauber region. 
There it is sufficient to only deform $k_s^+$ {\it or} $k_s^-$.

The deformations in Eqs.~\eqref{eq:glauber-deform-k-} and \eqref{eq:glauber-deform-k+} are only applied to the Glauber region, but not
necessary to the other soft subregions. One may devise a uniform deformation formula for the whole soft region, e.g.,
\beq[eq:glauber-deform]
	k_s^- \mapsto k_s^- + i \, \rho\pp{ \frac{2k_s^+ k_s^-}{k_{sT}^2} } \, v(k_s^-),
	\quad
	k_s^+ \mapsto k_s^+ - i \, \rho\pp{ \frac{2k_s^+ k_s^-}{k_{sT}^2} } \, v(k_s^+),
\eeq
where $\rho(x)$ has the property that $\rho(x) \simeq 1$ as $|x| \ll 1$ and $\rho(x) \to 0$ as $|x| \gtrsim 1$. One simple choice is
\beq
	\rho(x) = \frac{(\lambda/Q)^2}{x^2 + (\lambda/Q)^2}.
\eeq
After this deformation, the components $(k_s^+, k_s^-, k_{sT})$ are of the same order, restoring the uniform scaling [\eq{eq:soft-scaling}]. 
Then the soft approximations in Eqs.~\eqref{eq:soft-approx-k-in-A} and \eqref{eq:soft-approx-k-in-B} can be applied.

\subsection{Modified approximations}
\label{ssec:glauber-modify}
The soft approximations can be applied only after the contour deformation, so it is important that they do not introduce any poles that
obstruct the contour deformations in \eq{eq:glauber-deform}. 
Therefore, for the approximations in \eq{eq:soft-col-coupling-approx}, we need to carefully specify an $i\epsilon$ prescription for the 
soft poles introduced around 0. 
For soft gluon momenta $\hat{k}_{qs,i}$ and $\hat{k}_{\bar{q}s,j}$ flowing {\it into} $S$ from $C_q$ and $C_{\bar{q}}$, respectively, we 
modify \eq{eq:soft-col-coupling-approx} to
\beq[eq:soft-col-coupling-approx-modify]
	g^{\mu_i \nu_i} \mapsto \frac{\hat{k}_{qs,i}^{\mu_i} \, \bar{n}^{\nu_i} }{\hat{k}_{qs,i} \cdot \bar{n} + i\epsilon} 
	= \frac{\hat{k}_{qs,i}^{\mu_i} \, \bar{n}^{\nu_i} }{k_{qs,i} \cdot \bar{n} + i\epsilon},
	\quad\quad
	g^{\sigma_j \rho_j} \mapsto \frac{n^{\sigma_j} \, \hat{k}_{\bar{q}s, j}^{\rho_j} }{n \cdot \hat{k}_{\bar{q}s, j} + i\epsilon}
	= \frac{n^{\sigma_j} \, \hat{k}_{\bar{q}s, j}^{\rho_j} }{n \cdot k_{\bar{q}s, j} + i\epsilon}.
\eeq
Thus introduced poles for $k_s^{\pm}$ are all on the same half plane as those from the $C_{\bar{q}, q}$ propagators.

The collinear approximations in \eq{eq:g-approx} also introduce poles of the gluon momenta at 0. Even though they
are designed only for the collinear region, as we will see in \sec{sec:subtraction}, after applying those approximations, we extend the loop momenta
to all regions, including the soft and hard regions as well. 
The overlap with the soft region will be subtracted to avoid double counting. 
The subtraction term is obtained by first applying the soft approximation in 
Eqs.~\eqref{eq:soft-approx-k-in-A}\eqref{eq:soft-approx-k-in-B}\eqref{eq:soft-col-coupling-approx-modify},
and then applying the collinear approximation. 
Since the soft approximation is applied to the deformed contour, it is necessary that
the collinear approximation be compatible with such deformation when the same gluon momentum enters the soft region.
The same collinear gluon momentum $k_i$ entering $C_q$ from $H$ can also enter the soft region, 
where it flows from $C_{\bar{q}}$ into $S$ and has soft poles on the lower half plane for its plus momentum component.
Similarly, for the collinear gluon momentum $l_j$ entering $C_{\bar{q}}$ from $H$, we need to avoid the soft pole on the lower half plane
for its minus momentum component.
Therefore, we need to modify \eq{eq:g-approx} to
\begin{align} \label{eq:g-approx-modify}
	g^{\mu_i \nu_i} \, \mapsto \frac{n^{\mu_i} \, \hat{k}_i^{\nu_i} }{ \hat{k}_i \cdot n + i\epsilon} 
		= \frac{n^{\mu_i} \, \hat{k}_i^{\nu_i} }{ k_i \cdot n + i\epsilon}, \quad\quad
	g^{\sigma_j \rho_j} \, \mapsto \frac{\hat{l}_j^{\sigma_i} \, \bar{n}^{\rho_j} }{ \hat{l}_j \cdot \bar{n} + i\epsilon} 
		=  \frac{\hat{l}_j^{\sigma_i} \, \bar{n}^{\rho_j} }{ l_j \cdot \bar{n} + i\epsilon}.
\end{align}

In this way, the necessity of contour deformation to get out of the Glauber region dictates the $i\epsilon$ prescriptions in the soft and collinear
approximations. The direction of the deformation is determined only by the causal structure of the scattering process,
so are the $i\epsilon$ prescriptions, which will give correct causal properties for each factor in the factorization result.

\section{Subtraction formalism and factorization}
\label{sec:subtraction}

The leading power contribution of a certain amplitude or cross section does not {\it a priori} correspond to a factorized expression. 
The latter is motivated by the fact that the leading regions have distinct momentum scales, classified as the hard, the collinear, and the soft momenta, 
as given by the 
Libby-Sterman analysis. By choosing proper approximations as in \sec{sec:approx-region}, the different momentum scales detach so
as to imply a factorized expression. 

\begin{figure}[htbp]
	\centering
	\begin{tabular}{cc}
	\includegraphics[clip, trim={0 -7mm 0 0}, scale=0.65]{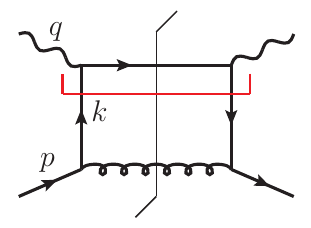} & \hspace{4em}
	\includegraphics[scale=0.65]{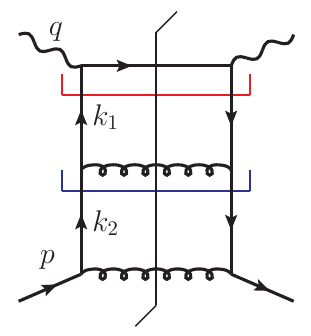} \\
	(a) & \hspace{4em} (b)
	\end{tabular}
	\caption{Examples of DIS diagrams for an elementary target. 
	(a) is the LO diagram, with one single leading region indicated by the red hooked line. 
	(b) is an NLO diagram, with two leading regions indicated by the red and blue hooked lines, respectively.
	}
	\label{fig:DIS-LO-NLO}
\end{figure}

As a necessary condition for factorization, the simplest nontrivial diagram that only has a single leading region with 
two distinct momentum scales should simply factorize as a result of the approximations, e.g., the leading-order DIS diagram as in \fig{fig:DIS-LO-NLO}(a).
If the whole amplitude or cross section only had one single region, then factorization would come as a direct result of proper approximations.
However, as a renormalizable gauge theory, each process in QCD has an infinite number of diagrams with arbitrarily complicated leading regions.
A proper treatment must take into account all possible regions of all diagrams, with careful avoidance of double counting between neighboring regions.
The factorization is then a highly nontrivial result, being very intricate and also fragile. 

In this section, we briefly review the subtraction formalism used in the treatment of multiple regions in the derivation of factorization. 
First, we will consider the two low-order DIS diagrams in \fig{fig:DIS-LO-NLO} and examine their factorization. 
Starting with the smallest region and building up larger and larger regions while
avoiding double counting of overlapping regions naturally motivate the subtraction method.  
Then after a formal discussion of the subtraction formalism, we demonstrate its application to the simplest case,
in a real physical situation, the DIS factorization in the light-cone gauge.

\subsection{A particular DIS diagram as an example}
\label{ssec:DIS-light-cone-ex}
Each leading region of a DIS diagram contains only two subgraphs, one hard subgraph $H$ connected to the virtual photon lines, and 
one collinear subgraph $C$ attaching to the external target. The two subgraphs are joined by a set of collinear lines whose propagators 
we include in the collinear subgraph. \fig{fig:DIS-LO-NLO} shows two diagrams for an elementary target, which we take as an
on-shell quark with a small mass $m$ to cut off collinear divergences, with no concern for confinement issues.
Taking the kinematics 
\beq
	p = \pp{ p^+, \frac{m^2}{2p^+}, \bm{0}_T }, \quad
	q = \pp{ - x p^+, \frac{Q^2}{2 x p^+}, \bm{0}_T }, \quad
\eeq
for the target and the photon, respectively. 
With $Q \gg m$, the lines in the hard subgraph have virtualities of order $Q^2$, and
those in the collinear subgraph have virtualities of order $m^2$. The leading regions are represented by the massless reduced diagrams, obtained
by taking $m \to 0$, which yields $p \to \hat{p} = (p^+, 0^-, \bm{0}_T)$.

In this section, we work in the light-cone gauge, where the gluon propagator numerator is 
\beq[eq:lc-g-propagator]
	-g^{\mu\nu} + \frac{n^{\mu} k^{\nu} + n^{\nu} k^{\mu}}{k \cdot n},
\eeq
with $n$ is defined in \eq{eq:n-nbar}.
This suppresses the longitudinal polarization so that the leading regions only have two quark or transversely polarized gluon lines 
joining $C$ to $H$. The approximation can be easily devised, following the spirit in \sec{sec:approx-region}, as
\begin{itemize}
\item
	for each collinear momentum $k_i$ flowing into the hard subgraph $H$, we approximate it by only keeping its plus component,
	\beq[eq:DIS-col-approx-lc]
		k_i \mapsto \hat{k}_i = (k_i \cdot n) \bar{n};
	\eeq
\item 
	for each quark line entering (leaving) $H$, insert the spinor projector 
	\beq[eq:DIS-spinor-proj]
		\P_n = \gamma^- \gamma^+ / 2 \quad (\Pb_n = \gamma^+ \gamma^- / 2);
	\eeq
\item 
	for each gluon line connecting $C$ to $H$, insert the Lorentz tensor $g_{\perp}^{\mu\nu}$ to project out the transverse polarization.
\end{itemize}
Denoting the effect of such approximation by
\beq[eq:DIS-T-approximator]
	H \otimes C \equiv \int [dk] H(k) \, C(k)  \; \mapsto \;
	\int [dk] H(k) \, \hat{T} \, C(k),
\eeq
where $k$ collectively denotes all the collinear momenta, and 
$\hat{T}$ acts on the integrand, inserting certain projectors between $H$ and $C$ and neglecting certain momenta in $H$ to its left.

The LO diagram $\Gamma_0$ in \fig{fig:DIS-LO-NLO}(a) has two leading regions, 
(1) one with $k = \alpha \, \hat{p}$ ($0 < \alpha < 1$) collinear to the target, so that the reduced diagram has one hard subgraph and one collinear
subgraph, separated by the red hooked line, which inserts necessary projectors and approximates on shell the collinear momenta passing it;
(2) and the other with $k^2 \sim \order{Q^2}$, so that the whole diagram is the hard subgraph, to which the external photon and quark lines attach.
To simplify the discussion in this section, we assume the second region is power suppressed by some nonperturbative effect, so that we only have the first
leading region.
Factorization then follows trivially from the fact that only $k^+$ flows through $H$, and this leads to
\beq[eq:DIS-LO-fact]
	C_R \Gamma_0 = T_R \Gamma_0 = \int \frac{d^4 k}{(2\pi)^4} H_0(k) \, \hat{T} \, C_0(k)
	= \int dk^+ H_0(\hat{k}) \bb{ \int \frac{dk^- d^2 \bm{k}_T}{(2\pi)^4} \P_n \, C_0(k)  \,\Pb_n },
\eeq
up to power suppressed contribution. 
Here we use the compound notation ``$C_R \Gamma_0$'' to refer to the leading contribution of the diagram $\Gamma_0$ from the region $R$.
$T_R$ is the approximator designed for this region; it acts on $\Gamma_0$ to extract the leading contribution. 
Since $\Gamma_0$ only has one leading region $R$, $C_R \Gamma_0$ is just equal to $T_R \Gamma_0$.
We will deal with the spin and color projectors in more details later in \sec{ssec:DIS-Feynman-lo-pdf}.

The NLO diagram $\Gamma_1$ in \fig{fig:DIS-LO-NLO}(b) is more complicated. We have two leading regions:
(1) region $R_1$: $k_1$ is in $H$ with a large virtuality, while $k_2 = \beta \hat{p}$ ($0 < \beta < 1$) is collinear, as indicated by the blue hooked line;
and
(2) region $R_2$: both $k_1$ and $k_2$ are collinear to the target, with 
\beq
	k_1 = \alpha k_2, \quad
	k_2 = \beta \hat{p}, \quad
	0 < \alpha, \beta < 1,
\eeq
which is indicated by the red hooked line in \fig{fig:DIS-LO-NLO}(b).
The topology of the region $R_1$ is defined in the momentum space as $k_1^2 \neq 0$ and $k_2 = \beta \hat{p}$. 
Apparently, its closure contains $R_2$ as a subset. This relation between the two leading regions is denoted as $R_1 > R_2$.
$R_1$ is greater than $R_2$ in the sense that it has more lines with hard virtualities, while $R_2$ has more lines in the collinear subgraph.

To move forward, we first define the diagram \fig{fig:DIS-LO-NLO}(b) as
\beq
	\Gamma_1 = \int \frac{d^4k_1}{(2\pi)^4} \frac{d^4k_2}{(2\pi)^4} \bb{ H_0(q; k_1) \cdot K_0(k_1, k_2) \cdot C_0(p; k_2)  }.
\eeq
The factor $H_0(q; k_1)$ includes the quark line on the top and the two photon vertices, $K_0(k_1, k_2)$ includes the propagators of $k_1$, 
the gluon line of $(k_2 - k_1)$, and their vertices, and $C_0(p; k_2)$ includes the rest of the diagram. These three factors are convoluted in
momenta $k_1$ and $k_2$, which has been explicitly written, and spinor and color indices, which are implicitly implied by the dot notation.

To extract the leading contribution from $R_2$, we can insert the approximator $\hat{T}$ between $H_0$ and $K_0$, formally written as
\beq
	C_{R_2} \Gamma_1 = T_{R_2} \Gamma_1 
	= \int \frac{d^4k_1}{(2\pi)^4} \frac{d^4k_2}{(2\pi)^4} \bb{ H_0(q; k_1) \hat{T} K_0(k_1, k_2) \cdot C_0(p; k_2)  },
\eeq
which projects $k_1$ in $H_0$ by $\hat{k}_1$ according to \eq{eq:DIS-col-approx-lc} and inserts the spinor projectors $\P_n$ and $\Pb_n$. 
This expression simply factorizes into
\beq[eq:DIS-NLO-fact-R2]
	C_{R_2} \Gamma_1 = \int dk^+ H_0(q; \hat{k})
		\bb{ \int \frac{dk^- d^2 \bm{k}_T}{(2\pi)^4} \frac{d^4k_2}{(2\pi)^4} \P_n \big( K_0(k, k_2) \cdot C_0(p; k_2) \big) \Pb_n },
\eeq
which adds to \eq{eq:DIS-LO-fact} with the same hard coefficient, but as an NLO correction to the collinear factor.

Now we consider the contribution from $R_1$. Naively, $C_{R_1} \Gamma_1$ is obtained by
\begin{align}
	C_{R_1} \Gamma_1 \sim T_{R_1} \Gamma_1
	& = \int \frac{d^4k_1}{(2\pi)^4} \frac{d^4k_2}{(2\pi)^4} \bb{ H_0(q; k_1) \cdot K_0(k_1, k_2) \, \hat{T} \, C_0(p; k_2)  } \nn\\
	& = \int \frac{d^4k_1}{(2\pi)^4} \frac{d^4k_2}{(2\pi)^4} \bb{ H_0(q; k_1) \cdot K_0(k_1, \hat{k}_2) \cdot \pp{ \P_n C_0(p; k_2) \Pb_n }  },
\end{align}
which seems to factorize into a hard factor $H_0(q; k_1) \cdot K_0(k_1, \hat{k}_2)$ and a collinear factor $\P_n C_0(p; k_2) \Pb_n$.
Now the hard factor includes the integration of $k_1$, which should be constrained to the hard region.  
However, this is technically hard to define, given also the need to deform contours when {\it unpinched} propagators become close to
the on-shell poles. It would be ideal to have the $k_1$ integration in the hard factor to extend to all regions. 
Then it can unavoidably reach the collinear region where $k_1 = \alpha \hat{k}_2$. This is still a leading region in the hard subgraph.
But such contribution has been included in the region $R_1$. 
This reflects a general fact that a larger region $R_1$ has overlap with smaller regions $R_2 < R_1$, 
such that the approximator $T_{R_1}$ alone is not sufficiently exclusive to extract only the contribution from $R_1$ when acting on the graph $\Gamma$ alone.
Therefore, when applying $T_{R_1}$, one should first subtract the contribution from smaller regions, such that
\begin{align}\label{eq:DIS-NLO-fact-R1}
	C_{R_1} \Gamma_1 & = T_{R_1} (\Gamma_1 - C_{R_2} \Gamma_1) = T_{R_1} (1 - T_{R_2}) \Gamma_1 	\nn\\
	& = \int \frac{d^4k_1}{(2\pi)^4} \frac{d^4k_2}{(2\pi)^4} \bb{ H_0(q; k_1) \, (1 - \hat{T}) \, K_0(k_1, k_2) \, \hat{T} \, C_0(p; k_2)  }	\nn\\
	& = \int dk^+ \bb{ \int \frac{d^4k_1}{(2\pi)^4} H_0(q; k_1) \, (1 - \hat{T}) \, K_0(k_1, \hat{k}) }
			\bb{ \int \frac{dk^- d^2 \bm{k}_T}{(2\pi)^4} \P_n C_0(p; k) \Pb_n },
\end{align}
where in the third line we gave the factorization expression, which adds onto \eq{eq:DIS-LO-fact} with the same collinear factor, but as
an NLO correction to the hard factor, defined as
\begin{align}
	&\int \frac{d^4k_1}{(2\pi)^4} H_0(q; k_1) \, (1 - \hat{T}) \, K_0(k_1, \hat{k})	\nn\\
	&\;\; = \int \frac{d^4k_1}{(2\pi)^4} H_0(q; k_1) \cdot K_0(k_1, \hat{k})
		- \int d k^+_1 H_0(q; \hat{k}_1) \cdot \bb{ \int\frac{dk_1^- d^2\bm{k}_{1T}}{(2\pi)^4} \P_n K_0(k_1, \hat{k}) \Pb_n}.
\end{align}

Adding the leading contribution from $R_1$ and $R_2$, we have the total leading-power contribution of $\Gamma_1$,
\begin{align}\label{eq:DIS-region-decomp}
	\sum_R C_R \Gamma_1 &\, = C_{R_2} \Gamma_1 + C_{R_1} \Gamma_1
	= T_{R_2} \Gamma_1 + T_{R_1} (1 - T_{R_2}) \Gamma_1	\nn\\
	&\, = T_{R_2} \Gamma_1 + T_{R_1} \Gamma_1 - T_{R_1} T_{R_2} \Gamma_1,
\end{align}
where in the second line, we have separated all terms of different approximator applications. 
By construction, $T_{R_2}$ gives a good approximation to $\Gamma_1$ when both $k_1$ and $k_2$ have low virtualities, but gives a poor
description when either of them is hard. 
(Recalling our assumption that the region with both momenta having hard virtualities is suppressed, so 
$T_{R_2}$ is bad when $k_1$ is hard but $k_2$ is collinear.)
Also, $T_{R_1}$ gives a good approximation to $\Gamma_1$ only when $k_1$ is hard and $k_2$ is collinear.
Now, in the region $R_2$, among the three terms in the second line of \eq{eq:DIS-region-decomp}, the first term $T_{R_2} \Gamma_1$
gives a good approximation of $\Gamma_1$. The other two terms combine into $T_{R_1}  (1 - T_{R_2}) \Gamma_1$, which is suppressed
in this region due to the $(1 - T_{R_2})$ factor that suppresses the low virtuality region of $k_1$. 
Since $T_{R_1}$ keeps $k_1$ unchanged, it does not affect such suppression.
In the region $R_1$, the second term gives a good approximation of $\Gamma_1$. The other two terms combine into
$(1 - T_{R_1})T_{R_2} \Gamma_1$, which is suppressed due to the factor $(1 - T_{R_1})$ that suppresses the hard virtuality region of $k_1$.

Now we examine how well \eq{eq:DIS-region-decomp} can approximate the graph $\Gamma_1$ by constructing the error term,
\beq[eq:DIS-error-2]
	r_1 \equiv \Gamma_1 - \sum_R C_R \Gamma_1 = (1 - T_{R_1}) (1 - T_{R_2}) \Gamma_1.
\eeq
The factor $(1 - T_{R_2})$ accounts for the error introduced by neglecting $k_{1T}$ with respect to $Q$, so gives
\beq
	(1 - T_{R_2}) \Gamma_1 = \order{\frac{k_{1T}}{Q} } \Gamma_1.
\eeq
The other factor $(1 - T_{R_1})$ accounts for the error introduced by neglecting $k_{2T}$ with respect to $k_{1T}$ and $Q$, so
\beq
	r_1 = \order{\frac{k_{2T}}{k_{1T}}, \frac{k_{2T}}{Q} } \order{\frac{k_{1T}}{Q} } \Gamma_1
	= \order{\frac{k_{2T}}{Q}, \frac{k_{1T} k_{2T}}{Q^2} } \Gamma_1.
\eeq
Since we always have $k_{2T} \simeq m \ll Q$ and $k_{1T} \lesssim Q$, the error is power suppressed,
\beq
	r_1 = \order{\frac{m}{Q}}\Gamma_1.
\eeq

In fact, the argument of the subtraction formalism [\eq{eq:DIS-region-decomp}] can start with the error term \eq{eq:DIS-error-2}. 
Successively applying $(1 - T_R)$ on $\Gamma_1$ must yield a power suppressed result, since it successively suppresses all
the leading regions. Therefore, we have, right from the beginning,
\beq[eq:DIS-remainder]
	r_1 \equiv (1 - T_{R_1}) (1 - T_{R_2}) \Gamma_1 = \order{\frac{m}{Q}} \Gamma_1.
\eeq
This can be reorganized as
\beq
	r_1 = (1 - T_{R_2}) \Gamma_1 - T_{R_1} (1 - T_{R_2}) \Gamma_1
		= \Gamma_1 - T_{R_2} \Gamma_1 - T_{R_1} (1 - T_{R_2}) \Gamma_1,
\eeq
so immediately gives the subtraction formula,
\beq
	\Gamma_1 = T_{R_2} \Gamma_1 + T_{R_1} (1 - T_{R_2}) \Gamma_1 + \order{\frac{m}{Q}},
\eeq
where the power suppressed term is from $r_1$. Such subtraction formalism systematically extracts the leading-power contributions
from all regions, with loop momenta extending to all regions and without double counting. 
Our analysis of the DIS diagrams up to NLO 
should motivate that summing over all regions and diagrams can lead to a factorization result, which we will discuss in \sec{ssec:DIS-light-cone}.

\subsection{Subtraction formalism}
\label{ssec:subtraction}
Generally, a diagram $\Gamma$ can contain multiple leading regions, $R_i$. 
In the Sudakov form factor example \fig{fig:sudakov-reduced-diagram}(b), different
regions of a particular diagram differ by having different lines or different numbers of lines in the hard, $A$-collinear, $B$-collinear, or soft subgraphs.
In the DIS example, different regions of the diagram $\Gamma_1$ in \fig{fig:DIS-LO-NLO}(b) 
differ by having different numbers of ladders in the hard of collinear subgraphs.
In each region $R_i$, we design suitable approximator $T_{R_i}$ that acts on the {\it integrand} 
and gives a proper approximation for the {\it integral} in that region.
The leading power contribution of each region $R$ is iteratively defined as
\begin{align}\label{eq:region-contrib}
	C_{R} \Gamma & = 
		\begin{dcases}
			T_R \Gamma, \quad &\mbox{if no region $R_i$ of $\Gamma$ is smaller than $R$;} \\
			T_R \pp{ \Gamma - \sum\nolimits_{R' < R}  C_{R'} \Gamma }, \quad & \mbox{otherwise.}
		\end{dcases}
\end{align}
And then summing over all regions gives an approximation to the original diagram $\Gamma$ up to power suppressed corrections,
\beq[eq:subtraction-formalism]
	\Gamma = \sum_R C_R \Gamma + {\rm p.s.}
\eeq
This is the general subtraction formalism of extracting the leading-power contribution from a diagram $\Gamma$.

No attempt will be given here to prove \eq{eq:subtraction-formalism}.%
\footnote{A formal argument can be found in \citep{Collins:2011zzd}.} 
We simply motivate it by examining a simple case where the leading regions are strictly nested, i.e., 
all the leading regions $R_i$ of any graph $\Gamma$ have the strict ordering $R_1 > R_2 > \cdots > R_n$.
This is true for diagrams with only two kinds of subgraphs, such as DIS diagrams which only have collinear and hard subgraphs 
(but not for Sudakov form factor which has four subgraphs).
Similar to \eq{eq:DIS-remainder}, by successively subtracting all the leading region contributions, the remainder
\beq[eq:ordered-remainder]
	r = (1 - T_{R_1}) (1 - T_{R_2}) \cdots (1 - T_{R_n}) \Gamma
\eeq
is power suppressed. Then by reorganizing \eq{eq:ordered-remainder}, we have
\begin{align}\label{eq:ordered-reorganize}
	\Gamma & = \Gamma - r + {\rm p.s.}	\nn\\
	& = \Gamma - (1 - T_{R_2}) \cdots (1 - T_{R_n}) \Gamma + T_{R_1} (1 - T_{R_2}) \cdots (1 - T_{R_n}) \Gamma + {\rm p.s.}	\nn\\
	& = \Gamma - (1 - T_{R_3}) \cdots (1 - T_{R_n}) \Gamma + T_{R_2} (1 - T_{R_3}) \cdots (1 - T_{R_n}) \Gamma \nn\\
	& \hspace{2em}
		+ T_{R_1} (1 - T_{R_2}) \cdots (1 - T_{R_n}) \Gamma + {\rm p.s.}	\nn\\
	& = T_{R_n} \Gamma + T_{R_{n-1}}(1 - T_{R_{n}}) \Gamma + \cdots 
		+ T_{R_1} (1 - T_{R_2}) \cdots (1 - T_{R_n}) \Gamma + {\rm p.s.}	\nn\\
	& = C_{R_n} \Gamma + C_{R_{n-1}} \Gamma + \cdots + C_{R_1} \Gamma + {\rm p.s.},
\end{align}
where in the last step we defined
\beq[eq:ordered-Ci]
	C_{R_i} = T_{R_i} (1 - T_{R_{i+1}}) \cdots (1 - T_{R_n}) \Gamma,
\eeq
which agrees with the general definition in \eq{eq:region-contrib}.

In a general Feynman diagram, the relations among all the leading regions form an ordered graph, 
starting from the largest region where all loop momenta
have hard virtualities, and ending at the smallest region where as many loop momenta as possible are 
in the soft (or collinear, when there is no soft subgraph) region.
Then the contribution $C_R \Gamma$ of any region $R$ has the formal structure,
\begin{align}\label{eq:region-subtraction-term}
	C_R \Gamma & = T_R \Gamma - \sum_{R' < R} T_R C_{R'} \Gamma	\nn\\
		& = T_R \Gamma - \sum_{R' < R} T_R T_{R'} (\Gamma - \sum_{R'' < R'} C_{R''} \Gamma )	\nn\\
		& = T_R \Gamma + T_R \sum_{R' < R} (- T_{R'}) \Gamma
				+ T_R \sum_{R' < R} (- T_{R'}) \sum_{R'' < R'} (- T_{R''}) \Gamma + \cdots	\nn\\
		& = T_R \Gamma + T_R \sum_{ \{ R'_i \} } \prod_i (-T_{R'_i}) \Gamma,
\end{align}
where in the last line the sum is over all possible nestings of regions smaller than $R$: $R > R'_1 > R'_2 > \cdots > R'_n$, 
and 
\beq
	\prod_i (-T_{R'_i}) = (-T_{R'_1})(-T_{R'_2}) \cdots (-T_{R'_n}),
\eeq
with the approximators for larger regions on the left. 
Each nesting can terminate at any region $R'_n < R$, not necessarily the smallest region. 

\eq{eq:region-subtraction-term} gives a general formula for the subtraction terms in $C_R \Gamma$. 
They are given by successively applying $-T_{R'}$ for smaller regions $R' < R$, and then applying $T_R$. 
When applying any $T_{R'}$, we treat the loop momenta as classified in the same way as the region $R'$,
and the approximator $T_{R'}$ works by deforming certain momentum contours, neglecting certain momentum components,
and inserting certain factors.
That is, any subtraction term in $C_R \Gamma$ is treated to have the same subgraph decomposition as $R$, and it is
to be approximated in the same way as $T_R \Gamma$. 
Since $R' < R$ has more lines in the collinear or soft subgraphs and/or fewer in the hard subgraph, the approximator $T_{R'}$
inserts certain approximation factors and truncates some momenta in the hard and/or collinear subgraphs of $R'$.
When applying $T_R$ on top of $T_{R'}$, some soft lines are upgraded to collinear lines 
and some collinear lines to hard lines, so then the approximator factors of $T_{R'}$ will reside within
the new hard and collinear subgraphs specified by $R$.
Therefore, graphically, $T_R \Gamma$ and the subtraction terms $\sum_{R' < R} T_R (-C_{R'} \Gamma)$ correspond to 
the same region graph, with the only difference being that the subtraction terms already have some ``sub-approximators'' applied within.

In order for the presence of subtraction not to affect the argument of the approximation $T_R$, especially the application of 
Ward identities that follows the soft and collinear approximators defined in 
Eqs.~\eqref{eq:soft-col-coupling-approx-modify} and \eqref{eq:g-approx-modify}, the same simplifications following $T_R \Gamma$
should also apply to the subtraction terms $\sum_{R' < R} T_R (-C_{R'} \Gamma)$.
So we require the approximator factors as in Eqs.~\eqref{eq:soft-col-coupling-approx-modify} and \eqref{eq:g-approx-modify}
not to cause gauge-violating structures.
Applying $T_{R}$ on $\Gamma$ allows to factorize soft or collinear gluons onto Wilson lines for any region $R$,
so does the $T_{R'}$ for smaller regions $R' < R$ in the subtraction terms of $C_R \Gamma$.
Then further applying $T_R$ on the Wilson lines resulted from $T_{R'}$ should also lead to a Wilson line structure.
To guarantee this, it is sufficient to require the use of Ward identities on a Wilson line to give back the same Wilson line.
On the other hand, since $T_{R'}$ is likely to involve contour deformations for some soft momenta,
it is necessary that the approximator $T_R$ does not introduce any factors contradicting such deformations
when those soft momenta are upgraded to collinear or hard momenta.
This explains the $i\epsilon$ choices in \eq{eq:g-approx-modify}.

\begin{figure}[htbp]
	\centering
	\begin{tabular}{cc}
	\includegraphics[clip, trim={0 -7mm 0 0}, scale=0.65]{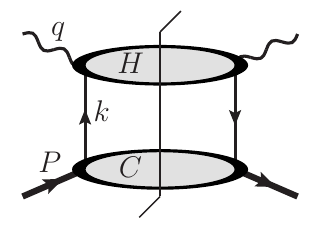} & \hspace{4em}
	\includegraphics[scale=0.65]{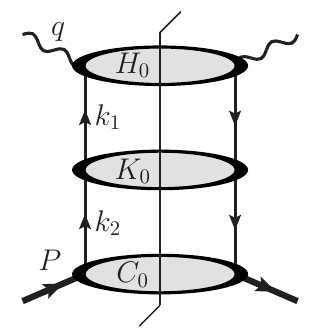} \\
	(a) & \hspace{4em} (b)
	\end{tabular}
	\caption{(a) A general leading region for a DIS cut diagram in light-cone gauge is divided into a hard subgraph $H$ and a collinear subgraph $C$, 
	joined by two collinear quark or gluon lines.
	(b) The ladder expansion for a certain DIS cut diagram is decomposed into a series of 2PI subdiagrams connected by two quark or gluon lines.
	The thick external lines represent hadron targets.
	}
	\label{fig:DIS-leading-ladder}
\end{figure}

\subsection{Factorization of DIS in light-cone gauge}
\label{ssec:DIS-light-cone}
Now we extend to all orders the discussion in \sec{ssec:DIS-light-cone-ex} of the DIS factorization in light-cone gauge, as a simple
application and illustration of the subtraction formalism in \sec{ssec:subtraction}. We will replace the elementary quark target by a physical
on-shell hadron target, such as a proton, of momentum $P$ and mass $M$.

In the light-cone gauge, the general leading region for DIS contains a hard subgraph and a collinear subgraph, which are joined by
two collinear quark or gluon lines, as shown in \fig{fig:DIS-leading-ladder}(a). 
For a given graph, a larger region has more lines in $H$ and fewer in $C$. This motivates the ladder expansion of a general DIS diagram,
as shown in \fig{fig:DIS-leading-ladder}(b), where each unit of $H_0$, $K_0$, and $C_0$ is two-particle-irreducible (2PI), which means
that they cannot be divided into two parts by only cutting two propagators, such that they cannot have further ladder expansion. 
We denote $H_0$, $K_0$, and $C_0$ as sums of all possible 2PI subgraphs for given external lines, each being a function of external momenta, spin indices,
and color indices. 
Then the sum of all DIS diagrams is given by
\begin{align}\label{eq:DIS-ladder}
	W & = D_0 + H_0 \cdot C_0 + H_0 \cdot K_0 \cdot C_0 + H_0 \cdot K_0 \cdot K_0 \cdot C_0  + \cdots	\nn\\
		& = D_0 + \sum_{n = 0}^{\infty} H_0 \cdot K_0^n \cdot C_0,
\end{align}
where $D_0$ is the minimal graph which is itself 2PI and has no ladder decomposition. 
We stress that each factor represents an all-order sum of 2PI {\it perturbative} diagrams, by the 
assumption (1) in \sec{ssec:fac-assumption}, 
with the hadron-parton vertex described by some hadron wavefunction.

By directly coupling the hadron to the virtual photon, $D_0$ has all lines being highly virtual, 
so it is power suppressed, by the same assumption (1) in \sec{ssec:fac-assumption}.

A graph $\Gamma_{n}$ with $n$ ladders ($n$ is the number of $K_0$ factor in \eq{eq:DIS-ladder}) has $n + 1$ leading regions, 
$ R_0 > R_1 > \cdots R_n$, with $R_i$ referring to the region with $i$ lower $K_0$ ladders belonging to the collinear subgraph.
Then following the discussion in Eqs.~\eqref{eq:ordered-remainder} - \eqref{eq:ordered-Ci}, 
the leading contribution of each region $R_i$ is 
\begin{align}
	C_{R_i} \Gamma_{n} &= T_{R_i} (1 - T_{R_{i+1}}) \cdots (1 - T_{R_{n}}) \Gamma_{n} 
		=  H_0 \cdot \bb{ (1 - \hat{T}) K_0 }^{n - i} \hat{T} \bb{ K_0 }^i \cdot C_0,
\end{align}
where $\hat{T}$ is defined in the same way as \eq{eq:DIS-T-approximator}.
This factorizes into a hard factor $H_0 \cdot [ (1 - \hat{T}) K_0 ]^{n - i}$ and a collinear factor $\P_n [ K_0 ]^i \cdot C_0 \Pb_n$,
similar to the low-order examples in Eqs.~\eqref{eq:DIS-LO-fact}\eqref{eq:DIS-NLO-fact-R2} and \eqref{eq:DIS-NLO-fact-R1}.
Summing over $i$ from $0$ to $n$ and then over $n$ from $0$ to $\infty$ amounts to summing over the ladders in the hard and 
collinear factors separately,
\begin{align}\label{eq:DIS-What}
	\hat{W} & = \sum_{n = 0}^{\infty} \sum_{i = 0}^n C_{R_i} \bb{ H_0 \cdot K_0^n \cdot C_0 }		\nn\\
		& = \sum_{n = 0}^{\infty} \sum_{i = 0}^n H_0 \cdot \bb{ (1 - \hat{T}) K_0 }^{n - i} \hat{T} \bb{ K_0 }^i \cdot C_0		\nn\\
		& = \sum_{i = 0}^{\infty} \sum_{j = 0}^{\infty} H_0 \cdot \bb{ (1 - \hat{T}) K_0 }^{i} \hat{T} \bb{ K_0 }^j \cdot C_0	\nn\\
		& = H_0 \cdot \frac{1}{1 - (1 - \hat{T}) K_0} \, \hat{T} \, \frac{1}{1 - K_0} \cdot C_0.
\end{align}
This factorizes into a hard factor
\beq[eq:DIS-hard]
	H(q; \hat{k}) = H_0 \cdot \frac{1}{1 - (1 - \hat{T}) K_0},
\eeq
and a collinear factor
\beq[eq:DIS-col]
	C(P, k) = \P_n \bb{ \frac{1}{1 - K_0} \cdot C_0 } \Pb_n.
\eeq
They are convoluted in the momentum $k$ and in color and spin indices. \eq{eq:DIS-What} approximates $W$ in \eq{eq:DIS-ladder}
with the remainder term
\beq
	r = W - \hat{W} = D_0 + \sum_{n = 0}^{\infty} H_0 \cdot \bb{ (1 - \hat{T}) K_0 }^n \cdot (1 - \hat{T}) C_0,
\eeq
where both terms are power suppressed. Therefore, we get the factorized result of the DIS cross section
\beq
	W = \int dk^+ H(q; \hat{k}) \bb{ \int \frac{dk^- d^2 \bm{k}_T}{(2\pi)^4} C(P, k) } + \order{\frac{M}{Q}}.
\eeq
We will deal with the spin and color connections later in \sec{ssec:DIS-Feynman-lo-pdf}.

Note that the sum over regions ($i$) and graphs ($n$) has been converted to two independent sums over each subgraph, 
which leads to the factorized expression. 
However, it is the subtraction of smaller regions from larger regions that separates different momentum scales. 
In this way, the hard factor defined in \eq{eq:DIS-hard} has removed all contribution from the regions where any of the loop momenta
become collinear. In terms of perturbative Feynman amplitudes, it is free from collinear singularity, and the corresponding
Feynman integrals are only sensitive to the hard scale $Q$, 
so we are allowed to use perturbative descriptions due to the asymptotic freedom, and it is
safe to neglect the parton masses and virtualities therein, as is encoded in the approximator $\hat{T}$.
The collinear factor defined in \eq{eq:DIS-col} collects all the pinch singularities in perturbative diagrams, 
so parton momenta in it are trapped in the low-virtuality regions. It is thus not perturbatively tractable, but 
the all-order sum in \eq{eq:DIS-col} can be {\it formally} defined nonperturbatively. 
Then the perturbative pinch singularities are interpreted to be reflecting the sensitivity to nonperturbative dynamics. 
Even though the result is obtained by perturbative diagram expansion, the overall sum, regardless of its convergence issue, 
is still assumed to reflect the correct reality, by the assumption (3) in \sec{ssec:fac-assumption}.
Its actual value can be obtained by fitting to experimental data, by virtue of its universality. 

In this way, the subtraction formalism together with the sum over regions and graphs factorizes the DIS cross section into hard and 
collinear factors, with the former only sensitive to dynamics at a hard scale $Q$, and the latter only to the nonperturbative soft scale $m$.
The separation of distinct scales is the essence of factorization.

\section{Ward identity and Wilson lines: DIS factorization in Feynman gauge}
\label{sec:ward-identity}

In the light-cone gauge, the gluon propagators are constrained to be physically polarized, 
and the leading region has a simple structure, as in \fig{eq:DIS-ladder} for the DIS. 
A simple application of the subtraction formalism directly yields the factorization formalism.
However, as indicated by \eq{eq:lc-g-propagator}, the light-cone gauge introduces extra poles
at $k \cdot n = k^+ = 0$. The leading regions of DIS have $k^+ \simeq \order{Q}$, without any pinch
singularity at $k^+ = 0$, so such poles do not pose severe trouble for the factorization argument.
However, in more complicated situations with also soft subdiagrams, this gives extra poles in the
soft region. Moreover, the $1/k\cdot n$ should be interpreted in the principle-value sense as
\beq
	\frac{1}{k\cdot n} = \frac{1}{2} \pp{ \frac{1}{k\cdot n + i \epsilon} + \frac{1}{k\cdot n - i \epsilon} },
\eeq
with both $i \epsilon$ prescriptions being present. 
Such poles severely forbid the factorization arguments in cases where a deformation of the $k^+$ contour
is needed, such as the Sudakov form factor and Drell-Yan process.
Therefore, in most general cases, one prefers to use Feynman gauge for the proof of factorization.
This avoids issues of unphysical poles, but introduces extra complications from unsuppressed 
longitudinally polarized gluons,
as discussed in \sec{ssec:power counting} and shown in \fig{fig:sudakov-reduced-diagram}(b) 
for the leading region of the Sudakov form factor. 

While the unphysical longitudinal polarization arises due to gluons being massless vector bosons as a result of
QCD being a gauge theory, they can be dealt with by gauge invariance arguments. As we shall see 
in this section, such gauge redundancy will be decoupled by Ward identities. However, contrary 
to what one might expect, such longitudinally polarized gluons are intermediate lines so gauge
invariance does not mean that they result in zero. Instead, they have specific irreducibility 
properties and will be collected into Wilson line structures.

In the following, we will first discuss the basic use of Ward identities for a certain region $R$ of the DIS diagram $\Gamma$
with subtraction, i.e., the term $T_R \Gamma$. 
The simple Wilson line structure shall make it evident that the use of Ward identities on a Wilson line gives back the same 
Wilson line. This will be used to take subtractions into account in approximating the region $R$, i.e., the term $C_R \Gamma$.
Finally, we will use the Ward identity argument on the Sudakov form factor to factorize the latter into soft, collinear, and hard
factors.

\subsection{DIS factorization at the lowest order: PDF and quark polarization}
\label{ssec:DIS-Feynman-lo-pdf}

\begin{figure}[htbp]
	\centering
	\begin{tabular}{cccc}
	\includegraphics[scale=0.65]{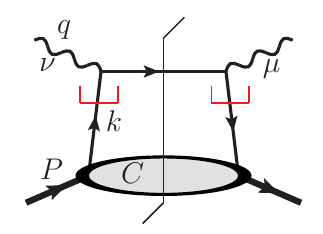} & 
	\includegraphics[scale=0.65]{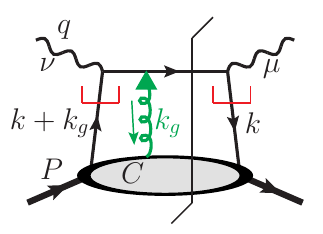} &
	\includegraphics[scale=0.65]{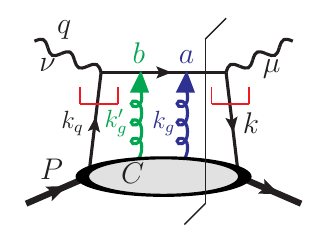} &
	\includegraphics[scale=0.65]{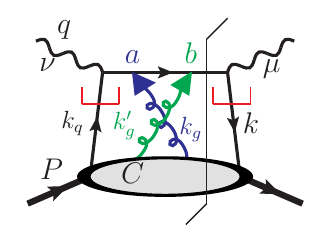} \\
	(a) & (b) & (c) & (d)
	\end{tabular}
	\caption{Leading-power leading-order sample DIS diagrams, initiated by a quark parton. 
	The quark parton can be accompanied by an arbitrary number of gluons with longitudinal polarizations.
	We take the convention that all the gluon momenta flow into the collinear subgraph.
	}
	\label{fig:DIS-LO-g}
\end{figure}

We first examine the LO DIS diagram $\Gamma_0$, given in \fig{fig:DIS-LO-g}(a).
It has no difference from the diagram in the light-cone gauge discussed in \sec{ssec:DIS-light-cone-ex}, 
with one single leading region $R$, 
and the same approximation leads to a factorization formula similar to \eq{eq:DIS-LO-fact}.
As a useful startup, in this section we continue the discussion in \eq{eq:DIS-LO-fact} to disentangle the quark spinor indices.

Note that in \fig{fig:DIS-LO-g}(a), the collinear subgraph $C$ includes all possible diagrams, so can be written as a Green function,
\beq[eq:C0-green]
	C_{\alpha\beta}(P, k) = \int d^4y\, e^{-i k \cdot y} 
		\vv{ P | \bar{\psi}_{\beta, j}(y) \psi_{\alpha, i}(0) | P },
\eeq
where $\alpha$ and $\beta$ are spinor indices and $i$ and $j$ are color indices.
The normal ordering of the fermion fields, instead of time ordering, is fixed because we are dealing with
a cut diagram. 
With the integration of $k^-$ and $\bm{k}_T$ in \eq{eq:DIS-LO-fact}, the coordinate separations $y^+$ and $\bm{y}_T$
are set to zero, so the collinear factor becomes
\beq[eq:F0-ab]
	F^0_{\alpha\beta}(P, k^+) 
	= \int \frac{dk^- d^2 \bm{k}_T}{(2\pi)^4} C_{\alpha\beta}(P, k) 
	= \int \frac{dy^-}{2\pi}\, e^{-i k^+ \, y^-} 
		\vv{ P | \bar{\psi}_{\beta, j}(0^+, y^-, \bm{0}_T) \psi_{\alpha, i}(0) | P }.
\eeq
Throughout this section, we use the superscript to refer to the number of collinear gluons attaching $C$ to $H$.
The spinor matrix in \eq{eq:F0-ab} can be expanded on the complete Dirac basis,
\beq[eq:f-a-b]
	F^0_{\alpha\beta}
	= S \, \mathds{1}_{\alpha\beta} + \wt{S} \pp{ \gamma_5}_{\alpha\beta} 
		+ V_{\mu} \pp{ \gamma^{\mu} }_{\alpha\beta} 
		+ A_{\mu} \pp{ \gamma_5 \gamma^{\mu} }_{\alpha\beta} 
		+ \frac{1}{2} T_{\mu\nu} \pp{ \sigma^{\mu\nu} }_{\alpha\beta}.
\eeq
Being sandwiched between $\P_n $ and $\Pb_n$, only three of the spinor structures are nonzero, 
\beq
	\P_n  (F^0_{\alpha\beta}) \Pb_n
	= V^+ \gamma^-  +  A^+ \gamma_5 \gamma^- - \sum_{i = 1, 2} T^{+i} \sigma^{-i}.
\eeq
The coefficients can be projected from \eq{eq:f-a-b} by tracing with certain gamma matrices.
This defines the three parton distribution functions (PDFs),
\bse\label{eq:PDFs-no-W}\begin{align}
	f^0(x) &= \Tr\bb{ \frac{\gamma^+}{2} F^0 }
		&&= \int \frac{dy^-}{4\pi}\, e^{-i k^+ \, y^-} 
		\vv{ P | \bar{\psi}_{j}(0^+, y^-, \bm{0}_T) \gamma^+ \psi_{i}(0) | P }, \\
	\Delta f^0(x) &= \Tr\bb{ \frac{\gamma^+\gamma_5}{2} F^0 }
		&&= \int \frac{dy^-}{4\pi}\, e^{-i k^+ \, y^-} 
		\vv{ P | \bar{\psi}_{j}(0^+, y^-, \bm{0}_T) \gamma^+\gamma_5 \psi_{i}(0) | P }, 	\\
	\delta_T^i f^0(x) &= \Tr\bb{ \frac{\gamma^+\gamma^i\gamma_5}{2} F^0 }
		&&= \int \frac{dy^-}{4\pi}\, e^{-i k^+ \, y^-} 
		\vv{ P | \bar{\psi}_{j}(0^+, y^-, \bm{0}_T) \gamma^+\gamma^i\gamma_5 \psi_{i}(0) | P },
\end{align}\ese
called unpolarized, polarized, and transversity PDFs, respectively,
which are dimensionless and boost invariant so only depends on $x = k^+ / P^+$.
The collinear factor in \eq{eq:DIS-LO-fact} then has the form 
\beq[eq:PDF-spinor-decomp]
	\bb{ \P_n  F^0(P, k) \Pb_n}_{\alpha\beta}
	= f^0(x) \, \bb{ \frac{\gamma^-}{2} }_{\alpha\beta} 
		+ \Delta f^0(x) \, \bb{ \frac{\gamma_5\gamma^-}{2} }_{\alpha\beta}
		+ \delta_T^i f^0(x) \, \bb{ \frac{\gamma^- \gamma^i \gamma_5}{2} }_{\alpha\beta}.
\eeq
Contracting with the hard part then converts \eq{eq:DIS-LO-fact} to the factorization formula in terms of
the PDFs,
\begin{align}
	C_R \Gamma^0
	& = \int \frac{d k^+}{k^+} H^{\mu\nu}_{0, \beta\alpha}(\hat{k}) 
		\bb{
			f^0(x) \, \frac{k^+\gamma^-}{2}
		+ \Delta f^0(x) \, \frac{\gamma_5\gamma^-k^+}{2}
		+ \delta_T^i f^0(x) \, \frac{k^+\gamma^- \gamma^i \gamma_5}{2}
		}_{\alpha\beta}	\nn\\
	& = \int \frac{d x}{x} f^0(x) \Tr\bb{ H^{\mu\nu}_0(\hat{k}) \, \frac{\slash{\hat{k}}}{2} 
			\big( 1 - \lambda^0(x) \gamma_5 
				- b_T^{0,i}(x) \gamma_5 \gamma^i  
			\big) },
\label{eq:DIS-LO-factorize-pdf}
\end{align}
where $\slash{\hat{k}} = k^+ \gamma^-$, and 
$( \lambda^0(x), b_T^{0,i}(x) ) = \big( \Delta f^0(x), \delta_T^{i} f^0(x) \big) / f^0(x)$
quantify the quark polarization state. 

The uncontracted hard part $H_0(\hat{k})$ can be explicitly obtained from 
\fig{fig:DIS-LO-g}(a),
\beq[eq:DIS-H-LO]
	H^{\mu\nu}_0(\hat{k})
	= \gamma^{\mu} \delta^+\big( (\hat{k} + q)^2 \big) (2\pi) \gamma \cdot (\hat{k} + q) \gamma^{\nu}
	= (2\pi) \delta(1 - x / x_B) \bb{ \gamma^{\mu} \gamma \cdot (\hat{k} + q) \gamma^{\nu} }
\eeq
where $q = (-x_B P^+, Q^2 / 2 x_B P^+, 0_T)$ is the photon momentum.
This is trivially color diagonal, so its contraction with \eq{eq:PDF-spinor-decomp} in color space,
which has been implicitly left open and entangled,
can be reduced into a $\delta_{ij}$ contracting with \eq{eq:PDF-spinor-decomp},
giving an unweighted sum of the color indices in \eq{eq:PDFs-no-W}.
Even so, the mere color sum does not make \eq{eq:PDFs-no-W} gauge invariant because the fermion 
fields are at different positions. As we will see later, the inclusion of longitudinally polarized gluons make
render the PDFs gauge invariant.

Tracing \eq{eq:DIS-H-LO} with the given Dirac matrices in \eq{eq:DIS-LO-factorize-pdf}, we have
\bse\begin{align}
	C^{\mu\nu}
	& = \Tr\bb{ H^{\mu\nu}_0(\hat{k}) \frac{\slash{\hat{k}}}{2} }	
	= 4 \pp{ \hat{k}^{\mu} - \frac{\hat{k} \cdot q}{q^2} q^{\mu} } 
		\pp{\hat{k}^{\nu} - \frac{\hat{k} \cdot q}{q^2} q^{\nu} }
		+ \pp{ q^2 g^{\mu\nu} - q^{\mu} q^{\nu} },	\\
	\Delta C^{\mu\nu}
	& = \Tr\bb{ H^{\mu\nu}_0(\hat{k}) \frac{\gamma_5\slash{\hat{k}}}{2} }	
	 = - 2 i \, \epsilon^{\mu\nu\alpha\beta} \hat{k}_{\alpha} q_{\beta},	\\
	\delta_T^i C^{\mu\nu}
	& = \Tr\bb{ H^{\mu\nu}_0(\hat{k}) \frac{\slash{\hat{k}}\gamma^i\gamma_5}{2} }	
	 = 0,
\end{align}\ese
with an overall coefficient $(2\pi) \delta(1 - x / x_B) \delta_{ij}$ omitted.

\subsection{Low-order DIS factorization example in Feynman gauge: one extra collinear gluon}
\label{ssec:DIS-Feynman-lo-1g}

In Feynman gauge, even without loop effects, the LO diagram in \fig{fig:DIS-LO-g}(a) can receive contribution
from diagrams like \fig{fig:DIS-LO-g}(b)--(d), where the quark enters the hard scattering together with 
an arbitrary number of gluons. 
To the leading power, only the plus components of these gluons' polarizations contribute, which we have 
termed longitudinal polarization.
Following \eq{eq:g-approx-modify}, we include in the definition of the approximator $\hat{T}$ that the contraction of 
a longitudinally polarized gluon of momentum $k_g$ to the hard part to the left of the cut is to be approximated as
\beq[eq:DIS-g-approx]
	H_{\rho}(k_g) \, g^{\rho\sigma} \, C_{\sigma}(k_g)
	\to 
	H_{\rho}(\hat{k}_g) \, \frac{\hat{k}_g^{\rho} n^{\sigma}}{k_g \cdot n + i \epsilon} C_{\sigma}(k_g),
\eeq
where we only left explicit the dependence on the gluon variables,
$k_g$ is taken to flow into the collinear subgraph $C$ and $\hat{k}_g$ is defined in \eq{eq:DIS-col-approx-lc}. 
To the right of the cut, we take gluon momenta to flow out of $C$ and reverse the $i \epsilon$ signs.
The sign of the $i \epsilon$ is in fact not crucial in DIS since there is only one collinear subgraph.
We fix it to be the same as the SIDIS.

Apparently, after applying $\hat{T}$, the structure $H(\hat{k}_g) \cdot \hat{k}_g$ allows to use Ward identity.
In the diagram \fig{fig:DIS-LO-g}(b), denoted as $\Gamma_1$, this gives 
\begin{align}
	H^{\mu\nu}_1(\hat{k}_g) \cdot \hat{k}_g
	& = \gamma^{\mu} \, (2\pi) \delta^+\pp{ (\hat{k} + q)^2 } \gamma\cdot (\hat{k} + q) \,
		( -ig t^a \gamma \cdot \hat{k}_g ) 
		\frac{i}{ \gamma \cdot (\hat{k} + \hat{k}_g + q) + i \epsilon}
		\gamma^{\nu},
\label{eq:H_1*k}
\end{align}
where $t^a = (t^a_{ji})$ is the SU(3) generator and carries the color dependence, with $i$ and $j$ corresponding
to the color indices in \eq{eq:PDFs-no-W}.
Using the familiar identity,
\begin{align}
	&\gamma\cdot (\hat{k} + q) \,(\gamma \cdot \hat{k}_g) \, \frac{1}{ \gamma \cdot (\hat{k} + \hat{k}_g + q) + i \epsilon}
	\nn\\
	& \hspace{2em}
	= \gamma\cdot (\hat{k} + q) \,
		\bb{ \gamma \cdot (\hat{k} + \hat{k}_g + q) - \gamma \cdot (\hat{k} + q) }
		\, \frac{1}{ \gamma \cdot (\hat{k} + \hat{k}_g + q) + i \epsilon}	\nn\\
	& \hspace{2em}
	= \gamma\cdot (\hat{k} + q) - (\hat{k} + q)^2 \, \frac{1}{ \gamma \cdot (\hat{k} + \hat{k}_g + q) + i \epsilon},
\label{eq:line-id-ex}
\end{align}
the second term vanishes due to the on-shell condition set by the cut line (or the $\delta$-function in \eq{eq:H_1*k}),
\eq{eq:H_1*k} becomes
\beq[eq:H*k-g1-ward]
	H^{\mu\nu}_1(\hat{k}_g) \cdot \hat{k}_g
	= H^{\mu\nu}_0(\hat{k}) \times  \bb{ i \, ( -ig t^a) },
\eeq
which differs from \eq{eq:DIS-H-LO} only by an extra color factor.
Combining with the remaining factors in \eq{eq:DIS-g-approx}, the first $i$ factor that comes from the numerator of 
the cancelled quark propagator combines with $1 / (k_g \cdot n + i \epsilon)$ to form a new eikonal propagator, 
and the color factor $( -ig t^a)$ detaches from the fermion line and forms a new eikonal vertex with $n^{\sigma}$.
So we have
\beq[eq:CR-G1]
	C_R \Gamma_1 
	= \int d k^+ H^{\mu\nu}_0(\hat{k}) \cdot
		\bb{
		\int \frac{dk^- d^2\bm{k}_T}{(2\pi)^4} \frac{d^4 k_g}{(2\pi)^4}
		\pp{ \frac{i}{k_g \cdot n + i \epsilon} \, ( -i g n^{\sigma} t^a) }
		\P_n  \, C^a_{\sigma}(P; k, k_g)\, \Pb_n
		},
\eeq
where both spinor and color indices are traced over, and the $a$ and $\sigma$ refer to the color and Lorentz
indices of the gluon, respectively. Since the dependence of $H$ on the gluon has been detached by Ward identity,
we include the integration of $k_g$ in the collinear part, together with the eikonal factors.

The collinear subgraph $C_{\sigma}(P; k, k_g)$ in \fig{fig:DIS-LO-g}(b) contains all possible disgrams
so can be converted to a Green function similar to \eq{eq:C0-green},
\beq
	C^a_{\sigma}(P; k, k_g)
	= \int d^4y \,d^4y_1\, e^{-i k \cdot y - i k_g \cdot y_1} 
		\vv{ P | \bar{\psi}_{\beta, j}(y) \T\bb{ A^a_{\sigma}(y_1) \psi_{\alpha, i}(0) }| P }
\eeq
where we suppressed the spinor indices of $C$, and the time ordering $\T$ is added because both the
quark and gluon are to the left of the cut in the amplitude.
Now the color factor in \eq{eq:CR-G1} combines with the gluon field, with the colors summed over.
By converting the eikonal propagator to
\beq[eq:eikonal-int]
	\frac{i}{k_g \cdot n + i \epsilon}
	 = \int_0^{\infty} d\lambda \, e^{ i \lambda (k_g \cdot n + i \epsilon) },
\eeq
the $k_g$ integration can be directly carried out and sets $y_1 = \lambda n$,
\begin{align}
	&\int \frac{d^4 k_g}{(2\pi)^4} 
		\pp{ \frac{i}{k_g \cdot n + i \epsilon} \, ( -i g n^{\sigma} t^a) }
		C^a_{\sigma}(P; k, k_g)		\nn\\
	& \hspace{1em}
	= \int_0^{\infty} d\lambda \int d^4y \,d^4y_1\, 
		e^{-i k \cdot y} 
		\int \frac{d^4 k_g}{(2\pi)^4} e^{- i k_g \cdot (y_1 - \lambda n)}
		\vv{ P | \bar{\psi}_{\beta}(y) \T\bb{-i g \, n \cdot A^a(y_1) \, t^a \,\psi_{\alpha}(0) }| P }	\nn\\
	& \hspace{1em}
	= \int_0^{\infty} d\lambda \int d^4y \, 
		e^{-i k \cdot y} 
		\vv{ P | \bar{\psi}_{\beta}(y) \bb{-i g \, n \cdot A^a(\lambda n) \, t^a \,\psi_{\alpha}(0) }| P },
\label{eq:eikonal-1g-n}
\end{align}
where the time ordering has been trivially removed since $\lambda > 0$. 
Further integrating out $k^-$ and $\bm{k}_T$ also sets $y$ on the light cone, and gives the collinear factor,
\begin{align}
	F^1_{\alpha\beta}
	= \int \frac{dy^-}{2\pi} e^{-i k^+ y^-} 
		\vv{ P | \bar{\psi}_{\beta}(y^-) \bb{ \int_0^{\infty} d\lambda \pp{-i g \, n \cdot A^a(\lambda n) \, t^a} }
			\psi_{\alpha}(0)| P }.
\label{eq:eikonal-1g-n-k-int}
\end{align}
The spinor matrix reduction is to be done in the same way as \eq{eq:PDF-spinor-decomp} after projecting with 
$\P_n $ and $\Pb_n$ so will not be repeated.
Since \eqs{eq:DIS-LO-factorize-pdf}{eq:CR-G1} have the same hard coefficient, we can combine them to define
a whole collinear factor,
\begin{align}
	(F^0 + F^1)_{\alpha\beta}
	= \int \frac{dy^-}{2\pi} e^{-i k^+ y^-} 
		\vv{ P | \bar{\psi}_{\beta}(y^-) \bb{ 1 -i g \int_0^{\infty} d\lambda \, n \cdot A^a(\lambda n) \, t^a }
			\psi_{\alpha}(0) | P }.
\label{eq:F0+F1}
\end{align}

\begin{figure}[htbp]
	\centering
	\begin{align*}
		\eqfig[0.65]{figures/DIS-LO-1g.pdf} + {\rm c.c.}
		\quad = \quad
		\eqfig[0.65]{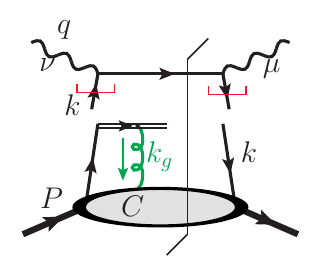} +
		\eqfig[0.65]{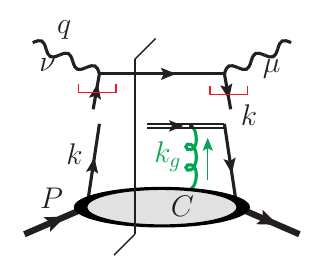}
	\end{align*}
	\caption{Results of applying Ward identity on \fig{fig:DIS-LO-g}(b) and its complex conjugate.
	The gluon momentum $k_g$ flows along the double line, which has the eikonal propagator the vertex as given in
	\eq{eq:CR-G1}.
	}
	\label{fig:DIS-LO-1g-w}
\end{figure}

Of course, at this order, one can also have the gluon to the right of the cut, which gives an eikonal line associated with
the field $\bar{\psi}$ at $y^-$. The result of the eikonal propagators and vertices can be represented graphically by the
double lines in \fig{fig:DIS-LO-1g-w}. The arrows on the double lines indicate the color flows in the fundamental 
representation, which is dictated by the fields $\psi$ and $\bar{\psi}$ on the ends.

\subsection{Low-order DIS factorization example in Feynman gauge: two extra collinear gluons}
\label{ssec:DIS-Feynman-lo-2g}

Now we consider the two diagrams in \fig{fig:DIS-LO-g}(c) and (d), which have two extra longitudinally polarized
gluons attaching the collinear subgraph to the hard part. The same approximation as \eq{eq:DIS-g-approx} applies
to both gluons so allows the use of Ward identity. The corresponding hard part is
\begin{align}
	&H^{2, ab}_{\rho\sigma} \, \hat{k}_g^{\rho} \, \hat{k}_g^{\prime\sigma}	
	= \gamma^{\mu}\, (2\pi) \delta^+\pp{ (\hat{k} + q)^2 } \gamma\cdot (\hat{k} + q) \,	\nn\\
	& \hspace{3em} \times
		\bb{ (-ig t^a \slash{\hat{k}}_g ) \,\frac{i}{\gamma\cdot (\hat{k} + \hat{k}_g + q) + i \epsilon}
			\, (-ig t^b \slash{\hat{k}}'_g ) \, \frac{i}{\gamma\cdot (\hat{k} + \hat{k}_g + \hat{k}'_g + q) + i \epsilon}
			\right.\nn\\
	& \hspace{4em} \left. + 
			(-ig t^b \slash{\hat{k}}'_g ) \, \frac{i}{\gamma\cdot (\hat{k} + \hat{k}'_g + q) + i \epsilon}
			\, (-ig t^a \slash{\hat{k}}_g ) \, \frac{i}{\gamma\cdot (\hat{k} + \hat{k}_g + \hat{k}'_g + q) + i \epsilon}
		} \gamma^{\nu},
\label{eq:DIS-LO-2g}
\end{align}
where the two terms are for \fig{fig:DIS-LO-g}(c) and (d), respectively,
$a$ and $b$ are the color indices of the gluons of momenta $k_g$ and $k'_g$, respectively, and both momenta
are taken to flow into the collinear subgraph $C$. 
We first perform Ward identity for the gluon $k_g$ and use the identities
\bse\label{eq:kg-feyn-id}\begin{align}
	\slash{\hat{k}}_g &= \gamma\cdot (\hat{k} + \hat{k}_g + q) - \gamma\cdot (\hat{k} + q),	\label{eq:kg-feyn-id1} \\
	&= \gamma\cdot (\hat{k} + \hat{k}_g + \hat{k}'_g + q) - \gamma\cdot (\hat{k} + \hat{k}'_g + q), \label{eq:kg-feyn-id2}
\end{align}\ese
for the two terms in \eq{eq:DIS-LO-2g}, respectively.
The cut line similarly sets the second term in \eq{eq:kg-feyn-id1} to zero, so we have
\beq[eq:ward-kg-1]
	(-ig)^2 (t^a t^b) i \, \slash{\hat{k}}'_g \, \frac{i}{\gamma\cdot (\hat{k} + \hat{k}_g + \hat{k}'_g + q) + i \epsilon}
\eeq
for the first term, and
\beq[eq:ward-kg-2]
	(-ig)^2 (t^b t^a) i \, \slash{\hat{k}}'_g \,
	\bb{ \frac{i}{\gamma\cdot (\hat{k} + \hat{k}'_g + q) + i \epsilon} 
		- \frac{i}{\gamma\cdot (\hat{k} + \hat{k}_g + \hat{k}'_g + q) + i \epsilon}
	}
\eeq
for the second term. Now \eq{eq:ward-kg-1} combines with the second term in \eq{eq:ward-kg-2} into
\beq[eq:ward-kg-1-22]
	(-ig)^2 [t^a, t^b] i \, \slash{\hat{k}}'_g \, \frac{i}{\gamma\cdot (\hat{k} + \hat{k}_g + \hat{k}'_g + q) + i \epsilon},
\eeq
while the first term in \eq{eq:ward-kg-2} allows a further use of Ward identity by noticing
$\slash{\hat{k}}'_g = \gamma\cdot (\hat{k} + \hat{k}'_g + q) - \gamma\cdot (\hat{k} + q)$, 
and gives
\beq[eq:ward-kg-21]
	(-ig)^2 (t^b t^a) i^2.
\eeq
Combining \eqs{eq:ward-kg-1-22}{eq:ward-kg-21} and inserting back to \eq{eq:DIS-LO-2g} gives
\begin{align}
	&H^{2, ab}_{\rho\sigma} \, \hat{k}_g^{\rho} \, \hat{k}_g^{\prime\sigma}	
	= \gamma^{\mu}\, (2\pi) \delta^+\pp{ (\hat{k} + q)^2 } \gamma\cdot (\hat{k} + q) \,	\nn\\
	& \hspace{6em} \times (-ig)^2  i^2
		\bb{  [t^a, t^b] \, \slash{\hat{k}}'_g \, \frac{1}{\gamma\cdot (\hat{k} + \hat{k}_g + \hat{k}'_g + q) + i \epsilon}
			+ t^b t^a
		} \gamma^{\nu}.
\label{eq:DIS-LO-2g-ward}
\end{align}

\begin{figure}[htbp]
	\centering
	\begin{align*}
		\eqfig[0.68]{figures/DIS-LO-2g1.pdf} + 
		\eqfig[0.68]{figures/DIS-LO-2g2.pdf}
		=
		\eqfig[0.68]{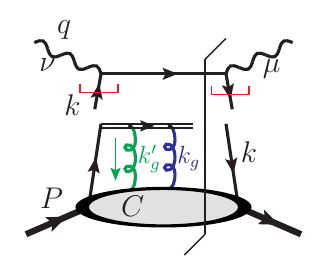} +
		\eqfig[0.68]{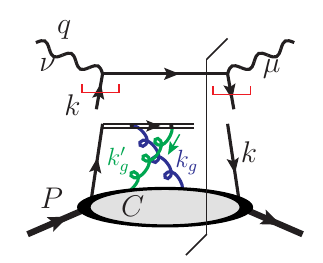}
	\end{align*}
	\caption{Results of applying Ward identity on \fig{fig:DIS-LO-g}(c) and (d).
	Both momenta $k_g$ and $k_g'$ flow into $C$ and the momentum flows on the double line satisfy momentum conservation.
	}
	\label{fig:DIS-LO-2g-w}
\end{figure}

In an Abelian gauge theory, the commutator $[t^a, t^b]$ vanishes, and \eq{eq:DIS-LO-2g-ward} already detaches 
the two gluons out of the hard part. In the non-Abelian gauge theory, $[t^a, t^b] = i f^{abc} t^c$ relates this term 
to tri-gluon coupling, which we will further explore in \sec{ssec:line-id}. 
Effectively, the use of Ward identity for the gluon of $k_g$ detaches it from the hard part and attaches it to the gluon
of $k_g'$, but the Ward-identity vertex is still $\slash{\hat{k}}'_g$, which cannot relate the two neighboring propagators
of momenta $(\hat{k} + q)$ and $(\hat{k} + \hat{k}_g + \hat{k}'_g+ q)$ by any identity similar to \eq{eq:kg-feyn-id},
so Ward identity cannot be simply applied.
The way out is to notice that $\hat{k}_g'$ has been projected on shell, so we rewrite it as
\beq
	\slash{\hat{k}}'_g = \gamma^- k_g^{\prime +}
		= \gamma^- (k_g + k_g^{\prime} )^+ \frac{k_g^{\prime +}}{(k_g + k_g^{\prime} )^+ + i \epsilon}
		= \gamma\cdot (\hat{k}_g + \hat{k}_g^{\prime} ) \bb{ \frac{k_g' \cdot n}{n \cdot (k_g + k_g^{\prime} ) + i \epsilon} }.
\eeq
The first factor allows the use of Ward identity, whereas the second factor modifies the eikonal identity associated with $k_g'$.
Then we can proceed with \eq{eq:DIS-LO-2g-ward} as
\begin{align}
	&H^{2, ab}_{\rho\sigma} \, \hat{k}_g^{\rho} \, \hat{k}_g^{\prime\sigma}	
	= H^{\mu\nu}_0(\hat{k}) \cdot (-ig)^2  i^2
		\bb{  [t^a, t^b] \, \frac{k_g' \cdot n}{n \cdot (k_g + k_g^{\prime} ) + i \epsilon}
			+ t^b t^a
		},
\label{eq:DIS-LO-2g-ward-2}
\end{align}
which again factorizes the dependence on the collinear gluons out of $H$ and gives the same hard part factor as 
\eqs{eq:DIS-H-LO}{eq:H*k-g1-ward}.
Combined with the remaining approximator factors in \eq{eq:DIS-g-approx} for both gluons, we have the overall eikonal factor as
\begin{align}
	(-ig n^{\rho}) (-ig n^{\sigma}) 
	\bb{ [t^a, t^b] \,  \frac{i}{n \cdot k_g + i \epsilon} \, \frac{i}{n \cdot (k_g + k_g^{\prime} ) + i \epsilon}
		+ t^b t^a \, \frac{i}{n \cdot k_g + i \epsilon} \,  \frac{i}{n \cdot k_g' + i \epsilon} 
	}.
\label{eq:eikonal-2g}
\end{align}
This can be further simplified by writing $[t^a, t^b] = t^a t^b - t^b t^a$, and using the simple eikonal identity,
\beq
	\frac{i}{n \cdot k_g + i \epsilon} \, \frac{i}{n \cdot k_g' + i \epsilon}
	=  \frac{i}{n \cdot k_g + i \epsilon} \, \frac{i}{n \cdot (k_g + k_g^{\prime} ) + i \epsilon}
		+ \frac{i}{n \cdot k_g' + i \epsilon} \, \frac{i}{n \cdot (k_g + k_g^{\prime} ) + i \epsilon},
\eeq
which converts \eq{eq:eikonal-2g} into a more symmetric form,
\begin{align}
	(-ig n^{\rho}) (-ig n^{\sigma}) 
	\bb{ t^a t^b \, \frac{i}{n \cdot k_g + i \epsilon} \, \frac{i}{n \cdot (k_g + k_g^{\prime} ) + i \epsilon}
		+ t^b t^a \, \frac{i}{n \cdot k_g' + i \epsilon} \, \frac{i}{n \cdot (k_g + k_g^{\prime} ) + i \epsilon}
	}.
\label{eq:eikonal-2g-sym}
\end{align}
This is graphically represented by \fig{fig:DIS-LO-2g-w}, where each double line has two propagators corresponding
to the two eikonal factors in each term in \eq{eq:eikonal-2g-sym}.

\eq{eq:eikonal-2g-sym} is to be multiplied on the collinear subgraph $C_{\rho\sigma}(P; k, k_g, k_g')$
and defines the collinear factor. But here for fixed momentum assignments of $k_g$ and $k_g'$, we do not
include diagrams in $C$ that differ only by exchanging the two gluons. Otherwise, one would have double counting
in \fig{fig:DIS-LO-g}(c) and (d).
After the use of Ward identity, the integrals of $k_g$ and $k_g'$ are completely within the collinear factor, so we
can reverse the color and momentum labels in the second term of \eq{eq:eikonal-2g-sym} (or the second diagram
on the right-hand side of \fig{fig:DIS-LO-2g-w}). This reduces \eq{eq:eikonal-2g-sym} to only the first term, but
now with the subgraph $C$ to include all possible diagrams. 
Then we can write it as a Green function,
\beq[eq:C-2g-green]
	C^{ab}_{\rho\sigma}(P; k, k_g, k_g')
	= \int d^4y \,d^4y_1\,d^4y_2\, e^{-i k \cdot y - i k_g \cdot y_1 - i k_g' \cdot y_2} 
		\vv{ P | \bar{\psi}_{\beta, j}(y) \T\bb{ A^a_{\rho}(y_1) A^b_{\sigma}(y_2) \psi_{\alpha, i}(0) }| P }
\eeq

The first term of \eq{eq:eikonal-2g-sym} can be converted to integrals by applying \eq{eq:eikonal-int} to both propagators,
\beq[eq:eikonal-int-2]
	\frac{i}{n \cdot k_g + i \epsilon} \, \frac{i}{n \cdot (k_g + k_g^{\prime} ) + i \epsilon}
	= \int_0^{\infty} d \lambda_1 \int_0^{\infty} d \lambda_2 \, \theta(\lambda_1 - \lambda_2) \,
		e^{i \lambda_1 n \cdot k_g} e^{i \lambda_2 n \cdot k_g'}.
\eeq
Combining this with \eq{eq:C-2g-green} and integrating over $k_g$ and $k_g'$ set $y_1 = \lambda_1 n$ and $y_2 = \lambda_2 n$.
Further integrating over $k^-$ and $\bm{k}_T$ also sets $y$ on the light cone. 
Finally, we have the form of the collinear factor,
\begin{align}
	F^2_{\alpha\beta}
	=& \int \frac{dk^- d^2\bm{k}_T}{(2\pi)^4} \, \frac{d^4k_g}{(2\pi)^4} \, \frac{d^4k_g'}{(2\pi)^4} \,
	(-ig n^{\rho} t^a) (-ig n^{\sigma} t^b) 	\nn\\
	& \hspace{3em} \times
		\bb{ \frac{i}{n \cdot k_g + i \epsilon} \cdot \frac{i}{n \cdot (k_g + k_g^{\prime} ) + i \epsilon} }
		C^{ab}_{\rho\sigma}(P; k, k_g, k_g')	\nn\\
	=& \int_0^{\infty} d \lambda_1 \int_0^{\infty} d \lambda_2 \, \theta(\lambda_1 - \lambda_2) \,
		\int \frac{dy^-}{2\pi} e^{-i k^+ y^-} \,	\nn\\
	& \hspace{3em} \times
		\vv{ P | \bar{\psi}_{\beta, j}(y^-) 
			\bb{ -ig \, n \cdot A^a(\lambda_1 n) t^a } \bb{ -ig \, n \cdot A^b(\lambda_2 n) t^b) } \psi_{\alpha, i}(0) | P },
\end{align}
where we have removed the time ordering as in \eq{eq:eikonal-1g-n}.
Note that both gluon fields are integrated along the light cone, and their order cannot be reversed due to the non-Abelian nature.
The $\theta(\lambda_1 - \lambda_2)$ function dictates the gluon field to the left to be also in front in the light-cone path along $n$,
so is equivalent to a path ordering, which is defined as
\begin{align}
	&\P \int_0^{\infty} d \lambda_1 \int_0^{\infty} d \lambda_2 \, O_1(\lambda_1) O_2(\lambda_2)	\nn\\
	&\hspace{2em}
	= \int_0^{\infty} d \lambda_1 \int_0^{\infty} d \lambda_2 
		\bb{ \theta(\lambda_1 - \lambda_2) O_1(\lambda_1) O_2(\lambda_2) 
			+ \theta(\lambda_2 - \lambda_1) O_2(\lambda_2) O_1(\lambda_1)},
\label{eq:path-ordering}
\end{align}
where $\P$ refers to path ordering operator.
Therefore, we write the collinear factor as
\beq[eq:F2-w]
	F^2_{\alpha\beta}
	= \frac{1}{2} \int \frac{dy^-}{2\pi} e^{-i k^+ y^-} \vv{ P | \bar{\psi}_{\beta, j}(y^-) 
		\cc{ \P \bb{ -i g \int_0^{\infty} d \lambda\, n \cdot A^a(\lambda n) t^a }^2 }_{ji} \psi_{\alpha, i}(0) | P }.
\eeq

\eq{eq:F2-w} combines with \eq{eq:F0+F1} to give
\begin{align}
	(F^0 + F^1 + F^2)_{\alpha\beta}
	= \int \frac{dy^-}{2\pi} e^{-i k^+ y^-} 
		\vv{ P | \bar{\psi}_{\beta}(y^-) \sum_{n = 0}^2 \frac{1}{n!} \P
			\bb{ -i g \int_0^{\infty} d\lambda \, n \cdot A^a(\lambda n) \, t^a }^n
			\psi_{\alpha}(0) | P }.
\label{eq:F0+F1+F2}
\end{align}
The same analysis can be done for the diagrams with the two gluons to the right of the cut, or with one gluon on both sides of the cut.
The structure of the result in \eq{eq:F0+F1+F2} easily motivates one to conjecture it to all orders,
\beq[eq:F-a-b-W-all-orders]
	F_{\alpha\beta}
	= \int \frac{dy^-}{2\pi} e^{-i k^+ y^-} 
		\vv{ P | \bar{\psi}_{\beta}(y^-) W^{\dag}(\infty, y^-; n) \,
			W(\infty, 0; n) \psi_{\alpha}(0) | P },
\eeq
with the colors summed over. 
Here $W(\infty, y; n) = (W_{ij})(\infty, y; n)$ is the straight Wilson line from $y$ to $\infty$ 
along the light-cone direction $n$,
\beq[eq:DIS-wilson-line]
	W(\infty, y; n)
	= \P \exp{ -i g \int_0^{\infty} d\lambda \, n \cdot A^a(y + \lambda n) \, t^a }.
\eeq
It points to the future because of the $i\epsilon$ choice in the approximator \eq{eq:DIS-g-approx}, as easily seen from \eq{eq:eikonal-int}.
The Wilson lines associated with the fermion fields render the parton distribution gauge invariant. Using the unitarity property of the 
Wilson line allows us to join the two infinitely long Wilson lines to convert to a finite one,
\beq
	F_{\alpha\beta}
	= \int \frac{dy^-}{2\pi} e^{-i k^+ y^-} 
		\vv{ P | \bar{\psi}_{\beta}(y^-) W(y^-, 0; n) \psi_{\alpha}(0) | P },
\eeq
with the gauge invariance being obvious. 
We will show the all-order derivation in the following.

\subsection{Elements of Ward identity: perturbative line identities}
\label{ssec:line-id}

\begin{figure}[htbp]
	\centering
	\begin{align*}
		\eqfig[0.55]{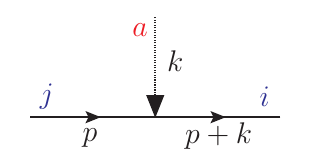}
		=
		\eqfig[0.55]{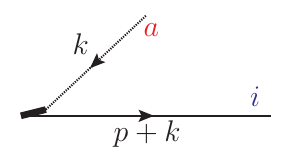} +
		\eqfig[0.55]{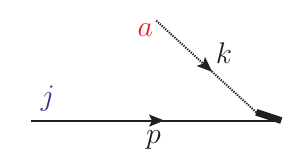}
	\end{align*}
	\caption{Ward identity element for the quark-gluon vertex. The dashed lines refer to ghosts, but the
	vertex on the left-hand side is a regular quark-gluon vertex, with the arrow on the end of the ghost line being
	the ghost momentum $k$ contracted with the vertex.
	}
	\label{fig:ward-q}
\end{figure}

As we have seen from the low-order examples in Secs.~\ref{ssec:DIS-Feynman-lo-1g} and \ref{ssec:DIS-Feynman-lo-2g}, 
Ward identity is mainly using the simple identity like \eqs{eq:line-id-ex}{eq:kg-feyn-id} and the successive cancellation.
For the quark process we have examined, the relevant line identity is shown in \fig{fig:ward-q}, which reads
\begin{align}
	\frac{i}{\slash{p} + \slash{k}} \, \pp{ -ig t^a_{ij} \slash{k} } \, \frac{i}{\slash{p}}
	= i \bb{ \frac{i}{\slash{p} + \slash{k}} \pp{ +ig t^a_{ij} } + \pp{ -ig t^a_{ij} } \frac{i}{\slash{p}} }.
\end{align}
Apart from the $i$ factor, the identity contains two special vertices, denoted by the thick diagonal lines in \fig{fig:ward-q},
with $(+ig t^a_{ij})$ for the vertex corresponding to a field $\bar{\psi}$, 
and $(-ig t^a_{ij})$ for the one corresponding to $\psi$.
The reason for denoting the arrowed gluon line in \fig{fig:ward-q} as a ghost line is that the reduced vertices on the 
right-hand side are related to the BRST variations of the quark fields.

\begin{figure}[htbp]
	\centering
	\begin{align*}
		\eqfig[0.55]{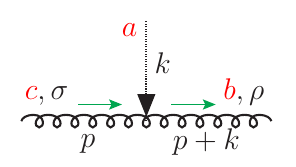}
		=
		\eqfig[0.55]{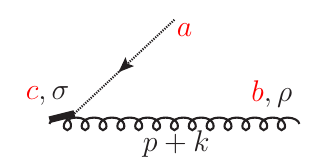} +
		\eqfig[0.55]{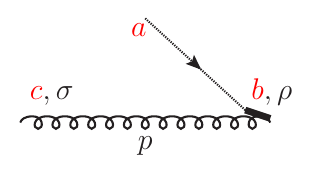} +
		\eqfig[0.55]{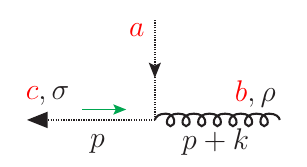} +
		\eqfig[0.55]{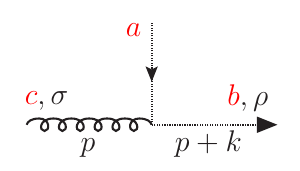}
	\end{align*}
	\caption{Ward identity element for the tri-gluon vertex.	
	}
	\label{fig:ward-g}
\end{figure}

Similarly, we can derive the Ward identity element for the tri-gluon vertex, as shown in \fig{fig:ward-g}, reading
\begin{align}
	& \frac{-i}{(p + k)^2} \, 
		\cc{ (-g f^{abc}) k_{\mu} 
			\bb{ (2k + p)^{\sigma} g^{\mu\rho} - (k + 2p)^{\mu} g^{\rho\sigma} + (p - k)^{\rho} g^{\mu\sigma} } 
		}
		\frac{-i}{p^2}	\nn\\
	& \hspace{1em}
	= i \bb{ \pp{ -g f^{acb} \, g^{\rho\sigma} } \cdot \frac{-i}{(p + k)^2}
		+ \pp{ -g f^{abc} \, g^{\rho\sigma} } \cdot \frac{-i}{p^2}	
		+ \bb{ -g f^{bca} \, (-p)^{\rho} } \cdot (-p)^{\sigma} \cdot \frac{i}{p^2} \frac{-i}{(p + k)^2}	\right.\nn\\
	& \hspace{6em} \left.
		+ \bb{ -g f^{cba} \, (p+k)^{\sigma} } \cdot (p+k)^{\rho} \cdot \frac{-i}{p^2} \frac{i}{(p + k)^2}
		},
\end{align}
term by term. We have written each term to make the special vertices clear. For the last two terms, a ghost line
with an arrow on the end is multiplied by a factor of its momentum; apart from this, it couples to other particles 
on this end in the same way as a gluon.
In this way, Ward identity relates gluon lines to ghost lines, as a special feature of non-Abelian gauge theory.
The resulted ghost lines with arrows allow iterative use of Ward identities until we reach the external legs of the
diagram.

\begin{figure}[htbp]
	\centering
	\begin{align*}
		\eqfig[0.55]{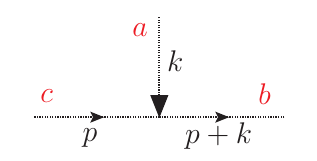}
		=
		\eqfig[0.55]{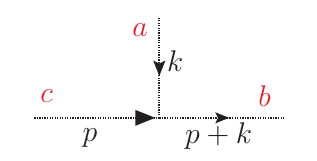} +
		\eqfig[0.55]{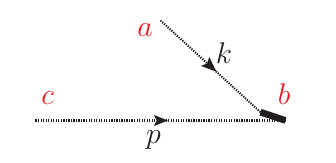}
	\end{align*}
	\caption{Ward identity element for the ghost-gluon vertex.
	}
	\label{fig:ward-ghost}
\end{figure}

A special case occurs for the gluon-ghost vertex. It is only proportional to the momentum of the outgoing ghost,
so does not have the form like a scalar QED vertex nor a regular vertex identity like \figs{fig:ward-q}{fig:ward-g}.
A yet useful identity is shown in \fig{fig:ward-ghost}, which reads
\beq
	\frac{i}{(p+k)^2} \bb{ ( -g f^{abc} ) k \cdot (p+k) } \frac{i}{p^2}
	= \frac{i}{(p+k)^2} \bb{ ( -g f^{cba} ) p \cdot (p+k) } \frac{i}{p^2}
		+ ( -g f^{abc} ) \frac{i}{p^2}.
\eeq
In a physical amplitude, ghosts only appear in closed loops (both clockwise and counterclockwise) with a minus sign, 
accompanied by the same gluon loops. When applying Ward identities to a gluon loop, the successive use of the last two terms
in \fig{fig:ward-g} converts the gluon loop to two ghost loops, each with the final end of the ghost line pointing back to itself.
Such ghost loops do not have a minus signs, and they cancel the regular ghost loops by use of the first term in \fig{fig:ward-ghost}.

\begin{figure}[htbp]
	\centering
	\begin{align*}
		\eqfig[0.55]{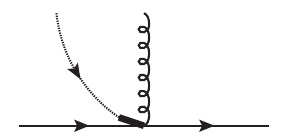} +
		\eqfig[0.55]{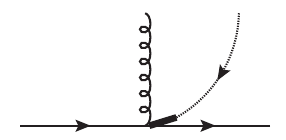} +
		\eqfig[0.55]{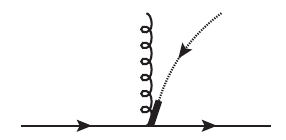}
		 = 0	
	\end{align*}	\\
	(a) \\
	\begin{align*}
		\eqfig[0.55]{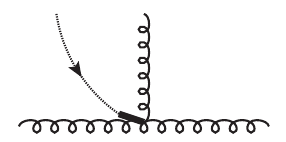} +
		\eqfig[0.55]{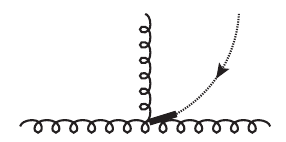} +
		\eqfig[0.55]{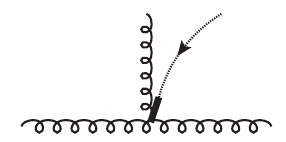} +
		\eqfig[0.55]{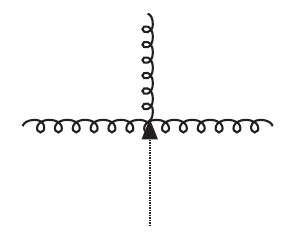}
		 = 0	
	\end{align*}	\\
	(b) 
	\begin{align*}
		\eqfig[0.55]{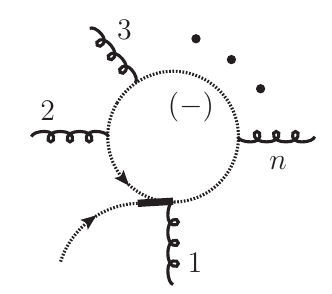} +
		\eqfig[0.55]{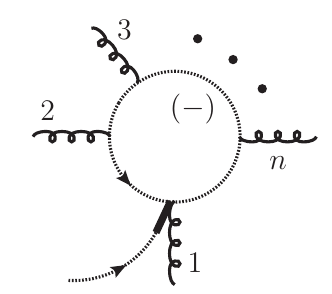} +
		\eqfig[0.55]{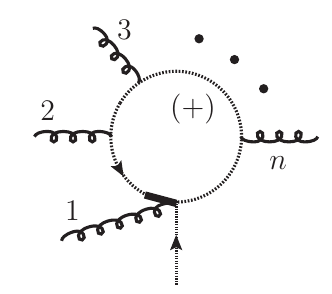} 
		 = 0	
	\end{align*}	\\
	(c) 
	\caption{Vertex identities for the use of Ward identity. 
	In (c), the first two diagrams arise from ghost loops so are multiplied by a $(-1)$ 
	while the third one arises from a gluon loop so does not have this factor.
	}
	\label{fig:qg-wand-cancel}
\end{figure}

In using Ward identities, we need to sum over all possible attachments of the same gluon whose vertex is multiplied by its momentum.
This results in a whole set of special vertices in \figs{fig:ward-q}{fig:ward-g}. 
The essence of Ward identity is the chain of cancellations among the neighboring vertices.
These are shown in \fig{fig:qg-wand-cancel}(a) and (b). For diagrams involving gluon loops, iterative use of Ward identities can convert
the gluon loops to ghost loops, without the minus signs. Then special vertices associated with ghost lines arise, like the third diagram 
in \fig{fig:qg-wand-cancel}(c). This is cancelled by Ward identity diagrams for accompanied ghost loops, with one example given in
\fig{fig:qg-wand-cancel}(c). The identity in \fig{fig:qg-wand-cancel}(a) is due to $[t^a, t^b] = i f^{abc}$, which is what we encountered in 
\eq{eq:ward-kg-1-22} when moving one gluon across another one. The remaining $[t^a, t^b]$ term there reflects the missing third diagram 
in \fig{fig:qg-wand-cancel}(a). Both the identities in \fig{fig:qg-wand-cancel}(b) and (c) are results of the Jacobi identity.

For a full Green function with only quark and gluon external lines, attaching a gluon line in all possible ways with the vertex contracted
with its momentum gives~\citep{Collins:2011zzd},
\begin{equation}
	\adjincludegraphics[valign=c, scale=0.7]{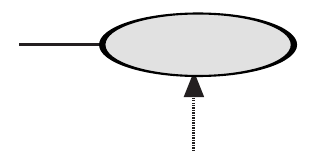}\quad
	\raisebox{0.81em}{$ = \;\; \displaystyle\sum$}
		\adjincludegraphics[valign=c, scale=0.7]{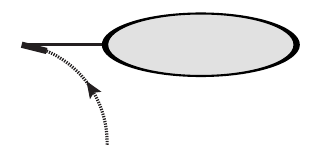}
	\raisebox{0.81em}{$ + \;\; \displaystyle\sum$}
		\adjincludegraphics[valign=c, scale=0.7]{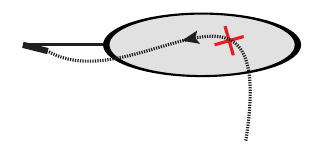}
		\raisebox{0.81em}{,}
\label{eq:ward-green-function}
\end{equation}
after successive use of the line identities. 
On the right-hand side of \eq{eq:ward-green-function}, we sum over the special vertex for each external line in the first term, 
which are given in \fig{fig:ward-q} for quarks and \fig{fig:ward-g} for gluons (without the last two terms).
When the gluon attaches to gluon lines, iterative use of the last two terms in \fig{fig:ward-g} replaces the gluon
line by a ghost line, which traverses the whole graph and join the external lines at special vertices, as given by
the second term on the right-hand side of \eq{eq:ward-green-function}.
This line is anchored in the internal graph, as indicated by the cross label.
For an external gluon line, we note that it can be completely converted to a ghost line with a vertex proportional to its momentum,
like the last two terms in \fig{fig:ward-g}. In this case, the special vertex on the right-hand side of \eq{eq:ward-green-function}
is only one single ghost line with an arrow. 

When the external lines are amputated on shell by the LSZ reduction formula, all the special vertices vanish because they do not
give the needed mass poles, so \eq{eq:ward-green-function} gives zero. For the special gluon vertex that is purely a ghost line
times its momentum, it becomes zero when contracting with a physical polarization vector.
Therefore, applying Ward identities on a physical amplitude, with all possible diagrams and attachments included, gives zero.

\subsection{All-order derivation of the Wilson line structure in DIS}
\label{ssec:DIS-Feynman-wilson}

Now we can come back to the low-order examples in \fig{fig:DIS-LO-g} that we dealt with in 
Sec.~\ref{ssec:DIS-Feynman-lo-pdf}--\ref{ssec:DIS-Feynman-lo-2g}. 
Even though we have projected on shell the collinear quark and gluons lines as external legs of the hard subgraph,
the use of Ward identity for the longitudinally polarized gluons does not give 0 because we are missing the diagrams
with the gluons directly attaching to the collinear quark or gluon lines. The hard subgraph is required to be one-particle-irreducible (1PI)
with respect to the collinear lines. Two merged collinear lines still have a low virtuality so belongs to the collinear subgraph 
before entering the hard subgraph. Such missing diagrams cause the remaining terms in \eqs{eq:H*k-g1-ward}{eq:DIS-LO-2g-ward},
which eventually turn into the Wilson line structure.

\begin{figure}[htbp]
	\centering
	\begin{tabular}{cc}
	\includegraphics[trim={-1cm 0 -1cm 0}, clip, scale=0.75]{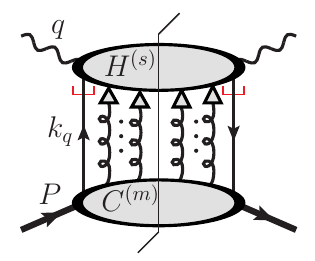} &
	\includegraphics[trim={-1cm 0 -1cm 0}, clip, scale=0.75]{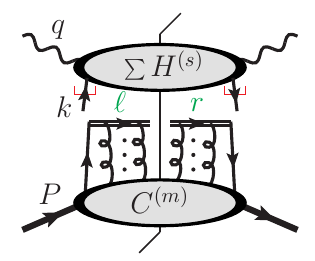} \\
	(a) & (b)
	\end{tabular}
	\caption{(a) A leading region of a DIS diagram. The two subgraphs $H^{(s)}$ and $C^{(m)}$ are joined by
	two quark lines, $\ell$ gluon lines to the left of the cut, and $r$ gluon lines to the right.
	The hooks on the quark lines refer to the on-shell projections by \eqs{eq:DIS-col-approx-lc}{eq:DIS-spinor-proj},
	and the arrows on the gluon lines denote the approximators in \eq{eq:DIS-g-approx}.
	(b) The result of Ward identity for the gluons and the Wilson structures. The upper blob contains all possible diagrams
	of $H^{(s)}$ at the same order, and the Wilson line contains all permutations of the gluons, while the collinear subdiagram 
	$C^{(m)}$ is fixed.
	}
	\label{fig:DIS-LR}
\end{figure}

With the idea of identifying the missing diagrams, we can give an all-order derivation of the Wilson line.
We will consider a particular region $R^{s, m}_{\ell, r}$ of a certain DIS diagram $\Gamma$, shown in \fig{fig:DIS-LR}(a), which is characterized by a hard subgraph 
$H^{(s)}$ and a collinear subgraph $C^{(m)}$ that are of perturbative orders at, respectively, $\order{(\alpha_s)^s}$ and $\order{(\alpha_s)^m}$ in QCD.
\footnote{One shall not confuse the label ``$s$'' for the (integer) order of $H^{(s)}$ from the subscript in the strong coupling $\alpha_s$.}
They are joined by two collinear quark lines and $(\ell + r)$ longitudinally polarized gluons, with $\ell$ to the left and $r$ to the right of the cut.
We will show that by summing all possible diagrams of $H^{(s)}$ and all possible attachments of the collinear gluons, Ward identity factorizes the gluons 
out of the hard part and recollects them onto two gauge links along the light-cone direction $n$, as shown in \fig{fig:DIS-LR}(b).

\begin{figure}[htbp]
	\centering
	\begin{tabular}{ccc}
	\includegraphics[trim={-1cm 0 -1cm 0}, clip, scale=0.75]{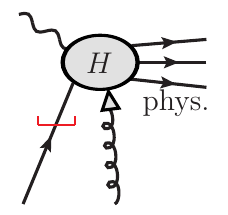} &
	\includegraphics[trim={-1cm 0 -1cm 0}, clip, scale=0.75]{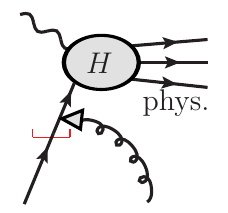} &
	\includegraphics[trim={-1cm 0 -1cm 0}, clip, scale=0.75]{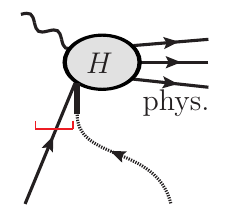} \\
	(a) & (b) & (c)
	\end{tabular}
	\caption{Relevant diagrams for one collinear gluon attachment. Both the gluon and ghost lines are amputated.
	}
	\label{fig:DIS-LR-1g}
\end{figure}

To start with, let us first consider the region $R^{s, m}_{1, 0}$ with $(\ell, r) = (1, 0)$, for which we show the relevant subdiagram in \fig{fig:DIS-LR-1g}(a),
in the form of a scattering amplitude. Only physical particles are included in the final state. They do not contribute to the Ward identity.
The quark parton line has been projected on shell when entering the hard part. Its full propagator has been amputated in $H$,
with an on-shell polarization vector effectively set by the inserted spinor projector $\P_n$,
\beq
	\P_n = \frac{\gamma^- \gamma^+}{2} = \frac{\slash{\hat{k}}_q \gamma^+}{2 k_q^+}
		= \sum_{\lambda} u_{\lambda}(\hat{k}_q) \bb{ \frac{\bar{u}_{\lambda}(\hat{k}_q) \, \gamma^+ }{2 k_q^+} },
\eeq
which vanishes when multiplied by $\slash{\hat{k}}_q$ on the left. 
The gluon approximator in \eq{eq:DIS-g-approx} allows to use Ward identity. Summing over all possible diagrams in $H$ and all possible 
attachments of the gluon, including the one in \fig{fig:DIS-LR-1g}(b), would render the physical amplitude to zero, by the result following 
\eq{eq:ward-green-function}. 

However, \fig{fig:DIS-LR-1g}(b) is not included in the region $R^{s, m}_{1, 0}$. Because both the quark and gluon are collinear, 
the merged quark line directly attaching to $H$ is also collinear. All the three lines are to be included in the collinear subgraph, so \fig{fig:DIS-LR-1g}(b)
belongs to the region $R^{s, m}_{0, 0}$. This diagram represents the missing term in the use of Ward identity.
It alone gives \fig{fig:DIS-LR-1g}(c) by the identity in \fig{fig:ward-q},
with the special vertex being $-i g t^a_{ij}$ (accompanied by an extra $(-i)$ factor), where $a$ is the color index
for the gluon, $i$ the color index of the quark directly connecting to $H$, and $j$ of the one to $C$.
Together with the eikonal factor in \eq{eq:DIS-g-approx}, we can convert the factorized gluon to a Wilson line structure like \eq{eq:eikonal-1g-n}.
But the discussion here applies to all orders of both $H$ and $C$, since we are using the general theorem of Ward identity, not restricted to the
special LO diagram as for \eq{eq:eikonal-1g-n}.

Note that we have been taking the convention of all gluon momenta flowing back to $C$ for simple eikonal propagators in \eq{eq:DIS-g-approx}, 
as explicitly done in Secs.~\ref{ssec:DIS-Feynman-lo-1g} and \ref{ssec:DIS-Feynman-lo-2g}. But in the explicit derivation of the Ward identity in 
\sec{ssec:line-id}, the gluon momenta flow into the graphs. Therefore, we have an extra minus sign, apart from the overall $i$ factor for each gluon.

In the literal consideration of \fig{fig:DIS-LR-1g}(b), both the quark and gluon momenta are projected on shell, causing a formal singularity,
\beq
	\frac{\gamma\cdot (\hat{k}_q + \hat{k}_g)}{(\hat{k}_q + \hat{k}_g)^2} (\gamma \cdot \hat{k}_g) \P_n,
\eeq
which is indefinite. For the use of Ward identity, we therefore first treat $\hat{k}_g$ as off the light cone, and then take the light-cone limit after 
cancelling the quark propagator, as
\beq[eq:on-shell-limit]
	\lim_{k_g \to \hat{k}_g} \frac{\gamma\cdot (\hat{k}_q + k_g)}{(\hat{k}_q + k_g)^2} (\gamma \cdot k_g) \P_n
	= \lim_{k_g \to \hat{k}_g} \frac{\gamma\cdot (\hat{k}_q + k_g)}{(\hat{k}_q + k_g)^2} 
		\bb{\gamma \cdot (\hat{k}_q + k_g) -  \gamma \cdot \hat{k}_q }\P_n
	= \P_n.
\eeq

The same analysis can be performed for the region with any arbitrary $\ell$, but the number of missing terms increases very rapidly as $\ell$
grows, along with a more careful successive use of \eq{eq:on-shell-limit}. 
Before taking on this journey, we note that all the missing terms have some (or all) of the gluon lines attach to the quark line or 
other gluon lines, so they have fewer gluon lines attach to the hard subgraph $H$. While the gluons attaching directly to the quark line can be 
made to the end of the quark line like \fig{fig:DIS-LR-1g}(c), those attaching to $H$ of a smaller number can be related to regions with 
smaller $\ell$ by induction. In the end, all the gluons are detached from $H$ and attached to the end of the quark line, in a complicated tower.
In this process, the details of $H$ are not so important. All it matters is that it is the external quark line that attaches to $H$ along with the gluons.
So the result would be the same if we replace $H$ by a single quark line, exactly like the LO examples in 
Secs.~\ref{ssec:DIS-Feynman-lo-1g} and \ref{ssec:DIS-Feynman-lo-2g}.

One can go further by noting that the quark line can even be replaced by a gauge link that attaches to the incoming quark-photon vertex on one end
and extends to $\infty$ on the other end. Because the Ward identity element for a gluon attaching to the gauge link works in the same way as
it attaches to a quark line, with the $\infty$ end corresponding to the on-shell external leg in the final state, summing all possible attachments of 
the longitudinally polarized gluons to the on-shell quark and gauge link yields zero. 
Then, when requiring the gluons to only attach to the gauge 
link, we can apply the same argument of missing terms, which will give the same result as attaching the gluons to $H$.
By choosing the gauge link along $n$, it has propagators and vertices like $i / (n \cdot k)$ and $-i g n^{\mu} t^a$, respectively.
So automatically, only the plus momenta flow through it, and the approximator in \eq{eq:DIS-g-approx} leaves it unchanged in a trivial way,
\beq
	\pp{ -i g \, n^{\mu} \, t^a } \cdot \frac{\hat{k}_g^{\mu} \, n^{\nu}}{k_g \cdot n + i \epsilon}
	= -i g \, n^{\nu} \, t^a.
\eeq
That is, applying Ward identity for all the gluons attaching to the gauge link returns the same gauge link.

In conclusion, after summing over all possible gluon attachments to the hard subgraph (which itself has included all possible diagrams 
for use of Ward identity), we can factorize out the gluons and reorganize them in a gauge link structure, 
\beq
	\displaystyle\sum_{H, \; \cc{i_1, \cdots, i_l}}
		\adjincludegraphics[valign=c, scale=0.7]{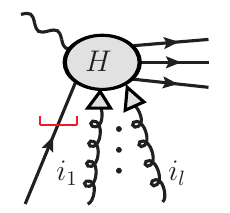}
	\quad = \quad \displaystyle\sum_{H, \; \cc{i_1, \cdots, i_l}}
		\adjincludegraphics[valign=c, scale=0.7]{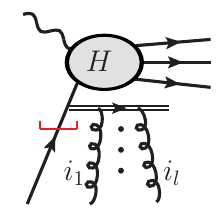}.
\label{eq:dis-hl-w}
\eeq
Since Ward identity receives no contribution from physical external legs, the sum over gluon attachments in \fig{fig:DIS-LR}(a) does not 
cross the cut, and so we get the same gauge link structure on both sides of the cut, as shown in \fig{fig:DIS-LR}(b), with $\ell$ gluons
on the left and $r$ on the right.
On either side, the gluons are summed over all possible permutations at give momentum assignments for a particular diagram $C^{(m)}$. 
Then as for \eq{eq:C-2g-green}, by including the gluon momentum integrations into the collinear factor, we can relabel the gluon momenta 
and colors to sum over the diagrams in $C^{(m)}$, with a fixed gluon attachment configuration. 

To put it formally, we have examined a particular region $R^{s, m}_{l, r}(\Gamma)$ of a particular diagram $\Gamma$, which can be written as
\begin{align}
	& H^{(s)}_{\ell, r}[\Gamma] \otimes C^{(m)}_{\ell, r}[\Gamma] = 
	\int \frac{d^4k}{(2\pi)^4} \, \bb{ \prod_{i = 1}^{\ell} \frac{d^4k^L_i}{(2\pi)^4} \, \prod_{j = 1}^{r} \frac{d^4k^R_j}{(2\pi)^4} }
		H^{(s)}_{\ell, r}[\Gamma]_{\cc{\mu_{i}}, \cc{\nu_{j}}}\pp{ q; k, \cc{k^{L}_i}, \cc{k^{R}_j} }\nn\\
	& \hspace{9em} \times
		\bb{ \prod_{i = 1}^{\ell} g^{\mu_i \rho_i} \prod_{j = 1}^{r} g^{\nu_j \sigma_j} }
		C^{(m)}_{\ell, r}[\Gamma]_{\cc{\rho_{i}}, \cc{\sigma_{j}}}\pp{P; k, \cc{k^{L}_i}, \cc{k^{R}_j} },
\end{align}
where the spinor and color indices are suppressed, $k$ is the total collinear momentum on either side of the cut, $\cc{k^L_i}_{i = 1}^{\ell}$ and 
$\cc{k^R_j}_{j = 1}^{\ell}$ are the two sets of collinear gluon momenta to the left and right of the cut, respectively.
Summing over all possible gluon attachments for all diagrams in the hard part factorizes the gluons out and recollects them
onto two gauge links,
\begin{align}
	& \sum_{\Gamma/R^{s, m}_{\ell, r}} T_R \cc{ H^{(s)}_{\ell, r}[\Gamma] \otimes C^{(m)}_{\ell, r}[\Gamma] }
	= \int dk^+ \, \bb{ \sum_{H} H^{(s)}_{0, 0} (q; \hat{k}) }	
		\int\frac{dk^- d^2\bm{k}_T}{(2\pi)^4} 
		\int \prod_{i = 1}^{\ell} \frac{d^4k^L_i}{(2\pi)^4} \prod_{j = 1}^{r} \frac{d^4k^R_j}{(2\pi)^4}	\nn\\
	& \hspace{3em} \times 
		\prod_{i = 1}^{\ell} \bb{ \, (-i g\, n^{\rho_i} \, t^{a_i} ) \, \frac{i}{n \cdot (k^L_1 + \cdots + k^L_i) + i \epsilon} }
		\prod_{j = 1}^{r} \bb{ (i g\, n^{\sigma_j} \, t^{b_j} ) \, \frac{-i}{n \cdot (k^R_1 + \cdots + k^R_j) - i \epsilon} }	\nn\\
	& \hspace{3em} \times 
		\bb{ \sum_C \P_n C^{(m)}_{\ell, r} \Pb_n}^{\cc{a_i}, \cc{b_j}}_{\cc{\rho_{i}}, \cc{\sigma_{j}}}\pp{P; k, \cc{k^{L}_i}, \cc{k^{R}_j} },
\label{eq:TR-factorize}
\end{align}
where the spinor and color indices are traced over, and in the second line, 
the product over $i$ puts smaller $i$'s to the left, while that over $j$ puts smaller $j$'s to the right.
On the left-hand side of \eq{eq:TR-factorize}, we sum over graphs $\Gamma$ with respect to the same region specification $R^{s, m}_{\ell, r}$.
On the right-hand side, the hard subgraph is summed over with no extra gluon attachments, $(\ell, r) = (0, 0)$, while the collinear
subgraph is summed over at the given order $m$ and with the given numbers of gluons, $\ell$ and $r$.

Then summing over $m$ converts the collinear factor into a Green function, by use of generalized \eqs{eq:eikonal-int}{eq:eikonal-int-2},
\beq[eq:F-a-b-l-r]
	F^{(\ell, r)}_{\alpha\beta}
	= \int \frac{dy^-}{2\pi} e^{-i k^+ y^-} 
		\vv{ P | \bar{\psi}_{\beta}(y^-) W^{(r) \dag}(\infty, y^-; n) \,
			W^{(\ell)}(\infty, 0; n) \psi_{\alpha}(0) | P },
\eeq
where the gauge links exactly reproduce the $\ell$- and $r$-th orders of the two Wilson lines in \eq{eq:F-a-b-W-all-orders}, respectively,
\bse\label{eq:wilson-line-order-l-r}\begin{align}
	W^{(\ell)}(\infty, 0; n) & = \frac{1}{\ell \,!} \, \P \bb{ -i g \int_0^{\infty} d\lambda \, n \cdot A^a(\lambda n) \, t^a }^{\ell}, \\
	W^{(r) \dag}(\infty, y^-; n) & = \frac{1}{r \,!} \, \Pb \bb{ i g \int_0^{\infty} d\lambda \, n \cdot A^a(y^- + \lambda n) \, t^a }^{r},
\end{align}\ese
with $\P$ and $\Pb$ standing for the path-ordering [\eq{eq:path-ordering}] and anti-path-ordering (obviously modified from \eq{eq:path-ordering}) 
operations, respectively.
Notably, the color flow of the partons is completely detached from the hard subgraph onto the gauge links. 
Thus in \eq{eq:F-a-b-l-r}, the colors are traced over. 
More subtleties about the color indices arise for exclusive processes and will be discussed in \ch{ch:exclusive}.

Further summing over $\ell$ and $r$ would convert \eq{eq:F-a-b-l-r} into the gauge-invariant form in \eq{eq:F-a-b-W-all-orders}, 
with the full Wilson line structures. However, doing so requires a sum over different regions and graphs. Without a careful treatment
of the overlaps among regions, such a sum gives unphysical results. So in the following, we include the subtraction of overlapping
regions, and then derive the full factorization result for the DIS.

\subsection{All-order factorization of DIS in Feynman gauge: with subtraction}
\label{ssec:DIS-Feynman-sub}

We have seen in \sec{ssec:DIS-Feynman-wilson} how the sum over graphs with respect to a given region allows use of Ward identity
to factorize the hard and collinear subgraphs, as summarized in \eq{eq:TR-factorize}. 
While this applies to any regions, one cannot simply proceed to sum over all regions and obtain factorization. 
Instead, the contribution from each region $R$ of a graph $\Gamma$ must contain subtractions of smaller regions.

As shown in \eq{eq:region-subtraction-term}, each of the subtraction terms is obtained for a nesting of regions
$R > R'_1 > R'_2 > \cdots > R'_n$, given by first applying the approximators $T_{R'}$ for smaller regions $R' < R$, 
and then applying the approximator $T_R$, i.e.,
\beq[eq:subtraction-term]
	\text{subtraction terms} = \sum_{\cc{R > R'_1 > R'_2 > \cdots > R'_n}} T_R (-T_{R'_1})(-T_{R'_2}) \cdots (-T_{R'_n}) \Gamma.
\eeq
In any region $R$, we can classify each loop momentum $k_i$ as belonging to either the hard region $H_R$ or the collinear region $C_R$.
Then a larger region $R' > R$ would upgrade some momenta from $C_R$ into $H_R$, and a smaller region $R'' < R$ would take some momenta
from $H_R$ into $C_R$. That is, $R' > R$ if and only if $H_{R'} \supset H_R$ and $C_{R'} \subset C_R$.
Therefore, in the successive applications of the region approximators in \eq{eq:subtraction-term}, an approximator that applies later only assigns
new collinear lines into the hard subgraph, but never takes lines out from the hard subgraph.
	
For the smallest region $R'_n$ in the nesting $R > R'_1 > R'_2 > \cdots > R'_n$, 
the graph $\Gamma$ can be decomposed into a hard subgraph $H_{R'_n}$ and 
a collinear subgraph $C_{R'_n}$, joined by a set of longitudinally polarized gluon and two quark lines.
This $H_{R'_n}$ is necessarily a subset of $H_R$, so as we sum over the hard subgraph $H_R$ in $T_R \Gamma$ 
with respect to the specified region decomposition of $R$, we automatically sum over $H_{R'_n}$ 
for each fixed region decomposition of $R'$.
As argued in \sec{ssec:DIS-Feynman-wilson}, the approximator $T_{R'_n}$ then allow to factorize 
these collinear lines out of $H_{R'_n}$ and collect them onto two gauge links. 
The factorized hard part $H_{R'_n}$ only depends on the approximated momenta of the quark lines.
And in the factorized collinear part $C_{R'_n}$, all momenta are as if they are unapproximated. 
 
Now we consider all graphs with the same region decomposition as $R$ and the same collinear subgraph as $C^{(m)}_{l,r}$
but different hard subgraphs $H^{(s)}_{l, r}$ at the same order $s$.
Among all the region nestings of these graphs that are of the same length as $R > R'_1 > R'_2 > \cdots > R'_n$, 
collectively denoted as
$\cc{R > R^i_1 > R^i_2 > \cdots > R^i_n}$, 
there are more than one of the smallest regions $R^i_n$ of those graphs 
with the same hard and collinear subgraph decompositions as $R'_n$
and the same collinear subgraphs $C_{R'_n}$, 
which has allowed us to sum over all graphs in $H_{R'_n}$ and gluon attachments to factorize the latter.
There are even more graphs that after factorizing the smallest regions have the same factorized hard subgraph as $H^0_{R'_n}$.
Working with this set of diagrams, their collinear factors $C_{R^i_n}$ all have two gauge links, 
but which may collect arbitrary numbers of gluons as allowed by the given perturbative order. 
When going to next regions $R^i_{n-1}$, some of the collinear lines become hard, and some collinear lines get into or connected to 
the new hard subgraphs, but they must all stay within the factorized collinear subgraphs $C_{R^i_n}$, not attached to the previously
factorized hard part $H^0_{R^i_n}$; otherwise, they would have stayed in or attached to $H_{R^i_n}$ when considering the region $R^i_n$.
Therefore, in the regions $R^i_{n-1}$, the hard subgraphs $H_{R^i_{n-1}}$ are connected to the two gauge links by arbitrary numbers of gluons, 
and the collinear subgraphs $C_{R^i_{n-1}}$ are joined to them by collinear gluon lines and two quark lines.

We only consider those of the diagrams that have the same collinear subgraph as $C_{R'_{n-1}}$, with the same number of collinear
gluons attaching to the hard subgraphs $H_{R^i_{n-1}}$, which can be arbitrary.
Again, by summing these diagrams, allowing the collinear gluons to also attach to the gauge links, 
we can factorize them out of the hard subgraphs onto two new gauge links.
The factorized hard factor $H^0_{R'_{n-1}}$ is summed over all possible graphs and only depends on two external amputated quark lines,
convoluted through their plus momentum with the collinear factor $C_{R'_{n-1}}$, which has certain number of gluons attached to its gauge
links.

Following the same analysis, we now consider all of those diagrams which are factorized into the same hard factors 
$H^0_{R'_{n}}$ and $H^0_{R'_{n-1}}$ (both being summed over all possible graphs at particular orders).
This can be similarly factorized for the region $R^i_{n-2}$. 
Continuing this analysis iteratively until we reach the region $R$, which can be similarly factorized by summing over the graphs.
In this way, we showed that
\begin{align}\label{eq:fac-subtraction}
	&\sum_{\Gamma / \{ R > R_1^i > \cdots > R_n^i \} }
		T_R \, T_{R_1^i} \, T_{R_2^i} \cdots T_{R_n^i} \, \Gamma	\nn\\
	&\hspace{4em}
	= C^{(m)}_{l, r} \otimes 
		\sum_{ \{H^0_{R_0}, H^0_{R_1}, \cdots, H^0_{R_n}\} } 
			H^0_{R_0} \otimes H^0_{R_1} \otimes H^0_{R_{2}} \cdots \otimes H^0_{R_{n}},
\end{align}
which is also true if multiplied by $(-1)^n$ as required by \eq{eq:subtraction-term}.
On the left-hand side, we sum over all the region nestings of length $(n + 1)$ of the graphs $\Gamma$ that are of the same order and 
have the same collinear subgraph $C^{(m)}_{l, r}$ for the region specified by $R$.
On the right-hand side, each hard factor $H^0_{R_{i}}$ has only two gauge links and two external amputated quark lines. They are
convoluted only via the plus momenta, and color and spinor indices are summed over within each factor.
As a fixed overall perturbative order $s$, we sum over all possible convolutions of the hard factors 
$\{H^0_{R_1}, H^0_{R_2}, \cdots, H^0_{R_n}\}$.

The main reason why \eq{eq:fac-subtraction} holds is that applying Ward identity on a gauge link works in the same way as on a single
quark line. So the Ward identity for $T_R$ applies equally with or without subtractions, and yields the same collinear factor with gauge links. 
Then we have, for the region $R^{s, m}_{l, r}$, summing over graphs factorizes $C_R \Gamma$ into the same form as \eq{eq:TR-factorize}, 
just with $T_R$ replaced by $C_R$, and $H^{(s)}_{0, 0}$ replaced by the subtracted hard factor,
\beq[eq:DIS-H-sub]
	H^{(s)}_{\rm sub} 
	= H^{(s)}_{0, 0} + \sum_{n = 1}^s (-1)^n 
		\sum_{\{H^0_{R_0}, H^0_{R_1}, \cdots, H^0_{R_n}\} } 
			H^0_{R_0} \otimes H^0_{R_1} \otimes H^0_{R_{2}} \cdots \otimes H^0_{R_{n}},
\eeq
with all possible subregion contribution removed at a fixed order $s$, for which one at most has $s$ nested regions so $n \leq s$.  

Note that in the subtracted version of \eq{eq:TR-factorize}, we sum over all possible diagrams in $H^{(s)}$ and $C^{(m)}_{\ell, r}$, 
with no contribution from smaller regions.
The same thing can be done for any other regions of any other graphs, and they yield the same factorization structure. 
Then we sum over all the regions of all graphs, which can be in turn converted to a sum over $s, m, \ell, r$ and $H^{(s)}$ and $C^{(m)}_{\ell, r}$,
\begin{align}
	\sum_{R, \Gamma} C_R \Gamma
	&= \sum_{\ell, r} \sum_{s, m} \sum_{C^{m}_{\ell, r}} \sum_{H^{s}_{\ell, r}}
		C_R \bb{ H^{(s)}_{\ell, r} \otimes C^{(m)}_{\ell, r} }	\nn\\
	&= \sum_{\ell, r} \sum_{s, m} \sum_{C^{m}_{\ell, r}} 
		H^{(s)}_{\rm sub} \otimes \bb{ (\text{gauge links})_{\ell, r} \cdot C^{(m)}_{\ell, r} }	\nn\\
	&= \sum_{\ell, r} \bb{ \sum_s H^{(s)}_{\rm sub} } \otimes F^{(\ell, r)}	\nn\\
	&= H_{\rm sub} \otimes F,
\label{eq:DIS-sum-over-R}
\end{align}
where in the last step we defined the all-order hard coefficient $H_{\rm sub} = \sum_s H^{(s)}_{\rm sub}$, and 
all-order parton distribution $F = \sum_{\ell, r} F^{(\ell, r)}$ in \eq{eq:F-a-b-W-all-orders}, 
with $F^{(\ell, r)}$ given in \eq{eq:F-a-b-l-r}.

In \eq{eq:DIS-sum-over-R}, the hard and collinear factors are only convoluted in the plus momentum.
The spinor indices are still entangled, and can be separated in the same way as \eq{eq:PDF-spinor-decomp}.
In this way, we proved to all orders that the DIS cross section can be factorized into 
parton distribution functions with hard coefficients.
The result takes the same form as \eq{eq:DIS-LO-factorize-pdf}, just with the ``0'' indices deleted. 
The all-order parton distribution functions take the same forms as \eq{eq:PDFs-no-W}, 
but with the ``0'' indices deleted and the Wilson lines inserted, as in \eq{eq:F-a-b-W-all-orders}.

The parton distribution functions have collected all the pinched propagators, so 
they have sensitive dependence on low-momentum-scale QCD dynamics and are nonperturbative.
The hard coefficients, on the other hand, have subtracted all the low-scale dependence and are free from pinch singularities.
The loop momenta therein always have (or can be freely deformed to have) high virtualities.
They are therefore only sensitive to the hard-scale dynamics, which is purely perturbative in QCD.
In this way, not only does factorization give a factorized form for the DIS cross section, 
but also each factor has a difference momentum scale dependence.

Moreover, the systematic factorization formalism gives field-theoretic operator definitions to the nonperturbative
parton distribution functions in \eq{eq:F-a-b-W-all-orders}. 
Similar factorization theorems can be derived for other processes, including SIDIS and Drell-Yan, 
and arrive at the same set of parton distribution functions. 
The operator definition thus allows to prove the universality of the nonperturbative set of functions.
This important fact equips factorization with predictive power, which follows by comparing those different processes
with the same set of PDFs but different perturbatively calculable hard coefficients.
Also, the operator definition implies that PDFs can be studied on their own, independent of the physical processes 
from which they are derived.
It is an ongoing and prospering effort in the literature to use nonperturbative methods like Lattice QCD to compute 
the PDFs, which are soon to reach an era when they can be in comparison with globally fit PDF results. 
See~\citep{Constantinou:2020hdm} for a review and references therein.

Throughout our discussion, we only considered the single-flavor case with the active parton being a quark. 
For the real case with several quark and antiquarks, each of our derivation steps like 
\eqss{eq:TR-factorize}{eq:fac-subtraction}{eq:DIS-sum-over-R}
needs to contain a sum over quark flavors, so the resultant factorization formula in
\eq{eq:DIS-LO-factorize-pdf} needs to be supplemented 
by a sum over quark and antiquark flavors, with a corresponding hard coefficient for each channel.
The same discussion holds for the most general case that also involve gluons.
For the gluon case, one can arrive at the same factorization result 
by a similar but more complicated derivation of the Wilson lines.
A thorough discussion with full details to all orders is still missing in the literature~\citep{Collins:2008sg}. 
We leave it to a future study and will not delve into it in this thesis.

\subsection{UV renormalization and factorization scale}
\label{ssec:dis-uv}

Now in the whole discussion of region analysis, we have explicitly considered only two regions: 
collinear (with $k_T \ll Q$) and hard (with $k_T \sim \O(Q)$). 
The original DIS diagram corresponds to a four-point (cut) Green function 
and has a superficial ultraviolet (UV) divergence degree $D = -1$.
This means that the region with $k_T \gg \O(Q)$ is power suppressed, 
so it is justified to only consider those two regions.
However, the factorization procedure that we laid out in the previous sections 
has implicitly introduced some artificial UV divergences, as we are going to treat now.

The definition of the PDF in \eq{eq:F-a-b-W-all-orders}
(and similarly the low-order ones in Eqs.~\eqref{eq:F0-ab} \eqref{eq:eikonal-1g-n-k-int}
and \eqref{eq:F2-w})
as a parton correlation function on the light cone results from including 
the integrations of $k^-$ and $\bm{k}_T$ within the collinear factor and extending them to infinity,
which is beyond the collinear region that it is supposed to capture. 
By only having a parton spinor vertex 
$\gamma^+$ (or $\gamma^+\gamma_5$ and $\gamma^+\gamma^i\gamma_5$) 
in \fig{fig:DIS-LR}(b), 
the PDF is essentially a three-point Green function,
with a superficial divergence degree $D = 0$,
Hence, the UV region with $k_T \gg \O(Q)$ in the PDF is not power suppressed 
but gives logarithmically divergent contribution. 
The formally defined collinear and hard factors in \eq{eq:DIS-sum-over-R} are thus both UV divergent.
However, by the subtraction formalism, the same UV divergences in the PDF are formally 
subtracted in the hard factor, so the overall convolution is free of UV divergences.
Even so, it is still physically significant to use UV finite definitions for both factors.
Therefore, we need to subtract the UV divergence in the PDF to define a renormalized PDF.

\begin{figure}[htbp]
\centering
	\begin{tabular}{ccc}
		\includegraphics[scale=0.65]{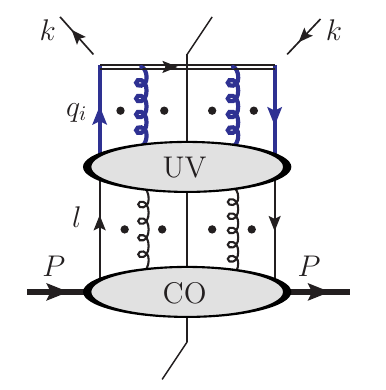} &
		\includegraphics[scale=0.65]{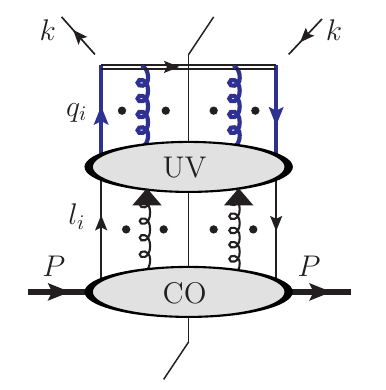} &
		\includegraphics[scale=0.65]{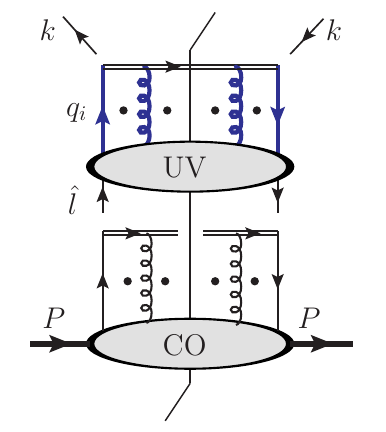}	\\
		(a) & (b) & (c) 
	\end{tabular}
	\caption{Separation of ultraviolet (UV) and collinear (CO) regions in a general diagram of PDF. 
	The thick blue lines have UV momenta flowing through, with the scaling in \eq{eq:uv scaling}.}
\label{fig:PDF-UVg}
\end{figure}

The renormalization procedure is very similar to the factorization analysis, and works in the following steps: 
\begin{enumerate}
\item [(1)] 
Treat the PDF as an independent amplitude, and separate its internal momenta as either UV or collinear.
This yields the same region analysis and the same subgraph decomposition.
But the hard subgraph is replaced by a UV subgraph, as shown in \fig{fig:PDF-UVg}(a),
and the hard scale $Q$ is replaced by a UV scale $\Lambda_{\rm UV}$ that is to be taken to infinity.
A momentum in the UV region has the scaling,
	\beq
		k_{\rm UV}^{\mu} 
			= \pp{k_{\rm UV}^+, k_{\rm UV}^-, \bm{k}_{\rm UV, T} } 
			\sim \pp{Q, \Lambda_{\rm UV}^2 / Q, \Lambda_{\rm UV}}.
		\label{eq:uv scaling}
	\eeq
\item [(2)]
In \fig{fig:PDF-UVg}(a), the collinear subgraph (CO) is connected to the UV subgraph (UV) 
by two quark lines and arbitrarily many gluon lines that are of longitudinal polarization.
Any other configuration is power suppressed by $1/\Lambda_{\rm UV}$ that will eventually vanish.
\item [(3)]
For each such region, the same approximations as Eqs.~\eqref{eq:DIS-col-approx-lc}\eqref{eq:DIS-spinor-proj} 
and \eqref{eq:DIS-g-approx} together with Ward identity factorize the collinear subgraph from the UV subgraph,
as shown in \fig{fig:PDF-UVg}(b)(c).
\item [(4)]
Still, subtractions of smaller region are needed for each region to avoid double counting. 
(In the language of renormalization, smaller regions yield subdivergences.)
After summing over all regions of all graphs, one would reach the same factorization structure as 
\eq{eq:DIS-sum-over-R}, 
but now the hard factor is replaced by the UV factor and has a similar expression as \eq{eq:DIS-H-sub}. 
In dimensional regularization with $d = 4 - 2\epsilon$, 
the UV factor is a function of $1/\epsilon$, depending on the subtraction schemes, and
the power suppressed term $1/\Lambda_{\rm UV}$ is not present.
\end{enumerate}

Schematically, \fig{fig:PDF-UVg}(c) together with subtractions leads to the factorization structure,
\begin{align}
	&\int \frac{d^4{\ell}}{(2\pi)^4}  \Tr{ \UV(k^+, \ell) \cdot \CO(\ell, p) } 	\nn\\
	& \hspace{2em} =  
		\int_{k^+}^{p^+} d \ell^+ \, \Tr{ \UV(k^+, \hat{\ell}) \frac{\gamma^-}{2} } \cdot 
		\bb{ \int \frac{d \ell^- d^2 \bm{\ell}_T}{(2\pi)^4} \Tr{ \frac{\gamma^+}{2} \CO(\ell, p)} } + \cdots		\nn\\
	& \hspace{2em} = 
		\int_x^1 \frac{d z}{z} \, \Tr{ \UV(k^+, \hat{\ell}) \frac{\ell^+ \gamma^-}{2} } \cdot 
		\bb{ \int \frac{d \ell^- d^2 \bm{\ell}_T}{(2\pi)^4} \Tr{ \frac{\gamma^+}{2} \CO(\ell, p)} } + \cdots			\nn\\
	& \hspace{2em} =
		\int_x^1 \frac{d z}{z} \, \UV(x/z) \cdot \bb{ \int \frac{d \ell^- d^2 \bm{\ell}_T}{(2\pi)^4} \Tr{ \frac{\gamma^+}{2} \CO(\ell, p)} } + \cdots,
\label{eq:UV factorize}
\end{align}
where $\cdots$ refers to other spinor structures and flavor channels, 
and Tr stands for the trace of Dirac spinor indices. 
The square bracket in the last line is the renormalized PDF with UV divergence removed. 

So, similar to how the region analysis of the cross section diagram in \fig{fig:DIS-LR}(a) leads to the 
factorization result in \eq{eq:DIS-sum-over-R},
the region analysis on the ``bare'' PDF results in a {\it multiplicative} renormalization,
Using the unpolarized PDF as an example, we have
\beq
	f_i^{\rm bare}(x, 1/\epsilon) = \sum_j \int_x^1 \frac{d z}{z} \, \big(Z^{-1}\big)_{ij}(x/z, 1/\epsilon; \alpha_s(\mu)) \, f_j(z, \mu),
\label{eq:uv divergence}
\eeq
where $(Z^{-1})_{ij} = \UV_{ij}$ collects all the UV divergences and $i$ and $j$ stand for the parton flavors.
Here we have explicitly indicated the UV divergence in the bare PDF, 
which is a polynomial of $\epsilon^{-1}$ in dimensional regularization.
Inverting \eq{eq:uv divergence} gives the renormalized PDF $f_i(x, \mu)$,
\beq
	f_i(x, \mu) = \sum_j \int_x^1 \frac{d \, z}{z} \, Z_{ij}(x/z, 1/\epsilon; \alpha_s(\mu)) \, f_j^{\rm bare}(z, 1/\epsilon),
\label{eq:pdf renormalize}
\eeq
where $Z_{ij}(x/z, 1/\epsilon; \mu)$ is the renormalization coefficient, being the inverse of UV divergence,
\beq
	\sum_j \int_x^1 \frac{d \, y}{y} \, Z_{ij}(x/y, 1/\epsilon; \alpha_s(\mu)) \, \big(Z^{-1}\big)_{jk}(y, 1/\epsilon; \alpha_s(\mu)) 
	= \delta_{ik} \, \delta(1-x).
\eeq

The renormalized PDF in \eq{eq:pdf renormalize} introduces a dependence on the {\it untraviolet} 
renormalization scale $\mu$ through $\alpha_s(\mu)$.
This scale is different from the UV renormalization scale in the QCD Lagrangian, but is
more of an artifact of the factorization. It is called the factorization scale.
The dependence on the factorization scale $\mu$ leads to a multiplicative evolution equation for the PDFs,
\beq
	\frac{d\, f_i(x, \mu)}{d \ln\mu^2} 
	= \sum_j \int_x^1 \frac{d z}{z} \, P_{ij}(x/z, \alpha_s(\mu)) \, f_j(z, \mu),
\label{eq:pdf-evolution}
\eeq
where the evolution kernel $P_{ij}(x/z)$ can be obtained by
\beq
	\frac{d }{d \ln\mu^2} Z_{ik}(z, 1/\epsilon, \alpha_s(\mu)) 
	= \sum_{j} \int_z^1 \frac{d y}{y} \, P_{ij}(z/y, \alpha_s(\mu)) \, Z_{jk}(y, 1/\epsilon, \alpha_s(\mu)),
\label{eq:kernel}
\eeq
which can be solved perturbatively order by order. 

Restoring the ``bare" notation in \eq{eq:DIS-sum-over-R} and substituting the renormalization expressions of PDFs
like \eq{eq:uv divergence} for the bare ones, we get the same factorization formula for the DIS cross section, 
but now in terms of the UV renormalized PDF and infrared (and UV) finite hard coefficient,
\beq[eq:DIS-factorize-ren]
	\sigma_{\rm DIS}(x_B, Q) = \sum_i \int_{x_B}^1 \frac{d x}{x} f_i(x, \mu) \, C_i\pp{\frac{x_B}{x}, \frac{Q}{\mu}; \alpha_s(\mu)}
		+ \mbox{polarized}
		+ \order{\LQCD / Q},
\eeq
which is similar to \eq{eq:DIS-sum-over-R}, but has an extra dependence on the factorization scale $\mu$.
In \eq{eq:DIS-factorize-ren} we only explicitly write the term for the unpolarized PDF,
whose hard coefficient has been relabeled as $C$;
the polarized PDF terms can be easily obtained like in \eq{eq:DIS-LO-factorize-pdf}. 
The renormalized hard coefficient $C_i$ is related to the bare one in \eq{eq:DIS-sum-over-R} through
\beq
	C_i(x, \mu) 
		= \sum_{j} \int_{x}^1 \frac{dy}{y} \, Z^{-1}_{ij}(x / y, 1/\epsilon; \alpha_s(\mu)) \, 
		C_j^{\rm bare}(y, 1/\epsilon).
\eeq

This ends our discussion of the DIS factorization. 
The UV finite PDFs now can be studied both theoretically and phenomenologically.
Parametrized at a base scale $\mu_0$, the PDFs can be evolved to any arbitrary scale via \eq{eq:pdf-evolution},
which allows them to be used in \eq{eq:DIS-factorize-ren} at the scale $\mu \sim \O(Q)$.
By comparing the experimentally measured cross section with the right-hand side in \eq{eq:DIS-factorize-ren},
one can obtain the PDFs that are best fit to experiments and then employ them to give predictions for 
the same experiment at a different energy or for other experiments.

\section{Factorization of Sudakov form factor}
\label{sec:sudakov-factorization}

Now we deal with the Sudakov form factor, whose leading region graph is shown in \fig{fig:sudakov-reduced-diagram}(b).
This is more complicated than the DIS factorization in \sec{sec:ward-identity} by having two collinear subgraphs
and one extra soft subgraph connected to them. 
We will show in this section that by summing over all regions and graphs,
the Sudakov form factor can be factorized into a hard factor, two collinear factors, and a soft factor.
As in \sec{sec:ward-identity}, we will first see how the region approximator $T_R$ alone can factorize each subdiagrams,
and then examine how subtractions of smaller regions will modify the factorization structure.

\subsection{Factorization without subtraction}
\label{ssec:sudakov-w/o-sub}

For any graph with the region decomposition in \fig{fig:sudakov-reduced-diagram}(b), 
we route each $q$ ($\bar{q}$) collinear loop momentum to flow between the hard subgraph $H$
and the collinear subgraph $C_q$ ($C_{\bar{q}}$) or completely within the latter,
and each soft momentum to circulate between $H$, $C_q$, $C_{\bar{q}}$, and the soft subgraph $S$,
or completely within $S$.
After the approximation [\eq{eq:sudakov-col-mom}] for collinear momenta flowing into $H$, and 
that [\eqs{eq:soft-approx-k-in-A}{eq:soft-approx-k-in-B}] for soft momenta flowing into $C_q$ or $C_{\bar{q}}$,
the soft momenta are completely zero in $H$.
So then the collinear approximation in \eq{eq:CHC-conv1}, with the necessary modification in \eq{eq:g-approx-modify},
allows an exact use of Ward identity for collinear gluons attaching to $H$. 

\begin{figure}[htbp]
	\centering
	\begin{tabular}{cc}
	\includegraphics[trim={0 0 -2em 0}, clip, scale=0.75]{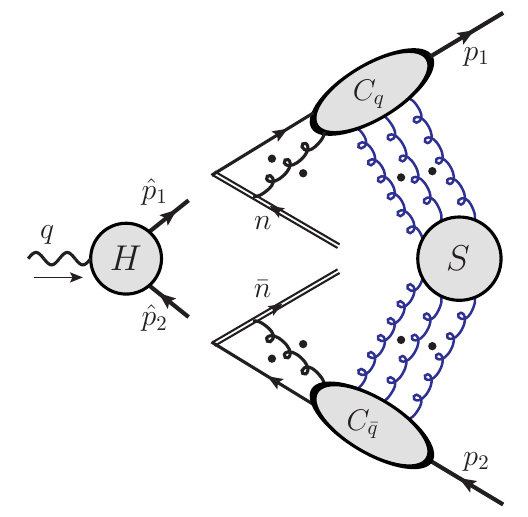} &
	\includegraphics[trim={-2em 0 -2em 0}, clip, scale=0.75]{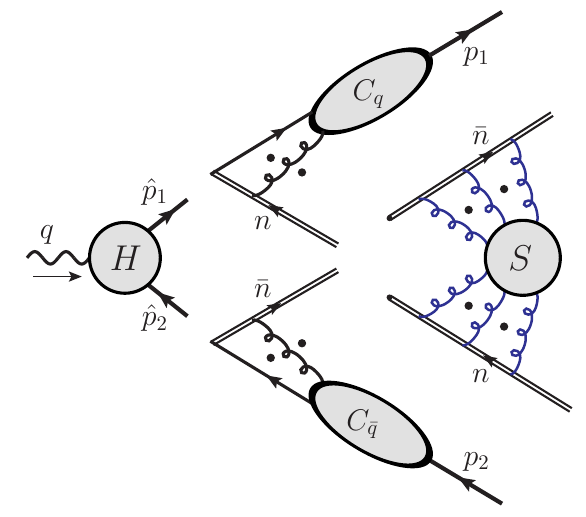}\\
	(a) & (b)
	\end{tabular}
	\caption{Factorization of the Sudakov form factor.
	(a) Factorization of collinear gluons out of the hard subgraph.
	(b) Factorization of soft gluons out of the collinear subgraphs.
	}
	\label{fig:sudakov-fac}
\end{figure}

This works in a way similar to the DIS factorization in \sec{sec:ward-identity}, for both collinear subgraphs.
For each collinear sector, the external lines of $H$ connected to the other collinear subgraph are projected on shell, 
so do not contribute to Ward identity. Then by the same argument as \sec{ssec:DIS-Feynman-wilson}, 
after summing over the diagrams in $H$ and all possible collinear gluon attachments 
with respect to the same region specification, 
the collinear gluons are factorized onto gauge links, as shown in \fig{fig:sudakov-fac}(a). 
The $q$-collinear gluons are collected by a gauge link along the $n$ direction, and 
the $\bar{q}$-collinear ones by a gauge link along the $\bar{n}$ direction.
By the $i\epsilon$ prescriptions in \eq{eq:g-approx-modify} with the collinear momenta flowing into $C_q$ or $C_{\bar{q}}$,
both gauge links point to the future.

The advantage of first factorizing collinear gluons is that they are organized onto Wilson line structures, which makes
the further factorization of soft gluons much simpler than the original graph in \fig{fig:sudakov-reduced-diagram}(b),
where the colors are more entangled among all the subgraphs. 
The factorized collinear subgraphs $C_q$ and $C_{\bar{q}}$ are connected Green functions with one external amputated on-shell
quark or antiquark line, one un-amputated fermion line corresponding the field $\bar{\psi}$ or $\psi$, and an arbitrary number of
gluon lines. They are attached by soft gluons in any arbitrary way. 
Those gluons are also required to be 1PI before entering the collinear subgraphs.

As given in \eq{eq:soft-c-c-coupling}, the region approximator is designed such that the soft gluon momenta are light-like when 
flowing into the collinear subgraphs. So exact use of Ward identity can be made by treating them as all on shell.
Since the Wilson lines organize the color flows such that all the off-shell collinear quark (antiquark) and gluon lines in 
$C_q$ ($C_{\bar{q}}$) together behave as a single quark (antiquark) field, applying Ward identity on the soft gluons 
yields the same result as if they are only attached to the end of the quark (antiquark) line. 
Then we get a similar result as \eq{eq:dis-hl-w}. 
Since it is the field $\bar{\psi}$ that corresponds to the $C_q$ subgraph, and 
$\psi$ to $C_{\bar{q}}$, the resultant Wilson lines collecting the soft gluons are respectively,
\begin{align}
	W^{(s)}(\infty, 0; \bar{n})	    \text{ on } C_q \text{ side;  }
	\quad
	W^{(\bar{s})\dag}(\infty, 0; n)	\text{ on } C_{\bar{q}} \text{ side, }  
\label{eq:sudakov-soft-wilson}
\end{align}
whose explicit forms are the same as \eq{eq:wilson-line-order-l-r},
at orders given by the number of the soft gluons, $s$ and $\bar{s}$, 
connecting to $C_q$ and $C_{\bar{q}}$, respectively.
In doing so, we are working at a particular order of $H$, $C_q$, $C_{\bar{q}}$, and $S$, and have summed over 
all possible graphs in the first three subgraphs.  

In this way, for a certain leading region $R$, specified by the specific order of each subgraph and
the numbers of collinear and soft gluons connecting these subgraphs, applying the approximator
$T_R$ and summing over the graphs factorizes these subgraphs, as shown in \fig{fig:sudakov-fac}(b).
The color flows in the following way:
\begin{align*}
	\mbox{antiquark in $C_{\bar{q}}$} 
	& \to 
	\mbox{out from the collinear Wilson line along $\bar{n}$ on the $\infty$ end}	\\
	&\to 
	\mbox{into $S$ from the $\infty$ end of the soft Wilson line along $n$}	\\
	&\to 
	\mbox{out from $S$ on the $0$ end of the soft Wilson line along $n$}	\\
	&\to 
	\mbox{into $H$ at the antiquark leg}
	\to
	\mbox{out from $H$ at the quark leg}	 \\
	&\to
	\mbox{into $S$ from the $0$ end of the soft Wilson line along $\bar{n}$}	\\
	& \to
	\mbox{out from $S$ on the $\infty$ end of the soft Wilson line along $\bar{n}$}	\\
	& \to
	\mbox{into the collinear Wilson line in $C_q$ along $n$ on the $\infty$ end}	\\
	& \to
	\mbox{out of $C_q$ at the quark line.}
\end{align*}
The hard subgraph only depends on the external two amputated on-shell quark legs, so is automatically 
a color singlet. This helps contracts the two soft Wilson lines on their $0$ ends. By summing over the graphs
in $S$, we can convert it into a Green function,
\beq[eq:sudakov-soft-factor-s]
	S^{(s, \bar{s})} = \frac{1}{N_c} \tr \, \vv{0 | W^{(s)}(\infty, 0; \bar{n}) \, W^{(\bar{s})\dag}(\infty, 0; n) | 0 }.
\eeq
As for \eqs{eq:TR-factorize}{eq:F-a-b-l-r}, this entails a relabeling of the soft gluon colors and momenta, 
avoiding a double counting of the soft gluon permutations.
Apparently, the unitarity of Wilson lines has helped us to identify $S^{(s, \bar{s})}$ also as a color singlet, 
so the color flows are disentangled between the collinear and soft subgraphs. 
Similarly, by summing over the graphs in $C_q$ and $C_{\bar{q}}$, we can write down their 
Green function definition,
\begin{align}
	C_q^{r}  = \vv{ q(p_1) | \bar{\psi}(0) W^{(r)\dag}(\infty, 0; n) | 0 } \P, \quad
	C_{\bar{q}}^{\bar{r}}  = \Pb \vv{ \bar{q}(p_2) | W^{(\bar{r})}(\infty, 0; n) \psi(0) | 0 },
\label{eq:sudakov-col-factor-r}
\end{align}
whose color dependence is also trivial, and where $r$ and $\bar{r}$ are the numbers of collinear gluons 
attaching to the Wilson lines. 
Here $\P = \Pb$ are the spinor projectors defined in \eqs{eq:sudakov-projector-P}{eq:sudakov-projector-Pb}.

We note that the conversions into Green functions are possible in 
\eqs{eq:sudakov-soft-factor-s}{eq:sudakov-col-factor-r}
because we have extended the collinear and the soft momenta into infinity after factorizing them out.
This, of course, goes out of the momentum regions where the approximator is designed for.
But as argued in \sec{sec:subtraction}, the imperfections of the approximated integrals at larger 
regions are taken care of by the subtractions when we consider those larger regions. 
Hence, to sum over all regions and graphs, we need to carefully subtract double-counting contribution 
from smaller regions.

\subsection{Rapidity divergence and modifications}
\label{ssec:sudakov-rapidity}

There is, however, a flaw in the derived factorization result in 
\eqs{eq:sudakov-soft-factor-s}{eq:sudakov-col-factor-r}.
As an explicit low-order calculation can show, the are divergences associated with Wilson lines
along light-like directions. These are because a gluon attached to such a Wilson line has no bound
on its rapidity, whose integration thus extends to infinity and leads to divergence.
So they are called rapidity divergence. 
In the factorization argument, light-like Wilson lines arise because we use the light-like vectors 
$n$ and $\bar{n}$ in the approximations \eqs{eq:soft-col-coupling-approx-modify}{eq:g-approx-modify}.

To cure this issue, we modify the vectors $n$ and $\bar{n}$ off the light cone,
\beq[eq:shift-off-lc]
	n \to u = n - e^{2y} \bar{n}, \quad
	\bar{n} \to \bar{u} = \bar{n} - e^{-2\bar{y}} n,
\eeq
where $\bar{y}$ and $y$ have large values and are approximating the rapidities of $\bar{n}$ and $n$, with 
$\bar{y} > 0$ and $y < 0$. 
The $u$ and $\bar{u}$ are to replace the $n$ and $\bar{n}$ in \eq{eq:soft-col-coupling-approx-modify}.
The minus signs are in order not to introduce extra poles for the soft momenta in \eq{eq:soft-col-coupling-approx-modify}.
This makes $u$ and $\bar{u}$ space-like.

To further require the soft approximator on a Wilson line to give back the same Wilson line, we require
\beq
	\hat{k}_{qs} \cdot \bar{u} = k_{qs} \cdot \bar{u}, \quad
	\hat{k}_{\bar{q}s} \cdot u = k_{\bar{q}s} \cdot u,
\eeq
for the soft momenta $k_{qs}$ and $k_{\bar{q}s}$ flowing into $C_q$ and $C_{\bar{q}}$, respectively.
This can be done by choosing 
\beq[eq:soft-momenta-modify]
	\hat{k}_{qs} = \frac{(k_{qs} \cdot \bar{u}) \, n}{\bar{u} \cdot n}
		= (k_{qs}^- - e^{-2\bar{y}} k_{qs}^+) \, n,
	\quad
	\hat{k}_{\bar{q}s} = \frac{(k_{\bar{q}s} \cdot u) \, \bar{n}}{u \cdot \bar{n}}
		= (k_{\bar{q}s}^+ - e^{2y} k_{\bar{q}s}^-) \, \bar{n},
\eeq
which are still along the lightcone directions $n$ and $\bar{n}$, and keep the important minus or plus components intact.
The modified approximations in \eqs{eq:shift-off-lc}{eq:soft-momenta-modify} differ from the original ones only by 
power-suppressed effects due to the parameters $y$ and $\bar{y}$. 
They still keep the soft gluons lightlike and on-shell when flowing into the collinear subgraphs, and vanishing in the hard subgraph.
So Ward identity can work equally and exactly, leading to modified soft factor in \eq{eq:sudakov-soft-factor-s-modify},
\beq[eq:sudakov-soft-factor-s-modify]
	S^{(s, \bar{s})}(\bar{y}, y) = \frac{1}{N_c} \tr \, \vv{0 | W^{(s)}(\infty, 0; \bar{u}) \, W^{(\bar{s})\dag}(\infty, 0; u) | 0 }.
\eeq

Similar modification should also be done for the collinear factors in \eq{eq:sudakov-col-factor-r}. 
But as we shall see shortly, 
subtractions of soft subregions from the collinear region cancel the rapidity divergences.

\subsection{Factorization with subtraction}
\label{ssec:sudakov-w-sub}

We examine some region $R^{r, \bar{r}}_{s, \bar{s}}$ that 
has $r$ ($\bar{r}$) collinear gluons connecting $H$ to $C_q$ ($C_{\bar{q}}$)
and $s$ ($\bar{s}$) soft gluons connecting $S$ to $C_q$ ($C_{\bar{q}}$), 
at a particular order for all these subgraphs,
and we include all such possible diagrams, denoted as $\Gamma/R^{r, \bar{r}}_{s, \bar{s}}$.

For a particular region $R$ of a certain graph $\Gamma$, a smaller region $R' < R$ can have some collinear lines of $R$ in $S$, 
or some hard lines of $R$ in the collinear subgraphs. 
Identifying the subgraphs for each region, together with subtractions of smaller
regions, each subgraph can include the same subgraph but with sub-approximators applied within. 
After summing over all the graphs in $\Gamma/R^{r, \bar{r}}_{s, \bar{s}}$, 
each subgraph is summed over all possible diagrams of its order, 
together with all possible (nested) subtractions. 
Since at each stage of the approximator, Ward identity is allowed to factorize collinear gluons out of the hard subgraph, and
soft gluons out of collinear subgraphs, each of the hard and the collinear factors involves factors of gauge links structures
like \eqs{eq:sudakov-soft-factor-s-modify}{eq:sudakov-col-factor-r}.
Thus even for the graph in \fig{fig:sudakov-reduced-diagram}(b) with subtractions in each subgraph, Ward identity can be 
applied equally to factorize different subgraphs. The resultant hard, collinear, and soft factors are at the corresponding orders,
but now with subtractions for smaller regions.

Since the soft subgraphs of any diagram in $\Gamma/R^{r, \bar{r}}_{s, \bar{s}}$ do not contain smaller regions, 
there is no further subtractions, and \eq{eq:sudakov-soft-factor-s-modify} is the final soft factor.

\begin{figure}[htbp]
	\centering
	\begin{tabular}{cc}
	\includegraphics[trim={0 0 -2em 0}, clip, scale=0.75]{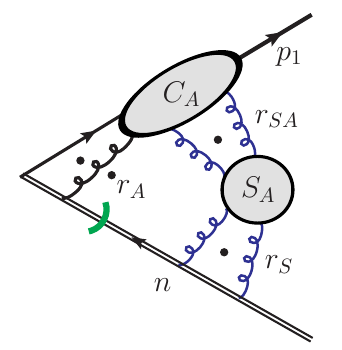} &
	\includegraphics[trim={-2em 0 -2em 0}, clip, scale=0.75]{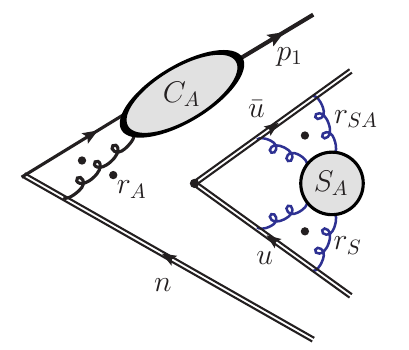}\\
	(a) & (b)
	\end{tabular}
	\caption{Factorization of the collinear subgraph $C_q$.
	(a) is one of its leading regions that has a collinear and a soft subgraphs within $C_q$.
	(b) is the result of factorizing the soft gluons onto Wilson lines.
	}
	\label{fig:sudakov-CA-fac}
\end{figure}

The collinear subgraphs, however, contain subtractions of smaller regions where some of the lines become soft.
For example, let us consider the factorized subgraph $C_q$ without subtraction. Extending its loop momenta to all regions
has converted it to the Green function expressions in \eq{eq:sudakov-col-factor-r}.
Now let us confine the loop momenta to the $q$-collinear region, and only allow them to reach the soft subregions.
Then this collinear subgraph $C_q$ can be further decomposed into two subgraphs, as shown in \fig{fig:sudakov-CA-fac}(a),
one collinear subgraph $C_A$ whose lines are collinear to $q$, and one soft subgraph $S_A$ collecting all the soft lines.

Each particular graph in $C_q$ can be decomposed as \eq{fig:sudakov-CA-fac}(a) in many ways, 
defining a whole set of regions. For each such region can be defined an approximator that inherits from 
the approximator for the whole graph in \fig{fig:sudakov-reduced-diagram}(b).
Since we only consider the two momentum regions, this approximator is denoted in \fig{fig:sudakov-CA-fac}(a)
by the thick green hooked line. To its left are the collinear momenta flowing along the Wilson line along $n$ with large
plus momentum components. 
The soft momenta are approximated as \eq{eq:soft-momenta-modify} when flowing into $C_A$ or across the 
hooked line, beyond which they are exactly zero on the Wilson line. 
Then, the same approximation allows us to factorize the soft gluons attached to $C_A$ to a Wilson line along $\bar{u}$,
while those attached to the Wilson line are directly separated from those collinear lines onto a separate Wilson line 
whose rapidity can be modified to $y$ without loss of the leading-power accuracy.
The result is shown in \fig{fig:sudakov-CA-fac}(b), for which we considered the specific case when 
$S_A$ is connected to $C_A$ via $r_SA$ soft gluon lines and to the Wilson line via $r_S$ ones.

\begin{figure}[htbp]
	\centering
	\begin{tabular}{cc}
	\includegraphics[trim={0 0 -2em 0}, clip, scale=0.75]{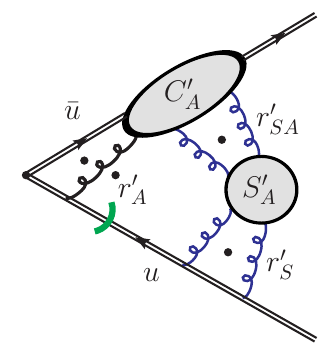} &
	\includegraphics[trim={-2em 0 -2em 0}, clip, scale=0.75]{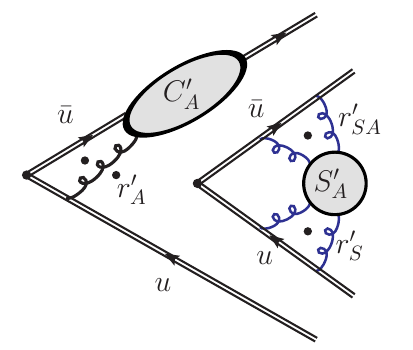}\\
	(a) & (b)
	\end{tabular}
	\caption{Factorization of the soft subgraph $S_A$ within the collinear subgraph $C_q$.
	(a) is one of its leading regions that has a collinear and a soft subgraphs within $S_A$, and
	(b) is the result of factorizing the soft gluons onto Wilson lines.
	}
	\label{fig:sudakov-SA-fac}
\end{figure}

In a certain nesting of regions that constitutes one subtraction term in the collinear subgraph $C_q$, 
a larger region would take some of the soft lines,
which have already been factorized into the $S_A$ factor in \fig{fig:sudakov-CA-fac}(b),
to collinear to $C_A$. 
This further decomposes $S_A$ into two subgraphs, $C_A'$ and $S_A'$, as shown in \fig{fig:sudakov-SA-fac}(a). 
The region approximator works in the same way, and we can factorize the soft gluons out of $C_A'$ onto 
Wilson lines of the same structure, one along $u$ and the other along $\bar{u}$.

This procedure can be carried on iteratively, with one minus sign for each iteration. 
For the $C_q$ at a certain order, summed over all graphs with the same number of gluons $r$,
the subtractions are on all possible such iterations, so we have
\begin{align}\label{eq:Cq-subtraction}
	C_q^{r, {\rm sub}}
	= C_q^{r} + 
		\sum_{n = 1}^r \sum_{s_1, \cdots, s_n} \, \sum_{r_0 + r_1 + \cdots r_n = r} \, (-1)^n \,
			C_q^{r_0} \cdot S^{s_1, r_1}(\bar{y}, y) \cdots S^{s_n, r_n}(\bar{y}, y),
\end{align}
where $s_i, r_i \geq 1$ for $i \geq 1$, 
and the sum of perturbative orders of all factors on the right-hand side is the same as $C_q^{r}$.
The $C_q^{r}$ is the unsubtracted collinear factor given in \eq{eq:sudakov-col-factor-r}.

The same subtraction procedure can be done for the other collinear factor $C_{\bar{q}}$, 
following the same idea of ``refactorization'', which leads a result similar to \eq{eq:Cq-subtraction}.
Then for the sum over diagrams in $\Gamma/R^{r, \bar{r}}_{s, \bar{s}}$, we achieve a factorization result,
\beq[eq:sudakov-factorization-fixed]
	\sum \Gamma/R^{r, \bar{r}}_{s, \bar{s}}
	= H^{\rm sub} \cdot C_q^{r, {\rm sub}} \cdot C_{\bar{q}}^{\bar{r}, {\rm sub}}
		\cdot S^{s, \bar{s}}(\bar{y}, y),
\eeq
where the hard factor with subtractions of smaller regions can be simply obtained by matching the left-hand side
onto the right-hand side with the definitions of the last three factors. 

Clearly, with careful subtraction included in \eq{eq:sudakov-factorization-fixed}, 
we can sum over graphs to all orders for each subgraph independently, 
and over $r$, $\bar{r}$, $s$, and $\bar{s}$. This is equivalent to a sum over all graphs
and all regions, so reproduces the complete leading-power contribution of the Sudakov form factor.
This converts the soft factor to a complete expression of Wilson lines,
\beq[eq:sudakov-soft-factor]
	S(\bar{y}, y) = \sum_{s, \bar{s} = 0}^{\infty} S^{s, \bar{s}}(\bar{y}, y) 
		= \frac{1}{N_c} \tr \, \vv{0 | W(\infty, 0; \bar{u}) \, W^{\dag}(\infty, 0; u) | 0 },
\eeq
where $s$ and $\bar{s}$ are either both 0 or both nonzero.
The sum over collinear subgraphs $C_q$ and $r$ in \eq{eq:Cq-subtraction} can be converted to independent
sums over all the indices, $n$, $s_i$, and $r_i$,
\begin{align}
	C_q^{\rm sub} & = \sum_{C_q, \, r} C_q^{r, {\rm sub}}
	= \sum_{C_q, \, r} C_q^{r} + 
		\sum_{n = 1}^{\infty} (-1)^n \cdot \sum_{s_i, \, r_i \geq 1} C_q^{r_0} \cdot S^{s_1, r_1}(\bar{y}, y) \cdots S^{s_n, r_n}(\bar{y}, y)
	\nn\\
	& =  \sum_{C_q, \, r}  C_q^{r} \cdot \bb{ 1 + \sum_{n = 1}^{\infty} (-1)^n \big( S(\bar{y}, y) - 1 \big)^n }	
	= \frac{C_q^{\rm unsub} }{ S(\bar{y}, y) },
\end{align}
where $C_q^{\rm unsub} \equiv \sum_{C_q, \, r} C_q^{r}$ sums over the graphs and $r$ in the unsubtracted collinear factor
in \eq{eq:sudakov-col-factor-r}.
In the second-to-last step we have used $S^{0, 0}(\bar{y}, y) = 1$.
To deal with rapidity divergences, we should have taken $n$ in \fig{fig:sudakov-CA-fac}(a) to be off light cone, so that
the collinear factor also has a rapidity cutoff $y$. Then we have the all-order collinear factors with subtractions,
\beq
	C_q^{\rm sub}(y_q, y) = \frac{C_q^{\rm unsub}(y_q, y)}{S(\bar{y}, y)}, \quad
	C_{\bar{q}}^{\rm sub}(\bar{y}, y_{\bar{q}}) = \frac{C_{\bar{q}}^{\rm unsub}(\bar{y}, y_{\bar{q}})}{S(\bar{y}, y)},
\eeq
where $y_q > 0$ and $y_{\bar{q}} < 0$ are the rapidities of external on-shell quark and antiquark, taken to be massive and physical.
Notably, the rapidity divergences associated with $y \to -\infty$ or $\bar{y} \to \infty$ are cancelled in both factors with subtraction.
Combined with the soft factor, we then have the all-order factorization result for the Sudakov form factor,
\beq[eq:sudakov-factorization]
	\Gamma_{\rm Sudakov}
	= H^{\rm sub} \cdot \frac{C_q^{\rm unsub}(y_q, y) \cdot C_{\bar{q}}^{\rm unsub}(\bar{y}, y_{\bar{q}})}{S(\bar{y}, y)}
		+ {\rm p.s.}
\eeq
Evidently, in this combination, the dependence on the artificially introduced rapidity cutoffs $y$ and $\bar{y}$ are also cancelled.
Now we see how including subtraction converts the soft factor into the denominator, contrary to what one might naively expected
from the structure of \fig{fig:sudakov-fac}(b).

To move forward, it helps to takes the limit $y \to -\infty$ and $\bar{y} \to \infty$. 
But this makes each factor in \eq{eq:sudakov-factorization} ill-defined so forbids a study of each of them alone.
Therefore, as introduced in~\citep{Collins:2011zzd}, one can define two reorganized collinear factors by noticing
the property $S(\bar{y}, y) = S(\bar{y} -  y)$ by Lorentz symmetry,
\begin{align}
	\wt{C}_q & = \wt{C}_q(y_q, y_n) 
		= C_q^{\rm unsub}(y_q, -\infty) \sqrt{ \frac{S(\infty, y_n)}{S(\infty, -\infty) \, S(y_n, -\infty)} }, 	\nn\\
	\wt{C}_{\bar{q}} &= \wt{C}_{\bar{q}}(y_n, y_{\bar{q}})
		= C_{\bar{q}}^{\rm unsub}(\infty, y_{\bar{q}}) \sqrt{ \frac{S(y_n, -\infty)}{S(\infty, -\infty) \, S(\infty, y_n)} },
\label{eq:sudakov-col-factors-reorganize}
\end{align}
with a new rapidity separator parameter $y_n$.
Then \eq{eq:sudakov-factorization} becomes
\beq[eq:sudakov-factorization-reorganize]
	\Gamma_{\rm Sudakov}
	= H^{\rm sub} \cdot \wt{C}_q(y_q, y_n) \cdot \wt{C}_{\bar{q}}(y_n, y_{\bar{q}}) + {\rm p.s.}
\eeq
Requiring invariance of \eq{eq:sudakov-factorization-reorganize} with respect to $y_n$ leads to an evolution equation
that allows to resum large logarithms. We will not cover this topic in this thesis.

Contrary to DIS, the Sudakov form factor is only a perturbative amplitude and does not directly correspond to a physical process. 
In QCD, quarks do not appear as observed particles, and massless external states cause infrared divergences; in our analysis,
we have implicitly added nonzero masses to the quarks as infrared regulators. The infrared divergences due to the massless
gluon are also implicitly regulated by methods like dimensional regularization which we have not specified carefully.

Because of this, the collinear factors in \eq{eq:sudakov-col-factors-reorganize}, albeit as nonperturbative as the PDFs,
are not to be obtained from fitting to experimental data.
The factorization analysis, however, serves as a very useful prototype for discussing factorization in more complicated 
{\it and physical} situations, including the single-hadron inclusive process in leptonic collisions, the SIDIS, and the Drell-Yan
processes. 
Another important ingredient needs to be developed there.
Because any observed particles in the final state induce pinch singularities, we may only talk about inclusive observables,
for which summing over all other unobserved particles allows us to employ the unitarity of QCD to show that all those
associated pinch singularities are cancelled.
This topic is not of direct relevance to the rest of this thesis, so will not be further reviewed.

The focus of the rest of this thesis is on exclusive processes, which are similar to the exclusive Sudakov form factor amplitude
but differ by concerning with only hadronic external states.
Similar decomposition of a region into hard, collinear, and soft subgraphs is to be carried out, with slightly different structures
for each factor. 
As we will show, while working with exclusive processes with a single hard scattering $H$, 
each hadron is connected to $H$ via a collinear set of parton lines that are very close to each other
so behave as color singlet as a whole, to the leading-power accuracy.
Soft gluons attached to them are therefore cancelled.
This mechanism of cancellation differs from those inclusive processes that use unitarity,
and is a direct result of color confinement.


\chapter{QCD Factorization of exclusive processes}
\label{ch:exclusive}


We have reviewed in \ch{ch:factorization} the main principles and methodology of QCD factorization, which applies to 
hadronic scattering processes with one hard scale $Q$ much greater than $\LQCD$. Normally, this would localize the interaction 
to become sensitive to the partonic degrees of freedom in the hadron(s). To the leading power in $\LQCD / Q$, only one 
parton enters
\footnote{A similar story holds for the inclusive hadron production process where one parton leaves the hard interaction
and initializes a jet of hadrons.}
the hard interaction. This breaks the incoming hadron into colored objects, which exchange soft gluons to neutralize the colors.
The final state is then a series of hard jets surrounded by soft hadrons. Any query about a specific soft hadron would touch the 
long-distance nonperturbative QCD dynamics and go beyond the control of a perturbative method. 
Hence in such situations it is more sensible to study inclusive observables, 
which ``inclusively sums over" (a fancier way of saying ``neglecting") anything else besides 
the directly observed hard particles or quantities.
Such processes are called inclusive processes, where
the unitarity sum cancels the soft gluon exchanges between different collinear sectors 
and establishes independent parton density functions or fragmentation functions.
The universality of those nonperturbative functions gives QCD factorization predictive power, and allows them to be measured
to reveal certain aspects of the hadron structures.

It should be noted that inclusiveness is not the absolutely necessary condition for well-defined observables in hadronic scattering,
but more of a practically convenient choice against our inability to deal with the nonperturbative soft regime. This is in contrast to
the soft divergences in QED: There the massless photons pose infrared divergences in both virtual and 
real processes, and it is only the inclusive observables that are well defined, 
given that arbitrarily soft photons can never be detected by an equipment of a finite size. 
In QCD, however, due to the color confinement, all final-state particles can in principle be captured by detectors --- 
no soft gluons elude the observation. 
Hence, it is in principle sensible to talk about the amplitude or cross section for producing a certain number of particles
of given types and momenta. Such processes are termed {\it exclusive processes}. 

In practice, including the context of this thesis, exclusive processes usually refer to a narrower class of processes 
in which hadrons are {\it unbroken}, given that 
the multiple soft radiations triggered by broken hadrons are easily intractable, 
both theoretically and experimentally. 
In this sense, we divide the exclusive processes to be discussed in this thesis into three types:
\begin{itemize}
\item
{\it large-angle scattering}, referring generally to a hadronic $2 \to n$ process 
in which the final-state hadronic particles are hard and well separated,
and no hadrons are found in the direction(s) of the hadron beam(s),
\item
{\it single-diffractive scattering}, which is similar to the previous case, but has one diffracted hadron in one of the hadron beam directions, and
\item 
{\it double-diffractive scattering}, which has one diffracted hadron in both hadron beam directions.
\end{itemize}
We only discuss at most $2\to 2$ processes for the large-angle scattering, with $n > 2$ a trivial generalization. 
The minimal configuration involves only one hadron, and 
the maximal one is a scattering of two hadrons into two other hadrons. 
At a high collision energy, a large-angle scattering happens at such a local region, characterized
by the inverse of the transverse momentum scale of the final-state particles,
that soft gluons communicating different collinear sectors at a long-distance scale are cancelled, 
leading the {\it amplitudes} to be factorized into hadron {\it distribution amplitudes} (DAs). 
For the single-diffractive scattering, we can similarly show the soft cancellation, and then the diffraction subprocess is factorized into generalized 
parton distributions (GPDs). 
For the double-diffractive scattering and beyond, however, we will show that soft gluons can be pinched in the Glauber region, 
which prohibits a factorization theorem from being derived.

By the exclusive nature of such processes, each hadron is connected to the hard scattering by at least two partons. 
This makes them more power suppressed than inclusive processes, 
by the counting rules in Table~\ref{tab:power-counting}. 
Hence, exclusive processes are more suitably studied at low energy scattering, 
while as the colliding energy increases, hadrons are more likely to break, leading to the inclusive
regime. Nevertheless, the universal parton correlation functions, the DAs and GPDs, obtained from the factorization of exclusive processes, provide
valuable information on the hadron structures complementary to the correlation functions obtained from inclusive processes, 
as will be discussed in more details in \ch{ch:GPD}.


\section{Large-angle exclusive meson scattering}
\label{sec:large-angle-meson}

We confine our discussion within large-angle $2\to 2$ exclusive meson scatterings,
\beq
	A(p_1) + B(p_2) \to C(q_1) + D(q_2),
\eeq
in which we always take $A$ as a meson.
These processes can be categorized according to the beam particle $B$, and we look at three types of processes: 
(1) electron-meson scattering, with $B = e^-$,
(2) photon-meson scattering, with $B = \gamma$, and
(3) meson-meson scattering, with $B = $ meson.
The factorization discussion can be trivially adapted to other processes with no mesons in the initial states or processes involving baryons.

\subsection{Large-angle electron-meson scattering}
\label{ssec:exclusive-em}
To the leading order (LO) in QED, the beam particle $B = e^-$ is scattered into the final state, so
we take $C = e^-$ as well. 
The other particle $D$ can be either a photon or a meson, which we now discuss sequentially.

\subsubsection{Single-meson process: $D = \gamma$}
\label{sssec:em2ea}
\begin{figure}[htbp]
	\centering
	\begin{tabular}{ccc}
	\includegraphics[scale=0.65]{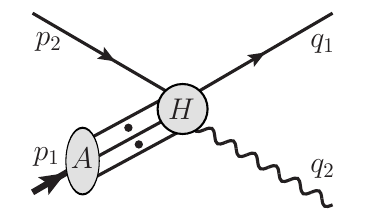} &
	\includegraphics[scale=0.65]{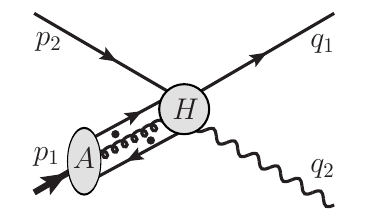} &
	\includegraphics[scale=0.65]{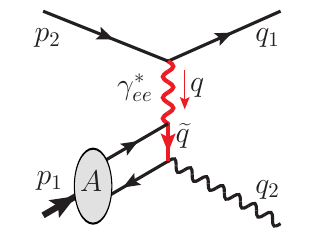} \\
	(a) & (b) & (c) 
	\end{tabular}
	\caption{(a) Reduced diagram for a general pinch surface of the exclusive scattering process 
	$\pi^0(p_1) + e(p_2) \to e(q_1) + \gamma(q_2)$. 
	The dots represent an arbitrary number of collinear lines.
	(b) is the leading region, where the dots alongside the gluon line represent an arbitrary number of collinear longitudinally polarized gluons.
	(c) is one LO diagram. Reversing the fermion arrow gives the other LO diagram.}
	\label{fig:em2ea}
\end{figure}

First, electric charge conservation constrains the meson $A$ to be neutral, so for simplicity, we take $A = \pi^0$ to be the charge-neutral pion.
The scattering 
\beq[eq:em2ea]
	\pi^0(p_1) + e(p_2) \to e(q_1) + \gamma(q_2)
\eeq
thus gives the $\pi$-$\gamma$ transition form factor~\citep{Lepage:1980fj}. 
As usual, we define the Mandelstam variables
\beq[eq:em2ea-mandelstam]
	s = (p_1 + p_2)^2, \quad
	t = (p_1 - q_1)^2, \quad
	u = (p_1 - q_2)^2.
\eeq
We work in the c.m.~frame, where $A$ always moves along the $+\hat{z}$ direction, 
and $e(q_1)$ has a transverse momentum $\vec{q}_T$.
In the limit $s \to \infty$ while $t / s$ and $u / s$ stay constant, i.e., $s\to \infty$ with $q_T / \sqrt{s}$ constant, 
the pion is connected to the hard scattering
via a set of collinear lines, as shown by the reduced diagram in \fig{fig:em2ea}(a). 
Following the Landau criterion in \sec{ssec:landau}, it represents a
general pinch surface in the parton loop momentum space that possibly gives mass divergences. 
The most general pinch surface here can also have arbitrarily
many soft lines connecting $A$ and $H$, but they are power suppressed by 
the general power counting rule in \sec{sec:power-counting}, so are neglected.

The power counting rules for the collinear lines, as discussed in \sec{sec:power-counting}, can be summarized as
(1) one collinear fermion line or transversely polarized gluon line is associated with a power $\lambda / Q$, where 
$Q = \order{q_T} = \order{\sqrt{s}}$ and $\lambda = \order{\LQCD} = \order{m_{\pi}, f_{\pi}}$, 
with $m_{\pi}$ and $f_{\pi}$ the pion mass and decay constant, respectively,
and 
(2) a longitudinally polarized gluon line is associated with a power $(\lambda / Q)^0$.
Hence, the leading region should have two collinear quark or 
transversely polarized gluon lines connecting $A$ to $H$, 
together with arbitrarily many longitudinally polarized gluons. 
The pure gluon channel violates isospin symmetry.
So we only have one type of leading regions,  
with one collinear subgraph and a hard subgraph, joined by a pair of quark lines and arbitrarily many gluon lines of longitudinal polarization, 
similar to the DIS shown in \fig{fig:DIS-LR}.
One example of the LO diagrams is shown in \fig{fig:em2ea}(c), where the scattered electron exchanges a highly-virtual photon $\gamma^*_{ee}$ with
the quark. The latter is then excited to a high virtuality. After a short lifetime, it annihilates with the antiquark to emit a real photon. Exchanging the roles of quark and 
antiquark gives the other LO diagram.

One can immediately notice the difference of exclusive processes from inclusive ones 
like \figs{fig:DIS-LR}{fig:DIS-LR-1g} that now a collinear subgraph in the amplitude 
is connected to the hard subgraph by at least two parton lines. 
This is because the hadron participating in the exclusive process must be a color singlet. 
To stay intact, they must only exchange a color-singlet state with the hard interaction. 
As a result, the leading power for the exclusive amplitude in \eq{eq:em2ea}
already counts as $(\lambda / Q)^1$.

The factorization works in a way similar to DIS treated in \sec{sec:ward-identity}. 
We include all the collinear propagators in the collinear subgraph $C$.
For each $k_i$ flowing in $H$, which scales as 
\beq[eq:e2em-col-scaling]
	k_i \sim (Q, \lambda^2/Q, \lambda),
\eeq 
we approximate it by only retaining the plus component,
\beq[eq:e2em-col-mom-approx]
	k_i \to \hat{k}_i = (k_i \cdot n) \bar{n},
\eeq
where we used the same light-like auxiliary vectors defined in \eq{eq:n-nbar}.
We project on shell the quark and antiquark lines external to $H$ by inserting the Dirac matrices, respectively,
\beq[eq:e2em-col-q-approx]
	\P_A = \frac{\gamma\cdot \bar{n} \gamma\cdot n}{2} = \frac{\gamma^- \gamma^+}{2}
	\quad \mbox{ and } \quad
	\Pb_A = \frac{\gamma\cdot n \gamma\cdot \bar{n}}{2} = \frac{\gamma^+ \gamma^-}{2}.
\eeq
Each gluon has its polarization dominantly proportional to its momentum, so we approximate its connection to $H$ by 
\beq[eq:e2em-col-g-approx]
	H_{\mu_i}(k_i) g^{\mu_i \nu_i} C_{\nu_i}(k_i) 
	\mapsto
	H_{\mu_i}(\hat{k}_i) \frac{\hat{k}_i^{\mu_i} n^{\nu_i} }{ k_i \cdot n - i\epsilon} C_{\nu_i}(k_i),
\eeq
for a particular gluon of momentum $k_i$ flowing {\it into} $H$. Since there is no soft region here, the $i\epsilon$ is not important; we added it only for convention.

\begin{figure}[htbp]
	\centering
		\begin{align*}
			\eqfig[0.5]{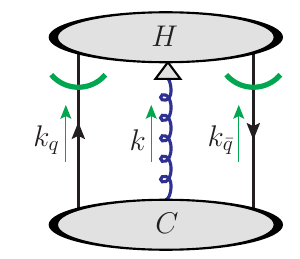}
			=
				- \eqfig[0.5]{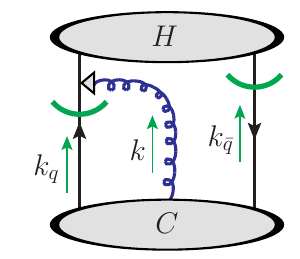}
				- \eqfig[0.5]{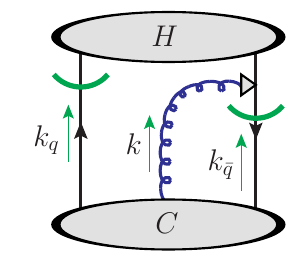}
			= \eqfig[0.5]{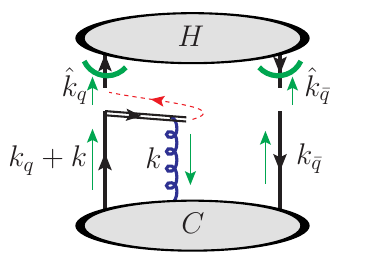} +
				\eqfig[0.5]{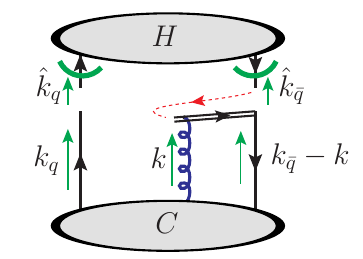} 
		\end{align*}
	\caption{Graphic representation of the two steps to detach a longitudinally polarized collinear gluon from the collinear subgraph $C$ to the hard subgraph $H$, 
	and reconnect it to corresponding gauge links of the $C$. The red thin dashed lines represent the color flows.
	}
\label{fig:em2ea-ward-1g}
\end{figure}

The approximator defined in Eqs.~\eqref{eq:e2em-col-mom-approx}--\eqref{eq:e2em-col-g-approx} 
can be collectively denoted as $\hat{T}$. 
It acts on one leading region $R$, which has the decomposition into a hard subgraph 
$H_n$ and and a collinear subgraph $C_n$ as in \fig{fig:em2ea}(b), 
of a certain diagram $\Gamma$, where $n$ is the number of gluons connecting 
$H_n$ and $C_n$ alongside the quark and antiquark lines. 
The leading-power contribution from the region $R$ can be obtained by applying $\hat{T}$, 
with the contribution from smaller regions subtracted,
\beq[eq:e2em-T]
	C_{R } \Gamma = H_n^{\rm sub} \cdot \hat{T} \cdot C_n.
\eeq
Since a smaller region with respect to $R$ necessarily has some lines in $H_n$ belong to the collinear region,
the subtraction in $C_{R } \Gamma$ only affects $H_n$.
In \eq{eq:e2em-T}, $H_n^{\rm sub}$ is the hard subgraph with subtraction for smaller regions,
and $\hat{T}$ acts to the left on $H_n^{\rm sub}$ by 
neglecting small components of collinear momenta (as specified in \eq{eq:e2em-col-mom-approx}) 
and inserting spinor and Lorentz projectors (as specified in \eqs{eq:e2em-col-q-approx}{eq:e2em-col-g-approx}).
$H_n^{\rm sub}$ can be further written in a form like \eq{eq:DIS-H-sub} after summing over graphs.

By our assumptions (1) and (3) in \sec{ssec:fac-assumption}, 
the amplitude of \eq{eq:em2ea} is given by the sum over all graphs.
Each region $R$ of a graph $\Gamma$ is uniquely specified by 
the hard subgraph $H$ and collinear subgraph $C$. 
Varying $H$ with a given $C$ or vice versa corresponds to a different graph. 
Summing over all regions of all diagrams is equivalent to summing over the subgraphs $H$ and $C$ individually.
And for each given subgraph $H$, the associated subtraction for smaller regions is also uniquely determined.
Now for a given $C_n$, we sum over all possible diagrams in $H_n$ at a given $n$ and perturbative order
{\it and} over all possible attachments of the collinear gluons onto the hard subgraph.
This, together with the $H(\hat{k}) \cdot \hat{k}$ structure after the approximation in \eq{eq:e2em-col-g-approx}, 
allows the use of Ward identity for the collinear gluons, 
which also equally applies to each subtracted term in $H_n^{\rm sub}$ 
(as illustrated in \sec{ssec:DIS-Feynman-sub}). 

Due to the presence of two quark lines, the Ward identity results in {\it two} gauge links that 
collect all the collinear gluons, different from the Sudakov form factor in \sec{sec:sudakov-factorization}.
This can be easily demonstrated for $n = 1$, as shown in \fig{fig:em2ea-ward-1g},
by the same method of finding the missing terms as in \fig{fig:DIS-LR-1g}.
Each quark is connected by a gauge link that goes along the lightcone direction $n$
to $\infty$ in the future, and the gluon can be connected to either one.
The same procedure can be applied inductively to an arbitrary $n$, and the result only depends on the colors
of the external quark legs. Therefore, as in \sec{ssec:DIS-Feynman-wilson}, we would reach the same Ward identity
result by attaching all the gluons to two gauge links.
The vertices and propagators along the gauge links are the same as those that can be obtained from the operator%
\footnote{\eq{eq:Wilson-line-operator-n} only specifies the momentum $k$ going out from the quark-to-Wilson-line vertex. 
The momentum from the antiquark-to-Wilson-line vertex is $p_1 - k$, determined by momentum conservation. }
\beq[eq:Wilson-line-operator-n]
	\int d^4y e^{i k \cdot x} \cc{ \bb{ \bar{\psi}_q(0) W^{\dag}(\infty, 0; n) }_j \bb{ W(\infty, y; n) \psi_q(y) }_i }^n,
\eeq
which has been expanded to the $n$-th order in $g$, and where $\psi_q$ is the quark field of flavor $q$, 
$W(\infty, y; n)$ is the Wilson line from $y$ to $\infty$ along $n$, as defined in \eq{eq:DIS-wilson-line},
and $i$, $j$ are color indices in the fundamental representation. 
As \eq{eq:F-a-b-l-r}, going from the diagrammatic expressions to \eq{eq:Wilson-line-operator-n} entails
a relabelling of gluon momenta and colors.
Different from \eq{eq:F-a-b-l-r} though, the expansion of \eq{eq:Wilson-line-operator-n} 
to the $n$-th order allows arbitrary numbers of gluons attached to either gauge link,
whereas \eq{eq:F-a-b-l-r} contains a cut line so the number of gluons assigned to each side is fixed.

The hard subgraph then only has two external quark lines, with momenta $k$ and $p_1 - k$, respectively. 
So we can write \eq{eq:e2em-T} as
\beq
	H^{\rm sub}(\hat{k}, \hat{p}_1 - \hat{k}) \otimes \bb{ \hat{T}_w \cdot C_n }(k; p_1),
\eeq
where $\hat{T}_w$ acting on $C_n$ is not different from $\hat{T}$, but just refers to the fact that Ward identity has been used to attach all collinear gluons onto gauge links
that only belong to $C_n$. Then summing over all possible diagrams for $H_n$ and $C_n$ gives a factorized result,
\beq[eq:em2ea-sum-over-regions]
	\sum_{R, \Gamma} C_{R } \Gamma 
	= \int \frac{d^4k}{(2\pi)^4}
		\H_{\beta\alpha; ji}(\hat{k}, \hat{p}_1 - \hat{k}) \,
		\C_{\alpha\beta; ij}(k; p_1),
\eeq
where we have left explicit the momentum convolution, and the dependence on colors and spinor indices.
In \eq{eq:em2ea-sum-over-regions}, 
$\H = \sum_{H} H^{\rm sub}$ is the subtracted hard subgraph with gluons factored out, 
and $\C = \hat{T}_w \sum_C C$ is the collinear subgraph, both being summed over all diagrams. 
Since there is no any subtraction involved in the collinear subgraph, we can write it as a matrix element form, 
by extending \eq{eq:Wilson-line-operator-n} to all orders,
\begin{align}\label{eq:em2ea-col-factor-0}
	\C_{\alpha\beta; ij}(k; p_1) 
	= & \P_{A, \alpha\alpha'} 
		\int d^4y e^{i k \cdot y} \langle 0 | \T \cc{ 
			\bb{ \bar{\psi}_{q, \beta'}(0) W^{\dag}(\infty, 0; n) }_j  
		\right.\nn\\
		& \hspace{8em}
		\left. \times
			\bb{ W(\infty, y; n) \psi_{q, \alpha'} (y) }_i 
		} | \pi^0(p_1) \rangle
		\Pb_{A, \beta'\beta},
\end{align}
where we have left explicit the dependence on colors and spinor indices, and 
have included the spinor projectors in \eq{eq:e2em-col-q-approx}. 
By construction, \eq{eq:em2ea-sum-over-regions} approximates the amplitude of 
\eq{eq:em2ea} at the leading power.

Now $\H$ is only convoluted with $\C$ by
(1) color indices $i$ and $j$,
(2) spinor indices $\alpha$ and $\beta$, and
(3) the plus component of the quark (or antiquark) momentum $k$.
Since the pion is color neutral, only the color singlet component of $\C_{\alpha\beta; ij}$ is nonzero, so we can define a gauge invariant factor $\C_{\alpha\beta}$
by summing over the color diagonal elements,
\beq[eq:em2ea-color]
	\C_{\alpha\beta; ij}(k; p_1)  = \frac{1}{N_c} \delta_{ij} \, \C_{\alpha\beta}(k; p_1).
\eeq
Similar to \sec{ssec:DIS-Feynman-lo-pdf},
by expanding $\C_{\alpha\beta}(k; p_1)$ in terms of the 16 independent Dirac matrices, 
we can see that only the $\gamma^-$, $\gamma_5\gamma^-$, and
$\sigma^{-\perp}$ components survive under the projection of $\P_A$ and $\Pb_A$. 
The pseudoscalar nature of $\pi^0$ further kills the $\gamma^-$ and
$\sigma^{-\perp}$ components, ending up with only one spinor structure,
\beq[eq:em2ea-spinor]
	 \C_{\alpha\beta}(k; p_1) 
	= \pp{ \frac{\gamma_5 \gamma\cdot \bar{n}}{2} }_{\alpha\beta} \C(k; p_1) 
\eeq
with the coefficient being a singlet in both color and spinor space,
\beq
	\C(k; p_1) 
	= \int d^4y e^{i k \cdot y} \langle 0 | \T \cc{ 
			\bb{ \bar{\psi}_{q, \alpha}(0) W^{\dag}(\infty, 0; n) }_i 
			\frac{\gamma\cdot n \gamma_5}{2} 
			\bb{ W(\infty, y; n) \psi_{q, \alpha} (y) }_i 
		} | \pi^0(p_1) \rangle.
\eeq
Finally, for the momentum convolution, since the hard part $\H$ only depends on $k \cdot n$, we insert into \eq{eq:em2ea-sum-over-regions} the factor
\beq[eq:identity-factor]
	1 = \int dx \, \delta\pp{ x - \frac{k \cdot n}{p_1 \cdot n} } 
	= (p_1 \cdot n)  \int dx \, \delta\pp{ x \, p_1 \cdot n - k \cdot n }
	= (p_1 \cdot n)  \int dx \int \frac{d\lambda}{2\pi} e^{i \lambda (x \, p_1 \cdot n - k \cdot n) }.
\eeq
Together with the color and spinor factors in Eqs.~\eqref{eq:em2ea-color} and \eqref{eq:em2ea-spinor}, 
the convolution in \eq{eq:em2ea-sum-over-regions} becomes
\begin{align}\label{eq:em2ea-factorize0}
	&\int dx \bb{ (p_1 \cdot n) \frac{1}{N_c} \delta_{ij} 
			\pp{ \frac{\gamma_5 \gamma\cdot \bar{n}}{2} }_{\alpha\beta}
			\H_{\beta\alpha; ji}(x\hat{p}_1, (1 - x) \hat{p}_1) 
		} \, \nn\\
		& \hspace{8em} 
		\times
		\bb{
			\int \frac{d\lambda}{2\pi} \int \frac{d^4k}{(2\pi)^4}
			e^{i \lambda (x \, p_1 \cdot n - k \cdot n) }
			\C(k; p_1)
		},
\end{align}
This completes the derivation of factorization for the amplitude,
\beq[eq:em2ea-factorize]
	\M_{\pi^0 e \to e \gamma} 
	= \sum_q \int dx \, D_{q / \pi^0}(x) \, \H_q(x; \vec{q}_T, s) + \order{\LQCD / q_T},
\eeq
where we have left explicit the sum over the quark flavor $q$, and changed the notation $\C$ to define the DA for $\pi^0$,
\begin{align}\label{eq:em2ea-DA}
	&D_{q / \pi^0}(x)
	= \int \frac{d\lambda}{2\pi} \int \frac{d^4k}{(2\pi)^4}
			e^{i \lambda (x \, p_1 \cdot n - k \cdot n) }
			\C(k; p_1)	\nn\\
	&\hspace{2em}= \int_{-\infty}^{\infty} \frac{d\lambda}{2\pi} 
		e^{i \lambda  x \, p_1 \cdot n }
		\langle 0 | \T \cc{ 
			\bb{ \bar{\psi}_{q}(0) W^{\dag}(\infty, 0; n) }
			\frac{\gamma\cdot n \gamma_5}{2} 
			\bb{ W(\infty, \lambda n; n) \psi_{q} (\lambda n) }
		} | \pi^0(p_1) \rangle \nn\\
	&\hspace{2em}= \int_{-\infty}^{\infty} \frac{d\lambda}{2\pi} 
		e^{i \lambda  x \, p_1 \cdot n }
		\langle 0 | \T \cc{ 
			\bar{\psi}_{q}(0)
			\frac{\gamma\cdot n \gamma_5}{2} 
			W(0, \lambda n; n) \psi_{q} (\lambda n)
		} | \pi^0(p_1) \rangle.
\end{align}

The integration of $k^-$ and $\vec{k}_T$ sets the operator on the light cone $n$, along which the operators have canonical commutation relations.
Then we can equivalently remove the time ordering in \eq{eq:em2ea-DA}.
It can also be shown by the analyticity properties of $\C(k; p_1)$ as a scattering amplitude under the integration of 
$k^-$~\citep{Diehl:1998sm}, following the assumption (3) in \sec{ssec:fac-assumption} and that the analyticity properties are 
the same as the corresponding perturbative Feynman diagrams.
This would allow the insertion of physical states,
\begin{align}\label{eq:em2ea-DA-X}
	D_{q / \pi^0}(x)
	& = \sum_X \int_{-\infty}^{\infty} \frac{d\lambda}{2\pi} 
		e^{i \lambda  x \, p_1 \cdot n }
		\langle 0 | 
				\bb{ \bar{\psi}_{q}(0) W^{\dag}(\infty, 0; n) }
		| X \rangle 
			\frac{\gamma\cdot n \gamma_5}{2} \nn\\
		& \hspace{5em} \times
		\langle X | 
			\bb{ W(\infty, \lambda n; n) \psi_{q} (\lambda n) }
		| \pi^0(p_1) \rangle \nn\\
	& = \sum_X \delta\pp{ p_X \cdot n - (1-x) p_1 \cdot n }
		\langle 0 | 
				\bb{ \bar{\psi}_{q}(0) W^{\dag}(\infty, 0; n) }
		| X \rangle 
			\frac{\gamma\cdot n \gamma_5}{2} \nn\\
		& \hspace{5em} \times
		\langle X | 
			\bb{ W(\infty, 0; n) \psi_{q} (0) }
		| \pi^0(p_1) \rangle,
\end{align}
where momentum conservation constrains the total plus momentum of the state $X$,
\beq
	p_X^+ = (1 - x) p_1^+.
\eeq
For $X$ to be a physical state, we must require $p_X^+ \geq 0$, so that $x \leq 1$.
On the other hand, by using the canonical commutation relation for $\psi$ and $\bar{\psi}$, \eq{eq:em2ea-DA-X} can also be written as
\begin{align}\label{eq:em2ea-DA-X2}
	D_{q / \pi^0}(x)
	& = - \sum_X \int_{-\infty}^{\infty} \frac{d\lambda}{2\pi} 
		e^{i \lambda  x \, p_1 \cdot n }
		\Tr\cc{
		\frac{\gamma\cdot n \gamma_5}{2}
		\langle 0 | 
			\bb{ W(\infty, \lambda n; n) \psi_{q} (\lambda n) }
		| X \rangle 
		\right. \nn\\
		& \hspace{6em} \left. \times
		\langle X | 
			\bb{ \bar{\psi}_{q}(0) W^{\dag}(\infty, 0; n) }
		| \pi^0(p_1) \rangle
		} \nn\\
	& = - \sum_X \delta\pp{ p_X \cdot n - x\, p_1 \cdot n }
		\Tr\cc{
		\frac{\gamma\cdot n \gamma_5}{2}
		\langle 0 | 
			\bb{ W(\infty, 0; n) \psi_{q} (0) }
		| X \rangle 
		\right. \nn\\
		& \hspace{6em} \left. \times
		\langle X | 
			\bb{ \bar{\psi}_{q}(0) W^{\dag}(\infty, 0; n) }
		| \pi^0(p_1) \rangle
		},
\end{align}
where $\Tr$ takes the spinor and traces.
Now we have
\beq[eq:em2ea-x-2]
	p_X^+ = x \, p_1^+ \geq 0,
\eeq
which requires $x \geq 0$.
Together, we must have $x \in [0, 1]$ for the DA to be nonzero. 
Therefore, we should constrain the $x$ integration in \eq{eq:em2ea-factorize} to be from 0 to 1.

Such constraint is not a mandatory condition inherent from factorization, 
but as a result of having the operator on light cone in collinear factorization
and setting $x = k^+ / p_1^+$ on the real axis.
This then causes a problem of endpoint singularity. 
We note that the above approximations $\hat{T}$ defined in 
Eqs.~\eqref{eq:e2em-col-mom-approx}--\eqref{eq:e2em-col-g-approx} 
is true only for the scaling in \eq{eq:e2em-col-scaling}, 
which corresponds to the pinch surface whose surrounding region gives the leading-power contribution to the amplitude. 
In principle, one should keep the scaling $k_i^+ \sim \order{Q}$ throughout the factorization analysis. 
Nevertheless, in the result of factorization, \eq{eq:em2ea-factorize}, the variable $x$ is integrated from $0$ to $1$, 
so that we have to include the region where one of the active partons has momentum $k_i^+ \ll Q$. 
Perturbatively, this does not lead to a pinch, so we should have deformed the contour of $k_i^+$ by $\order{Q}$ 
to make the associated propagator in the hard subgraph to have high virtuality. 
For example, as shown later in \eq{eq:DVCS-hel-amps-hard},
the LO hard coefficient contains a term that is proportional to $1/(x - i\epsilon)Q^2$ which becomes soft as $x\to 0$, 
and we should deform the contour of $x$ to the lower half complex plane to make $\Im x \sim \order{1}$. 
Similar issue arises as $x\to 1$. 
However, since the DA only has support in $x\in[0,1]$, such deformation is forbidden by the end points of the $z$ integration. 
Therefore, the validity of the DA factorization in \eq{eq:em2ea-factorize} needs to be supplemented with an additional assumption 
that the end point region should be strongly suppressed by the DA, which we refer to as {\it soft-end suppression}. 
This situation could be improved by the Sudakov suppression factor introduced in \citep{Li:1992nu}.
Further work is needed on this issue.
 
So far, we have been working with the bare DA and hard coefficient, 
without caring for the possible UV divergences introduced by the approximator $\hat{T}$.
The original amplitude $\M_{\pi^0 e \to e \gamma}$ contains no UV divergence. 
But the approximator $\hat{T}$ short-circuits the integration of $k^-$ and 
$\vec{k}_T$ into $\C$ in \eq{eq:em2ea-factorize0}, and extends the integration to infinity. 
This introduces an (artificial) UV divergence.
Since the hard coefficient $\H$ is defined with subtraction of smaller regions, 
which are themselves factorized in the same way the DA is factorized (see \eq{eq:DIS-H-sub}),
whatever UV divergences introduced by $\hat{T}$ in the DA have been compensated by the subtraction in $\H$. 
In this way, both the DA and $\H$ contain UV divergences, which cancel each other and make up a finite convolution result in \eq{eq:em2ea-factorize}. 
Nevertheless, it would be nice to define a {\it renormalized} DA by taking off the UV divergences therein. 

Similar to the PDF renormalization in \sec{ssec:dis-uv}, the DA is also renormalized multiplicatively,
\beq[eq:em2ea-uv-ren-DA]
	D_{q/\pi^0}(x, \mu) = \int_0^1 dz \, Z(x, z; \alpha_s(\mu); \epsilon^{-1} ) D^{\rm bare}_{q/\pi^0}(z; \epsilon^{-1} ),
\eeq
with an invertible renormalization coefficient,
\beq[eq:em2ea-uv-div]
	D^{\rm bare}_{q/\pi^0}(x; \epsilon^{-1}) 
	= \int_0^1 dz \, Z^{-1}(x, z; \alpha_s(\mu); \epsilon^{-1}) D_{q/\pi^0}(z, \mu).
\eeq
This introduce a factorization scale $\mu$ dependence in the renormalized DA, 
which will lead to an evolution equation (not to be discussed in this thesis).
We note that due to the lack of gluon channel, the DA renormalization 
has no mixing between the quark and gluon or between different quark flavors.
Restoring the ``bare" notation in \eq{eq:em2ea-factorize} and substituting 
\eq{eq:em2ea-uv-div} for the bare DA, we get the same factorization formula for
the amplitude, but now in terms of 
the UV renormalized DA and infrared (and UV) finite hard coefficient,
\beq[eq:em2ea-factorize-ren]
	\M_{\pi^0 e \to e \gamma} 
	= \sum_q \int_0^1 dx \, D_{q / \pi^0}(x, \mu) \, \H_q\pp{x; \frac{\vec{q}_T}{\sqrt{s}}, \frac{q_T}{\mu}}  
		+ \order{\LQCD / q_T},
\eeq
which looks like \eq{eq:em2ea-factorize} but has an extra factorization scale dependence. 
The renormalized hard coefficient is related to the bare one by
\beq
	\H_q\pp{x; \frac{\vec{q}_T}{\sqrt{s}}, \frac{q_T}{\mu}} 
	= \int_0^1 dz \, Z^{-1}(x, z; \alpha_s(\mu); \epsilon^{-1}) \, \H^{\rm bare}(z; \vec{q}_T, s; \epsilon^{-1}).
\eeq

In this way, we finished proving the factorization for the amplitude of \eq{eq:em2ea}. 
We have not only obtained the operator definition for the pion DA, 
given in Eqs.~\eqref{eq:em2ea-DA} and renormalized in \eqref{eq:em2ea-uv-ren-DA},
but also provided a practical procedure for calculating the hard coefficient to all perturbative orders. 
By projecting the pion state in \eq{eq:em2ea-factorize-ren} to an on-shell parton-pair state, 
we can expand both sides order by order and obtain the hard coefficient at each order by an iterative matching.

\subsubsection{Double-meson process: $D = $ meson}
\label{sssec:em2em}
Now we discuss the electron induced meson production. 
Similarly, the electromagnetic current does not change the flavor of the meson, 
so for concreteness we take $A = D = \pi^+$, with the scattering of other mesons generalized in a trivial way. 
The process 
\beq
	\pi^+(p_1) + e(p_2) \to e(q_1) + \pi^+(q_2)
\label{eq:em2em}
\eeq
probes the electromagnetic pion form factor. 
The kinematics are also defined as in \eq{eq:em2ea-mandelstam}. We work in the c.m.~frame with the $\pi^+(p_1)$ along $+\hat{z}$ direction, 
under the limit $q_T \gg m_{\pi}$ and $q_T / \sqrt{s} = \order{1}$.

\begin{figure}[htbp]
	\centering
	\begin{tabular}{cc}
	\includegraphics[scale=0.65]{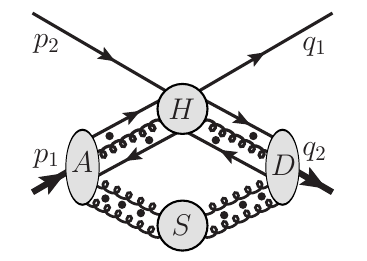} &
	\includegraphics[scale=0.65]{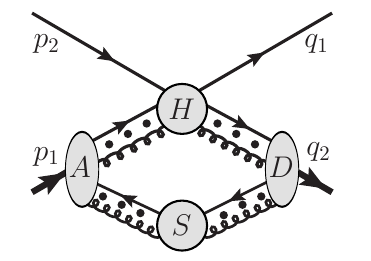}  \\
	(a) & (b)
	\end{tabular}
	\caption{Leading regions of the exclusive scattering process in \eq{eq:em2em}.}
	\label{fig:em2em}
\end{figure}

\begin{figure}[htbp]
	\centering
	\begin{tabular}{cc}
	\includegraphics[scale=0.65]{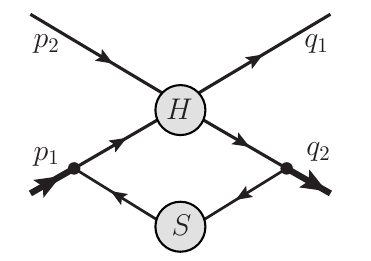} &
	\includegraphics[scale=0.65]{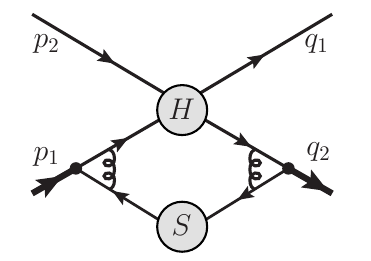}  \\
	(a) & (b)
	\end{tabular}
	\caption{Two examples for the leading region (b) in \fig{fig:em2em}.}
	\label{fig:em2em-2}
\end{figure}

Following the same procedure, we can list the leading region diagrams for the meson production amplitude, shown in \fig{fig:em2em}.
Immediately, one can notice the differences from the real photon production process discussed above:
\begin{enumerate}
\item[(1)]
there are two collinear subgraphs now, which are connected by an extra soft subgraph, and
\item[(2)]
there are two kinds of leading regions, shown in Figs.~\ref{fig:em2em}(a) and~\ref{fig:em2em}(b), which we denote as region (a) and region (b). 
For region (b), only one active quark parton enters the hard interaction, and the other one is soft and only transmits the needed quantum number.
\end{enumerate}
Region (b) raises some theoretical difficulty for factorization argument. 
However, we note that such leading regions are obtained based on the soft scaling in \eq{eq:soft-scaling} with $\lambda_S = \order{\lambda^2 / Q}$.
In this region, a soft parton has virtuality of order $\lambda^4 / Q^2$, well below the nonperturbative threshold, 
so we consider such ultrasoft region to be cut off by nonperturbative dynamics.
As argued in \sec{ssec:pinch-surface}, considering $\lambda_S \lesssim \lambda^2 / Q$ is more important for inclusive processes where
we replace the sum over final-state hadrons by the sum over final-state on-shell partons. 
Here we are dealing with exclusive processes, where all the partons are directly connected to hadrons, 
so we confine our discussion within $\lambda_S \gtrsim \lambda$,
for which the power counting rules are given by the last two columns of Table~\ref{tab:power-counting}.
We note the suppression from having soft momenta flowing through more than one collinear lines.
This constrains the diagrams for region (b) to be at very low order due to the continuity of fermion lines, 
while for region (a), we must require the soft gluons to attach to the collinear lines right before they enter the hard part.

While it is likely not well defined, the lowest-order diagram for region (b) can be conceived as in \fig{fig:em2em-2}(a), where the two quark lines directly attach to 
the ``pion wavefunction''. In this case, the two collinear lines have virtualities $\lambda Q$, while the soft subgraph has a power counting $\lambda^3$, so it gives
the power counting $\lambda / Q$ in total, which is one power higher than the counting $(\lambda/Q)^2$ for region (a). 
However, this assumes the bare quark-pion coupling to scale as $1$. 
In the kinematical regime when the pion is highly boosted, it is hardly conceivable that all the
pion momentum is carried by one of the two valence partons. 
So we add into the {\it soft-end suppression assumption} made for the $e\pi^0 \to e \gamma$ process in \sec{sssec:em2ea}
that diagrams like \fig{fig:em2em}(b) receive a high enough suppression 
from the nonperturbative hadron wavefunction such that they are power suppressed compared
to the region in \fig{fig:em2em}(a). 
This assumption is supported by high-order QCD corrections. 
As shown in \fig{fig:em2em-2}(b), when there are gluon connections between the soft and collinear partons,
the whole diagram becomes power suppressed compared to \fig{fig:em2em}(a), 
by the counting rule in Table~\ref{tab:power-counting}.
We leave a detailed study to future work.
For now, we simply note that the soft-end suppression brings the leading regions down to the one in \fig{fig:em2em}(a).

To simplify the following discussion, we note that by virtue of the large $q_T$, 
one can always boost to the frame where $A$ 
is moving along $+z$ direction and $D$ is moving along $-z$ direction, 
as was done in \citep{Collins:1981ta, Nayak:2005rt}, 
which brings the discussion similar to the Sudakov form factor in \sec{sec:sudakov-factorization}. 
This can be achieved in a covariant way by defining two sets of light-cone vectors
\begin{align}\label{eq:em2em-aux-vectors}
	w_A^{\mu} = \frac{1}{\sqrt{2}}\pp{1, \hat{z} }, \quad
	\bar{w}_A^{\mu} = \frac{1}{\sqrt{2}}\pp{1, -\hat{z} },
	\quad
	w_D^{\mu} = \frac{1}{\sqrt{2}}\pp{1, \hat{w} }, \quad
	\bar{w}_D^{\mu} = \frac{1}{\sqrt{2}}\pp{1, -\hat{w} },
\end{align}
where $\hat{w} = (\sin\theta \cos\phi, \sin\theta \sin\phi, \cos\theta)$ is the direction of the final-state meson $D$.
Then any momentum four-vector $r$ can be expanded in the $w_A$-$w_D$ frame as
\beq[eq:frame]
	r^{\mu} = r^+ \, w_A^{\mu} + r^- \, w_D^{\mu} + r_{T}^{\mu},
\eeq
where $r^{\pm} = ( r \cdot w_{D,A} ) / ( w_A\cdot w_D )$ are the longitudinal components, and $w_A \cdot w_D \sim \order{1}$ does not affect the power counting. Under this notation, we have
\beq
	r^2 = 2 \, r^+ r^- w_A \cdot w_D - \bm{r}_{T}^2,
\eeq
where $\bm{r}_{T}^2 = - g_{\mu\nu} r_{T}^{\mu} r_{T}^{\nu}$. The $A$-collinear momentum $k_A$ and $D$-collinear momentum $k_D$ have dominant components along $w_A$ and $w_D$, respectively,
\begin{align}\label{eq:cov scaling}
	k_A^{\mu} &= \pp{k_A^+, k_A^-, \bm{k}_{A,T}}_{AD} \sim (Q, \lambda^2 / Q, \lambda),	\nn\\
	k_D^{\mu} &= \pp{k_D^+, k_D^-, \bm{k}_{D,T}}_{AD} \sim (\lambda^2 / Q, Q, \lambda),
\end{align}
where the subscript ``$AD$" refers to light-front coordinates in the $w_A$-$w_D$ frame. 
A soft momentum $k_s$ exchanged between the $A$- and $D$-collinear subgraphs is in the central rapidity region with respect to 
the $w_A$-$w_D$ frame, so we have
\beq[eq:em2em soft]
	k_s^{\mu} = \pp{k_s^+, k_s^-, \bm{k}_{s,T}}_{AD} \sim (\lambda_S, \lambda_S, \lambda_S),
\eeq
with $\lambda_S$ varying between $\lambda^2 / Q$ and $\lambda$.
In the following discussion of this section, we will stay in this frame and omit the subscripts ``$AD$".

As noted in \sec{sec:glauber-region}, however, the Glauber region of the soft gluons requires special care, 
where the soft momentum $k_s$ has the scaling
\beq[eq:em2em-glauber]
	k_s^{\rm Glauber} \sim (\lambda^2 / Q, \lambda^2 / Q, \lambda).
\eeq
Similar to the Sudakov form factor case detailed in \sec{ssec:glauber-deform}, 
there is no pinch that traps the soft momentum in the Glauber region. 
So we can deform the contour to stay away from the Glauber region.
For a soft momentum $k_s$ flowing from $A$ into $S$ and then into $D$, 
its minus component only receives poles from the $A$-collinear lines, 
which all lie on the upper half plane, whereas its plus component only has poles from the $D$-collinear lines 
which also lie on the upper half plane. 
Hence, in this region, we deform the contour as 
\beq[eq:em2em-deform]
	k_s^+ \mapsto k_s^+ - i \, v(k_s^+), \quad
	k_s^- \mapsto k_s^- - i \, v(k_s^-),
\eeq
where $v(k_s^{\pm})$ is a positive real function defined in \sec{ssec:glauber-deform}. 
Such deformation deforms the Glauber momenta back to the uniform soft scaling in \eq{eq:em2em soft}. 
Then we can define the approximator $\hat{T}$ for a leading region $R$:
\begin{enumerate}
\item [(a)]
For a soft momentum $k_{SA}$ ($k_{SD}$) flowing in $A$ ($D$), we approximate it by
\beq[eq:em2em-s-mom]
	k_{SA} \mapsto \hat{k}_{SA} = \frac{k_{SA} \cdot w_A}{w_A \cdot w_D} w_D, \quad
	k_{SD} \mapsto \hat{k}_{SD} = \frac{k_{SD} \cdot w_D}{w_A \cdot w_D} w_A.
\eeq
\item [(b)]
For a soft momentum $k_{SA}$ flowing from $A$ {\it into} $S$, we include its propagator in $S$ and approximate its coupling with $A$ by
\beq[eq:em2em-s-A]
	C_{A, \mu}(k_A; k_{SA}) \, g^{\mu\nu} \, S_{\nu}(k_{SA})
		\mapsto
		C_{A, \mu}(k_A; \hat{k}_{SA}) \, \frac{\hat{k}_{SA}^{\mu} w_A^{\nu} }{k_{SA} \cdot w_A - i \epsilon} \, S_{\nu}(k_{SA}),
\eeq
where $k_A$ stands for some $A$-collinear momentum, and 
the $i\epsilon$ prescription makes the artificially introduced pole at $k_{SA}^- = 0$ on the upper half plane, compatible with the needed deformation
in \eq{eq:em2em-deform}. 
In \eq{eq:em2em-s-A}, we used $C_A$ and $S$ to refer to the collinear subgraph $A$ and the soft subgraph, respectively.
\item [(c)]
For a soft momentum $k_{SD}$ flowing from $D$ {\it into} $S$, we include its propagator in $S$ and approximate its coupling with $D$ by
\beq[eq:em2em-s-D]
	C_{D, \mu}(k_D; k_{SD}) \, g^{\mu\nu} \, S_{\nu}(k_{SD})
		\mapsto
		C_{D, \mu}(k_D; \hat{k}_{SD}) \, \frac{\hat{k}_{SD}^{\mu} w_D^{\nu} }{k_{SD} \cdot w_D + i \epsilon} \, S_{\nu}(k_{SD}),
\eeq
Note that we flipped the soft momentum flow relative to that in \eq{eq:em2em-deform}.
Similarly to \eq{eq:em2em-s-A}, here $C_D$ refers to the collinear subgraph $D$.
\item [(d)]
For an $A$ ($D$) collinear momentum $k_{AH}$ ($k_{DH}$) flowing in $H$, we approximate it by
\beq[eq:em2em-col-mom]
	k_{AH} \mapsto \hat{k}_{AH} = (k_{AH} \cdot \bar{w}_A) w_A, \quad
	k_{DH} \mapsto \hat{k}_{DH} = (k_{DH} \cdot \bar{w}_D) w_D.
\eeq
Here we project $k_{AH}$ ($k_{DH}$) against $\bar{w}_A$ ($\bar{w}_D$), instead of $w_D$ ($w_A$), 
such that after factoring the collinear subgraphs out of $H$, 
each collinear subgraph is independent of one another. 
Such replacement keeps the leading momentum components, so does not affect the leading-power accuracy.
\item [(e)]
For a collinear gluon attaching $A$ to $H$, its polarization is dominantly longitudinal. 
We include its propagator in $C_A$ and approximate its coupling with $H$ by
\beq[eq:em2em-col-A-g]
	H_{\mu}(k_H; k_{AH}) \, g^{\mu\nu} \, C_{A, \nu}(k_{AH}) 
		\mapsto
		H_{\mu}(k_H; \hat{k}_{AH}) \, \frac{\hat{k}_{AH}^{\mu} \bar{w}_A^{\nu}}{k_{AH} \cdot \bar{w}_A + i \epsilon} \, 
		C_{A, \nu}(k_{AH}) ,
\eeq
where $k_H$ is some hard momentum in $H$, and we take $k_{AH}$ to flow from $H$ {\it into} $A$. 
This introduces a pole at $k_{AH} \cdot \bar{w}_A = 0$. 
The $i \epsilon$ is introduced to make it compatible with the deformation in 
\eq{eq:em2em-deform}, as explained in \sec{ssec:glauber-modify}. 
The same momentum $k_{AH}$ can reach the soft region, where it flows from $S$ into $A$, 
through $H$, into $B$, and back to $S$.
The deformation in \eq{eq:em2em-deform} is then adapted to
\beq
	\Delta k_{AH}^S = +i \, \order{\lambda} (w_A + w_D),
\eeq
which deforms the denominator $k_{AH} \cdot \bar{w}_A$ by 
\beq 
	\Delta k^S_{AH} \cdot \bar{w}_A = +i \, \order{\lambda},
\eeq
into the upper half plane. So we need the $+i\epsilon$ prescription in \eq{eq:em2em-col-A-g}.
This will lead to a future-pointing Wilson line along $\bar{w}_A$.
\item [(f)]
For a collinear gluon attaching $D$ to $H$, we include its propagator in $C_D$ and approximate its coupling with $H$ by
\beq[eq:em2em-col-D-g]
	H_{\mu}(k_H; k_{DH}) \, g^{\mu\nu} \, C_{D, \nu}(k_{DH}) 
		\mapsto
		H_{\mu}(k_H; \hat{k}_{DH}) \, \frac{\hat{k}_{DH}^{\mu} \bar{w}_D^{\nu}}{k_{DH} \cdot \bar{w}_D - i \epsilon} \, 
		C_{D, \nu}(k_{DH}) ,
\eeq
where we take $k_{DH}$ to flow from $H$ {\it into} $D$. The $i \epsilon$ is introduced in a similar way to \eq{eq:em2em-col-A-g}.
This will lead to a past-pointing Wilson line along $\bar{w}_D$.
\item [(g)]
For the quark and antiquark lines entering $H$ from $A$, we insert the spinor projectors
\beq[eq:em2em-spinor-A]
	\P_A = \frac{\gamma\cdot w_A \, \gamma\cdot \bar{w}_A}{2}, \quad
	\Pb_A = \frac{\gamma\cdot \bar{w}_A \, \gamma\cdot w_A}{2},
\eeq
respectively.
For the quark and antiquark lines leaving $H$ to $D$, we insert the spinor projectors
\beq[eq:em2em-spinor-D]
	\Pb_D = \frac{\gamma\cdot \bar{w}_D \, \gamma\cdot w_D}{2}, \quad
	\P_D = \frac{\gamma\cdot w_D \, \gamma\cdot \bar{w}_D}{2},
\eeq
respectively.
\end{enumerate}

A region $R$ for a graph $\Gamma$ is specified by the set of collinear and soft gluons (and the two pairs of collinear quark lines by default); 
any other lines belong to the hard subgraph $H$. We denote the graph contribution in such a region as
\beq[eq:em2em-convolute]
	H_{n_1, n_2} \otimes C_{A, n_1; m_1} \otimes C_{B, n_2; m_2} \otimes S_{m_1, m_2},
\eeq
where $n_1$ and $n_2$ are the number of collinear gluons connecting $H$ to $A$ and $B$, respectively, 
and $m_1$ and $m_2$ are the number of soft gluons connecting $S$ to $A$ and $B$, respectively. 
The symbol $\otimes$ refers collectively to the momentum convolutions and color and spinor contractions.
For the same graph $\Gamma$, there may be smaller regions $R'$ than $R$, which have fewer lines in $H$ and/or $C_{A, D}$, and/or more lines in $S$.
The contribution from the region $R$ is then extracted by applying $\hat{T}$ after the subtraction of smaller region contributions,
\beq[eq:em2em-region-contrib]
	C_{R} \Gamma = \hat{T} \pp{ \Gamma - \sum_{R' < R} C_{R'} \Gamma },
\eeq
where the contribution $C_{R'} \Gamma$ is obtained by iterative use of \eq{eq:em2em-region-contrib}. 
The subtraction terms in \eq{eq:em2em-region-contrib} also have $\hat{T}$ acted in front, just as in \eq{eq:region-subtraction-term}. 
They are obtained by treating the lines in the same way as in $R$, with certain lines belonging to $H$, certain lines to $A$, etc.,
even though the approximators for $R'$ have been applied that treat those lines in some other (smaller) regions.
Therefore, the subtraction terms in \eq{eq:em2em-region-contrib} have the same structure as \eq{eq:em2em-convolute}, with different factors
$H$, $C_A$, and $C_D$, but the same $S$.
Those subtraction terms can be uniquely determined once $R$ is specified.

\begin{figure}[htbp]
	\centering
	\begin{tabular}{cc}
	\includegraphics[trim={-1em -2em -1em 0}, clip, scale=0.5]{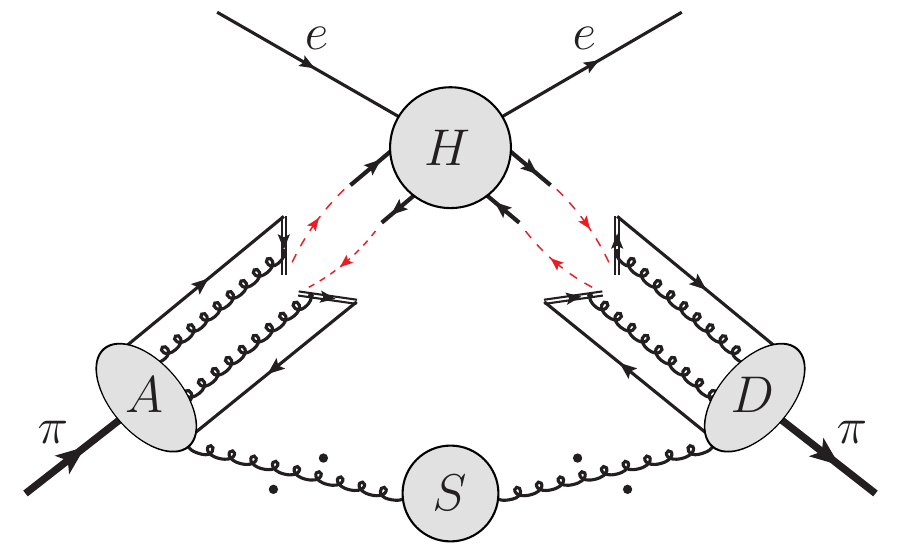} &
	\includegraphics[trim={-1em 0 -1em 0}, clip, scale=0.5]{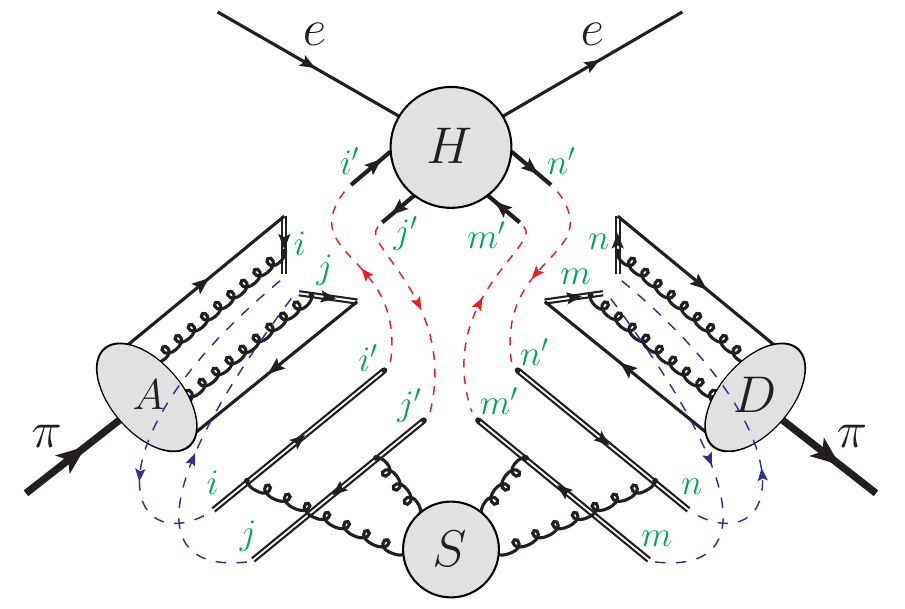}  \\
	(a) & (b)
	\end{tabular}
	\caption{Factorization of collinear gluons out of the hard subgraph (a) 
	and also soft gluons out of the collinear subgraphs (b) 
	for the process in \eq{eq:em2em}. 
	The red and blue thin dashed lines refer to the color flows. 
	The external thick quark lines of the hard subgraph $H$ are amputated.
	The numbers of gluons in all sectors can be arbitrary.}
	\label{fig:em2em-fac}
\end{figure}

The approximator $\hat{T}$ modifies certain momenta and inserts some Lorentz and spinor projectors 
in a way that the use of Ward identity for soft and collinear gluons is exact.
This applies to both $\hat{T}\Gamma$ and the subtracted terms $C_{R'} \Gamma$ 
(which is obtained after applying their approximators) in \eq{eq:em2em-region-contrib}. 
Acting $\hat{T}$ on the latter further modifies the momenta and introduces projectors that makes a further use of Ward identity exact.
Therefore, after applying $\hat{T}$ in \eq{eq:em2em-region-contrib}, we can use Ward identity for both $\Gamma$ and the subtracted terms
in an exact way.
Then we sum over all possible diagrams with the same region specification as \eq{eq:em2em-convolute}, 
with $H$ having a fixed order $N_h$ of $\alpha_s$.
Among them, the sum of those with the same subgraphs $A$, $D$ and $S$ 
but different $H$ and collinear gluon attachments
allows the use of Ward identity for the $A$ and $D$ collinear gluons. 
This factorizes the collinear gluons out of $H$, and simplifies \eq{eq:em2em-convolute} to
\beq[eq:em2em-factor-H]
	H^{(N_h)} \otimes \hat{T}_w 
		\bb{ C_{A, n_1; m_1} \otimes S_{m_1, m_2} \otimes C_{D, n_2; m_2} },
\eeq
where $\hat{T}_w$ is the same as $\hat{T}$ but just refers to the fact that the collinear gluons are now collected by two pairs gauge links.
This is shown graphically in \fig{fig:em2em-fac}(a).
\eq{eq:em2em-factor-H} applies to both terms in \eq{eq:em2em-region-contrib}, so the factorized result also extends to the subtracted factors,
\beq[eq:em2em-factor-H-sub]
	H_{\rm sub}^{(N_h)} \otimes \hat{T}_w 
		\bb{ C_{A, n_1; m_1} \otimes S_{m_1, m_2} \otimes C_{D, n_2; m_2} }_{\rm sub},
\eeq
Now $H^{(N_h)}$ is only specified by two pairs of external amputated collinear quarks, at a given order $N_h$ of $\alpha_s$. 
The sum over $H$ is then independent of the other factors, and each given $H$ determines uniquely the subtracted terms within. 
So the sum over $H$ yields the partially factorized result,
\beq[eq:em2em-factor-H-sum]
	H_{\rm sub} \otimes \hat{T}_w 
		\bb{ C_{A, n_1; m_1} \otimes S_{m_1, m_2} \otimes C_{D, n_2; m_2} }_{\rm sub},
\eeq
which applies for any values of $n_1$ and $n_2$. 
Here 
\beq
	H_{\rm sub} 
		= \sum_{N_h} \sum_i \bb{ H_i^{(N_h)} - (\mbox{subtraction for smaller regions}) },
\eeq
with $H_i^{(N_h)}$ denoting the $i$-th graph at $N_h$-th order for the two pairs of external collinear quarks.
Since smaller regions in $H$ refer to some of the lines becoming collinear or soft, their contributions are obtained by 
iterative use of the approximator $\hat{T}_w$, which can be refactorized as soft and collinear factors.

Then we sum over all subdiagrams for $A$ and $D$ at a given order 
$N_a$ and $N_d$ of $\alpha_s$, respectively, and for fixed $n_1$ and $n_2$.
This allows the use of Ward identities for the soft gluons to factorize them out of the collinear subgraphs. 
Again, this applies to both $A$ and $D$ themselves and the subtracted terms therein, so we have
\beq[eq:em2em-factor-S]
	H_{\rm sub} \otimes \bb{ C^{(N_a)}_{A, n_1; \rm sub} \otimes S_{m_1, m_2} \otimes C^{(N_d)}_{D, n_2; \rm sub} },
\eeq
which is graphically shown in \fig{fig:em2em-fac}(b), where the soft gluons are collected by two pairs of gauge links, 
one along $w_A$ and the other along $w_D$.
Then we can sum over $n_1$, $n_2$, $N_a$, and $N_d$ independently, 
which converts the two collinear factors into matrix element definitions,
\bse\label{eq:em2em-col-factors}\begin{align}
	C^{\rm unsub}_{A, \alpha\beta, ij}(k)
	&\, = \int d^4 z \, e^{ik \cdot z} \P_{A, \alpha\alpha'}  \, 
		\langle 0 | \T \cc{ \bb{ W(\infty, z; \bar{w}_A) \psi_{1, \alpha'}(z) }_i  \right.
	\nn\\
	&\hspace{10em} \left. \times
	\bb{ \bar{\psi}_{2, \beta'}(0) W^{\dag}(\infty, 0; \bar{w}_A) }_j 
	} 
	| \pi^+ \rangle 
	\Pb_{A, \beta'\beta}, \\
	C^{\rm unsub}_{D, \gamma\delta, mn}(l)
	&\, = \int d^4 z \, e^{-il \cdot z} \P_{D, \gamma\gamma'}  \, \langle \pi^+ | \T \cc{ \bb{ W(-\infty, 0; \bar{w}_D) \psi_{2, \gamma'}(0) }_m  \right.
	\nn\\
	&\hspace{10em} \left. \times
	\bb{ \bar{\psi}_{1, \delta'}(z) W^{\dag}(-\infty, z; \bar{w}_D) }_n
	} 
	| 0 \rangle 
	\Pb_{D, \delta'\delta},
\end{align}\ese
where the subscript ``unsub'' means that these factors have not included the subtraction for smaller regions (where some of the lines go soft), 
$(\alpha, \beta, \gamma, \delta)$ are spinor indices, $(i, j, m, n)$ are color indices in the fundamental representation,
and we keep the general notations $\psi_1$ and $\psi_2$, which are $u$ and $d$ quark fields for $\pi^+$.
	
The sum over all possible soft subdigrams and over $m_1$ and $m_2$ 
can be done independently and converts it into a matrix element definition,
\begin{align}\label{eq:em2em-soft-factor}
	S_{i' i, j' j; m' m, n' n} 
	&\, = \langle 0 | \T \cc{ W_{i' i}(0, -\infty; w_A) W^{\dag}_{j j'}(0, -\infty; w_A) \right. \nn\\
	& \hspace{6em} \left. \times
	W^{\dag}_{m' m}(\infty, 0; w_D) W_{n n'}(\infty, 0; w_D)  } |0 \rangle,
\end{align}
where the color indices $(i, j, m, n)$ match the ones for the collinear factors in \eq{eq:em2em-col-factors}, and $(i', j', m', n')$ are to contract with those
of the hard factor $H_{i'j', m'n'}$. The soft subgraph contains no subtraction for smaller regions, so \eq{eq:em2em-soft-factor} is the final result. 

Now for the same reason as \eq{eq:em2ea-color}, the collinear factors in \eq{eq:em2em-col-factors} are color singlets, such that
\beq[eq:em2em-col-singlet]
	C^{\rm unsub}_{A, \alpha\beta, ij}(k) = \frac{1}{N_c} \delta_{ij} \, C^{\rm unsub}_{A, \alpha\beta}(k), \quad
	C^{\rm unsub}_{D, \gamma\delta, mn}(l) = \frac{1}{N_c} \delta_{mn} \, C^{\rm unsub}_{D, \gamma\delta}(l),
\eeq
and their contraction with the soft factor renders the latter into an identity matrix,
\begin{align}\label{eq:em2em-soft-cancel}
	\delta_{ij} \, S_{i' i, j' j; m' m, n' n} \delta_{mn} 
	= \delta_{i'j'} \delta_{m'n'},
\end{align}
by unitarity of the Wilson lines.
Hence all exchanges of soft gluons are canceled. 
This also applies to the subtraction terms in $A$ and $D$ subgraphs. 
Those smaller regions where some of the gluon lines turn soft are canceled order by order after summing over graphs. 
Therefore, the unsubtracted collinear factors in \eq{eq:em2em-col-factors} are the same as the subtracted ones.

We note that the choices of lightlike vectors for the soft approximations in Eqs.~\eqref{eq:em2em-s-A} and \eqref{eq:em2em-s-D}
could have introduced rapidity divergences if the soft factor does not reduce to unity. 
In that case, one needs to choose some non-lightlike vectors to use in the soft approximations, 
as in \sec{ssec:sudakov-rapidity}, which does not affect the result that $S = 1$.
The cancellation of soft gluons for the exclusive processes is a direct result of the scattered particles being color singlets, 
which itself is the consequence of color confinement. 
This is in contrast to inclusive processes, where soft cancellation is a result of unitarity, due to the sum over final states.
If we also compare QCD to non-confined gauge theories like QED, the latter have bare charges that can emit and absorb soft (and/or collinear)
gauge bosons (photons for QED), which can introduce corresponding divergences to the amplitudes. A finite cross section is achieved only
after a proper sum over the final (and/or initial) states, by virtue of unitarity. 
Hence, exclusive processes are only well defined for a confined gauge theory like QCD, but not for non-confined ones, where only inclusive 
processes are sensibly defined.

After a similar spinor and momentum decomposition as in Eqs.~\eqref{eq:em2ea-spinor} and \eqref{eq:identity-factor}, 
we get the factorized expression for the amplitude of \eq{eq:em2em},
\beq[eq:em2em-factorize]
	\M_{\pi^+ e \to e \pi^+} = \int_0^1 dx \, dy \, D_{u / \pi^+}(x, \mu) \bar{D}_{u / \pi^+}(y, \mu) \H\pp{ x, y; \frac{\vec{q}_T}{\sqrt{s}}, \frac{q_T}{\mu} },
\eeq
where we have used the multiplicative renormalization to convert each factor into the renormalized one. 
Here the two bare DAs are defined as
\begin{align}\label{eq:em2em-DA-def}
	D^{\rm bare}_{u / \pi^+}(x) 
	& = 
	\int \frac{d\lambda}{4\pi} \, e^{i \lambda x p_1 \cdot \bar{w}_A} \, 
	\langle 0 | \T \cc{ \bar{\psi}_2(0) \, \gamma\cdot \bar{w}_A \gamma_5 \, W(0, \lambda \bar{w}_A; \bar{w}_A) \psi_1(\lambda \bar{w}_A) }
	| \pi^+(p_1) \rangle, \nn\\
	\bar{D}^{\rm bare}_{u / \pi^+}(y) 
	& = 
	\int \frac{d\lambda}{4\pi} \, e^{-i \lambda y q_2 \cdot \bar{w}_D} \, 
	\langle \pi^+(q_2)  | \T \cc{ \bar{\psi}_1(\lambda \bar{w}_D) \, \gamma\cdot \bar{w}_D \gamma_5 \,
		W(\lambda \bar{w}_D, 0; \bar{w}_D) \psi_2(0) }
	| 0\rangle ,
\end{align}
where the time ordering can be deleted, as in \eq{eq:em2ea-DA}. 
It can be easily shown that the DA value does not depend on the momentum direction of the hadron, 
and the DA for the final-state pion differs from the initial-state one by a complex conjugate,
\beq
	\bar{D}_{u / \pi^+}(x) = \bb{ D_{u / \pi^+}(x) }^*,
\eeq
which applies to both the bare and renormalized DAs.
The hard coefficient is defined as the scattering of two pairs of collinear $[q_1\bar{q}_2]$ states that form color singlets,
\beq
	\H\pp{ x, y; \frac{\vec{q}_T}{\sqrt{s}}, \frac{q_T}{\mu} }
	= \frac{1}{N_c^2} \bb{ \pp{ \frac{\gamma_5 \gamma\cdot w_A}{2} }_{\alpha\beta} 
		H_{\beta \alpha, \delta\gamma}\pp{ x, y; \frac{\vec{q}_T}{\sqrt{s}}, \frac{q_T}{\mu} }
		\pp{ \frac{\gamma_5 \gamma\cdot w_D}{2} }_{\gamma\delta} },
\eeq
with subtraction for smaller region contributions.

\eq{eq:em2em-factorize} can be readily extended to other mesons and baryons, just with a proper change of
the DA and hard coefficients~\citep{Lepage:1980fj}.

\subsubsection{Choice of Glauber deformation for the double-meson process}
\label{sssec:em2em-deformation}

In the discussion of \sec{sssec:em2em}, the contour deformation to get the soft gluon momentum $k_s$ out of the Glauber region 
is symmetric with $k_s^+$ and $k_s^-$, as was employed for the Sudakov form factor in \sec{ssec:glauber-deform}. 
This is, nevertheless, not the unique choice~\citep{Collins:2004nx}, as it is sufficient to get rid of the Glauber region 
as long as $| k_s^+ k_s^- | \gtrsim | k_{sT}^2 |$. 
By examining the contour of $k_s^+$, we note that while all the $k_s^+$ poles from the $D$-collinear lines are of $\order{\lambda^2 / Q}$ 
and lie on the same half plane, 
the poles from the $A$-collinear lines and soft lines are of order $Q$ in the Glauber region.
Hence one may choose to only deform the contour of $k_s^+$, but now by a magnitude of $\order{Q}$,
\beq[eq:em2em-ks+-deform-Q]
	k_s^+ \mapsto k_s^+ + i \, \order{Q},
\eeq
when $k_s$ flows from $D$ into $S$. 
This deforms a Glauber gluon momentum into the $A$-collinear region with the scaling $(Q, \lambda^2 / Q, \lambda)$, 
and then one can perform usual approximations and apply Ward identities for the rest of the soft gluon momenta. 
In this way, although the Glauber region will not be treated accurately by the soft approximation, it will be by the collinear approximation.

The soft gluons factorized from $D$ are attached to two Wilson lines along $w_D$, 
and the $A$-collinear longitudinally polarized gluons are collected by two Wilson lines along $\bar{w}_A$; 
both of the two sets of Wilson lines point to the future.
Since we do not deform the contour of $k_s^-$, it does not matter what $i\epsilon$ prescription we assign to the approximator $1/k_s^-$; 
the $+i\epsilon$ choice leads to same result%
\footnote{Here $k_s$ is the same as in \eq{eq:em2em-ks+-deform-Q}, flowing from $D$ to $S$ and then to $A$.} 
as the symmetric deformation in \sec{sssec:em2em}, 
with soft Wilson lines along $w_A$ and collinear Wilson lines along $\bar{w}_D$ both pointing from/to the past, 
but the $-i\epsilon$ choice would have both point to the future.

Similarly, one may also choose to only deform $k_s^-$ as $k_s^- \mapsto k_s^- - i\order{Q}$ 
when it flows out of $A$-collinear lines into $S$, and then the $i\epsilon$ prescription for $k^+$ is not important 
as long as every $k_s^-$ is associated with the same prescription as in $1/(k_s^- + i\epsilon)$. 

This gives some freedom in choosing the suitable $i\epsilon$ prescriptions to achieve universal definitions 
for the soft factor and collinear factors when compared with other processes~\citep{Collins:2004nx}. 
Within collinear factorization framework, the soft factor cancels no matter what prescription is used, 
and the Wilson lines associated with the collinear factors also become straight lines on the light cone 
due to unitarity of the Wilson lines, so that universality is a trivial property in the collinear factorization for exclusive processes. 
However, such freedom as in \eq{eq:em2em-ks+-deform-Q} is necessary for the factorization of diffractive processes, 
as we will discuss in \sec{sssec:dvmp}.

\subsection{Large-angle photon-meson scattering}
\label{ssec:exclusive-am}
For photon-meson scattering, the beam particle $B$ stands for a photon. The final-state particles $CD$ can take
(1) $(CD) = (l^+l^-)$, 
(2) $(CD) = (\gamma\gamma)$, 
(3) $(CD) = (\gamma, \mbox{meson})$, or
(4) $(CD) = (\mbox{meson}, \mbox{meson})$.
The first three cases do not raise new issues in factorization, which we will briefly address.
The last case, however, requires a generalization of our factorization argument 
for the electron-meson scattering in \sec{ssec:exclusive-em}.

\subsubsection{Single-meson process: $(CD) = (l^+l^-)$ or $(\gamma\gamma)$.}
\label{sssec:am2ll}
For the dilepton or diphoton production, the color structure does not differ from the pion-photon transition in \sec{sssec:em2ea}. 
The leading region in QCD thus takes the same form as \fig{fig:em2ea}(b) with a mere change of external lines. 

\begin{figure}[htbp]
	\centering
	\begin{tabular}{cc}
	\includegraphics[trim={-1em 0 -1em 0}, clip, scale=0.75]{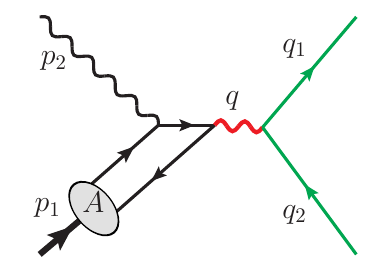} &
	\includegraphics[trim={-1em 0 -1em 0}, clip, scale=0.75]{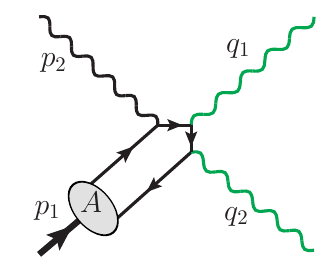}  \\
	(a) & (b)
	\end{tabular}
	\caption{LO diagrams for photoproduction of dileptons (a) and diphotons (b).}
	\label{fig:am2ll-am2aa-LO}
\end{figure}

At LO in QED, the dilepton production happens via the decay of a timelike virtual photon of invariant mass $Q = m_{ll}$, 
as shown in \fig{fig:am2ll-am2aa-LO}(a), so this process is probing the {\it timelike} meson-photon transition form factor. 
This can only happen for a charge-neutral meson with an even charge-conjugation parity (C-even), such as $\pi^0$. 
In this case, the hard scale for factorization is provided by the high virtuality $Q$, which is ensured by the condition of a large $q_T$. 
However, the factorization holds as long as $Q \gg \LQCD$, even when $q_T$ is small. 
Therefore, we have a factorization formula as \eq{eq:em2ea-factorize-ren}, with the same DA definition and a proper change of the hard coefficient.
The power suppressed correction is now $\order{\LQCD / m_{ll}}$.

In contrast, for the diphoton production, all the three photons directly couple to the quark line, 
as shown by the LO diagram in \fig{fig:am2ll-am2aa-LO}(b).
In this case, the hard scale is necessarily provided by the large $q_T$. 
The same factorization cannot extend to the forward kinematic region.
Now that the meson is coupled to three photons, such processes can only happen to 
charge-neutral C-odd mesons, such as the $\rho$ vector meson.
Since we neglect the quark masses in the hard part, the collinear $q\bar{q}$ state from the meson must have zero helicity, 
so the vector meson must be longitudinally polarized. 
The factorization formula therefore is extended from \eq{eq:em2ea-factorize-ren} to
\beq[eq:am2aa-factorize-ren]
	\M_{\rho_L \gamma \to \gamma \gamma} 
	= \sum_q \int_0^1 dx \, D_{q / \rho_L}(x, \mu) \, \H_q\pp{x; \frac{\vec{q}_T}{\sqrt{s}}, \frac{q_T}{\mu}}  + \order{\LQCD / q_T}.
\eeq
Here the bare DA is defined as
\begin{align}\label{eq:am2aa-DA}
	D^{\rm bare}_{q / \rho_L}(x)
	= \int_{-\infty}^{\infty} \frac{d\lambda}{2\pi} 
		e^{i \lambda  x \, p_1 \cdot n }
		\langle 0 | \T \cc{ 
			\bar{\psi}_{q}(0)
			\frac{\gamma\cdot n}{2} 
			W(0, \lambda n; n) \psi_{q} (\lambda n)
		} | \rho_L(p_1) \rangle,
\end{align}
and the bare hard coefficient is the scattering $[q\bar{q}](\hat{p}_1) + \gamma(p_2) \to \gamma(q_1) + \gamma(q_2)$, defined as
\beq
	\H^{\rm bare}_q\pp{x; \frac{\vec{q}_T}{\sqrt{s}}} 
	= (p_1 \cdot n) \frac{1}{N_c} \delta_{ij} 
			\pp{ \frac{\gamma\cdot \bar{n}}{2} }_{\alpha\beta}
			\H_{\beta\alpha; ji}(x\hat{p}_1, (1 - x) \hat{p}_1),
\eeq
up to subtractions of smaller regions.

\subsubsection{Double-meson process: $(CD) = (\gamma, \mbox{meson})$.}
\label{sssec:am2am}
\begin{figure}[htbp]
	\centering
	\begin{tabular}{cc}
	\includegraphics[scale=0.8]{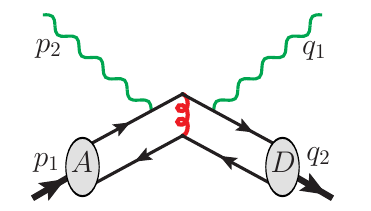} &
	\includegraphics[scale=0.8]{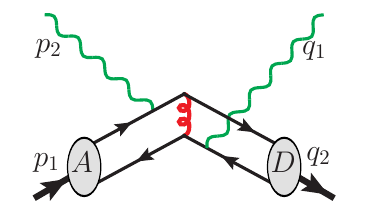}  \\
	(a) & (b)
	\end{tabular}
	\caption{LO diagrams for the photoproduction of photon-meson pairs.}
	\label{fig:am2am-LO}
\end{figure}

The photon-meson pair production has the same color structure as 
the elastic electron-meson scattering in \sec{sssec:em2em}, 
so the leading region differs from \fig{fig:em2em} only by changing 
the external electron lines by photon lines. 
By the same argument (including the soft-end suppression assumption), 
we can obtain a factorization formula like \eq{eq:em2em-factorize}.
However, we need to note that now the LO diagrams have both 
external photons attach to the quark lines, resulting in
three propagators in the hard part, as shown in \fig{fig:am2am-LO}. 
The calculation of the hard coefficient thus becomes more involved, 
but it also reveals a richer structure.

While the two photon lines can attach to the same quark line as in \fig{fig:am2am-LO}(a), they
may also attach to different quark lines as in \fig{fig:am2am-LO}(b).
In the first case, all the propagators in the hard part are either connected to two external on-shell lines 
or amputated parton lines on one of its ends,
whereas in the second case the gluon propagator is not.
As a result, the hard coefficient has an (unpinched) pole in the middle of the $(x, y)$ integration in the DA convolution. 
This introduces an imaginary part to the amplitude even at LO. 
More importantly, as will be elaborated in \ch{ch:GPD}, 
the $(x, y)$ and $q_T$ dependencies within the hard coefficient cannot be separated. 
Their entanglement will lead to a nontrivial sensitivity to the $x$ dependence of the DA.

\subsubsection{Triple-meson process: $(CD) = (\mbox{meson}, \mbox{meson})$. Symmetric deformation.}
\label{sssec:am2mm-sym}

\begin{figure}[htbp]
	\centering
	\begin{tabular}{ccc}
		\includegraphics[scale=0.75]{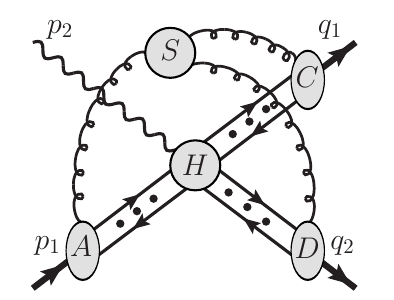} &
		\includegraphics[scale=0.75]{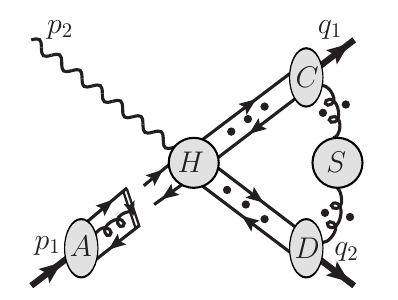} &
		\includegraphics[scale=0.75]{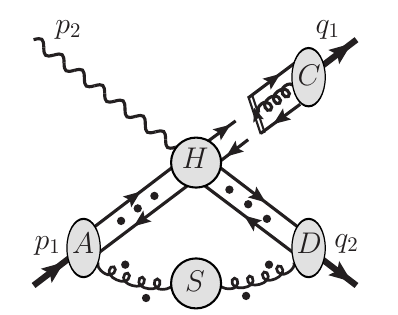} \\
		(a) & (b) & (c)
	\end{tabular}
	\caption{(a) Leading-region graph for the photoproduction of light meson pairs. 
	There can be any numbers of soft gluons connecting $S$ to each collinear subgraph. 
	The regions with $S$ connecting to one or more collinear subgraphs via quark lines or transversely polarized gluon lines are omitted. 
	Depending on the quantum numbers, the collinear quark lines may be replaced by transversely polarized gluon lines. 
	The dots represent arbitrary numbers of longitudinally polarized collinear gluons.
	(b) The result of factorizing the collinear subgraph $A$ out of the hard subgraph $H$, with the soft gluons coupled to $A$ canceled.
	(c) The result of factorizing the collinear subgraph $C$ out of the hard subgraph $H$, with the soft gluons coupled to $C$ canceled.}
	\label{fig:am2mm}
\end{figure}

The process 
\beq[eq:am2mm]
	M_A(p_1) + \gamma(p_2) \to M_C(q_1) + M_D(q_2)
\eeq 
has three hadrons in three different directions, among which arbitrary soft
gluon exchanges can happen. The leading region is shown in \fig{fig:am2mm}(a). The regions where the soft subgraph is connected to any two 
collinear subgraphs via quark lines or transversely polarized gluon lines are omitted, which are power suppressed by the soft-end suppression.

In such a three-meson process, each of the final-state mesons exchanges soft gluons with both initial-state and final-state mesons. 
This makes it difficult to uniformly deform the contour to move the soft gluons away from the Glauber region.
To put forward the formal discussion, 
we work in the c.m.~frame and define some auxiliary vectors by extending \eq{eq:em2em-aux-vectors},
\begin{align}\label{eq:auxiliary-A-C-D}
	w_A^{\mu} = \frac{1}{\sqrt{2}}(1, \hat{z}), \quad
	\bar{w}_A^{\mu} = \frac{1}{\sqrt{2}}(1, -\hat{z}),	\quad
	w_C^{\mu} = \frac{1}{\sqrt{2}}(1, \hat{n}) = \bar{w}_D^{\mu}, \quad
	\bar{w}_C^{\mu} = \frac{1}{\sqrt{2}}(1, -\hat{n}) = w_D^{\mu},
\end{align}
where $\hat{z}$ and $\hat{n}$ are normalized three-vectors along the directions of the initial-state meson $M_A$ and final-state meson $M_C$. 
Basically, $w_{A,C,D}$ are the light-cone vectors along the directions of meson $A$, $C$ and $D$, respectively, 
and the corresponding vectors with bars refer to the conjugate light-cone vectors along the opposite directions. 

The essential point is that any soft gluon momentum $k_s$ can be routed to only flow through two collinear subgraphs. 
For this, we introduce the notation $k_s^{(ij)}$ to be a soft gluon momentum that flows from the collinear subgraph $i$ into $S$, 
then proceeds into the collinear subgraph $j$, passes through the hard subgraph $H$, and finally returns to $i$. 
Apparently, we have $k_s^{(ij)} = -k_s^{(ji)}$, with $i, j = A, C, D$ and $i \neq j$.

When considering the soft gluon momentum $k_s^{(ij)}$, we expand it in the $w_i$-$w_j$ frame as defined in \eq{eq:frame},
\footnote{While we may define the plus and minus components in each $w_i$-$w_j$ frame like Eqs.~\eqref{eq:frame}--\eqref{eq:cov scaling}, 
having multiple such frames makes the notation cumbersome, so we stick to the covariant notations.}
\beq[eq:soft-kij-expand]
	k_s^{(ij)} = w_i \frac{k_s^{(ij)}\cdot w_j}{w_i \cdot w_j} + w_j \frac{k_s^{(ij)}\cdot w_i}{w_i \cdot w_j} + k_{sT}^{(ij)} \,,
\eeq
where all the three terms on the right-hand side are of the same size, $\order{\lambda_S}$.
When it flows in the collinear subgraph $i$, whose momenta are dominantly along $w_i$, 
the $k_s^{(ij)}$ can be approximated by only retaining the $w_j$ component, 
\beq[eq:soft kij]
	k_s^{(ij)} \simeq \hat{k}_s^{(ij)} = w_j \frac{k_s^{(ij)}\cdot w_i}{w_i\cdot w_j}	\,.
\eeq
Moreover, the coupling of this soft gluon to the collinear subgraph $J^i$ can be approximated as
\begin{align}\label{eq:soft coupling ij}
	J^i_{\mu}(k_i, k_s^{(ij)}) \, g^{\mu\nu} \, S_{\nu}(k_s^{(ij)})		
		\simeq 
		J^i_{\mu}(k_i, \hat{k}_s^{(ij)}) \, \frac{\hat{k}_s^{(ij)\mu} \, w_i^{\nu}}{k_s^{(ij)} \cdot w_i} \, S_{\nu}(k_s^{(ij)})	\,,
\end{align}
because it is the specific component of $g^{\mu\nu}$ given by $w_j^{\mu} w_i^{\nu} / w_i \cdot w_j$ that provides the dominant contribution. 
In \eq{eq:soft coupling ij}, $k_i$ stands for some collinear momentum in the subgraph $i$.
This approximation will allow the use of Ward identity to factorize the soft gluons out of the collinear subgraphs.
\footnote{We should note that the argument given here is equivalent to \citep{Collins:1981ta, Nayak:2005rt} 
that boost into the rest frame of two collinear subgraphs. The underlying reason is that any two distinct collinear subgraphs 
are well separated in rapidity; in the language here, it is $w_i \cdot w_j \simeq \order{1}$.}

While this is a good approximation for the central soft region, it is not for the Glauber region in which 
\beq[eq:glauber ij]
	| k_s^{(ij)}\cdot w_i | \, | k_s^{(ij)}\cdot w_j | \ll | k_{sT}^{(ij)} |^2 \, w_i \cdot w_j \,.
\eeq
Now because all the collinear lines in the subgraph $i$ or $j$ only give poles for $k_s^{(ij)}\cdot w_i$ or $k_s^{(ij)}\cdot w_j$
on the same half complex plane, the integration contour of $k_s^{(ij)}$ is not pinched in the Glauber region, 
and a proper deformation can get it out of the Glauber region.
However, if we take the symmetric deformation as in \eq{eq:glauber-deform}, we need to deform in opposite ways the soft momenta coupling $A$ to $C$
and those coupling $D$ to $C$. Specifically, for $k_s^{(AC)}$, it receives poles on the upper half plane for both the component 
$k_s^{(AC)}\cdot w_A$ from $A$-collinear lines, and $k_s^{(AC)}\cdot w_C$ from $C$-collinear lines. So we need to deform its contour as
\beq[eq:deform-AC]
	k_s^{(AC)} \mapsto k_s^{(AC)} - i \order{\lambda} w_C - i \order{\lambda} w_A,
\eeq
following the expansion in \eq{eq:soft-kij-expand}. On the other hand, the soft momentum $k_s^{(DC)}$ has poles for $k_s^{(DC)} \cdot w_D$ on the lower
half plane, so the deformation is
\beq[eq:deform-DC]
	k_s^{(DC)} \mapsto k_s^{(DC)} + i \order{\lambda} w_C - i \order{\lambda} w_D.
\eeq
While such sign difference is for different soft momentum attachments so does not pose any difficulty like a Glauber pinch, 
it does imply that a $C$-collinear longitudinally polarized gluon $k_C$ has soft subtraction terms with different contour deformation. 

If we take $k_C$ to flow from $H$ to $C$, then we approximate $k_C$ by 
\beq
	k_C \to \hat{k}_C = (k_C \cdot \bar{w}_C) w_C
\eeq
in $H$, and its coupling to $H$ by
\beq[eq:am2mm-col-approx]
	H_{\mu}(k_H, k_C) \, g^{\mu\nu} \, J^C_{\nu}(k_C)
		\simeq 
		H_{\mu}(k_H, \hat{k}_C) \,
		\frac{\hat{k}_C^{\mu} \, \bar{w}_C^{\nu} }{k_C \cdot \bar{w}_C} \,
		J^C_{\nu}(k_C)	\,.
\eeq
This introduces a pole at $k_C \cdot \bar{w}_C = 0$, which can potentially obstruct the deformations 
in Eqs.~\eqref{eq:deform-AC} and \eqref{eq:deform-DC} in the soft subtraction terms,
as explained in \sec{ssec:glauber-modify}.
Even though we are approximating the collinear region, 
which does not suffer from the Glauber region problem, 
\eq{eq:am2mm-col-approx} is applied to the whole diagram with deformed contours. 
Furthermore, the same gluon $k_C$ considered in \eq{eq:am2mm-col-approx} 
can also go into the soft region, attaching to $A$- or $D$-collinear subgraph, 
for which we will change the notation $k_C$ to $k_C^{(A)}$ or $k_C^{(D)}$,
whose contribution has already been included in the soft approximations defined in \eq{eq:soft coupling ij}. 
A subtraction is needed from \eq{eq:am2mm-col-approx} 
to avoid such double counting, which is obtained by first applying the 
soft approximation [\eq{eq:soft coupling ij}] and then applying the collinear approximation [\eq{eq:am2mm-col-approx}]. 
Since the subtraction mixes the collinear and soft approximations for the same gluons, 
and the latter require deformation of contours, we do need the $i\epsilon$ prescription 
in \eq{eq:am2mm-col-approx} not to obstruct the contour deformations in Eqs.~\eqref{eq:deform-AC} and \eqref{eq:deform-DC}. 
The latter would need to deform the denominator in \eq{eq:am2mm-col-approx} by
\beq
	\Delta( k_C^{(A)} \cdot \bar{w}_C ) =  - i \order{\lambda} w_C \cdot \bar{w}_C - i \order{\lambda} w_A \cdot \bar{w}_C = - i \order{\lambda},
\eeq
when $k_C$ reaches the soft region and attaches to $A$, or by
\beq
	\Delta( k_C^{(D)} \cdot \bar{w}_C ) =  + i \order{\lambda} w_C \cdot \bar{w}_C - i \order{\lambda} w_D \cdot \bar{w}_C = + i \order{\lambda},
\eeq
when $k_C$ reaches the soft region and attaches to $D$.
Therefore, there is not a uniform $i\epsilon$ choice for the collinear gluon approximation [\eq{eq:am2mm-col-approx}] to respect the soft deformations in the 
corresponding subtraction terms.

To avoid this difficulty, we note that the collinear subgraph $A$ only couples to 
final-state mesons by the soft gluons. So we can first factorize $A$ out of the 
the hard part. To do that, we approximate all $A$-collinear momenta $k_A$ by
\beq
	k_A \to \hat{k}_A = (k_A \cdot \bar{w}_A) w_A
\eeq
when they flow in $H$. We insert proper spinor or Lorentz projectors for the collinear quark or transversely polarized gluon lines.
The coupling of each $A$-collinear longitudinally polarized gluon to $H$ is approximated by
\beq[eq:am2mm-A-col-approx]
	H_{\mu}(k_H, k_A) \, g^{\mu\nu} \, J^A_{\nu}(k_A)
		\simeq 
		H_{\mu}(k_H, \hat{k}_A) \,
		\frac{\hat{k}_A^{\mu} \, \bar{w}_A^{\nu} }{k_A \cdot \bar{w}_A} \,
		J^A_{\nu}(k_A)	\,.
\eeq
By taking $k_A$ to flow from $H$ into $A$, the soft subtraction terms contain the regions
$k_A = k_A^{(C)} = k_s^{(CA)}$ and $k_A = k_A^{(D)} = k_s^{(DA)}$.
These require the deformations
\beq
	\Delta k_A^{(C)} = + i\order{\lambda} (w_A + w_C), \quad
	\Delta k_A^{(D)} = + i\order{\lambda} (w_A + w_D),
\eeq
which change the denominator $k_A \cdot \bar{w}_A$ in the collinear approximation [\eq{eq:am2mm-A-col-approx}] by
\beq
	\Delta (k_A^{(C)} \cdot \bar{w}_A) = + i\order{\lambda}, \quad
	\Delta (k_A^{(D)} \cdot \bar{w}_A) = + i\order{\lambda},
\eeq
respectively. 
Therefore, it is possible to introduce a $+i\epsilon$ prescription to \eq{eq:am2mm-A-col-approx} to be compatible with such deformations.
This leads to future-pointing Wilson lines along $\bar{w}_A$ to collect the $A$-collinear longitudinal gluons.

In contrast, it is easy to choose the $i\epsilon$ prescriptions for all soft gluons to make the approximation in \eq{eq:soft coupling ij} 
compatible with the deformations. We choose $-i\epsilon$ when $i = A$ and $+i\epsilon$ when $i = C$ or $D$.	
As a result, the soft gluons attached to $A$ will be collected by a pair of Wilson lines along $w_A$ that come from the past infinity, 
and those attached to $C$ ($D$) by a pair of Wilson lines along $w_C$ ($w_D$) that go to the future infinity.

Then following the same procedure as \sec{sssec:em2em}, we can factorize the $A$ subgraph out of $H$, and soft gluons out of $A$. 
This Ward-identity argument applies equally to the approximated region itself and 
to the subtracted smaller regions, to which the same approximator applied. 
Then because the meson $M_A$ is a color singlet state, 
the same soft cancellation happens as Eqs.~\eqref{eq:em2em-col-singlet} and \eqref{eq:em2em-soft-cancel}.
That is, the two infinitely long Wilson lines associated with the $A$ subgraph are joined to a connected one with a finite length, and 
the soft gluons coupling to $A$ are canceled. 
Although the argument in Eqs.~\eqref{eq:em2em-col-singlet} and \eqref{eq:em2em-soft-cancel} is for the 
whole Wilson lines to all orders, it applies to each finite perturbative order as well. 

The result is shown in \fig{fig:am2mm}(b). The remaining gluons only couple to $C$ and $D$, which are both in the final state. 
Then the symmetric deformation to get the gluons out of Glauber region works in the same way as the Sudakov form factor in \eq{eq:glauber-deform}, namely,
\beq
	k_s^{(CD)} \mapsto k_s^{(CD)} + i \order{\lambda} (w_D - w_C).
\eeq
The following factorizations of collinear subgraphs and soft gluons work a similar way to the Sudakov form factor, so will not be repeated here.
The resultant soft Wilson lines are canceled in the same way as those coupling to $A$, as a result of $M_C$ and $M_D$ being color neutral mesons. 

Therefore, we end up with the factorization result of the amplitude,
\begin{align}\label{eq:am2mm-factorize}
	\M_{M_A \gamma \to M_C M_D} & = \sum_{i,j,k} \int_0^1 dx\, dy\, dz \, 
		D_{i/A}(x, \mu) \, H_{i \gamma \to j k}\pp{ x, y, z; \frac{\vec{q}_T}{\sqrt{s}}, \frac{q_T}{\mu} } \nn\\
	& \hspace{8em} \times 
			\bar{D}_{j/C}(y, \mu) \, \bar{D}_{k/D}(z, \mu),
\end{align}
where the sum is over all possible parton flavors, and we have used the multiplicative renormalization of DAs to write each factor as the renormalized one. 
The hard coefficient $H$ is the scattering of the photon off a collinear pair of on-shell massless partons $i$, into two pairs of partons $j$ and $k$. 
Note that the soft cancellation applies also to the subtraction terms in $A, C, D$, and $H$, so the unsubtracted DA factors are the same as the subtracted ones, 
and the $H$ only contains collinear subtractions.

\subsubsection{Triple-meson process: $(CD) = (\mbox{meson}, \mbox{meson})$. Asymmetric deformation.}
\label{sssec:am2mm-asym}
The factorization procedure for the process in \eq{eq:am2mm} is based on the symmetric deformation.
Its feasibility relies on there being only one collinear subgraph $A$ in the initial state, which only exchanges soft gluons 
with final-state particles. The strategy does not apply to the process in \eq{eq:mm2mm} that involves
two widely separated mesons in both initial and final states. On the other hand, we cannot extend the proof to the 
single-diffractive case where the meson $M_A$ is replaced by a diffracted hadron $h$, which enters the hard interaction 
with $\gamma$, but also produces another hadron $h'$ in the nearly forward direction. 
As we will see in \sec{sssec:dvmp}, there exists a kinematic region where the 
momentum component $k_s \cdot w_A$ is pinched in the
Glauber region when the soft gluon $k_s$ is exchanged between the diffracted hadron and final-state mesons. 
It then forbids the symmetric deformation such as \eq{eq:deform-AC}. 
Therefore, a different approach is needed for extending the proof. 
Now, we explore the possibility of asymmetric deformation.

Given the need to allow generalization of the factorization proof to the single-diffractive process, 
we choose not to deform the contour of $k_s^{(Aj)} \cdot w_A$ when a soft momentum $k_s^{(Aj)}$ flows in the $A$-collinear subgraph,
and will instead try to factorize soft interactions from the collinear subgraphs $C$ and $D$.

The needed deformations can be motivated by examining a single soft gluon exchange between different collinear subgraphs. 
We first consider the collinear subgraph $C$ that has one soft gluon $k_s^{(CA)}$ and $k_s^{(CD)}$ exchange 
with the $A$-collinear subgraph and $D$-collinear subgraph, respectively. 
Since $k_s^{(CA)}$ flows in $C$ in the same direction as the $C$-collinear lines, 
the poles of $k_s^{(CA)} \cdot w_C$ are all on the lower half plane, so we deform the contour of $k_s^{(CA)}$ by
\beq[eq:deform CA]
	k_s^{(CA)} \to k_s^{(CA)} + i \, w_A \, \order{Q}  \,,
\eeq
when it is in the Glauber region, similar to \eq{eq:em2em-ks+-deform-Q}. 
Similarly, we deform the contour of $k_s^{(CD)}$ by
\beq[eq:deform CD]
	k_s^{(CD)} \to k_s^{(CD)} + i \, w_D \, \order{Q}  \,.
\eeq
Such deformations move the soft gluon momenta from the Glauber region all the way into the $A$ or $D$ collinear region,
which will be properly treated by collinear approximations.

In order for the approximator in \eq{eq:soft coupling ij} not to obstruct such deformations, we modify it to
\bse\label{eq:soft coupling C}
\begin{align}
	&J^C_{\mu}(k_C, k_s^{(CA)}) \, g^{\mu\nu} \, S_{\nu}(k_s^{(CA)})				
	\simeq 
		J^C_{\mu}(k_C, \hat{k}_s^{(CA)}) \, 
			\frac{\hat{k}_s^{(CA)\mu} \, w_C^{\nu}}{k_s^{(CA)} \cdot w_C + i\epsilon} \, 
			S_{\nu}(k_s^{(CA)})	\,,	\label{eq:soft coupling CA}	\\
	&J^C_{\mu}(k_C, k_s^{(CD)}) \, g^{\mu\nu} \, S_{\nu}(k_s^{(CD)})		
	\simeq 
		J^C_{\mu}(k_C, \hat{k}_s^{(CD)}) \, 
			\frac{\hat{k}_s^{(CD)\mu} \, w_C^{\nu}}{k_s^{(CD)} \cdot w_C + i\epsilon} \, 
			S_{\nu}(k_s^{(CD)})	\,, \label{eq:soft coupling CD}
\end{align}
\ese
where only the relevant arguments are written explicitly.
Both approximations in \eq{eq:soft coupling C} have the structure
\begin{align}
	J^C_{\mu}(k_C, k_s) \, g^{\mu\nu} \, S_{\nu}(k_s)			
	\simeq 
		J^C_{\mu}(k_C, \hat{k}_s) \, 
		\frac{\hat{k}_s^{\mu} \, w_C^{\nu}}{k_s \cdot w_C + i\epsilon} \, 
		S_{\nu}(k_s)	\,,
\end{align}
where the structure $\hat{k}_s^{\mu} \, J^C_{\mu}(k_C, \hat{k}_s)$ allows the use of Ward identity in a uniform way, no matter which other collinear subgraph $k_s$ flows through. The $+ i\epsilon$ choice will lead to future-pointing soft Wilson lines.

Now we consider the collinear longitudinally polarized gluons attaching $C$ to $H$. Similarly, the approximation can be obtained by examining a single gluon, whose momentum $k_C$ flows from $H$ into $C$ and can be expanded in the $w_C$-$\bar{w}_C$ frame,
\beq
	k_C = w_C \, (k_C \cdot \bar{w}_C) + \bar{w}_C \, (k_C \cdot w_C) + k_{C,T} \, ,
\eeq
where among the three terms on the right, the $w_C$ component dominates and scales as $\order{Q}$. Then we approximate $k_C$ in $H$ by
\beq
	k_C \to  \hat{k}_C = w_C \, (k_C \cdot \bar{w}_C) \,,
\eeq
and the coupling of the collinear gluon to $H$ by
\begin{align}\label{eq:collinear approx C}
	H_{\mu}(k_H, k_C) \, g^{\mu\nu} \, J^C_{\nu}(k_C)	
		\simeq 
		H_{\mu}(k_H, \hat{k}_C) \,
		\frac{\hat{k}_C^{\mu} \, \bar{w}_C^{\nu} }{k_C \cdot \bar{w}_C - i\epsilon} \,
		J^C_{\nu}(k_C)	\,,
\end{align}
where only the relevant argument dependence is written explicitly and $k_H$ stands for some hard momentum in $H$. 

When applying \eq{eq:collinear approx C} to the whole graph with deformed contours,
the same gluon $k_C$ can go into the soft region, 
which we notate as $k_C^{(A)}$ or $k_C^{(D)}$
when it attaches to the $A$- or $D$-collinear subgraph. 
Then the deformations in Eqs.~\eqref{eq:deform CA} and \eqref{eq:deform CD} are adapted to
\footnote{Note that now the soft momentum direction is reversed compared to the convention of 
$k_s^{(CA)}$ and $k_s^{(CD)}$, which are used in Eqs.~\eqref{eq:deform CA} and \eqref{eq:deform CD}.}
\beq
	\Delta k_C^{(A)} = -i \, w_A \, \order{Q},
	\quad
	\Delta k_C^{(D)} = -i \, w_D \, \order{Q},
\eeq
implying that the denominator in \eq{eq:collinear approx C} needs to be compatible with the deformations
\begin{align}
	\Delta k_C^{(A)} \cdot \bar{w}_C &= -i \, (w_A\cdot \bar{w}_C) \, \order{Q} = -i  \, \order{Q}	\,,	\nn\\
	\Delta k_C^{(D)} \cdot \bar{w}_C &= -i \, (w_D\cdot \bar{w}_C) \, \order{Q} = 0	\,.
\end{align}
This explains the $- i\epsilon$ choice in \eq{eq:collinear approx C}. 
After applying Ward identity, it leads to collinear Wilson lines pointing to the past.

Eqs.~\eqref{eq:soft coupling C} and \eqref{eq:collinear approx C} constitute the needed approximations related to the collinear subgraph $C$.
Even though we only considered a single soft or collinear gluon connection, they generalize to multiple gluon connections in an obvious way: one just applies \eq{eq:soft coupling C} to every soft gluon connecting $C$ to $A$ or $D$, and \eqref{eq:collinear approx C} to every collinear longitudinally polarized gluon connecting $H$ to $C$. Then by applying suitable on-shell projections to the $C$-collinear quark lines or transversely polarized gluon lines, and summing over all possible attachments of the collinear gluons, we can factorize the collinear longitudinally polarized gluons out of the hard part $H$ onto two Wilson lines along $\bar{w}_C$ pointing to the past, and the soft gluons out of $C$ onto two Wilson lines along $w_C$ pointing to the future.

Similar to the discussion in \sec{sssec:em2em}, although choosing the lightlike $w_C$ 
in the soft approximation \eq{eq:soft coupling C} causes rapidity divergences,
it does not affect our conclusion that the soft gluons eventually cancel as a result of exclusiveness.
Remedying this superficial flaw with a non-lightlike vector $n_C$ is straightforward and shall lead to the same result.

The subsequent argument follows the same line as Secs.~\ref{sssec:em2em} and \ref{sssec:am2mm-sym}. 
By the color neutrality of $M_C$, the soft gluons factorized out of $C$ are canceled order by order, which is proved by identifying the
Wilson line structure that they form. 
This reduces the graph in \fig{fig:am2mm}(a) to the partially factorized one in \fig{fig:am2mm}(c), 
in which only the two collinear subgraphs $A$ and $D$ are coupled to the hard subgraph $H$, 
and the soft subgraph $S$ is only coupled to $A$ and $D$ subgraphs.

With the $C$-collinear subgraph factorized out, the leading-region graph in \fig{fig:am2mm}(c) is 
similar to that in \fig{fig:em2em}(a), whose factorization is discussed in \sec{sssec:em2em}, 
with the asymmetric deformation in \sec{sssec:em2em-deformation}. 
Again, in the treatment of the soft region, one only needs to deform the contour of soft gluon $k_s^{(DA)}$ by 
\beq
	k_s^{(DA)} \to k_s^{(DA)} + i \, w_A \, \order{Q},
\eeq
regardless of the poles of $k_s^{(DA)} \cdot w_A$ provided by the $A$-collinear propagators. By the same argument as for the $C$ subgraph, the soft gluons coupling to $D$ are canceled, and the $D$ subgraph is factorized out of $H$ into the DA for $M_D$. Then the soft gluons are only coupled to the $A$ subgraph and no longer pinched. They can then be deformed into the $A$-collinear region and grouped into a part of $A$-collinear subgraph, which can be further factorized from $H$ into the DA of $M_A$. 

The soft cancellation applies equally to the subtracted terms of smaller regions, so this procedure leads to the same factorization in \eq{eq:am2mm-factorize}.
Even though the Wilson lines associated with the collinear factors point to different directions from the ones in \sec{sssec:am2mm-sym} with symmetric
deformations, due to the cancellation of soft gluons, the Wilson line pair for each collinear factor 
join together into a finite-length Wilson line, with the segments pointing to infinity canceled. 
The resultant DA definitions are therefore universal and do not depend on the specific deformation ways. 
This is a property of collinear factorization.

\subsection{Large-angle meson-meson scattering}
\label{ssec:exclusive-mm}
For meson-meson scattering, the beam particle $B$ is also a meson. The final-state particles $C$ and $D$ can take
(1) $(CD) = (l^+l^-)$, 
(2) $(CD) = (\gamma\gamma)$, 
(3) $(CD) = (\gamma, \mbox{meson})$, or
(4) $(CD) = (\mbox{meson}, \mbox{meson})$.
The first three cases do not raise new issues in factorization, which we will briefly remark on, 
and the last case only requires a simple generalization of our factorization argument for the photon-meson scattering in \sec{sssec:am2mm-asym}.

\begin{figure}[htbp]
	\centering
	\begin{tabular}{cc}
	\includegraphics[scale=0.8]{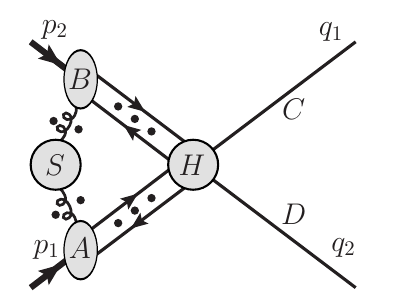} &
	\includegraphics[scale=0.8]{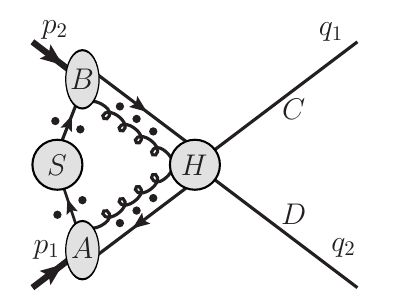}  \\
	(a) & (b)
	\end{tabular}
	\caption{Leading regions of the processes in \eqs{eq:mm2ll}{eq:mm2aa},
	with the final-state lines being electrons or photons. The quark lines can be replaced by transversely polarized
	gluons, and the dots refer to arbitrary numbers of gluon lines with longitudinally polarized gluons.}
	\label{fig:mm2ll}
\end{figure}

\subsubsection{Double-meson process: $(CD) = (l^+l^-)$ or $(\gamma\gamma)$.}
\label{sssec:mm2ll}
The processes 
\beq[eq:mm2ll]
	M_A(p_1) + M_B(p_2) \to l^+(q_1) + l^-(q_2)
\eeq 
and 
\beq[eq:mm2aa]
	M_A(p_1) + M_B(p_2) \to \gamma(q_1) + \gamma(q_2)
\eeq 
have similar color structures as the meson electroproduction (discussed in \sec{sssec:em2em}) 
and photoproduction (discussed in \sec{sssec:am2am}), respectively,
except only that both mesons are now in the initial state. 
They thus have similar leading regions, as shown in \fig{fig:mm2ll}.
As usual, the region (b) is assumed to be power suppressed by the soft-end suppression.

At LO in QED, the dilepton production happens via the production and decay of a timelike virtual photon of invariant mass $Q = m_{ll}$. 
An example is $\pi^+ \pi^- (\to \gamma^*) \to l^+ l^-$. 
This property means that it is the invariant mass $Q$ that provides the hard scale for factorization, 
regardless of the transverse momentum $q_T$ of the leptons, similar to the dilepton photoproduction in \sec{sssec:am2ll}.
On the other hand, in the diphoton production, both photons directly attach to the quark parton lines, 
and the large $q_T$ is necessary for factorization. 
Examples are $\pi^+ \pi^- \to \gamma \gamma$ or $\pi^0 \pi^0 \to \gamma \gamma$.

Factorization for the region (a) works in a similar way to the meson electroproduction and photoproduction discussed before.
One can use either symmetric or asymmetric deformation to avoid the Glauber region. They give different soft and collinear Wilson lines
in intermediate steps, but result in the same soft cancellation and the same collinear factor definitions, as a property of collinear factorization.
The asymmetric deformation is particularly important for later generalization to single-diffractive scattering. 
For a soft gluon momentum $k_s$ flowing from $B$ to $A$, we expand it as
\beq
	k_s = \frac{k_s \cdot w_A}{w_A \cdot w_B} w_B + \frac{k_s \cdot w_B}{w_A \cdot w_B} w_A + k_{sT},
\eeq
with $w_A$ defined as in \eq{eq:auxiliary-A-C-D} and $w_B = \bar{w}_A$ in the c.m.~frame, and deform its contour by
\beq
	k_s \mapsto k_s - i \order{Q} w_A.
\eeq
This then determines all necessary $i\epsilon$ prescriptions for the soft and collinear approximations.
	
In the end, we get a factorization formula for the scattering amplitude,
\begin{align}
	&\M_{M_A M_B \to l^+ l^- / \gamma\gamma}	
	= \sum_{i, j} \int_0^1 dx \, dy \, D_{i/A}(x, \mu) \, D_{j/B}(y, \mu) \,
		H_{ij \to l^+ l^- / \gamma\gamma}\pp{ x, y; \frac{\vec{q}_T}{\sqrt{s}}, \frac{q_T}{\mu} },
\end{align}
where the DAs are for initial-state meson annihilations, and the hard coefficient $H$ is for the scattering of two pairs of collinear partons $i$
and $j$ into $l^+ l^-$ or $\gamma\gamma$.
We have used their multiplicative renormalization to convert each factor to renormalized ones, which introduces the factorization scale $\mu$.

\begin{figure}[htbp]
	\centering
	\begin{tabular}{cc}
	\includegraphics[scale=0.8]{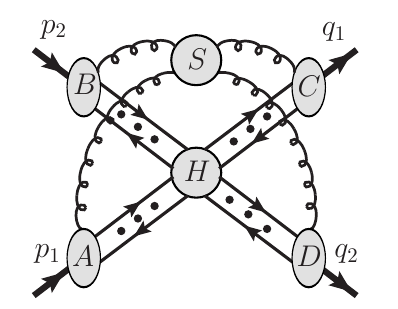} &
	\includegraphics[scale=0.8]{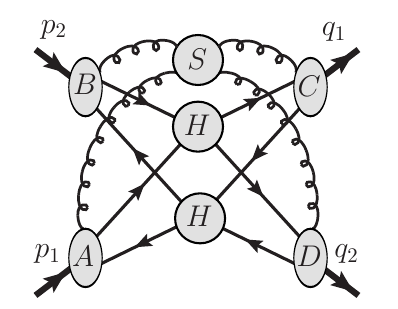}  \\
	(a) & (b)
	\end{tabular}
	\caption{Leading regions of the process in \eq{eq:mm2mm}. 
	(a) has one single hard scattering, and (b) has two hard scatterings.}
	\label{fig:mm2mm}
\end{figure}

\subsubsection{Triple-meson process: $(CD) = (\gamma, \mbox{meson})$.}
\label{sssec:mm2am}

The triple-meson process 
\beq[eq:mm2am]
	M_A(p_1) + M_B(p_2) \to \gamma(q_1) + M_D(q_2)
\eeq 
exactly resembles the meson pair photoproduction in \eq{eq:am2mm}, treated in
Secs.~\ref{sssec:am2mm-sym} and \ref{sssec:am2mm-asym}, just with the exchange of the photon and one meson.
As there, both symmetric and asymmetric deformations are applicable to deal with the Glauber region.
The symmetric one starts with the factorization of $D$-collinear subgraph, which reduces the leading region to 
\fig{fig:mm2ll}, discussed in \sec{sssec:mm2ll}.
The asymmetric deformation keeps intact the soft gluon momentum components flowing 
along the $A$-collinear subgraph, but deforms the other
components in $B$ or $D$ collinear subgraphs by an order of $Q$.
The soft gluons cancel in both cases as a result of the mesons being color neutral. 
Finally, the amplitude is factorized into
\begin{align}
	\M_{M_A M_B \to \gamma M_D} = & \sum_{i, j, k} \int_0^1 dx \, dy \, dz \, D_{i/A}(x, \mu) D_{j/B}(y, \mu)
	\nn\\
	&\hspace{6em} \times
		H_{ij \to \gamma k}\pp{ x, y, z; \frac{\vec{q}_T}{\sqrt{s}}, \frac{q_T}{\mu} }
		\bar{D}_{k/D}(z, \mu).
\end{align}

\subsubsection{Quadruple-meson process: $(CD) = (\mbox{meson}, \mbox{meson})$.}
\label{sssec:mm2mm}
The quadruple-meson process 
\beq[eq:mm2mm]
	M_A(p_1) + M_B(p_2) \to M_C(q_1) + M_D(q_2)
\eeq 
has two collinear sectors in both initial and final states. 
The leading region is shown in \fig{fig:mm2mm}(a).
The symmetric deformation out of Glauber region does not trivially apply, as explained in \sec{sssec:am2mm-sym}.
So we will simply employ the asymmetric deformation in \sec{sssec:am2mm-asym}.

For a soft gluon momentum $k_s^{(Cj)}$ flowing from the collinear subgraph $C$ to some other one $j$, we expand it as
\beq
	k_s^{(Cj)} = \frac{k_s^{(Cj)} \cdot w_C}{w_C \cdot w_j} w_j + \frac{k_s^{(Cj)} \cdot w_j}{w_C \cdot w_j} w_C + k_{sT}.
\eeq
When it flows through the $C$-collinear subgraph, we approximate it by
\beq
	k_s^{(Cj)} \mapsto \hat{k}_s^{(Cj)} = \frac{k_s^{(Cj)} \cdot w_C}{w_C \cdot w_j} w_j.
\eeq
This component receives poles from the $C$-collinear lines, which are all on the lower half plane.
So we choose to deform its contour by
\beq[eq:mm2mm-deform]
	k_s^{(Cj)} \mapsto k_s^{(Cj)} + i \order{Q} w_j.
\eeq
This determines the $i\epsilon$ prescription in the approximation of its coupling to the $C$-collinear subgraph,
\beq
	J_{\mu}^C(k_C, k_s^{(Cj)}) \, g^{\mu\nu} \, S_{\nu}(k_s^{(Cj)})
		\mapsto
		J_{\mu}^C(k_C, \hat{k}_s^{(Cj)}) 
			\, \frac{\hat{k}_s^{(Cj)\, \mu} w_C^{\nu}}{k_s^{(Cj)}\cdot w_C + i \epsilon}
			\, S_{\nu}(k_s^{(Cj)}).
\eeq
This will allow the use of Ward identity to factorize soft gluons out of $J^C$ onto a pair of Wilson lines along $w_C$ pointing into the future.

For a $C$-collinear momentum $k_C$, we expand it as
\beq
	k_C = (k_C \cdot w_C) \, \bar{w}_C + (k_C \cdot \bar{w}_C) \, w_C + k_{C, T},
\eeq
and only keep the large component $k_C \cdot \bar{w}_C$ in the hard part $H$,
\beq
	k_C \mapsto \hat{k}_C = (k_C \cdot \bar{w}_C) \, w_C.
\eeq
The coupling of a collinear longitudinally polarized gluon $k_C$ to $H$ is approximated by
\beq
	H_{\mu}(k_H; k_C) \, g^{\mu\nu} \, J^C_{\nu}(k_C)
	\mapsto
		H_{\mu}(k_H; \hat{k}_C) \, 
			\frac{\hat{k}_C^{\mu} \bar{w}_C^{\nu}}{k_C \cdot \bar{w}_C - i \epsilon} \, 
			J^C_{\nu}(k_C),
\eeq
for $k_C$ to flow from $H$ into $J^C$. The $- i \epsilon$ is uniquely determined to be 
compatible with the deformation in \eq{eq:mm2mm-deform},
given the need of soft subtraction. 
This will lead to a pair of past-pointing Wilson lines along $\bar{w}_C$ 
to collect all $C$-collinear gluons with longitudinal polarization.

After factorizing the $C$-collinear subgraph from $H$, and soft gluons out of $C$, one can easily identify the soft Wilson lines as an identity by the 
color neutrality of the meson $M_C$. This soft cancellation applies to both the approximated region and the subtracted smaller regions. 
Thus we have a factorized DA for $M_C$, whose unsubtracted version is the same as the soft-subtracted one, convoluted with the rest of the 
graph. It then has the same color structure as the triple-meson process in \eq{eq:mm2am} and factorizes in the same manner.
Eventually, we have the factorized expression for the amplitude,
\footnote{Note the symbol $D$ has been used to denote both the DA and the particle $D$ in the $2\to2$ scattering, which should not cause confusion.}
\begin{align}
	\M_{M_A M_B \to M_C M_D} & = \sum_{i,j,k,l} \int_0^1 dx \, dy \, dz \, dw \, D_{i/A}(x, \mu) D_{j/B}(y, \mu) \nn\\
	& \hspace{6em}
	\times H_{ij \to kl}\pp{ x, y, z, w; \frac{\vec{q}_T}{\sqrt{s}}, \frac{q_T}{\mu} }
		\bar{D}_{k/C}(z, \mu) \bar{D}_{l/D}(w, \mu),
\end{align}
where the sum is over all parton flavors and their spin structures, and the hard coefficient $H$ is the scattering of two pairs of collinear partons $i$ and $j$
into another two pairs $k$ and $l$. Again, the hard coefficient contains subtraction of collinear regions for each of the four mesons, and we have used
the multiplicative renormalization of DAs to convert all factors into renormalized ones, which introduced the factorization scale $\mu$.

The leading regions that contain soft quark or physically polarized gluon lines to directly couple any of the collinear subgraphs to the soft subgraph
are assumed to be power suppressed, by the same soft-end suppression assumption that applies to all the processes discussed before.
However, for the quadruple-meson scattering process, there is one different type of regions that count at a more leading power. 
This is given by the reduced diagram in \fig{fig:mm2mm}(b) that has two separated hard scattering subgraphs. 
Discussion of such multiple hard scattering case is beyond the scope of this thesis, for which we refer to \citep{Landshoff:1974ew, Botts:1989kf}.
In this thesis, we assume that all processes are dominated by one single hard scattering.

\section{Single-diffractive hard exclusive processes}
\label{sec:SDHEP}

\begin{figure}[htbp]
	\centering
	\begin{tabular}{cc}
	\includegraphics[trim={-2em 0 -2em 0}, clip, scale=0.75]{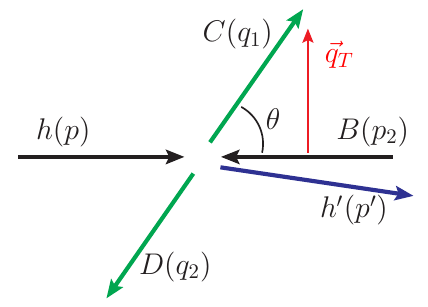} &
	\includegraphics[trim={-2em 0 -2em 0}, clip, scale=0.75]{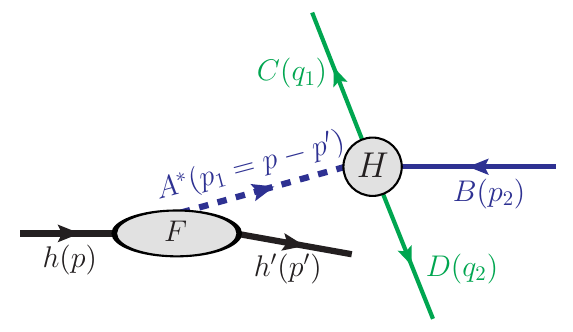}  \\
	(a) & (b)
	\end{tabular}
	\caption{(a) Illustration of the kinematic configuration of the SDHEP in the lab frame. 
	(b) The two-stage paradigm of the SDHEP.}
	\label{fig:sdhep}
\end{figure}

Now we generalize the $2\to2$ large-angle meson scattering processes by 
allowing one extra hadron $h'$ in the final state along the direction of one of the
initial-state hadrons $h$. The extra hadron $h'$ is the {\it diffraction} of the initial-state hadron $h$. 
To allow perturbative QCD study, we further require a hard scale in the scattering process, 
so we take the two particles $C$ and $D$ in the final state to have hard transverse momenta with respect to
the collision axis. Thus the minimal configuration we study is a generic $2\to3$ process that we call {\it single-diffractive hard exclusive process} (SDHEP),
\beq[eq:sdhep]
	h(p) + B(p_2) \to h'(p') + C(q_1) + D(q_2),
\eeq 
where $h$ of momentum $p$ is the hadron we would like to study, 
$B$ of momentum $p_2$ is a colliding lepton, photon or meson, 
and $C$ and $D$ of momentum $q_1$ and $q_2$, respectively, are two final-state particles, which can be a lepton, photon or meson, with large transverse momenta, 
\beq[eq:hard qT]
	q_{1T}\sim q_{2T} \gg \sqrt{-t}\, ,
\eeq 
with $t \equiv (p-p')^2$. 
In the lab frame with $h$ along $+\hat{z}$ and $B$ along $-\hat{z}$, the scattering configuration is illustrated in \fig{fig:sdhep}(a).
	
\begin{figure*}[htbp]
\def\sc{1.3}
\centering
	\begin{align*}
		&\adjincludegraphics[valign=c, scale=0.7]{figures/sdhep.pdf}
	 	\; \Scale[\sc]{=}
	 	\adjincludegraphics[valign=c, scale=0.7]{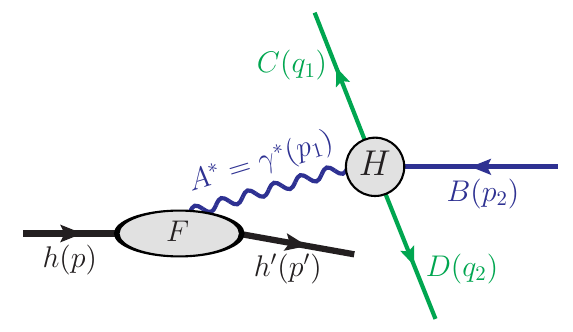} \;\Scale[\sc]{+} \\
		& \hspace{1.8em} \Scale[\sc]{+}
	 	\adjincludegraphics[valign=c, scale=0.7]{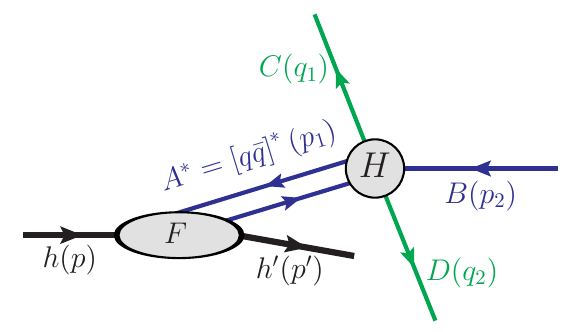} \;\Scale[\sc]{+} 
	 	\adjincludegraphics[valign=c, scale=0.7]{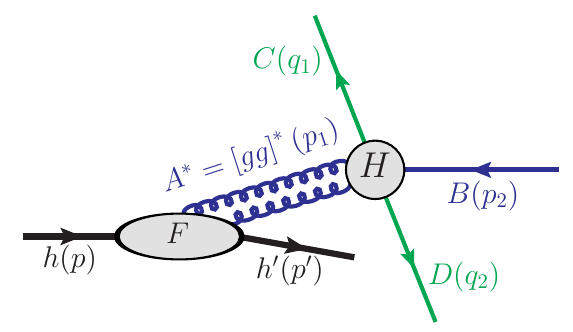} \Scale[\sc]{+\cdots}
	\end{align*}
	\caption{The representation of the SDHEP in terms of all possible exchanged channels 
	of the virtual state $A^*(p_1)$ between the single-diffractive $h\to h'$ transition [\eq{eq:diffractive}] 
	and the $2\to 2$ hard exclusive process [\eq{eq:hard 2to2}]. 
	The two gluons in $gg$ channel have physical polarizations. 
	The $q\bar{q}$ and $gg$ channels can be accompanied by arbitrary numbers of 
	collinear longitudinally polarized gluons. 
	The ``$\cdots$" refers to the channels with more than two physically polarized partons, 
	which are power suppressed compared to the two-parton case.}
\label{fig:decomposition}
\end{figure*}

There are two distinct scales involved in the SDHEP, one soft scale $\sqrt{-t}$ characterizing the diffraction subprocess, and one hard scale $Q = \order{q_T}$ characterizing the
production of the particles $C$ and $D$. Then the SDHEP can be pictured as a two-stage process, as shown in \fig{fig:sdhep}(b),
being a combination of a diffractive production of a single long-lived state $A^*(p_1)$,
\beq[eq:diffractive]
	h(p) \to A^*(p_1) + h'(p'), \quad \mbox{with } p_1 = p - p',
\eeq
and a hard exclusive $2\to2$ scattering between the two nearly head-on states $A^*(p_1)$ and $B(p_2)$,
\beq[eq:hard 2to2]
	A^*(p_1) + B(p_2) \to C(q_1) + D(q_2).
\eeq
In the c.m.~frame of $A^*$ and $B$, as a necessary condition for factorization, the transverse momentum $q_T$ of $C$ or $D$ is required to be much greater than the invariant mass of $A^*$ or $B$. 

The $2\to 2$ hard exclusive process $H$ in \fig{fig:sdhep}(b) takes place at a short distance $1/Q \ll 1/\LQCD\sim 1\sim\fm$ 
and is sensitive to the partonic structure of the exchanged state  $A^*(p_1)$. 
The scattering amplitude of the SDHEP should include a sum of all possible partonic states, as illustrated in \fig{fig:decomposition}, 
which can be schematically described as
\beq[eq:channels]
	\mathcal{M}_{hB\to h'CD} = \sum_{n = 1}^{\infty} \sum_{f_n} F_{h\to h'}^{f_n}(p, p') \otimes C_{f_n B \to CD},
\eeq
where $n$ and $f_n$ represent the number and flavor content of the particles in the exchanged state $A^*$, respectively,
$F_{h\to h'}^{f_n}(p, p')$ is a ``form factor'' responsible for the $h\to h'$ transition, and 
$C_{f_n B \to CD}$ denotes the scattering amplitude of the hard part $H$, 
along with the sum running over all possible exchanged states characterized by $n$ and $f_n$. 
Throughout the discussion in this thesis, 
we retain the scattering amplitude $C_{f_n B \to CD}$ at the lowest order in the QED coupling constant 
for given exchanged state $f_{n}$ and particle types of $B$, $C$, and $D$, 
while investigating contributions from QCD to all orders in its coupling constant.

For $n = 1$, the only possible case is a virtual photon exchange, i.e., $f_1 = \gamma^*$, 
which resembles the Bethe-Heitler process for the DVCS (see~\citep{Ji:1996nm} for example). 
Rather than probing the partonic structure of $h$, this channel only gives an access to the 
electromagnetic form factor of $h$ evaluated at a relatively soft scale $t$.  
As discussed below, the $\gamma^*$-mediated subprocess gives a 
``superleading-power" background for the $n\geq 2$ channels, 
and should not be disregarded even if suppressed by higher-order QED couplings, 
unless symmetry considerations prohibit it. 
The scattering amplitude of the SDHEP should be expanded in inverse powers of the hard scale, and then 
followed by a perturbative factorization for the leading-power contribution 
(and subleading-power contribution if needed, see, e.g.,~\citep{Kang:2014tta}). 
If the $n=1$ subprocess is forbidden (as discussed below), 
then the scattering amplitude of the SDHEP starts with $n = 2$ subprocesses.

For $n = 2$, we can have QCD subprocesses with $f_2 = [q\bar{q}']$ or $[gg]$. This gives the leading-power contribution that, 
as shown in the following subsections, can be factorized into GPDs with corresponding hard coefficients.  
The channels with $n \geq 3$ belong to high-twist subprocesses that are suppressed by powers of $\sqrt{-t}/Q$ 
and will be neglected in the following analysis.

\subsection{General discussion of the $\gamma^*$-mediated channel}
\label{ssec:n=1}

Before providing the detailed arguments for QCD factorization of SDHEPs, initiated by a lepton, photon or meson beam, respectively, 
in the next three subsections, 
we first give a general discussion for the $\gamma^*$-mediated hard subprocesses, 
corresponding to the $n = 1$ channel in \eq{eq:channels}, independent of the particle types of $B, C$ and $D$.  
More detailed discussions for specific processes will be given in later subsections.

One difference between the $n = 1$ and $n \geq 2$ subprocesses is that the virtual photon momentum 
is fully determined by the diffraction of the hadron $h$. The amplitude of the $\gamma^*$-mediated subprocess 
can be trivially factorized into the electromagnetic form factor of $h$,
\begin{align}\label{eq:n1 FF}
	\M^{(1)} 
	&= -\frac{e}{t} \, \langle h'(p') | J^{\mu}(0) | h(p) \rangle \cdot \langle C(q_1) D(q_2) | \pp{ -i e J_{\mu}(0) } | B(p_2) \rangle	\nn \\
	&\equiv -\frac{e}{t} \, F^{\mu}(p, p') \, \H_{\mu}(p_1, p_2, q_1, q_2),
\end{align}
where the superscript ``$(1)$'' refers to the contribution to the SDHEP amplitude from the $n = 1$ channel, 
and $J^{\mu} = \sum_{i \in q} e_i \, \bar{\psi}_i \gamma^{\mu} \psi_i$ is the electromagnetic current of quarks, 
summing over their flavors ``$i$'' and weighted by their fractional charges $e_i$, 
normalized such that $e_u = 2/3$ and $e_d = -1/3$. 
In the second step we defined the hard factor $\H_{\mu}$ that includes the QED coupling $-ie$, 
and the electromagnetic form factor,
\begin{align}
	F^{\mu}(p, p') = \langle h'(p') | J^{\mu}(0) | h(p) \rangle	
		= F_1^h(t) \, \bar{u}(p') \gamma^{\mu} u(p) - F_2^h(t) \, \bar{u}(p')\frac{i\sigma^{\mu\nu}p_{1\nu}}{2m_h} u(p),
\label{eq:EM-form-factor}
\end{align}
which has the leading component $F^+ \sim \order{Q}$ as the $h$-$h'$ system is highly boosted along the $z$ direction. 
However, when this component is contracted with $\H_{\mu}$, which scales as $\order{Q^0}$ for each component, we have
\begin{align}
	&F^+ \H^- = \frac{1}{p_1^+} \, F^+ \pp{ p_1^+ \, \H^-}	
		= \frac{1}{p_1^+}  \, F^+ \pp{p_1 \cdot \H + \bm{p}_{1T}\cdot \bm{\H}_{T} - p_1^- \H^+ },
\end{align}
where in the bracket, the first term vanishes by the Ward identity of QED, 
and the other two scale as $\sqrt{-t}$ and $t/p_1^+$, respectively. 
So the leading power of $F \cdot \H$ scales as $\sqrt{-t}$ and is given by the transverse polarization of the virtual photon. 
Therefore, the power counting of $\M^{(1)}$ is of the order $1 / \sqrt{-t}$, which is higher than the $n = 2$ channel by one power of $Q / \sqrt{-t}$. 

One caution should be noted that it is not appropriate to only keep $p_1^+$ in the amplitude $\H_{\mu}(p_1, p_2, q_1, q_2)$ 
because the approximation introduces an error of order $\sqrt{-t}/Q$. While this is power suppressed 
comparing to the leading contribution from the $n = 1$ channel, it could scale at the same order as the contribution 
from the $n = 2$ channel since both of them have the power counting $1/Q$. By neglecting all the $n \geq 3$ channels, 
our approximation to the full SDHEP amplitude is up to the error at $\order{\sqrt{-t}/Q^2}$, so that the $1/Q$ part 
should be kept as exact when evaluating the contribution from the $n=1$ channel.
This will be explicitly demonstrated for the single-diffractive real photon electroproduction in \sec{sec:dvcs-sdhep}.

There is one further subtlety when the $\gamma^*$-mediated subprocess involves light mesons in $\H$. 
The conventional practice is to factorize it into meson distribution amplitudes (DAs).
While this is true to the leading power at $\order{1/\sqrt{-t}}$, it neglects the power correction of 
$\order{\LQCD/Q} \cdot \order{1/\sqrt{-t}} = \order{1/Q}$, which is of the same order as the $n = 2$ channels, i.e., the GPD channels. 
Keeping the exact $1/Q$ contribution thus requires the subleading-power (or, twist-3) factorization for the 
$\gamma^*$-mediated subprocess that involves any mesons, which is beyond the scope of this thesis.

There are two cases in which the $\gamma^*$-channel is forbidden. The first is for a flavor-changing channel 
with $h \neq h'$ that cannot be achieved by the electromagnetic interaction, like the pion-nucleon scattering processes 
in \citep{Berger:2001zn, Qiu:2022bpq} which can involve the proton-neutron transition. 
The second case is for particular combinations of the particle types of $B, C$ and $D$ 
that mandate $\H_{\mu}(p_1, p_2, q_1, q_2) = 0$ by some symmetries. 
Apart from these two cases, we should generally include the $\gamma^*$-mediated subprocess. 

For example, for the photoproduction of diphotons considered in~\citep{Pedrak:2017cpp}, 
one should include the $\gamma^*$ channel that involves photon-photon scattering in $\H_{\mu}$. 
Even though this is suppressed by $\alpha_{\rm em}$ compared to the GPD subprocess that 
corresponds to the $n = 2$ channel, the $\gamma^*$ channel at $n=1$ is power enhanced by $Q/ \sqrt{-t}$. 
In such cases, we need to carefully compare the contributions from both channels,  
and to develop an experimental approach to remove the background due to the $\gamma^*$ channel 
in order to extract GPDs from the experimental data. 
One common approach by using azimuthal correlations will be discussed 
in Secs.~\ref{sec:SDHEP-frame} and \ref{sec:dvcs-sdhep}.

\subsection{SDHEP with a lepton beam}
\label{ssec:sdhep-lepton}

For single-diffractive hard exclusive electroproduction processes, we have $B = C = e$. 
The other particle $D$ can be a photon $\gamma$ or a light meson $M_D$. 
Both of these two processes allow the $\gamma^*$-initialized channel with $n = 1$. 
For the $n = 2$ channel, $D = \gamma$ leads to the 
deeply virtual Compton scattering (DVCS)~\citep{Ji:1996nm, Radyushkin:1997ki}, 
and $D = \mbox{meson}$ corresponds to the 
deeply virtual meson production (DVMP)~\citep{Brodsky:1994kf, Frankfurt:1995jw}. 
Both processes have been proved to be factorized into GPDs~\citep{Collins:1998be, Collins:1996fb}. 
Here, we will switch the theoretical perspective from~\citep{Collins:1998be, Collins:1996fb} 
by fitting them into the general SDHEP framework. 
The proof follows the two-stage paradigm depicted in Eqs.~\eqref{eq:diffractive}--\eqref{eq:channels}. 
This approach incorporates the $\gamma^*$-initialized $n=1$ channel naturally, and for the $n = 2$ channel, 
it leads to a direct analogy to the large-angle meson scattering processes in \eq{eq:hard 2to2} by 
having $A^*$ being some meson state carrying the quantum number of the $[q\bar{q}']$ or $[gg]$ state. 
Our strategy for the proof follows a two-step process introduced in~\citep{Qiu:2022bpq, Qiu:2022pla}: 
(1) justify the factorization for a simpler $2\to 2$ hard exclusive process in \eq{eq:hard 2to2}, which has been done in \sec{sec:large-angle-meson}, and 
(2) extend the factorization to the full SDHEP in \eq{eq:sdhep} by addressing extra complications, 
including especially the difficulty from Glauber gluons. 
As expected, we will reproduce the proofs in~\citep{Collins:1998be, Collins:1996fb}.

\begin{figure}[htbp]
\centering
	\begin{tabular}{ccc}
		\includegraphics[scale=0.7]{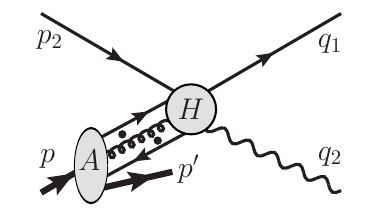} &
		\includegraphics[clip, trim={0 -2mm 0 0}, scale=0.7]{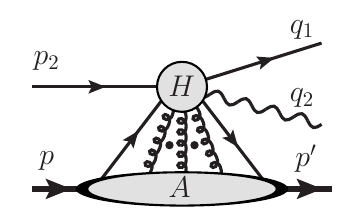} &
		\includegraphics[scale=0.7]{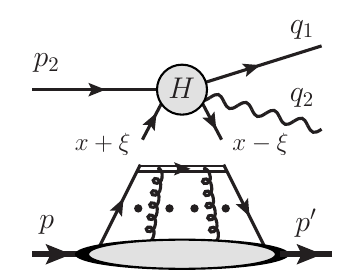} \\
		(a) & (b) & (c)
	\end{tabular}
\caption{Leading-region graphs of the DVCS for the (a) ERBL region and (b) DGLAP region of the GPD. 
	(c) illustrates the result after factorizing the collinear subgraph out of $H$ into a GPD.
	The two quark lines can be replaced by two transverse gluon lines. }
\label{fig:dvcs}
\end{figure}

\subsubsection{Real photon production: $D = \gamma$}
\label{sssec:dvcs}

For $n = 1$, this gives the Bethe-Heitler process, and the amplitude $\H^{\mu}$ in \eq{eq:n1 FF} 
is the scattering amplitude of $\gamma^*(p_1) + e(p_2) \to e(q_1) + \gamma(q_2)$ with $q_{1T}^2 \gg |p_1^2|=|t|$.

For $n = 2$, the state $A^*$ can be either a collinear $q\bar{q}$ or $gg$ pair, which interacts with the electron beam 
by exchanging a virtual photon $\gamma_{ee}^*$ with momentum $q = p_2 - q_1$, similar to \fig{fig:em2ea}(c).
This channel is therefore referred to as deeply virtual Compton scattering (DVCS).
The $[q\bar{q}]$ and $[gg]$ state can be accompanied by an arbitrary number of longitudinally polarized collinear gluons. 
The traditional treatments are all carried out in the Breit frame of the virtual photon 
$\gamma_{ee}^*$ and hadron beam $h$~\citep{Collins:1998be, Collins:1996fb}. 
Here, we follow the kinematic setup of the SDHEP in~\eq{eq:sdhep} to work in the 
c.m.~frame of the initial-state hadron and electron with the hadron along the $z$ axis.
The requirement of a high virtuality $Q^2 = -q^2$ for the $\gamma_{ee}^*$ is equivalent to 
the requirement of hard transverse momenta $q_T$ for the final-state electron and photon in this frame, 
since $Q^2 \propto q_T^2$. 
Hence, the virtual photon $\gamma_{ee}^*$ has a short lifetime and belongs to the hard part, 
and therefore we have the leading-region diagrams as in \fig{fig:dvcs}(a)(b).

Due to the presence of the diffraction, now we have two types of leading regions. 
In the first region [\fig{fig:dvcs}(a)], all the $h$-collinear parton lines go into the hard scattering $H$ with positive plus momenta. 
This region is greatly similar to the leading region of meson scattering, e.g., \fig{fig:em2ea}(b). 
We call it ERBL region. 
In the other region [\fig{fig:dvcs}(b)], however, we also have some of the $h$-collinear parton lines go out of $H$ with positive plus momenta
and merge with the beam remnants to form the diffracted hadron $h'$.
This region is called DGLAP region. 
It has no analogy in the large-angle meson scattering so represents a new feature of the diffractive scattering.
In the DGLAP region, not only do we have long-lived propagating partons lines connecting the collinear subgraph to the hard subgraph $H$, but
also have long-lived remnant particles propagating along the collinear direction of $h$ as spectators of the hard interaction. 
Therefore, one has an opened color object lasting for a long time throughout the whole scattering.
Soft gluons can be exchanged between the spectators and colored lines along other collinear directions. 
This will lead to the problem of Glauber pinch that we will discuss in detail below. 

Luckily, the Glauber region does not cause any issue for the real photon electroproduction process here, since there is only one
collinear subgraph and no soft subgraph. 
Therefore, the factorization proof can be directly built on that of the corresponding meson scattering process treated in \sec{sssec:em2ea}.

For both the ERBL and DGLAP regions, the collinear momenta $k_i^{\mu}$ are pinched for 
their minus components if $\sqrt{-t}  \ll p_1^+ \sim q_T$. 
Introducing the scaling variable $\lambda =\sqrt{-t} \ll q_T$, the collinear momentum scaling is the same as in \eq{eq:e2em-col-scaling}. 
And then the same approximations as in Eqs.~\eqref{eq:e2em-col-mom-approx}--\eqref{eq:e2em-col-g-approx} 
can be made to factorize the collinear subgraph from the hard subgraph for quark-initiated processes.
For the diffractive scattering, one no longer has isospin symmetry to forbid the gluon-initiated channel, 
so the leading region in \fig{fig:dvcs} contains an extra case when all collinear parton lines are gluons. 
Then we replace each gluon coupling by
\beq
	H_{\mu}(k_i; k_H) \, g^{\mu\nu} \, C_{\nu}(k_i)
		=
		H_{\mu}(k_i; k_H) \, (K^{\mu\nu}(k_i, n) + G^{\mu\nu}(k_i, n)) \, C_{\nu}(k_i),
\eeq
with
\beq
	K^{\mu\nu}(k_i, n) = \frac{k_i^{\mu} n^{\nu}}{k_i \cdot n - i \epsilon}, \quad
	G^{\mu\nu}(k_i, n) = g^{\mu\nu} - \frac{k_i^{\mu} n^{\nu}}{k_i \cdot n - i \epsilon}.
\eeq
Note that no replacement of $k_i \to \hat{k}_i$ as \eq{eq:e2em-col-mom-approx} has been made in $H$.
A gluon with its coupling replaced by the $K$ ($G$) factor is called a $K$-gluon ($G$-gluon). 
When all or all but one gluons are $K$-gluons, we get a super-leading power contribution.
The region that has two $G$-gluons with all the others being $K$-gluons corresponds to the leading power.
When there are three or more $G$-gluons, one receives a power suppression.
As demonstrated in \citep{Collins:2008sg}, the super-leading power contribution is cancelled, 
but those regions still give nonzero contribution at leading power, which combines with 
the leading regions to give the full leading-power contribution.

After use of Ward identities and sum over regions and graphs, the collinear lines are factorized out of the hard part,
as in \eq{eq:em2ea-sum-over-regions},
\begin{align} \label{eq:eh2eah-factorize}
	 \M^{(2)}_{he \to h' e\gamma} 
	&= \int \frac{d^4k}{(2\pi)^4} \bb{ 
		\sum_q 
			\H^q_{\beta\alpha; ji}(\hat{k}, \hat{p}_1 - \hat{k}) \,
			\C^q_{\alpha\beta; ij}(k; p, p') 
			\right. \nn\\
		& \hspace{13.5em} \left.
			+ \H^g_{\nu\mu; ba}(\hat{k}, \hat{p}_1 - \hat{k}) \,
			\C^g_{\mu\nu; ab}(k; p, p') 
		},
\end{align}
up to terms suppressed by powers of $\lambda / q_T$, 
where the superscript ``(2)'' refers to the contribution to the SDHEP amplitude from the $n = 2$ channel.
We have included the contributions from both quark and gluon channels.
The collinear factor $\C^q$ for the quark parton differs from \eq{eq:em2ea-col-factor-0} only in the external hadron states,
\begin{align}\label{eq:eh2eah-col-factor-q0}
	\C^q_{\alpha\beta; ij}(k; p, p') 
	= & \P_{A, \alpha\alpha'} 
		\int d^4y \, e^{i k \cdot y} \, \langle h'(p') | \T \cc{ 
			\bb{ \bar{\psi}_{q, \beta'}(0) W^{\dag}(\infty, 0; n) }_j  
		\right.\nn\\
		& \hspace{10em}
		\left. \times
			\bb{ W(\infty, y; n) \psi_{q, \alpha'} (y) }_i 
		} | h(p) \rangle \, 
		\Pb_{A, \beta'\beta},
\end{align}
but now we are allowed to have more spin structures,
\beq[eq:eh2eah-col-factor-q0-expand]
	\C^q_{\alpha\beta; ij}(k; p, p') 
	= \frac{\delta_{ij}}{N_c} \bb{ 
			\C^{q, +}(k; p, p') \frac{\gamma^-}{2}
			+ \wt{\C}^{q, +}(k; p, p') \frac{\gamma_5\gamma^-}{2}
			+ \sum_{\ell = 1, 2} \C_{\perp}^{q, +\ell}(k; p, p') \frac{\sigma^{\ell -}}{2}
		}_{\alpha\beta},
\eeq
with each factor defined as
\begin{align}\label{eq:eh2eah-col-factor-q1}
	& \pp{ \C^{q, +}, \wt{\C}^{q, +}, \C_{\perp}^{q, +\ell} }(k; p, p') = 
	\int d^4y \, e^{i k \cdot y} \, \langle h'(p') | \T \cc{ 
			\bar{\psi}_{q}(0) W^{\dag}(\infty, 0; n)
		\right.\nn\\
		& \hspace{15em}
		\left. \times 
			\pp{ 
				\frac{\gamma^+}{2}, \frac{\gamma^+\gamma_5}{2}, \frac{\sigma^{+\ell}}{2}
			}
			W(\infty, y; n) \psi_{q} (y)
		} | h(p) \rangle.
\end{align}
The collinear factor for the gluon parton is
\begin{align}\label{eq:eh2eah-col-factor-g0}
	\C^{g,\mu\nu}_{ab}(k; p, p') 
	= & \frac{1}{k^+ (k - p_1)^+}
		\int d^4y \, e^{i k \cdot y} \, \langle h'(p') | \T \cc{ 
			\bb{ G^{+\nu}(0) W_A^{\dag}(\infty, 0; n) }_b  
		\right.\nn\\
		& \hspace{12em}
		\left. \times
			\bb{ W_A(\infty, y; n) G^{+\mu}(y) }_a 
		} | h(p) \rangle,
\end{align}
where $G_a^{\mu\nu} = \partial^{\mu} A^{\nu}_a - \partial^{\nu} A^{\mu}_a - g f^{abc} A_b^{\mu} A_c^{\nu}$ is the gluon field strength tensor, and 
$W_A$ is the Wilson line in the adjoint representation,
\beq[eq:wilson-line-a]
	W_{A, ab}(\infty, y; n)= \P \exp\cc{ - g \int_0^{\infty} d\lambda \, n_{\mu} \, A^{\mu}_c(y + \lambda n) \, (f^{cab}) },
\eeq
obtained by replacing $t^a$ in \eq{eq:DIS-wilson-line} by $T^a_A = -i (f^{abc})$. 
Due to the antisymmetry of $G_a^{\mu\nu}$, only the components of $\C^{g,\mu\nu}_{ab}$ with $\mu, \nu = 1, 2$ are nonzero, so 
$\C^{g,\mu\nu}_{ab}$ also has four independent Lorentz structures, similar to \eq{eq:eh2eah-col-factor-q0-expand},
\beq[eq:eh2eah-col-factor-g0-expand]
	\C^{g, ij}_{ab}
	= \frac{\delta_{ab}}{2(N_c^2 - 1)} \bb{
			\C^{g, \bar{i} \bar{i}} \delta^{ij}
			+ \pp{ \C^{g, ij} - \C^{g, ji} }
			+ \pp{ \C^{g, ij} + \C^{g, ji} - \C^{g, \bar{i} \bar{i}} \delta^{ij} }
		}, 
\eeq
where the repeated index $\bar{i}$ is summed over, and the quantities 
$\C$ without color subscripts $(a, b)$ have already included a trace over them. 
We may rewrite \eq{eq:eh2eah-col-factor-g0-expand} by abusing the 
notations of Pauli matrices $\sigma_k = (\sigma_k^{ij})$ ($k = 1, 2, 3$),
\beq[eq:GPD-Cg-pauli]
	\C^{g, ij}_{ab}
	= \frac{\delta_{ab}}{2(N_c^2 - 1)} \bb{
			\C_0 \, \delta^{ij} + \sum_{k = 1}^3 \C_k \, \sigma_k^{ij},
		},
\eeq
with the coefficients determined as
\begin{align}
	\C_0 & = \tr\pp{\C^g} = \C^{g, 11} + \C^{g, 22}, 
	&& \C_2 = \tr\pp{\C^g \sigma_2} = -i (\C^{g, 21} - \C^{g, 12}), \nn\\
	\C_1 & = \tr\pp{\C^g \sigma_1} = \C^{g, 12} + \C^{g, 21}, 
	&& \C_3 = \tr\pp{\C^g \sigma_3} = \C^{g, 11} - \C^{g, 22},
\label{eq:GPD-Cgi-pauli}
\end{align}
where the ``tr'' is over the Lorentz indices $(i, j)$.

Define the kinematics associated with the collinear factors,
\beq[eq:eh2eah-kin]
	P = (p + p') / 2,
	\quad
	\Delta = p_1 =  p - p',
	\quad
	\xi = \frac{(p - p')^+}{(p + p')^+},
	\quad
	k^+ = (x + \xi) P^+, 
\eeq
and so 
\beq
	p^+ = (1 + \xi) P^+, \quad
	p^{\prime+} = (1 - \xi) P^+, \quad
	(k - p_1)^+ = (x - \xi) P^+.
\eeq 
Then because only the plus parton momentum flows in $H$, 
the momentum integration in \eq{eq:eh2eah-factorize} can be reduced to a mere convolution in $k^+$,
or in $x$. 
This then factorizes the whole amplitude into GPDs that capture the infrared sensitivity,
\begin{align} \label{eq:eh2eah-factorize-x}
	 \M^{(2)}_{he \to h' e\gamma} 
	=&\, \sum_f \int_{-1}^1 dx 
			\bb{ F^f(x, \xi, t) H^f(x, \xi)
				+ \wt{F}^f(x, \xi, t) \wt{H}^f(x, \xi)	\right.\nn\\
			& \left. \hspace{10em}
				+ \sum\nolimits_{i = 1, 2} F_{T}^{f, i}(x, \xi, t) H_T^{f, i}(x, \xi)
			},
\end{align}
which sums over the parton flavors $f = q, g$, as illustrated in \fig{fig:dvcs}(c).
We have defined the quark and gluon GPDs, obtained by integrating 
Eqs.~\eqref{eq:eh2eah-col-factor-q1} and \eqref{eq:GPD-Cgi-pauli}
over $\bm{k}_T$ and $k^-$,
\begin{align}\label{eq:eh2eah-GPDs}
	&\pp{ F^q, \wt{F}^q, F_{T}^{q, 1}, F_{T}^{q, 2} }(x, \xi, t) 
		= \int \frac{dy^-}{4\pi} e^{i (x + \xi) P^+ \, y^-} \\
	& \hspace{8em} \times
			\langle h'(p') | \T \cc{ 
			\bar{\psi}_{q}(0) W(0, y^-; n)
			\pp{ \gamma^+, \gamma^+\gamma_5, \sigma^{+1}, \sigma^{+2} } 
			\psi_{q} (y^-)
		} | h(p) \rangle,  \nn\\
	&\pp{ F^g, \wt{F}^g, F_{T}^{g, 1}, F_{T}^{g, 2} }(x, \xi, t) 
		= \int \frac{dy^-}{2\pi P^+} e^{i (x + \xi) P^+ \, y^-} \\
	& \hspace{8em} \times
			\pp{ \delta^{ij}, -\sigma_2^{ij}, \sigma_3^{ij}, \sigma_1^{ij} }
			\langle h'(p') | \T \cc{ 
			G^{+j}(0) W_A(0, y^-; n) 
			G^{+i} (y^-)
		} | h(p) \rangle \nn
\end{align}
for the unpolarized, (longitudinally) polarized, and transversity ones. 
The corresponding hard coefficients are,
\begin{align}
	&\pp{ H^q, \wt{H}^q, H_{T}^{q, 1}, H_{T}^{q, 2} }(x, \xi) 
		= \frac{1}{2N_c}
			\pp{ \gamma^-, \gamma_5\gamma^-, \sigma^{1-}, \sigma^{2-} }_{\alpha\beta}
			H^q_{\beta\alpha; \bar{i}\bar{i}}(\hat{k}, \hat{p}_1 - \hat{k}) ,  \\
	&\pp{ H^g, \wt{H}^g, H_{T}^{g, 1}, H_{T}^{g, 2} }(x, \xi) 
		= \frac{1}{2(N_c^2 - 1)} \frac{1}{x^2 - \xi^2}
			\pp{ \delta^{ij}, -\sigma_2^{ij}, \sigma_3^{ij}, \sigma_1^{ij} }
			H^g_{ji; \bar{a}\bar{a}}(\hat{k}, \hat{p}_1 - \hat{k}),	\nn
\end{align}
where the spinor indices $(\alpha, \beta)$ and transverse Lorentz indices $(i, j)$ are summed over.
Note that the factor $1 / (x^2 - \xi^2)$ in the gluon hard coefficient does not raise problems for the 
$x$ integration at $x = \pm \xi$ because such poles are introduced by the artificial use of the field strength
tensor in the gluon GPD definition. The latter contains zeros at $x = \pm \xi$, which cancel the poles at the
hard coefficients.

For the DVCS, transversity GPDs do not contribute because the massless parton approximation
renders the corresponding hard coefficients to vanish. Then \eq{eq:eh2eah-factorize-x} only has the first line.
By a similar argument as Eqs.~\eqref{eq:em2ea-DA-X}--\eqref{eq:em2ea-x-2}, the time ordering can be dropped~\citep{Diehl:1998sm} 
in \eq{eq:eh2eah-GPDs}, allowing for insertion of physical states.
Then we have the support conditions for the GPDs,
\beq[eq:GPD-x-range]
	p^+ - k^+ = (1 - x) P^+ \geq 0, \quad
	p^+ - (p_1 - k)^+ = (1 + x) P^+ \geq 0.
\eeq
Thus GPDs are only nonzero when $x \in [-1, 1]$, which explains the integration range in \eq{eq:eh2eah-factorize-x}.
Due to the amplitude nature, GPDs are (non-local) matrix elements between two {\it pure} hadron state.
There is no way for hadron spin averages to come in before we square the amplitude. 
All possible GPDs should be kept unless forbidden by symmetries.
This is different from the collinear factorization of inclusive processes such as DIS, 
for which the polarization state of partons is dependent on the target spins state,
and polarized PDFs are nonzero only when the targets are polarized.
Moreover, we note that the flavor sum in \eq{eq:eh2eah-factorize-x} is only over all possible quark flavors and gluon, not antiquarks.
Because the $x$ is integrated from $-1$ to $1$, there is no need to introduce antiquark GPDs separately.

Due to the absence of a soft subgraph, the collinear factors in \eq{eq:eh2eah-GPDs} do not need further subtraction.
The factorization result in \eq{eq:eh2eah-factorize-x} is obtained mainly by use of Ward identities, which applies equally to the leading
region $R$ under approximation and to the subtracted smaller regions $R' < R$, effectively contained in $H$. 
The hard coefficients in \eq{eq:eh2eah-factorize-x} contain subtractions of smaller regions when some lines become collinear,
which can be dealt with recursively using the same factorization procedure.

However, the GPDs defined in \eq{eq:eh2eah-GPDs}, as well as the corresponding hard coefficients due to collinear subtractions, 
contain artificial UV divergences, as a result of short circuiting the $\bm{k}_T$ integration in the collinear factors. 
They hence need additional renormalization. 
Similar to PDFs in \sec{ssec:dis-uv}, GPDs can also be multiplicatively renormalized.
This can be used to convert each factor in \eq{eq:eh2eah-factorize-x} to a renormalized version, with the same factorization structure.
All the renormalized factors depend additionally on the factorization scale $\mu$, 
\begin{align} \label{eq:eh2eah-factorize-ren}
	 \M^{(2)}_{he \to h' e\gamma} 
	=&\, \sum_f \int_{-1}^1 dx 
			\bb{ F^f(x, \xi, t; \mu) H^f(x, \xi; \mu)
				+ \wt{F}^f(x, \xi, t; \mu) \wt{H}^f(x, \xi; \mu)	
			},
\end{align}
which implies a set of evolution equations that can be used to improve the factorization predictivity.

Compared to the corresponding DA factorization in \sec{sssec:em2ea},
the soft parton issue can also arise here, i.e., some of the parton momenta may have $k_i^+ \ll Q$, 
which violates the scaling in \eq{eq:e2em-col-scaling}, and thus the corresponding approximations. 
This is termed the ``breakpoint" issue in~\citep{Collins:1996fb}. 
However, since the region $k_i^+ \sim 0 \ll Q$ is not pinched, we can deform the contour of $k^+$ integration 
by $k^+ \mapsto k^+ \pm i \order{Q}$~\citep{Collins:1996fb}. 
This deformation is allowed because the breakpoint only lies on the boundary 
between the ERBL and DGLAP regions, but not at the GPD end points. 

Perturbatively, the soft parton singularity appears in \eq{eq:eh2eah-factorize-ren} at $x = \pm \xi$. 
As an example, the LO DVCS hard coefficient, to be calculated in \eq{eq:dvcs-coefs-x},
contains terms that are proportional to $1/(x\pm \xi \mp i\varepsilon)$.
We can deform the $x$ contour to avoid the poles at $\mp \xi$.
This is achieved in practical calculations by 
\beq[eq:principle value]
	\frac{1}{x \pm \xi \mp i \varepsilon} = P \frac{1}{x \pm \xi} \pm i \pi \, \delta(x \pm \xi),
\eeq
where $P$ denotes principal-value integration.

\subsubsection{Light meson production: $D = \mbox{meson}$}
\label{sssec:dvmp}
Similarly, the single-diffractive hard electroproduction of a light meson $M_D$ can be built on the large-angle meson electron scattering process in \sec{sssec:em2em}.
We keep the same definitions in Eqs.~\eqref{eq:em2em-aux-vectors}--\eqref{eq:em2em-glauber}, and use the same approximations in 
Eqs.~\eqref{eq:em2em-s-mom}--\eqref{eq:em2em-spinor-D} 
except the Eqs.~\eqref{eq:em2em-s-A} and \eqref{eq:em2em-col-A-g}, which will be explained below.
We will rely on the asymmetric deformation introduced in \sec{sssec:em2em-deformation} 
which has been extensively used in Secs.~\ref{sssec:am2mm-asym} and \ref{sssec:mm2mm}.

\begin{figure}[htbp]
\centering
	\begin{tabular}{cc}
		\includegraphics[clip, trim={-1em 0 -1em 0}, scale=0.75]{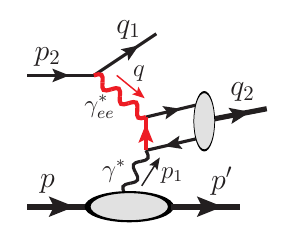} &
		\includegraphics[clip, trim={-1em 0 -1em 0}, scale=0.75]{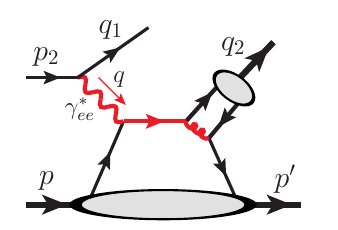}	\\
		(a) & (b) 
	\end{tabular}
\caption{Examples of LO diagrams for the light meson production in the SDHEP with an electron beam, for 
	(a) the $n = 1$ channel and (b) the $n = 2$ channel for $[q\bar{q}']$ case, 
	where the red thick lines indicate those with a hard $q_T$ flow and high virtualities.}
\label{fig:dvmp-lo}
\end{figure}

First, the $n = 1$ $\gamma^*$-initialized channel exists for a neutral meson production, which gives the subprocess
\beq[eq:eh2emh-n1]
	\gamma^*(p_1) + e(p_2) \to e(q_1) + M_D(q_2).
\eeq
One LO diagram is shown in \fig{fig:dvmp-lo}(a).
The slightly off-shell photon $\gamma^*(p_1)$ scatters with the highly virtual photon $\gamma^*_{ee}(q = p_2 - q_1)$ to produce the meson $M_D$. 
\eq{eq:eh2emh-n1} is just the reversed process of the large-angle real photon production in electron-meson scattering, discussed in \sec{sssec:em2ea},
although now the photon $\gamma^*$ is virtual. 
As in \sec{sssec:em2ea}, we can also factorize the amplitude of the process [\eq{eq:eh2emh-n1}] into the
DA of $M_D$ to the leading power of $m_D / q_T$, similar to \eq{eq:em2ea-factorize-ren}. 
As noted in \sec{ssec:n=1}, however, this approximation is only valid at leading power of the process in \eq{eq:eh2emh-n1}, 
which is of one power higher (super-leading) 
than the $n = 2$ GPD channel of our main interest. A more consistent treatment needs to 
factorize the process in \eq{eq:eh2emh-n1} to the subleading power, which is
beyond the scope of this thesis. 
Alternatively, one may choose to parametrize the amplitude by the $\gamma^* \gamma^*_{ee} \to M_D$ form factor, without use of factorization.
The $n = 1$ channel would be forbidden for the production of a charged meson like $\pi^{\pm}$, 
or of a neutral meson with odd $C$ parity, such as $\rho$ and $J/\psi$.

\begin{figure}[htbp]
\centering
	\begin{tabular}{cc}
		\includegraphics[clip, trim={-1em 0 -1em 0}, scale=0.75]{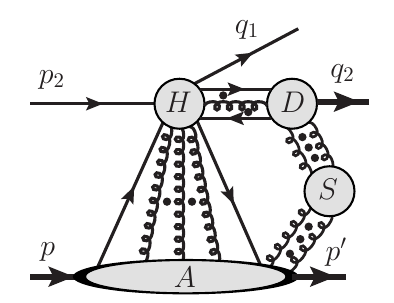} &
		\includegraphics[clip, trim={-1em 0 -1em 0}, scale=0.75]{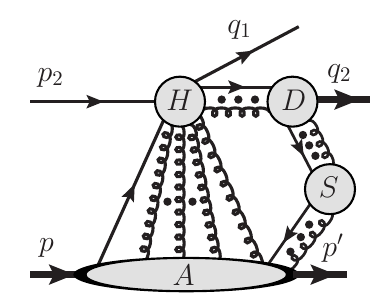} \\
		(a) & (b)
	\end{tabular}
	\caption{Leading-region graphs (a)(b) for producing a light meson from the SDHEP with a lepton beam. 
	Depending on the quantum numbers, the quark lines may be replaced by transversely polarized gluon lines.
	(c) is the result after factorizing it into the DA and GPD.}
\label{fig:eh2emh-leading}
\end{figure}

For the $n = 2$ channel, the diffracted hadron $h$ can exchange a collinear $[q\bar{q}']$ or $[gg]$ state with the hard scattering. 
The latter only holds when $M_D$ is charge neutral. 
One leading-order diagram for the quark channel is shown in \fig{fig:dvmp-lo}(b).
The hard electron scattering still happens by exchanging a highly virtual photon $\gamma^*_{ee}$, 
and so this (sub)process is referred to as deeply virtual meson production (DVMP).
The leading regions are shown in \fig{fig:eh2emh-leading}(a) and \fig{fig:eh2emh-leading}(b). 
In region (b) physically polarized quarks or gluons are attaching the collinear subgraphs to the soft subgraph; it is power suppressed 
by the soft-end suppression with respect to the meson wavefunction, as explained in \sec{sssec:em2em}.

\begin{figure}[htbp]
\centering
	\begin{tabular}{ccc}
		\includegraphics[clip, trim={0 -4pt 0 0}, scale=0.7]{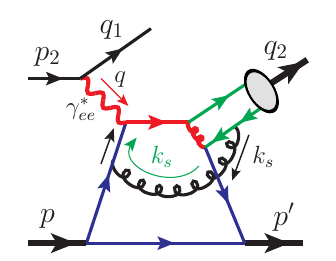} &
		\includegraphics[clip, trim={0 -4pt 0 0}, scale=0.7]{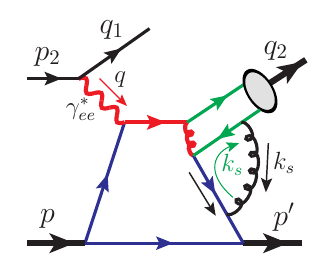} &
		\includegraphics[scale=0.7]{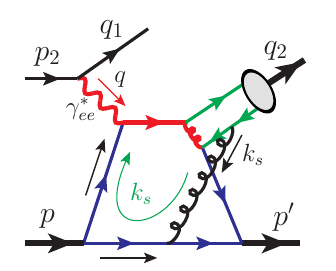}	\\
		(a) & (b) & (c)
	\end{tabular}
	\caption{Three example diagrams illustrating the soft gluon exchange between the collinear subgraphs along the diffractive hadron 
	and the final-state meson, for the DGLAP region of the GPD in a simple model theory. The green thin curved lines indicate the soft momentum flows.}
\label{fig:dvmp-glauber}
\end{figure}

As for the DVCS in \sec{sssec:dvcs}, the diffractive kinematics introduces the extra DGLAP region, compared to the
meson scattering case in \sec{sssec:em2em}. 
While this does not cause problems for the DVCS, it does lead to obstacles in factorizing soft gluons out of the $A$-collinear subgraph. 
This is illustrated in a simple model theory in \fig{fig:dvmp-glauber}, where we have indicated the chosen soft momentum flows 
by the thin curved arrowed lines. We make the following observations:
\begin{enumerate}
\item[(1)]
	DGLAP region has active collinear parton lines both before and after the hard interactions, and the soft gluons can attach to both, 
	as shown in Figs.~\ref{fig:dvmp-glauber}(a) and~\ref{fig:dvmp-glauber}(b). 
	With the soft momentum flows as indicated, attaching to the initial-state collinear parton gives a pole of 
	$k_s^-$ at $\order{\lambda^2}/Q - i\epsilon$, 
	while the final-state one gives a pole of 
	$k_s^-$ at $\order{\lambda^2}/Q + i\epsilon$;
\item[(2)]
	DGLAP region also has some spectator partons going in the forward direction. 
	When the soft gluon attaches to the spectator lines, as shown in \fig{fig:dvmp-glauber}(c), 
	it flows both in the same and opposite directions as the target-collinear lines, so that one single diagram gives both 
	$\order{\lambda^2} / Q \pm i\epsilon$ poles for $k_s^-$ contour.
	\footnote{Rerouting the soft momentum flow can change the situation (1) such that it also flows through the spectators and leads to both kinds of poles.}
\end{enumerate}
Diagrams like \fig{fig:dvmp-glauber}(c) pinch the $k_s^-$ contour at small values, such that for a Glauber gluon 
with the momentum scaling as in \eq{eq:em2em-glauber}, one cannot deform the $k_s^-$ contour to get out of the Glauber region, 
as was allowed by the corresponding $2\to2$ meson scattering in \eq{eq:em2em-deform}. 
While the diagrams in Figs.~\ref{fig:dvmp-glauber}(a) and~\ref{fig:dvmp-glauber}(b) do not directly cause pinch in the Glauber region, 
they cannot be trivially dealt with, either. Note that factorizing the soft gluons from the $A$-collinear lines requires to first deform soft gluons out of the 
Glauber region and then apply Ward identities. Even though we can deform the $k_s^-$ contour to get out the Glauber region for both diagrams,
the deformation directions are opposite. 
For \fig{fig:dvmp-glauber}(a), we need to replace the gluon coupling by
\beq
	J^A_{\mu}(k_s, k_A) \, g^{\mu\nu} \, S_{\nu}(k_s) 
	\mapsto
	J^A_{\mu}(\hat{k}_s, k_A) \, \frac{\hat{k}_s^{\mu} w_A^{\nu}}{k_s \cdot w_A + i \epsilon},
\eeq
whereas for \fig{fig:dvmp-glauber}(b), we need to flip the $i\epsilon$ sign. 
This would forbid use of Ward identity for the soft gluons, since different terms do not combine and cancel. 
This feature is closely related to the existence of Glauber pinch for $k_s^-$ in \fig{fig:dvmp-glauber}(c).

The way out is to note that all the soft $k_s^+$ poles come from the $D$-collinear lines, and lie on the lower half plane when $k_s$ flows from $D$ into $S$.
One may thus deform $k_s^+$ as 
$k_s^+ \mapsto k_s^+ + i \order{Q}$ 
while keeping $k_s^-$ contour unchanged, as was done in \eq{eq:em2em-ks+-deform-Q}.
While it is a free choice for the $2\to2$ hard exclusive scattering, this deformation is necessary here
due to the pinch in the DGLAP region of the diffractive process, and it moves all Glauber gluon momenta to the $A$-collinear region. 
For the same reason as discussed in \sec{sssec:em2em-deformation}, 
the $i\epsilon$ prescription for $k^-$ does not matter so it can be chosen in an arbitrary but consistent way.

\begin{figure}[htbp]
\centering
	\begin{tabular}{cc}
		\includegraphics[clip, trim={-3em 0 0 0}, scale=0.75]{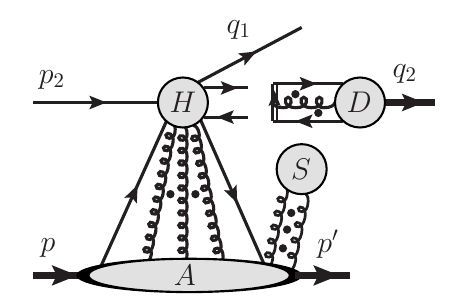}  &
		\includegraphics[clip, trim={-4em 0 0 0}, scale=0.75]{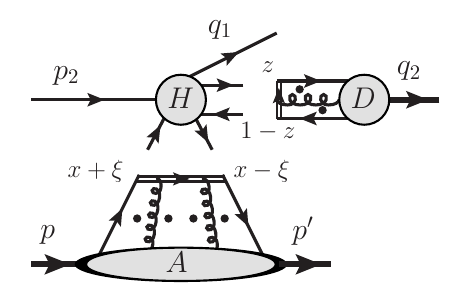} \\
		(a) & (b)
	\end{tabular}
	\caption{(a) Factorization of soft subgraph from the collinear subgraph of the final-state meson.
	(b) Factorization of the $A$-collinear subgraph out of the hard subgraph into GPD.}
\label{fig:eh2emh-factorize}
\end{figure}

Then by a similar discussion to \sec{sssec:am2mm-asym}, we can first factorize the $D$-collinear subgraph out of $H$, and 
soft gluons out of $D$. 
The same line of arguments applies here for the neutrality of meson $D$, the soft cancellation, and that the pair of
collinear Wilson lines associated with $D$ is joined into a finite-length Wilson line along $\bar{w}_D$. 
This applies to both the approximated region in \fig{fig:eh2emh-leading}(a) and smaller regions for subtraction, and reduces the leading region to 
\fig{fig:eh2emh-factorize}(a). 
Then by only attaching to the collinear subgraph $A$, soft gluons are no longer pinched. 
Because all the $k_s^+$ poles are of order $Q$, one may deform the $k_s^+$ contour by order $Q$ to make it a $A$-collinear momentum.
We can thus group the soft subgraph into the $A$-collinear subgraph. 
Then \fig{fig:eh2emh-factorize}(a) is exactly similar to \fig{fig:dvcs}(a)(b) for the single-diffractive real photon electroprodcution, and we can follow the same
procedure to factorize the collinear subgraph associated with the diffracted hadron out of $H$ into the GPD. 

Finally, we achieve the factorization of the amplitude,
\begin{align}\label{eq:eh2emh-factorize}
	\mathcal{M}^{(2)}_{he \to h' e M_D} 
	= &\, \sum_{i, j} \int_{-1}^1 dx \int_0^1 dz \, F_{hh'}^i(x, \xi, t; \mu) C_{ie \to ej}(x, \xi; z; \bm{q}_T, \mu) \, \bar{D}_{j/D}(z, \mu),
\end{align}
up to $1/q_T$ power suppressed terms, as diagrammatically shown in \fig{fig:eh2emh-factorize}(b).
The hard coefficient is a scattering of a collinear and on-shell parton pair $i$ along $w_A$ 
off the electron into another collinear and on-shell parton pair $j$ along $w_D$.
It contains collinear subtractions from both the GPD $F_{hh'}^i$ and DA $\phi_{j/D}$, but the latter two do not contain further soft subtractions, as a feature of
collinear factorization. We have used the multiplicative renormalizations of GPD and DA to convert each factor to a renormalized one, which introduces
a factorization scale $\mu$ and the associated evolution equations. The sum over $i$ and $j$ runs over all possible flavors and spin structures.

For example, the charged pion $\pi^+$ production $p e \to n e \pi^+$ only supports the channel $i = j = [u \bar{d}]$, which gives the factorization formula,
\begin{align}\label{eq:eh2emh-factorize-pi+}
	\mathcal{M}^{(2)}_{pe \to n e \pi^+} 
	= &\, \int_{-1}^1 dx \int_0^1 dz \, \wt{F}_{pn}^u(x, \xi, t; \mu) 
		\, C_{[u\bar{d}] e \to e [u\bar{d}]}(x, \xi; z; \bm{q}_T, \mu) 
		\, \bar{D}_{u/\pi^+}(z, \mu).
\end{align}
To the leading order of QED, the hard coefficient is only nonzero for the polarized GPD $\wt{F}_{pn}^u$ due to the QED Ward identity,
\beq
	C_{[u\bar{d}] e \to e [u\bar{d}]} \propto \bar{u}(q_1) \gamma_{\mu} u(p_2) \frac{-i g_{\mu\nu}}{q^2} (\hat{p}_1 + \hat{q}_2)^{\nu},
\eeq
which requires a $\gamma$ from the GPD. The bare flavor-changing GPD $\wt{F}_{pn}^u$ is defined as
\begin{align}\label{eq:eh2ehh-GPDs}
	\wt{F}_{pn}^{u, {\rm bare}}(x, \xi, t) 
		= \int \frac{dy^-}{4\pi} e^{i (x + \xi) P^+ \, y^-} 
			\langle n(p') |
			\bar{\psi}_{d}(0) W(0, y^-; n)
			\gamma^+\gamma_5
			\psi_{u} (y^-)
		| p(p) \rangle.
\end{align}
The neutral pion $\pi^0$ production $p e \to p e \pi^0$, on the other hand, supports both quark and gluon channels,
\begin{align}\label{eq:eh2emh-factorize-pi0}
	\mathcal{M}^{(2)}_{p e \to p e \pi^0} 
	= &\, \sum_{i = u, d, g} \sum_{j = u, d}
	\int_{-1}^1 dx \int_0^1 dz \, \wt{F}_{p}^i(x, \xi, t; \mu) 
		\, C_{[i\bar{i}] e \to e [j\bar{j}]}(x, \xi; z; \bm{q}_T, \mu) 
		\, \bar{D}_{j/\pi^0}(z, \mu).
\end{align}
The flavor-changing GPDs can be related to the flavor-conserving ones by isospin symmetry~\citep{Mankiewicz:1997aa},
\beq[eq:GPD-isospin]
	\wt{F}^u_{pn}(x, \xi, t) = \wt{F}^d_{np}(x, \xi, t) 
		= \wt{F}^u_{p}(x, \xi, t) - \wt{F}^d_{p}(x, \xi, t) = \wt{F}^d_{n}(x, \xi, t) - \wt{F}^u_{n}(x, \xi, t),
\eeq
which also applies to $F$.

\subsubsection{Extending to virtual photon or heavy quarkonium production}
\label{sssec:e-virtual-photon}

The DVCS and DVMP differ in how the observed particle couples to the hard interaction: 
the photon of the DVCS couples directly to the hard collision 
while the light meson of DVMP couples to the hard collision via two collinear partons. 
The factorization proof for the DVCS should apply equally to the case of producing a virtual photon 
$\gamma^*_f$ with high $q_T$ and low virtuality $Q'^2$ that decays into a pair of charged leptons. 
Even if $q_T \gg Q'$, there is no large logarithm of $q_T / Q'$ that spoils perturbation theory, 
contrary to the inclusive process~\citep{Berger:2001wr}, 
because such logarithms are associated with diagrams' collinear sensitivity, 
which require two collinear parton lines to connect the low mass virtual photon to the hard part, 
which is suppressed by one power of $Q'/q_T$ compared to the direct photon attachment. 
In contrast, the DVMP amplitude has large logarithms of $q_T / m_D$, 
due to the long-distance evolution of the collinear parton lines. 
Such logarithms are incorporated by the evolution equation associated with 
the factorization formula in Eqs.~\eqref{eq:em2em-factorize} and \eqref{eq:eh2emh-factorize}.

For a virtual photon $\gamma^*_f$ with its virtuality $Q'$ of the same order as $q_T$ 
(but, sufficiently away from masses of heavy quarkonia), it should belong to the short-distance hard part, 
and the whole process becomes $e^- + h \to h' + 2e^- + e^+$. 
This is no longer a $2\to 3$ SDHEP-type process, 
but we can still relate it to the SDHEP type by considering the kinematic regime
where one of the final-state electrons has a high transverse momentum $q_T$, 
balanced by the other $e^+e^-$ pair, which also has a large invariant mass $Q' \sim q_T$. 

First of all, the $\gamma^*$-mediated channel at $n = 1$ is allowed, with the hard scattering $e^- + \gamma^* \to 2e^- + e^+$. 
Second, the $n = 2$ channel does not unambiguously lead to the double DVCS (DDVCS) process~\citep{Guidal:2002kt} 
because it is not possible to distinguish which of the final-state electrons comes from the scattering of the initial-state electron.  
By labeling the final-state electrons and positron as $\pp{e^-_1, e^-_2, e^+}$, 
we find that a single configuration of $\pp{e^-_1, e^-_2, e^+}$ could correspond to 
both high-$Q'$ and low-$Q'$ processes. Specifically, let us consider the following three kinematic cases:
\begin{enumerate}
\item[(1)]
	All the $(e^-_1, e^-_2, e^+)$ have high transverse momenta, of order $q_T \gg \sqrt{-t}$, 
	and the two invariant masses $( m_{e^-_1 e^+}, m_{e^-_2 e^+})$ are large, of the same order of $q_T$. 
	This case leads unambiguously to DDVCS, and the factorization of DVCS can be trivially generalized here. 
	But one needs to consider both diagrams with either $e^-_1$ or $e^-_2$ coming from the decay of the virtual photon $\gamma^*_f$.
\item[(2)]
	All the $(e^-_1, e^-_2, e^+)$ have high transverse momenta, of order $q_T \gg \sqrt{-t}$, 
	but one of the invariant lepton-pair masses, say $m_{e^-_1 e^+}$, is much less than $q_T$, 
	and the other pair has a large invariant mass, i.e., $q_T \sim m_{e^-_2 e^+} \gg m_{e^-_1 e^+}$. 
	In this case, one can have 
	(a) $(e^-_1, e^+)$ comes from the decay of a low-virtuality $\gamma^*_f$, and 
	(b) $(e^-_2, e^+)$ comes from the decay of a high-virtuality $\gamma^*_f$. 
	While both correspond to the DDVCS processes, it is the case (a) with a low-mass electron pair that contributes at a leading power.
\item[(3)]
	$(e_2^-, e^+)$ have high transverse momenta, of order $q_T \gg \sqrt{-t}$, 
	and $e_1^-$ has a low transverse momentum, much less than $q_T$. 
	Automatically, we have both $(m_{e^-_1 e^+}, m_{e^-_2 e^+})$ to be large. 
	This gives two different cases: 
	(a) $e_1^-$ comes from the diffraction of the initial-state electron, 
	which gives out a quasireal photon $\gamma^*_{ee}$ that scatters with the diffractive hadron $h$ 
	and produces a highly virtual photon $\gamma^*_f$ that decays into the $(e_2^-, e^+)$ pair; 
	(b) $e_2^-$ comes from the hard scattering of the initial-state electron, 
	whose interaction with the diffractive hadron $h$ produces a highly virtual photon $\gamma^*_f$ 
	with a high transverse momentum, which decays into the $(e_1^-, e^+)$ pair.
	Now only the case (b) corresponds to the DDVCS process, and case (a) gives a subprocess of 
	(quasi)real photon scattering with the hadron, whose factorization will be proved later in \sec{sssec:ah2llh-ah2aah}. 
	While both subprocesses are factorizable, it is the subprocess (a) that gives the leading power contribution.
\end{enumerate}
Of course, if the virtual photon $\gamma_f^*$ decays into a lepton pair of other flavors, 
like a $\mu^+\mu^-$ pair, then it unambiguously leads to the DDVCS process and can be factorized in the same way as the DVCS.

When the $\gamma_f^*$ virtuality $Q'$ becomes much greater than $q_T$, one starts entering the two-scale regime. 
Whether there will be large logarithms of $Q'/q_T$ that requires a new factorization theorem 
to be developed is not a trivial problem based on our analysis so far. We leave that discussion to the future.

For a heavy quarkonium production, unfortunately, it is not obvious that the factorization in \sec{sssec:dvmp} can be easily generalized. 
The key points to the factorization are
\begin{enumerate}
\item[(i)]
	there is a pinch singularity that forces a collinear momentum to have the scaling in \eq{eq:e2em-col-scaling}, with a leading component and two smaller components;
\item[(ii)]
	soft gluons can be factorized from the collinear lines.
\end{enumerate}
The exclusive production of a heavy quarkonium naturally has the most contribution from 
producing a heavy quark pair with an invariant mass $M_H\sim 2m_Q$, 
where $m_Q \gg \LQCD$ is the heavy quark mass. 
Since the corresponding heavy quark GPD in $h$-$h'$ transition is suppressed, 
we do not suffer from the extra region like \fig{fig:eh2emh-leading}(b). 
When the transverse momentum $q_T$ of the heavy quarkonium is much greater than $m_Q$, 
the heavy quark can be thought of as the active parton line associated with the observed particle $D$ in \fig{fig:eh2emh-leading}(a), 
and the heavy quarkonium is attached to the hard part by a pair of nearly collinear heavy quark lines, whose momenta scale as
\beq
	k_Q \sim \pp{\lambda_Q ^2, 1, \lambda_Q } q_T,
	\quad
	\mbox{with } \lambda_Q = m_Q / q_T,
\eeq
when the heavy quarkonium moves along the minus direction. 
This pinches the plus momentum components to be small, and for a soft gluon $k_s$ attached to such heavy quark lines, 
one may keep only the $k_s^+$ component, which allows us to factorize the soft gluon out of the collinear lines. 
Hence, for $q_T \gg m_Q \gg \LQCD$, one can still factorize the heavy quarkonium production amplitude into the 
heavy quarkonium DA, up to the error of $\order{m_Q / q_T}$. 
See \citep{Kang:2014tta} for a similar discussion of the inclusive production of a heavy quarkonium.

When $m_Q \sim q_T\gg \LQCD$, the error estimated above becomes ${\cal O}(1)$, 
which invalidates the factorization into heavy quarkonium DA. 
However, if $M_H/2 - m_Q \ll m_Q \sim q_T$, the formation of the heavy quarkonium 
from the produced heavy quark pair might be treated in terms of the color singlet model~\citep{Einhorn:1975ua, Chang:1979nn, Berger:1980ni} 
or the velocity expansion of nonrelativistic QCD with color singlet long-distance matrix elements~\citep{Bodwin:1994jh}.  
For this exclusive production, the soft gluon interaction from the diffractive hadron 
with the heavy quark pair at $q_T\sim m_Q\gg \LQCD$ is expected to be suppressed by powers of $m_Qv/q_T\sim v$ 
with $v$ being the heavy quark velocity in the quarkonium's rest frame.  
More detailed study for the heavy quarkonium production when $q_T \lesssim m_Q$ will be presented in a future publication.

\subsection{SDHEP with a photon beam}
\label{ssec:sdhep-photon}
For single-diffractive hard exclusive photoproduction processes, we have $B = \gamma$. 
The other particles $C$ and $D$ can be two elementary particles, one elementary particle and one light meson, or two light mesons. 
So we consider the three cases,
(1) massive dilepton $(CD) = (l^+l^-)$~\citep{Berger:2001xd, CLAS:2021lky} 
	or diphoton $(\gamma\gamma)$ production~\citep{Pedrak:2017cpp, Grocholski:2021man, Grocholski:2022rqj}, 
(2) real photon and light meson pair $(CD) = (\gamma M_D)$ production~
\citep{Boussarie:2016qop, Duplancic:2018bum, Duplancic:2022ffo, Duplancic:2023kwe, Qiu:2023mrm},
and 
(3) light meson pair $(CD) = (M_C M_D)$ production~\citep{ElBeiyad:2010pji}. 
These are similar to the large-angle photon-meson scattering treated in \sec{ssec:exclusive-am}.
In this section, we generalize the factorization arguments there to the corresponding single-diffractive processes,
following the same two-stage paradigm as the single-diffractive lepton-hadron scattering in \sec{ssec:sdhep-lepton}.

\subsubsection{Dilepton or diphoton production: $(CD) = (l^+l^-)$ or $(\gamma\gamma)$}
\label{sssec:ah2llh-ah2aah}
Both production processes allow the $\gamma^*$-mediated $n=1$ subprocesses. 
For the dilepton production, we have the partonic process $\gamma \gamma^* \to l^+ l^-$, 
starting at $\order{e^2}$ in terms of the QED coupling $e$, 
while we have $\gamma\gamma^* \to \gamma\gamma$ for the diphoton production, starting at $\order{e^4}$. 
Since this $\gamma^*$-mediated $n=1$ channel has a power enhancement of 
$\order{q_T/\sqrt{-t}}$ compared to the $n = 2$ channel, it cannot be simply neglected even though 
its scattering amplitude might require a higher power in QED coupling. 
A careful quantitive comparison in size between $\gamma^*$-mediated $n=1$ and 
GPD-sensitive $n=2$ subprocesses is needed in practical evaluation.

\begin{figure}[htbp]
\centering
	\begin{tabular}{cc}
		\includegraphics[trim={-2em 0 -2em 0}, clip, scale=0.75]{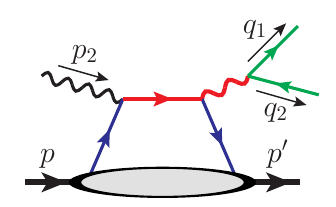} &
		\includegraphics[trim={-2em 0 -2em 0}, clip, scale=0.75]{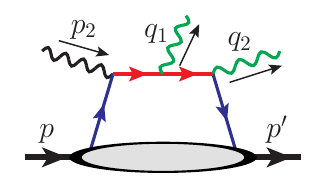}	\\
		(a) & (b) 
	\end{tabular}
	\caption{Examples of leading-order diagrams in the $n = 2$ (GPD) channel for the 
	single-diffractive hard exclusive photoproduction of massive (a) dilepton and (b) diphoton processes.}
\label{fig:ah2llh-ah2aah-LO}
\end{figure}

For $n = 2$ channel, these two processes share the same color structure as the DVCS, 
and thus the same leading-region graphs in \fig{fig:dvcs} with a proper change of the external lines, 
because $B$, $C$, and $D$ are all elementary colorless particles.
The argument for factorization into GPDs works in the same way as for the DVCS in Sec.~\ref{sssec:dvcs} 
and will not be repeated here. 
The process with $(CD) = (l^+l^-)$ happens by producing a timelike photon $\gamma^{\prime*}$ 
in the exclusive $\gamma h \to \gamma^{\prime*} h'$ process followed by the decay $\gamma^{\prime*} \to l^+l^-$, 
which is thus called timelike Compton scattering (TCS), as shown in \fig{fig:ah2llh-ah2aah-LO}(a). 
For the process with $(CD) = (\gamma\gamma)$, all the three photons couple to the quark lines, 
as illustrated in \fig{fig:ah2llh-ah2aah-LO}(b).
In both processes, it is the high $q_T$ that provides the hard scale for factorizability, 
by creating high virtualities through the invariant mass of the virtual photon in the dilepton case 
or having the $q_T$ flow through the quark lines in the diphoton case.

\begin{figure}[htbp]
\centering
	\begin{tabular}{cc}
		\includegraphics[trim={-2em 0 -2em 0}, clip, scale=0.7]{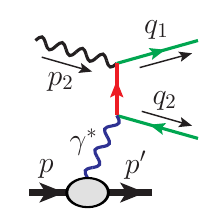} &
		\includegraphics[trim={-2em 0 -2em 0}, clip, scale=0.7]{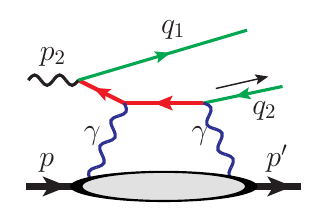}	\\
		(a) & (b) 
	\end{tabular}
\caption{(a) The sample diagram for the $\gamma^*$-mediated channel of the photoproduction of a massive lepton pair, 
where the internal lepton propagator (in red) has a hard virtuality only when $q_T$ is large. 
(b) At large $m_{ll}$ but small $q_T$, the forward scattering diagrams with two photon exchanges 
between the diffractive hadron and the quasireal lepton can become important and 
compete with the TCS mechanism in \fig{fig:ah2llh-ah2aah-LO}(a).}
\label{fig:ah2llh a}
\end{figure}

It is important to note that in general, the requirement of a high invariant mass 
for the pair of particles $(CD)$ is not the same as requiring a hard $q_T$. 
This is similar to the large-angle photon-meson scattering in \sec{sssec:am2ll}.
For the TCS, it is the invariant mass of the lepton pair $m_{ll}$ that provides the hard scale for the partonic collision, 
and hence keeping $m_{ll}$ large is sufficient for TCS to be factorized into GPD, 
independent of the magnitude of $q_T$ of the observed lepton. 
However, a hard $q_T$ is needed to guarantee the $\gamma^*$-mediated $n = 1$ subprocess 
$\gamma\gamma^* \to l^+ l^-$ to be a hard scattering process, as illustrated in \fig{fig:ah2llh a}(a). 
If $q_T$ is too low, then this amplitude introduces another enhancement factor of $\order{m_{ll} / q_T}$, 
in addition to the $m_{ll} / \sqrt{-t}$ enhancement of the $n = 1$ channel, 
as correctly pointed out in \citep{Berger:2001xd}. 
Then, this could allow other subprocesses to happen that may compete with the TCS subprocess in magnitude. 
For example, one may have an $n = 2$ channel mediated by $f_2 = [\gamma\gamma]$, 
as shown in \fig{fig:ah2llh a}(b), which is suppressed by $e^2$ and one power of $\sqrt{-t} / m_{ll}$ 
compared to the $n = 1$ channel, but is still one power $\order{m_{ll} / q_T}$ higher than the TCS channel. 
The relative order comparison is then too complicated to be obvious, 
and the extraction of GPDs from the TCS amplitude becomes even harder. 

\begin{figure}[htbp]
\centering
	\includegraphics[scale=0.7]{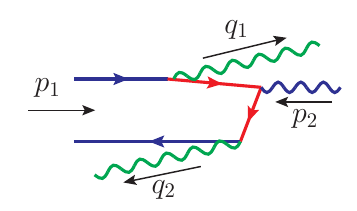}
\caption{A sample diagram for the photoproduction of diphoton process at low $q_T$, 
where the photon $q_1$ is radiated collinearly by the incoming quark.}
\label{fig:ah2aah-col}
\end{figure}

On the other hand, if $q_T$ is too low in the diphoton production process, 
some quark lines could have low virtualities of order $q_T$, 
as the photons could be radiated from the quark lines (see \fig{fig:ah2aah-col}) 
almost collinearly, introducing the long-distance physics into the ``hard probe'', 
which invalidates our factorization arguments.

\subsubsection{Real photon and light meson pair production: $(CD) = (\gamma M_D)$}
\label{sssec:ah2amh}
For $(CD) = (\gamma M_D)$ with $M_D$ being a light meson, the $n = 1$ channel 
corresponds to the subprocess $\gamma^* \gamma \to \gamma M_D$. 
This is forbidden for a charged meson like $\pi^{\pm}$, as considered in \citep{Duplancic:2018bum}, 
or for a neutral meson with even $C$-parity, like $\pi^0$, $\eta$, etc. 
In the high-$q_T$ scattering, the $n = 1$ amplitude can be factorized into the DA of $M_D$.  

The $n = 2$ channel has the same color structure as the DVMP process in \sec{sssec:dvmp}, 
and the leading region is also as in \fig{fig:eh2emh-leading} just with the proper change 
of the external electron lines by photon lines. The argument for factorization then 
works in the same way, and is not to be repeated here. 
For the same reason as the diphoton production process in the previous subsection, 
we emphasize the necessity of the hard transverse momentum $q_T$, 
which is not equivalent to requiring a large invariant mass of the $\gamma M_D$ pair.

\subsubsection{Light meson pair production: $(CD) = (M_C M_D)$}
\label{sssec:ah2mmh}
The single-diffractive photoproduction with $(CD) = (M_C M_D)$ differs from the 
electroproduction of a light meson in \sec{sssec:dvmp} by having one more hadron in the final state. 
This leads to one more collinear subgraph in another direction but does not make the factorization proof very different. 
As for the DVMP, generalizing the proof of the corresponding meson scattering in \sec{sssec:am2mm-sym} to the 
diffractive case encounters the trouble of Glauber pinch for gluons attaching to the diffracted hadron.
As a result, we will need to use the asymmetric contour deformation in \sec{sssec:am2mm-asym}.

First, the $n = 1$ channel is given by the subprocess $\gamma^* \gamma \to M_C M_D$, 
which may or may not happen depending on the quantum numbers of $M_C$ and $M_D$. 
This was considered first in \citep{Brodsky:1981rp}. 
The $t$-channel crossing process $M_A \gamma \to \gamma M_D$ is briefly discussed in \sec{sssec:am2am}.
The time-reversal process $M_A + M_B \to \gamma \gamma$ was also studied in \citep{Qiu:2022bpq}. 
Its amplitude can be factorized into the DAs of $M_C$ and $M_D$, as a simple generalization of the factorization
proof for the process in \sec{sssec:em2em}.

\begin{figure}[htbp]
\centering
	\begin{tabular}{cc}
		\includegraphics[clip, trim={-1em 0 -1em 0}, scale=0.7]{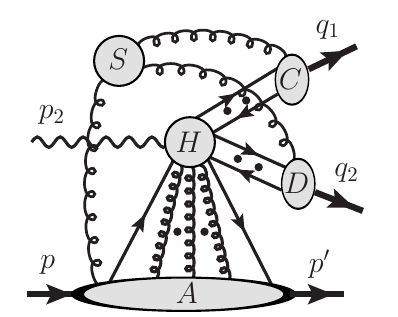} &
		\includegraphics[clip, trim={-1em 0 -1em 0}, scale=0.7]{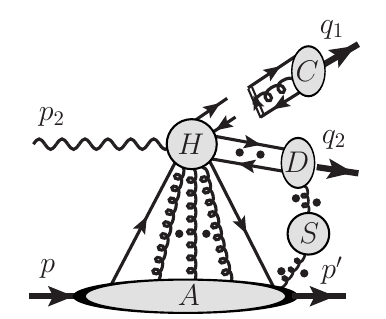} 	\\
		(a) & (b) 
	\end{tabular}
	\caption{(a) Leading-region graphs for the single-diffractive hard photoproduction of a light meson pair. 
	(b) is obtained as an intermediate step after factorizing the $C$-collinear subgraph out of the hard subgraph $H$ 
	and soft subgraph $S$.}
\label{fig:ah2mmh-leading}
\end{figure}

For the $n = 2$ channel, the leading regions is shown in \fig{fig:ah2mmh-leading}(a), 
where all lines in the hard part ``$H$'' are off shell by order of the hard scale $Q\sim q_T$, 
which effectively makes the contribution from attaching soft gluons to $H$ power suppressed. 
There could be additional leading regions in which one or more of the collinear subgraph 
is connected to the soft subgraph by one quark or transversely polarized gluon line, 
while connecting to the hard subgraph by the other quark or transversely polarized gluon line. 
Following the same assumption that such soft end point region is strongly suppressed 
by the nonperturbative QCD dynamics from the meson distribution amplitude, 
we neglect them and consider only the leading regions in \fig{fig:ah2mmh-leading}(a).

Extending the factorization of the meson scattering process to the corresponding single-diffractive process is trivial. 
The only complication arises from the extra DGLAP region in the single-diffractive channel of the hadron $h\to h'$, 
which causes the momentum $k_s$ of the soft gluon coupling to the $A$-collinear subgraph to be pinched 
in the Glauber region for its component $k_s \cdot w_A$, as explained in Sec.~\ref{sssec:dvmp}. 
This makes the use of symmetric deformation as in \sec{sssec:am2mm-sym} not possible.
But the asymmetric deformation strategy in \sec{sssec:am2mm-asym} applies here with no change, 
because we never deformed the contour of $k_s \cdot w_A$ when $k_s$ flows through the $A$-collinear subgraph. 
The important step of factorizing the $C$-collinear subgraph is shown in \fig{fig:ah2mmh-leading}(b). 
In the end, the diffractive amplitude is factorized into the hadron GPD and meson DAs,
\begin{align}\label{eq:p3 factorize}
	&\mathcal{M}^{(2)}_{h\gamma \to h' M_C M_D} 
	= \sum_{i, j, k} \int_{-1}^1 dx \int_0^1 d z_C \, dz_D F_{hh'}^i(x, \xi, t; \mu) 
	\nn\\
	&\hspace{10em}\times
	C_{i\gamma \to jk}(x, \xi; z_C, z_D; \bm{q}_T, \mu) \, \phi_{j/C}(z_C, \mu) \, \phi_{k/D}(z_D, \mu) ,
\end{align}
up to $1/q_T$ power suppressed terms, where the sum over $i$, $j$, and $k$ runs over all possible flavors and spin structures.

\subsection{SDHEP with a meson beam}
\label{ssec:sdhep-meson}
For the SDHEP with a meson beam, we have $B$ being some meson $M_B$, 
which is usually a pion or kaon. Similar to the case with a photon beam, 
we consider three cases for the particles $C$ and $D$: 
(1) massive dilepton $(CD) = (l^+l^-)$ or diphoton $(\gamma\gamma)$ production; 
(2) real photon and light meson pair $(CD) = (\gamma M_D)$ production; and 
(3) light meson pair $(CD) = (M_C M_D)$ production. 
The dilepton and diphoton production processes have been studied in 
\citep{Berger:2001zn, Qiu:2022bpq}, respectively, and their factorizations are similar to the DVMP process. 
The processes (2) and (3) have not been considered in the literature. 
In this section, we address the factorization of these processes in the framework of the SDHEP within the two-stage paradigm.

\subsubsection{Massive dilepton or diphoton production: $(CD) = (l^+l^-)$ or $(\gamma\gamma)$}
\label{sssec:mh2llh-mh2aah}
The SDHEPs of massive dilepton and diphoton productions are
\beq[eq:mh2llh]
	h(p) + M_B(p_2) \to h'(p') + l^-(q_1) + l^+(q_2),
\eeq
and 
\beq[eq:mh2aah]
	h(p) + M_B(p_2) \to h'(p') + \gamma(q_1) + \gamma(q_2),
\eeq
respectively. 
Both processes have $C$ and $D$ being colorless elementary particles, 
and they are similar to the meson production in the SDHEP with a lepton beam in \sec{sssec:dvmp} 
and the meson-photon pair production in the SDHEP with a photon beam in \sec{sssec:ah2amh}, respectively.  
The difference comes from switching the final-state meson with the initial-state lepton or photon. 
The argument for the factorization works in essentially the same way, 
with only a slight change due to the meson being in the initial state instead of final state. 
In reality, only charged light meson beams such as $\pi^{\pm}$ or $K^{\pm}$ are readily accessible in experiments, 
so we will consider only those beams. 
Then charge conservation implies a flavor change of the diffractive hadron, i.e., $h' \neq h$, 
which forbids the $\gamma^*$-mediated $n = 1$ channel.
Therefore, the leading-power contributions to the amplitudes in Eqs.~\eqref{eq:mh2llh} and \eqref{eq:mh2aah} 
start with the $n = 2$ channels, which are factorized into the GPDs associated with the hadron transition $h \to h'$, 
as in~\citep{Berger:2001zn, Qiu:2022bpq}.

For the process in \eq{eq:mh2llh}, at the lowest order in QED, 
the high-$q_T$ lepton pair is produced via a timelike photon $\gamma_{ll}^*$ 
with a high virtuality $Q \sim \order{q_T}$, 
when it is sufficiently away from the resonance region of a heavy quarkonium. 
This process can hence be referred to as exclusive Drell-Yan process~\citep{Berger:2001zn}.
It is this highly virtual photon that couples directly to the parton lines from the $h$-$M_B$ interaction, 
whose virtuality $Q$ provides the hard scale that localizes the parton interactions. 
This is sufficient for the factorization argument. 
Furthermore, due to the lack of $\gamma^*$-mediated $n=1$ subprocess, 
the requirement of the high invariant mass for the lepton-pair is a sufficient condition for factorization, 
allowing us to release the high $q_T$ requirement, 
which is contrary to the requirement for the lepton-pair production in the SDHEP with a photon beam, 
as discussed in Sec.~\ref{sssec:ah2llh-ah2aah}.

In contrast, the process in \eq{eq:mh2aah} has the two final-state photons 
directly couple to the parton lines, and the hard scale is solely provided 
by their high transverse momentum $q_T$, 
which is both the sufficient and necessary condition for collinear factorization. 
In the low-$q_T$ regime, one starts to have two widely separated scales in the same process, 
$q_T^2\ll \hat{s} = (p-p'+p_2)^2$, 
just as the photoproduction of diphoton process in \sec{sssec:ah2llh-ah2aah}, 
the factorization for which needs further study.

\subsubsection{Real photon and light meson pair production: $(CD) = (\gamma M_D)$}
\label{sssec:mh2amh}
Now we consider the process
\beq[eq:mh2amh]
	h(p) + M_B(p_2) \to h'(p') + \gamma(q_1) + M_D(q_2),
\eeq
which differs from the photoproduction of a meson pair process in Sec.~\ref{sssec:ah2mmh} 
by switching the initial-state photon with one of the final-state mesons. 
The $n = 1$ channel corresponds to the subprocess $\gamma^*(p_1) + M_B(p_2) \to \gamma(q_1) + M_D(q_2)$, 
which has been discussed in \sec{sssec:am2am}. 
Depending on the quantum numbers of $M_B$ and $M_D$, this channel may or may not be present. 
The amplitude can be factorized into the DAs of $M_B$ and $M_D$.

The amplitude of $n = 2$ channel can be factorized into a GPD and two DAs, 
whose proof can be adapted from \sec{sssec:am2mm-asym} with straightforward modifications: 
one can first factorize the $D$-collinear subgraph and the soft gluons attached to it, 
and then do the same thing for $B$, which is sufficient to complete the proof. 

\begin{figure}[htbp]
\centering
	\includegraphics[scale=0.7]{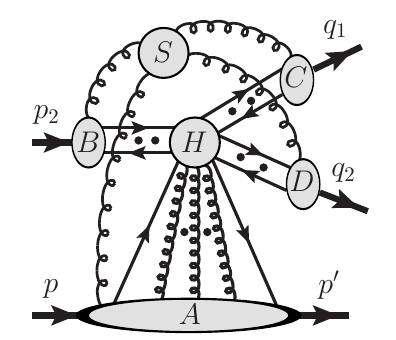}
\caption{Leading-region graphs for the single-diffractive hard mesoproduction of two mesons. 
There can be any numbers of soft gluons connecting $S$ to each collinear subgraph. 
Depending on the quantum numbers, the quark lines may be replaced by transversely polarized gluon lines. 
The dots represent arbitrary numbers of longitudinally polarized collinear gluons.}
\label{fig:mh2mmh-leading}
\end{figure}

\subsubsection{Light meson pair production: $(CD) = (M_C M_D)$}
\label{sssec:mh2mmh}
Now we consider the process
\beq[eq:mh2mmh]
	h(p) + M_B(p_2) \to h'(p') + M_C(q_1) + M_D(q_2),
\eeq
whose corresponding $2\to2$ hard meson scattering is discussed in \sec{sssec:mm2mm}.
The $n = 1$ channel, $\gamma^*(p_1) + M_B(p_2) \to M_C(q_1) + M_D(q_2)$,
which may or may not contribute depending on the quantum numbers, can be analyzed in the same way as
the photon-meson scattering in Secs.~\ref{sssec:am2mm-sym} and \ref{sssec:am2mm-asym}.
The $n = 2$ channel has leading regions shown in \fig{fig:mh2mmh-leading}, 
under the assumptions of strong soft-end suppression 
{\it and} a single hard scattering in which all the parton lines are off shell by the hard scale. 
Compared to the meson pair photoproduction process in \sec{sssec:ah2mmh}, 
there is one more collinear subgraph in the initial state, and factorization works with a simple generalization. 
In \fig{fig:mh2mmh-leading} one does not deform the contours of soft gluon momenta $k_s$ 
for their components $k_s \cdot w_A$ when they flow in the $A$-collinear subgraph. 
We first factorize $C$, $D$, and $B$ from $H$ sequentially, together with the soft gluons attached to them, 
and then group the soft gluons into the $A$-collinear subgraph to 
complete the proof in a way similar to \sec{sssec:ah2mmh}.  
Consequently, the amplitude of the diffractive process in \eq{eq:mh2mmh} can be factorized into the GPD and DAs,
\begin{align}\label{eq:mh2mmh-factorize}
	&\mathcal{M}^{(2)}_{h M_B \to h' M_C M_D} 
	= \sum_{i, j, k, l} \int_{-1}^1 dx \int_0^1 dz_B \, dz_C \, dz_D  
		F_{hh'}^i(x, \xi, t; \mu) \, \phi_{j/B}(z_B, \mu)\nn\\
	&\hspace{6em}\times
		C_{i j \to kl}(x, \xi; z_B, z_C, z_D; \bm{q}_T, \mu) \, \phi_{k/C}(z_C; \mu) \, \phi_{l/D}(z_D; \mu) ,
\end{align}
up to $1/q_T$ power suppressed terms, 
where where the sum over $i$, $j$, and $k$ runs over all possible flavors and spin structures, 
and the hard coefficient $C_{i j \to kl}(x, \xi; z_B, z_C, z_D; \bm{q}_T, \mu)$ 
can be calculated as the scattering of two collinear parton pairs $i$ and $j$ into another two pairs $k$ and $l$.

\section{Further discussion on single diffractive processes}
\label{sec:discussion}

In this section, we give a few general remarks on the properties of SDHEPs, and their factorizability and sensitivities for extracting GPDs.

\subsection{Two-stage paradigm and factorization}
\label{ssec:discussion two stage}
We have presented the arguments to prove the factorization of SDHEPs 
with different colliding beams and different types of final-state particles. 
Our proofs follow a unified two-stage approach by taking advantage of the unique feature of SDHEPs, 
which can be effectively separated into two stages, 
as specified in Eqs.~\eqref{eq:diffractive} and~\eqref{eq:hard 2to2}. 
By requiring $q_T \gg \sqrt{-t}$, we effectively force the exchanged state $A^*$ 
between the single diffractive transition of $h\to h'$ and the hard exclusive $2\to 2$ 
scattering to be a low-mass and long-lived state in comparison to the timescale~$\sim {\cal O}(1/q_T)$ 
of the hard exclusive process, and effectively reduce the SDHEP into two stages: 
single diffractive (SD) + hard exclusive (HE) with the quantum interference 
between these two subprocesses suppressed by powers of $\sqrt{-t}/q_T$. 
As emphasized earlier, requiring large transverse momenta for the final-state particles 
$C$ and $D$ is not equivalent to requiring a large invariant mass of them, 
$m_{CD}\gg \sqrt{-t}$; the latter does not necessarily guarantee a hard collision.

This two-stage paradigm gives a unified picture for the microscopic mechanism of the SDHEPs, 
described in \eq{eq:channels} and \fig{fig:decomposition}. It accounts for the 
$\gamma^*$-mediated $n=1$ channel in a coherent framework, 
which is usually regarded as a ``byproduct" of the GPD channel in the literature 
and can be easily forgotten but which is in fact one power higher than the GPD channel 
and should be incorporated unless it is forbidden by some quantum number conservation. 

Furthermore, this two-stage paradigm leads to a simple methodology 
for proving factorization of the SDHEPs in \eq{eq:sdhep}, in particular, for the $n = 2$ channel. 
By treating the long-lived exchanged state $A^*$ as a ``meson'' capturing the quantum number of $h\to h'$ transition, 
we make the corresponding scattering $A^* + B \to C + D$ effectively a $2\to 2$ exclusive process 
with a single hard scale, whose factorization is relatively easier to prove. 
In this way, the factorization proof of the SDHEP can focus on its differences from the $2\to 2$ hard exclusive process.

\begin{figure}[htbp]
\centering
	\begin{tabular}{cc}
		\includegraphics[clip, trim={-1em 0 -1em 0}, scale=0.7]{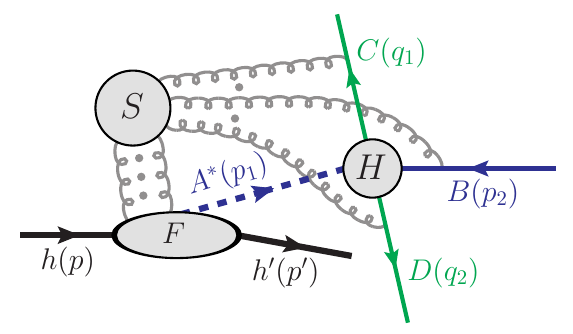} &
		\includegraphics[clip, trim={-1em 0 -1em 0}, scale=0.7]{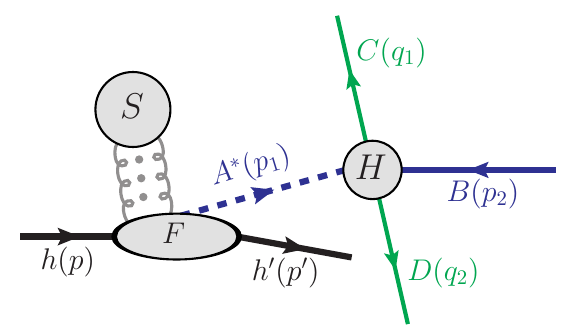} \\
		(a) & (b) 
	\end{tabular}
\caption{(a) SDHEP in the general case, with all possible soft gluon connections.
	(b) The result of soft cancellation in (a). The cancellation of the soft gluons in the $2\to 2$ hard exclusive scattering 
	implies the same cancellation of the soft gluons that couple to $B$, $C$, and/or $D$.}
\label{fig:sdhep soft A}
\end{figure}

The only difference between the factorization of the $2\to 2$ hard exclusive process 
and the full SDHEP is that the GPD channel supports both ERBL and DGLAP regions, 
and a Glauber pinch can exist for the DGLAP region. However, since we only have one diffractive hadron, 
only one component $k_s\cdot w_A$ of the soft gluon momentum $k_s$ is pinched in the Glauber region. 
The factorizability of the corresponding $2 \to 2$ exclusive process implies that 
soft gluons coupling to $B$, $C$, and/or $D$ are canceled, which applies equally to the situation of SDHEPs. 
The rest of the soft gluons only couple to the diffracted hadron and can be grouped into the 
collinear subgraph of the diffractive hadron $h\to h'$; see \fig{fig:sdhep soft A} as an illustration. 
The factorization of soft gluons leads to the independence among different collinear subgraphs, 
and help to establish the factorization of the collinear subgraph of the diffractive hadron into a universal GPD, 
and the other collinear subgraphs into universal meson DAs.

\subsection{Assumptions for the exclusive factorization}
\label{ssec:assumption}

The keys to collinear factorization are the cancellation of soft subgraphs that 
connect to different collinear subgraphs and the factorization of all collinear subgraphs 
from the infrared-safe short-distance hard part.

The first assumption that we made is that the leading active quark lines 
or transversely polarized gluon lines from the mesons must be coupled to the hard interaction, 
but not to the soft subgraph, for which we effectively assume that we could get an 
additional suppression from the expected end point behavior of meson wave function, 
when one of the active quarks (or gluons) has a soft momentum, 
which we have referred to as the soft-end suppression. 
The result of this assumption is that, to the leading power, the soft subgraph is 
only connected to collinear subgraphs by gluon lines that are longitudinally polarized, 
for which Ward identity can be applied to factorize them onto Wilson lines. 
The soft Wilson lines are only connected to the rest of the graph by colors, 
and can be disentangled and factorized from the collinear subgraphs because 
the collinear subgraphs are in color singlet states, which is an important feature of exclusive processes. 
Consequently, the soft cancellation for the factorization of SDHEPs is 
very different from typical soft cancellation for the factorization of inclusive processes~\citep{Collins:1989gx}.

Another consequence of the soft-end suppression is that we are allowed to 
constrain the light-cone parton momenta of the mesons on the real axis 
and arrive at a definition of meson DA, $\phi(z)$ with $0 < z < 1$, 
as argued at the end of \sec{sssec:dvcs}.

This assumption was also applied to most factorizations of exclusive processes 
involving high-momentum mesons, notably for the pion form factor and large-angle production processes; 
see the review~\citep{Brodsky:1989pv}. 
Even though the soft-end region was conjectured to be Sudakov suppressed in~\citep{Brodsky:1989pv}, 
which is more than the power suppression taken as our assumption, 
a more extensive discussion on this issue is still lacking in the literature.

The second assumption that we implicitly made is that there is only one single hard interaction 
in which all the parton lines are effectively off shell by the hard scale. 
This applies especially to the mesoproduction of a meson pair process in 
Secs.~\ref{sssec:mm2mm} and \ref{sssec:mh2mmh}. 
It is well known that the exclusive hadron-hadron scattering into large-angle hadrons can happen 
via multiple hard interactions, which has an enhanced power counting with respect to 
the single hard interaction~\citep{Landshoff:1974ew, Botts:1989kf}. 
We have shown the factorization for the hard exclusive $2\to 2$ meson-meson scattering 
and the corresponding SDHEP with a meson beam for the single hard interaction case. 
Within the two-stage paradigm, it is unclear to us whether the factorization of the 
large-angle meson-meson scattering via multiple hard interactions can imply 
a corresponding factorization for the SDHEP with a meson beam; it is left for future study.

One may also consider representing $A^*$ as a sum over virtual hadronic states, 
instead of the expansion in terms of partonic states like $[q\bar{q}']$ and $[gg]$. 
However, the exchanged state $A^*$ in the SDHEP enters a hard collision, 
which has a resolution scale $1/Q$ much smaller than the typical hadronic scale, 
and therefore it is the partonic degrees of freedom inside the virtual hadronic state or the diffractive hadron that are probed. 
For example, the leading-power contribution from a virtual hadronic state 
should also be mediated by two active parton lines, just as in 
Figs.~\ref{fig:em2ea}(b), \ref{fig:em2em}(a), \ref{fig:am2mm}(a), \ref{fig:mm2ll}(a), and~\ref{fig:mm2mm}(a), 
along with the same short-distance hard part as the $n=2$ 
partonic channel in connection with GPDs.
In principle, to this power, one should add all the two-parton-mediated contributions 
from all possible virtual hadronic states of the same diffractive hadron, 
which could possibly recover the full contributions from the corresponding GPDs of the same hadron, 
but, only from their ERBL region. GPDs also contain the DGLAP region, 
which cannot be covered by the subprocesses mediated by virtual hadronic states.
The approach of taking out a virtual meson $A^*$ from the $h\to h'$ transition, 
described by some form factor $F^{A}_{h\to h'}(t)$, followed by extracting two parton lines 
via its distribution amplitude, should also be captured by the GPD of $h\to h'$ transition in a more general sense.
The choice to represent $A^*$ by a single virtual meson state, like the Sullivan process, is therefore an additional approximation. 
On the other hand, the expansion in terms of the number of partons, $n$, is an expansion in powers of $1/Q$.

\subsection{Non-factorizability of double diffractive processes}
\label{ssec:double-diffractive}

From the procedure for proving factorization in the two-stage paradigm, 
it is easy to understand the importance of the {\it single} diffraction for factorizability of the exclusive process. 
The whole difficulty from the diffraction is the DGLAP region that 
pinches one component of the soft gluon momentum in the Glauber region, 
and we get away with it by only deforming the other components associated with other mesons. 
After factorizing out all the other mesons, 
the rest of the soft gluons are only coupled to the diffracted hadron and can be grouped together into this hadron's GPDs.

\begin{figure}[htbp]
\centering
		\includegraphics[scale=0.75]{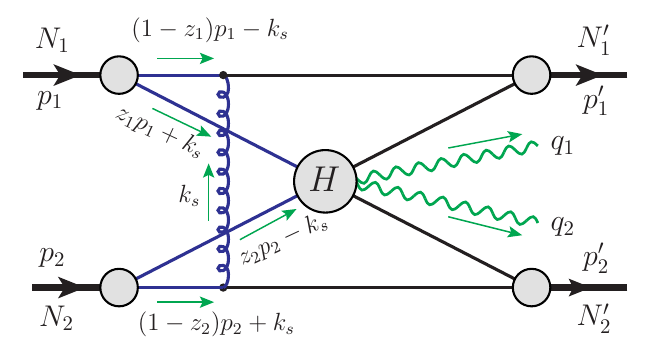}
	\caption{Diphoton production in a double diffractive hard exclusive scattering process 
	between two head-on hadrons $N_1$ and $N_2$ along the $z$ axis.}
\label{fig:double diffractive}
\end{figure}

If we consider the double diffractive process, as shown in \fig{fig:double diffractive}, 
the soft gluon $k_s$ exchanged between the remnants along opposite directions 
is pinched in the Glauber region for both $k_s^+$ and $k_s^-$, and thus 
no deformation can be done to get it out. As a result, this process cannot be factorized, 
even if we do have a hard scale provided by the transverse momentum $q_T$ of the final-state photon pair.

Similar conclusion holds for the {\it inclusive} diffractive processes~\citep{Soper:1997gj, Collins:1997sr}. 
The observation of the diffracted hadron anchors the inclusive sum over the final state 
and forbids the use of unitarity to cancel the Glauber gluon exchanges. 
While the soft gluon momentum can be deformed out of the Glauber region 
for single diffractive inclusive processes~\citep{Collins:1997sr}, in a similar way 
to the exclusive processes discussed in this thesis, it does not work for 
inclusive diffractive hadron-hadron scattering~\citep{Landshoff:1971zu, Henyey:1974zs, Cardy:1974vq, DeTar:1974vx, Collins:1992cv, Soper:1997gj}.

This phenomenon is very similar to the factorization of Drell-Yan process 
at high twists \citep{Qiu:1990xxa, Qiu:1990xy}, where the hadron connected 
by more than two active partons to the hard part is analogous to the diffracted hadron here. 
Even though the extra transversely polarized gluon lines at a high twist 
may be confused by soft gluons and endangers factorization, 
this is still factorizable as one can first factorize soft gluons out of the other hadron at the leading twist, 
similar to the procedure for the single diffractive process here that 
we first factorize the soft gluons out of the other mesons. 
This can only be done at the {\it first} subleading twist for which one of the two hadrons 
still has a twist-2 PDF involved, and so the Drell-Yan process is not factorizable 
beyond the first nonvanishing subleading twist, similar to the nonfactorizability of double diffractive processes.

\subsection{Comparison to high-twist inclusive processes}
\label{ssec:high-twist}

\begin{figure}[htbp]
\centering
	\begin{tabular}{cc}
		\includegraphics[clip, trim={-1em -1mm -1em 0}, scale=0.75]{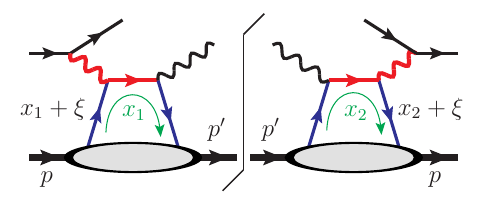} &
		\includegraphics[clip, trim={-1em 0 -1em 0}, scale=0.75]{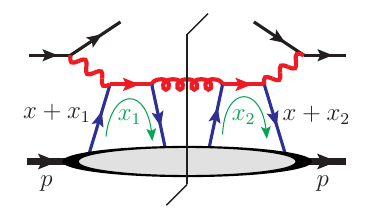}	\\
		(a) & (b) 
	\end{tabular}
	\caption{Sample leading-order cut diagrams for 
	(a) DVCS amplitude squared and 
	(b) inclusive DIS cross section at twist-4. 
	The red thick lines indicate the hard parts, and the blue lines are collinear partons.}
\label{fig:t4}
\end{figure}

\begin{figure}[htbp]
\centering
	\begin{tabular}{cc}
		\includegraphics[clip, trim={-1em 0 -1em 0}, scale=0.75]{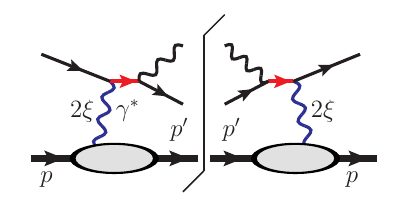}	&
		\includegraphics[clip, trim={-1em 0 -1em 0}, scale=0.75]{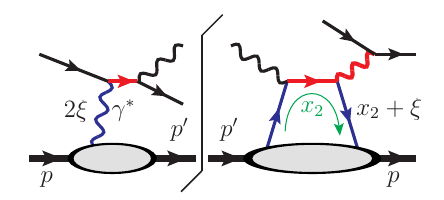}	\\
		(a) & (b) 
	\end{tabular}
	\caption{Sample cut diagrams of the amplitude squared of the real photon electroproduction process for 
	(a) the $\gamma^*$-mediated channel, and 
	(b) the interference between the $\gamma^*$-mediated channel and GPD channel. 
	The red thick lines indicate the hard parts, and the blue lines are collinear partons or photons.}
\label{fig:t23}
\end{figure}

The factorization of exclusive processes at the amplitude level shares 
many common features with the inclusive process factorization at a high twist. 
Taking the leading-order DVCS amplitude as an example, 
we show the amplitude square as a cut diagram in \fig{fig:t4}(a), 
which is compared with one of the leading-order diagrams of the inclusive DIS at twist-4 in \fig{fig:t4}(b). 
They only differ in that the cut line for the DVCS forces an exclusive final state. 
Both diagrams have two collinear parton lines connecting the hadron-collinear subgraph to the hard part, 
in both the amplitude to the left of the cut and conjugate amplitude to the right. 
In this sense, the DVCS amplitude squared corresponds to a twist-4 contribution 
to the cross section of the real photon electroproduction process. 
On the other hand, the amplitude squared of the $n = 1$ channel for the 
$\gamma^*$-mediated subprocess corresponds to a twist-2 contribution (see \fig{fig:t23}(a)), 
and the interference between the $n = 1$ and $n = 2$ channels corresponds to a twist-3 contribution (see \fig{fig:t23}(b)).

In the DVCS amplitude in \fig{fig:t4}(a), the two partons carry momenta 
$(x_1 + \xi) P^+$ and $(x_1 - \xi) P^+$ (following the directions indicated by the curved arrow), 
with $x_1$ integrated in $[-1, 1]$. In its conjugate amplitude, the two partons carry momenta 
$(x_2 \pm \xi) P^+$ with $x_2$ integrated in the same range. 
Similarly, for the twist-4 DIS diagram in \fig{fig:t4}(b), the amplitude part has two collinear partons with momenta 
$(x+x_1)p^+$ and $x_1 \, p^+$, with $x_1$ integrated in $[-1, 1-x]$. 
The conjugate amplitude part has two collinear partons with momenta 
$(x+x_2)p^+$ and $x_2 \, p^+$, with $x_2$ integrated in the same range. 
In both cases, the $x_1$ and $x_2$ integrations are not related and to be integrated independently. 
Only the total momentum of the two partons, which is 
$2\xi P^+$ for the DVCS and $x p^+$ for the twist-4 DIS, is observable, whose dependence is probed by the experiment.

On the other hand, there are soft breakpoint poles of $x_1$ (or $x_2$), 
given by the situations when one of the two partons has zero momentum, 
which is $x_1 = \pm \xi$ for DVCS and $x_1 = 0$ or $-x$ for twist-4 DIS. 
However, those poles are not pinched and they happen at the middle part of the $x_1$ integration range. 
As a result, we can deform the contour of $x_1$ to avoid them, 
just as discussed around \eq{eq:principle value}. This situation is contrary to the DA factorization, 
for which the soft poles happen at the end points of the DA integration and cannot be deformed away, 
which requires us to make the soft-end suppression assumption in \sec{ssec:assumption}.


\chapter{Generalized parton distributions}
\label{ch:GPD}


The generalized parton distributions (GPDs) resulting from factorization of single-diffractive exclusive scattering processes are important
nonperturbative parton correlation functions that reveal many aspects of the confined partonic structures of hadrons. 
By their universal operator definitions, GPDs can be studied by themselves. 
Their values can be obtained by nonperturbative calculation methods like Lattice QCD~\citep{Ji:2013dva, Chen:2019lcm, Alexandrou:2020zbe, Lin:2020rxa, Lin:2021brq, Hashamipour:2022noy, Bhattacharya:2022aob}, 
which will not be discussed in this thesis, 
or by fitting to experimental data by virtue of the factorization theorems discussed in \sec{ch:exclusive}. 
Nevertheless, the exclusive nature of the GPD factorization poses substantial challenges for the fitting programs, making the 
extraction of GPDs, especially their $x$ dependence, from experimental data, extremely difficult. 
This is our focus in this section.
First, we will first review some important properties of GPDs, especially their roles in unveiling the hadron structures.
Then we will lay out the phenomenological framework for the single-diffractive hard exclusive processes (SDHEPs).
We will see that the two-stage paradigm gives a clear description of the azimuthal correlations that arise from different spin structures 
of the GPDs.
As an illustration, we will discuss the most popular process, deeply virtual Compton scattering (DVCS), within this framework.
Similar to many other processes, it can probe GPDs only up to a few moments. 
This information is far from enough to map out the full $x$ dependence of GPDs.
To resolve this issue, following a general discussion on the $x$ sensitivity of GPD processes, 
we will introduce a type of processes that can provide enhanced sensitivity to the $x$ dependence, 
and demonstrate how well they can help determine the latter. 
We will close this section by proposing a global analysis of all types of observables that can be used for the task of determining GPDs.


\section{GPD properties}
\label{sec:gpd-property}

\subsection{Definitions and spin dependence}
\label{ssec:gpd-def}
As remarked below \eq{eq:GPD-x-range}, the GPDs defined in \eq{eq:eh2eah-GPDs} contain full dependence on the hadron spin states.
This dependence shall be separated by decomposing the matrix elements into independent form factors,
\bse\label{eq:GPD-def-q}\begin{align}
	F^q(x, \xi, t) & = 
		\int \frac{dy^-}{4\pi} e^{-i x P^+ \, y^-}
			\langle p', \alpha' |
			\bar{\psi}_{q}\pp{ y^- / 2} \, 
			\gamma^+ \, 
			\psi_{q} \pp{ -y^- / 2}
		| p, \alpha \rangle	\nn\\
	& = \frac{1}{2 P^+} \bar{u}(p', \alpha')
		\bb{ H^q(x, \xi, t) \gamma^+ 
			- E^q(x, \xi, t) \frac{i \sigma^{+\alpha} \Delta_{\alpha}}{2m}
		} u(p, \alpha),	\label{eq:GPD-def-F-q}\\
	\wt{F}^q(x, \xi, t) & = 
		\int \frac{dy^-}{4\pi} e^{-i x P^+ \, y^-}
			\langle p', \alpha' |
			\bar{\psi}_{q}\pp{ y^- / 2} \, 
			\gamma^+ \gamma_5 \, 
			\psi_{q} \pp{ -y^- / 2}
		| p, \alpha \rangle	\nn\\
	& = \frac{1}{2 P^+} \bar{u}(p', \alpha')
		\bb{ \wt{H}^q(x, \xi, t) \gamma^+ \gamma_5 
			- \wt{E}^q(x, \xi, t) \frac{\gamma_5 \Delta^+}{2m}
		} u(p, \alpha),
\end{align}\ese
where we take the hadron states as protons for definiteness, and use the kinematic convention in \eq{eq:eh2eah-kin}. 
$\alpha$ and $\alpha'$ explicitly denote the spin states.
The $\Delta$ differs from the usual convention~\citep{Diehl:2003ny} by a sign, 
which has been compensated by the minus sign in front of $E$ and $\wt{E}$ such that the GPDs are the same.
We have dropped the time ordering and omitted the Wilson lines.
The decomposition is done following Lorentz covariance, parity invariance, and Dirac matrix properties.
The same decomposition applies to gluon GPDs,
\bse\label{eq:GPD-def-g}\begin{align}
	F^g(x, \xi, t) & = 
		\sum\nolimits_{i, j}\delta^{ij}
		\int \frac{dy^-}{2\pi P^+} e^{-i x P^+ \, y^-}
			\langle p', \alpha' |
			G^{+i} \pp{ y^- / 2} \, 
			G^{+j} \pp{ -y^- / 2}
		| p, \alpha \rangle	\nn\\
	& = \frac{1}{2 P^+} \bar{u}(p', \alpha')
		\bb{ H^g(x, \xi, t) \gamma^+ 
			- E^g(x, \xi, t) \frac{i \sigma^{+\alpha} \Delta_{\alpha}}{2m}
		} u(p, \alpha),	\label{eq:GPD-def-F-g}\\
	\wt{F}^g(x, \xi, t) & = 
		\sum\nolimits_{i, j} ( -i \epsilon_T^{ij} )
		\int \frac{dy^-}{2\pi P^+} e^{-i x P^+ \, y^-}
			\langle p', \alpha' |
			G^{+i} \pp{ y^- / 2} \, 
			G^{+j} \pp{ -y^- / 2}
		| p, \alpha \rangle	\nn\\
	& = \frac{1}{2 P^+} \bar{u}(p', \alpha')
		\bb{ \wt{H}^g(x, \xi, t) \gamma^+ \gamma_5 
			- \wt{E}^g(x, \xi, t) \frac{\gamma_5 \Delta^+}{2m}
		} u(p, \alpha).
\end{align}\ese
It is the scalar coefficients $H^{q,g}$, $\wt{H}^{q,g}$, $E^{q,g}$, $\wt{E}^{q,g}$ that are usually referred to as GPDs,
which are constrained to be real functions that are even in $\xi$.
In this thesis we loosely refer to both these coefficients and $F$'s, $\wt{F}$'s as GPDs. 
There are also form factor decompositions for the transversity GPDs, but we will not discuss them in this thesis.

In the GPD definitions, the parton spin states are dictated by the spinor or tensor projectors, 
$\gamma^+$ and $\gamma^+\gamma_5$, or $\delta^{ij}$ and $-i \epsilon_T^{ij}$,
whereas the proton spin structures are selected by the different form factors. 
It is, however, not straightforward to quantitatively describe them.
First, the partons in GPDs are not on-shell, but instead we have integrated out their transverse and minus momentum components
(see the discussion above \eq{eq:eh2eah-GPDs}).
Second, the proton states are not both along the $z$ direction, and one can even go to a frame 
where both $p$ and $p'$ are not along the $z$ direction.
On the other hand, the parton states in the hard scattering have been projected to be on-shell along the $z$ direction, and their 
spin states can be chosen as the helicities.
To unify the whole picture, we now introduce the concept of light-cone helicity state.

\subsubsection{Transverse boost and light-cone helicity}
\label{sssec:transverse-boost}
A {\it transverse boost} $\Lambda(\bm{v})$ is a special Lorentz transformation that takes a momentum $k$ to
$k^{\prime \mu} = \Lambda^{\mu}{}_{\nu}(\bm{v}) k^{\nu}$ by 
\beq[eq:trans-boost]
	k^{\prime +} = k^+, \quad
	\bm{k}'_T = \bm{k}_T + \sqrt{2} k^+ \bm{v}, \quad
	k^{\prime -} = k^- + \sqrt{2} \bm{k}_T \cdot \bm{v} + k^+ v^2,
\eeq
where $\bm{v} = (v_1, v_2)$ is a transverse vector and $v^2 = v_1^2 + v_2^2$.
This keeps the plus momentum invariant but shifts the transverse momentum 
(the $k^-$ transformation is determined by requiring $k^2$ to be invariant).
The transformation matrix $\Lambda(\bm{v})$ can be written in the Cartesian coordinate system as
\beq[eq:trans-boost-mat]
	\Lambda(\bm{v}) 
	= \pp{ \Lambda^{\mu}{}_{\nu}(\bm{v}) }
	= \begin{pmatrix}
		1 + v^2 / 2 & v_1 & v_2 & v^2 / 2 \\
		v_1 & 1 & 0 & v_1 \\
		v_2 & 0 & 1 & v_2 \\
		-v^2 / 2 & -v_1 & -v_2 & 1 - v^2 / 2
	\end{pmatrix}
	\simeq
	1 + 
	\begin{pmatrix}
		0 & v_1 & v_2 & 0 \\
		v_1 & 0 & 0 & v_1 \\
		v_2 & 0 & 0 & v_2 \\
		0 & -v_1 & -v_2 & 0
	\end{pmatrix}.
\eeq
In the last step, we have taken the small $v$ approximation and thrown away higher-power terms. 
This helps identify the transverse boost generators $\bm{T} = (T_1, T_2)$ with the usual boost and rotation generators,
$\bm{K} = (K_1, K_2, K_3)$ and $\bm{J} = (J_1, J_2, J_3)$, 
\beq[eq:trans-boost-gen]
	T_1 = K_1 + J_2, \quad
	T_2 = K_2 - J_1.
\eeq
The transverse boost can then be written as
\beq[eq:trans-boost-exp]
	\Lambda(\bm{v}) = e^{-i \bm{T} \cdot \bm{v} },
\eeq
which induces the transverse boost operator $\hat{U}(\bm{v}) = \exp( -i \hat{\bm{T}} \cdot \bm{v} )$ that acts on the quantum Hilbert space.

Using \eq{eq:trans-boost-gen} and the Poincare algebra, we can get $[T_1, T_2] = 0$ 
and work out their commutation relations with the momentum operator $\hat{P}^{\mu}$,
\beq[eq:trans-gen-mom]
	[\hat{T}^i, \hat{P}^+] = 0, \quad
	[\hat{T}^i, \hat{P}^-] = - \sqrt{2} \, i \, \hat{P}^i, \quad
	[\hat{T}^i, \hat{P}^j] = -i \, \delta^{ij} \sqrt{2} \, \hat{P}^+,
\eeq
where $i, j = 1, 2$ are the transverse indices, and we take $\hat{T}^i = \hat{T}_i$. 
By defining $\hat{P}^{\mu}(\bm{v}) = \hat{U}(\bm{v}) \hat{P}^{\mu} \hat{U}^{-1}(\bm{v})$, we have
\beq
	\frac{\partial}{\partial v^i} \hat{P}^{\mu}(\bm{v})
	= \hat{U}(\bm{v}) (-i) \bb{ \hat{T}^i, \hat{P}^{\mu} }  \hat{U}^{-1}(\bm{v}),
\eeq
which gives
\beq
	\frac{\partial}{\partial v^i} \hat{P}^{+}(\bm{v}) = 0, \quad
	\frac{\partial}{\partial v^i} \hat{P}^{j}(\bm{v}) = - \delta^{ij} \sqrt{2} \hat{P}^+(\bm{v}), \quad
	\frac{\partial}{\partial v^i} \hat{P}^{-}(\bm{v}) = - \sqrt{2} \hat{P}^i(\bm{v}).
\eeq
This gives the solution
\beq
	\hat{U}(\bm{v}) \hat{P}^{\mu} \hat{U}^{-1}(\bm{v})
	= \pp{ \hat{P}^+, \hat{P}^- - \sqrt{2} \hat{\bm{P}}_T \cdot \bm{v} + \hat{P}^+ v^2, \hat{\bm{P}}_T - \sqrt{2} \hat{P}^+ \bm{v} }.
\eeq
A one-particle state can be specified by its plus and transverse momentum, $|k^+, \bm{k}_T \rangle$, 
with its minus momentum component determined as $k^- = (k^2 + k_T^2) / (2k^+)$.
After acting on it a transverse boost operation, we have
\begin{align}
	\hat{P}^{\mu} \hat{U}(\bm{v}) |k^+, \bm{k}_T \rangle
	&\, = \hat{U}(\bm{v}) \hat{P}^{\mu}(-\bm{v}) |k^+, \bm{k}_T \rangle	\nn\\
	& \, = \pp{ k^+, k^- + \sqrt{2} \bm{k}_T \cdot \bm{v} + k^+ v^2, \bm{k}_T + \sqrt{2} k^+ \bm{v}} \hat{U}(\bm{v}) |k^+, \bm{k}_T \rangle.
\end{align}
That is,
\beq[eq:trans-boost-state]
	\hat{U}(\bm{v}) |k^+, \bm{k}_T \rangle = |k^+, \bm{k}_T + \sqrt{2} k^+ \bm{v} \rangle
\eeq
realizes the same momentum transformation as in \eq{eq:trans-boost}. Together with the normalization
\beq
	\vv{ k^+, \bm{k}_T | k^{\prime +}, \bm{k}'_T } = (2\pi)^3 \, (2 k^+) \, \delta(k^+ - k^{\prime +}) \, \delta^{(2)}(\bm{k}_T - \bm{k}'_T),
\eeq
\eq{eq:trans-boost-state} facilitates the unitary representation of $\hat{U}(\bm{v})$ in the Hilbert space. 
The helicity quantum number will be specified in the following.

The usual helicity state $|\bm{k}, \lambda \rangle$ is defined by transforming from the basic reference state 
$|k_0 \hat{z}, \lambda \rangle$ by first boosting along the $z$ direction such that it has the same energy as $|\bm{k}, \lambda \rangle$,
and then rotating around the $y$ and $z$ axes to reach the momentum $\bm{k}$ (see \ch{ch:Poincare} for details),
\beq
	|\bm{k}, \lambda \rangle \equiv U(R_z(\phi)) U(R_y(\theta)) U(\Lambda_z(\beta)) |k_0 \hat{z}, \lambda \rangle,
\eeq
where $\theta$ and $\phi$ are the polar and azimuthal angles of $\bm{k}$, respectively.
For such helicity state, the spin quantization axis is the momentum direction, which transforms as we rotate the momentum.
A rotation around the $z$ axis thus transforms $|\bm{k}, \lambda \rangle$ in a trivial way 
(see the discussion below \eq{eq:lorentz-rep-massive-R} in \ch{ch:Poincare}),
\beq
	U(R_z(\alpha)) |\bm{k}, \lambda \rangle = |R_z(\alpha)\bm{k}, \lambda \rangle,
\eeq
without any phase signifying a spin component along the $z$ direction.

In a similar way, we define the light-cone helicity state $|k^+, \bm{k}_T, \lambda \rangle$ by transforming from the basic reference state
$|k_0^+, \bm{0}_T, \lambda \rangle$. First boost along $z$ to reach the plus momentum $k^+$. Then perform a transverse boost with
$\bm{v} = \bm{k}_T / (\sqrt{2} k^+)$, that is,
\beq[eq:light-cone-helicity-state]
	|k^+, \bm{k}_T, \lambda \rangle = U\pp{ \frac{\bm{k}_T}{\sqrt{2} k^+} } |k^+, \bm{0}_T, \lambda \rangle.
\eeq
This applies to both massless and massive particle states.
\footnote{The only exception is for massless states moving along the $-z$ direction, which have zero plus momentum 
so cannot be achieved by \eq{eq:light-cone-helicity-state}.}
Since the transverse boosts form an Abelian subgroup, acting a transverse boost on $|k^+, \bm{k}_T, \lambda \rangle$ only changes the 
momentum component $\bm{k}_T$, but keeps $k^+$ and $\lambda$ invariant.
Also, by noting $[\bm{T}, K_3] = i \bm{T}$ and thus 
\beq
	e^{-i K_3 \beta} \, \bm{T} \, e^{i K_3 \beta} = e^{-\beta} \bm{T},
\eeq
the light-cone helicity state transforms under a longitudinal boost as
\begin{align}\label{eq:trans-boost-z}
	e^{-i K_3 \beta} |k^+, \bm{k}_T, \lambda \rangle 
	&\, = \pp{ e^{-i K_3 \beta} e^{-i \bm{T} \cdot \bm{v}} e^{i K_3 \beta} } e^{-i K_3 \beta} |k^+, \bm{0}_T, \lambda \rangle	\nn\\
	&\, = e^{-i \bm{T} \cdot ( e^{-\beta} \bm{v} ) } |e^{\beta} k^+, \bm{0}_T, \lambda \rangle\nn\\
	&\, = |e^{\beta} k^+, \bm{k}_T, \lambda \rangle,
\end{align}
which keeps the light-cone helicity invariant. 
Similarly, under a rotation around the $z$ direction, the state transforms as
\begin{align}\label{eq:trans-rot-z}
	e^{-i J_3 \alpha} |k^+, \bm{k}_T, \lambda \rangle 
	&\, = \pp{ e^{-i J_3 \alpha} e^{-i \bm{T} \cdot \bm{v}} e^{i J_3 \alpha}  } e^{-i J_3 \alpha}  |k^+, \bm{0}_T, \lambda \rangle	\nn\\
	&\, = e^{-i \lambda \alpha} e^{-i \bm{T} \cdot [ R_z(\alpha) \bm{v} ] } |k^+, \bm{0}_T, \lambda \rangle \nn\\
	&\, = e^{-i \lambda \alpha} |k^+, R_z(\alpha) \bm{k}_T, \lambda \rangle,
\end{align}
which is obtained by using $[J_3, T_1] = i T_2$, $[J_3, T_2] = -i T_1$ and
\beq
	e^{-i J_3 \alpha} \, \bm{T} \, e^{i J_3 \alpha} = R^{-1}_z(\alpha) \, \bm{T}.
\eeq
Eqs.~\eqref{eq:trans-boost-z} and \eqref{eq:trans-rot-z} establish the light-cone helicity $\lambda$ with a physical interpretation
of the spin component along the $z$ direction. 

We note the difference between the light-cone helicity state and the canonical spin state. The latter only applies to a massive state
and is defined by boosting the basic reference state $|0, s \rangle$ along the momentum direction,
\beq
	|\bm{k}, j, s \ra = e^{-i \bm{K} \cdot \bm{\beta} } |0, j, s \rangle
	= U(R_3(\phi) R_2(\theta)) e^{-i K_3 \beta} U^{-1}(R_3(\phi) R_2(\theta)) |0, j, s \rangle,
\eeq
where $\bm{k}$ is along the $(\theta, \phi)$ direction, $\bm{\beta} = \bm{k} / k^0$ and $\beta = |\bm{\beta}|$. 
This is related to the helicity state by
\beq[eq:canonical-helicity-relation]
	|\bm{k}, j, s \ra = e^{i s \phi} \sum_{\lambda} d^j_{s \lambda}(\theta) \,  |\bm{k}, j, \lambda \ra .
\eeq
Under a rotation around $z$, this also transforms as
$|\bm{k}, j, s \ra \to e^{-i s \alpha } |\bm{k}, j, s \ra$. 
But under a boost along $z$, $s$ is not kept invariant. 
By using Eqs.~\eqref{eq:little-group-chi} and \eqref{eq:boost-transformation-massive}, it transforms into
\begin{align}
	&U(\Lambda_z(\beta)) |\bm{k}, j, s \ra
	= e^{i s \phi} \sum_{\lambda, \lambda'} d^j_{s \lambda}(\theta) \, |\Lambda_z(\beta)\bm{k}, j, \lambda' \ra \, d^j_{\lambda' \lambda}\pp{ \chi(\beta, k) } \nn\\
	&\hspace{2em} 
	= e^{i s \phi} \sum_{\lambda} d^j_{s \lambda}\big( \theta - \chi(\beta, k) \big) |\Lambda_z(\beta) \bm{k}, j, \lambda \ra \nn\\
	&\hspace{2em} 
	= \sum_{s'} \bb{ e^{i s \phi} \, d^j_{ss'}\big( \theta - \chi(\beta, k) - \theta'(\Lambda k) \big) \, e^{-i s' \phi} } |\Lambda_z(\beta) \bm{k}, j, s' \ra 	\nn\\
	&\hspace{2em} 
	= \sum_{s'} |\Lambda_z(\beta) \bm{k}, j, s' \ra \, D^j_{s's}\big( \phi, \theta'(\Lambda k) - \theta + \chi(\beta, k), \phi \big)
\label{eq:boost-canonical}
\end{align}
where in the third step we used the inverse of \eq{eq:canonical-helicity-relation}.
In \eq{eq:boost-canonical}, $\Lambda_z(\beta)$ does not change the azimuthal angle of $\bm{k}$, 
but changes its polar angle to $\theta'(\Lambda k)$.
This can be easily verified not to equal $\theta - \chi(\beta, k)$. 
Therefore, the canonical spin component $s$ shall only be interpreted as the spin component along the $z$ direction in the rest frame, 
but not in the boosted frame.

\subsubsection{Light-front quantization}
\label{sssec:lf-quantization}
One may designate the light-cone helicity states defined in \eq{eq:light-cone-helicity-state} into field decomposition and define single-particle creation and annihilation operators.
For a fermion field, the amplitude for annihilating a state at some space-time point $x$ is given by
\beq[eq:lf-fermion-anni]
	\la 0 | \psi(x) | k^+, \bm{k}_T, \lambda \ra
	 = u_{\lambda}(k^+, \bm{k}_T) e^{-i k \cdot x},
\eeq 
which defines the spinor $u_{\lambda}(k^+, \bm{k}_T)$ associated with this state. 
By using the definition in \eq{eq:light-cone-helicity-state}, the left-hand side of \eq{eq:lf-fermion-anni} becomes
\begin{align}
	& \la 0 | \psi(x) U(\bm{v}) | k^+, \bm{0}_T, \lambda \ra
	= \la 0 | U^{-1}(\bm{v}) \, \psi(x) \, U(\bm{v}) | k^+, \bm{0}_T, \lambda \ra 	\nn\\
	&\hspace{2em}
	= S(\bm{v}) \la 0 | \psi\big( \Lambda^{-1}(\bm{v}) x \big) | k^+, \bm{0}_T, \lambda \ra
	= S(\bm{v}) u_{\lambda}(k^+, \bm{0}_T) e^{-i k \cdot x},
\end{align}
which therefore gives the definition for the spinor, in a similar way to the state definition,
\beq[eq:lf-fermion-spinor-def]
	u_{\lambda}(k^+, \bm{k}_T) = S(\bm{v}) \, u_{\lambda}(k^+, \bm{0}_T).
\eeq
Here $S(\bm{v})$ is the Lorentz group representation for the Dirac spinor associated with the transverse boost 
$\Lambda(\bm{v})$. It can be easily solved by using the generator definitions in \eq{eq:trans-boost-gen}, 
and gives the explicit spinor definitions,
\begin{align}\label{eq:lc-spinor-u}
	u_{+}(k^+, \bm{k}_T)
	= \bb{ \sqrt{2} k^+ }^{1/2}
	\begin{pmatrix}
	 \frac{m}{\sqrt{2} k^+} 
		\begin{pmatrix} 
	 		1 \\ 0 
		\end{pmatrix} \\
	 	\begin{pmatrix} 
	 		1 \\ 
			\frac{k_T \, e^{i \phi} }{\sqrt{2} k^+}
		\end{pmatrix}
	\end{pmatrix},
\quad
	u_{-}(k^+, \bm{k}_T)
	= \bb{ \sqrt{2} k^+ }^{1/2}
	\begin{pmatrix}
		\begin{pmatrix} 
	 		-\frac{k_T \, e^{-i \phi} }{\sqrt{2} k^+} \\ 
			1
		\end{pmatrix}	\\
		\frac{m}{\sqrt{2} k^+} 
		\begin{pmatrix} 
	 		0 \\ 1 
		\end{pmatrix}
	\end{pmatrix},
\end{align}
where we took $\bm{k}_T = k_T (\cos\phi, \sin\phi)$. The spinors for antiparticles can be obtained in the same way, 
or by simply using charge conjugation relation $v_{\lambda}(k^+, \bm{k}_T) = i \gamma^2 u^*_{\lambda}(k^+, \bm{k}_T)$,
\begin{align}\label{eq:lc-spinor-v}
	v_{+}(k^+, \bm{k}_T)
	= \bb{ \sqrt{2} k^+ }^{1/2}
	\begin{pmatrix}
	 	\begin{pmatrix} 
	 		\frac{k_T \, e^{-i \phi} }{\sqrt{2} k^+} \\ 
			-1
		\end{pmatrix}	\\
		\frac{m}{\sqrt{2} k^+} 
		\begin{pmatrix} 
	 		0 \\ 1 
		\end{pmatrix}
	\end{pmatrix},
\quad
	v_{-}(k^+, \bm{k}_T)
	= \bb{ \sqrt{2} k^+ }^{1/2}
	\begin{pmatrix}
		\frac{m}{\sqrt{2} k^+} 
		\begin{pmatrix} 
	 		1 \\ 0 
		\end{pmatrix} \\
	 	\begin{pmatrix} 
	 		-1 \\ 
			-\frac{k_T \, e^{i \phi} }{\sqrt{2} k^+}
		\end{pmatrix}
	\end{pmatrix}.
\end{align}
It is straightforward to verify that the light-cone helicity spinors satisfy the usual normalization relations,
\begin{align}
	&\bar{u}_{\lambda}(k^+, \bm{k}_T) \, u_{\lambda'}(k^+, \bm{k}_T) = 2m \delta_{\lambda \lambda'},
	\quad
	\bar{v}_{\lambda}(k^+, \bm{k}_T) \, v_{\lambda'}(k^+, \bm{k}_T) = -2m \delta_{\lambda \lambda'},
	\nn\\
	&u^{\dag}_{\lambda}(k^+, \bm{k}_T) \, u_{\lambda'}(k^+, \bm{k}_T) = 2 E \delta_{\lambda \lambda'},
	\quad
	v^{\dag}_{\lambda}(k^+, \bm{k}_T) \, v_{\lambda'}(k^+, \bm{k}_T) = 2 E \delta_{\lambda \lambda'},
	\nn\\
	& \bar{u}_{\lambda}(k^+, \bm{k}_T) \, \gamma^{\mu} \, u_{\lambda'}(k^+, \bm{k}_T)
	= \bar{v}_{\lambda}(k^+, \bm{k}_T) \, \gamma^{\mu} \, v_{\lambda'}(k^+, \bm{k}_T)
	= 2 k^{\mu} \delta_{\lambda\lambda'},
\end{align}
with $E = (k^+ + k^-) / \sqrt{2}$,
the orthogonality,
\beq
	\bar{u}_{\lambda}(k^+, \bm{k}_T) \, v_{\lambda'}(k^+, \bm{k}_T)
	= \bar{v}_{\lambda}(k^+, \bm{k}_T) \, u_{\lambda'}(k^+, \bm{k}_T)
	= 0,
\eeq
and the sum rules,
\beq
	\sum_{\lambda} u_{\lambda}(k^+, \bm{k}_T) \, \bar{u}_{\lambda}(k^+, \bm{k}_T) 
	= \slash{k} + m,
	\quad
	\sum_{\lambda} v_{\lambda}(k^+, \bm{k}_T) \, \bar{v}_{\lambda}(k^+, \bm{k}_T) 
	= \slash{k} - m.
\eeq

The same procedure can apply to a vector particle state, which is annihilated by the vector field by
\beq[eq:lf-vector-anni]
	\la 0 | A^{\mu}(x) | k^+, \bm{k}_T, \lambda \ra
	 = \epsilon^{\mu}_{\lambda}(k^+, \bm{k}_T) e^{-i k \cdot x},
\eeq 
where the polarization vector $\epsilon^{\mu}_{\lambda}(k^+, \bm{k}_T)$ is defined by the 
transverse boost in a similar way to \eq{eq:lf-fermion-spinor-def},
\beq[eq:lf-vector-polv-def]
	\epsilon^{\mu}_{\lambda}(k^+, \bm{k}_T)
	 = \Lambda^{\mu}{}_{\nu}(\bm{v}) \, \epsilon^{\nu}_{\lambda}(k^+, \bm{0}_T).
\eeq
For massless vector bosons, the basic polarization vector is transverse,
$
\epsilon^{\mu}_{\lambda}(k^+, \bm{0}_T) = (0, \bm{\epsilon}_{T}, 0),
$
where $\bm{\epsilon}_{T} = (\epsilon^1_{T}, \epsilon^2_{T})$ is the transverse part.
With the help of \eq{eq:trans-boost-mat}, 
this then gives the polarization vector for a general momentum,
\beq
	\epsilon^{\mu}_{\lambda}(k^+, \bm{k}_T)
	= \pp{ 
		\frac{\bm{k}_T \cdot \bm{\epsilon}_{T}}{\sqrt{2} k^+}, 
		\bm{\epsilon}_{T},
		-\frac{\bm{k}_T \cdot \bm{\epsilon}_{T}}{\sqrt{2} k^+}
		}
	= \pp{ 0^+, \frac{\bm{k}_T \cdot \bm{\epsilon}_{T}}{k^+}, \bm{\epsilon}_{T} }_{\rm lc},
\eeq
where the last expression is in light-front coordinates. 
We note that it has the same transverse component as the basic vector 
$\epsilon^{\mu}_{\lambda}(k^+, \bm{0}_T)$.

With the fixed definitions of the spinors and polarization vectors for arbitrary momenta, 
one can decompose the fields in terms of the light-cone helicity state creation and annihilation operators.
For a fermion field, one has
\beq[eq:lf-fermion-field-expan]
	\psi(x) = \sum_{\lambda}
		\int \frac{d k^+ d^2\bm{k}_T}{(2\pi)^3 \sqrt{2k^+}}
		\bb{
			b(k^+, \bm{k}_T, \lambda) \, u_{\lambda}(k^+, \bm{k}_T) \, e^{-i k\cdot x}
			+ 
			d^{\dag}(k^+, \bm{k}_T, \lambda) \, v_{\lambda}(k^+, \bm{k}_T) \, e^{i k\cdot x}
		},
\eeq
where $b(k^+, \bm{k}_T, \lambda)$ and $d(k^+, \bm{k}_T, \lambda)$ respectively 
annihilate a fermion and an anti-fermion with momentum $(k^+, \bm{k}_T)$ and 
light-cone helicity $\lambda$, and the integration of $k^+$ is from $0$
to $\infty$. And for a massless vector field, one has
\beq[eq:lf-vector-field-expan]
	A^{\mu}(x) = \sum_{\lambda}
		\int \frac{d k^+ d^2\bm{k}_T}{(2\pi)^3 \sqrt{2k^+}}
		\bb{
			a(k^+, \bm{k}_T, \lambda) \, \epsilon^{\mu}_{\lambda}(k^+, \bm{k}_T) \, e^{-i k\cdot x}
			+ 
			a^{\dag}(k^+, \bm{k}_T, \lambda) \, \epsilon^{\mu *}_{\lambda}(k^+, \bm{k}_T) \, e^{i k\cdot x}
		},
\eeq
where $\lambda = \pm 1$, and 
we have taken $A$ to be a Hermitian field, as is the case for photons and gluons.
The operators $a(k^+, \bm{k}_T, \lambda)$ and $a^{\dag}(k^+, \bm{k}_T, \lambda)$ respectively
annihilate and create a vector boson with momentum $(k^+, \bm{k}_T)$ and 
light-cone helicity $\lambda$. 
Then Eqs.~\eqref{eq:lf-fermion-anni} and \eqref{eq:lf-vector-anni} can be realized by 
Eqs.~\eqref{eq:lf-fermion-field-expan} and \eqref{eq:lf-vector-field-expan} provided
the single-particle state definitions,
\beq[eq:light-cone-helicity]
	| k^+, \bm{k}_T, \lambda \ra
	= \sqrt{2k^+} \, a^{\dag}(k^+, \bm{k}_T, \lambda) |0\ra,
\eeq
etc., and the commutation relations,
\beq[eq:lf-commutation]
	\bb{ a(k^+, \bm{k}_T, \lambda), a^{\dag}(k^{\prime +}, \bm{k}'_T, \lambda') }
	= (2\pi)^3 \, \delta(k^+ - k^{\prime +}) \, \delta^{(2)} (\bm{k}_T - \bm{k}'_T) \,
		\delta_{\lambda \lambda'},
\eeq
and similar anticommutation relations for the fermion operators.

Nevertheless, when converting such commutators among the creation and annihilation operators 
into the canonical commutators or anticommutators among the fields,
one does not get the ``naturally conjectured" equal-$x^+$ commutation relations, but instead,
at the last step one has to use $\int dk^+ d^2\bm{k}_T / 2 k^+ = \int d^3\bm{k} / 2 E_k$ to get the 
equal-time commutation relations. 
The existence of nonzero $\bm{k}_T$ and mass $m$ in the spinors and polarization vectors forbids the 
derivation of an equal-$x^+$ commutation relation. 
The use of light-cone helicity states does not embed itself into a simply covariant formalism.

To overcome this problem, we introduce the light-front quantization.
Instead of taking equal-time commutation relations plus time evolutions,
one take, right in the beginning, equal-$x^+$ commutation relations, 
and evolve everything with respect to $x^+$, under the ``Hamiltonian" $P_+$,
\beq
	i \frac{\partial}{\partial x^+} O(x^+) = [O(x^+), P_+].
\eeq
Then one shall immediately notice from the QCD Lagrangian that there are some ``bad" field components 
that are non-evolving and dynamically dependent on other components, 
and some ``good" dynamically independent field components. 
In the light-cone gauge $A^+ = 0$, the good fields components are
\beq[eq:lf-good-field]
	\psi_G(x) = \frac{\gamma^- \gamma^+}{2} \psi(x), \quad
	\bar{\psi}_G(x) = \bar{\psi}(x) \frac{\gamma^+ \gamma^-}{2}, \quad
	A^{\perp}(x) = (A^1(x), A^2(x)),
\eeq
where the color indices are omitted.

The field decompositions for the good fields are particularly simple. We notice that the spinor projector 
$\gamma^- \gamma^+ / 2$ takes all the spinors in Eqs.~\eqref{eq:lc-spinor-u} and \eqref{eq:lc-spinor-v} 
to their lightlike versions,
\begin{align}
	&\frac{\gamma^- \gamma^+}{2} 
		u_{\lambda}(k^+, \bm{k}_T; m)
		= u_{\lambda}(k^+, \bm{0}_T; 0)
		\equiv u_{\lambda}(k^+),	
	\nn\\
	&\frac{\gamma^- \gamma^+}{2} 
		v_{\lambda}(k^+, \bm{k}_T; m)
		= v_{\lambda}(k^+, \bm{0}_T; 0)
		\equiv v_{\lambda}(k^+).
\end{align}
Similarly, the transverse components of the polarization vectors are reduced to 
$\epsilon^{\mu}_{\lambda}(k^+, \bm{0}_T)$, for both massless and massive vector particles.
Then, the decompositions in Eqs.~\eqref{eq:lf-fermion-field-expan} and \eqref{eq:lf-vector-field-expan}
become,
\bse\label{eq:lf-field-expan-m0}\begin{align}
	\psi_G(x) = \sum_{\lambda}
		\int \frac{d k^+ d^2\bm{k}_T}{(2\pi)^3 \sqrt{2k^+}}
		\bb{
			b(k^+, \bm{k}_T, \lambda) \, u_{\lambda}(k^+) \, e^{-i k \cdot x}
			+ 
			d^{\dag}(k^+, \bm{k}_T, \lambda) \, v_{\lambda}(k^+) \, e^{i k\cdot x}
		},	
	\label{eq:lf-fermion-field-expan-m0}\\
	A^{i}(x) = \sum_{\lambda}
		\int \frac{d k^+ d^2\bm{k}_T}{(2\pi)^3 \sqrt{2k^+}}
		\bb{
			a(k^+, \bm{k}_T, \lambda) \, \epsilon^{i}_{\lambda}(k^+) \, e^{-i k\cdot x}
			+ 
			a^{\dag}(k^+, \bm{k}_T, \lambda) \, \epsilon^{i *}_{\lambda}(k^+) \, e^{i k\cdot x}
		},	
	\label{eq:lf-vector-field-expan-m0}
\end{align}\ese
where $i = 1, 2$ and $x = (0^+, x^-, \bm{x}_T)$ is at the zero light-front time.
Now there is no place for $k^-$ to come in, and the good field components satisfy the equal-$x^+$ 
commutation relations,
\begin{align}
	&\cc{ \psi_G(x^+, x^-, \bm{x}_T), \bar{\psi}_G(x^+, x^{\prime -}, \bm{x}'_T) }
	 = \frac{\gamma^-}{2} \, \delta(x^- - x^{\prime -}) \, \delta^{(2)}(\bm{x}_T - \bm{x}'_T), \\
	&\bb{ A^{i}(x^+, x^-, \bm{x}_T), \partial_- A^j(x^+, x^{\prime -}, \bm{x}'_T) }
	 = \frac{i}{2} \, \delta^{ij} \, \delta(x^- - x^{\prime -}) \, \delta^{(2)}(\bm{x}_T - \bm{x}'_T).
\end{align}

\subsubsection{Parton spin structure}
\label{sssec:GPD-parton-spin}
The parton spin structure is best understood in the light cone gauge $A^+ = 0$. 
The presence of the Wilson lines in the covariant gauge obscures the parton picture.

In the quark GPD definitions [\eq{eq:GPD-def-q}], 
the quark fields sandwich a $\gamma^+$ matrix. Since 
\beq
	\gamma^+ = 
		\pp{ \frac{\gamma^+ \gamma^-}{2} } \gamma^+ \pp{ \frac{\gamma^- \gamma^+}{2} },
\eeq
both the quark and antiquark fields are projected to be the good field components,
\beq
	\bar{\psi} \gamma^+ (1, \gamma_5) \psi = \bar{\psi}_G \gamma^+ (1, \gamma_5) \psi_G.
\eeq
Similarly, in the light-cone gauge, the gluon fields in the gluon GPD definitions [\eq{eq:GPD-def-g}]
only have the transverse components, so are also good field components. 
Thus we can decompose the fields according to \eq{eq:lf-field-expan-m0}, with the partons 
interpreted as carrying light-cone helicities. Note that in this picture, the creation and annihilation 
operators are for on-shell partons, which may or may not be massless but whose light-cone helicity
states are the same as the massless parton helicity states moving along the $z$ direction. 

Inserting \eq{eq:lf-fermion-field-expan-m0} into \eq{eq:GPD-def-q}, the light-cone operators can be expanded as,
\begin{align}
	&\int \frac{dy^-}{4\pi} e^{-i x P^+ \, y^-} \bar{\psi}_{q}( y^- / 2) \, \gamma^+ \,(1, \gamma_5) \, \psi_{q}( -y^- / 2 )	\nn\\
	&\hspace{1em}
	= \sum_{\lambda\lambda'} \int\frac{dk^+ dk^{\prime +} d^2\bm{k}_T d^2\bm{k}'_T}{(2\pi)^6} \theta(k^+) \theta(k^{\prime +}) 
	\nn\\
	& \hspace{4em} \times \bb{
			b^{\dag}(k^{\prime +}, \bm{k}'_T, \lambda') \, b(k^+, \bm{k}_T, \lambda) 
				\pp{ \delta_{\lambda'\lambda}, \sigma^3_{\lambda'\lambda} }
				\delta\big(2 x P^+ - (k^+ + k^{\prime +})\big)	\right.\nn\\
	& \hspace{5em} 		
		+ d(k^{\prime +}, \bm{k}'_T, \lambda') \, d^{\dag}(k^+, \bm{k}_T, \lambda) 
				\pp{ \delta_{\lambda'\lambda}, -\sigma^3_{\lambda'\lambda} }
				\delta\big(2 x P^+ + (k^+ + k^{\prime +})\big)	\nn\\
	& \hspace{5em} 		
		+ b^{\dag}(k^{\prime +}, \bm{k}'_T, \lambda') \, d^{\dag}(k^+, \bm{k}_T, \lambda) 
				\pp{ -\sigma^1_{\lambda'\lambda}, -i \sigma^2_{\lambda'\lambda} }
				\delta\big(2 x P^+ + (k^+ - k^{\prime +})\big)	\nn\\
	& \hspace{5em} 		
		\left. + d(k^{\prime +}, \bm{k}'_T, \lambda') \, b(k^+, \bm{k}_T, \lambda) 
				\pp{ -\sigma^1_{\lambda'\lambda}, i\sigma^2_{\lambda'\lambda} }
				\delta\big(2 x P^+ - (k^+ - k^{\prime +})\big)
		},
\label{eq:lc-expansion-qGPD-0}
\end{align}
where we used the explicit results for the light-cone helicity spinor algebra,
\begin{align}
	\bar{u}_{\lambda'}(k^{\prime +}) \gamma^+ (1, \gamma_5) u_{\lambda}(k^+)
		&= 2\sqrt{k^+ k^{\prime +}} \pp{ \delta_{\lambda'\lambda}, \sigma^3_{\lambda'\lambda} }, \nn\\
	\bar{v}_{\lambda'}(k^{\prime +}) \gamma^+ (1, \gamma_5) v_{\lambda}(k^+)
		&= 2\sqrt{k^+ k^{\prime +}} \pp{ \delta_{\lambda'\lambda}, -\sigma^3_{\lambda'\lambda} }, \nn\\
	\bar{u}_{\lambda'}(k^{\prime +}) \gamma^+ (1, \gamma_5) v_{\lambda}(k^+)
		&= 2\sqrt{k^+ k^{\prime +}} \pp{ -\sigma^1_{\lambda'\lambda}, -i \sigma^2_{\lambda'\lambda} }, \nn\\
	\bar{v}_{\lambda'}(k^{\prime +}) \gamma^+ (1, \gamma_5) u_{\lambda}(k^+)
		&= 2\sqrt{k^+ k^{\prime +}} \pp{ -\sigma^1_{\lambda'\lambda}, i\sigma^2_{\lambda'\lambda} },
\end{align}
with abuse of the Pauli matrix notations as in \eq{eq:GPD-Cg-pauli}.
Now in each term of \eq{eq:lc-expansion-qGPD-0}, we label the momentum $k^+$ as $(x + \xi) P^+$ when it corresponds 
to an annihilated quark or $-(x + \xi) P^+$ when it corresponds to a created antiquark. The momentum $k^{\prime +}$
can be determined by the $\delta$-function. This gives
\begin{align}
	&\int \frac{dy^-}{4\pi} e^{-i x P^+ \, y^-} \bar{\psi}_{q}( y^- / 2) \, \gamma^+ \,(1, \gamma_5) \, \psi_{q}( -y^- / 2 )	\nn\\
	&\hspace{0.2em}
	= \sum_{\lambda\lambda'} P^+ \int\frac{dx d^2\bm{k}_T d^2\bm{k}'_T}{(2\pi)^6}
	\nn\\
	& \hspace{2em} \times \bb{
			b^{\dag}\big((x-\xi)P^+, \bm{k}'_T, \lambda'\big) \, b\big((x+\xi)P^+, \bm{k}_T, \lambda\big) 
				\pp{ \delta_{\lambda'\lambda}, \sigma^3_{\lambda'\lambda} }
				\theta(x + \xi) \theta(x - \xi)	 \right.\nn\\
	& \hspace{3em} 		
		+ d^{\dag}\big(-(x+\xi)P^+, \bm{k}_T', \lambda'\big) \, d\big((\xi - x)P^+, \bm{k}_T, \lambda\big)
				\pp{ -\delta_{\lambda'\lambda}, \sigma^3_{\lambda'\lambda} }
				\theta(\xi - x) \theta(-x - \xi)		\nn\\
	& \hspace{3em} 		
		+ b^{\dag}\big((x-\xi)P^+, \bm{k}'_T, \lambda'\big) \, d^{\dag}\big(-(x+\xi)P^+, \bm{k}_T, \lambda\big)
				\pp{ -\sigma^1_{\lambda'\lambda}, -i \sigma^2_{\lambda'\lambda} }
				\theta(x - \xi) \theta(-x - \xi)	\nn\\
	& \hspace{3em} 		
		\left. + d\big((\xi - x)P^+, \bm{k}'_T, \lambda'\big) \, b\big((x+\xi)P^+, \bm{k}_T, \lambda\big)
				\pp{ -\sigma^1_{\lambda'\lambda}, i\sigma^2_{\lambda'\lambda} }
				\theta(\xi - x) \theta(x + \xi)
		},
\label{eq:lc-expansion-qGPD}
\end{align}
where the $\theta$-functions arise from the constraints $k^+ > 0$ and $k^{\prime+} > 0$.
In the second term, we used $d d^{\dag} = - d^{\dag} d$ to reverse the operator order and relabeled the momenta and helicities.
This results in an extra minus sign to the unpolarized antiquark GPD $F^{\bar{q}}$
but the correct sign for the polarized antiquark GPD $\wt{F}^{\bar{q}}$ to have the helicity polarization interpretation.
Depending on the sign of $\xi$, only three terms of \eq{eq:lc-expansion-qGPD} can survive. 
For the case $\xi > 0$, the term $b^{\dag} d^{\dag}$ that corresponds to the creation of a $[q\bar{q}]$ pair is not allowed.
For the remaining three terms, 
\begin{itemize}
\item 
	when $x > \xi$, it is the $b^{\dag} b$ term that 
	annihilates a quark with light-cone momentum fraction $(x + \xi)$ and helicity $\lambda$, 
	and then inserts back a quark with with light-cone momentum fraction $(x - \xi)$ and helicity $\lambda' = \lambda$. 
	The GPD $F^q$ simply adds the two helicity states so is unpolarized, 
	whereas $\wt{F}^q$ takes the difference and thus corresponds to the helicity polarization;
\item 
	when $x < - \xi$, it is the $d^{\dag} d$ term that 
	annihilates an antiquark with light-cone momentum fraction $(\xi - x)$ and helicity $\lambda$, 
	and then inserts back an antiquark with with light-cone momentum fraction $(- \xi - x)$ and helicity $\lambda' = \lambda$. 
	Similarly, $F^q$ is unpolarized and $\wt{F}^q$ is helicity polarized;
\item
	when $-\xi < x < \xi$, it is the $db$ term that
	annihilates a pair of quark and antiquark, with light-cone momentum fractions $(\xi \pm x)$ and helicities $(\lambda, \lambda')$ respectively.
	Due to the $\sigma^1$ or $\sigma^2$ structure, we either have $(\lambda, \lambda') = (+, -)$ or $(-, +)$ so that the 
	$|q_{\lambda} \bar{q}_{\lambda'}\ra$ state has zero net helicity. 
	The GPD $F^q$ adds the two helicity configurations, $|q_+ \bar{q}_-\ra + |q_- \bar{q}_+\ra$, so is unpolarized,
	while $\wt{F}^q$ takes the difference, $|q_+ \bar{q}_-\ra - |q_- \bar{q}_+\ra$, and is polarized.
\end{itemize} 
When inserting \eq{eq:lc-expansion-qGPD} between the hadron states $\la p' | \cdot | p \ra$, momentum conservation yields two more 
$\delta$-functions for $x$ and $\bm{k'}_T$ to kill the corresponding integrations, leaving us with only $\bm{k}_T$ integration.
It also constrains $x$ within $[-1, 1]$, as argued around \eq{eq:GPD-x-range}.

Due to the different physical interpretations of GPDs at different $x$, we call the region with $\xi < |x| < 1$ DGLAP region, which contains 
two subregions, one for quark and the other for antiquark, and the region with $|x| < \xi$ ERBL region.

The gluon GPDs have similar decompositions as \eq{eq:lc-expansion-qGPD}, which we will not repeat here but refer to \citep{Diehl:2003ny} for details.

\subsubsection{Proton spin structure}
\label{sssec:GPD-proton-spin}
In Eqs.~\eqref{eq:GPD-def-q} and \eqref{eq:GPD-def-g}, each GPD is defined for a certain parton
spin structure.  
The form factor decomposition for each GPD corresponds to different proton spin structure.
Following the discussion above \sec{sssec:transverse-boost}, we also describe the proton spin
using light-cone helicity states. 
With the notation $\Gamma_{\alpha, \alpha'} = (2 P^+)^{-1} \bar{u}(p', \alpha') \Gamma u(p,\alpha)$ 
and using the explicit spinor forms in \eq{eq:lc-spinor-u}, we have
\begin{align}
	\bb{ \gamma^+ }_{++} = \bb{ \gamma^+ }_{--} = \sqrt{1 - \xi^2}, & \quad
	\bb{ \gamma^+ }_{+-} = \bb{ \gamma^+ }_{-+} = 0, \nn\\
	\bb{ \gamma^+ \gamma_5 }_{++} = - \bb{ \gamma^+  \gamma_5}_{--} = \sqrt{1 - \xi^2}, &\quad
	\bb{ \gamma^+ \gamma_5 }_{+-} = \bb{ \gamma^+  \gamma_5}_{-+} = 0,
\end{align}
for the helicity non-flipping structures, and
\begin{align}
	& \bb{ \frac{-i \sigma^{+\alpha} \Delta_{\alpha} }{2m} }_{++} 
		= \bb{ \frac{-i \sigma^{+\alpha} \Delta_{\alpha} }{2m} }_{--} 
		= \frac{- \xi^2}{\sqrt{1 - \xi^2}}, \nn\\
	& \bb{ \frac{-i \sigma^{+\alpha} \Delta_{\alpha} }{2m} }_{+-} 
		= -\bb{ \frac{-i \sigma^{+\alpha} \Delta_{\alpha} }{2m} }^*_{-+} 
		= - e^{i \phi_{\Delta}} \cdot \frac{\sqrt{t_0 - t}}{2m}, \nn\\
	& \bb{ \frac{-\gamma_5 \Delta^+ }{2m} }_{++} 
		= -\bb{ \frac{-\gamma_5 \Delta^+ }{2m} }_{--} 
		= \frac{- \xi^2}{\sqrt{1 - \xi^2}}, \nn\\
	& \bb{ \frac{-\gamma_5 \Delta^+ }{2m} }_{+-} 
		= \bb{ \frac{-\gamma_5 \Delta^+ }{2m} }^{*}_{-+} 
		= -\xi \, e^{i \phi_{\Delta}} \cdot \frac{\sqrt{t_0 - t}}{2m},
\end{align}
for the helicity flipping structures, where $t_0 = -4 \xi^2 m^2 / (1 - \xi^2)$ is the maximum value of $t$
at a given $\xi$. 
Here we describe the diffraction of $p \to p'$ by using the azimuthal angle $\phi_{\Delta}$ of $\Delta$.
Making explicit the proton helicity labels in the GPDs in Eqs.~\eqref{eq:GPD-def-q} and \eqref{eq:GPD-def-g} as
$F_{\alpha\alpha'}$ and $\wt{F}_{\alpha\alpha'}$, we have
\begin{align}
	&F_{++} = F_{--} = \sqrt{1 - \xi^2} \bb{ H - \frac{\xi^2}{1 - \xi^2} E}, 
	&&F_{+-} = - F_{-+}^* = - e^{i \phi_{\Delta}} \cdot \frac{\sqrt{t_0 - t}}{2m} \, E, \nn\\
	&\wt{F}_{++} = -\wt{F}_{--} = \sqrt{1 - \xi^2} \bb{ \wt{H} - \frac{\xi^2}{1 - \xi^2} \wt{E}}, 
	&&\wt{F}_{+-} = \wt{F}_{-+}^* = - \xi e^{i \phi_{\Delta}} \cdot \frac{\sqrt{t_0 - t}}{2m} \, \wt{E},
\label{eq:GPD-proton-spin-structure}
\end{align}
which applies to both quark and gluon GPDs.

In this way, the GPDs $H$ and $\wt{H}$ are associated with proton helicity non-flipping channels, whereas
the GPDs $E$ and $\wt{E}$ are with proton helicity flipping ones. 
Since we are dealing with parton helicity non-flipping GPDs, the proton helicity flipping breaks the
light-cone angular momentum conservation by one unit.
This is compensated by the {\it linear} power of a nonzero $\Delta_T$, which is described in the lab frame 
by the phase $e^{i \phi_{\Delta}}$ and the factor
\beq
	\sqrt{t_0 - t} = \sqrt{\frac{1 + \xi}{1 - \xi}} \Delta_T.
\eeq

The information contained in the GPDs $E$ and $\wt{E}$ can only be probed by the exclusive diffractive processes.
In most experiments, the diffracted proton spin is not observed. 
If we only consider the GPD channels,
then the unpolarized proton scattering cross section will depend on $E$ (or $\wt{E}$) through their squares, or their
interference with $H$ and $\wt{H}$, which is, however, suppressed by the small $\xi^2$.
By having the initial-state proton to be transversely polarized, the associated azimuthal asymmetry observables will 
have leading dependence on $E$ (or $\wt{E}$) through its product with the $H$ (or $\wt{H}$) GPDs, 
as we will see in Secs.~\ref{sec:diphoton} and \ref{sec:photoproduction}.
If we also have the $\gamma^*$-mediated channel at $n = 1$, like the Bethe-Heitler process, its interference with
the GPD channels can also offer linear dependence on both $H$ (or $\wt{H}$) and $E$ (or $\wt{E}$), as we will 
see in \sec{sec:dvcs-sdhep}.

\subsection{Moments and sum rules}
\label{ssec:gpd-moments}
Because GPDs only have support in $x \in [-1, 1]$, taking the $x$ moments converts them
into the matrix elements of local twist-2 operators,
\begin{align}\label{eq:x-moments-Fq}
	& \int_{-1}^1 dx \, x^n \, F^q(x, \xi, t) = \int_{-\infty}^{\infty} dx x^n F^q(x, \xi, t)	\nn\\
	& \hspace{2em}
		= \int_{-\infty}^{\infty} dx \int \frac{dy^-}{4\pi} 
	 		\bb{ \pp{ \frac{i}{P^+} \frac{\partial}{\partial y^-} }^n e^{-i x P^+ \, y^-} }
			\langle p', \alpha' |
			\bar{\psi}_{q}\pp{ y^- / 2}
			\gamma^+ \, 
			\psi_{q} \pp{ -y^- / 2}
		| p, \alpha \rangle	\nn\\
	& \hspace{2em}
		= \int \frac{dy^-}{4\pi}  \int_{-\infty}^{\infty} dx \,
			e^{-i x P^+ \, y^-} \pp{ \frac{-i}{P^+} \frac{\partial}{\partial y^-} }^n
			\langle p', \alpha' |
			\bar{\psi}_{q}\pp{ y^- / 2} 
			\gamma^+ \, 
			\psi_{q} \pp{ -y^- / 2}
		| p, \alpha \rangle	\nn\\
	& \hspace{2em}
		= \int \frac{dy^-}{4\pi} (2\pi) \delta(P^+ \, y^-)
			\pp{ \frac{1}{P^+} }^n
			\langle p', \alpha' |
			\bar{\psi}_{q}\pp{ y^- / 2}
			\gamma^+ \, 
			(i \lrp^+)^n
			\psi_{q} \pp{ -y^- / 2}
		| p, \alpha \rangle	\nn\\
	& \hspace{2em}
		= \frac{1}{2(P^+)^{n+1}} 
			\langle p', \alpha' |
			\bar{\psi}_{q}\pp{0}
			\gamma^+ \, 
			(i \lrp^+)^n
			\psi_{q} \pp{0}
		| p, \alpha \rangle,
\end{align}
where $\lrp^+ = (\rp^+ - \lp^+)/2$ will become the covariant derivative
$\lrD^+ = (\rD^+ - \lD^+)/2$ once the Wilson line is included. 
Similar relations apply to the other GPDs,
\begin{align}\label{eq:x-moments-Fq}
	& \int_{-1}^1 dx \, x^n \, \wt{F}^q(x, \xi, t) 
		= \frac{1}{2(P^+)^{n+1}} 
			\langle p', \alpha' |
			\bar{\psi}_{q}\pp{0}
			\gamma^+ \gamma_5 \, 
			(i \lrD^+)^n
			\psi_{q} \pp{0}
		| p, \alpha \rangle,	\nn\\
	& \int_{-1}^1 dx \, x^{n-1} \, F^g(x, \xi, t) 
		= \frac{1}{(P^+)^{n+1}} \delta^{ij}
			\langle p', \alpha' |
			G^{+i}\pp{0}
			(i \lrD_A^+)^{n-1}
			G^{+j}\pp{0}
		| p, \alpha \rangle, \nn\\
	& \int_{-1}^1 dx \, x^{n-1} \, \wt{F}^g(x, \xi, t) 
		= \frac{1}{(P^+)^{n+1}} (-i \epsilon_T^{ij})
			\langle p', \alpha' |
			G^{+i}\pp{0}
			(i \lrD_A^+)^{n-1}
			G^{+j}\pp{0}
		| p, \alpha \rangle,
\end{align}
where we weight the gluon GPDs by $x^{n-1}$ instead of $x^n$ such that the local twist-2 operators have spin $(n+1)$,
similar to the quark case.
	
The off-forward matrix elements of the twist-2 operators can be decomposed into independent form factors based on
Lorentz covariance, parity, and time reversal symmetries. Taking $+$ for all the Lorentz indices then leads to important
polynomiality properties for the GPDs,
\bse\label{eq:H-moments}\begin{align}
	\int_{-1}^1 dx \, x^n \, H^q(x, \xi, t)
		= \sum_{i = 0, 2, \cdots}^n (2\xi)^i \, A^q_{n+1, i}(t) + {\rm mod}(n, 2) (2\xi)^{n+1} \, C^q_{n+1}(t), \\
	\int_{-1}^1 dx \, x^n \, E^q(x, \xi, t)
		= \sum_{i = 0, 2, \cdots}^n (2\xi)^i \, B^q_{n+1, i}(t) - {\rm mod}(n, 2) (2\xi)^{n+1} \, C^q_{n+1}(t), \\
	\int_{-1}^1 dx \, x^{n-1} \, H^g(x, \xi, t)
		= \sum_{i = 0, 2, \cdots}^n (2\xi)^i \, A^g_{n+1, i}(t) + {\rm mod}(n, 2) (2\xi)^{n+1} \, C^g_{n+1}(t), \\
	\int_{-1}^1 dx \, x^{n-1} \, E^g(x, \xi, t)
		= \sum_{i = 0, 2, \cdots}^n (2\xi)^i \, B^g_{n+1, i}(t) - {\rm mod}(n, 2) (2\xi)^{n+1} \, C^g_{n+1}(t), 
\end{align}\ese
for the unpolarized GPDs, and
\bse\label{eq:Ht-moments}\begin{align}
	\int_{-1}^1 dx \, x^n \, \wt{H}^q(x, \xi, t)
		= \sum_{i = 0, 2, \cdots}^n (2\xi)^i \, \wt{A}^q_{n+1, i}(t), \\
	\int_{-1}^1 dx \, x^n \, \wt{E}^q(x, \xi, t)
		= \sum_{i = 0, 2, \cdots}^n (2\xi)^i \, \wt{B}^q_{n+1, i}(t), \\
	\int_{-1}^1 dx \, x^{n-1} \, \wt{H}^g(x, \xi, t)
		= \sum_{i = 0, 2, \cdots}^n (2\xi)^i \, \wt{A}^g_{n+1, i}(t), \\
	\int_{-1}^1 dx \, x^{n-1} \, \wt{E}^g(x, \xi, t)
		= \sum_{i = 0, 2, \cdots}^n (2\xi)^i \, \wt{B}^g_{n+1, i}(t).
\end{align}\ese
That is, the $x$ moments of GPDs reduce to even polynomials of $\xi$. 
For unpolarized GPDs, the maximum power of $\xi$ is equal to the spin of the corresponding twist-2 operator,
whereas for polarized GPDs, it is the spin minus 1.

The low-order moments are related to the matrix elements of physical currents that can be probed in experiments.
For $n = 0$, the twist-2 quark operators become the electric and axial currents,
\beq
	J^{\mu}_q = \bar{\psi}_q \gamma^{\mu} \psi_q, \quad
	J^{5\mu}_q = \bar{\psi} \gamma^{\mu} \gamma_5 \psi,
\eeq
which can be accessed experimentally through electromagnetic and weak interactions, respectively,
giving the Dirac and Pauli form factors,
\beq
	\la p' | J^{\mu}_q(0) | p \ra 
		= \bar{u}(p') 
			\bb{ F_1^q(t) \, \gamma^{\mu} - 
				F_2^q(t) \, \frac{i \, \sigma^{\mu \alpha} \Delta_{\alpha}}{2m}
			} u(p),
\eeq
and the axial and pseudoscalar form factors,
\beq
	\la p' | J^{5\mu}_q(0) | p \ra 
		= \bar{u}(p') 
			\bb{ g_A^q(t) \, \gamma^{\mu} \gamma_5 - 
				g_P^q(t) \, \frac{\gamma_5 \Delta^{\mu}}{2m}
			} u(p).
\eeq
Taking $\mu = +$ then relates them to the form factors of GPDs in 
Eqs.~\eqref{eq:H-moments} and \eqref{eq:Ht-moments},
\begin{align}
	A_{1, 0}^q(t) &= \int_{-1}^1 dx \, H^q(x, \xi, t) = F_1^q(t), \quad
	&&B_{1, 0}^q(t) = \int_{-1}^1 dx \, E^q(x, \xi, t) = F_2^q(t), \nn\\
	\wt{A}_{1, 0}^q(t) &= \int_{-1}^1 dx \, \wt{H}^q(x, \xi, t) = g_A^q(t), \quad
	&&\wt{B}_{1, 0}^q(t) = \int_{-1}^1 dx \, \wt{E}^q(x, \xi, t) = g_P^q(t).
\label{eq:1st-moments-form-factor}
\end{align}
Since twist-2 gluon operators start from spin 2, there are no corresponding relations for gluon GPDs.

For $n = 1$, the spin-2 twist-2 operators are just the energy momentum tensor~\citep{Polyakov:2018zvc},
\begin{align}
	\int_{-1}^1 dx \, x\, F^q(x, \xi, t) &= \frac{1}{2(P^+)^2} 
		\langle p', \alpha' |
			\bar{\psi}_{q}\pp{0}
			\gamma^+ \, 
			(i \lrD^+)
			\psi_{q} \pp{0}
		| p, \alpha \rangle	\nn\\
	& = \frac{1}{2(P^+)^2} \langle p', \alpha' | T^{++}_q(0) | p, \alpha \rangle,	\nn\\
	\int_{-1}^1 dx \, F^g(x, \xi, t) &= \frac{1}{(P^+)^2} 
		\langle p', \alpha' |
			G^{+\mu}\pp{0}
			G_{\mu}{}^{+}\pp{0}
		| p, \alpha \rangle	\nn\\
	& = \frac{1}{(P^+)^2} \langle p', \alpha' | T^{++}_g(0) | p, \alpha \rangle,
	\label{eq:2nd-moments-EMT}
\end{align}
where the energy momentum tensors are
\begin{align}
	T^{\mu\nu}_q &= \frac{1}{2} \bar{\psi}_q \pp{i \, \lrD^{\mu} \gamma^{\nu} + i \, \lrD^{\nu} \gamma^{\mu} } \psi_q
		- g^{\mu\nu} \bar{\psi}_q \pp{ i \gamma \cdot \lrD - m_q } \psi_q, \nn\\
	T^{\mu\nu}_g &= G^{a, \mu\rho} G^{a}{}_{\rho}{}^{\nu} + \frac{1}{4} g^{\mu\nu} (G^a_{\rho\sigma})^2.
\end{align}
The latter can be decomposed into the so-called gravitational form factors,
\begin{align}
	\langle p', \alpha' | T^{\mu\nu}_i(0) | p, \alpha \rangle
	& \, = \bar{u}(p', \alpha') 
			\left[ A_i(t) \, \frac{\gamma^{(\mu} P^{\nu)}}{2} - B_i(t) \, \frac{i P^{(\mu} \sigma^{\nu) \rho} \Delta_\rho}{4 m}
				\right.\nn\\
	& \hspace{7em} \left. 
			+ D_i(t) \, \frac{\Delta^\mu \Delta^\nu - g^{\mu \nu} \Delta^2}{4 m}
			+m \, \bar{c}_i(t) \, g^{\mu \nu}\right] u(p, \alpha),
\end{align}
where $i = q, g$, and we used the notation $a^{(\mu} \, b^{\nu)} = a^{\mu} \, b^{\nu} + a^{\nu} \, b^{\mu}$.
Taking $\mu = \nu = +$ gives
\begin{align}
	\langle p', \alpha' | T^{++}_i(0) | p, \alpha \rangle
	& \, = P^+ \, \bar{u}(p', \alpha') 
			\left[ \pp{ A_i(t) + \xi^2 D_i(t) } \gamma^+ 
			\right.\nn\\
	& \hspace{8em} \left. 
				- \pp{ B_i(t) - \xi^2 D_i(t) } \frac{i \sigma^{+ \rho} \Delta_\rho}{2 m}
			\right] u(p, \alpha).
\label{eq:T++-form-factors}
\end{align}
Comparing this with Eqs.~\eqref{eq:2nd-moments-EMT}\eqref{eq:GPD-def-F-q} and \eqref{eq:GPD-def-F-g}, 
we have the sum rules,
\begin{align}\label{eq:GPD-moments-GFF}
	&\int_{-1}^1 dx \, x\, H^q(x, \xi, t) = A_q(t) + \xi^2 \, D_q(t),
	&&\int_{-1}^1 dx \, x\, E^q(x, \xi, t) = B_q(t) - \xi^2 \, D_q(t), \nn\\
	&\int_{-1}^1 dx \, H^g(x, \xi, t) = 2 \pp{A_g(t) + \xi^2 \, D_g(t) },
	&&\int_{-1}^1 dx \, E^g(x, \xi, t) = 2 \pp{B_g(t) - \xi^2 \, D_g(t) },
\end{align}
which relate the gravitational form factors to the GPD moments.
While the former cannot be easily measured in experiments, the latter can in principle be measured 
(or calculated in lattice QCD) and give a probe to the energy momentum tensor. 
This can uncover certain global dynamic properties inside the proton. 

In the case of PDFs, $p = p'$, $\Delta = 0$, and $\xi = t = 0$. Then \eq{eq:T++-form-factors} reduces only to the $A_i$ factor,
\begin{align}
	\langle p, \alpha' | T^{++}_i(0) | p, \alpha \rangle
	& \, = p^+ \, A_i(0) \, \bar{u}(p, \alpha') \gamma^+  u(p, \alpha)
	= 2 (p^+)^2 \, A_i(0) \, \delta_{\alpha\alpha'},
\label{eq:T++-form-factors-PDF}
\end{align}
and \eq{eq:GPD-moments-GFF} reduces to the forward limit which only gives access to the total momentum fraction $A_i(0)$ of each parton flavor,
\begin{align}\label{eq:PDF-moments-EMT}
	&\int_{-1}^1 dx \, x\, f^q(x)  = \int_{0}^1 dx \, x\, \pp{ f^q(x) + f^{\bar{q}}(x) } = A_q(0), \nn\\
	&\int_{-1}^1 dx \, x \, f^g(x) = 2 \int_0^1 dx \, x \, f^g(x) = 2 A_g(0).
\end{align}
Accessing the same operators as the PDFs, the GPDs have the capability of probing the gravitational form factors because 
they are associated with off-forward kinematics, which opens up the dependence on $\xi$ and $t$ and the related form factors.

By combining the moments of $H$ and $E$ in \eq{eq:GPD-moments-GFF}, the $D$ terms cancel and we get
the sum rule,
\begin{align}
	&\int_{-1}^1 dx \, x\, \pp{ H^q(x, \xi, t) + E^q(x, \xi, t) }
		= A_q(t) + B_q(t)
		= 2 J_q(t),	\nn\\
	&\int_{-1}^1 dx \, \pp{ H^g(x, \xi, t) + E^g(x, \xi, t) }
		= 2 \pp{ A_g(t) + B_g(t) }
		= 4 J_g(t),
	\label{eq:A-B-J}
\end{align}
where the $J_a(t)$ form factor is related to the angular momentum sum of the parton $a$, normalized by
\beq
	\sum_a J_a(0) = \frac{1}{2}.
\eeq
\eq{eq:A-B-J} then gives the angular momentum sum rules for partons inside the proton~\citep{Ji:1996ek},
\begin{align}
	J_q &= \frac{1}{2} \lim_{t\to 0} \int_{-1}^1 dx \, x\, \big( H^q(x, \xi, t) + E^q(x, \xi, t) \big), \nn\\
	J_g &= \frac{1}{4} \lim_{t\to 0} \int_{-1}^1 dx \, \big( H^g(x, \xi, t) + E^g(x, \xi, t) \big).
	\label{eq:ang-momentum-sum-rule}
\end{align}
Therefore, the measurement of GPDs, especially the construction of the moments of their $x$ distributions,
gives important handles to the partonic dynamics inside a hadron.

\subsection{Two-scale nature and hadron tomography}
\label{ssec:tomography}
The relation of GPDs to form factors [\eq{eq:1st-moments-form-factor}] 
encodes a great aspect of internal hadron structures, as pointed out by~\citep{Burkardt:2000za, Burkardt:2002hr}.
On the one hand, \eq{eq:1st-moments-form-factor} can be viewed as a decomposition of form factors in the $x$ space. In the limit of $\xi \to 0$,
$x$ is the momentum fraction of the active quark. Since the Fourier transform of the form factors with respect to $t$ gives a spatial distribution
of the quarks, the Fourier transform of the GPDs with respect to $t$ in the limit of $\xi \to 0$ should give a decomposition of the spatial image 
of the partons in the $x$ space.
On the other hand, in the forward limit, the electric current operator projects out the electric charge $Q_p$ of the proton, while 
in the off-forward case, the extra $t$ dependence maps out a spatial distribution of that quantum number. 
Similarly, since the (non-local light-cone) GPD operator projects out the $x$ distribution in the forward limit, i.e., the PDF,
it should further map out a spatial distribution of the PDF for each given value of $x$ in the off-forward case. 
To put it more formally, we expect
\beq[eq:gpd-tomography]
	f(x, \bm{b}_T) 
	= \int \frac{d^2 \bm{\Delta}_T}{(2\pi)^2} \, e^{i \bm{\Delta}_T \cdot \bm{b}_T} \,
		H(x, \xi = 0, t = -\Delta_T^2)
\eeq
to be the parton number density in the three-dimensional space of the longitudinal momentum fraction $x$ 
and transverse spatial position $\bm{b}_T$. 
Now we give a detailed derivation of \eq{eq:gpd-tomography}.

To discuss the transverse spatial distribution, we need to first localize the proton states in the coordinate space.
The easiest way is to construct a transverse position eigenstate from the light-cone helicity state in the momentum space
as \eq{eq:light-cone-helicity-state},
\beq
	| p^+, \bm{b}_T, \lambda \rangle
		= \int \frac{d^2\bm{p}_T}{(2\pi)^2} \, e^{-i \bm{p}_T \cdot \bm{b}_T} \, | p^+, \bm{p}_T, \lambda \rangle,
\eeq
which is normalized by
\beq
	\langle p^{\prime +}, \bm{b}'_T, \lambda' | p^+, \bm{b}_T, \lambda \rangle
		= 2\pi \, (2p^+) \, \delta_{\lambda \lambda'} \, \delta(p^+ - p^{\prime +}) \, \delta^{(2)}( \bm{b}_T - \bm{b}'_T ).
\eeq
Such state can be created by the operator
\beq
	b^{\dag}(k^+, \bm{b}_T, \lambda)
		 = \int \frac{d^2\bm{k}_T}{(2\pi)^2} \, e^{-i \bm{k}_T \cdot \bm{b}_T} \, b^{\dag}(k^+, \bm{k}_T, \lambda),
\eeq
with the commutation relation,
\begin{align}
	\big\{ b(k^+, \bm{b}_T, \lambda), b^{\dag}(k^{\prime +}, \bm{b}'_T, \lambda') \big\}
		= 2\pi \, \delta_{\lambda \lambda'} \, \delta(k^+ - k^{\prime +}) \, \delta^{(2)}( \bm{b}_T - \bm{b}'_T ),	
	\nn\\
	\big\{ b(k^+, \bm{b}_T, \lambda), b(k^{\prime +}, \bm{b}'_T, \lambda') \big\}
		= \big\{ b^{\dag}(k^+, \bm{b}_T, \lambda), b^{\dag}(k^{\prime +}, \bm{b}'_T, \lambda') \big\}
		= 0.
\end{align}
From this one can define a parton number operator
\beq
	\hat{N} = \sum\nolimits_{\lambda} \int \frac{d k^+ d^2 \bm{b}_T}{2\pi} \, b^{\dag}(k^+, \bm{b}_T, \lambda) \, b(k^+, \bm{b}_T, \lambda),
\eeq
which is normalized properly to give a particle number interpretation since
\beq[eq:commute-N]
	[ b(k^+, \bm{b}_T, \lambda), \hat{N} ] = b(k^+, \bm{b}_T, \lambda), \quad
	[ b^{\dag}(k^+, \bm{b}_T, \lambda), \hat{N} ] = -b^{\dag}(k^+, \bm{b}_T, \lambda).
\eeq
Then the number density operator for partons with light-cone helicity $\lambda$ and longitudinal momentum fraction $x$
while situated at the transverse distance $\bm{b}_T$ from the proton center is,
\beq
	\frac{d\hat{N}}{dx \, d^2\bm{b}_T} = \frac{P^+}{2\pi} \, b^{\dag}(k^+, \bm{b}_T, \lambda) \, b(k^+, \bm{b}_T, \lambda),
\eeq
where $x = k^+ / P^+$ with $P^+$ the proton momentum.

With all these defined, the parton density in the $x$ and $\bm{b}_T$ space is
\beq[eq:f-3D-0]
	f(x, \bm{b}_T)
		= \frac{P^+}{2\pi} \, \frac{\langle P^+, \bm{0}_T, \alpha' | 
			b^{\dag}(k^+, \bm{b}_T, \lambda') \, b(k^+, \bm{b}_T, \lambda)
			| P^+, \bm{0}_T, \alpha \rangle
			}{ \langle P^+, \bm{0}_T | P^+, \bm{0}_T \rangle},
\eeq
where we also allow off-diagonal parton and proton helicities,
and the infinite normalization in the denominator can formally resolve the infinity in the numerator,
\beq
	\langle P^+, \bm{0}_T | P^+, \bm{0}_T \rangle
	= 2\pi \, (2 P^+) \delta(0^+) \delta^{(2)}(\bm{0}_T)
	= 2 P^+ \int dx^- \int \frac{d^2\bm{p}_T}{(2\pi)^2}.
\eeq
The parton creation and annihilation operators can be related to the fermion fields through
\begin{align}
	b(k^+, \bm{b}_T, \lambda) 
		&\, = \frac{1}{\sqrt{2k^+}} \int dx^- \, e^{i k^+ x^-} \bar{u}(k^+, \lambda) \gamma^+ \psi_G(x^-, \bm{b}_T), \nn\\
	b^{\dag}(k^+, \bm{b}_T, \lambda)
		&\, = \frac{1}{\sqrt{2k^+}} \int dx^- \, e^{-i k^+ x^-} \bar{\psi}_G(x^-, \bm{b}_T) \gamma^+ u(k^+, \lambda),
\label{eq:extracting-b-operator}
\end{align}
where the ``light-cone time'' $x^+$ is taken to 0.
The numerator in \eq{eq:f-3D-0} has the expression
\begin{align}
	&\langle P^+, \bm{0}_T, \alpha' | O(x^-, \bm{b}_T) | P^+, \bm{0}_T, \alpha \rangle	
		= \int \frac{d^2 \bm{p}_T d^2 \bm{p}'_T}{(2\pi)^4} 
			\langle P^+, \bm{p}'_T, \alpha' | O(x^-, \bm{b}_T) | P^+, \bm{p}_T, \alpha \rangle
	\nn\\
	&\hspace{2em} 
		= \int \frac{d^2 \bm{p}_T d^2 \bm{p}'_T}{(2\pi)^4} \, e^{-i(\bm{p}'_T - \bm{p}_T) \cdot \bm{b}_T} \,
			\langle P^+, \bm{p}'_T, \alpha' | O(0^-, \bm{0}_T) | P^+, \bm{p}_T, \alpha \rangle,
\end{align}
where the operator $O(x^-, \bm{b}_T)$ is at the coordinate $(0^+, x^-, \bm{b}_T)$.
Then inserting \eq{eq:extracting-b-operator} to \eq{eq:f-3D-0} gives the numerator 
\begin{align}
	\langle \cdots \rangle
	& \, = \frac{1}{2k^+} \int \frac{d^2 \bm{p}_T d^2 \bm{p}'_T}{(2\pi)^4} \, e^{-i(\bm{p}'_T - \bm{p}_T) \cdot \bm{b}_T}
		\int dx^- dx^{\prime -} e^{-i k^+(x - x')^-}
	\nn\\
	& \hspace{3em} \times
		\langle P^+, \bm{p}'_T, \alpha' | 
			\bar{\psi}_G(x^- - x^{\prime-}, \bm{0}_T) \gamma^+ u(k^+, \lambda')
			\bar{u}(k^+, \lambda) \gamma^+ \psi_G(0, \bm{0}_T)
		| P^+, \bm{p}_T, \alpha \rangle
	\nn\\
	& \, = \bb{ \int \frac{d^2 \bm{p}'_T}{(2\pi)^2} \int dx^{\prime-}}
		\frac{1}{2k^+} 
		\int \frac{d^2 \bm{\Delta}_T}{(2\pi)^2} e^{i \bm{\Delta}_T \cdot \bm{b}_T}
		\int dx^- e^{-i k^+ x^-}
	\nn\\
	& \hspace{3em} \times
		\langle P^+, -\bm{\Delta}_T, \alpha' | 
			\bar{\psi}_G(x^-, \bm{0}_T) \gamma^+ u(k^+, \lambda')
			\bar{u}(k^+, \lambda) \gamma^+ \psi_G(0, \bm{0}_T)
		| P^+, \bm{0}_T, \alpha \rangle,
\label{eq:tomography-num}
\end{align}	
where in the second step we performed a transverse boost to set the $p_T$ of the initial proton state to zero.
The operator is left invariant under such a boost because 
only the minus coordinate component is nonzero and the Lorentz index of the Dirac matrix is plus.
Now we make a simplification by only considering the diagonal parton helicity elements, $\lambda = \lambda' = \pm 1/2$;
the off-diagonal ones are related to transversity GPDs beyond the scope of our discussion.
Then we can use
\beq
	u(k^+, \pm) \bar{u}(k^+, \pm)
	= P_{\pm} \sum\nolimits_{\lambda} u(k^+, \lambda) \bar{u}(k^+, \lambda)
	= k^+ \, P_{\pm} \gamma^-,
\eeq
where $P_{\pm} = (1 \pm \gamma_5) / 2$.
Inserting this back to \eq{eq:tomography-num}, with the other factors included in \eq{eq:f-3D-0} gives
\begin{align}
	&f^{\pm}_{\alpha\alpha'}(x, \bm{b}_T)
	= \int \frac{d^2 \bm{\Delta}_T}{(2\pi)^2} e^{i \bm{\Delta}_T \cdot \bm{b}_T}
	\nn\\
	&\hspace{6.5em}\times
		\int \frac{dx^-}{4\pi} e^{-i k^+ x^-} 
		\langle P^+, -\bm{\Delta}_T, \alpha' | 
			\bar{\psi}(x^-, \bm{0}_T) \gamma^+ P_{\pm} \psi(0, \bm{0}_T)
		| P^+, \bm{0}_T, \alpha \rangle.
\label{eq:f-3D-1}
\end{align}
The second line in \eq{eq:f-3D-1} is exactly the GPD at $\xi = 0$, but we have allowed for arbitrary parton and proton helicities.

For a general proton spin state described by the spin vector $\bm{s} = (s_1, s_2, \lambda_p)$, we have the 
three-dimensional quark number density,
\beq
	f(x, \bm{b}_T) = \int \frac{d^2 \bm{\Delta}_T}{(2\pi)^2} \, e^{i \bm{\Delta}_T \cdot \bm{b}_T}
		\, \rho_{\alpha\alpha'}(\bm{s}) F_{\alpha\alpha'}(x, -\Delta_T^2),
\eeq
and helicity density,
\beq
	\Delta f(x, \bm{b}_T) = \int \frac{d^2 \bm{\Delta}_T}{(2\pi)^2} \, e^{i \bm{\Delta}_T \cdot \bm{b}_T}
		\, \rho_{\alpha\alpha'}(\bm{s}) \wt{F}_{\alpha\alpha'}(x, -\Delta_T^2),
\eeq
where $\rho(\bm{s}) = (1 + \bm{\sigma}\cdot \bm{s}) / 2$ is the proton spin density matrix 
and $(F, \wt{F})_{\alpha\alpha'}(x, -\Delta_T^2) \equiv (F, \wt{F})_{\alpha\alpha'}(x, \xi = 0, t = -\Delta_T^2)$
are the GPDs defined in \eq{eq:GPD-def-q} at $\xi = 0$ and with the subscripts denoting the proton helicities.
According to \eq{eq:GPD-proton-spin-structure}, we have
\begin{align}
	F_{++} &= F_{--} = H(x, 0, -\Delta_T^2), &
	F_{+-} &= -F_{-+}^* = - e^{i \phi_{\Delta}} \frac{\Delta_T}{2m} E(x, 0, -\Delta_T^2), \nn\\
	\wt{F}_{++} &= -\wt{F}_{--} = \wt{H}(x, 0, -\Delta_T^2), &
	\wt{F}_{+-} &= \wt{F}^*_{-+} = 0,
\end{align}
where $\Delta_T = \abs{\bm{\Delta}_T}$ for definiteness. This then gives
\begin{align}
	f(x, \bm{b}_T) &\, = \int \frac{d^2 \bm{\Delta}_T}{(2\pi)^2} \, e^{i \bm{\Delta}_T \cdot \bm{b}_T}	
		\bb{ H(x, 0, -\Delta_T^2) - \frac{i}{2m} (s_1 \Delta_2 - s_2 \Delta_1) E(x, 0, -\Delta_T^2) }
	\nn\\
	&\, = f_H(x, \bm{b}_T) - \frac{1}{2m} \bb{ \bm{s}_T \times \nabla_{\bm{b}_T} f_E(x, \bm{b}_T) }^z
\end{align}
for the unpolarized parton density,
where $f_{H, E}$ are the Fourier transforms of $H$ and $E$, respectively. Note that 
both $H$ and $E$ are rotationally invariant with respect to $\bm{\Delta}_T$, so 
$f_{H, E}(x, \bm{b}_T) = f_{H, E}(x, b^2_T)$ and 
\beq
	\nabla_{\bm{b}_T} f_E(x, \bm{b}_T)
	= \pp{ \nabla_{\bm{b}_T} b_T^2 } \frac{\partial f_E(x, b^2_T)}{\partial b_T^2}
	= 2 \bm{b}_T \frac{\partial f_E(x, b^2_T)}{\partial b_T^2}.
\eeq
Hence, 
\beq
	f(x, \bm{b}_T) 
	= f_H(x, b^2_T) - \frac{\pp{ \bm{s}_T \times \bm{b}_T }^z}{m} \frac{\partial f_E(x, b^2_T)}{\partial b_T^2},
\eeq
so that the unpolarized proton contains an isotropic parton density in the transverse plane whereas
a nonzero transverse spin $\bm{s}_T$ yields an asymmetry in the perpendicular direction, 
whose magnitude is quantified by the gradient of the GPD $f_E$ with respect to $b_T^2$.
The parton helicity distribution, on the other hand, is only governed by the GPD $\wt{H}$,
\beq[eq:tomography-fs]
	\Delta f(x, \bm{b}_T) = \lambda_p \int \frac{d^2 \bm{\Delta}_T}{(2\pi)^2} \, e^{i \bm{\Delta}_T \cdot \bm{b}_T}
		\wt{H}(x, 0, -\Delta_T^2)
	= \lambda_p \, f_{\wt{H}}(x, b^2_T),
\eeq
which is isotropic and has no distortion from the proton's transverse spin.

The derivation from \eq{eq:f-3D-0} to \eq{eq:tomography-fs} is general and applies to the antiquark and gluon cases as well.
The generalization to transversely polarized partons is also straightforward. 
In this way, we showed that the GPDs encode transverse spatial parton images at each slice of the 
longitudinal momentum fraction $x$. This is sometimes referred to as QCD tomography, or proton imaging.
Compared with PDFs, such tomography is due to the extra scale $t$, which is a nonperturbative soft scale. 
In the SDHEPs (and other processes) for probing GPDs, one has two separated scales ---
one hard scale $Q = q_T$ that localizes the interaction to identify partons,
and one soft scale $t$ that gives an extra handle for probing the confined parton images.
Such two-scale property is also true for TMDs, where one also has a hard scale $Q$ for probing parton degrees of freedom,
but instead of $t$, one has a low transverse momentum scale $Q_T \ll Q$ to probe the (nonperturbative) 
transverse partonic motion. TMDs and GPDs thus provide complementary information about the parton dynamics.

\section{The SDHEP kinematics and observables}
\label{sec:SDHEP-frame}

\begin{figure}[htbp]
	\centering
		\includegraphics[scale=0.8]{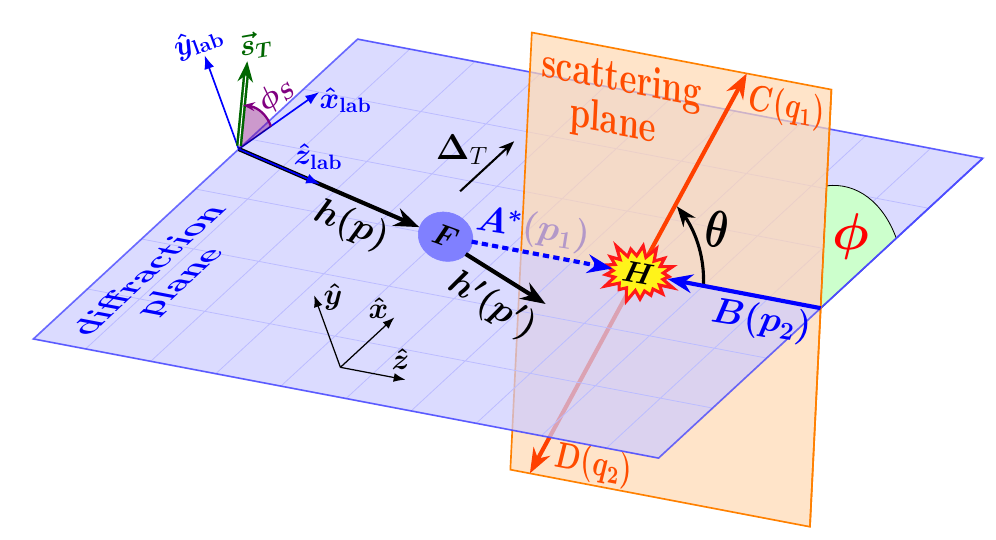}
	\caption{
		The two frames for describing the SDHEP process. 
		The Lab frame $\hat{x}_{\rm Lab}$-$\hat{y}_{\rm Lab}$-$\hat{z}_{\rm Lab}$
		is the c.m.~frame of the colliding $h$ and $B$ beams, with $\hat{z}_{\rm Lab}$ along the direction of $h$
		and $\hat{x}_{\rm Lab}$ along the diffraction direction, $\bm{\Delta}_T$.
		The SDHEP frame $\hat{x}$-$\hat{y}$-$\hat{z}$ is the c.m.~frame of $A^*$ and $B$, 
		with $\hat{z}$ along the $A^*$ momentum direction and $\hat{x}$ on the diffraction plane along the $\bm{\Delta}_T$ in the Lab frame. 
		$F$ denotes the (nonperturbative) diffraction process $h\to h' + A^*$, which happens in the blue plane (``diffraction plane"), 
		and $H$ denotes the hard interaction between $A^*$ and $B$ to produce $C$ and $D$, 
		which happens in the orange plane (``scattering plane"). The two planes form an angle of $\phi$ 
		and intersect at the collision axis between $A^*$ and $B$. 
	}
\label{fig:sdhep-frame}
\end{figure}

By the two-stage paradigm described at the beginning of \sec{sec:SDHEP}, we introduce a two-step description
of the SDHEP kinematics. 
First, the Lab frame is naturally chosen as the c.m.~frame of the colliding beams $h(p)$ and $B(p_2)$, with 
$h$ along the $\hat{z}_{\rm Lab}$ direction. 
The kinematics of the diffraction subprocess [\eq{eq:diffractive}] is fully captured by the momentum transfer vector
$\Delta = p_1 = p - p'$ through its invariant mass $t = \Delta^2$, azimuthal angle $\phi_{\Delta}$, and the variable $\xi$
related to its leading momentum component, $\xi = \Delta^+ / (p + p')^+$. 
Once the diffraction subprocess (and thereby the c.m.~energy of the $2\to2$ hard scattering subprocess [\eq{eq:hard 2to2}]) 
is determined, we perform a Lorentz transformation to the SDHEP frame to describe the $2\to2$ subprocess, 
defined as the c.m.~frame of $A^*B$ with $A^*$ along the $\hat{z}$ axis, 
in terms of the polar and azimuthal angles $(\theta, \phi)$ of $C(q_1)$ (or $D(q_2)$).
This is shown in \fig{fig:sdhep-frame}, where the diffraction subprocess [\eq{eq:diffractive}] happens in the blue plane (diffraction plane), 
and the hard scattering subprocess [\eq{eq:hard 2to2}] happens in the orange plane (scattering plane). 
The $\hat{x}$ axis lies on the diffraction plane and points to the same direction as $\bm{\Delta}_T$ in the Lab frame, 
and the $\hat{y}$ axis is perpendicular to the diffraction plane, determined by the right-hand rule. 

In a fixed Lab frame coordinate system, the $\phi_{\Delta}$ distribution is nontrivial only 
when $h$ has a transverse spin to break the azimuthally rotational invariance.
To simplify the kinematic description, we choose the $\hat{x}_{\rm Lab}$ axis of the Lab frame to be along the 
diffraction direction, $\hat{x}_{\rm Lab} = \bm{\Delta}_T / \Delta_T$, and $\hat{y}_{\rm Lab} = \hat{z}_{\rm Lab} \times \hat{x}_{\rm Lab}$
is determined accordingly. This choice varies event by event, and trades the azimuthal angle $\phi_{\Delta}$ of the diffraction in a fixed Lab
frame for the azimuthal angle $\phi_S$ of the transverse spin $\bm{s}_T$ of $h$ in the varying $\hat{x}_{\rm Lab}$-$\hat{y}_{\rm Lab}$ frame.
This is also illustrated in \fig{fig:sdhep-frame} but it needs to be emphasized that the 
$\hat{x}_{\rm Lab}$-$\hat{y}_{\rm Lab}$-$\hat{z}_{\rm Lab}$ coordinate system together with $\bm{s}_T$ and $\phi_S$
should be considered to be in the Lab frame, not in the c.m.~frame of the hard scattering (that is, the SDHEP frame).

The c.m.~energy square of the $2\to2$ hard scattering is
\beq[eq:sdhep-shat]
	\hat{s} = (\Delta + p_2)^2 = t + m_2^2 + 2 (\Delta^+ p_2^- + \Delta^- p_2^+) 
		= t + m_2^2 + 2 \Delta^+ p_2^- + \frac{(t + \Delta_T^2) m_2^2}{2 \Delta^+ p_2^-},
\eeq
which is written in terms of the kinematic variables in the Lab frame. 
Here $m_2$ is the mass of the beam particle $B$, which may be a lepton, photon, or light meson.
As elaborated in \ch{ch:exclusive}, 
the factorizability into GPD (and light meson DA) necessarily requires
$\hat{s} \gtrsim q_T^2 \gg t, m_2^2$, so that we can approximate $\hat{s}$ as
\beq[eq:sdhep-shat-approx]
	\hat{s} \simeq 2 \Delta^+ p_2^-,
\eeq
up to error of order $\order{(t, m_2^2)/q_T^2}$.
This can be related to the overall c.m.~energy square,
\beq[eq:sdhep-s-approx]
	s = (p + p_2)^2 = m^2 + m_2^2 + 2p^+p_2^- + \frac{m^2 m_2^2}{2p^+p_2^-}
		= 2p^+p_2^- + \order{\frac{m^2, m_2^2}{s}},
\eeq 
by
\beq[eq:sdhep-s-hat-approx]
	\hat{s} \simeq 2\pp{\frac{2 \, \xi}{1 + \xi}p^+}p_2^- \simeq \frac{2 \, \xi}{1 + \xi} s.
\eeq

Going from the Lab frame to the SDHEP frame is most easily done when we neglect the mass of particle $B$ 
(which is within the error of factorization) and approximate its momentum by $\hat{p}_2 = (0, p_2^-, 0_T) = p_2^- \, n$. 
In both frames, this momentum is along the minus light-cone direction, so that
the Lorentz transformation connecting the two frames can be constructed as a Lorentz transformation $S$
that leaves the lightlike vector $n$ invariant, followed by a boost along $\hat{z}$ to enter the c.m.~frame of $A^* B$.
Since an arbitrary vector $r$ transforms under $S$ with its plus component unchanged, 
$(Sr)^+ = (Sr) \cdot n = (Sr) \cdot (Sn) = r \cdot n = r^+$,
$S$ is exactly a transverse boost introduced in \eq{eq:trans-boost}, which can be parametrized by a two-dimensional 
transverse vector $\bm{v}_T$. 
By requiring $S$ to transform $\Delta = (\Delta^+, (t + \Delta_T^2)/2\Delta^+, (\Delta_T, 0))$ in the Lab frame
to $(S\Delta) = (\Delta^+, t/2\Delta^+, 0_T)$ in the SDHEP frame gives the solution
$\bm{v}_T = \pp{-\Delta_T/\sqrt{2} \Delta^+, 0}$, where we have used the fact that the $\hat{x}_{\rm lab}$ is chosen along the 
direction of $\bm{\Delta}_T$.
Therefore, the transformation $S$ takes any vector $r^{\mu}$ to
\beq[eq:lorentz-trans-sdhep]
	r = (r^+, r^-, r_x, r_y) \to 
	S \cdot r = \pp{ r^+, r^- - r_x \frac{\Delta_T}{\Delta^+} + \frac{r^+}{2} \pp{ \frac{\Delta_T}{\Delta^+} }^2, r_x - \Delta_T \frac{r^+}{\Delta^+}, r_y},
\eeq
where we have written explicitly $\bm{r}_T = (r_x, r_y)$. 
Following $S$, we may approximate $\Delta$ by $\hat{\Delta} = \Delta^+ \bar{n}$ 
and perform a trivial boost along $\hat{z}$ to make $\hat{\Delta}$ and $\hat{p}_2$ have the same energy.

The fact that it is the transverse boost that connects the Lab and the SDHEP frame means that
GPD definitions in Eqs.~\eqref{eq:GPD-def-q} and \eqref{eq:GPD-def-g} are the same in both frames,
so are the GPDs $H$, $\wt{H}$, $E$, and $\wt{E}$. 
As a result, one can use the same factorization formulas in the two frames, making the use of the
SDHEP frame rather convenient for describing the $2\to2$ hard scattering subprocess. 
We note that the transformation from the Lab to the SDHEP frame can also be achieved by 
boosting along $-\bm{p}'$~\citep{Berger:2001xd}, but expressing in terms of the transverse boost elucidates the 
light-cone kinematics more clearly.

Now we give the differential cross section formula for SDHEPs in terms of the two-stage description. 
Given that the amplitude $\M$ can be written as convolutions of GPDs and hard coefficients equally in both frames,
the cross section $\sigma$ is 
\beq
	\sigma = \frac{1}{2s} 
		\int \frac{d^3\bm{p'}}{(2\pi)^3 2 E_{p'}} \frac{d^3\bm{q_1}}{(2\pi)^3 2 E_{q_1}} \frac{d^3\bm{q_2}}{(2\pi)^3 2 E_{q_2}}
		\, \overline{|\M|^2}
		\, (2\pi)^4 \delta^{(4)}(p + p_2 - p' - q_1 - q_2),
\eeq
where $\overline{|\M|^2}$ is the amplitude squared, averaged over the initial-state spins with corresponding density matrices
and summed over final-state spins. Rewriting $d^3\bm{p'} / (2 E_{p'})$ as $d^4p' \, \delta^+(p'^2 - m^2)$ and 
trading $p'$ for $\Delta$ gives
\begin{align}\label{eq:sdhep-xsec-0}
	\sigma = &\, \frac{1}{2s} 
		\int \frac{d^4\Delta}{(2\pi)^3} \delta^+((p - \Delta)^2 - m^2)	\nn\\
		& \hspace{2em} \times \int \frac{d^3\bm{q_1}}{(2\pi)^3 2 E_{q_1}} \frac{d^3\bm{q_2}}{(2\pi)^3 2 E_{q_2}}
		\, \overline{|\M|^2}
		\, (2\pi)^4 \delta^{(4)}(\Delta + p_2 - q_1 - q_2),
\end{align}
where the first line describes the diffraction kinematics while the second line is for the $2\to2$ hard scattering.
Taking advantage of the Lorentz invariance of each integration measure, the delta functions, and the amplitude squared,
we express the first line in the Lab frame and the second line in the SDHEP frame. This gives
\begin{align}\label{eq:ps-diffraction}
	\int \frac{d^4\Delta}{(2\pi)^3} \delta^+((p - \Delta)^2 - m^2)
	= \frac{d|t| \, d\xi \, d\phi_S}{2(2\pi)^3 (1 + \xi)^2},
\end{align}
for the diffraction part, and, by approximating $\Delta$ and $p_2$ by $\hat{\Delta}$ and $\hat{p}_2$, 
\begin{align}\label{eq:ps-hard}
	\int \frac{d^3\bm{q_1}}{(2\pi)^3 2 E_{q_1}} \frac{d^3\bm{q_2}}{(2\pi)^3 2 E_{q_2}}
		(2\pi)^4 \delta^{(4)}(\hat{\Delta} + \hat{p}_2 - q_1 - q_2)
	= \frac{d\cos\theta \, d\phi}{8(2\pi)^2},
\end{align}
for the hard-scattering part, which can be equally expressed in terms of the transverse momentum $\bm{q}_T$
of $C$ or $D$.
Note that in \eq{eq:ps-diffraction}, we have traded $\phi_{\Delta}$ for $\phi_S$ since we choose an event-by-event
$\hat{x}_{\rm Lab}$ axis along the direction of $\bm{\Delta}_T$, which makes the azimuthal angle of the hadron's 
transverse spin $\bm{s}_T$, which is fixed in the lab setting, vary event by event.
Combining Eqs.~\eqref{eq:sdhep-xsec-0}\eqref{eq:ps-diffraction} and \eqref{eq:ps-hard} gives the fully differential cross
section,
\beq[eq:sdhep-xsec]
	\frac{d\sigma}{d|t| \, d\xi \, d\phi_S \, d\cos\theta \, d\phi}
	= \frac{\overline{|\M|^2}}{ (4\pi)^5 \, (1 + \xi)^2 \, s }.
\eeq

Building the $2\to2$ hard scattering kinematics on top of the diffraction plane, the angle $\phi$ describes the angular correlation 
between the diffraction and the hard collision. Its distribution is solely determined by the spin states of $A^*$ and $B$. 
If we denote the helicities of $A^*$ and $B$ by $\lambda_A$ and $\lambda_B$, respectively, then the $\phi$ dependence of the hard scattering amplitude is captured by a phase factor, $e^{i\pp{\lambda_A - \lambda_B} \phi}$. 

For the $n = 1$ channel, $A^* = \gamma^*$ has three helicity states $(+1, 0, -1)$. 
For the $n = 2$ channel, the quark GPDs have three possible helicities $\lambda_A^q = 0$ or $\pm 1$, 
where $\lambda_A^q = 0$ has two independent contributions from the unpolarized and polarized GPDs, 
while $\lambda_A^q = \pm 1$ is given by the two transversity GPDs. 
Similarly, the gluon GPDs also have three helicities $\lambda_A^g = 0$ or $\pm 2$, with 
$\lambda_A^g = 0$ receiving contributions from both the unpolarized and polarized GPDs 
and $\lambda_A^g = \pm 2$ from the two transversity GPDs.

Because of the exclusive nature, the SDHEP cross section can receive contributions from the interferences among any two of $A^* = \gamma^*, [q\bar{q}']$ and $[gg]$ channels as well as their different polarization states.
Schematically, the scattering amplitude can be written as
\beq[eq:sdhep-M-decomp-A-phi]
	\M \sim \sum_{A^*} e^{i\pp{\lambda_A - \lambda_B} \phi} F_{A^*} \otimes C_{A^*B \to CD}(\hat{s}, \theta),
\eeq
in accordance with \eq{eq:channels}, where $F_{A^*}$ is the hadron structure function associated with the diffraction $h \to h' + A^*$, 
and $C$ is the corresponding hard scattering amplitude, with the $\phi$ dependence factored out.
Squaring the amplitude in \eq{eq:sdhep-M-decomp-A-phi} can cause interference of any two different channels $A^*$.
Having a transversely polarized beam $B$ can also induce interference of its different helicity states.
The interference between $(\lambda_A, \lambda_B)$ and $(\lambda_A^{\prime}, \lambda_B^{\prime})$ leads to the azimuthal correlations
\beq
	\cos(\Delta\lambda_A - \Delta\lambda_B)\phi,\;
	\mbox{ and/or }\;
	\sin(\Delta\lambda_A - \Delta\lambda_B)\phi,
\eeq
depending on details of the interaction, where $\Delta\lambda_{A, B} = \lambda_{A, B} - \lambda_{A, B}^{\prime}$. Extracting different trigonometric components of the azimuthal distribution is a great way to disentangle different GPD contributions, in a way similar to using the angular modulations
in the SIDIS to extract different TMDs~\citep{Bacchetta:2006tn}. 
Similarly, the angular distribution of the lepton pair in the Drell-Yan process~\citep{Lam:1978pu} was studied to capture richer structures of QCD dynamics than the production rate alone.

\section{DVCS as an SDHEP}
\label{sec:dvcs-sdhep}

The simplest SDHEP example is the real photon electroproduction in \sec{sssec:dvcs}, 
which gives the DVCS at $n = 2$. Usually calculations of the hard coefficients
are carried out in the c.m.~frame of the parton pair and the virtual photon $\gamma^*_{ee}$. 
Since $\gamma^*_{ee}$ has a hard virtuality and short lifetime, distinguishing the amplitudes according to its helicity
somewhat obscures the underlying physics. 
On the other hand, in the $n = 1$ channel, the virtual photon $\gamma^*$ 
that connects the diffracted hadron and the hard part has a long lifetime. 
This gives it the physical significance of being a ``quasi-real'' particle. 
Hence we may discuss the hard scattering amplitudes in terms of its helicity state. 
With these considerations, in this section, we adopt the setup outlined in \sec{sec:SDHEP-frame} and 
reformulate the DVCS within the SDHEP frame.

\subsection{Hard coefficients of the DVCS}
\label{ssec:hard-dvcs-calc}
\begin{figure}[htbp]
\centering
	\includegraphics[trim={-1em 0 -2em 0}, clip, scale=0.75]{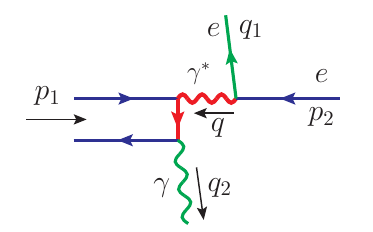}	
	\includegraphics[trim={-2em 0 -1em 0}, clip, scale=0.75]{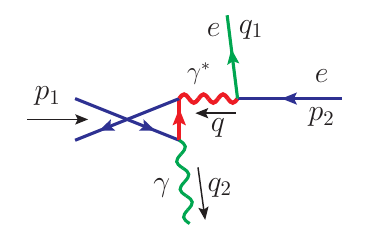}	
	\caption{The LO diagrams to the hard exclusive subprocess of the DVCS, initialized by the state $f_2 = [q\bar{q}]$. 
	The red thick lines indicate the propagators with high virtualities and thus belong to the hard part.}
\label{fig:dvcs-lo}
\end{figure}

For the $n = 2$ channel, the DVCS amplitude can be factorized into GPDs, as given in \eq{eq:eh2eah-factorize-ren},
\begin{align}
	\M_{\alpha_1 \alpha_2 \lambda_2}
		= \sum_q \int_{-1}^1 dx \bb{ F^q(x, \xi, t) \, C_{\alpha_1 \alpha_2 \lambda_2}(x, \xi; \hat{s}, \theta, \phi)
			+ \wt{F}^q(x, \xi, t) \, \wt{C}_{\alpha_1 \alpha_2 \lambda_2}(x, \xi; \hat{s}, \theta, \phi) },
\label{eq:dvcs-factorize-M}
\end{align}
where $\hat{s}$ is given in \eq{eq:sdhep-shat}, 
$\theta$ and $\phi$ are defined in \fig{fig:sdhep-frame} as the final-state electron direction, and
$\alpha_1$, $\alpha_2$, and $\lambda_2$ are helicities of the initial-state electron, 
final-state electron and photon {\it in the SDHEP frame}, respectively.
Following discussions in \sec{sec:SDHEP-frame}, the GPDs take the same values in the SDHEP frame
as in the Lab frame.  
We work at LO, where only quark GPDs contribute and we suppress the factorization scale dependence.
The hard coefficients are given by the scattering of a collinear $[q\bar{q}]$ pair of zero helicity
with an electron, 
\beq[eq:dvcs-hard]
	[q\bar{q}](\hat{p}_1, 0) + e(p_2, \alpha_1) \to e(q_1, \alpha_2) + \gamma(q_2, \lambda_2),
\eeq 
as given by the two diagrams in \fig{fig:dvcs-lo}.
The momentum $\hat{p}_1$ is projected on shell as in \eq{eq:e2em-col-mom-approx}. 
The SDHEP frame is the c.m.~frame of the hard scattering in \eq{eq:dvcs-hard}.
By introducing two auxiliary light-like vectors $w$ and $\bar{w}$,
\beq[eq:dvcs-w-wbar]
	\bar{w} = \frac{1}{\sqrt{2}} \pp{ 1, \vec{q}_1 / |\vec{q}_1|}, \quad
	w = \frac{1}{\sqrt{2}} \pp{ 1, -\vec{q}_1 / |\vec{q}_1|} = \frac{1}{\sqrt{2}} \pp{ 1, \vec{q}_2 / |\vec{q}_2| },
	\quad
	w \cdot \bar{w} = 1,
\eeq
the kinematics can be described as
\beq
	\hat{p}_1 = \sqrt{\hat{s}/2} \,\bar{n}, \quad
	p_2 = \sqrt{\hat{s}/2} \,n, \quad
	q_1 = \sqrt{\hat{s}/2} \,\bar{w}, \quad
	q_2 = \sqrt{\hat{s}/2} \,w,
\eeq
with $n$ and $\bar{n}$ being the same as \eq{eq:n-nbar} and the scalar products,
\beq
	w \cdot n = \bar{w} \cdot \bar{n} = (1 - \cos\theta) / 2, \quad
	w \cdot \bar{n} = \bar{w} \cdot n = (1 + \cos\theta) / 2.
\eeq

As indicated by \eq{eq:dvcs-hard} and \fig{fig:dvcs-lo}, 
we calculate the hard coefficients by thinking of both the quark and antiquark as entering the hard interaction, 
carrying momenta $(\xi + x) \hat{P}$ and $(\xi - x) \hat{P}$, respectively. The results can be directly extrapolated to the full
$x$ range $[-1, 1]$ by keeping all related $i\epsilon$ prescriptions explicitly. 
Following the two-stage paradigm, it is convenient to make analogy between the DVCS and the corresponding meson process
in \sec{sssec:em2ea}, such that it is natural to introduce a variable change $x \to z = (x + \xi) / (2 \xi)$. 
Then the two quarks carry momenta $z \hat{p}_1$ and $(1-z) \hat{p}_1$, respectively. 
The hard coefficients $C$ and $\wt{C}$ are then obtained by contracting the amputated parton lines with 
$\gamma \cdot \hat{p}_1 / 2$ and $\gamma_5\gamma \cdot \hat{p}_1 / 2$ (with an additional $1/2\xi$ factor), for the unpolarized
and polarized GPDs, respectively.

The P-even hard coefficient $C$ is,
\begin{align}
	2 \xi C_{\alpha_1 \alpha_2 \lambda_2}
	= -\frac{i e_q^2 e^3}{q^2} \bar{u}_2 \gamma_{\mu} u_1
		\bigg\{ \frac{1}{k_1^2 + i \epsilon} 
			\Tr\bb{\frac{\slash{\hat{p}}_1}{2} \slash{\epsilon}^*_{\lambda_2} \slash{k}_1 \gamma^{\mu} }
		+ \frac{1}{k_2^2 + i \epsilon} 
			\Tr\bb{\frac{\slash{\hat{p}}_1}{2} \gamma^{\mu} \slash{k}_2  \slash{\epsilon}^*_{\lambda_2}  }
		\bigg\},
\label{eq:dvcs-amp-0}
\end{align}
where $e_q$ is the electric charge of the quark $q$ with $e_u = 2/3$ and $e_d = -1/3$, and we have taken the electron charge to be $-1$.
Here $u_1 = u(p_2, \alpha_1)$ and $u_2 = u(q_1, \alpha_2)$ are the initial- and final-state electron spinors, respectively, 
$\epsilon_{\lambda_2} = \epsilon_{\lambda_2}(q_2)$ is the final-state photon polarization vector,
and we have the following internal momentum definitions,
\begin{align}
	q & = p_2 - q_1, &
	q^2 & = -2 p_2 \cdot q_1 = - \hat{s} \, (1 + \cos\theta) / 2, \nn\\
	k_1 & = q_2 - (1-z) \hat{p}_1, &
	k_1^2 &= -2(1-z) \hat{p}_1 \cdot q_2 = - (1-z) \, \hat{s} \, (1 + \cos\theta) / 2, \nn\\
	k_2 & = z \hat{p}_1 - q_2, &
	k_2^2 &= -2 z \hat{p}_1 \cdot q_2 = - z \, \hat{s} \, (1 + \cos\theta) / 2.
\end{align}
One can immediately notice that the two terms in the curly bracket in \eq{eq:dvcs-amp-0} are related to each other by a 
charge conjugation and $z \to 1 - z$, up to an overall minus sign, so we have
$C_{\alpha_1 \alpha_2 \lambda_2}(z) = -C_{\alpha_1 \alpha_2 \lambda_2}(1 - z)$,
or 
\beq
	C_{\alpha_1 \alpha_2 \lambda_2}(x, \xi) = -C_{\alpha_1 \alpha_2 \lambda_2}(-x, \xi)
\eeq
in terms of the original GPD variable $x$.
This property goes beyond LO and ensures that only the charge-conjugation-even (C-even) unpolarized GPD component
\beq[eq:C-even-F]
	F^+(x, \xi, t) \equiv F(x, \xi, t) - F(-x, \xi, t)
\eeq 
is probed by the DVCS.
Evaluating the fermion traces, we have
\begin{align}
	2 \xi C_{\alpha_1 \alpha_2 \lambda_2}
		= -\frac{i e_q^2 e^3}{q^2} \pp{ \bar{u}_2 \gamma_{\mu} u_1 \, \epsilon^*_{\lambda_2,\nu} }
			\bb{ g^{\mu\nu} - \frac{\bar{n}^{\mu} w^{\nu} + \bar{n}^{\nu} w^{\mu}}{\bar{n} \cdot w} }
			\bb{ \frac{1}{1 - z - i \epsilon} - \frac{1}{z - i \epsilon} }.
\label{eq:dvcs-ce-tensor}
\end{align}

Similarly, replacing $\slash{p}_1$ in \eq{eq:dvcs-amp-0} by $\gamma_5\slash{p}_1$ gives the P-odd hard coefficient,
\begin{align}
	2 \xi \wt{C}_{\alpha_1 \alpha_2 \lambda_2}
		= -\frac{i e_q^2 e^3}{q^2} \pp{ \bar{u}_2 \gamma_{\mu} u_1 \, \epsilon^*_{\lambda_2,\nu} }
			\frac{-i \, \epsilon^{\bar{n} w \mu \nu}}{\bar{n} \cdot w}
			\bb{ \frac{1}{1 - z - i \epsilon} + \frac{1}{z - i \epsilon} },
\label{eq:dvcs-co-tensor}
\end{align}
which is invariant under $z \to 1 - z$ (or equivalently, $x \to -x$) and thus probes the C-even polarized GPD component
\beq[eq:C-even-Ft]
	\wt{F}^+(x, \xi, t) \equiv \wt{F}(x, \xi, t) + \wt{F}(-x, \xi, t),
\eeq 
a property that goes beyond LO.

Inserting explicit forms for the spinors and polarization vector into Eqs.~\eqref{eq:dvcs-ce-tensor} and \eqref{eq:dvcs-co-tensor},
we have the helicity amplitudes in the SDHEP frame,
\bse\label{eq:DVCS-hel-amps-hard}\begin{align}
	2 \xi C_{\pm\pm\pm} & = -e_q^2 e^3 \sqrt{\frac{2}{\hat{s}}} \cdot \bb{ \frac{1}{1 - z - i \epsilon} - \frac{1}{z - i \epsilon} } 
		\cdot \frac{4 \cos(\theta/2)}{(1 + \cos\theta)^2} \, e^{\mp i \phi / 2}, 
	\\
	2 \xi C_{\pm\pm\mp} & = +e_q^2 e^3 \sqrt{\frac{2}{\hat{s}}} \cdot \bb{ \frac{1}{1 - z - i \epsilon} - \frac{1}{z - i \epsilon} } 
		\cdot \frac{4 \cos(\theta/2)}{(1 + \cos\theta)^2} \, \sin^2(\theta/2) \, e^{\mp i \phi / 2},
	\\
	2 \xi \wt{C}_{\pm\pm\pm} & = \mp e_q^2 e^3 \sqrt{\frac{2}{\hat{s}}} \cdot \bb{ \frac{1}{1 - z - i \epsilon} + \frac{1}{z - i \epsilon} } 
		\cdot \frac{1}{\cos(\theta/2)} \, e^{\mp i \phi / 2},
	\\
	2 \xi \wt{C}_{\pm\pm\mp} & = \mp e_q^2 e^3 \sqrt{\frac{2}{\hat{s}}} \cdot \bb{ \frac{1}{1 - z - i \epsilon} + \frac{1}{z - i \epsilon} } 
		\cdot \frac{ \sin^2(\theta/2) }{\cos(\theta/2)} \, e^{\mp i \phi / 2}.
\end{align}\ese
All the other helicity amplitudes are zero.
We see that the $z$ dependence factors out of the $\theta$ and $\phi$ dependence, while the latter are experimental observables.
In terms of $x$, the $z$ factor becomes
\beq[eq:dvcs-coefs-x]
	\frac{1}{2\xi} \bb{ \frac{1}{1 - z - i \epsilon} \mp \frac{1}{z - i \epsilon} } 
	= \frac{1}{\xi - x - i \epsilon} \mp \frac{1}{\xi + x - i \epsilon} ,
\eeq
which gives the (negative) ``zeroth GPD moments'',
\beq[eq:dvcs-GPD-moments]
	-F^+_0(\xi, t) \equiv - \int_{-1}^1 dx \, \frac{F^+(x, \xi, t)}{x - \xi + i \epsilon}, \quad
	-\wt{F}^+_0(\xi, t) \equiv - \int_{-1}^1 dx \, \frac{\wt{F}^+(x, \xi, t)}{x - \xi + i \epsilon},
\eeq
when convoluting with $F$ and $\wt{F}$, respectively.
Then the DVCS amplitudes in \eq{eq:dvcs-factorize-M} are
\bse\label{eq:dvcs-helicity-amplitudes}\begin{align}
	\M^{\rm DVCS}_{\pm\pm\pm}
	& = e^3 \sqrt{\frac{2}{\hat{s}}} e^{\mp i \phi / 2} \sum_q e_q^2 
		\bb{ F^{q, +}_0(\xi, t) \frac{4 \cos(\theta/2)}{(1 + \cos\theta)^2}
			\pm \wt{F}^{q, +}_0(\xi, t) \frac{1}{\cos(\theta/2)} },	\\
	\M^{\rm DVCS}_{\pm\pm\mp}
	& = e^3 \sqrt{\frac{2}{\hat{s}}} \sin^2(\theta/2) e^{\mp i \phi / 2} \sum_q e_q^2 
		\bb{ -F^{q, +}_0(\xi, t) \frac{4 \cos(\theta/2)}{(1 + \cos\theta)^2}
			\pm \wt{F}^{q, +}_0(\xi, t) \frac{ 1 }{\cos(\theta/2)} } ,
\end{align}\ese
which are written as helicity amplitudes for the initial-state electron and final-state electron and photon, 
constituting four independent helicity structures.
The diffracted hadron helicities are implicitly encoded in the GPDs.

\subsection{Bethe-Heitler process}
\label{ssec:bh-eh2eah}

As we noted in \sec{ssec:n=1} and confirmed in \eq{eq:dvcs-helicity-amplitudes}, the DVCS amplitudes, which 
correspond to the $n = 2$ channel for the real photon electroproduction process, have the power counting $\O(1/Q)$,
with $Q \sim q_T \sim \sqrt{\hat{s}}$ being the hard scale. 
In contrast, the $\gamma^*$-mediated channel at $n = 1$, i.e., the Bethe-Heitler (BH) process,
counts at a more leading power, $\O(1/\sqrt{|t|})$, as discussed in \sec{ssec:n=1}.
So it contributes more to the overall cross section. 
In this section, we calculate the BH amplitudes {\it in the SDHEP frame}, where the $\gamma^*$ carries momentum $p_1$
and collides with the electron in the c.m.~frame,
\beq[eq:dvcs-bh]
	\gamma^*(p_1, \lambda) + e(p_2, \alpha_1) \to e(q_1, \alpha_2) + \gamma(q_2, \lambda_2),
\eeq
where the helicity index $\lambda$ of $\gamma^*$ is to be specified in the following. 
One difference from the DVCS amplitude calculation is that here we need to keep $p_1$ as exact because neglecting its
virtuality may induce an error of order $1/Q$ in the amplitude, which is suppressed with respect to the BH amplitude, 
but not to the DVCS. By denoting $\hat{s}$ as the unapproximated c.m.~energy square of the hard
collision system, we have the kinematics,
\begin{align}
	p_1 &= \frac{\sqrt{\hat{s}}}{2} \pp{ \frac{\hat{s} + t}{\hat{s}}, 0, 0, \frac{\hat{s} - t}{\hat{s}} },
	&p_2 &= \frac{\sqrt{\hat{s}}}{2} \pp{ \frac{\hat{s} - t}{\hat{s}}, 0, 0, - \frac{\hat{s} - t}{\hat{s}} },	\nn\\
	q_1 &= \frac{\sqrt{\hat{s}}}{2} \pp{ 1, \bm{q}_1 / |\bm{q}_1| },
	&q_2 &= \frac{\sqrt{\hat{s}}}{2} \pp{ 1, -\bm{q}_1 / |\bm{q}_1| }.
\end{align}

As discussed in \sec{ssec:n=1}, the BH amplitude reduces to the electromagnetic form factor of the hadron, 
as given in Eqs.~\eqref{eq:n1 FF} and \eqref{eq:EM-form-factor}. 
In the SDHEP frame, the amplitude structure simplifies by use of Ward identities for both $F$
and $\H$,
\beq[eq:BH-ward]
	F\cdot p_1 = F^+ p_1^- + F^- p_1^+ = 0, 
	\quad
	\H \cdot p_1 = \H^+ p_1^- + \H^- p_1^+ = 0, 
\eeq
such that $F^+ \H^- = F^- \H^+$.
By the power counting $(F^+, F^-, \bm{F}_T) \sim (Q, t/Q, \sqrt{|t|})$, we have
$(\H^+, \H^-, \bm{\H}_T) \sim (1, t/Q^2, 1)$ in the SDHEP frame, instead of the superficial 
power counting $\H^{\mu} \sim \order{1}$.
Then the scalar product of $F$ and $\H$ is
\beq[eq:F-H-SDHEP]
	F \cdot \H = 2F^+\H^- - \bm{F}_T \cdot \bm{\H}_T
	= 2 (F\cdot n)(\bar{n} \cdot \H) - 
		\sum\nolimits_{\lambda = \pm} \bb{ F \cdot \epsilon_{\lambda}^*(p_1)} \bb{\epsilon_{\lambda}(p_1) \cdot \H}.
\eeq
Although this is only valid in the SDHEP frame, it equips the virtual photon with well-defined polarization states. 
The first term on the right-hand side of \eq{eq:F-H-SDHEP} corresponds to a longitudinally polarized photon state, whose polarization vector 
$\epsilon_0(p_1) \equiv \bar{n}$
is contracted with $\H$. 
Along with the $1/t$ factor from the photon propagator, it contributes to the amplitude at the power 
$1/t \times t/Q = 1/Q$, the same as the GPD channel. So when calculating the hard coefficient 
$\bar{n} \cdot \H$,
we may keep only the leading power in $\sqrt{|t|} / Q$.
The second term in \eq{eq:F-H-SDHEP} corresponds to two transverse polarization states, with the polarization vectors defined as
\beq[eq:ga*-pol-v]
	\epsilon_{\pm}^{\mu}(p_1) = (0, \mp 1, - i, 0) / \sqrt{2}.
\eeq
It contributes to the amplitude with a power counting of $1/t \times \sqrt{|t|} \sim 1 / \sqrt{|t|}$, which is one power higher than the GPD channel.
Hence we need to keep the hard coefficient $\H_\lambda \equiv \epsilon_{\lambda} \cdot \H$ up to the subleading power in $\sqrt{|t|} / Q$.

The above discussion of the $\gamma^*$ channel is generic to all processes. 
For the real photon electroproduction, the subprocess in \eq{eq:dvcs-bh} is an elementary scattering.
As we will see through explicit calculations, the $t$ dependence in 
$\epsilon_{\pm} \cdot \H$ arises from kinematic effects. 
Since there is no singularity associated with $t\to 0$ in $\H$, the subleading term in $\H_\pm$ starts at $t / Q^2$, 
which is power suppressed with respect to the GPD channel. 
Thus, it is a valid approximation to also neglect $t$ in the transverse amplitudes.

\begin{figure}[htbp]
\centering
	\includegraphics[clip, trim={-1em 0 -2em 0}, scale=0.75]{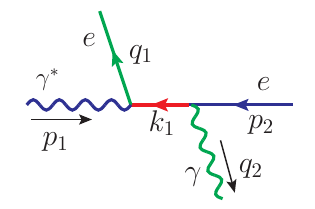}	
	\includegraphics[clip, trim={-2em 0 -1em 0}, scale=0.75]{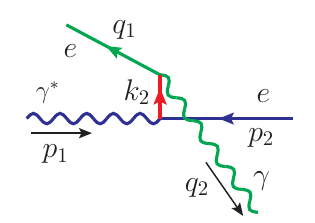}	
	\caption{The LO diagrams for the BH subprocess, initialized by the virtual photon state.}
\label{fig:bh-lo}
\end{figure}

The hard scattering diagrams for the BH amplitude are shown in \fig{fig:bh-lo}. 
Denoting these hard coefficients as 
$\H_{\lambda, \alpha_1 \alpha_2 \lambda_2} \equiv \epsilon_{\lambda} \cdot \H_{\alpha_1 \alpha_2 \lambda_2}$ with 
$\lambda = \pm1, 0$ being the helicity of the virtual photon $\gamma^*$, and
$\alpha_1$, $\alpha_2$, and $\lambda_2$ being helicities of the initial-state electron, final-state electron and photon,
respectively. Explicitly, we have
\begin{align}
	\H_{\lambda, \alpha_1 \alpha_2 \lambda_2}
	& = -i e^2 \, \bar{u}_2 \bb{ \frac{1}{k_1^2} \slash{\epsilon}_{\lambda} \slash{k}_1 \slash{\epsilon}^*_{\lambda_2} 
		+ \frac{1}{k_2^2} \slash{\epsilon}^*_{\lambda_2} \slash{k}_2 \slash{\epsilon}_{\lambda} } u_1,
\end{align}
where $k_1 = p_2 - q_2$, $k_2 = p_1 + p_2$, and $u_1$, $u_2$, and $\epsilon_{\lambda_2}$ are the same as in \eq{eq:dvcs-amp-0}.
The helicity amplitudes for the longitudinal photon polarization are
\beq
	\H_{0, \pm\pm\mp}
	= -2e^2 \frac{t}{s} \frac{1}{\sqrt{1 - t/s}} \cos(\theta/2) e^{\mp i \phi / 2}
	= -2e^2 \frac{t}{s} \cos(\theta/2) e^{\mp i \phi / 2} + \order{t^2 / s^2},
\eeq
which start at $\order{t/s}$ as argued below \eq{eq:BH-ward}. In the second step we only kept the leading terms.
The transverse photon polarizations give
\begin{align}
	\H_{\pm, \pm\pm\pm}
		& =  \mp \frac{2e^2}{\sin(\theta/2)} \sqrt{1 - \frac{t}{s}} \, e^{\pm i \phi / 2}
		&& = \mp \frac{2e^2}{\sin(\theta/2)} \, e^{\pm i \phi / 2} + \order{t /s}, \nn\\
	\H_{\pm, \pm\pm\mp}
		& =  \mp \frac{e^2 (1 + \cos\theta)}{\sin(\theta/2)} \frac{t}{s} \frac{1}{\sqrt{1 - t/s}} \, e^{\pm i \phi / 2}
		&& = 0 + \order{t /s}, \nn\\
	\H_{\pm, \mp\mp\pm}
		& =  \pm 2e^2 \sin(\theta/2) \frac{1}{\sqrt{1 - t/s}} \, e^{\pm 3 i \phi / 2}
		&& = \pm 2e^2 \sin(\theta/2) \, e^{\pm 3 i \phi / 2} + \order{t /s},
\end{align}
which scale as $\O(1)$, with subleading terms of $\order{t/s}$.
All other helicity amplitudes are zero.

Together with Eqs.~\eqref{eq:n1 FF} and \eqref{eq:F-H-SDHEP}, we can determine the helicity amplitudes for the BH process,
\bse\label{eq:BH-helicity-amplitudes}\begin{align}
	\M^{\rm BH}_{\pm\pm\pm} &\, = \mp \frac{2e^3}{t} \, F_{\pm} \, \frac{e^{\pm i \phi/2}}{\sin(\theta/2)}  + \order{\sqrt{|t|} / Q^2 }, \\
	\M^{\rm BH}_{\pm\pm\mp} &\, = \frac{4e^3}{\hat{s}} \, F_0 \, \cos(\theta/2) e^{\mp i \phi/2}
		\mp \frac{2e^3}{t} \, F_{\mp} \, \sin(\theta/2) e^{\mp 3 i \phi/2}   + \order{\sqrt{|t|} / Q^2}.
\end{align}\ese
Similar to \eq{eq:dvcs-helicity-amplitudes}, there are four independent helicity amplitudes.
Each of them are obtained by summing over the helicity of the intermediate virtual photon $\gamma^*$. 
The diffracted hadron helicities are implicitly encoded in the form factors 
$F_0 = F \cdot n = F^+$ and $F_{\pm} = F \cdot \epsilon^*_{\pm}$.

\subsection{Combining the DVCS and Bethe-Heitler processes}
\label{ssec:bh-dvcs}
The full amplitude of the single-diffractive real photon electroproduction is obtained by the channel decomposition,
with the first two powers given by the BH and DVCS amplitudes,
\beq[eq:dvcs-amp-LP-NLP]
	\M = \M^{\rm BH} + \M^{\rm DVCS} + \order{\sqrt{|t|} / Q^2}.
\eeq
As we have seen in Secs.~\ref{ssec:hard-dvcs-calc} and \ref{ssec:bh-eh2eah}, 
the leading-power contribution starts at $\O(1/\sqrt{|t|})$ and is given by the transversely polarized photon channel in the BH amplitude,
\begin{align}\label{eq:dvcs-LP-amplitudes}
	\M^{\rm LP}_{\alpha \alpha' \pm \pm \pm}
		= \pp{ F_{\alpha\alpha'} \cdot \epsilon^*_{\pm} } \A^T_{\pm \pm \pm}, 
	\quad
	\M^{\rm LP}_{\alpha \alpha' \pm \pm \mp}
		= \pp{ F_{\alpha\alpha'} \cdot \epsilon^*_{\mp} } \A^T_{\pm \pm \mp},
\end{align}
where we have made explicit the dependence on the hadron helicities $\alpha$ and $\alpha'$ 
which is encoded in the EM form factor $F_{\alpha\alpha'}$,
and the reduced hard scattering amplitudes $\A^T$ can be easily matched by comparing with \eq{eq:BH-helicity-amplitudes}.
The next-to-leading power contribution is at $\O(1/Q)$, given by the DVCS amplitude and the longitudinally polarized photon channel in the BH amplitude,
\bse\label{eq:dvcs-NLP-amplitudes}\begin{align}
	\M^{\rm NLP}_{\alpha \alpha' \pm\pm\pm}
	& = \F_{\alpha\alpha'} \, \G_{\pm \pm \pm} + \Ft_{\alpha\alpha'} \, \Gt_{\pm \pm \pm} ,	\\
	\M^{\rm NLP}_{\alpha \alpha' \pm\pm\mp}
	& = \F_{\alpha\alpha'} \, \G_{\pm \pm \mp} + \Ft_{\alpha\alpha'} \, \Gt_{\pm \pm \mp}
		+ \pp{ F_{\alpha\alpha'} \cdot n } \A^L_{\pm \pm \mp}.
\end{align}\ese
Here we defined the shorthand notations for the GPD convolutions,
\begin{align}
	\F_{\alpha\alpha'} & = \F_{\alpha\alpha'}(\xi, t) \equiv \sum\nolimits_q e_q^2 F^{q, +}_{0, \alpha \alpha'}(\xi, t)
		= \frac{1}{2P^+} \bar{u}(p', \alpha') \bb{ \H \, \gamma^+ - \E \, \frac{i \sigma^{+\Delta}}{2m} } u(p, \alpha), 
	\nn\\
	\Ft_{\alpha\alpha'} & = \Ft_{\alpha\alpha'}(\xi, t) \equiv \sum\nolimits_q e_q^2 \wt{F}^{q, +}_{0, \alpha \alpha'}(\xi, t)
		= \frac{1}{2P^+} \bar{u}(p', \alpha') \bb{ \Ht \, \gamma^+\gamma_5 - \Et \, \frac{\gamma_5 \Delta^+}{2m} } u(p, \alpha),
\end{align}
with similar definitions for the (complex-valued) GPD moments $\H$, $\E$, $\Ht$, and $\Et$, 
\beq[eq:dvcs-GPD-form-factors]
	(\H, \E, \Ht, \Et) \equiv \sum\nolimits_q e_q^2 ( H^+_0, E^+_0, \wt{H}^+_0, \wt{E}^+_0 )(\xi, t).
\eeq 
The associated hard coefficients $\G$ and $\Gt$ in \eq{eq:dvcs-NLP-amplitudes} can be obtained by matching 
with \eq{eq:dvcs-helicity-amplitudes}.
The longitudinal BH channel only contributes to the helicity amplitudes $\M^{\rm NLP}_{\alpha \alpha' \pm\pm\mp}$
and its hard coefficients $\A^L$ can be easily obtained from \eq{eq:BH-helicity-amplitudes}.

We allow arbitrary polarization states of the hadron and electron beams in the Lab frame. They can be introduced by
averaging over the initial-state hadron and electron spins by the density matrices 
$\rho^N_{\alpha\bar{\alpha}}$ and $\rho^e_{\alpha_e\bar{\alpha}_e}$. 
In the Lab frame, they can be written in terms of the spin Bloch vectors
$\bm{s}^N = (\bm{s}_T, \lambda_N)$ and $\bm{s}^e = (\bm{s}^e_T, \lambda_e)$. 
Since we neglect the electron mass, $\bm{s}^e_T$ cannot enter the amplitude square in this process, 
and $\lambda_e$ is Lorentz invariant. 
The initial-state hadron spin average can be performed in a Lorentz covariant way by using the spin 4-vector $S^{\mu}$ and using 
\beq
	\sum_{\alpha, \bar{\alpha}} u(p, \alpha) \rho^N_{\alpha \bar{\alpha}} (\bm{s}) \bar{u} \left(p, \bar{\alpha}\right)
	=\frac{1}{2}(\slash{p} + m) \left(1 + \frac{\gamma_5 \Slash{S}}{m}\right),
\eeq
where $S^{\mu}$ transforms covariantly and takes the explicit form $S^{\mu}_{\rm rest} = m (0, \bm{s})$ in the hadron rest frame in terms of 
Cartesian coordinates. Transforming it to the Lab frame, we have
\beq
	S^{\mu}_{\rm Lab} = \pp{ \lambda_N p^+, - \lambda_N \frac{m^2}{2p^+}, m \, \bm{s}_T }_{\rm lc},
\eeq
in the lightfront coordinates, together with the diffracted hadron momenta,
\beq[eq:Lab-p-p']
	p^{\mu}_{\rm Lab} = \pp{ p^+, \frac{m^2}{2 p^+}, \bm{0} },
	\quad
	p^{\prime\mu}_{\rm Lab} = \pp{ \frac{1 - \xi}{1 + \xi} p^+, \frac{1 + \xi}{1 - \xi} \frac{m^2 + \Delta_T^2}{2p^+}, - \bm{\Delta}_T }.
\eeq
The polarization vectors of the virtual photon in \eq{eq:ga*-pol-v} are written in the SDHEP frame, and transform into
\beq[eq:Lab-e]
	\epsilon^{\mu}_{{\rm Lab}, \pm} = \epsilon^{\mu} + \frac{\bm{\Delta}_T \cdot \bm{\epsilon}_T}{\Delta^+} n^{\mu}
\eeq
in the Lab frame by the inversion of \eq{eq:lorentz-trans-sdhep}. 
In both Eqs.~\eqref{eq:Lab-p-p'} and \eqref{eq:Lab-e}, we note that $\bm{\Delta}_T$ is a transverse 2-component vector in the Lab frame,
taken along the $x$ direction, $\bm{\Delta}_T = (\Delta_T, 0)$.

The leading power of the amplitude square starts at $\O(1/t)$, given by the square of the transverse BH channels in \eq{eq:dvcs-LP-amplitudes},
\begin{align}
	\overline{|\M|_{\rm LP}^2} &\, = \rho^N_{\alpha\bar{\alpha}} \, \rho^e_{\alpha_e \bar{\alpha}_e} \,
		\M^{\rm LP}_{\alpha \alpha' \alpha_e \alpha_e' \lambda} \,
		\M^{{\rm LP} *}_{\bar{\alpha} \alpha' \bar{\alpha}_e \alpha_e' \lambda} 	\nn\\
	& \, = \rho^e_{++} \pp{ | \A^T_{+++} |^2 \, \vv{ F \cdot \epsilon^*_+ F^* \cdot \epsilon_+ }
									+ | \A^T_{++-} |^2 \, \vv{ F \cdot \epsilon^*_- F^* \cdot \epsilon_- } }	\nn\\
	& \hspace{1.2em}
			+ \rho^e_{--} \pp{ | \A^T_{---} |^2 \, \vv{ F \cdot \epsilon^*_- F^* \cdot \epsilon_- }
									+ | \A^T_{--+} |^2 \, \vv{ F \cdot \epsilon^*_+ F^* \cdot \epsilon_+ } },
\end{align}
where the repeated helicity indices are summed over and we introduced the notation 
\beq[eq:vFF]
	\vv{ F \cdot \epsilon^*_{\lambda} F^* \cdot \epsilon_{\lambda'} }
	\equiv 
	\rho^N_{\alpha\bar{\alpha}} \pp{ F_{\alpha \alpha'} \cdot \epsilon^*_{\lambda} } 
		\pp{ F_{\bar{\alpha} \alpha'}^* \cdot \epsilon_{\lambda'} }
\eeq
for the hadron spin average. From \eq{eq:BH-helicity-amplitudes}, we see that
\begin{equation*}
	| \A^T_{+++} |^2 = | \A^T_{---} |^2 = \pp{ \frac{2e^3}{t} }^2 \frac{1}{\sin^2(\theta/2)}, \quad
	| \A^T_{++-} |^2 = | \A^T_{--+} |^2 = \pp{ \frac{2e^3}{t} }^2 \sin^2(\theta/2),
\end{equation*}
such that
\begin{align}\label{eq:dvcs-M2-LP-0}
	\overline{|\M|_{\rm LP}^2} &\, = 
		\pp{ \frac{2e^3}{t} }^2 
		\bb{ \frac{1}{\sin^2(\theta/2)} \pp{ B_0 + \lambda_e B_1 }
			+ \sin^2(\theta/2) \pp{ B_0 - \lambda_e B_1 }
		},
\end{align}
where
\begin{align}
	B_0 &\,= \frac{1}{2} \pp{ \vv{ F \cdot \epsilon^*_+ F^* \cdot \epsilon_+ } + \vv{ F \cdot \epsilon^*_- F^* \cdot \epsilon_- } }	\nn\\
		&\, = -\pp{ 2m^2 + \frac{1 - \xi^2}{2 \xi^2} t } \pp{ F_1^2 - \frac{t}{4m^2} F_2^2 } 
			- t (F_1 + F_2)^2, 
	\nn\\
	B_1 &\,= \frac{1}{2} \pp{ \vv{ F \cdot \epsilon^*_+ F^* \cdot \epsilon_+ } - \vv{ F \cdot \epsilon^*_- F^* \cdot \epsilon_- } }	\nn\\
		&\, = -(F_1 + F_2) \cc{ \lambda_N \bb{ F_1 \pp{ \frac{4 \xi m^2}{1 + \xi} + \frac{t}{\xi} } + t \, F_2}
			+ \frac{\bm{s}_T \cdot \bm{\Delta}_T }{2m} \bb{ 4m^2 F_1 + \frac{1 + \xi}{\xi} \, t \, F_2 } },
\end{align}
with $\bm{s}_T \cdot \bm{\Delta}_T = s_T \Delta_T \cos\phi_S$ evaluated in the Lab frame.
Organizing \eq{eq:dvcs-M2-LP-0} in terms of the polarization parameters, we have
\begin{align}\label{eq:dvcs-M2-LP}
	\overline{|\M|_{\rm LP}^2} &\, = \pp{ \frac{2e^3 m }{t} }^2 
		\Sigma^{\rm LP}_{UU} \bb{ 1 + \lambda_e \lambda_N A^{\rm LP}_{LL} + \lambda_e s_T \, \frac{\Delta_T \cos\phi_S}{2m} A^{\rm LP}_{TL} },
\end{align}
with the dimensionless polarization parameters,
\bse\label{eq:dvcs-pol-LP}\begin{align}
	\Sigma^{\rm LP}_{UU} & \, 
		= \bb{ \frac{1}{\sin^2(\theta/2)} + \sin^2(\theta/2) } 
			\bb{ \pp{ \frac{1 - \xi^2}{2 \xi^2} \frac{-t}{m^2} - 2 } \pp{ F_1^2 - \frac{t}{4m^2} F_2^2 } 
				- \frac{t}{m^2} (F_1 + F_2)^2 },	\\
	A^{\rm LP}_{LL} & \, 
		= \Sigma_{UU}^{-1} \bb{ \frac{1}{\sin^2(\theta/2)} - \sin^2(\theta/2) } 
			(F_1 + F_2) \bb{ F_1 \pp{ \frac{-t}{\xi m^2} - \frac{4 \xi}{1 + \xi} } - \frac{t}{m^2} \, F_2},	\\
	A^{\rm LP}_{TL} & \, 
		= \Sigma_{UU}^{-1} \bb{ \frac{1}{\sin^2(\theta/2)} - \sin^2(\theta/2) } 
			(F_1 + F_2) \bb{ -4 F_1 + \frac{1 + \xi}{\xi} \, \frac{-t}{m^2} \, F_2 }.
\end{align}\ese
Note that the unpolarized part $\Sigma_{UU}$ is positive definite by the kinematic constraint $-t \geq 4 \xi^2 m^2 / (1 - \xi^2)$.
As evident form \eq{eq:BH-helicity-amplitudes}, at the leading power, 
different $\gamma^*$ helicities are associated with different helicity structures of the hard scattering amplitude. 
By neglecting the electron mass, there is no interference between the right-handed and left-handed $\gamma^*$ states,
and thus there is no nontrivial $\phi$ correlation between the diffraction and scattering planes. 

The subleading power of the amplitude square is of order $\O(1/Q \sqrt{|t|})$, given by the interference of the transverse BH amplitudes
with the longitudinal BH amplitude and DVCS amplitude in \eq{eq:dvcs-LP-amplitudes},
\begin{align}
	&\overline{|\M|_{\rm NLP}^2} = 2 \Re\bb{ \rho^N_{\alpha\bar{\alpha}} \, \rho^e_{\alpha_e \bar{\alpha}_e} \,
		\M^{\rm NLP}_{\alpha \alpha' \alpha_e \alpha_e' \lambda} \,
		\M^{{\rm LP} *}_{\bar{\alpha} \alpha' \bar{\alpha}_e \alpha_e' \lambda} 
		}	\nn\\
	& \hspace{1em} 
		= 2 \rho^e_{++} \Re\bigg\{
			\big[ \G_{+++} \, \vv{ \F F^* \cdot \epsilon_+ } 
				+ \Gt_{+++} \, \vv{ \Ft F^* \cdot \epsilon_+ } 
			\big] \A^{T *}_{+++}
				\nn\\
	& \hspace{4em}
			+ \big[ \G_{++-} \vv{ \F F^* \cdot \epsilon_- }
				+ \Gt_{++-} \vv{ \Ft F^* \cdot \epsilon_- } 
				+ \A^L_{++-} \vv{ F\cdot n \, F^* \cdot \epsilon_-}
				\big] \A^{T *}_{++-}
			\bigg\}
				\nn\\
	& \hspace{1.2em}
			+ 2 \rho^e_{--} \Re\bigg\{
				\big[ \G_{---} \, \vv{ \F F^* \cdot \epsilon_- } 
					+ \Gt_{---} \, \vv{ \Ft F^* \cdot \epsilon_- } 
				\big] \A^{T *}_{---}	\nn\\
	& \hspace{4em}
			+ \big[ \G_{--+} \vv{ \F F^* \cdot \epsilon_+ }
				+ \Gt_{--+} \vv{ \Ft F^* \cdot \epsilon_+ } 
				+ \A^L_{--+} \vv{ F\cdot n \, F^* \cdot \epsilon_+}
				\big] \A^{T *}_{--+}
			\bigg\},
\end{align}
where we introduced notations similar to \eq{eq:vFF}, e.g., 
\beq
	\vv{ \F F^* \cdot \epsilon_{\pm} } 
	\equiv 
	\rho^N_{\alpha\bar{\alpha}} \pp{ \F_{\alpha \alpha'} } 
		\pp{ F_{\bar{\alpha} \alpha'}^* \cdot \epsilon_{\pm} }.
\eeq
Using the explicit forms of the hard scattering amplitudes, we have
\begin{align}\label{eq:dvcs-M2-NLP-0}
	&\overline{|\M|_{\rm NLP}^2} = 
		\frac{4 e^6}{-t} \sqrt{\frac{2}{\hat{s}}}
		\bigg\{ 
			\frac{4}{s_\theta (1 + c_\theta)}
			\Re\bb{ e^{-i \phi} \rho^e_{++} \vv{ \F F^* \cdot \epsilon_+ } - e^{i \phi} \rho^e_{--} \vv{ \F F^* \cdot \epsilon_- } }	\nn\\
	&\hspace{8em}
			+ \frac{2}{s_\theta}
			\Re\bb{ e^{-i \phi} \rho^e_{++} \vv{ \Ft F^* \cdot \epsilon_+ } + e^{i \phi} \rho^e_{--} \vv{ \Ft F^* \cdot \epsilon_- } } \nn\\
	&\hspace{8em}
			- \frac{s_\theta (1-c_\theta)}{(1+c_\theta)^2}
			\Re\bb{ e^{i \phi} \rho^e_{++} \vv{ \F F^* \cdot \epsilon_- } - e^{-i \phi} \rho^e_{--} \vv{ \F F^* \cdot \epsilon_+ } } \nn\\
	&\hspace{8em}
			+ \frac{s_\theta (1-c_\theta)}{2(1+c_\theta)}
			\Re\bb{ e^{i \phi} \rho^e_{++} \vv{ \Ft F^* \cdot \epsilon_- } + e^{-i \phi} \rho^e_{--} \vv{ \Ft F^* \cdot \epsilon_+ } } \nn\\
	&\hspace{8em}
			+ \frac{s_\theta}{2\xi} \frac{1}{P^+}
			\Re\bb{ e^{i \phi} \rho^e_{++} \vv{ F\cdot n \, F^* \cdot \epsilon_- } - e^{-i \phi} \rho^e_{--} \vv{ F\cdot n\, F^* \cdot \epsilon_+ } }
		\bigg\}.
\end{align}
Clearly, it is composed of the interference of the left- or right-handed $\gamma^*$ in the BH channel with the unpolarized GPD, polarized GPD, and
the longitudinally polarized $\gamma^*$ from the BH channel. The latter three all correspond to exchanges of helicity-0 states, so there will be 
$\cos\phi$ and $\sin\phi$ correlations. Furthermore, the hadron spin average introduces dependence on $\bm{s}_T$, which enters linearly through
$\bm{s}_T \cdot \bm{\Delta}_T \propto \cos\phi_S$ or $\bm{s}_T \times \bm{\Delta}_T \propto \sin\phi_S$ and contributes nontrivial azimuthal 
diffractive distributions. In this way, we organize \eq{eq:dvcs-M2-NLP-0} according to the spin and azimuthal dependence,
\begin{align}\label{eq:dvcs-M2-NLP}
	&\overline{|\M|_{\rm NLP}^2} = 
		\frac{4 e^6 \, m}{-t} \sqrt{\frac{1}{\hat{s}}} \,
		\bb{
			\pp{ \Sigma_{UU}^{\rm NLP} + \lambda_e \lambda_N \Sigma_{LL}^{\rm NLP} } \cos\phi 
			+ \pp{ \lambda_e \Sigma_{UL}^{\rm NLP} + \lambda_N \Sigma_{LU}^{\rm NLP} }\sin\phi 	\right.\nn\\
	&\hspace{10em}\left.
			+ s_T \pp{ \Sigma_{TU, 1}^{\rm NLP} \cos\phi_S \sin\phi + \Sigma_{TU, 2}^{\rm NLP} \sin\phi_S \cos\phi }
				\right.\nn\\
	&\hspace{10em}\left.
			+ \lambda_e s_T \pp{ \Sigma_{TL, 1}^{\rm NLP} \cos\phi_S \cos\phi + \Sigma_{TL, 2}^{\rm NLP} \sin\phi_S \sin\phi }
		},
\end{align}
with the polarization coefficients
\bse\label{eq:dvcs-pol-NLP}\begin{align}
	\Sigma_{UU}^{\rm NLP} &=
		\frac{\Delta_T}{2m} \frac{1 + \xi}{\xi} 
		\bb{ \frac{2s_\theta}{\xi} \pp{ F_1^2 - \frac{t}{4m^2} F_2^2 }
			- \xi \frac{4 + (1 - c_\theta)^2}{s_\theta} (F_1 + F_2) \Re\Ht	\right.\nn\\
		& \hspace{6em} \left.
			-\frac{2}{s_\theta} \pp{ \frac{4 + (1 - c_\theta)^2}{1 + c_\theta} }
				\pp{ F_1 \Re\H - \frac{t}{4m^2} F_2 \Re\E }
		},	\\
	\Sigma_{LL}^{\rm NLP} &=
		- \frac{\Delta_T}{2m}
		\cc{ 2(F_1 + F_2) 
			\bb{ s_\theta \pp{ \frac{1 + \xi}{\xi} F_1 + F_2 }
				+ \frac{3 - c_\theta}{s_\theta} \pp{ (1 + \xi) \Re\H + \xi \Re\E }
			}
			\right.\nn\\
		& \hspace{6em} \left.
			+ s_\theta \frac{3 - c_\theta}{1 - c_\theta} 
				\bb{ \frac{1 + \xi}{\xi} F_1 \Re\Ht - 
					\pp{ \xi F_1 + (1 + \xi) \frac{t}{4m^2} F_2 } \Re\Et 
				}
		},	\\
	\Sigma_{UL}^{\rm NLP} &=
		- \frac{\Delta_T}{2m} \frac{1 + \xi}{\xi} \frac{2 (3 - c_\theta)}{s_\theta}
		\bb{ F_1 \Im\H - \frac{t}{4m^2} F_2 \Im\E + \xi \, c^2_{\theta/2} \, (F_1 + F_2) \Im\Ht }, 
	\\
	\Sigma_{LU}^{\rm NLP} &=
		- \frac{\Delta_T}{2m} \pp{ \frac{4 + (1 - c_\theta)^2}{s_\theta} }
		\bb{ \frac{F_1 + F_2}{c_{\theta/2}^2} \pp{ (1 + \xi) \Im\H + \xi \Im\E }
			+ \frac{1 + \xi}{\xi} F_1 \Im\Ht		\right.\nn\\
		& \hspace{6em} \left.
			- \pp{ \xi F_1 + \frac{t}{4m^2} (1 + \xi) F_2 } \Im\Et
		},	\\
	\Sigma_{TU, 1}^{\rm NLP} &=
		\frac{4 + (1 - c_\theta)^2}{s_\theta}
		\cc{ \frac{F_1 + F_2}{c^2_{\theta/2}} 
			\bb{ \xi \Im\H + \pp{ \frac{\xi^2}{1 + \xi^2} + \frac{t}{4m^2} } \Im\E }
			\right.\nn\\
		& \hspace{2em} \left.
			- \pp{ \xi F_1 - \frac{1 - \xi^2}{\xi} \frac{t}{4m^2} F_2 } \Im\Ht
			- \bb{ \frac{\xi^2}{1 + \xi} F_1 + \frac{t}{4m^2} (F_1 + \xi F_2) } \Im\Et
		},	\\
	\Sigma_{TU, 2}^{\rm NLP} &=
		\frac{4 + (1 - c_\theta)^2}{s_\theta}
		\cc{ \xi (F_1 + F_2) \pp{ \Im\Ht + \frac{t}{4m^2} \Im\Et }
			\right.\nn\\
		& \hspace{2em} \left.
			- \frac{1}{c^2_{\theta/2}} 
			\bb{ \pp{\xi F_1 + \frac{1 - \xi^2}{\xi} \frac{-t}{4m^2} F_2 } \Im\H
				+ \pp{ \pp{ \xi + \frac{t}{4\xi m^2} } F_1 + \frac{\xi t}{4m^2} F_2 } \Im\E
			}
		},	\\
	\Sigma_{TL, 1}^{\rm NLP} &=
		2(F_1 + F_2)
		\cc{ \frac{3 - c_\theta}{s_\theta} 
			\bb{ \xi \Re\H + \pp{ \frac{\xi^2}{1 + \xi} + \frac{t}{4m^2} } \Re\E }
			\right.\nn\\
		& \hspace{8em} \left.
			+ s_\theta \bb{ F_1 + \pp{ \frac{\xi}{1 + \xi} + \frac{t}{4\xi m^2} } F_2 }
		} 
		\nn\\
		& \hspace{2em}
			- \frac{4 - (1 - c_\theta)^2}{s_\theta}
			\cc{ 
				\bb{ \xi F_1 - \frac{t}{4m^2} \frac{1 - \xi^2}{\xi} F_2 } \Re\Ht
				\right.\nn\\
		& \hspace{10em} \left.
				+ \bb{ \pp{ \frac{\xi^2}{1 + \xi} + \frac{t}{4m^2} } F_1
					+ \frac{\xi t}{4 m^2} F_2 } 
					\Re\Et
			},	\\
	\Sigma_{TL, 2}^{\rm NLP} &=
	 	(F_1 + F_2)
		\cc{ 2\pp{ F_1 + \frac{t}{4m^2} F_2 } s_\theta
			- \xi \frac{4 - (1 - c_\theta)^2}{s_\theta}
			\pp{ \Re\Ht + \frac{t}{4m^2} \Re\Et }
		}	\nn\\
		& \hspace{2em}
			+ 2 \frac{3 - c_\theta}{s_\theta}
			\cc{ 
				\bb{ \xi F_1 - \frac{1 - \xi^2}{\xi} \frac{t}{4m^2} F_2 } \Re\H
				+ \bb{ \pp{ \xi + \frac{t}{4 \xi m^2} } F_1 + \frac{\xi t}{4m^2} F_2 } \Re\E
			}.
\end{align}\ese

Combining the leading power (LP) in \eq{eq:dvcs-M2-LP} and next-to-leading power (NLP) in \eq{eq:dvcs-M2-NLP} 
contribution and inserting them into \eq{eq:sdhep-xsec}, 
we have the differential cross section for the real photon electroproduction process,
\begin{align}
	\frac{d\sigma}{d|t| \, d\xi \, d\phi_S \, d\cos\theta \, d\phi}
	& = \frac{1}{(2\pi)^2} \frac{d\sigma^{\rm unpol}}{d|t| \, d\xi \, d\cos\theta}
		\cdot \bb{ 1 + \lambda_e \lambda_N A^{\rm LP}_{LL} + \lambda_e s_T \, \frac{\Delta_T \cos\phi_S}{2m} A^{\rm LP}_{TL} 
		\right.\nn\\
		&\hspace{4em} \left.
			+ \pp{ A_{UU}^{\rm NLP} + \lambda_e \lambda_N A_{LL}^{\rm NLP} } \cos\phi 
			+ \pp{ \lambda_e A_{UL}^{\rm NLP} + \lambda_N A_{LU}^{\rm NLP} }\sin\phi 	\right.\nn\\
		&\hspace{4em}\left.
			+ s_T \pp{ A_{TU, 1}^{\rm NLP} \cos\phi_S \sin\phi + A_{TU, 2}^{\rm NLP} \sin\phi_S \cos\phi }
				\right.\nn\\
		&\hspace{4em}\left.
			+ \lambda_e s_T \pp{ A_{TL, 1}^{\rm NLP} \cos\phi_S \cos\phi + A_{TL, 2}^{\rm NLP} \sin\phi_S \sin\phi }
		},
\label{eq:dvcs-LP-NLP-xsec}
\end{align}
where 
\beq
	\frac{d\sigma^{\rm unpol}}{d|t| \, d\xi \, d\cos\theta}
	= \frac{\lambda_e^3}{(1 + \xi)^2} \frac{m^2}{s \, t^2} \Sigma^{\rm LP}_{UU}
\eeq
is the unpolarized differential cross section with the azimuthal dependence integrated out, 
the LP polarization parameters $\Sigma^{\rm LP}_{UU}$, $A^{\rm LP}_{LL}$, and $A^{\rm LP}_{TL}$ are given in \eq{eq:dvcs-pol-LP},
and the NLP ones are obtained from \eq{eq:dvcs-pol-NLP} by
\begin{align}
	& (A_{UU}^{\rm NLP}, A_{LL}^{\rm NLP}, A_{UL}^{\rm NLP}, A_{LU}^{\rm NLP}, 
		A_{TU, 1}^{\rm NLP}, A_{TU, 2}^{\rm NLP}, A_{TL, 1}^{\rm NLP}, A_{TL, 2}^{\rm NLP})	\nn\\
	& \hspace{3em} 
	= \frac{-t}{m \sqrt{\hat{s}}} \frac{1}{\Sigma^{\rm LP}_{UU}}
		(\Sigma_{UU}^{\rm NLP}, \Sigma_{LL}^{\rm NLP}, \Sigma_{UL}^{\rm NLP}, \Sigma_{LU}^{\rm NLP}, 
		\Sigma_{TU, 1}^{\rm NLP}, \Sigma_{TU, 2}^{\rm NLP}, \Sigma_{TL, 1}^{\rm NLP}, \Sigma_{TL, 2}^{\rm NLP}),
\end{align}
with $\hat{s}$ determined from the hard-scattering kinematics or approximated by \eq{eq:sdhep-s-hat-approx}, the error of which
is power suppressed. 
Evidently, the NLP does not change the event rate, but only introduce azimuthal $\cos\phi$ and $\sin\phi$ modulations. 
Measuring the latter offers an efficient way to determine the GPDs, with knowledge of the form factors $F_{1, 2}$ from other experiments.
Notably, GPDs enter in a linear way, and we have in total 8 GPD factors together with 8 NLP polarization parameters,
with $A_{UU}^{\rm NLP}$, $A_{LL}^{\rm NLP}$, $A_{TL, 1}^{\rm NLP}$, and $A_{TL, 2}^{\rm NLP}$ depending on their real parts, 
and $A_{UL}^{\rm NLP}$, $A_{LU}^{\rm NLP}$, $A_{TU, 1}^{\rm NLP}$, and $A_{TU, 2}^{\rm NLP}$ on their imaginary parts.
In principle, the DVCS can fully determine the GPD moments $\H$, $\E$, $\Ht$, and $\Et$, for both real and imaginary parts. 
Especially, their imaginary parts can directly constrain the GPD values at $x = \pm\xi$.

By neglecting the three-parton channels in \eq{eq:channels}, the amplitude in \eq{eq:dvcs-amp-LP-NLP} is valid up to 
the error of $\sqrt{|t|} / Q^2$, and the cross section in \eq{eq:dvcs-LP-NLP-xsec} is accurate up to the error of $1/Q^2$. 
The square of the DVCS amplitude contributes at $\order{1/Q^2}$ which mixes with the interference 
between $\M^{\rm BH}$ and the power suppressed term $\mathcal{O}(\sqrt{|t|} / Q^2)$ in \eq{eq:dvcs-amp-LP-NLP}, 
so will not be included.

\subsection{Comments on related processes}
\label{ssec:dvcs-others}
As evident from the definitions in Eqs.~\eqref{eq:dvcs-GPD-moments} and \eqref{eq:dvcs-GPD-form-factors}, 
the DVCS can constrain the GPDs only up to their moments. There are three associated limitations: 
(1) only C-even GPD combinations are probed,
(2) the quark flavors cannot be disentangled,
and
(3) the $x$ dependence cannot be fully extracted, except the values at $x = \pm \xi$.
The first limitation is because the hard part of the DVCS only has two external photons, which automatically selects
C-even parton combinations. By including more processes like the photoproduction of two real photons~\citep{Pedrak:2017cpp, 
Grocholski:2021man, Grocholski:2022rqj}, one can gain exclusive access to C-odd GPD combinations. 
This process also gives different GPD flavor combinations by being proportional to the cubes of the quark electric charges,
instead of squares as the DVCS.
Similarly, in the DVMP, the quark charge enters linearly and gives another independent way to disentangle the quark flavors.
By selecting different produced mesons, we may also gain sensitivity to different GPD combinations, including 
flavor changing GPDs.

Nevertheless, the $x$ dependence cannot be accessed by simply including multiple processes. 
For all the above-mentioned processes, the dependence of the amplitudes on the GPDs is through the moments like 
\eq{eq:dvcs-GPD-moments}, at least to the leading perturbative order.
Even with a complete separation of flavors and charge conjugation combinations, only knowing the moment for each GPD
is not sufficient to map out the full $x$ dependence. As will be further discussed in the following sections, new types of 
processes with enhanced sensitivity to the GPD $x$ dependence are needed.

\section{Sensitivity to $x$-dependence of GPDs: a general discussion}
\label{sec:x-sensitivity}

As we have seen in \sec{sec:dvcs-sdhep}, the DVCS and similar processes only probe the GPDs via certain moments, 
which are not sufficient to determine the full $x$ dependence. 
In this section, we give a systematic discussion on the sensitivity of different SDHEPs to the $x$-dependence of GPDs.
As shown in \eq{eq:sdhep-M-decomp-A-phi}, the $x$ dependence is probed by the hard part, which is a function of the
kinematic variables $\hat{s}$, $q_T$ or $\theta$, and $\phi$. 
While $\hat{s}$ is determined by $\xi$ in \eq{eq:sdhep-s-hat-approx} and the $\phi$ dependence is determined by the spin
structures as in \eq{eq:sdhep-M-decomp-A-phi}, which can be useful to disentangle different GPD components,
only the $q_T$ or $\theta$ dependence can be closely connected to the $x$ dependence of GPD convolutions.

Here we are considering the $x$-sensitivity from the tree-level hard part $C(x, Q)$, 
where $Q$ is the external observable(s) not associated with the diffractive hadron;
\footnote{
Even though the GPD variable $\xi$ is also in the hard coefficient $C$ and is directly observable 
from the diffracted hadron momentum, we do not consider it to be included in $Q$, 
but instead it always comes with $x$ and is suppressed in $C(x, Q)$.}
in the context of SDHEP kinematics in \sec{sec:SDHEP-frame}, $Q$ can be $q_T$ or $\theta$ of the final-state particle $C$ or $D$.
We consider the two types of sensitivity:
\begin{enumerate}
\item[(I)] \emph{Moment-type sensitivity}: 
	$C(x, Q)$ factorizes into an $x$-dependent part and $Q$-dependent part,
	\beq[eq:hard coeff factorize]
		C(x, Q) = G(x) \, T(Q).
	\eeq
	In this case, the measurement of the $Q$ distribution, which is fully captured by the predictable $T(Q)$, 
	does not help in probing the $x$-dependence of GPDs, and all the sensitivity is in the moment-type quantity
	\beq[eq:GPD-moment-G]
		\int_{-1}^1 dx \, G(x) \, F(x, \xi, t).
	\eeq
	We call a process with only moment-type sensitivity a type-I process.
\item[(II)] \emph{Enhanced sensitivity}: 
	$C(x, Q)$ does not factorize, in the sense of \eq{eq:hard coeff factorize}. 
	Then, the distribution of $Q$ depends on the detailed $x$ distribution in the GPD. 
	To some extent, $Q$ is the ``conjugate variable" of $x$, and they are related in the amplitude
	\beq[eq:enhanced]
		\mathcal{M}(Q) \sim \int_{-1}^1 dx \, C(x, Q) \, F(x, \xi, t)
	\eeq
	through the transformation kernel $C(x, Q)$, which is, in general, not invertible, of course.
	We call a process with enhanced sensitivity a type-II process.
\end{enumerate}

Only having moment-type sensitivity is far from enough, even with next-to-leading-order hard coefficients and evolution effects included~\citep{Bertone:2021yyz}, as also confirmed in practical fits of GPDs~\citep{Diehl:2004cx, Hashamipour:2020kip, Hashamipour:2021kes, Guo:2022upw}. Given the complicated functional dependence of the GPD on $x$ plus its entanglement 
with $\xi$ and $t$ variables, one should have as much enhanced sensitivity as possible while 
having as many independent moment constraints. Among the processes that have been studied in the literature, only the DDVCS~\citep{Guidal:2002kt}, photoproduction of photon-meson pair~\citep{Boussarie:2016qop, Duplancic:2018bum, Duplancic:2022ffo, Duplancic:2023kwe, Qiu:2023mrm}, 
and meson-production of diphoton~\citep{Qiu:2022bpq} processes are type-II processes, and all the other processes~\citep{Ji:1996nm, Radyushkin:1997ki, Brodsky:1994kf, Frankfurt:1995jw, Berger:2001xd, Berger:2001zn, Pedrak:2017cpp} belong to type I.

\begin{figure}[htbp]
\centering
	\begin{tabular}{cc}
		\includegraphics[trim={0 -0.3cm -0.2cm 0}, clip, scale=1]{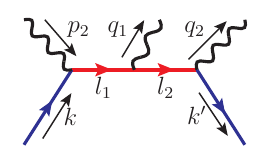}	&
		\includegraphics[trim={-0.3cm 0 0.35cm 0 0}, clip, scale=1]{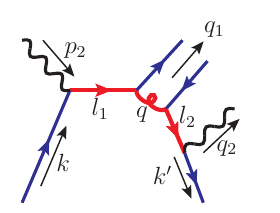}	\\
		(a) & (b) 
	\end{tabular}
	\caption{Sample diagrams for the hard scattering of the single diffractive 
	(a) photoproduction of diphoton process, and (b) photoproduction of photon-meson pair process. 
	The red thick lines indicate the propagators in the hard part, and 
	the blue lines are amputated parton lines that are put on shell and massless.}
\label{fig:hard}
\end{figure}

A careful examination of the denominator structure of the LO hard part 
of the partonic scattering can help understand and identify the difference 
in the $x$-sensitivity between these two types of processes.
The type-I processes have one common feature that every internal propagator 
can be made to have one end connect to two on-shell massless external lines, 
whether the external line is an amputated parton line or a real massless particle. 

Take the photoproduction of diphoton process, with one of its hard scattering diagrams 
in \fig{fig:hard}(a), as an example, the propagator of momentum $l_1$ is connected to 
an amputated parton line of on-shell momentum $k=(x+\xi)\hat{P}$ and the incoming photon line 
of momentum $p_2$, while the propagator of momentum $l_2$ is connected to 
an amputated parton line of momentum $k'=(x-\xi)\hat{P}$ and the outgoing photon line
of momentum $q_2$. In the c.m.~frame of the hard exclusive collision as defined in 
\fig{fig:sdhep-frame}, we have
\begin{align}\label{eq:kin in}
	\hat{P}^{\mu} &= \pp{P^+, 0^-, \bm{0}_T}, \quad
	p_2^{\mu} = \pp{0^+, p_2^-, \bm{0}_T} ,	\quad 
	\Delta^+  = p_1^+ = 2\xi P^+ = p_2^- = \sqrt{\hat{s}/2}	\,,	
\end{align}
and the final-state momenta $q_1$ and $q_2$, which define the hard scale $q_T$, 
\bse\label{eq:kin out}\begin{align}
	q_1^{\mu} &= \frac{\sqrt{\hat{s}}}{2} (1, \bm{n})
		= \pp{ \sqrt{ \frac{\hat{s}}{2} } \frac{1 + \cos\theta}{2}, 
			\sqrt{ \frac{\hat{s}}{2} } \frac{1 - \cos\theta}{2}, \bm{q}_T }_{\rm lc}	,	\\
	q_2^{\mu} & = \frac{\sqrt{\hat{s}}}{2} (1, -\bm{n}) 
		= \pp{ \sqrt{ \frac{\hat{s}}{2} } \frac{1 - \cos\theta}{2}, 
			\sqrt{ \frac{\hat{s}}{2} } \frac{1 + \cos\theta}{2}, -\bm{q}_T }_{\rm lc},	
\end{align}\ese
where we present them first in terms of Cartesian coordinates with $\bm{n}$ 
being a unit spatial vector defined as $\vec{q}_1/|\vec{q}_1|$
and then in light-front coordinates, and we also introduced the polar angle $\theta$ 
to represent $q_T \, ( = \sqrt{\hat{s}} \, \sin\theta / 2)$. 

With all external momenta defined in Eqs.~(\ref{eq:kin in}) and (\ref{eq:kin out}), 
we can express the virtuality of the internal momentum $l_1$ as
\beq[eq:l1]
	l_1^2 = 2 k \cdot p_2 = 2 (x+\xi) \hat{P} \cdot p_2  = \frac{x+\xi}{2\xi} \, \hat{s} \equiv x_{\xi}  \, \hat{s} \,,
\eeq
where $x_{\xi} = (x + \xi) / 2\xi$ is the same as the $z$ variable defined in \sec{ssec:hard-dvcs-calc}.
Similarly, we have the virtuality of the other internal momentum $l_2$ as
\beq[eq:l2]
	l_2^2 = 2 k^{\prime} \cdot q_2 = 2(x-\xi) \hat{P} \cdot q_2 
		= x_{\xi}^{\prime} \cdot \cos^2(\theta/2) \,  \hat{s} \,,
\eeq
where $x_{\xi}^{\prime} = (x - \xi) / 2\xi = x_{\xi} - 1$. 
And then the hard coefficient of the diagram \fig{fig:hard}(a) takes a factorized form, 
\begin{align}\label{eq:hard diphoton}
	 C(x, \xi, \cos\theta) 
		 \propto \frac{1}{(l_1^2 + i\varepsilon) (l_2^2 + i\varepsilon) }
		 \propto \bb{ \frac{1}{(x_{\xi}+i\varepsilon)(x_{\xi}^{\prime} + i\varepsilon)} } \cdot \frac{1}{\cos^2(\theta/2)}
\end{align}
in which the dependence on $\theta$ (or equivalently, $q_T$) is factorized from the 
momentum fraction $x$ of the relative momentum of the active $[q\bar{q}]$ pair. 

This is an immediate consequence of having the internal propagator directly 
connected to two external on-shell massless particles.
Generally, as a result of connecting to two on-shell massless lines with momenta $r_1$ and $r_2$, 
the virtuality of the internal propagator is just a product $r_1 \cdot r_2$, which simply factorizes 
into a GPD-$x$ (or DA-$z$) dependent factor and a factor that depends on the 
external observable such as $\theta$ in \eq{eq:l2}.
This example also indicates that the poles of $x$ take place at $x_{\xi} = 0$ or $x_{\xi}^{\prime} = 0$, 
that is, $x = \pm \, \xi$, which are at the boundary points between the DGLAP and ERBL regions.

In contrast, a type-II process has at least one internal line in the hard part that cannot be made to have either end connect to two on-shell massless lines. We take the photoproduction of a photon-meson pair as an example, for which one hard scattering diagram is shown in \fig{fig:hard}(b). The kinematics is the same as in Eqs.~\eqref{eq:kin in} and~\eqref{eq:kin out}, and two of the propagators, $l_1$ and $l_2$, are the same as the previous diphoton production example, given in Eqs.~\eqref{eq:l1} and~\eqref{eq:l2}.

However, the gluon propagator $q$ is connected to $l_1$ on one end and $l_2$ on the other end, 
both of which are not on shell. Letting the outgoing quark line along $q_1$ have its momentum $z q_1$, we have the gluon momentum,
\beq
	q = k + p_2 - z q_1 
		= (x + \xi) \hat{P} + p_2 - z q_1	\,,
\eeq
which has the virtuality
\beq[eq:gluon aM]
	q^2 = \hat{s} \bb{ x_{\xi} \pp{1 - z \sin^2(\theta/2)} - z \cos^2(\theta/2) } \,.
\eeq
This leads to a hard coefficient that does not take a simple factorized form to separate the $(x_{\xi}, z)$ dependence from the observable $\theta$, and therefore the distribution of $\theta$ contains extra sensitivity to the shape of $x$ and $z$ in the GPD and DA, respectively.

Compared to \eq{eq:hard diphoton}, the gluon propagator in \eq{eq:gluon aM} leads to some new poles of $x$, at
\beq
	x_{\xi} = \frac{z \cos^2(\theta/2)}{1 - z \sin^2(\theta/2)} \in [0, 1],
	\;
	\mbox{ for }
	z \in [0, 1],
\eeq
which corresponds to $x\in [-\xi, \xi]$, and thus lies in the ERBL region. 
These are not pinched poles, so do not pose any theoretical obstacles, 
but are just the regions where we need to deform the contour of $x$ to avoid them. 

Switching the roles of $p_2$ and $q_1$ in \fig{fig:hard}(b), we can get the 
diphoton mesoproduction process, as the kinematically crossing counterpart of 
the photon-meson pair photoproduction. The same gluon now has the momentum
\beq
	q' = k - q_1 + (1-z) p_2 = (x + \xi) \hat{P} - q_1 + (1-z) p_2,
\eeq
with the virtuality
\beq
	q^{\prime 2} = \hat{s} \bb{ x_{\xi} \pp{ \cos^2(\theta/2) - z } - (1 - z) \cos^2(\theta/2) },
\eeq
which equally does not factorize. It gives another new pole of $x$ at
\beq
	x_{\xi} = \frac{\cos^2(\theta/2) \, (1 - z)}{ \cos^2(\theta/2) - z } \in [1, \infty) \cup (-\infty, 0],
	\;
	\mbox{ for }
	z \in [0, 1],
\eeq
which corresponds to $|x| \geq \xi$ and lies in the DGLAP region. 
This process therefore differs from the photoproduction one
by giving complementary sensitivity to the $x$ dependence.

Similarly, in \fig{fig:hard}(a), if we make the photon $q_2$ virtual in the diphoton production process, the photon momenta in \eq{eq:kin out} will become
\begin{align}\label{eq:kin out m}
	q_1^{\mu} = \frac{\sqrt{\hat{s}}}{2} (1-\zeta) (1, \bm{n})	\,, 	\quad
	q_2^{\mu} = \frac{\sqrt{\hat{s}}}{2} (1+\zeta, -(1-\zeta) \bm{n}) 	\,,
\end{align}
where $\zeta = Q^{\prime 2} / \hat{s}$ with $Q^{\prime 2} = q_2^2$ being the virtuality of the photon $q_2$ that decays into a lepton pair. Then the propagator $l_2$ becomes
\beq[eq:l2 virtual a a]
	l_2^2 = \hat{s} \cc{ x_{\xi}^{\prime} \, \cos^2(\theta/2) + \zeta \bb{ 1 + x_{\xi}^{\prime} \, \sin^2(\theta/2) } }	\,,
\eeq
which differs from \eq{eq:l2} by having an additional term proportional to $\zeta$ that introduces an extra scale dependence. By varying $\zeta$ and $\theta$, one can get extra sensitivity to the $x$-dependence of the GPD. This is the same mechanism that gives the enhanced $x$-sensitivity as the DDVCS process~\citep{Guidal:2002kt}.
This propagator [\eq{eq:l2 virtual a a}] leads to a new pole of $x$ at
\beq
	x_{\xi}^{\prime} = \frac{- \zeta}{\cos^2(\theta/2) + \zeta \sin^2(\theta/2)} \in [-1, -\zeta],
	\;
	\mbox{ for } \theta \in [0, \pi],
\eeq
that is $x \in [-\xi, (1-2\zeta)\xi] \subset [-\xi, \xi]$, which is again inside the ERBL region.

By comparison, the type-I processes are usually topologically or kinematically simpler than the type-II processes, 
so their theoretical analysis and hard coefficient calculations are usually easier. 
The type-II processes introduce enhanced sensitivity to the $x$ dependence by having 
extra scale dependence that entangles with the $x$ flow. 
For the three type-II examples we have just examined, 
the photon-meson pair production and the diphoton production processes 
differ from the DVMP process by having one extra photon attaching to the active parton lines, 
while the virtual photon production process differs from the real photon production process by 
having an extra scale $Q'$ which is in turn achieved by having that photon decay into {\it two} leptons.
In general, extra scale dependence is introduced by more complicated topology,
\footnote{
Here, we consider virtual or massive particles as having more complicated topology 
than real massless particles, even in the case when the mass scale is not associated with virtual particle decay.} 
which is usually the necessary condition for enhanced sensitivity.

One important role that the SDHEP plays is that it sets a template for listing a number of processes, 
which we have categorized according to the beam types. 
We have shown the proof of factorization in a general sense. 
Within this framework, one shall study as many independent processes as possible, 
which should in turn constrain the $x$ dependence of GPDs as much as possible.

\section{Shadow GPDs}
\label{sec:shadow}

Only having type-I processes has an intrinsic problem that it is always possible to construct a function $S_G(x, \xi, t)$, 
called a shadow GPD~\citep{Bertone:2021yyz}, 
which has a vanishing moment in \eq{eq:GPD-moment-G} and forward limit,
\beq
	\int_{-1}^1 dx \, S_G(x, \xi, t) \, G(x, \xi) = 0, \quad
	S_G(x, 0, 0) = 0.
\eeq
Without including evolution effects and high-order corrections, shadow GPDs can never be disentangled from the ``regular'' GPDs in type-I processes.
From the discussion below \eq{eq:hard diphoton}, the kernels $G(x, \xi)$ at LO for type-I processes are limited to $1/(x - \xi \pm i \epsilon)$
and $1/(x + \xi \pm i \epsilon)$.
\footnote{In the most general case, they can also be multiplied by powers of $x$ and $\xi$, whose integrals with the GPDs can be reduced to simple
moments by use of identities like $x / (x - \xi + i \epsilon) = 1 + \xi / (x - \xi + i \epsilon)$.}
Hence it is straightforward to construct analytic shadow GPDs at LO.

We define the shadow GPDs $S(x, \xi)$ as having null forward limits and 
moment integrals in \eq{eq:dvcs-GPD-moments}, while having the same polynomiality and time reversal properties as normal GPDs. 
That is, we require
\beq[eq:sd gpd property]
	S(x, -\xi) = S(x, \xi),
	\quad
	S(\pm 1, \xi) = 0,
	\quad
	S(x, 0) = 0,
	\quad
	S(\pm \xi, \xi) = 0,
	\quad
	\int_{-1}^1 dx \frac{S(x, \xi)}{x - \xi} = 0,
\eeq
and the $(n+1)$-th moment of $S(x, \xi)$ to be an even polynomial of $\xi$ of at most $n$-th order,
\beq[eq:polynomiality]
	\int_{-1}^1 dx \, x^n \, S(x, \xi) = \sum_{i = 0, 2, \cdots}^n (2\xi)^i \, S_{n+1, i} \,.
\eeq
Note that we have dropped the $t$ dependence in $S$, which may be introduced~\citep{Moffat:2023svr} 
to relax the small $\xi$ suppression (due to $S(x, 0) = 0$) in \eq{eq:sd gpd property},
and a possible $\xi^{n+1}$ term in \eq{eq:polynomiality} which is associated with the $D$-term. 
We will construct a shadow $D$-term separately below.
As we have seen in \sec{ssec:hard-dvcs-calc} and will also see in Secs.~\ref{ssec:diphoton-amplitudes} and \ref{ssec:photo-amplitudes},
it is either the C-even or C-odd GPD combination that enters the scattering amplitudes,
so we require $S(x, \xi)$ to be either odd or even in $x$, when it is to be added to the GPDs $H(x, \xi, t)$ and $E(x, \xi, t)$ or 
$\wt{H}(x, \xi, t)$ and $\wt{E}(x, \xi, t)$.
This has allowed us to leave out the condition $\int_{-1}^1 dx \, S(x, \xi) / (x + \xi) = 0$ in \eq{eq:sd gpd property}
from which it can be inferred.
Besides, we also require the first moment of the shadow GPD to vanish 
since that can be constrained by the electromagnetic form factor measurements (see \eq{eq:1st-moments-form-factor}),
i.e., 
\beq[eq:sgpd-1st-moment]
	\int_{-1}^1 dx \, S(x, \xi) = S_{1, 0} = 0.
\eeq

The conditions in \eq{eq:sd gpd property} lead to some general constraints on the shadow GPDs.
In low energy scattering such as at JLab, the accessible $\xi$ values are small, $\xi \ll 1$.
The zeros at $x = \pm \xi$ then severely constrain the shadow GPD values in the ERBL region, 
which can only grow up to a certain power of $\xi$.
In this case, the integral in \eq{eq:sgpd-1st-moment} and the last equation in \eq{eq:sd gpd property} 
mainly receive contributions from the DGLAP region, which must be highly suppressed.
As a result, the shadow GPDs must have extra zeros in the DGLAP region, but not necessarily in the ERBL region.

As we shall see later, such oscillation will strongly suppress the contributions 
from the DGLAP regions of shadow GPDs to the special integrals 
in \eqs{eq:diphoton-special-int}{eq:photo-special-int}, which we have briefly discussed in \sec{sec:x-sensitivity}.
Even though it is the special integrals that distinguish the type-II processes from type-I ones by producing
$\cos\theta$ distributions that 
depend sensitively on the GPDs and 
have capability of disentangling shadow GPDs,
the oscillation of the latter at small $\xi$ makes them difficult to be probed.
As a consequence, in the examples to be shown in Secs.~\ref{ssec:diphoton-numerical} and~\ref{ssec:photo-numerical}, 
large weights are needed to produce significant effects.

In contrast, at a larger $\xi$, the ERBL region can become stronger, and such constraints no longer exist.

To construct specific models for shadow GPDs, we choose the following ansatz,
\beq[eq:ansatz]
	\hat{S}(x, \xi; n, a, b; c) = K_0 \, \xi^2 \, x^a \, (x^2 - \xi^2) \, (1 - x^2)^b \cdot Q_{2n}(x, c),
\eeq
where $a \geq 0$ and $b, n > 0$ are integers, and $Q_{2n}(x, c) = 1 + c \, x^2 + \cdots + q_{2n}(c) \, x^{2n}$ 
is an even $x$ polynomial of $(2n)$-th order.
This parametrization automatically satisfies the first four conditions in \eq{eq:sd gpd property}.
Since it is only a fourth order polynomial of $\xi$, the polynomiality condition can also be readily satisfied.
We have fixed the power of $(x^2 - \xi^2)$ to be unity; 
a higher power further suppresses the ERBL region 
and would lead to an even smaller impact on the integrals in \eqs{eq:diphoton-special-int}{eq:photo-special-int}.
For given $a$ and $b$, we choose $n$ to be the minimum integer such that \eq{eq:ansatz} satisfies all the conditions in 
Eqs.~\eqref{eq:sd gpd property}\eqref{eq:polynomiality} and \eqref{eq:sgpd-1st-moment}.
The single parameter $c$ is allowed to tune the shape of the shadow GPD. 
For any given choice, we fix the normalization $K_0$ (which is independent of $\xi$) 
such that $\int_{-1}^1 dx \, [ \hat{S}(x, \xi; n, a, b; c) ]^2 = 1$ when $\xi = 0.1$.

We choose the GK model as the standard GPDs, $H_0(x, \xi, t)$ and $\wt{H}_0(x, \xi, t)$, and vary them by adding shadow GPDs
to the $u$ quark GPD.
For the unpolarized GPD, since the $H^+$ that will enter the special integrals in \eqs{eq:diphoton-special-int}{eq:photo-special-int} 
is an odd function of $x$, we choose
$n = 3$ and $(a, b) = (1, 2)$ or $(1, 6)$.
The parameter $c$ is chosen to maximize the integral in \eq{eq:sgpd-1st-moment} from the DGLAP region, 
which gives $c = -11$ or $-17$, respectively.
That is, we defined two shadow GPDs,
\beq[eq:shadow-S12]
	S_1(x, \xi) = \hat{S}(x, \xi; 3, 1, 2; -11), \quad
	S_2(x, \xi) = \hat{S}(x, \xi; 3, 1, 6; -17),
\eeq
which, added to the GK model $H^u_0$, make up two other models, 
\beq[eq:GPD-model-Hh-12]
	H^u_{i}(x, \xi, t) = H^u_0(x, \xi, t) + S_{i}(x, \xi), 
	\quad
	\mbox{ for } 
	i = 1, 2.
\eeq
Similarly, for the polarized GPD, we choose $n = 3$ and $(a, b) = (0, 2)$ or $(0, 6)$, and $c = -24$ or $-40$. 
This gives two other shadow GPDs, 
\beq[eq:shadow-St12]
	\wt{S}_1(x, \xi) = \hat{S}(x, \xi; 3, 0, 2; -24), \quad
	\wt{S}_2(x, \xi) = \hat{S}(x, \xi; 3, 0, 6; -40),
\eeq
and GPD models with 
\beq[eq:GPD-model-Ht-12]
	\wt{H}^u_{i}(x, \xi, t) = \wt{H}^u_0(x, \xi, t) + \wt{S}_{i}(x, \xi), 
	\quad
	\mbox{ for } 
	i = 1, 2.
\eeq
The GPDs of other flavors are kept unchanged from the GK model.

\begin{figure}[htbp]
	\centering
		\includegraphics[scale=0.68]{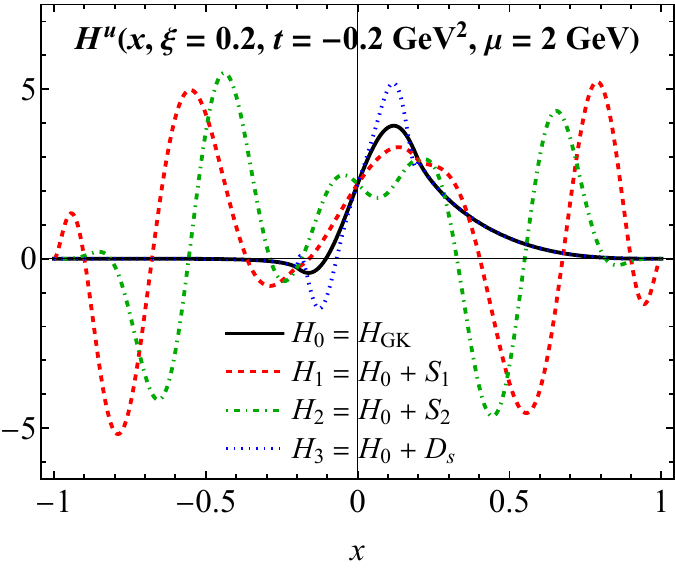}
		\includegraphics[scale=0.68]{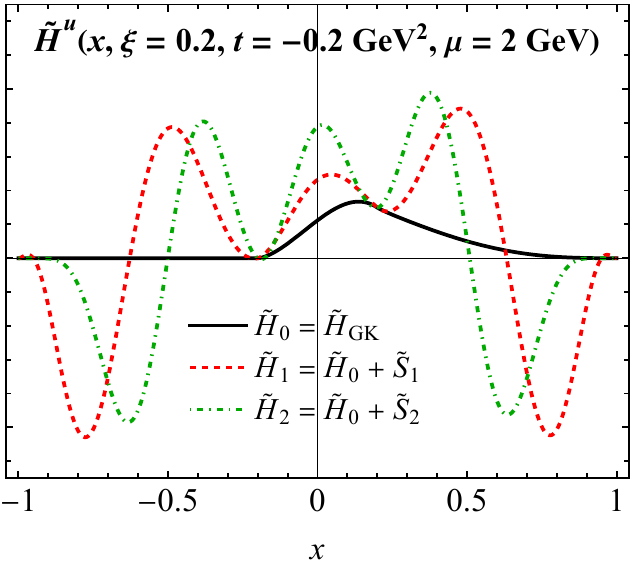}	\\ \vspace{2mm}
		\includegraphics[scale=0.698]{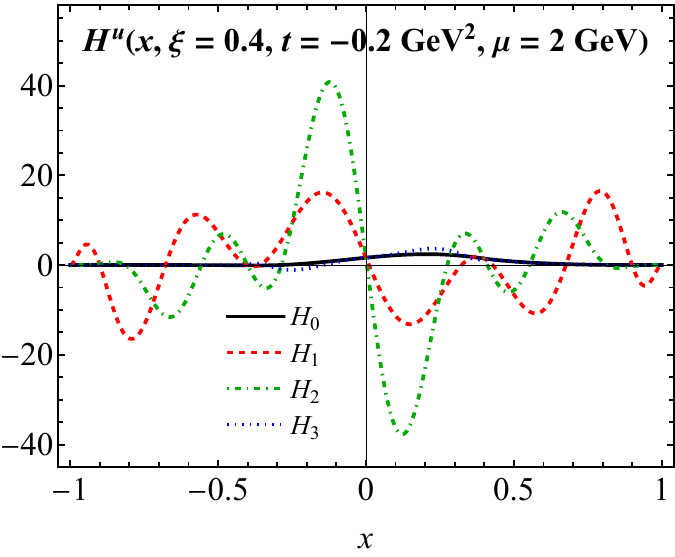}
		\includegraphics[trim={0 0 -2.5mm 0}, clip, scale=0.698]{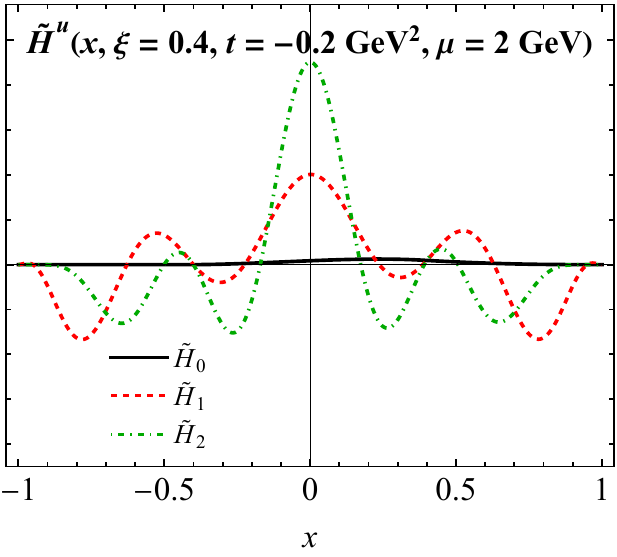}
	\caption{GPD models given in Eqs.~\eqref{eq:GPD-model-Hh-12}\eqref{eq:GPD-model-Ht-12} and \eqref{eq:GPD-model-Hh-3}.
	The upper row is for a small $\xi = 0.2$ while the lower row is for a large $\xi = 0.4$. 
	In the lower row, the thick black lines are the GK model, 
	which are small compared to the shadow GPDs and hard to notice in the figure.}
\label{fig:GPD-shadow}
\end{figure}

For the unpolarized GPD, an additional term proportional to $\mathrm{mod}(n, 2) (2\xi)^{n+1}$ 
can exist on the right hand side of \eq{eq:polynomiality}, just as in \eq{eq:H-moments}.
This comes from the $D$-term in the double distribution representation,
\beq
	H^q(x, \xi, t) = \int_{-1}^1 d\beta \int_{-1 + |\beta|}^{1 - |\beta|} d\alpha \, \delta(x - \beta - \xi \alpha) \, f^q(\beta, \alpha, t)
		+ \operatorname{sgn}(\xi) \, D^q(x / \xi, t) \, \theta\pp{ \xi^2 - x^2 } \,,
\eeq
where $D^q(x, t)$ is an odd function of $x$. 
Now we can construct a shadow GPD specifically for this term.
Similarly, to retain the conditions in \eq{eq:sd gpd property}, we drop the $t$ dependence and 
choose the ``shadow $D$-term'' $D_s(x)$ 
such that
\beq[eq:ds property]
	D_s(-x) = - D_s(x),
	\quad
	D_s(1) = 0,
	\quad
	\int_{-1}^1 dx \, \frac{D_s(x)}{x - 1} = 0,
\eeq
where the subscript ``s'' is to remind that this $D$-term is to be part of the shadow GPD, 
but not to be the requirement for the $D$-terms in the normal GPDs.
Note that since the $D$-term automatically disappears in the forward limit, 
its magnitude does not necessarily suffer from the suppression when $\xi$ is small.
Because of the last condition in \eq{eq:ds property}, a shadow $D$-term cannot be 
probed by the dispersion relation in the DVCS data~\citep{CLAS:2007clm, CLAS:2015uuo}, 
but it can modify the $D$-term in the gravitational form factor.
We choose the ansatz for the shadow $D$-term
\beq[eq:ds-ex]
	D_s(x) = J_0 \, x \, (1 - x^2) \cdot \pp{1 + c \, x^2 - \frac{7}{15}(3c + 5) x^4 } \, \theta(1 - x^2) \, ,
\eeq
with $c = 50$ and the normalization factor $J_0$ chosen to make $\int_{-1}^1 dx \, D_s^2(x) = 1$. 
Adding this to the $u$ quark GPD $H^u_0$ gives another GPD model,
\beq[eq:GPD-model-Hh-3]
	H^u_{3}(x, \xi, t) = H^u_0(x, \xi, t) + D_s(x / \xi).
\eeq

These GPD models are shown in \fig{fig:GPD-shadow} for the $u$ quark at a small $\xi = 0.2$ and a large $\xi = 0.4$, 
with a fixed $t = -0.2~\GeV^2$ and evolution scale $\mu = 2~\GeV$, for the GK model. 
As expected, at a small $\xi$, the shadow GPDs are small in the ERBL region, being dominated by the DGLAP region,
whereas at a larger $\xi = 0.4$, 
one can immediately notice that the shadow GPDs (not the shadow $D$-term) 
scale up with $\xi$ very rapidly and the ERBL region becomes dominant.

\section{Single diffractive hard exclusive diphoton mesoproduction}
\label{sec:diphoton}

Now let us study the single diffractive hard exclusive diphoton production in nucleon-pion collisions, 
\beq[eq:diphoton-process]
	N(p) + \pi(p_2) \to N'(p')+ \gamma(q_1)+\gamma(q_2),
\eeq
which was introduced in \citep{Qiu:2022bpq}.
Here $N$ can be a proton ($p$) or a neutron ($n$) and $\pi$ can be $\pi^-$ or $\pi^+$, 
making up various exclusive processes, such as 
$p\,\pi^-\to n\gamma\gamma$ and $n\,\pi^+\to p\gamma\gamma$,   
and those that could be measured with a pion beam at J-PARC~\citep{Aoki:2021cqa} and AMBER~\citep{Adams:2018pwt} experiments. 
The pion beam can also be replaced by a kaon beam and makes up more processes.
The exclusive process, $p\,\pi^-\to n\gamma\gamma$, could be made analogous to the $\pi^+\pi^-$ collision by thinking of the $p\to n$ transition 
as taking a virtual $\pi^+$ out of the proton, carrying momentum $\Delta = p - p'$ and colliding with $\pi^-$ to produce two hard photons exclusively. 

As discussed in \sec{sssec:mh2llh-mh2aah}, the charged meson beam requires a flavor change of the nucleon and forbids
the $\gamma^*$-mediated channel at $n = 1$. The leading contribution to the scattering amplitude of 
\eq{eq:diphoton-process} thus comes from the quark-antiquark pair exchange at $n = 2$ (the gluon pair exchange
is also forbidden by the charge conservation), 
which is factorized into nucleon transition GPDs and the pion DA, as argued in \sec{sssec:mh2llh-mh2aah}.
Taking the $p\,\pi^-\to n\gamma\gamma$ process as an example, the factorization formula of the amplitude is
\begin{align}\label{eq:diphoton-factorize}
	\M_{\lambda_1\lambda_2} = &\, 
		\int_{-1}^1 dx \int_0^1 dz \bb{ 
			F^u_{pn}(x, \xi, t)  \, \wt{C}_{\lambda_1\lambda_2}(x, \xi, z; \hat{s}, \theta, \phi)
			\right.\nn\\
			& \left. \hspace{8em}
			+ \wt{F}^u_{pn}(x, \xi, t) \, C_{\lambda_1\lambda_2}(x, \xi, z; \hat{s}, \theta, \phi)
		} D_{d/\pi^-}(z),
\end{align}
where we have suppressed the factorization scale, and $\lambda_1$ and $\lambda_2$ are the helicities of the final-state
two photons in the SDHEP frame. The polarized nucleon transition GPD $\wt{F}^u_{pn}(x, \xi, t)$ is defined in
\eq{eq:eh2ehh-GPDs}, and the unpolarized one $F^u_{pn}(x, \xi, t)$ can be similarly obtained by removing the $\gamma_5$.
Both take form factor decompositions similar to \eq{eq:GPD-def-q},
\bse\label{eq:pn-GPD}\begin{align}
	F^u_{pn}(x, \xi, t) & = 
		\frac{1}{2 P^+} \bar{u}(p', \alpha')
		\bb{ H^u_{pn}(x, \xi, t) \gamma^+ 
			- E^u_{pn}(x, \xi, t) \frac{i \sigma^{+\alpha} \Delta_{\alpha}}{2m}
		} u(p, \alpha),	\label{eq:pn-GPD-F}\\
	\wt{F}^u_{pn}(x, \xi, t) & = 
		\frac{1}{2 P^+} \bar{u}(p', \alpha')
		\bb{ \wt{H}^u_{pn}(x, \xi, t) \gamma^+ \gamma_5 
			- \wt{E}^u_{pn}(x, \xi, t) \frac{\gamma_5 \Delta^+}{2m}
		} u(p, \alpha),
\end{align}\ese
thereby defining the flavor transition GPDs $H^u_{pn}$, $E^u_{pn}$, $\wt{H}^u_{pn}$, and $\wt{E}^u_{pn}$, 
which have the same isospin relations to the flavor-diagonal GPDs as \eq{eq:GPD-isospin},
\beq
	H^u_{pn}(x, \xi, t) = H^u_p(x, \xi, t) - H^d_p(x, \xi, t),
\eeq
and similarly for $E^u_{pn}$, $\wt{H}^u_{pn}$, and $\wt{E}^u_{pn}$. 
The DA $D_{d/\pi^-}(z)$ is similarly defined as \eq{eq:em2em-DA-def}, 
\beq
	D_{d/\pi^-}(z)
	= \int \frac{dy^+}{4\pi} \, e^{i z p_2^- y^+} \, 
	\langle 0 | \bar{u}(0) \gamma^- \gamma_5 \, \Phi(0, y^+ \bar{n}; \bar{n}) \, d(y^+) | \pi^-(p_2) \rangle,
\eeq
which can be related to the DA $D_{u/\pi^+}(z)$ in \eq{eq:em2em-DA-def} by isospin symmetry,
\beq
	D_{d/\pi^-}(z) = -D_{u/\pi^+}(z) = - \frac{i f_{\pi}}{2} \phi(z).
\eeq
Similar factorization formulae can be written for the $n\,\pi^+\to p\gamma\gamma$ and other processes, with 
the GPDs replaced by $F^d_{np}(x, \xi, t)$ and $\wt{F}^d_{np}(x, \xi, t)$,
which are equal to $F^u_{pn}(x, \xi, t)$ and $\wt{F}^u_{pn}(x, \xi, t)$ by isospin symmetry,
and the DA by $D_{u/\pi^+}(z)$.

\subsection{Calculation of the hard coefficients}
\label{ssec:diphoton-hard}
The hard coefficients $\wt{C}_{\lambda_1\lambda_2}$ and $C_{\lambda_1\lambda_2}$ are the helicity amplitudes
of the collision between two collinear quark-antiquark pairs, 
$[q_1 \bar{q}_2]$ from the diffracted nucleon which carries light-like momenta $(\xi \pm x) P^+ \bar{n}$, respectively,
and $[q_2 \bar{q}_1]$ from the annihilated pion carrying light-like momenta $z p_2^- n$ and $(1 - z) p_2^- n$ respectively. 
For the $p \pi^-$ collision, we have $(q_1, q_2) = (u, d)$, while for the $n \pi^+$ collision, we have $(q_1, q_2) = (d, u)$.
In both processes, $\wt{C}_{\lambda_1\lambda_2}$ is obtained by contracting the amputated 
$[q_1 \bar{q}_2]$ legs with the spinor projector $\gamma\cdot\bar{n} / 2$, and 
$C_{\lambda_1\lambda_2}$ with $\gamma_5 \gamma\cdot\bar{n} / 2$. 
The amputated $[q_1 \bar{q}_2]$ legs are contracted with $\gamma_5 \gamma\cdot n / 2$ in any cases.
In this way, the hard coefficient $\wt{C}$ associated with the unpolarized GPD $F$ is parity-odd (P-odd),
while the $C$ associated with the polarized GPD $\wt{F}$ is parity-even (P-even).
The hard coefficients are then obtained by averaging over the colors of each quark pair, and then multiplied by
an extra factor $\hat{s} / 2$.

\begin{figure}[htbp]
	\centering
	\begin{tabular}{cccc}
		\includegraphics[scale=0.7]{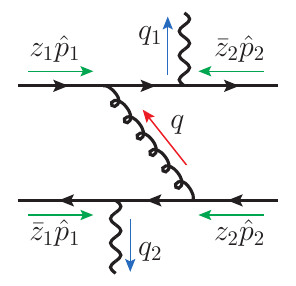} &
		\includegraphics[scale=0.7]{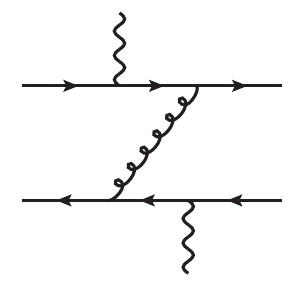} &
		\includegraphics[scale=0.7]{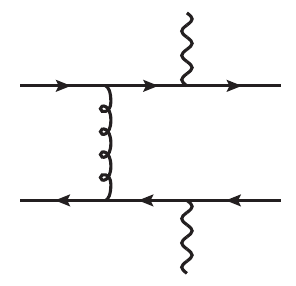} &
		\includegraphics[scale=0.7]{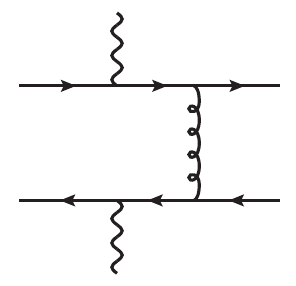} \\
		($A_1$) & ($A_2$) & ($A_3$) & ($A_4$)
	\end{tabular}
	\begin{tabular}{ccc}
		\includegraphics[scale=0.7]{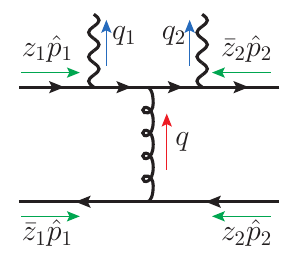} &
		\includegraphics[trim={0 -0.68cm 0 0}, clip, scale=0.7]{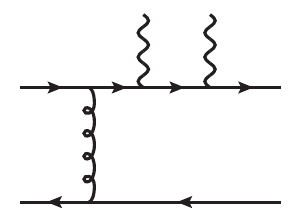} &
		\includegraphics[trim={0 -0.68cm 0 0}, clip, scale=0.7]{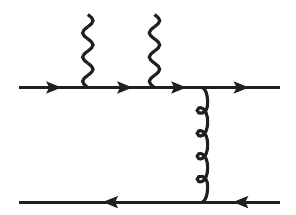} \\
		($B_1$) & ($B_2$) & ($B_3$)
	\end{tabular}
	\begin{tabular}{ccc}
		\includegraphics[scale=0.7]{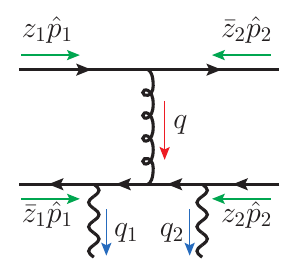} &
		\includegraphics[trim={0 0 0 -0.68cm}, clip, scale=0.7]{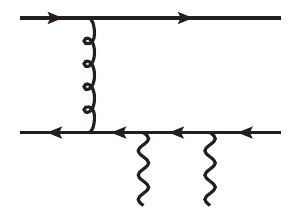} &
		\includegraphics[trim={0 0 0 -0.68cm}, clip, scale=0.7]{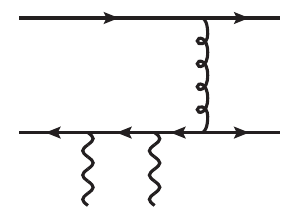} \\
		($B_4$) & ($B_5$) & ($B_6$)
	\end{tabular}
	\caption{Hard scattering diagrams for the diffractive nucleon-pion scattering into a photon pair. 
	The two incoming quark lines on the left are from the diffracted nucleon, carrying momenta 
	$z_1 \hat{p}_1$ and $\bar{z}_1 \hat{p}_1 \equiv (1-z_1) \hat{p}_1$, respectively, 
	where $\hat{p}_1 = (\Delta \cdot n) \bar{n}$.
	The two incoming quark lines on the right are to form the annihilated pion, carrying momenta 
	$z_2 \hat{p}_2$ and $\bar{z}_2 \hat{p}_2 \equiv (1-z_2) \hat{p}_2$, respectively, 
	where $\hat{p}_2 = (p_2 \cdot \bar{n})n$.
	The variables $z_1$ and $z_2$ are related to $x$ and $z$ by $z_1 = (x + \xi) / 2\xi$ and $z_2 = z$ (see the text).
	Another set of diagrams are also to be included by switching the two photon lines, giving 20 diagrams in total.}
	\label{fig:diphoton-diagrams}
\end{figure}

Note that we are taking the convention for both partons from the diffracted nucleon to enter the hard collisions.
This makes the calculation of the hard coefficients simpler and the two-stage paradigm more transparent.
By a variable change $z_1 = (x + \xi) / 2 \xi$, the $[q_1 \bar{q}_2]$ pair carries momenta
$z_1 p_1^+ \bar{n}$ and $(1 - z_1) p_1^+ \bar{n}$. To ease the notation, we also denote $z_2 = z$.
In the c.m.~frame of the hard collision, we take
\beq
	p_1^+ = p_2^- = \sqrt{\frac{\hat{s}}{2}} = \sqrt{ \frac{\xi \, s}{1 + \xi} } \, .
\eeq
The hard-scattering diagrams are shown in \fig{fig:diphoton-diagrams}. They are classified into two types:
($A$) there are 8 {\it type-$A$} diagrams for which the two photon lines are attached to two different fermion lines,
which are denoted as ($A_1$)--($A_4$), and ($A_1'$)--($A_4'$) that are obtained by switching the two photons,
and 
($B$) there are 12 {\it type-$B$} diagrams for which both photon lines are attached to one single fermion line, where
($B_1$)--($B_3$) and ($B_1'$)--($B_3'$) have both photons attached to the $q_1$ quark line that carries electric charge 
$e_1 e$, 
and 
($B_4$)--($B_6$) and ($B_4'$)--($B_6'$) have them attached to the $q_2$ quark line that carries electric charge 
$e_2 e$.

By defining two auxiliary light-like vectors $w$ and $\bar{w}$ as 
\beq[eq:diphoton-w-wbar]
	\bar{w} = \frac{1}{\sqrt{2}} \pp{ 1, \vec{q}_1 / |\vec{q}_1|}, \quad
	w = \frac{1}{\sqrt{2}} \pp{ 1, -\vec{q}_1 / |\vec{q}_1|} = \frac{1}{\sqrt{2}} \pp{ 1, \vec{q}_2 / |\vec{q}_2| },
	\quad
	w \cdot \bar{w} = 1,
\eeq
we can write the photon momenta as
\beq
	q_1 = (q_1 \cdot w) \bar{w} = \bar{w} \, \sqrt{\hat{s} / 2}, \quad
	q_2 = (q_2 \cdot \bar{w}) w = w \, \sqrt{\hat{s} / 2},
\eeq
with the scalar products,
\beq
	w \cdot n = \bar{w} \cdot \bar{n} = (1 - \cos\theta) / 2, \quad
	w \cdot \bar{n} = \bar{w} \cdot n = (1 + \cos\theta) / 2.
\eeq

For type-$A$ diagrams, the intermediate gluon propagator carries the hard transverse momentum $q_T$,
\beq[eq:diphoton-g-mom]
	q_A = -z_1 \hat{p}_1 - (1 - z_2) \hat{p}_2 + q_1
	= - \sqrt{\frac{\hat{s}}{2}} \pp{ z_1 \bar{n} + (1 - z_2) n - \bar{w} },
\eeq
with a virtuality,
\beq[eq:diphoton-g2-A]
	q_A^2 = - \frac{\hat{s}}{2} \big\{ 
		z_1 z_2 + (1 - z_1) (1 - z_2) - \cos\theta \bb{ z_1 z_2 - (1 - z_1) (1 - z_2) }
		\big\}.
\eeq
Switching the two photons changes $q_1$ and $\bar{w}$ in \eq{eq:diphoton-g-mom} by $q_2$ and $w$, respectively,
and results in a similar gluon virtuality,
\beq[eq:diphoton-g2-A']
	q_{A'}^{2} = - \frac{\hat{s}}{2} \big\{ 
		z_1 z_2 + (1 - z_1) (1 - z_2) + \cos\theta \bb{ z_1 z_2 - (1 - z_1) (1 - z_2) }
		\big\}.
\eeq
The type-$A$ gluons have special propagators whose dependence on the external observable $\theta$ cannot be
simply factored out, so they induce enhanced sensitivity to the $x$ dependence of GPDs, following the criterion in
\sec{sec:x-sensitivity}. 
In contrast, the type-$B$ diagrams have simple gluon propagators whose virtualities are
\beq
	q_B^2 = (1 - z_1) z_2 \hat{s}, \quad
	q_B^{\prime2} = z_1 (1 - z_2) \hat{s},
\eeq
for diagrams $(B_1, B_1', \cdots, B_3, B_3')$ and $(B_4, B_4', \cdots, B_6, B_6')$, respectively. 
These simply factorize into factors that only depend on $z_1$ and $z_2$, but not $\theta$ at all, following the
structure of \eq{eq:hard coeff factorize}. Similar are the quark propagators for all diagrams. 
As a result, the type-$B$ diagrams only yield moment-type sensitivity.

Denoting the hard coefficients as
\bse\label{eq:diphoton-Cmn}\begin{align}
	C_{\lambda_1\lambda_2}(z_1, z_2; \hat{s}, \theta) 
		&= C_{\mu\nu}(z_1, z_2; \hat{s}, n, \bar{n}, w, \bar{w}) \, 
			\epsilon^{\mu*}_{\lambda_1}(q_1) \epsilon^{\nu*}_{\lambda_2}(q_2),	\\
	\wt{C}_{\lambda_1\lambda_2}(z_1, z_2; \hat{s}, \theta) 
		&= \wt{C}_{\mu\nu}(z_1, z_2; \hat{s}, n, \bar{n}, w, \bar{w}) \, 
			\epsilon^{\mu*}_{\lambda_1}(q_1) \epsilon^{\nu*}_{\lambda_2}(q_2),
\end{align}\ese
the $\gamma_5$ structures in the corresponding fermion spinor traces determine that $\wt{C}_{\mu\nu}$
contains one antisymmetric Levi-Civita tensor, while $C_{\mu\nu}$ does not. 
The Ward identities
\beq
	\bar{w}_{\mu} \wt{C}^{\mu\nu} = \wt{C}^{\mu\nu} w_{\nu}
	= \bar{w}_{\mu} C^{\mu\nu} = C^{\mu\nu} w_{\nu}
	= 0
\eeq
then allows us to decompose them into gauge-invariant tensor structures,
\bse\label{eq:diphoton-tensor-decomp}\begin{align}
	C^{\mu\nu} &= C_+ \, \pp{ - g_{\perp}^{\mu\nu} } 
		+ C_- \, \pp{ - g_{\perp}^{\mu\nu} + 2 \frac{ \bar{n}_{\perp}^{\mu} \bar{n}_{\perp}^{\nu} }{ \bar{n}_{\perp}^2} }
		+ C_w \, \bar{n}_{\perp}^{\mu} w^{\nu} + C_{\bar{w}} \, \bar{w}^{\mu} \bar{n}_{\perp}^{\nu}
		+ C_{\bar{w}w} \, \bar{w}^{\mu} w^{\nu}, \\
	\wt{C}^{\mu\nu}
		&= i \wt{C}_+ \, \pp{ 
			\frac{ \bar{n}_{\perp}^{\mu} \, \epsilon^{\nu \bar{n}_{\perp} w \bar{w}} 
				- \epsilon^{\mu \bar{n}_{\perp} w \bar{w}} \bar{n}_{\perp}^{\nu} 
				}{ \bar{n}_{\perp}^2 } 
			}
		+ i \wt{C}_- \, \pp{ 
			\frac{ \bar{n}_{\perp}^{\mu} \, \epsilon^{\nu \bar{n}_{\perp} w \bar{w}} 
				+ \epsilon^{\mu \bar{n}_{\perp} w \bar{w}} \bar{n}_{\perp}^{\nu} 
				}{ \bar{n}_{\perp}^2 } 
			}	\nn\\
		& \hspace{1em}
		+ i \wt{C}_{\bar{w}} \, \bar{w}^{\mu} \, \epsilon^{\nu \bar{n}_{\perp} w \bar{w}} 
		+ i \wt{C}_w \, \epsilon^{\mu \bar{n}_{\perp} w \bar{w}} w^{\nu},
\end{align}\ese
where we defined
\beq[eq:diphoton-gperp]
	g_{\perp}^{\mu\nu} = g^{\mu\nu} - w^{\mu} \bar{w}^{\nu} - \bar{w}^{\mu} w^{\nu},
	\quad
	\bar{n}_{\perp}^{\mu} = g_{\perp}^{\mu\nu} \bar{n}_{\nu}.
\eeq
When contracting with the polarization vectors as \eq{eq:diphoton-Cmn}, the terms in \eq{eq:diphoton-tensor-decomp}
that are proportional to $\bar{w}^{\mu}$ or $w^{\nu}$ vanish. 
The helicity amplitudes are purely determined by the first two tensor structures,
\begin{align}
	C_{++} = C_{--} = \frac{N}{\hat{s}} \, C_+, &\quad
	C_{+-} = C_{-+} = \frac{N}{\hat{s}} \, C_-,	\nn\\
	\wt{C}_{++} = - \wt{C}_{--} = \frac{N}{\hat{s}} \, \wt{C}_+, &\quad
	\wt{C}_{+-} = - \wt{C}_{-+} = \frac{N}{\hat{s}} \, \wt{C}_-,
\label{eq:diphoton-helicity-structure}
\end{align}
where $N = 2i e^2 g^2 C_F / N_c$ and the four independent hard coefficients are 
\begin{align}
	&2\xi C_{+} =
		\pp{e_1-e_2}^2 \frac{2}{s_{\theta}^2} \cdot
			\P\frac{z_1 z_2 + (1-z_1)\, (1-z_2)}{ z_1 z_2 (1-z_1) (1-z_2)}
		\nn\\
	& \hspace{1em} 
		+ \frac{2 i \pi}{s_{\theta}^2} \cdot
			\bb{ (e_1^2-e_2^2)
				\pp{ \frac{\delta(z_1)}{z_2} - \frac{\delta(1-z_1)}{1-z_2} }
				- 2 e_1 e_2 
				\pp{ \frac{\delta(z_1)}{z_2} + \frac{\delta(1-z_1)}{1-z_2} }
			},
\label{eq:diphoton-C+}		\\
	&2\xi C_{-}  = 
 		\pp{e_1-e_2}^2 \frac{2}{ s_{\theta}^2 } 
			\cdot \P \frac{z_1 + z_2 - 2z_1 z_2}{z_1 z_2(1-z_1)(1-z_2)} 	
		+ (e_1^2 - e_2^2) \P \frac{z_1 - z_2}{z_1 z_2(1-z_1)(1-z_2)} 
	\nn\\
	& \hspace{1em} 
		+ \P \frac{2 e_1 e_2}{z_1 z_2 (1-z_1) (1-z_2) } \cdot
		\frac{ \pp{ z_1(1-z_1) + z_2(1-z_2) } \pp{ z_1 z_2 + (1-z_1)(1-z_2) } }{
			\pp{ 2 z_1 z_2 + (1 - c_\theta) (1-z_1-z_2) } 
			\pp{ 2 z_1 z_2 + (1 + c_\theta) (1 - z_1 - z_2) }}
	\nn\\
	& \hspace{1em} 
	 	- i\pi \bigg\{ 
			\pp{e_1-e_2}^2 \pp{\frac{\delta (1-z_1)}{z_2} + \frac{\delta (z_1)}{1-z_2}}
			- (e_1^2 - e_2^2) \frac{2}{s_{\theta}^2} \pp{ \frac{\delta (z_1)}{1-z_2} - \frac{\delta(1-z_1)}{z_2} }
			\nn\\
			& \hspace{4em}
			+ 2 e_1 e_2 \bigg[
				\frac{1 + s_{\theta}^2}{s_{\theta}^2}
					\pp{ \frac{\delta(1-z_1)}{z_2} + \frac{\delta (z_1)}{1-z_2} }
					+ \frac{1}{s_{\theta}^2} \times
					\nn\\
					& \hspace{6em} 
					\times
					\bigg( \sgn{c^2_{\theta/2} - z_2} \, \delta(z_1 - \rho(z_2)) 
						\bigg( \frac{c_{\theta}}{z_2(1-z_2)} - \frac{1}{c^2_{\theta/2} - z_2} \bigg)
						 \nn\\
						& \hspace{7.5em}
						+ \sgn{z_2 - s^2_{\theta/2}} \, \delta(z_1 - \wt{\rho}(z_2)) 
						\bigg( \frac{c_{\theta}}{z_2(1-z_2)} - \frac{1}{z_2 - s^2_{\theta/2}} \bigg)
					\bigg)
			\bigg]
		\bigg\},
\label{eq:diphoton-C-}		\\
	&2\xi\Ct_+ = 
		\pp{e_1-e_2}^2 \frac{ -2 }{ s_{\theta}^2 } \cdot
 			\P\frac{1 - z_1-z_2}{z_1\, z_2\, (1-z_1)(1-z_2)}
	\nn\\
	& \hspace{1em}
		-\frac{2 \pi i}{ s_{\theta}^2 }  \cdot \bb{
			(e_1^2 - e_2^2) \pp{ \frac{\delta(1-z_1)}{ 1 - z_2} + \frac{\delta (z_1)}{z_2} }
			- 2 e_1 e_2  \pp{ \frac{\delta (z_1)}{z_2}  -  \frac{\delta(1-z_1)}{ 1 - z_2} }
		},
\label{eq:diphoton-Ct+}	\\
	&2\xi\Ct_- =
 		\pp{e_1-e_2}^2 \frac{2c_{\theta}}{ s_{\theta}^2 } 
			\cdot \P \frac{z_1 - z_2}{z_1 z_2 (1-z_1)(1-z_2)}
		\nn\\
	& \hspace{1em}
		+ 
			\P \frac{2 e_1 e_2 \, c_{\theta}}{z_1 z_2 (1-z_1) (1-z_2) } \cdot
			\frac{ (z_1-z_2)(1-z_1-z_2)^2 }{ 
				\pp{ 2 z_1 z_2 + (1 - c_\theta) (1-z_1-z_2) } 
				\pp{ 2 z_1 z_2 + (1 + c_\theta) (1-z_1-z_2) } 
			} \nn\\
	&\hspace{1em} 
			- \frac{2 \pi i }{s_{\theta}^2} 
			\bigg\{ 
				(e_1^2 - e_2^2) c_\theta 
					\pp{ \frac{\delta(1-z_1)}{z_2} + \frac{\delta (z_1)}{1-z_2} }
				- e_1 e_2 
				\bb{ 
					c_\theta \pp{ \frac{\delta (z_1)}{1-z_2} - \frac{\delta(1-z_1)}{z_2} }
					\right.\nn\\
					&\hspace{4em}\left.
					+ \frac{z_1 - z_2}{z_2 (1 - z_2)}
					\pp{ 
						\sgn{ c^2_{\theta/2} - z_2} \delta(z_1-\rho(z_2)) - \sgn{s^2_{\theta/2} - z_2} \delta(z_1-\wt{\rho}(z_2)) 
					}
				}
			\bigg\},	
\label{eq:diphoton-Ct-}	
\end{align}
where $\P$ indicates that the hard coefficients should be understood in the sense of principle-value integration 
for $z_1$ (or $x$), when convoluted with the GPD and DA, and we used the notation 
$(c_{\theta}, s_{\theta}, c_{\theta/2}, s_{\theta/2}) = (\cos\theta, \sin\theta, \cos(\theta/2), \sin(\theta/2))$.

The special gluon propagators in the type-$A$ diagrams (cf. Eqs.~\eqref{eq:diphoton-g2-A}\eqref{eq:diphoton-g2-A'}) 
introduce new poles of $z_1$ in addition to 0 and 1,
\bse\label{eq:diphoton-special-poles-z}\begin{align}
	\rho(z_2) &\, = \frac{(1 + \cos\theta)(1 - z_2) }{1 + \cos\theta - 2 z_2} 
		= \frac{\cos^2(\theta/2)(1 - z_2) }{\cos^2(\theta/2) -  z_2},
	\\
	\wt{\rho}(z_2) & \, = \frac{(1 - \cos\theta)(1 - z_2) }{1 - \cos\theta - 2 z_2} 
		= \frac{\sin^2(\theta/2)(1 - z_2) }{\sin^2(\theta/2) -  z_2}
		= 1 - \rho(1 - z_2),
\end{align}\ese
which have small imaginary parts by the $i\epsilon$ prescription,
\beq
	i \epsilon \, \sgn{ z_2 - \cos^2(\theta/2)},
	\quad
	i \epsilon \, \sgn{ z_2 - \sin^2(\theta/2)},
\eeq
respectively. In terms of $x = \xi (2 z_1 - 1)$, \eq{eq:diphoton-special-poles-z} translates to the special poles,
\bse\label{eq:diphoton-special-poles-x}\begin{align}
	x_p(\xi, z, \theta)
		& = \xi \cdot \bb{ \frac{(1+\cos\theta)(1-z) + (1-\cos\theta) z}{(1+\cos\theta)(1-z) - (1-\cos\theta) z} }
		= \xi \cdot \bb{ \frac{1 - z + \tan^2(\theta / 2) z}{1 - z - \tan^2(\theta / 2) z} },
	\\
	\wt{x}_p(\xi, z, \theta)
		& = \xi \cdot \bb{ \frac{\tan^2(\theta / 2)(1 - z) + z}{\tan^2(\theta / 2)(1 - z) - z} }
		= - x_p(\xi, 1 - z, \theta).
\end{align}\ese
As $z$ goes from $0$ to $1$, the two poles go from $\xi$ to $\infty$ on the lower half complex $x$ plane, and
then from $-\infty$ to $-\xi$ on the upper half plane. They then cover the whole DGLAP regions of GPDs.

We immediately see from Eqs.~\eqref{eq:diphoton-C+}--\eqref{eq:diphoton-Ct-} that 
the $e_1^2$ or $e_2^2$ proportional terms, which come from type-$B$ diagrams, 
only carry moment-type sensitivity, 
while the $e_1 e_2$ terms coming also from type-$A$ diagrams carry enhanced $z_1$ (or $x$) sensitivity. 
We have organized Eqs.~\eqref{eq:diphoton-C+}--\eqref{eq:diphoton-Ct-} in terms of 
$(e_1-e_2)^2$, $(e_1^2 - e_2^2)$, and $e_1 e_2$,
such that if we had a neutral pion beam, $e_1 = e_2 = e_u$ or $e_d$, and then the moment-type terms proportional
to $(e_1-e_2)^2$ or $(e_1^2 - e_2^2)$ would be cancelled, further enhancing the $x$-sensitivity in $e_1 e_2$ terms.

Charge conjugation symmetry amounts to $(z_1, z_2) \leftrightarrow (1-z_1, 1-z_2)$ 
(or equivalently, $(x, z) \leftrightarrow (-x, 1-z)$) and $e_1 \leftrightarrow e_2$.
Due to the presence of two $\gamma_5$ matrices in the evaluation of $C_{\pm}$, the charge-conjugation-even (C-even)
$(e_1-e_2)^2$ and $e_1 e_2$ are manifestly invariant under $(z_1, z_2) \leftrightarrow (1-z_1, 1-z_2)$,
while the charge-conjugation-odd (C-odd) $(e_1^2 - e_2^2)$ term flips a sign. 
Given the symmetry property $D(z) = D(1-z)$ of the DA, the first two terms are probing 
the C-even GPD combination $\wt{F}^+(x, \xi, t) \equiv \wt{F}(x, \xi, t) + \wt{F}(-x, \xi, t)$, whereas
the $(e_1^2 - e_2^2)$ term is probing the C-odd GPD combination 
$\wt{F}^-(x, \xi, t) \equiv \wt{F}(x, \xi, t) - \wt{F}(-x, \xi, t)$. 
Similarly, since there is only one $\gamma_5$ matrix in the evaluation of $\wt{C}_{\pm}$, the C-even 
$(e_1-e_2)^2$ and $e_1 e_2$ terms flip signs under $(z_1, z_2) \leftrightarrow (1-z_1, 1-z_2)$,
while the C-odd $(e_1^2 - e_2^2)$ term is manifestly invariant. 
As a result, the first two terms are probing the C-even GPD combination 
$F^+(x, \xi, t) \equiv F(x, \xi, t) - F(-x, \xi, t)$, whereas
the $(e_1^2 - e_2^2)$ term is probing the C-odd GPD combination 
$F^-(x, \xi, t) \equiv F(x, \xi, t) + F(-x, \xi, t)$.

The $p\pi^-$ and $n\pi^+$ channels differ from each other by changing $(e_1, e_2) = (e_u, e_d)$ to $(e_d, e_u)$, 
which does not affect the $(e_1-e_2)^2$ and $e_1 e_2$ terms, but flips the signs of the $(e_1^2 - e_2^2)$ terms.
So combining both channels helps distinguish C-even and C-odd GPD components.

\subsection{Helicity amplitudes}
\label{ssec:diphoton-amplitudes}
The convolutions of the hard coefficients in Eqs.~\eqref{eq:diphoton-C+}--\eqref{eq:diphoton-Ct-} 
with the GPD and DA can be simplified by using symmetry property of the DA.
Specifically, by introducing the notation
\begin{align}
	C_{\alpha}^{[\wt{F}]} &\, \equiv \int_{-1}^1 dx \int_0^1 dz \, 
		\wt{F}(x, \xi, t) \, D(z) \, C_{\alpha}(x, \xi; z; \theta),
	\nn\\
	\wt{C}_{\alpha}^{[F]} &\, \equiv \int_{-1}^1 dx \int_0^1 dz \, 
		F(x, \xi, t) \, D(z) \, \wt{C}_{\alpha}(x, \xi; z; \theta),
	\label{eq:C-conv-short}
\end{align}
with $\alpha$ being any single or double helicity indices,
we have
\bse\label{eq:diphoton-convolution}\begin{align}
	C_+^{[\wt{F}]} = &
		- \frac{2 D_0}{\sin^2\theta} \cc{
			\pp{e_1-e_2}^2 \, \wt{F}^+_0(\xi, t)
			+ i\pi \bb{
				(e_1^2-e_2^2) \, \wt{F}^-(\xi, \xi, t)
				+ 2 e_1 e_2 \, \wt{F}^+(\xi, \xi, t)
			}
		},
\label{eq:diphoton-M+}	\\
	C_-^{[\wt{F}]}  = &
		- (e_1 - e_2)^2 \, D_0 \bb{ \frac{2}{\sin^2\theta} \, \wt{F}^+_0(\xi, t) 
				+ i \pi \wt{F}^+(\xi, \xi, t)
		}	\nn\\
	& - (e_1^2-e_2^2) \, D_0 \bb{ \wt{F}^-_0(\xi, t) + \frac{2i\pi}{\sin^2\theta} \wt{F}^-(\xi, \xi, t) }
		\nn\\
	& + e_1 e_2	\cc{
		\int_0^1dz \frac{D(z)}{z(1-z)}
			\bb{ \frac{1+\cos\theta-2z}{\sin^2\theta} - \frac{1}{1+\cos\theta-2z} } 
			\cdot I[\wt{F}^+; \xi, t, z, \theta]
		\right.
		\nn\\
		&\left. \hspace{4em} 
		- \frac{2 D_0}{\sin^2\theta} \cdot 
		\bb{ \wt{F}^+_0(\xi, t)
	 		+ i \pi \, (1 + \sin^2\theta) \, \wt{F}^+(\xi, \xi, t)
		}
	},
\label{eq:diphoton-M-}	\\
	\wt{C}_+^{[F]} = &
		-\frac{2D_0}{\sin^2\theta} \cc{
			(e_1 - e_2)^2 \, F^+_0(\xi, t)
			+ i \pi \bb{
				(e_1^2-e_2^2) \, F^-(\xi, \xi, t)
				+ 2 e_1 e_2 \, F^+(\xi, \xi, t)
			}
		},
\label{eq:diphoton-Mt+}	\\
	\wt{C}_-^{[F]} = &
		- \frac{2\cos\theta}{\sin^2\theta} \cdot D_0 \bb{
			(e_1 - e_2)^2 \, F^+_0(\xi, t)
			+ i \pi \, (e_1^2-e_2^2) \, F^-(\xi, \xi, t)
		}
		\nn\\
		&+ e_1 e_2 \cc{
			\int_0^1dz \frac{D(z)}{z(1-z)}
				\bb{ \frac{1+\cos\theta-2z}{\sin^2\theta} + \frac{1}{1+\cos\theta-2z} } \cdot
				I[F^+; \xi, t, z, \theta]
				\right.\nn\\
				&\left. \hspace{4em} - \frac{2 \cos\theta \, D_0}{\sin^2\theta} \cdot 
				\bb{ 
					F^+_0(\xi, t) + i \pi \, F^+(\xi, \xi, t)
				}
		},
\end{align}\ese
where we defined the “zeroth moments” of the DA and GPDs,
\beq[eq:GPD-DA-moments]
	D_0 \equiv \int_0^1 \frac{dz \, D(z)}{z}, 
	\quad
	\F_0(\xi, t) \equiv \P \int_{-1}^1 \frac{dx \, \F(x, \xi, t)}{x - \xi},
\eeq
and the special GPD integral,
\beq[eq:diphoton-special-int]
	I[\F; \xi, t, z, \theta] 
		\equiv 
		\int_{-1}^{1}dx \frac{\F(x, \xi, t)}{x - x_p(\xi, z, \theta) + i \epsilon \, \sgn{\cos^2(\theta/2) - z} },
\eeq
where $\F$ can take any GPD function such as $F^{\pm}$, $\wt{F}^{\pm}$, $H^{\pm}$, etc..
The special integral in \eq{eq:diphoton-special-int} yields a function of $z$ that depends sensitively on the $x$ distribution
of the GPD. On a further integration of $z$ under the profiling of a given DA, this maps out a distribution of $\theta$ that
contains enhanced sensitivity to the GPD $x$ dependence. Since the pole $x_p$ lies on the DGLAP region, the enhanced
sensitivity is mainly on that region.

\subsection{Cross section and single nucleon spin asymmetry}
\label{ssec:diphoton-cross-section}
First, we write \eq{eq:diphoton-factorize} explicitly as
\beq[eq:diphoton-amplitude-factorize]
	\M_{\lambda_1\lambda_2} = \frac{1}{2 P^+} \bar{u}(p', \alpha')
		\bb{ \Ct_{\lambda_1\lambda_2}^{[H]} \, \gamma^+ 
			- \Ct_{\lambda_1\lambda_2}^{[E]} \, \frac{i \sigma^{+\alpha} \Delta_{\alpha}}{2m}
			+ C_{\lambda_1\lambda_2}^{[\wt{H}]} \, \gamma^+ \gamma_5
			- C_{\lambda_1\lambda_2}^{[\wt{E}]} \, \frac{\gamma_5 \Delta^+}{2m}
		} u(p, \alpha),
\eeq
which is written in the SDHEP frame and
where the subscripts $(\lambda_1, \lambda_2)$ refer to the two final-state photon helicities.
As argued in \sec{sec:SDHEP-frame}, because the Lorentz transformation connecting the Lab and SDHEP frames is a transverse boost,
the factorization formula [\eq{eq:photo-factorize}] can be written equally in the Lab and SDHEP frames, 
and the corresponding GPDs $F$, $\wt{F}$, $H$, $E$, $\wt{H}$, and $\wt{E}$ take the same values in both frames.
This has allowed us to write the coefficients $C$ and $\wt{C}$, which contain the kinematics of the final state, in the SDHEP frame.

Now, we omit the photon helicities in the notations,
sum over the initial-state nucleon spin, and average over the initial-state nucleon spin using the density matrix,
\beq
	\rho_{\alpha\alpha'}(\bm{s}_T, \lambda) = \frac{1}{2} ( 1 + \bm{s} \cdot \bm{\sigma} )
	= \frac{1}{2} 	
		\begin{pmatrix}
			1 + \lambda & s_x - i s_y \\
			s_x + i s_y & 1 - \lambda
		\end{pmatrix},
\eeq
where $\bm{s} = (s_x, s_y, \lambda) = (s_T \cos\phi_S, s_T \sin\phi_S, \lambda)$ is the spin 
Bloch vector of the initial-state nucleon. The spin average can be conveniently done by introducing a covariant spin vector 
$S^{\mu}$ via
\beq
	\sum_{\alpha, \alpha'} u(p, \alpha) \, \rho_{\alpha \alpha'}(\bm{s}_T, \lambda) \, \bar{u}(p, \alpha')
	= \frac{1}{2}(\gamma\cdot p + m) \pp{ 1 + \gamma_5 \gamma\cdot S / m },
\eeq
with $S^{\mu} = (\lambda p^+, - \lambda m^2 / (2p^+), m \bm{s}_T)$ in the light-front coordinates in the Lab frame.
It is important to notice that $S$ enters the calculation at most in a linear way.
The evaluation of the spinor algebra thus involves three Lorentz invariants associated with $S$,
\begin{align}
	& p \cdot S = 0, \quad
	p' \cdot S = \frac{\lambda}{2} \pp{ -t + \frac{4 \xi m^2}{1 + \xi} } + m \, \bm{s}_T \cdot \bm{\Delta}_T	, \nn\\
	&\epsilon^{n p p' S} = m p^+ (\bm{s}_T \times \bm{\Delta}_T)^z 
		= - m p^+ s_T \Delta_T \sin\phi_S.
\end{align}
Then the nucleon-spin averaged amplitude square can be expressed as
\begin{align}\label{eq:diphoton-M2}
	\overline{\abs{\M}^2} = \langle \M \M^* \rangle_{NN'}
		= \A_0 + \lambda \, \A_1 + \frac{\bm{s}_T\cdot \bm{\Delta}_T}{2 m} \, \A_2^1 
			+ \frac{(\bm{s}_T\times \bm{\Delta}_T)^z}{2 m} \, \A_2^2,
\end{align}
where the photon helicities are left open but suppressed. 
The nucleon spin independent part $\A_0$ is
\begin{align}\label{eq:diphoton-A0}
& \A_0 
	= (1 - \xi^2) \pp{ \wt{C}^{[H]} \wt{C}^{[H] *} + C^{[\wt{H}]} C^{[\wt{H}] *} }
		- \pp{ \xi^2 + \frac{t}{4m^2}} \wt{C}^{[E]} \wt{C}^{[E] *} 	
		- \frac{\xi^2 t}{4m^2} C^{[\wt{E}]} C^{[\wt{E}] *}	
	\nn\\
& \hspace{4em}
		- \xi^2 \pp{ \wt{C}^{[H]} \wt{C}^{[E] *} + \wt{C}^{[E]} \wt{C}^{[H] *}
			+ C^{[\wt{H}]} C^{[\wt{E}] *} + C^{[\wt{E}]} C^{[\wt{H}] *} 
		}.
\end{align}
The nucleon helicity dependent part $\A_1$ is
\begin{align}\label{eq:diphoton-A1}
& \A_1 
	= (1 - \xi^2) \pp{ \wt{C}^{[H]} C^{[\wt{H}] *} + C^{[\wt{H}]} \wt{C}^{[H] *} }
		- \xi \pp{ \frac{t}{4m^2} + \frac{\xi^2}{1 + \xi} }
			\pp{ \wt{C}^{[E]} C^{[\wt{E}] *} + C^{[\wt{E}]} \wt{C}^{[E] *} }
	\nn\\
& \hspace{4em}
	- \xi^2 \pp{ \wt{C}^{[H]} C^{[\wt{E}] *} + C^{[\wt{E}]} \wt{C}^{[H] *} 
		+ C^{[\wt{H}]} \wt{C}^{[E] *} + \wt{C}^{[E]} C^{[\wt{H}] *} }.
\end{align}
The nucleon transverse spin dependent parts $\A_2^1$ and $\A_2^2$ are
\bse\label{eq:diphoton-A2}\begin{align}
& \A_2^1 
	= (1 + \xi) \, \bb{ C^{[\wt{H}]} \wt{C}^{[E] *} + \wt{C}^{[E]} C^{[\wt{H}] *}
		- \xi \pp{ \wt{C}^{[H]} C^{[\wt{E}] *} + C^{[\wt{E}]} \wt{C}^{[H] *} } }
	\nn\\
& \hspace{4em}
	- \xi^2 \, \pp{ \wt{C}^{[E]} C^{[\wt{E}] *} + C^{[\wt{E}]} \wt{C}^{[E] *}}, 	\\
& \A_2^2
	= i \, (1 + \xi) \, \bb{ \wt{C}^{[H]} \wt{C}^{[E] *} - \wt{C}^{[E]} \wt{C}^{[H] *}
		- \xi \pp{ C^{[\wt{H}]} C^{[\wt{E}] *} - C^{[\wt{E}]} C^{[\wt{H}] *} }}.
\end{align}\ese
In Eqs.~\eqref{eq:diphoton-A0}--\eqref{eq:diphoton-A2}, the omitted photon helicity indices take 
$(\lambda_1, \lambda_2)$ in the helicity amplitudes $C$ and $\wt{C}$, 
and $(\lambda_1', \lambda_2')$ in the complex conjugate ones $C^*$ and $\wt{C}^*$.

The values of $\xi$ and $t$ are limited in a certain experiment for a fixed collision energy. By introducing 
an upper cut for $t$, say $|t| \le 0.2~\GeV^2$, we have $\xi \leq 0.218$, 
which suppresses all the $E$ and $\wt{E}$ related terms in $\A_0$ and $\A_1$ by at least a factor of $0.048$.
Neglecting the $\xi^2$ and $t$ suppressed terms, we have the approximations for 
Eqs.~\eqref{eq:diphoton-A0}--\eqref{eq:diphoton-A2},
\bse\label{eq:diphoton-Ai-approx}\begin{align}
	\A_0 & = (1 - \xi^2) \pp{ \wt{C}^{[H]} \wt{C}^{[H] *} + C^{[\wt{H}]} C^{[\wt{H}] *} } + \mathcal{O}(\xi^2, \, t/m^2),\\
	\A_1 & = (1 - \xi^2) \pp{ \wt{C}^{[H]} C^{[\wt{H}] *} + C^{[\wt{H}]} \wt{C}^{[H] *} } + \mathcal{O}(\xi^2, \, \xi t/m^2),\\
	\A_2^1 & = (1 + \xi) \bb{ C^{[\wt{H}]} \wt{C}^{[E] *} + \wt{C}^{[E]} C^{[\wt{H}] *}
			- \xi \pp{ \wt{C}^{[H]} C^{[\wt{E}] *} + C^{[\wt{E}]} \wt{C}^{[H] *} } } + \mathcal{O}(\xi^2), \\
	\A_2^2 & = i \, (1 + \xi) \, \bb{ \wt{C}^{[H]} \wt{C}^{[E] *} - \wt{C}^{[E]} \wt{C}^{[H] *}
		- \xi \pp{ C^{[\wt{H}]} C^{[\wt{E}] *} - C^{[\wt{E}]} C^{[\wt{H}] *} }},
\end{align}\ese
so that the leading effect of the unpolarized cross section $\A$ and helicity asymmetry $\A_1$ is from the GPDs 
$H$ and $\wt{H}$, whereas that of the transverse spin asymmetries $\A_2^{1,2}$ also receives contribution from 
the GPDs $E$ and $\wt{E}$.

The helicity amplitude structure in \eq{eq:diphoton-helicity-structure} encodes various polarization degrees of freedom 
of the two photons. These are, unfortunately, not measured at a collider detector. The only polarization observable is of
the initial-state nucleon. Tracing over the two photon helicities in \eq{eq:diphoton-Ai-approx}, we get
\bse\label{eq:diphoton-Ai-avg}\begin{align}
	\A_0 & = 2 (1 - \xi^2) \pp{\frac{N}{\hat{s}} }^2 \pp{ | C_+^{[\wt{H}]} |^2 + | C_-^{[\wt{H}]} |^2 + | \wt{C}_+^{[H]} |^2 + | \wt{C}_-^{[H]} |^2 }
				+ \mathcal{O}(\xi^2, \, t/m^2),\\
	\A_2^2 & = -4(1 + \xi) \pp{\frac{N}{\hat{s}} }^2 \Im{ \wt{C}_+^{[H]} \wt{C}_+^{[E] *} + \wt{C}_-^{[H]} \wt{C}_-^{[E] *} 
				- \xi \pp{ C_+^{[\wt{H}]} C_+^{[\wt{E}] *} + C_-^{[\wt{H}]} C_-^{[\wt{E}] *} } },\\
	\A_1 & = \A_2^1 = 0.
\end{align}\ese
Without observing the photon polarizations, the only nucleon spin asymmetry is with respect to its transverse spin
component perpendicular to the diffraction plane, which gives a $\sin\phi_S$ correlation.

Inserting \eq{eq:diphoton-M2} to \eq{eq:sdhep-xsec}, with the explicit forms in \eq{eq:diphoton-Ai-avg}, and integrating
over the trivial $\phi$ dependence, we have the differential cross section of the diphoton production,
\begin{align}
	\frac{d\sigma}{d|t| \, d\xi \, d\phi_S \, d\cos\theta}
	= \frac{1}{2\pi} 
		\frac{d\sigma}{d|t| \, d\xi \, d\cos\theta}
		\bb{ 1 + s_T \, A_N(t, \xi, \cos\theta) \, \sin\phi_S },
\end{align}
where
\beq[eq:diphoton-unpol-xsec]
	\frac{d\sigma}{d|t| \, d\xi \, d\cos\theta} 
	= 2\pi \pp{ \alpha_e \alpha_s \frac{C_F}{N_c} }^2 \frac{1 - \xi^2}{\xi^2 \, s^3}
		\Sigma_{U}
\eeq
is the unpolarized differential cross section, with
\beq[eq:diphoton-sigma-U]
	\Sigma_{U} = | C_+^{[\wt{H}]} |^2 + | C_-^{[\wt{H}]} |^2 + | \wt{C}_+^{[H]} |^2 + | \wt{C}_-^{[H]} |^2 
		+ \mathcal{O}(\xi^2, \, t/m^2),
\eeq
and 
\beq
	A_N = \frac{\Delta_T / m }{(1 - \xi) \Sigma_{U}}
		\Im{ \wt{C}_+^{[H]} \wt{C}_+^{[E] *} + \wt{C}_-^{[H]} \wt{C}_-^{[E] *} 
				- \xi \pp{ C_+^{[\wt{H}]} C_+^{[\wt{E}] *} + C_-^{[\wt{H}]} C_-^{[\wt{E}] *} } }
\eeq
is the single transverse spin asymmetry (SSA) of the initial-state nucleon. Here $\Delta_T$ can be determined from 
$t$ and $\xi$ by
\beq
	\Delta_T = \frac{\sqrt{-(1 - \xi^2)t - 4 \xi^2 m^2}}{1 + \xi}.
\eeq
To the leading accuracy, the unpolarized cross section allows to probe the GPDs $H$ and $\wt{H}$, while
the SSA gives an opportunity to probe the GPD $E$ and $\wt{E}$, both with enhanced $x$-sensitivity.

\subsection{Numerical results}
\label{ssec:diphoton-numerical}

Now we present some numerical results for the diphoton production process, especially on the
enhanced $x$-sensitivity. 
We consider the pion beam as accessed at the J-PARC~\citep{Aoki:2021cqa} or AMBER~\citep{Adams:2018pwt} experiment,
with an energy $E_{\pi} = 20$ or $150~\GeV$, respectively.
Both facilities only probe unpolarized nucleon targets, 
so only give the differential cross section observable in \eq{eq:diphoton-unpol-xsec},
which receives contributions from the GPDs $H$ and $\wt{H}$.
For simplicity, we fix the DA to the asymptotic form with $\phi(z) = 6z(1-z)$, 
and renormalization and factorization scales to $2~\GeV$, turning off QCD evolution effects.

\begin{figure}[htbp]
	\centering
		\includegraphics[scale=0.665]{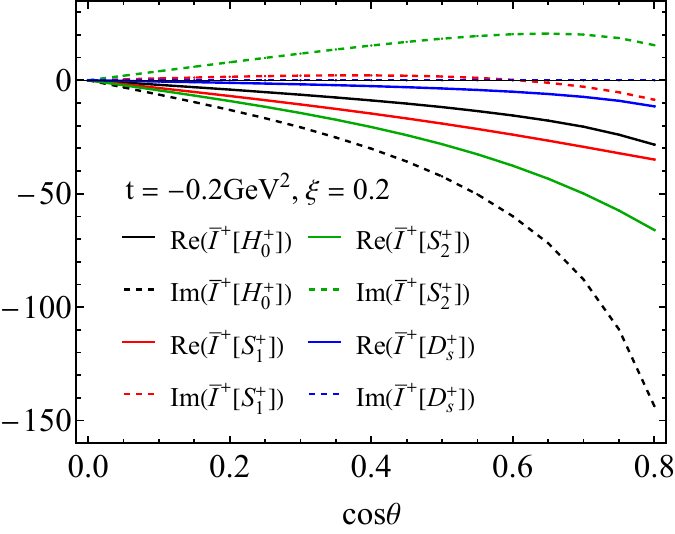}\hspace{1em}
		\includegraphics[scale=0.65]{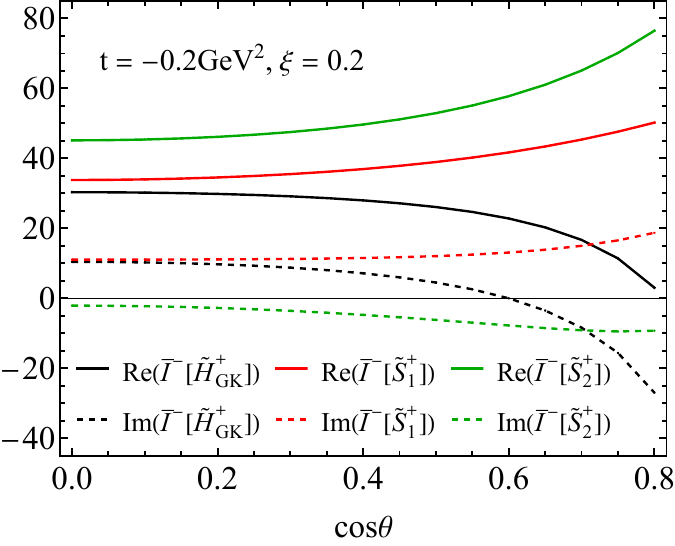}	
	\caption{The special integral in \eq{eq:diphoton-profiled-z-int}, evaluated for the $u$ quark GPDs in the GK model and
	the shadow GPDs defined in Eqs.~\eqref{eq:shadow-S12}~\eqref{eq:shadow-St12} and \eqref{eq:GPD-model-Hh-3}.
	On the left are $a = 1$ for the unpolarized GPD $H$ and associated shadow GPDs, while
	on the right are $a = -1$ for the polarized GPD $\wt{H}$ and associated shadow GPDs.
	The superscripts ``$+$'' refer to C-even components defined as \eqs{eq:C-even-F}{eq:C-even-Ft}.
	By symmetry, $\bar{I}^+$ is odd with respect to $\cos\theta$ and $\bar{I}^-$ is even, 
	so only the parts with $\cos\theta > 0$ are shown.
	}
\label{fig:diphoton-special-int}
\end{figure}

The main thing of our interest is how the special integral in \eq{eq:diphoton-special-int} helps to
distinguish the GPDs from shadow GPDs. 
Since this integral is to be convoluted with the DA,
we further define the profiled special integral,
\beq[eq:diphoton-profiled]
	\bar{I}^a[\phi, \F; \xi, t, \theta]
	= \int_0^1dz \frac{\phi(z)}{z(1-z)}
			\bb{ \frac{1+\cos\theta-2z}{\sin^2\theta} + \frac{a}{1+\cos\theta-2z} } \cdot
			I[\F; \xi, t, z, \theta],
\eeq
where we have temporarily removed the constant coefficient in the DA,
and $a = 1$ corresponds to $\F = H^+$ and $-1$ to $\F = \wt{H}^+$.
With the asymptotic DA, the $z$ variable in \eq{eq:diphoton-profiled} can be readily integrated to give
\begin{align}
	\bar{I}^a[\phi, \F; \xi, t, \theta] = \int_{-1}^{1}\dd{x} \F(x, \xi, t) K^a(x, \xi, \cos\theta),
\label{eq:diphoton-profiled-z-int}
\end{align}
with the kernel $K^a(x, \xi, \cos\theta)$ being
\begin{align}
	K^a(x, \xi, c) 
	&= 6 \cc{ a \, L(x, \xi, c)
		+ \frac{\xi^2(1-c^2)}{(x - \xi c)^2} 
			\bb{ L(x, \xi, c) + \frac{1}{\xi(1-c^2)} + \frac{c (x - \xi c)}{\xi^2 (1-c^2)^2}
			}	\right.\nn\\
	&\hspace{6em} \left.
		+ \, i \pi \cdot \theta(x^2 - \xi^2) \, {\rm sgn}(x) \cdot \frac{1}{-2(x - \xi c)}
		\bb{ \frac{\xi^2 (1 - c^2)}{(x - \xi c)^2} + a }
		},
\label{eq:diphoton-x-kernel}
\end{align}
where $c \equiv \cos\theta$ and 
\beq
	L(x, \xi, c) = \frac{1}{-2(x - \xi c)} \ln\abs{\frac{(1 - c)(x + \xi)}{(1 + c)(x - \xi)}}.
\eeq
At given $t$ and $\xi$, \eq{eq:diphoton-x-kernel} can be integrated for different GPDs to yield
distributions of $\cos\theta$.
In \fig{fig:diphoton-special-int} we show results of the special integral in \eq{eq:diphoton-profiled-z-int} for 
both the ``standard'' GPDs in the GK model and the shadow GPDs and shadow $D$-term. 
An interesting feature can be immediately noticed: 
On the left-hand side, the imaginary part of the integral $\bar{I}^+$ for $H^+_0$ is much greater than
all the others and the large oscillation in the DGLAP region of the shadow GPDs diminish 
their integrals a lot, especially for the imaginary parts;
whereas on the right-hand side, different components of the integration results are of similar orders,
and the shadow GPD integrals supersede the GK model ones.

\begin{figure}[htbp]
	\centering
		\includegraphics[scale=0.65]{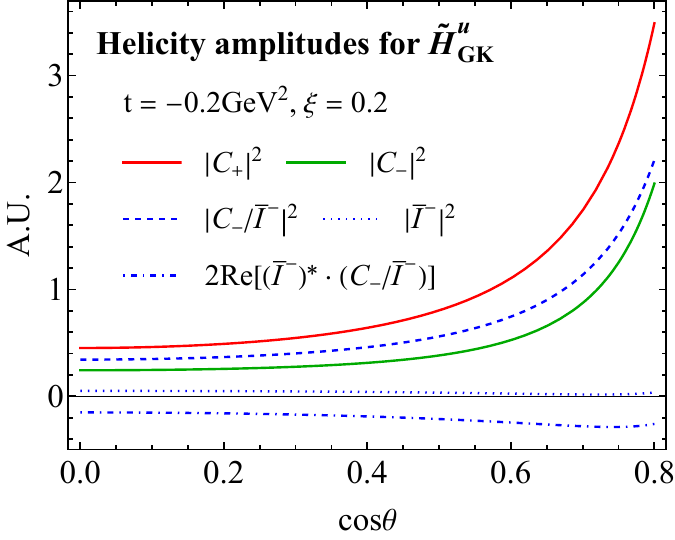}\hspace{1em}
		\includegraphics[scale=0.65]{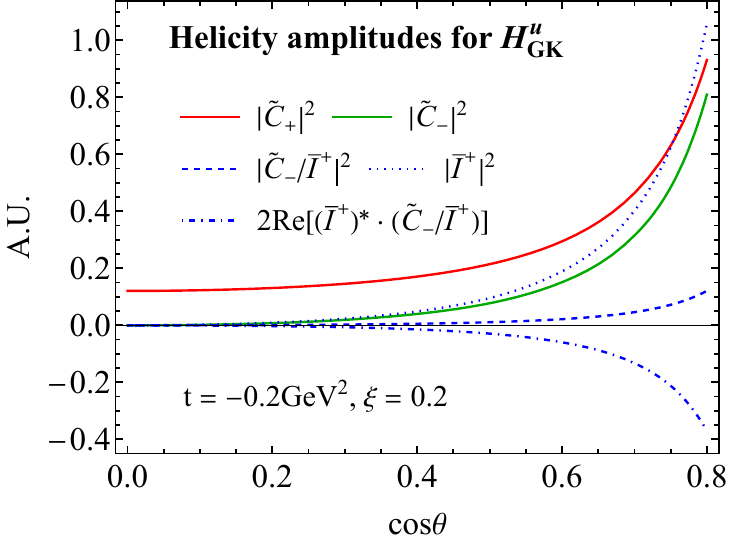}	
	\caption{Helicity amplitudes squared of (in arbitrary scales) $C_{\pm}^{[\wt{H}]}$ and $\wt{C}_{\pm}^{[H]}$ 
	in \eq{eq:diphoton-convolution} evaluated at $t = -0.2~\GeV^2$ and $\xi = 0.2$
	for the $u$ quark GPD $\wt{H}^u_{\rm GK}$ and $H^u_{\rm GK}$ of the GK model, respectively.
	The amplitudes $C_{-}^{[\wt{H}]}$ and $\wt{C}_{-}^{[H]}$ contain the special integrals 
	$\bar{I}^-[\wt{H}]$ and $\bar{I}^+[H]$, and we have displayed three separate contributions.
	On the left figure, $C_- / \bar{I}^-$ refers to the amplitude $C_{-}^{[\wt{H}]}$ with contribution from
	$\bar{I}^-[\wt{H}]$ removed, and 
	on the right figure, $\wt{C}_- / \bar{I}^+$ removes the contribution from $\bar{I}^+[H]$.
	}
\label{fig:diphoton-hel-amps}
\end{figure}

When combined into the whole amplitudes, the special integrals are further weighted by $e_1 e_2$,
which is $-2/9$ for a charged pion reaction and further suppresses their contribution. 
In \fig{fig:diphoton-hel-amps}, we show the helicity amplitudes (squared) in \eq{eq:diphoton-convolution}
for the GK model, with explicitly shown the contributions from the special integrals, including
both their interference with the other contributions. 
Both figures are in arbitrary scales, but the relative sizes are exact among all the curves and between the two figures.
So the P-even amplitudes (left) are larger than the P-odd ones (right).
In both figures, the helicity-conserving amplitudes are greater than the helicity-flipping ones.
For the P-even amplitude $C_-^{[\wt{H}]}$, the contribution from the special integral is much smaller than the rest, 
while for the P-odd one $\wt{C}_-^{[H]}$, they are of similar orders.
In both cases, the special integrals have destructive interference with the other terms.

\begin{figure}[htbp]
	\centering
		\includegraphics[scale=0.68]{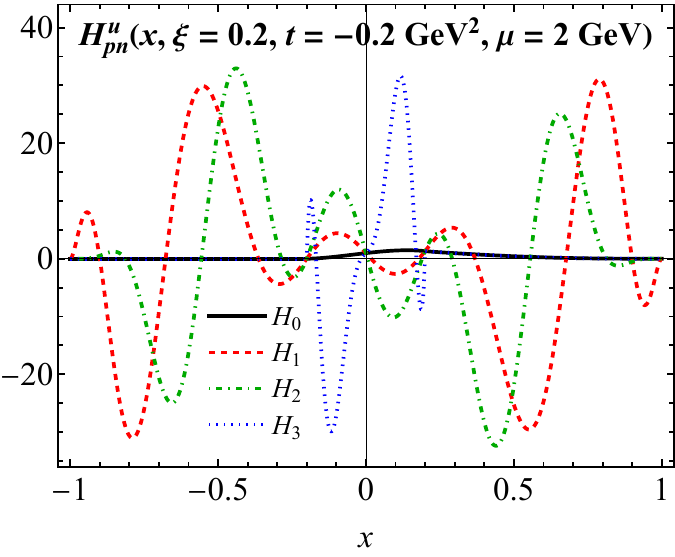}
		\includegraphics[scale=0.68]{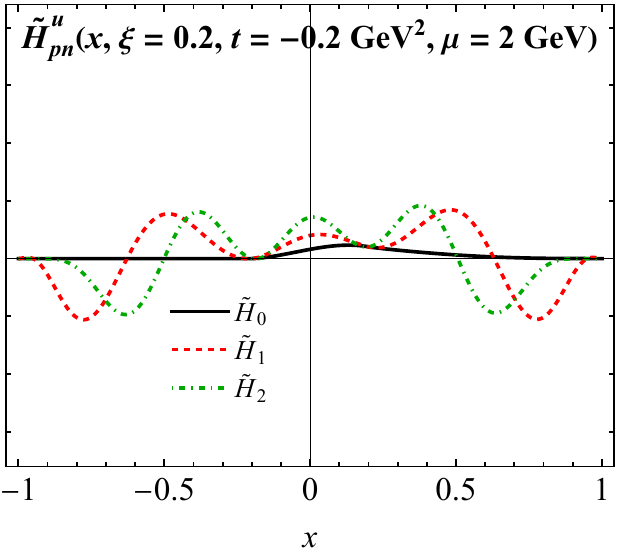}
	\caption{Flavor transition GPD models used for the diphoton production in $p\pi^-$ collision.}
\label{fig:diphoton-GPD}
\end{figure}

\begin{figure}[htbp]
	\centering
		\includegraphics[scale=0.65]{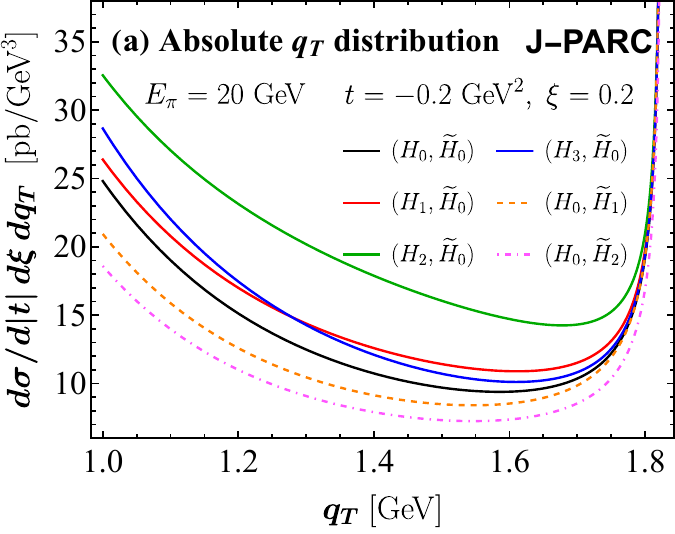} \quad
		\includegraphics[scale=0.65]{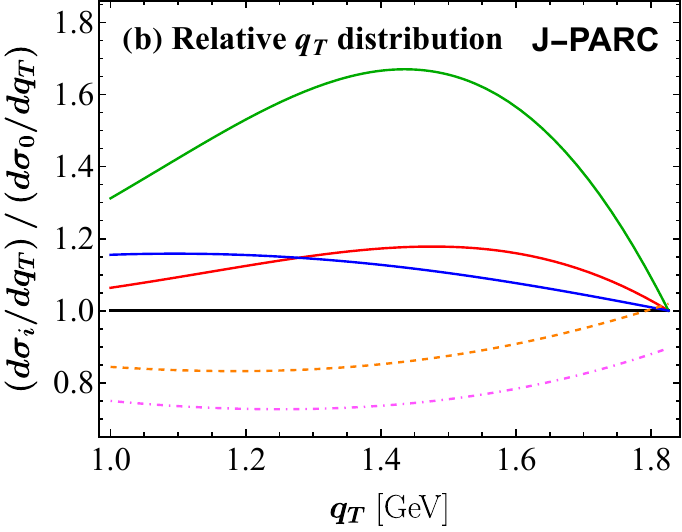}
	\caption{Unpolarized differential cross section in the transverse momentum $q_T$,
	for the diphoton diphoton production in $p\pi^-$ collision at the J-PARC energy $E_{\pi} = 20~\GeV$,
	evaluated for different GPD models in \fig{fig:diphoton-GPD} at $t = -0.2~\GeV^2$ and $\xi = 0.2$.
	(a) is the absolute $q_T$ distribution, and (b) exhibits the distribution ratio of each model to the GK model.}
\label{fig:diphoton-qt-20}
\end{figure}

\begin{figure}[htbp]
	\centering
		\includegraphics[scale=0.65]{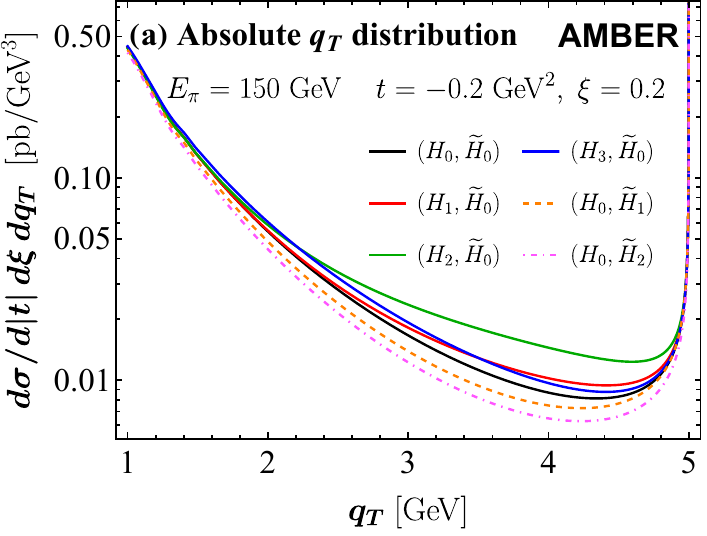} \quad
		\includegraphics[scale=0.65]{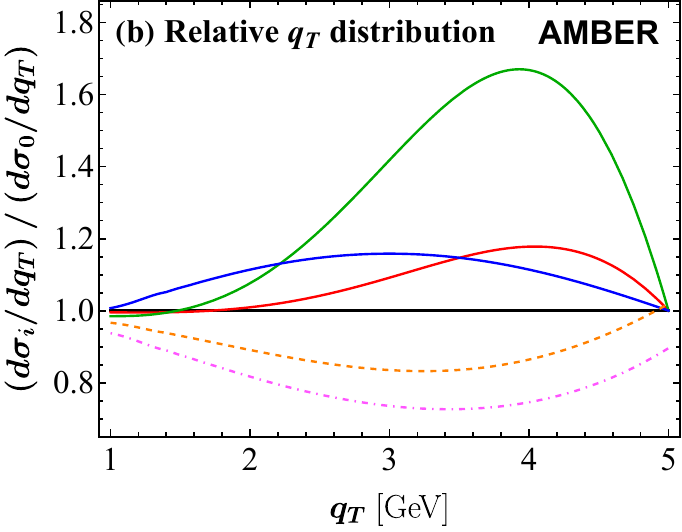}
	\caption{Same as \fig{fig:diphoton-qt-20}, but for the AMBER energy $E_{\pi} = 150~\GeV$. 
	The vertical axis of (a) is in log scale.}
\label{fig:diphoton-qt-150}
\end{figure}

Therefore, the capability of using the special integrals to distinguish different GPDs is limited.
Especially, given the fact that we only have the unpolarized cross section observable in \eq{eq:diphoton-sigma-U}
as the sum over all the amplitude squares,
which causes the large ``background'' contribution from $C_+^{[\wt{H}]}$ and $\wt{C}_+^{[H]}$.
To demonstrate the distinguishing power of the diphoton process, 
we weight the shadow GPDs by 6 in \eq{eq:GPD-model-Hh-12} and 2 in \eq{eq:shadow-St12}, and 
weight the shadow $D$-term by 24 in \eq{eq:GPD-model-Hh-3}, as a liberal choice.
This results in the flavor transition GPD models to be used for the $p\pi^-$ scattering.

The typical pion beam energy is $20~\GeV$ at the J-PARC experiment~\citep{Aoki:2021cqa}
and $150~\GeV$ at the AMBER experiment~\citep{Adams:2018pwt}.
Since the photons in the final state are identical particles, there is symmetry between the forward and backward regions,
so it is convenient to convert the $\cos\theta$ distribution in \eq{eq:diphoton-unpol-xsec} to a $q_T$ distribution,
which gives a Jacobian peak at $q_T = \sqrt{\hat{s}}/2$.
The results are shown in 
\figs{fig:diphoton-qt-20}{fig:diphoton-qt-150} for the J-PARC and AMBER energies, respectively,
for the GPD models in \fig{fig:diphoton-GPD} at the phase space point 
$(t, \xi) = (-0.2~\GeV^2, 0.2)$, including both the absolute and the relative distributions.
Importantly, both the absolute rates and relative shapes of the $q_T$ distributions are altered by the added shadow GPDs.
By the chosen weights, each one alters the $q_T$ distribution by a comparable amount.
Only by measuring the unpolarized $q_T$ distributions, 
one is more sensitive to the polarized GPD $\wt{H}$ than the unpolarized GPD $H$.
Since it requires a much larger weight for the shadow $D$-term to reach the same impact as the others, the diphoton process
is more sensitive to the DGLAP region than the ERBL region, as we have anticipated below \eq{eq:diphoton-special-int}.
Clearly, at a higher collision energy at AMBER, the cross section at each phase space point is much lower than that at J-PARC,
whereas the kinematic coverage is much wider and gives similar sensitivity in terms of the relative $q_T$ distribution shapes.

\section{Single diffractive hard exclusive photon-meson pair photoproduction}
\label{sec:photoproduction}

In this section, we study the crossing process of \eq{eq:diphoton-process},
the single diffractive hard exclusive photon-pion pair production in nucleon-photon collisions,
\beq[eq:photo-process]
	N(p) + \gamma(p_2) \to N'(p') + \pi(q_1) +  \gamma(q_2),
\eeq
which was studied in \citep{Boussarie:2016qop, Duplancic:2018bum, Qiu:2022pla, Duplancic:2022ffo, Duplancic:2023kwe, Qiu:2023mrm}.
The final-state pion can be replaced by any other light meson, and $N'$ does not have to be restricted to nucleons.
Similar to the process in \eq{eq:diphoton-process}, the $\gamma^*$-mediated channel is forbidden by charge conservation
for charged pion production or by the charge parity in the hard scattering for neutral pion production.
In the SDHEP kinematic region [\eq{eq:hard qT}], the leading configuration of the process [\eq{eq:photo-process}] happens
via an exchange of a collinear parton pair.
Following the discussion in \sec{sssec:ah2amh}, the amplitude of \eq{eq:photo-process} can be factorized into
the GPDs for the hadron transition $N \to N'$, a DA for the formation of the final-state pion, and 
perturbatively calculable coefficients,
\begin{align}\label{eq:photo-factorize}
	\M_{ N \gamma_{\lambda} \to N' \pi \gamma_{\lambda'} }^{[F, \wt{F}]}
	&\,= \sum_{f,f'} \int_{-1}^{1} dx \int_0^1 dz \, \bb{ 
			F_{NN'}^f(x, \xi, t) \, \widetilde{C}^{ff'}_{\lambda\lambda'}(x, \xi, z; \hat{s}, \theta, \phi)   \right.\nn\\
	&\hspace{8em}\left.	
			  + \wt{F}_{NN'}^f(x, \xi, t) \, C^{ff'}_{\lambda\lambda'}(x, \xi, z; \hat{s}, \theta, \phi)
		} \bar{D}_{f'/\pi}(z),
\end{align}
where $f=[q\bar{q}]$ and $[gg]$ for quark and gluon GPDs, respectively, if $N'=N$, 
or $f=[q\bar{q}']$ for transition GPDs with $N \neq N'$, and correspondingly, 
$f'=[q\bar{q}]$ or $[q\bar{q}']$ for the pion DA $\bar{D}_{f'/\pi}$.  
The hard coefficients $C^{ff'}_{\lambda\lambda'}$ and $\widetilde{C}^{ff'}_{\lambda\lambda'}$ 
are helicity amplitudes for the photon scattering off a collinear on-shell parton pair $f$ 
with $\lambda$ and $\lambda'$ denoting the photon helicities in the SDHEP frame [\fig{fig:sdhep-frame}].
The correction to the factorization in \eq{eq:photo-factorize} is suppressed by powers of $|t|/q_T^2 \ll 1$.

As we will see shortly, similar to the process [\eq{eq:diphoton-process}], 
in the crossing process [\eq{eq:photo-process}], 
the transverse momentum, or equivalently, the polar angle $\theta$, 
of the pion or final-state photon, provides an additional sensitive handle to the $x$ dependence of GPDs. 
However, there are some major differences from the diphoton production process 
that highlight the photoproduction one:
\begin{itemize}
\item the crossing kinematics provides an enhanced $x$-sensitivity mainly in the ERBL region, 
complementary to the diphoton production;
\item while only charged pion beams are accessible for the diphoton production, 
one can readily select a neutral pion product in the photoproduction process; 
this makes it not restricted to the flavor transition GPDs, and, more importantly, 
will further enhance the $x$-sensitivity due to a cancellation of certain ``moment" terms; and
\item the polarization of the initial-state photon beam can be easily controlled, 
as can be realized in the JLab Hall D facility~\citep{GlueX:2017zoo}. 
This allows the study of various polarization asymmetries, enabling to disentangle different types of GPDs.
\end{itemize}

\subsection{Calculation of the hard coefficients}
Similar to \sec{ssec:diphoton-hard},
the hard coefficients $\wt{C}^{ff'}_{\lambda_1\lambda_2}$ and $C^{ff'}_{\lambda_1\lambda_2}$ are the helicity amplitudes
of the collision $f(\hat{p}_1) + \gamma(p_2) \to f'(\hat{q}_1) + \gamma(q_2)$, 
where $\hat{p}_1 = (\Delta \cdot n) \bar{n}$ with $n=(0^+,1^-,{\bf 0}_T)$ and $\bar{n}=(1^+,0^-,{\bf 0}_T)$, 
and $\hat{q}_1 = (q_1 \cdot w)\bar{w}$ with $w$ and $\bar{w}$ defined as in \eq{eq:diphoton-w-wbar}.
For the quark GPD channel, $f = f' = [q_1 \bar{q}_2]$, 
with the two quarks $q_1$ and $\bar{q}_2$ from the diffracted nucleon carrying light-like momenta 
$(\xi \pm x) P^+ \bar{n}$, respectively,
and those for the produced pion carrying light-like momenta $z \hat{q}_1$ and $(1 - z) \hat{q}_1$, respectively. 
We neglect the gluon GPD in this thesis.

\begin{figure}[htbp]
	\centering
	\begin{tabular}{cccc}
		\includegraphics[scale=0.7]{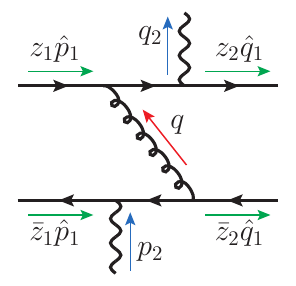} &
		\includegraphics[scale=0.7]{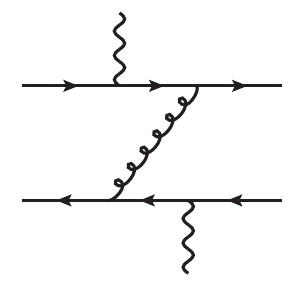} &
		\includegraphics[scale=0.7]{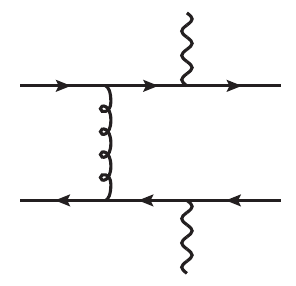} &
		\includegraphics[scale=0.7]{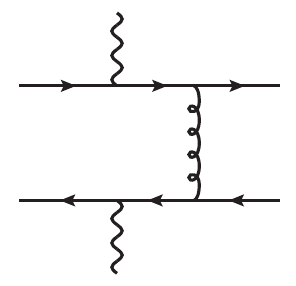} \\
		($A_1$) & ($A_2$) & ($A_3$) & ($A_4$)
	\end{tabular}
	\begin{tabular}{ccc}
		\includegraphics[scale=0.7]{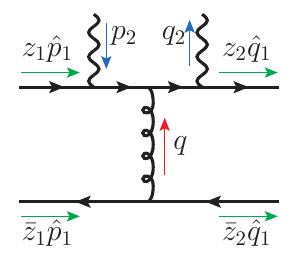} &
		\includegraphics[trim={0 -0.68cm 0 0}, clip, scale=0.7]{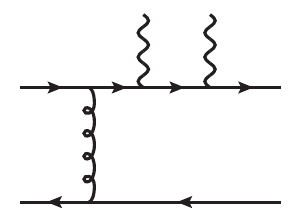} &
		\includegraphics[trim={0 -0.68cm 0 0}, clip, scale=0.7]{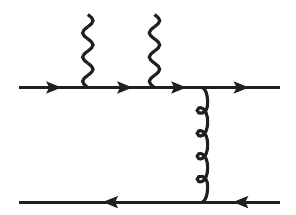} \\
		($B_1$) & ($B_2$) & ($B_3$)
	\end{tabular}
	\begin{tabular}{ccc}
		\includegraphics[scale=0.7]{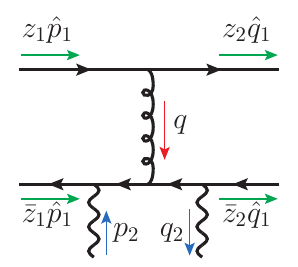} &
		\includegraphics[trim={0 0 0 -0.68cm}, clip, scale=0.7]{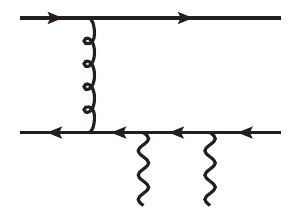} &
		\includegraphics[trim={0 0 0 -0.68cm}, clip, scale=0.7]{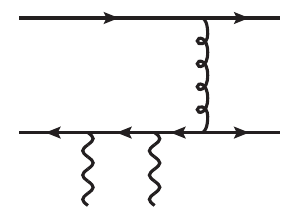} \\
		($B_4$) & ($B_5$) & ($B_6$)
	\end{tabular}
	\caption{Hard scattering diagrams for the photon-proton scattering into a photon-pion pair. 
	The two incoming fermion lines on the left are from the diffracted nucleon, carrying momenta 
	$z_1 \hat{p}_1$ and $\bar{z}_1 \hat{p}_1 \equiv (1-z_1) \hat{p}_1$, respectively. 
	The two outgoing fermion lines on the right are to form the produced pion, carrying momenta 
	$z_2 \hat{q}_1$ and $\bar{z}_2 \hat{q}_1 \equiv (1-z_2) \hat{q}_1$, respectively. 
	The variables $z_1$ and $z_2$ are related to $x$ and $z$ by $z_1 = (x + \xi) / 2\xi$ and $z_2 = z$ (see the text).
	Another set of diagrams are also to be included by switching the two photon lines, giving 20 diagrams in total.}
	\label{fig:photo-diagrams}
\end{figure}

For the $\pi^+$ production, we have $(q_1, q_2) = (u, d)$, while for the $\pi^-$ collision, we have $(q_1, q_2) = (d, u)$.
For the $\pi^0$ production, we have $q_1 = q_2 = u$ or $d$.
The hard coefficients $\widetilde{C}_{\lambda\lambda'}$ ($C_{\lambda\lambda'}$) are obtained from the diagrams 
in \fig{fig:photo-diagrams} by amputating the parton lines associated with the diffracted proton and produced pion, 
and contracting them with $\gamma \cdot \bar{n} / 2$ ($\gamma_5 \gamma \cdot \bar{n} / 2$) 
and $\gamma_5 \gamma \cdot \bar{w} / 2$, respectively, for the unpolarized (longitudinally polarized) GPD.
So similar to \sec{ssec:diphoton-hard}, the hard coefficient $\wt{C}$ associated with the unpolarized GPD $F$ is parity-odd (P-odd),
while the $C$ associated with the polarized GPD $\wt{F}$ is parity-even (P-even).
The hard coefficients are then obtained by averaging over the colors of each quark pair, and then multiplied by
an extra factor $\hat{s} / 2$.

Similar to \sec{ssec:diphoton-hard}, we take the convention for both partons from the diffracted nucleon to enter the hard collisions
and introduce the variable change $z_1 = (x + \xi) / 2 \xi$ and $z_2 = z$. Then the incoming $[q_1 \bar{q}_2]$ pair carries momenta
$z_1 \hat{p}_1$ and $(1 - z_1) \hat{p}_1$, respectively. In the c.m.~frame of the hard collision, we take 
\beq
	\hat{p}_1 = \sqrt{\frac{\hat{s}}{2}} \, \bar{n}, \quad
	p_2 = \sqrt{\frac{\hat{s}}{2}} \, n, \quad
	\hat{q}_1 = \sqrt{\frac{\hat{s}}{2}} \, \bar{w}, \quad
	q_2 = \sqrt{\frac{\hat{s}}{2}} \, w.
\eeq
Similar to \fig{fig:diphoton-diagrams}, the diagrams in \fig{fig:photo-diagrams} are classified into 8 type-$A$ ones and 12 type-$B$ ones.
The kinematic crossing of \eq{eq:diphoton-process} introduces different intermediate type-$A$ gluon propagators,
\beq[eq:photo-g-mom]
	q_A = -z_1 \hat{p}_1 + z_2 \hat{q}_1 + q_2
	= - \sqrt{\frac{\hat{s}}{2}} \pp{ z_1 \bar{n} - z_2 \bar{w} - w },
\eeq
and $q_{A'} = q_A(q_2 \to -p_2, w\to -n)$ obtained by switching the two photons in \fig{fig:photo-diagrams}.
They have virtualities different from the ones in \fig{fig:diphoton-diagrams},
\bse\label{eq:photo-g2-A}\begin{align}
	q_A^2 &= \hat{s} \bb{ 
		(1 - z_1) z_2 - \cos^2(\theta/2) z_1 (1 - z_2)
		},	\\
	q_{A'}^{2} &= \hat{s} \bb{ 
		z_1 (1 - z_2) - \cos^2(\theta/2) (1 - z_1) z_2
		},
\end{align}\ese
which contain non-factorizable $\theta$ dependence and 
thereby induce enhanced sensitivity to the $x$ dependence of GPDs, following the criterion in
\sec{sec:x-sensitivity}. 
The type-$B$ diagrams have simple gluon propagators with virtualities
\beq
	q_B^2 = - (1 - z_1) (1 - z_2) \hat{s} \sin^2(\theta/2), \quad
	q_B^{\prime2} = -z_1 z_2 \hat{s} \sin^2(\theta/2),
\eeq
for diagrams $(B_1, B_1', \cdots, B_3, B_3')$ and $(B_4, B_4', \cdots, B_6, B_6')$, respectively. 
Their $\theta$ dependence similarly factorizes out of $z_1$ and $z_2$, following the
structure of \eq{eq:hard coeff factorize}, and similarly for the quark propagators of all diagrams. 
So the type-$B$ diagrams only yield moment-type sensitivity.

Similar to \eq{eq:diphoton-Cmn}, we denote the hard coefficients in \eq{eq:photo-factorize} as
\bse\label{eq:photo-Cmn}\begin{align}
	C_{\lambda_1\lambda_2}(z_1, z_2; \hat{s}, \theta, \phi) 
		&= C_{\mu\nu}(z_1, z_2; \hat{s}, n, \bar{n}, w, \bar{w}) \, 
			\epsilon^{\mu}_{\lambda_1}(p_2) \epsilon^{\nu*}_{\lambda_2}(q_2),	\\
	\wt{C}_{\lambda_1\lambda_2}(z_1, z_2; \hat{s}, \theta, \phi) 
		&= \wt{C}_{\mu\nu}(z_1, z_2; \hat{s}, n, \bar{n}, w, \bar{w}) \, 
			\epsilon^{\mu}_{\lambda_1}(p_2) \epsilon^{\nu*}_{\lambda_2}(q_2),
\end{align}\ese
neglecting the superscripts $ff'$.
By the parity property, $\wt{C}_{\mu\nu}$ contains one antisymmetric Levi-Civita tensor, while $C_{\mu\nu}$ does not. 
Now, the Ward identities
\beq
	n_{\mu} \wt{C}^{\mu\nu} = \wt{C}^{\mu\nu} w_{\nu}
	= n_{\mu} C^{\mu\nu} = C^{\mu\nu} w_{\nu}
	= 0
\eeq
motivate us to expand the tensor structure in the light-cone basis formed by $n$ and $w$. 
Any vector $V^{\mu}$ can be decomposed in this basis as
\beq
	V^{\mu} = \frac{(V \cdot n) w^{\mu} + (V \cdot w) n^{\mu} }{n \cdot w} + V_{\perp}^{\mu},
\eeq
where the transverse component can be projected by acting on $V$ the tensor projector,
\beq
	g_{\perp}^{\mu\nu} = g^{\mu\nu} - \frac{w^{\mu} n^{\nu} + n^{\mu} w^{\nu}}{n \cdot w}.
\eeq
Then we can decompose $C_{\mu\nu}$ and $\wt{C}_{\mu\nu}$ into gauge-invariant tensor structures as,
\bse\label{eq:photo-tensor-decomp}\begin{align}
	C^{\mu\nu} &= C_- \, \pp{ -g_{\perp}^{\mu\nu} } 
		+ C_+ \, \pp{ -g_{\perp}^{\mu\nu} + 2 \frac{ \bar{n}_{\perp}^{\mu} \bar{n}_{\perp}^{\nu} }{ \bar{n}_{\perp}^2 } }
		+ C_n \, n^{\mu} \bar{n}_{\perp}^{\nu}  + C_w \, \bar{n}_{\perp}^{\mu} w^{\nu}
		+ C_{nw} \, n^{\mu} w^{\nu}, \\
	\widetilde{C}^{\mu\nu} 
	&= i \widetilde{C}_{-} \pp{\frac{ \bar{n}_{\perp}^{\mu} \varepsilon^{\nu n w \bar{n}_{\perp}} 
			+ \varepsilon^{\mu n w \bar{n}_{\perp}} \bar{n}_{\perp}^{\nu} }{n\cdot w \, \bar{n}_{\perp}^2} }
	+ i \widetilde{C}_{+} \pp{ \frac{\bar{n}_{\perp}^{\mu} \varepsilon^{\nu n w \bar{n}_{\perp}} 
			- \varepsilon^{\mu n w \bar{n}_{\perp}} \bar{n}_{\perp}^{\nu} }{n\cdot w \, \bar{n}_{\perp}^2} }
		\nn\\
	& \hspace{1em}
	+ i \widetilde{C}_n \, n^{\mu} \varepsilon^{\nu n w \bar{n}_{\perp}} 
	+ i \widetilde{C}_w \, \varepsilon^{\mu n w \bar{n}_{\perp}} w^{\nu},
\end{align}\ese
where we defined
$\bar{n}_{\perp}^{\mu} = g_{\perp}^{\mu\nu} \bar{n}_{\nu}$,
similarly to \eq{eq:diphoton-gperp}.
When contracting with the polarization vectors as \eq{eq:photo-Cmn}, the terms in \eq{eq:photo-tensor-decomp}
that are proportional to $n^{\mu}$ or $w^{\nu}$ vanish. 
The helicity amplitudes are purely determined by the first two tensor structures,
\begin{align}
	C_{\pm \pm}(z_1, z_2; \hat{s}, \theta, \phi) = \frac{N}{\hat{s}} e^{\mp i \phi} C_+(z_1, z_2; \theta), 
	&\quad
	C_{\pm \mp}(z_1, z_2; \hat{s}, \theta, \phi) = \frac{N}{\hat{s}} e^{\mp i \phi} C_-(z_1, z_2; \theta), 
	\nn\\
	\widetilde{C}_{\pm \pm}(z_1, z_2; \hat{s}, \theta, \phi) = \pm \frac{N}{\hat{s}} e^{\mp i \phi} \widetilde{C}_{+}(z_1, z_2; \theta),
	&\quad
	\widetilde{C}_{\pm \mp}(z_1, z_2; \hat{s}, \theta, \phi) = \pm \frac{N}{\hat{s}} e^{\mp i \phi} \widetilde{C}_{-}(z_1, z_2; \theta),
\label{eq:photo-helicity-structure}
\end{align}
where $N = 2i e^2 g^2 C_F / N_c$ and the four independent hard coefficients are 
\begin{align}
&
2 \xi C_{+} = 
	-(e_1-e_2)^2 \cdot
		t^2_{\theta/2} \,
			\P\frac{(1 - z_1) (1 - z_2) + z_1 z_2}{ 2 z_1 z_2 (1-z_1) (1-z_2) } 
	+ \frac{e_1^2 - e_2^2}{ s^{2}_{\theta/2} } 
		\cdot \P\frac{z_1 + z_2 - 1}{z_1 z_2 (1 - z_1) (1 - z_2)}
\nn\\
&\hspace{1em}
	+ \frac{e_1 e_2}{2} \, \P 
	\frac{s^2_{\theta/2} }{z_1 z_2 (1-z_1) (1-z_2)}
	\times
\nn\\
&\hspace{4em}
	\times
	\frac{ \pp{ z_1 (1 - z_2) + (1 - z_1) z_2 } \pp{ z_1 (1 - z_1) + z_2(1 - z_2) }}{
		\big( (1 - z_1) z_2 - c^2_{\theta/2} \, z_1 (1 - z_2) \big)
		\big( z_1 (1 - z_2) - c^2_{\theta/2} \, (1 - z_1) z_2 \big) 
	}
\nn\\
&\hspace{1em}
	+ i \pi \bigg\{
		(e_1-e_2)^2 \, \frac{2 c_{\theta}}{s^2_{\theta}}
			\pp{ \frac{ \delta(1 - z_1) }{1 - z_2} + \frac{\delta(z_1)}{z_2} }
		+ \frac{e_1^2 - e_2^2}{2} \pp{ 1 + s^{-2}_{\theta/2}}
			\pp{ \frac{\delta(1 - z_1)}{1 - z_2} - \frac{\delta(z_1)}{z_2} }
\nn\\
&\hspace{4em}
	- \frac{e_1 e_2 }{2} 
				\pp{ t^2_{\theta/2} - 2 s^{-2}_{\theta/2} }
   				\pp{ \frac{\delta(1 - z_1)}{1 - z_2} + \frac{\delta(z_1)}{z_2} }
			- \frac{e_1 e_2}{2 \,c^2_{\theta/2} \, z_2 \, (1 - z_2)} \times
\nn\\
&   
		\hspace{5em} \times 
		\bb{ 
			\pp{ \frac{z_1}{z_2} + c^2_{\theta/2} \frac{z_2}{z_1} }
				\delta\pp{ z_1 - \rho(z_2) }
			+ \pp{ c^2_{\theta/2} \frac{z_1}{z_2} + \frac{z_2}{z_1} }
				\delta\pp{ z_1 - \widetilde{\rho}(z_2) }
		}
	\bigg\} \,,
\label{eq:photo-C+}
\\
&
2 \xi C_{-} = 
	-(e_1 - e_2)^2 \, t^2_{\theta/2} \,
		\P\frac{z_1 (1 - z_2)  + (1 - z_1) z_2}{2 z_1 z_2 (1-z_1) (1-z_2)} 
\label{eq:photo-C-}
\\
&\hspace{1em}
	- i \pi \bigg\{
		\bigg( \frac{(e_1 - e_2)^2}{2c^2_{\theta/2}} + \frac{e_1 e_2 }{c^2_{\theta/2}} \bigg)
			\pp{ \frac{\delta(1 - z_1)}{z_2} + \frac{\delta(z_1)}{1 - z_2} }
		+ \frac{e_1^2 - e_2^2}{2} 
			\pp{ \frac{\delta(z_1)}{1 - z_2} - \frac{\delta(1 - z_1)}{z_2} }
	\bigg\} \,,
\nn\\
&
2 \xi \widetilde{C}_{+} =
	- \frac{(e_1 - e_2)^2}{2}  
		\pp{ 1 + c^{-2}_{\theta/2} }
			\, \P\frac{z_1 + z_2 - 1}{z_1 z_2 (1 - z_1) (1 - z_2)}
\label{eq:photo-Ct+}
\\
&	\hspace{1em}
	+ \frac{e_1 e_2}{2} \, \P
		\frac{1 + c^2_{\theta/2}}{z_1 z_2 (1-z_1) (1-z_2) }
		\times
\nn\\
&\hspace{4em}
	\times
		\frac{ (z_1 + z_2 - 1) (z_1 - z_2)^2 }{ 
			\big( (1 - z_1) z_2 - c^2_{\theta/2} \, z_1 (1 - z_2) \big)
			\big( z_1 (1 - z_2) - c^2_{\theta/2} \, (1 - z_1) z_2 \big)
		}	
\nn\\
&\hspace{1em}
	+ i \pi \bigg\{
		\bigg[ \frac{2 (e_1 - e_2)^2}{s^2_{\theta}} 
			+ e_1 e_2 \bigg( \frac{4}{s^2_{\theta}} - \frac{t^2_{\theta/2}}{2} \bigg)
		\bigg]
		\pp{ \frac{\delta(z_1)}{z_2} - \frac{\delta (1 - z_1)}{1 - z_2} }
		- \frac{e_1 e_2}{2} t^2_{\theta/2} \times
		\nn\\
&\hspace{4em}
		\frac{z_1 + z_2 - 1}{z_2 \, (1 - z_2)}
			\big[ 
				\delta \pp{ z_1 - \rho(z_2) } 
				+ \delta \pp{ z_1 - \widetilde{\rho}(z_2) }
			\big]
		+ \frac{e_1^2 - e_2^2}{2 \, t^2_{\theta/2}} \, 
			\pp{ \frac{\delta(1 - z_1)}{1 - z_2} + \frac{\delta(z_1)}{z_2} }
	\bigg\} \,,
\nn\\
&
2 \xi \widetilde{C}_{-} =
	(e_1 - e_2)^2 \,
		t^2_{\theta/2} \,
			\P \frac{z_1 - z_2}{2 z_1 z_2 (1-z_1) (1-z_2)} 
\label{eq:photo-Ct-}
\\
&\hspace{1em}
	- i \pi \bigg\{
		\bigg(
			\frac{(e_1 - e_2)^2}{2 c^2_{\theta/2}}  
			+ \frac{e_1 e_2 }{c^2_{\theta/2}} 
		\bigg)
			\pp{ \frac{\delta(z_1)}{1 - z_2} - \frac{\delta(1 - z_1)}{z_2} }
		+ \frac{e_1^2 - e_2^2}{2} 
			\pp{ \frac{\delta(z_1)}{1 - z_2} + \frac{\delta(1 - z_1)}{z_2} }
	\bigg\} \,,
\nn
\end{align}
where we introduced $t_{\theta/2} = \tan(\theta/2)$ and 
used the same notations for $\P$ and $(c_{\theta}, s_{\theta}, c_{\theta/2}, s_{\theta/2})$
as in Eqs.~\eqref{eq:diphoton-C+}--\eqref{eq:diphoton-Ct-}.

The special gluon propagators in the type-$A$ diagrams (cf. \eq{eq:photo-g2-A}) 
introduce special poles of $z_1$,
\bse\label{eq:photo-special-poles-z}\begin{align}
	\rho(z_2) &\, = \frac{z_2}{z_2 + \cos^2(\theta/2) (1 - z_2)},
	\\
	\wt{\rho}(z_2) & \, = \frac{\cos^2(\theta/2) \, z_2}{1 - z_2 + \cos^2(\theta/2) z_2}
		= 1 - \rho(1 - z_2),
\end{align}\ese
which both lie between 0 and 1, and have small positive and negative imaginary parts by the $i\epsilon$ prescription, respectively.
In terms of $x = \xi (2 z_1 - 1)$, \eq{eq:photo-special-poles-z} translates to the special poles,
\bse\label{eq:photo-special-poles-x}\begin{align}
	x_p(\xi, z, \theta)
		& = \xi \cdot \bb{ \frac{z - \cos^2(\theta/2)(1-z)}{z + \cos^2(\theta/2)(1-z)} },
	\label{eq:photo-special-pole-x-a}\\
	\wt{x}_p(\xi, z, \theta)
		& = \xi \cdot \bb{ \frac{\cos^2(\theta/2) z - (1-z)}{\cos^2(\theta/2) z + (1-z)} }
		= - x_p(\xi, 1 - z, \theta),
\end{align}\ese
which cross the whole ERBL region as $z$ goes from $0$ to $1$, complementary to the diphoton production process 
(see \eq{eq:diphoton-special-poles-x}).

Similar to Eqs.~\eqref{eq:diphoton-C+}--\eqref{eq:diphoton-Ct-} for the diphoton process, 
we have organized Eqs.~\eqref{eq:photo-C+}--\eqref{eq:photo-Ct-} in terms of 
$(e_1-e_2)^2$, $(e_1^2 - e_2^2)$, and $e_1 e_2$,
where the first two kinds of terms only carry moment-type sensitivity, 
whereas the $e_1 e_2$ terms carry enhanced $z_1$ (or $x$) sensitivity. 
For a neutral pion production channel, $e_1 = e_2$, and then the first two kinds of terms are cancelled, 
which further enhances the $x$-sensitivity in the $e_1 e_2$ terms.
By the charge conjugation symmetry, the $(e_1-e_2)^2$ and $e_1 e_2$ terms are probing C-even GPDs
$\wt{F}^+$ and $F^+$, whereas the $(e_1^2 - e_2^2)$ terms are probing C-odd GPDs $\wt{F}^-$ and $F^-$.
Also, the two processes 
$p \gamma \to n \pi^+\gamma$ and $p \gamma \to n \pi^+\gamma$ 
are related by isospin symmetry, which is broken by the $(e_1^2 - e_2^2)$ terms.
Combining both channels then can help distinguish C-odd GPD components from C-even ones.
The same is true for the two processes 
$p \gamma \to p \pi^0 \gamma$ and $n \gamma \to n \pi^0 \gamma$.

\subsection{Helicity amplitudes}
\label{ssec:photo-amplitudes}
The convolutions of the hard coefficients in Eqs.~\eqref{eq:photo-C+}--\eqref{eq:photo-Ct-}
with the GPD and DA can be simplified by using symmetry property of the DA.
With the same notation in \eq{eq:C-conv-short}, we have
\bse\label{eq:photon-convolution}\begin{align}
	C_+^{[\wt{F}]} = &
	\pp{e_1-e_2}^2 \, \bar{D}_0
			\bb{  
				\frac{1}{2} t^2_{\theta/2} \cdot \wt{F}^+_0(\xi, t)
				+ \frac{2 i \pi  \cos\theta}{\sin^2\theta} \cdot \wt{F}^+(\xi, \xi, t)
			}	\label{eq:photo-M+}	\\
	& - (e_1^2-e_2^2) \, \bar{D}_0
			\bb{
				\frac{2}{1-\cos\theta} \cdot \wt{F}^-_0(\xi, t)
				- \frac{i \pi}{2} \cdot \frac{3-\cos\theta}{1-\cos\theta} \cdot \wt{F}^-(\xi, \xi, t)
			} 	\nn\\
	& - \frac{e_1 e_2}{2}
			\cc{
				\int_0^1 \frac{dz \bar{D}(z)}{z(1-z)}
					\bb{ \frac{1}{c^2_{\theta/2} + z \, s^2_{\theta/2}} + \frac{c^2_{\theta/2} + z \, s^2_{\theta/2}}{c^2_{\theta/2}} } 
					 \cdot J[\wt{F}^+; \xi, t, z, \theta]
			\right.
			\nn\\
	&\hspace{4em}\left.
			- \bar{D}_0
			\bb{ t^2_{\theta/2} \, \wt{F}^+_0(\xi, t)  
				- i\pi \pp{ t^2_{\theta/2} - \frac{2}{s^2_{\theta/2}} }
					\wt{F}^+(\xi, \xi, t) 
			}
	},
\nn\\
	C_-^{[\wt{F}]}  = &
	\frac{\pp{e_1-e_2}^2}{2} \, \bar{D}_0
		\bb{
			t^2_{\theta/2} \cdot \wt{F}^+_0(\xi, t)
			- \frac{i \pi}{c^2_{\theta/2}} \cdot \wt{F}^+(\xi, \xi, t)
		}	\nn\\
	& + i \pi \bar{D}_0 
		\bb{ \frac{e_1^2-e_2^2}{2} \cdot \wt{F}^-(\xi, \xi, t)
			- \frac{e_1 e_2}{c^2_{\theta/2}} \cdot \wt{F}^+(\xi, \xi, t)
		},
\label{eq:photo-M-}	\\
	\wt{C}_+^{[F]} = &
		\pp{e_1-e_2}^2  \, \bar{D}_0
		\bb{ 
			\frac{3 + \cos\theta}{ 2\,(1+\cos\theta) } \cdot F^+_0(\xi, t)
			- \frac{2 i \pi  }{\sin^2\theta} \cdot F^+(\xi, \xi, t)
		} 
	\label{eq:photo-Mt+}	\\
	& + (e_1^2-e_2^2) \cdot \frac{i \pi }{2}  
		\cdot  \frac{1+ \cos\theta}{1- \cos\theta} \cdot \bar{D}_0 \cdot F^-(\xi, \xi, t) 	\nn\\
	& + \frac{e_1 e_2}{2}
		\cc{ 
			\int_0^1 \frac{dz \bar{D}(z)}{z(1-z)} 
			\bb{ \frac{1}{c^2_{\theta/2} + z \, s^2_{\theta/2}} - \frac{c^2_{\theta/2} + z \, s^2_{\theta/2}}{c^2_{\theta/2}} } 
				 \cdot J[F^+; \xi, t, z, \theta] 
			\right.
			\nn\\
	&\hspace{4em} \left.
			+ \bar{D}_0
			\bb{
				\frac{3+\cos\theta}{1+\cos\theta} \cdot F^+_0(\xi, t)
				- i \pi \pp{ \frac{8}{\sin^2\theta} - t^2_{\theta/2}}
					F^+(\xi, \xi, t)
			}
		},
	\nn\\
	\wt{C}_-^{[F]} = &
	- \frac{\pp{e_1-e_2}^2}{2} \, \bar{D}_0 \,
		\bb{
			t^2_{\theta/2} \cdot F^+_0(\xi, t) - \frac{i \pi }{c^2_{\theta/2}} F^+(\xi, \xi, t)
		}	\nn\\
	& - i \pi \bar{D}_0 
		\bb{ \frac{e_1^2-e_2^2}{2} \cdot F^-(\xi, \xi, t)
			- \frac{e_1 e_2}{c^2_{\theta/2}} \cdot F^+(\xi, \xi, t)
		},
\label{eq:photo-Mt-}
\end{align}\ese
where we used the same moment notations as \eq{eq:GPD-DA-moments},
and defined the special GPD integral,
\beq[eq:photo-special-int]
	J[\F; \xi, t, z, \theta] 
		\equiv 
		\int_{-1}^{1}dx \frac{\F(x, \xi, t)}{x - x_p(\xi, z, \theta) + i \epsilon},
\eeq
where $\F$ can take any GPD function such as $F^{\pm}$, $\wt{F}^{\pm}$, $H^{\pm}$, etc., 
and the pole $x_p$ is given in \eq{eq:photo-special-pole-x-a}.
The special integral will map out a distribution of $\theta$ that contains enhanced sensitivity to the 
GPD $x$ dependence. Since the pole $x_p$ lies on the ERBL region, the enhanced sensitivity is 
mainly in that region.

\subsection{Cross section and polarization asymmetries}
The treatment of the nucleon spins works in the same way as \sec{ssec:diphoton-cross-section} 
for the diphoton production process. We first write \eq{eq:photo-factorize} explicitly as
\beq[eq:photo-amplitude-factorize]
	\M_{\lambda\lambda'} = \frac{1}{2 P^+} \bar{u}(p', \alpha')
		\bb{ \Ct_{\lambda\lambda'}^{[H]} \, \gamma^+ 
			- \Ct_{\lambda\lambda'}^{[E]} \, \frac{i \sigma^{+\alpha} \Delta_{\alpha}}{2m}
			+ C_{\lambda\lambda'}^{[\wt{H}]} \, \gamma^+ \gamma_5
			- C_{\lambda\lambda'}^{[\wt{E}]} \, \frac{\gamma_5 \Delta^+}{2m}
		} u(p, \alpha),
\eeq
which is written in the SDHEP frame,
with the subscripts $\lambda$ and $\lambda'$ referring to the photon helicities.
Following the same derivation, we get the same 
Eqs.~\eqref{eq:diphoton-M2}--\eqref{eq:diphoton-Ai-approx},
just with different photon helicity labels that are left implicit.

Now we sum over the final-state photon helicity and average over the initial-state one by the density matrix,
\beq[eq:photon-denmtx-sdhep]
	\rho^{\gamma}_{\lambda\bar{\lambda}} 
	= \frac{1}{2}
		\begin{pmatrix}
		1 + \lambda_{\gamma} & - \zeta \, e^{2i \phi_{\gamma} }  \\
		- \zeta \, e^{-2i \phi_{\gamma} } & 1 - \lambda_{\gamma}
		\end{pmatrix},
\eeq
where $\lambda_{\gamma}$ is the helicity of the photon beam, and $\zeta > 0$ is the linear polarization degree, along 
the azimuthal direction $\phi_{\gamma}$ in the SDHEP frame. 
Introducing the shorthand notation,
\beq
	\vv{C_1^{[\F_1]} C_2^{[\F_2]} } \equiv 
		\sum_{\lambda \bar{\lambda} \lambda'} 
			C^{[\F_1]}_{1, \lambda\lambda'} \rho^{\gamma}_{\lambda \bar{\lambda}} C^{[\F_2]}_{2, \bar{\lambda}\lambda'},
\eeq
where $C_{i}$ can stand for $C$ or $\wt{C}$ and $\F_i$ for any GPD,
we have the photon polarization averaged result for each amplitude component in \eq{eq:diphoton-Ai-approx},
\bse\label{eq:photo-pol-avg}\begin{align}
	\vv{C^{[\F_1]} C^{[\F_2] *} }
	& = \pp{ \frac{N}{\hat{s}} }^2 \bb{ 
			C_+^{[\F_1]} C_+^{[\F_2] *} + C_-^{[\F_1]} C_-^{[\F_2] *} 	\right.\nn\\
			&\hspace{6em}\left.
			- \zeta \cos2(\phi - \phi_{\gamma}) \pp{ C_+^{[\F_1]} C_-^{[\F_2] *} + C_-^{[\F_1]} C_+^{[\F_2] *} }
			},
	\\
	\vv{\wt{C}^{[\F_1]} \wt{C}^{[\F_2] *} }
	& = \pp{ \frac{N}{\hat{s}} }^2 \bb{ 
			\wt{C}_+^{[\F_1]} \wt{C}_+^{[\F_2] *} + \wt{C}_-^{[\F_1]} \wt{C}_-^{[\F_2] *} 	\right.\nn\\
			&\hspace{6em}\left.
			+ \zeta \cos2(\phi - \phi_{\gamma}) \pp{ \wt{C}_+^{[\F_1]} \wt{C}_-^{[\F_2] *} + \wt{C}_-^{[\F_1]} \wt{C}_+^{[\F_2] *} }
			},
	\\
	\vv{C^{[\F_1]} \wt{C}^{[\F_2] *} }
	& = \pp{ \frac{N}{\hat{s}} }^2 \bb{ \lambda_{\gamma} \pp{ C_+^{[\F_1]} \wt{C}_+^{[\F_2] *} + C_-^{[\F_1]} \wt{C}_-^{[\F_2] *}  }	\right.\nn\\
			&\hspace{6em}\left.
			- i \zeta \sin2(\phi - \phi_{\gamma}) \pp{ C_+^{[\F_1]} \wt{C}_-^{[\F_2] *} + C_-^{[\F_1]} \wt{C}_+^{[\F_2] *} }
			},
	\\
	\vv{\wt{C}^{[\F_1]} C^{[\F_2] *} }
	& = \pp{ \frac{N}{\hat{s}} }^2 \bb{ 
			\lambda_{\gamma} \pp{ \wt{C}_+^{[\F_1]} C_+^{[\F_2] *} + \wt{C}_-^{[\F_1]} C_-^{[\F_2] *}  }		\right.\nn\\
			&\hspace{6em}\left.
			+ i \zeta \sin2(\phi - \phi_{\gamma}) \pp{ \wt{C}_+^{[\F_1]} C_-^{[\F_2] *} + \wt{C}_-^{[\F_1]} C_+^{[\F_2] *} }
			}.
\end{align}\ese
The the result of \eq{eq:diphoton-Ai-approx} with the photon polarization averaged is,
\bse\label{eq:photo-Ai-avg}\begin{align}
	\A_0 & = (1 - \xi^2) \pp{ \frac{N}{\hat{s}} }^2 
		\cc{  |C_+^{[\wt{H}]}|^2 + |C_-^{[\wt{H}]}|^2 +  |\wt{C}_+^{[H]}|^2 + |\wt{C}_-^{[H]}|^2
				\right.\nn\\
		&\left. \hspace{2em}
			+ 2 \zeta \cos2(\phi - \phi_{\gamma})
				\Re\bb{ \wt{C}_+^{[H]} \wt{C}_-^{[H] *} - C_+^{[\wt{H}]} C_-^{[\wt{H}] *}
				}
		}  + \mathcal{O}(\xi^2, \, t/m^2),	\\
	\A_1 & = 2(1 - \xi^2) \pp{ \frac{N}{\hat{s}} }^2 
		\cc{  \lambda_{\gamma} \Re
				\bb{ \wt{C}_+^{[H]} C_+^{[\wt{H}] *} + \wt{C}_-^{[H]} C_-^{[\wt{H}] *}
				}	\right.\nn\\
		&\left. \hspace{2em}
			- \zeta \sin2(\phi - \phi_{\gamma})
				\Im \bb{ \wt{C}_+^{[H]} C_-^{[\wt{H}] *} + \wt{C}_-^{[H]} C_+^{[\wt{H}] *}
				}
		} + \mathcal{O}(\xi^2, \, \xi t/m^2),	\\
	\A_2^1 & = 2(1 + \xi) \pp{ \frac{N}{\hat{s}} }^2 
		\cc{  \lambda_{\gamma} \Re
				\bb{ \wt{C}_+^{[E]} C_+^{[\wt{H}] *} + \wt{C}_-^{[E]} C_-^{[\wt{H}] *}
					-\xi \pp{ \wt{C}_+^{[H]} C_+^{[\wt{E}] *} + \wt{C}_-^{[H]} C_-^{[\wt{E}] *} 
					}
				}	\right. \nn\\
		&\left. \hspace{2em}
			- \zeta \sin2(\phi - \phi_{\gamma})
				\Im \bb{ \wt{C}_+^{[E]} C_-^{[\wt{H}] *} + \wt{C}_-^{[E]} C_+^{[\wt{H}] *}
					-\xi \pp{ \wt{C}_+^{[H]} C_-^{[\wt{E}] *} + \wt{C}_-^{[H]} C_+^{[\wt{E}] *}
					}
				}
		} \nn\\
		& \hspace{2em}
		+ \mathcal{O}(\xi^2), \\
	\A_2^2 & = - 2(1 + \xi) \pp{ \frac{N}{\hat{s}} }^2
		\cc{  \Im
				\bb{ \wt{C}_+^{[H]} \wt{C}_+^{[E] *} + \wt{C}_-^{[H]} \wt{C}_-^{[E] *}
					- \xi \pp{ C_+^{[\wt{H}]} C_+^{[\wt{E}] *} + C_-^{[\wt{H}]} C_-^{[\wt{E}] *} 
					}
				}	\right. \nn\\
		&\left. \hspace{2em}
			+ \zeta \cos2(\phi - \phi_{\gamma})
				\Im \bb{ \wt{C}_+^{[H]} \wt{C}_-^{[E] *} + \wt{C}_-^{[H]} \wt{C}_+^{[E] *}
					+ \xi \pp{ C_+^{[\wt{H}]} C_-^{[\wt{E}] *} + C_-^{[\wt{H}]} C_+^{[\wt{E}] *} 
					}
				}
		},
\end{align}\ese
Inserting \eq{eq:diphoton-M2} to \eq{eq:sdhep-xsec}, with the explicit forms in \eq{eq:photo-Ai-avg}, 
we have the differential cross section of the photoproduction process,
\begin{align}\label{eq:photo-xsec-differential}
	&\frac{\dd\sigma}{\dd |t| \, \dd\xi \, \dd\phi_S \, \dd\cos\theta \, \dd\phi}
	= \frac{1}{(2\pi)^2} \frac{\dd\sigma}{\dd |t| \, \dd\xi \, \dd\cos\theta }
		\bigg\{ 1 +  \lambda_N \lambda_{\gamma} A_{LL}	
				+ \zeta A_{UT} \cos2(\phi - \phi_{\gamma})
				\nn\\
	&\hspace{6em}
				+ \lambda_N \zeta A_{LT} \sin2(\phi - \phi_{\gamma})
			+ \frac{s_T\Delta_T}{m (1 - \xi)} \bigg[ 
					 A_{TU} \sin\phi_S
					+ \lambda_{\gamma} A_{TL} \cos\phi_{\Delta} \nn\\
					&\hspace{6em}
					+ \zeta A_{TT}^1 \cos\phi_{S} \sin2(\phi - \phi_{\gamma})
					+ \zeta A_{TT}^2 \sin\phi_{S} \cos2(\phi - \phi_{\gamma})
				\bigg]
	\bigg\}.
\end{align}
where $\lambda_N$ in place of $\lambda$ in \eq{eq:diphoton-M2} is the helicity of the initial-state nucleon
\beq[eq:photo-unpol-xsec]
	\frac{d\sigma}{d|t| \, d\xi \, d\cos\theta} 
	= \pi \pp{ \alpha_e \alpha_s \frac{C_F}{N_c} }^2 \frac{1 - \xi^2}{\xi^2 \, s^3}
		\Sigma_{UU}
\eeq
is the unpolarized differential cross section, with
\beq
	\Sigma_{UU} = | C_+^{[\wt{H}]} |^2 + | C_-^{[\wt{H}]} |^2 + | \wt{C}_+^{[H]} |^2 + | \wt{C}_-^{[H]} |^2 
		+ \mathcal{O}(\xi^2, \, t/m^2).
\eeq
The presence of the beam polarizations give rise to various asymmetries,
\bse\label{eq:photo-pol-asymmetry}\begin{align}
	A_{LL}
	& = 2 \, \Sigma_{UU}^{-1} \Re\bb{
				\wt{C}_+^{[H]} C_+^{[\wt{H}] *} + \wt{C}_-^{[H]} C_-^{[\wt{H}] *}
			}, \\
	A_{UT}
	& = 2 \, \Sigma_{UU}^{-1} \Re\bb{ 
				\wt{C}_+^{[H]} \wt{C}_-^{[H] *} - C_+^{[\wt{H}]} C_-^{[\wt{H}] *}
			}, \\
	A_{LT}
	& = -2 \, \Sigma_{UU}^{-1} \Im \bb{
				\wt{C}_+^{[H]} C_-^{[\wt{H}] *} + \wt{C}_-^{[H]} C_+^{[\wt{H}] *}
			}, \\
	A_{TU}
	& = \Sigma_{UU}^{-1} \Im \bb{
				\wt{C}_+^{[H]} \wt{C}_+^{[E] *} + \wt{C}_-^{[H]} \wt{C}_-^{[E] *}
					- \xi \pp{ C_+^{[\wt{H}]} C_+^{[\wt{E}] *} + C_-^{[\wt{H}]} C_-^{[\wt{E}] *} 
					}
			}, \\
	A_{TL}
	& = \Sigma_{UU}^{-1} \Re
				\bb{ \wt{C}_+^{[E]} C_+^{[\wt{H}] *} + \wt{C}_-^{[E]} C_-^{[\wt{H}] *}
					-\xi \pp{ \wt{C}_+^{[H]} C_+^{[\wt{E}] *} + \wt{C}_-^{[H]} C_-^{[\wt{E}] *} 
					}
			}, \\
	A_{TT}^1
	& = -\Sigma_{UU}^{-1} \Im 
				\bb{ \wt{C}_+^{[E]} C_-^{[\wt{H}] *} + \wt{C}_-^{[E]} C_+^{[\wt{H}] *}
					-\xi \pp{ \wt{C}_+^{[H]} C_-^{[\wt{E}] *} + \wt{C}_-^{[H]} C_+^{[\wt{E}] *}
					}
			}, \\
	A_{TT}^2
	& = \Sigma_{UU}^{-1} \Im 
				\bb{ \wt{C}_+^{[H]} \wt{C}_-^{[E] *} + \wt{C}_-^{[H]} \wt{C}_+^{[E] *}
					+ \xi \pp{ C_+^{[\wt{H}]} C_-^{[\wt{E}] *} + C_-^{[\wt{H}]} C_+^{[\wt{E}] *} 
					}
			},
\end{align}\ese
which are approximations of the full expressions up to the errors suppressed by $\xi^2$ or $t/m^2$.
The $A_{LL}$ can be measured through the double helicity asymmetry,
\beq
	A_{LL}(t, \xi, \cos\theta)
	= \frac{1}{\lambda_N \lambda_{\gamma}}
		\frac{
			\sigma(\lambda_N, \lambda_{\gamma}) - \sigma(\lambda_N, -\lambda_{\gamma})
		}{
			\sigma(\lambda_N, \lambda_{\gamma}) + \sigma(\lambda_N, -\lambda_{\gamma})
		},
\eeq
where $\sigma$ stands for the differential cross section in $(t, \xi, \cos\theta)$ with $\phi_S$ and $\phi$ integrated out.
All the other asymmetries can be measured through the azimuthal modulations.

To the leading accuracy, the differential cross section without $s_T$ allows to probe the GPDs $H$ and $\wt{H}$, 
whereas with a nonzero $s_T$, we also gain access to probe the GPD $E$ and $\wt{E}$, both with enhanced $x$-sensitivity.

\subsection{Numerical results}
\label{ssec:photo-numerical}

Now we present the numerical results for the photoproduction process on the enhanced $x$-sensitivity.
We target our analysis toward the JLab Hall D experiment, with 
a photon beam $E_{\gamma} = 9~\GeV$ of arbitrary polarization 
and a proton target that can be unpolarized or longitudinally polarized.
With a potential upgrade to JLab $22~\GeV$, a photon beam $E_{\gamma} = 22~\GeV$ can also be accessed.
Similarly, we fix the DA to the asymptotic form with $\phi(z) = 6z(1-z)$, 
and renormalization and factorization scales to $2~\GeV$, turning off QCD evolution effects.

\begin{figure}[htbp]
	\centering
	\includegraphics[scale=0.65]{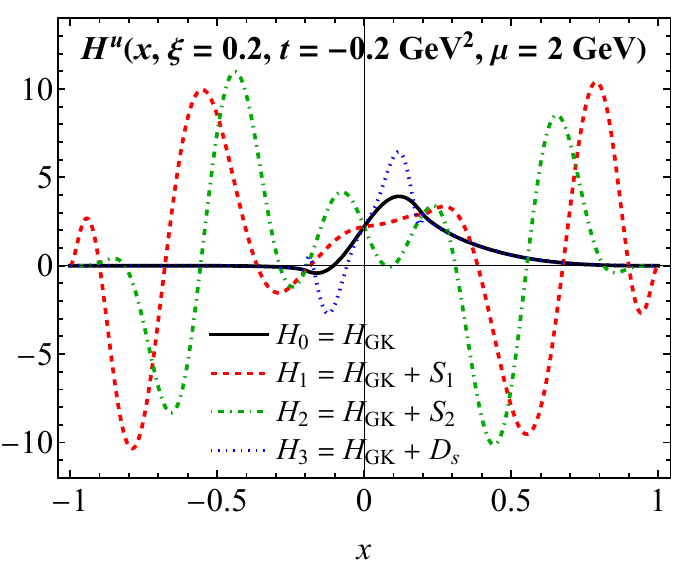}
	\includegraphics[scale=0.65]{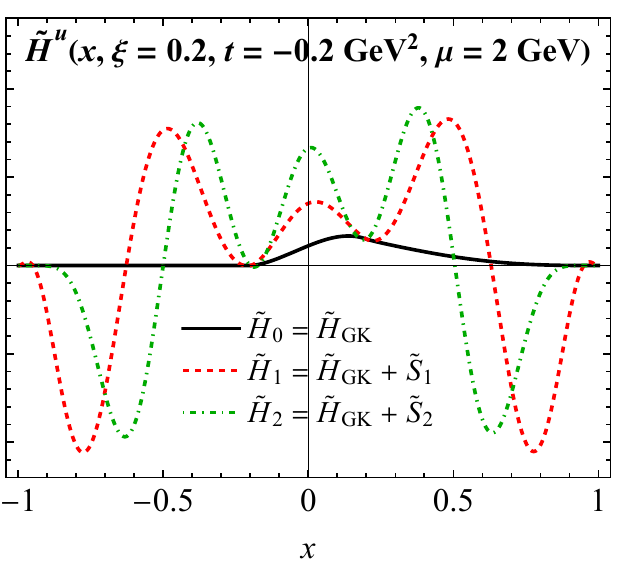} 
	\caption{Choices of the $u$-quark GPD models for the photoproduction process.}
	\label{fig:photo-gpds}
\end{figure}

As we shall see here, this photoproduction process gives better sensitivity to GPDs, 
so we lower the weights of both the shadow GPDs and $D$-term to 2 in 
Eqs.~\eqref{eq:GPD-model-Hh-12}\eqref{eq:GPD-model-Ht-12} and \eqref{eq:GPD-model-Hh-3},
in contrast to the diphoton process in \sec{ssec:diphoton-numerical}. 
The resultant GPD models for the $u$ quark are displayed in \fig{fig:photo-gpds}.

\begin{figure}[htbp]
	\centering
	\includegraphics[scale=0.75]{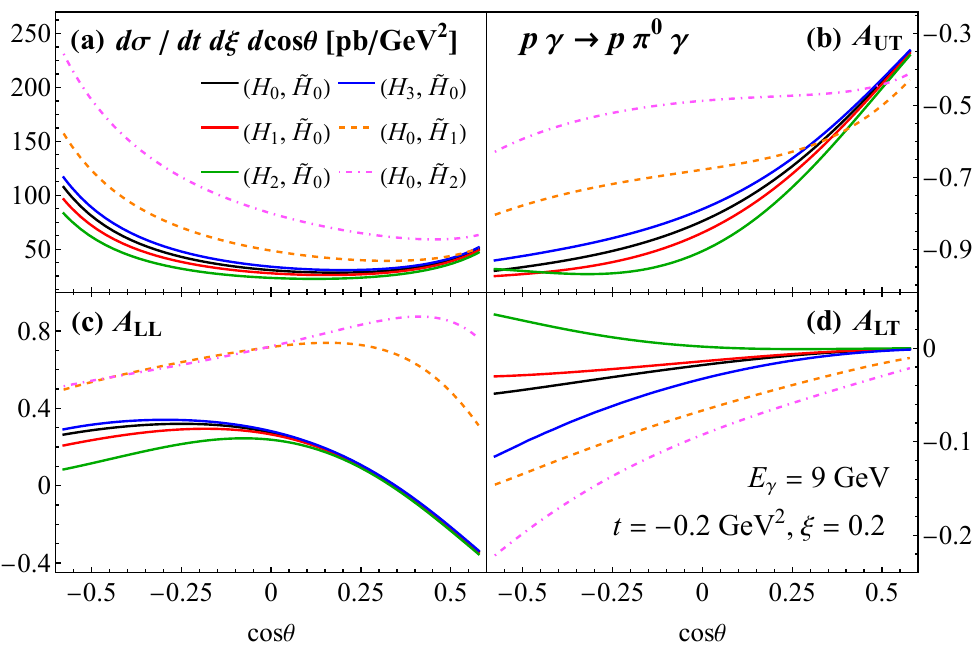}
	\caption{Unpolarized rate (a) and polarization asymmetries (b)-(d) 
	as functions of $\cos\theta$ at $(t, \xi) = (-0.2~\GeV^2, 0.2)$, 
	using different GPD sets as given in \fig{fig:photo-gpds}.
	}
	\label{fig:pi0-distributions}
\end{figure}

\begin{figure}[htbp]
	\centering
	\includegraphics[scale=0.75]{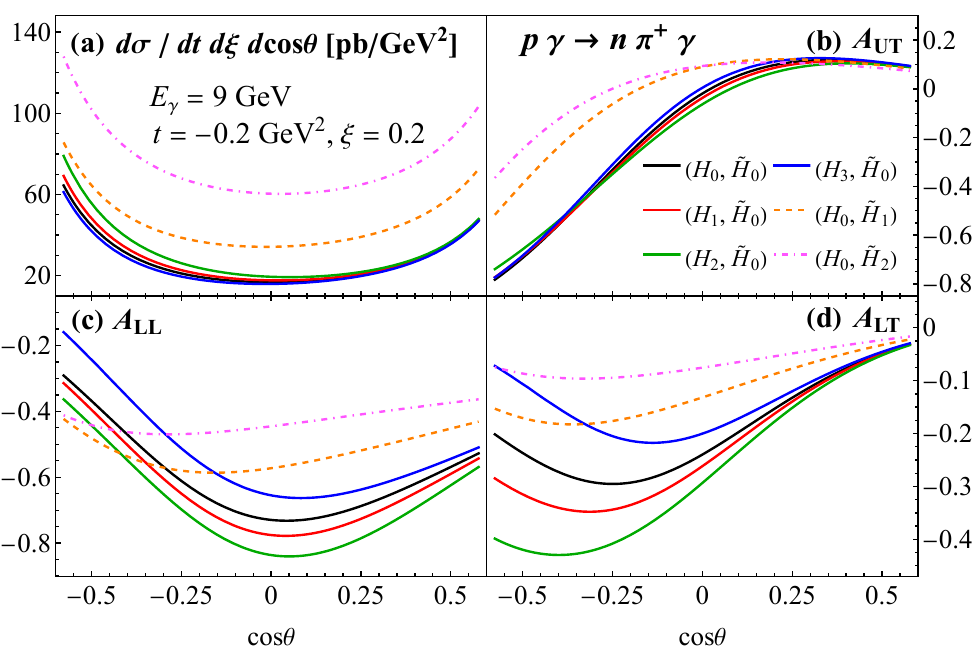}
	\caption{Same as \fig{fig:pi0-distributions}, but for the $p \gamma \to n \pi^+ \gamma$ process.}
	\label{fig:pi+distributions}
\end{figure}

\begin{figure}[htbp]
	\centering
	\includegraphics[scale=0.75]{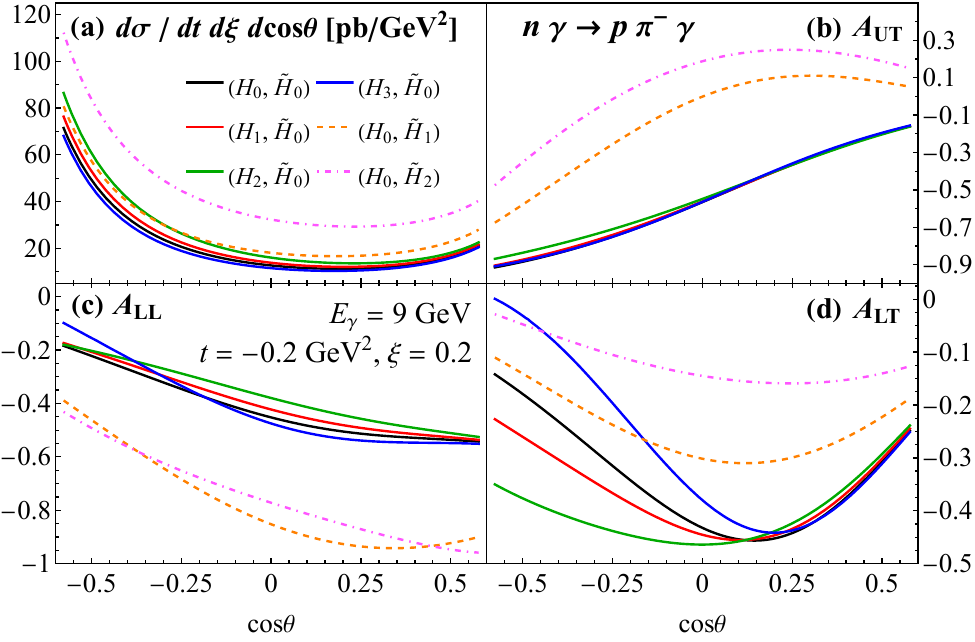}
	\caption{Same as \fig{fig:pi0-distributions}, but for the $n \gamma \to p \pi^- \gamma$ process.}
	\label{fig:pi-distributions}
\end{figure}

In \fig{fig:pi0-distributions}, we show the unpolarized differential cross section in \eq{eq:photo-unpol-xsec} 
together with the various asymmetries in \eq{eq:photo-pol-asymmetry} for $\pi^0$ production 
as a function of its polar angle $\theta$ in the SDHEP frame at $E_{\gamma} = 9~\GeV$.  
Since the amplitudes $C_-^{[\wt{H}]}$ and $\wt{C}_-^{[H]}$ only depend on GPDs through their moments 
(\eq{eq:photon-convolution}), they are not visible to the shadow GPDs.
On the other hand, the the amplitudes $C_+^{[\wt{H}]}$ and $\wt{C}_+^{[H]}$ have nontrivial
$\cos\theta$ dependence through the special GPD integral in \eq{eq:photo-special-int}.
Therefore, GPDs with different $x$-dependence lead to different rate and asymmetries.
In particular, the $A_{LT}$ is sensitive to the imaginary parts of the amplitudes, which are generated in
the ERBL region, and has better sensitivity to the shadow $D$-term than the other three observables 
as shown in \fig{fig:pi0-distributions}.

As for the diphoton production process in \sec{ssec:diphoton-numerical},
the oscillation of shadow GPDs in the DGLAP region generally causes a big cancellation in 
their contributions to the amplitudes, 
while the sensitivity is more positively correlated with the GPD magnitude in the ERBL region.
The shadow $\wt{S}_i$ associated with the $x$-dependence of the polarized GPD $\wt{H}$ 
gives bigger contribution to the amplitude $C_+^{[\wt{S}]}$ than $S_i$ to $\wt{C}_+^{[S]}$ 
due to charge symmetry property, so can be better probed. 

Contrary to diphoton production process, now we have various polarization observables that
have linear dependence on the helicity-flipping amplitudes, 
such as the $A_{UT}$ and $A_{LT}$ in \eq{eq:photo-pol-asymmetry},
thereby on the special integrals. 
The helicity-conserving amplitudes, which are blind to shadow GPDs,
now do not just come as large backgrounds, but are large coefficients of the special integrals 
to enhance the discriminative power. 
This therefore gives better opportunities to probe the GPDs.

\begin{figure}[htbp]
	\centering
	\includegraphics[scale=0.75]{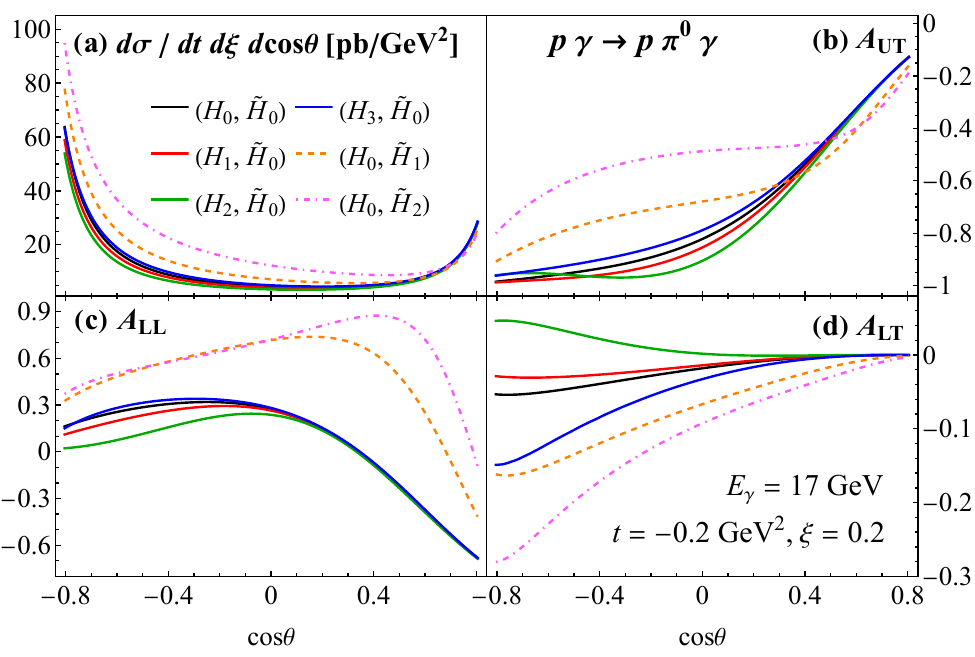}
	\caption{Same as \fig{fig:pi0-distributions}, but with $E_{\gamma} = 17~\GeV$. 
	}
	\label{fig:pi0-distributions-17}
\end{figure}

\begin{figure}[htbp]
	\centering
	\includegraphics[scale=0.75]{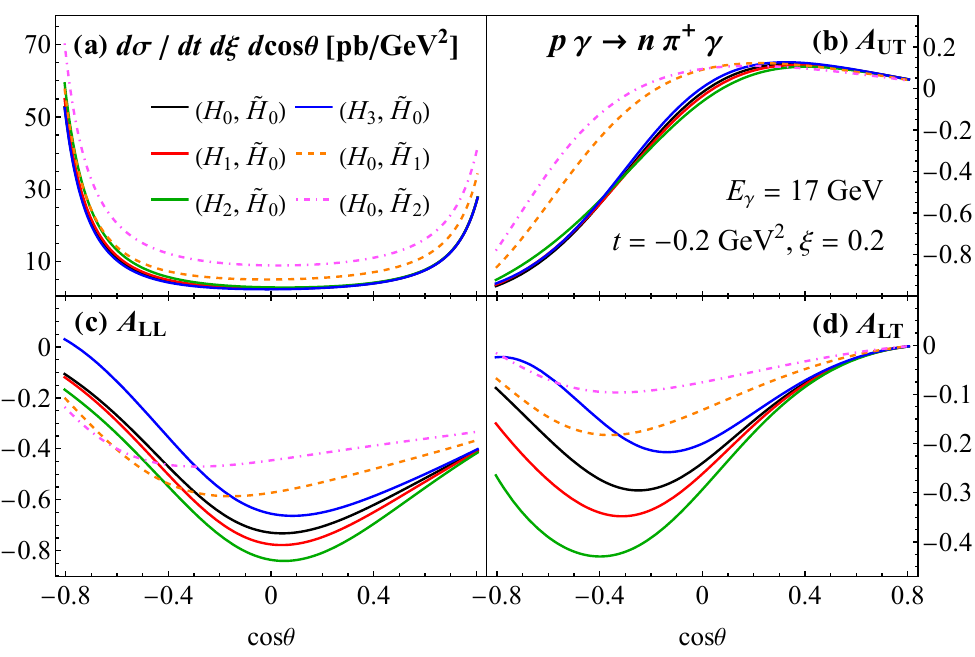}
	\caption{Same as \fig{fig:pi+distributions}, but with $E_{\gamma} = 17~\GeV$.}
	\label{fig:pi+distributions-17}
\end{figure}

\begin{figure}[htbp]
	\centering
	\includegraphics[scale=0.75]{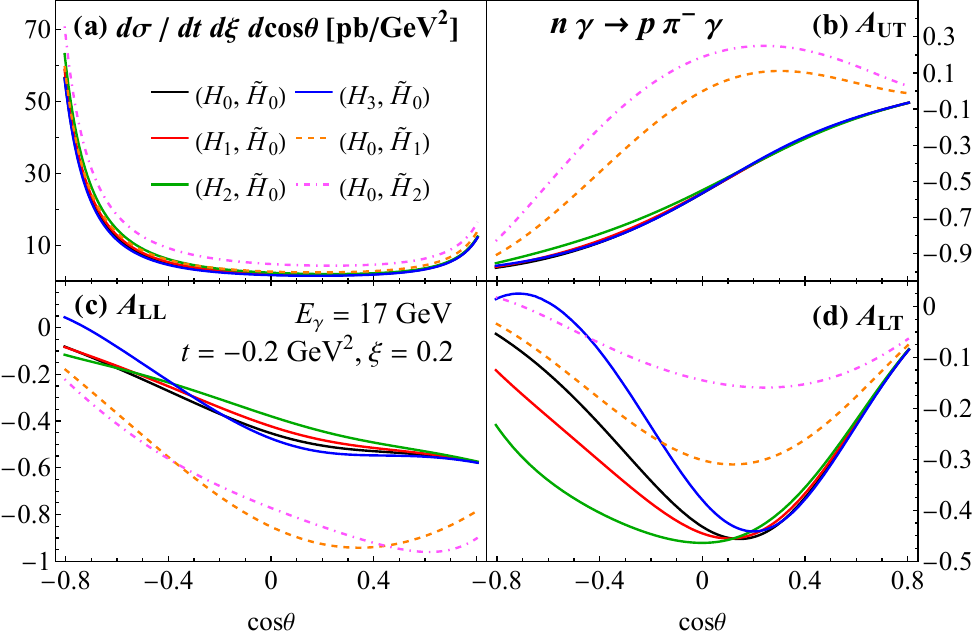}
	\caption{Same as \fig{fig:pi-distributions}, but with $E_{\gamma} = 17~\GeV$.}
	\label{fig:pi-distributions-17}
\end{figure}

Furthermore, for the neutral pion production, we can eliminate terms proportional to 
$(e_1-e_2)^2$ or $(e_1^2-e_2^2)$ in the amplitudes [\eq{eq:photon-convolution}] since $e_1 = e_2$,
which effectively removes a good number of moment-type terms, 
giving the maximum amount of entanglement and the most 
sensitivity to GPDs' $x$-dependence.
In \fig{fig:pi+distributions}, we present the same study 
for the $p \gamma \to n \pi^+ \gamma$ process. With different flavor combination, 
it provides different $x$-sensitivity.  
The $n \gamma \to p \pi^- \gamma$ process gives a similar result in \fig{fig:pi-distributions}, 
but with a smaller production rate.
Although the charged pion channels have less sensitivity in the absolute distributions, they 
yield greater polarization asymmetry values than the neutral pion channel so can provide
equally significant and complementary results.

For ungraded JLab energy, the photon beam with $E_{\gamma} = 17~\GeV$ yields similar results, 
shown in \fig{fig:pi0-distributions-17}. 
The rate at each phase space point decreases as the energy increases,
but the kinematic coverage becomes wider, 
giving a complementary opportunity to probe the $x$ dependence via wider $\cos\theta$ distributions.

As demonstrated in Figs.~\ref{fig:pi0-distributions}--\ref{fig:pi-distributions-17}, 
both the production rate and asymmetries are sizable and measurable, 
making the photoproduction process uniquely different from DVCS and other processes 
in terms of its enhanced sensitivity for extracting the $x$-dependence of GPDs.


\newpage
\chapter{Summary and Outlook}
\label{ch:summary-qcd}

One of the dominant features of QCD is that colors are fully entangled and confined within hadrons, 
which makes the internal structure of hadrons by no means like the atomic structure, 
where electrons are bound to nucleus in a sparsely distributed space. 
On the contrary, inside a hadron, quarks and gluons are densely distributed and strongly tied together. 
As a result, it is less useful to describe the hadronic structure using the concept of wavefunction in non-relativistic quantum mechanics. 
Instead, the study of hadrons' internal dynamics is to use parton correlation functions, 
which are the expectation values of a set of parton fields in a hadron state. 
A full understanding of partonic hadron structure can be obtained by knowing all possible parton correlation functions. 

However, the correlation functions are by definition nonperturbative and require experimental measurement, given the lack of
a full nonperturbative calculation method. The connection of the correlation functions to experimental observables is given by
QCD factorization theorems. At a hard scattering process involving hadrons, one can show that the scattering cross section or amplitude
can be factorized into certain parton correlation functions with perturbatively calculable hard coefficients, to the leading power of the hard scale. 
Depending on the specific type of processes, one end up with different type of parton correlation functions, which have operator definitions,
can be studied on their own, and uncover different aspects of the hadronic structures.

For inclusive processes, the factorization of their cross sections leads to the forward parton distribution functions, 
which capture the one-dimensional longitudinal parton correlation on the light cone within a fast-moving hadron, 
and transverse-momentum-dependent parton distribution functions, 
which in addition capture the parton correlations in the transverse plane. 
Both distributions correspond to cut diagrams, and are expressed as the diagonal matrix elements of parton operators.

Exclusive processes, on the other hand, are factorized at amplitude level into new types of parton correlation functions, 
among which are the meson or baryon distribution amplitudes that play the role of hadron wavefunctions on the light cone,
and the generalized parton distributions (GPDs), which form the main part of this thesis. 
Among others, the GPDs entail three-dimensional parton pictures in the space of parton momentum fraction $x$ and 
transverse position $\bm{b}_T$. We have shown a general class of $2 \to 3$ processes in \sec{sec:SDHEP}, 
the single diffractive hard exclusive processes, whose amplitudes can be factorized into GPDs and which can provide
useful experimental probes to GPDs.

While two of the three variables $(x,\xi,t)$ of GPDs are directly related to the measured momenta of the diffractive hadron, 
$p - p'$, it is the relative momentum fraction $x$ of the two exchanged partons, $[q\bar{q}']$ or $[gg]$, 
between the diffractive hadron and the hard probe that is the most difficult one to 
extract from the experimental measurement, while it is the most important one to define the slices of the hadron's spatial tomography.
We have systematically examined the sensitivity of various SDHEPs for extracting the $x$-dependence of GPDs in \sec{sec:x-sensitivity},
and divided the sensitivity into two types: moment type and enhanced type.  
We argued that the requirement for enhanced sensitivity on $x$ is to have at least one internal propagator 
in the hard part that is not connected to two on-shell massless external lines on either of its ends, 
which usually requires observing more than one external particle that comes out of the hard scattering.
We gave two example processes, the hard diphoton production in single-diffractive pion-nucleon collision, and
single-diffractive photoproduction of a hard photon-pion pair. These two processes give complementary enhanced sensitivity to the 
$x$ dependence of GPDs, which were demonstrated by using the shadow GPDs.
d\
Given both the theoretical and experimental difficulties to unambiguously extract the $x$-dependence of GPDs, 
one should not only study as many independent GPD-related processes as possible, 
but also identify more processes that yield enhanced sensitivity to the $x$ dependence of GPDs.
With a generic factorization proof, the SDHEP can serve as a framework to identify and 
categorize all specific processes for the study of GPDs. 
In this thesis, we categorized these processes in terms of the type of the beam colliding with the diffractive hadron.  
With the two-stage paradigm of the SDHEP, we are well motivated for the search of new processes for extracting GPDs, 
and in particular, their $x$-dependence.

\part{Single Transverse Polarization Phenomena at High-Energy Colliders}
\label{ch:spin-hep}

\chapter{Introduction}
\label{ch:intro-spin}

Spins are unique features of quantum mechanics, as a product of quantum Lorentz symmetry~\citep{Wigner:1939cj, Weinberg:1995mt}.
At high-energy colliders such as the Large Hadron Collider (LHC), however, spin phenomena are relatively rarely discussed, because
(1) the LHC is an unpolarized proton-proton collider, so usually it does not produce polarized particles; and
(2) the detectors of high-energy colliders only record the energy and momentum information, but do not measure the spins, 
so even if a particle is produced polarized, the spin information will be lost.
Both obstacles can be overcome. 
First, the Standard Model (SM) contains parity-violating weak interactions, so particles can be produced with net spin polarizations
along their momentum directions, or net helicities.
Furthermore, even without parity violations, there can be significant transverse polarizations produced even at unpolarized colliders.
In both contexts, the polarization refers to a single particle, with all other particles' spins unobserved, so it belongs to the regime of
single polarization production, similar to the discussion of single spin asymmetry at polarized colliders.
Second, even though the high-energy detectors do not directly measure spins, if the polarized particle is unstable and decays into
other particles, its polarization information will be imprinted on the kinematic distributions, especially angular distributions, of the
decay products. This is because the polarization of the mother particle breaks the spatial rotational invariance, and so it leads to certain
angular distributions, which can be determined by rotation group properties.

The same story holds for high-energy quarks and gluons produced in the hard collisions. 
Due to the asymptotic freedom, such particles are produced as (quasi-)free particles, with well defined polarization properties.
But as they travel away from each other, the color interaction among them becomes stronger and stronger, and eventually 
turns each fast-moving quark or gluon into a jet of hadrons. It may be argued that such hadronization process will wash out all the 
original parton spin information. But it is more presumably motivated from the high-energy jetty event structures that only soft
gluons are exchanged among the hard partons to neutralize colors~\citep{Collins:1992kk}. 
Perturbatively, soft gluon exchanges do not change the spins of hard partons, 
so we can expect the polarization of the quark or gluon produced from the hard collision to be preserved when it
fragments into a jet. As a result, the angular distribution of the jet constituents will reflect the polarization state of the parton
that initiates the jet.

Therefore, it is equally feasible to study spin phenomena at high-energy colliders as well as at low-energy experiments. 
This leads to much more observables than the pure production rates. 
Especially, as we will elaborate in this thesis, the transverse polarization corresponds to the quantum interference
between different helicity states. Such information would be lost had one not measured the decay distributions.
The spin-sensitive observables hence provide new tests on the interaction structures of the SM.

This rest of this thesis is devoted to the study of single transverse polarization phenomena at such high-energy colliders as the LHC.
Historically, such study dates back to 1976 when it was discovered at Fermilab that the inclusively produced $\Lambda^0$ hyperon
in hadron collisions had a substantial transverse polarization~\citep{Bunce:1976yb, Heller:1978ty}.
This triggered a number of both experimental and theoretical studies until today.
Among the early theoretical works was Ref.~\citep{Kane:1978nd}, where it was realized that the single transverse spin 
of a quark is an infrared-safe observable in Quantum Chromodynamics (QCD), which can be calculated perturbatively 
by virtue of the asymptotic freedom. Following the observation that only the transverse spin component perpendicular to
the scattering plane is allowed by parity conservation, the authors argued that this must be sourced by the imaginary part 
of the interference between a helicity-conserving and a helicity-flipping amplitudes. Therefore, one necessarily
requires a nonzero quark mass to flip the quark helicity {\it and} a threshold effect at loop level to generate a nonzero phase.
So then in the scaling limit, the quark polarization is suppressed by $\alpha_s \, m_q / \sqrt{s}$, 
where $\alpha_s$ is the strong coupling due to the loop effect, $m_q$ is the quark mass, and $\sqrt{s}$ is the scattering energy.

Although this means the single transverse spin of a strange quark produced at high-energy collisions would be too small
to explain the observed large $\Lambda^0$ polarization~\citep{Dharmaratna:1989jr}, it does imply the possibility of having
a largely polarized top quark~\citep{Kane:1991bg}, which is the heaviest quark in the SM and whose polarization could be
a new probe for new physics. Any deviation from the SM prediction, especially a nonzero transverse spin within the production
plane, could indicate the existence of a new interaction or even $CP$ violation.

One advantage of the transverse spin is that it leads to a nontrivial azimuthal correlation of the decay products 
with the spin direction, as a result of breaking the rotational invariance. Since a transverse spin is the interference
between different helicity states, $\lambda_1$ and $\lambda_2$, the specific correlation form can be easily 
obtained from rotational properties as $\cos(\lambda_1 - \lambda_2) \phi$ and/or $\sin(\lambda_1 - \lambda_2) \phi$,
with $\phi$ characterizing the overall azimuthal direction of the decay products. 
Such correlations can be readily measured to determine the value of the transverse polarization.
Unlike the helicity polarization that leads to a forward-backward asymmetry for the decay products with respect to the 
momentum direction of the mother particle, the azimuthal correlations resulting from the transverse polarization
stay invariant when the polarized particle is boosted. This makes them a source of new jet substructures for boosted
objects. However, due to the spin-half nature, the azimuthal correlations associated with a transversely polarized quark
are $\cos\phi$ and/or $\sin\phi$, the observation of which requires to identify the flavor of the decay products. 
For example, in a jet initialized by a transversely polarized $u$ quark, one may be observing the correlation of a charged
pion $\pi^+$ with the polarization direction.

On the other hand, a gluon can also be produced in high-energy collisions with a linear polarization, as was noticed 
around the same time as the transverse quark spin~\citep{Brodsky:1978hw}. 
Contrary to the latter, though, the linear gluon polarization does not suffer from the mass and high-order suppression,
and can in principle be produced at leading order with a large magnitude
~\citep{Brodsky:1978hw, Olsen:1979fp, Devoto:1979jm, Devoto:1979fq, DeGrand:1980yp, Petersson:1980cn, 
Olsen:1981ws, Devoto:1981vh, Korner:1981qj, Olsen:1983mm, Hara:1988uj, 
Jacobsen:1990jp, Groote:1997vg, Groote:1998qt, Groote:2002qc, Yu:2022kcj}. 
Since gluons are spin-one massless particles, their linear polarization is the interference between a $+1$ and a $-1$ 
helicity states, with a helicity flip by two units. Hence, they will leave $\cos2\phi$ and/or $\sin2\phi$ azimuthal 
correlations in the fragmented jets. Such correlations are invariant under $\phi \to \phi + \pi$ so do not require 
distinguishing the particle flavors, but instead they will be reflected as an azimuthal anisotropy in the energy deposition.
Observation of such polarized gluon jet substructure could be easier than for the polarized quark ones. 
As we will show, this can serve as a new tool to probe $CP$-violating interactions.

Similar effects also apply to massive vector bosons like the $W$ and $Z$, which can also be produced with a linear
polarization when they carry a nonzero transverse momentum. Such phenomena have actually been noticed all
along when one studies the angular functions of the Drell-Yan pair in their rest frame~\citep{Lam:1978pu}.
However, one may still gain some insights when framing in terms of linear polarizations. Especially, as one goes to
the boosted regime, a $W$ or $Z$ may be produced with a very high transverse momentum such that their decay
products are highly collimated. In particular, when they decay hadronically, 
it may not be easily determined whether they are QCD jets or are indeed from the heavy boson decays, 
and one cannot simply reconstruct the rest frame for each event. 
Carefully designed jet substructure observables must be employed to tag the boosted objects. 
Then the angular function decomposition back in the rest frame loses its advantage, 
but the azimuthal correlation substructures due to the linear polarization retain 
their simplicity and can be used to tag the observed jets.

Linearly polarized gluons can not only be produced from hard collisions, but also can exist ubiquitously elsewhere,
such as from heavy meson decay~\citep{Brodsky:1978hw, Koller:1980fk, Robinett:1990qt} 
and from parton showering~\citep{DeGrand:1980yp}. 
In particular, it has been noticed that a linearly polarized gluon can be emitted in the shower of an unpolarized parton 
and lead to nontrivial $\cos2\phi$ correlations~\citep{Chen:2020adz, Chen:2021gdk, Karlberg:2021kwr, Hamilton:2021dyz}.
The reason for this is that a $1 \to 2$ splitting in the boosted parton showering defines a plane and allows a linear
polarization along or perpendicular to the plane.
This resembles the gluon Boer-Mulders function in the transverse-momentum-dependent QCD factorization
~\citep{Mulders:2000sh, Nadolsky:2007ba, Boer:2010zf, Catani:2011kr, Sun:2011iw, Qiu:2011ai, Boer:2011kf}, for which
a linearly polarized gluon distribution can exist in an unpolarized hadron target when the gluon carries a nonzero 
transverse momentum.

Again, similar effects can be extended to massive vector bosons, which can come from the decay of a boosted heavy object
like a top quark or Higgs boson~\citep{Yu:2021zpe}. For the same reason, the intermediate vector boson can carry a linear
polarization and then decay into light particles preferentially along the direction parallel or perpendicular to the polarization
direction. This leads to a more complicated azimuthal correlation in the original boosted heavy particle. 
The minimal configuration is a $1 \to 3$ decay. When extended to the hadronic decay mode, the intermediate linearly polarized 
vector boson gives rise to an azimuthally inhomogeneous energy deposition pattern that makes the whole ``fat'' jet
more circular or planar. Such phenomena could be measured as a precision test of the SM and probe for new physics.

The rest of this thesis is organized as the following.
First, to lay the foundation of the polarization study, I will review in \ch{ch:Poincare} the definitions of the spin states 
and their Lorentz transformation properties which are governed by the corresponding little group, 
mainly following the discussion in~\citep{Weinberg:1995mt}.
Along the line will be derived the explicit little group forms for some important cases that will be used in later sections.
Then in \ch{ch:pol-fermion}, I will discuss the transverse spins of quarks, using the top quark as a main example. 
The discussion is mainly as an introduction for the vector boson polarization in the following chapter, 
with most being known in the literature. A brief comparison between the single quark polarization and the quark spin-spin
correlations is given at the end of this chapter.
Next, \ch{ch:pol-vboson} is devoted to the study of linear vector boson polarization at the LHC. 
This forms the main part of the rest of this thesis. 
We will first discuss the linear polarization of a gluon as produced directly from a hard collision. 
This discussion leads to the definition of a polarized gluon jet function, which provides a concrete procedure for measuring
the gluon polarization at the LHC. As will be explained, such measurement will provide a sensitive probe for possible 
$CP$-violating effects.
Then we will discuss the linear polarization of a vector boson that comes from the decay of a boosted heavy object. 
The focus will be on a boosted top quark that decays into a bottom quark and a $W$ boson, which further decays into 
a lepton pair or quark pair. We will give a physical argument of why the $W$ boson can be linearly polarized in the boosted
regime. The derivation will clarify it as a general phenomenon that a boosted $1\to3$ decay system can exhibit such 
azimuthal correlation if it is mediated by a vector boson.
Finally, in \ch{ch:spin-summary}, we conclude our discussion and present the outlook.

\chapter{Poincare group representation and little group transformation}
\label{ch:Poincare}


In this chapter, we review the \Poincare group representation and the associated little group, following Ch.~2.5 of \citep{Weinberg:1995mt}.


\section{General formalism}
\label{sec:poincare-general}

Setting the \Poincare symmetry as the fundamental spacetime symmetry, 
we identify states that can transform into each other under
a \Poincare transformation as belonging to the same particle species. 
The \Poincare symmetry transformation acts on the coordinate space as
\beq[eq:Poincare-x]
	x^{\mu} \to x^{\prime\mu} = \Lambda^{\mu}{}_{\nu} \, x^{\nu} + a^{\mu},
\eeq
defined to make $g_{\mu\nu} \, dx^{\mu} \, dx^{\nu}$ invariant, with 
$g_{\mu\nu} = {\rm diag}\{1, -1, -1, -1\}$.
This defines the Lorentz transformation $\Lambda^{\mu}{}_{\nu}$ and the translation $a^{\mu}$,
such that 
$g_{\mu\nu} \, \Lambda^{\mu}{}_{\rho} \, \Lambda^{\nu}{}_{\sigma} = g_{\rho\sigma}$.
\eq{eq:Poincare-x} induces a unitary operator $U(\Lambda, a)$ on the Hilbert space, 
the whole set of which forms the \Poincare group and satisfies
\beq
	U(\Lambda, a) \, U(\Lambda', a') = U(\Lambda \Lambda', a + \Lambda a'),
\eeq
and
\beq
	U(1, 0) = 1, \quad
	U^{\dag}(\Lambda, a) = U^{-1}(\Lambda, a) = U(\Lambda^{-1}, -\Lambda^{-1} a).
\eeq
We also define the Lorentz transformation $U(\Lambda) \equiv U(\Lambda, 0)$, 
whose set forms the Lorentz group, as a unitary subgroup of the \Poincare group.

By the translation properties, we label each single-particle state by its momentum $p^{\mu}$ and 
some internal quantum number collectively denoted as $\sigma$,
\beq
	\hat{P}^{\mu} | p, \sigma \rangle = p^{\mu} | p, \sigma \rangle,
\eeq
where $\hat{P}^{\mu}$ is the momentum operator, defined as the generator of the translation group,
\beq
	U(1, a) = e^{i \hat{P} \cdot a}.
\eeq
The momenta of two states can be related to each other under Lorentz transformation 
only if they have the same mass $m^2 = p^{\mu} p_{\mu}$ and, if $m^2 > 0$, sign of $p^0$.
Since 
\begin{align*}
	&\hat{P}^{\mu} \, U(\Lambda) | p, \sigma \rangle 
		= U(\Lambda) \bb{ U^{-1} (\Lambda) \hat{P}^{\mu} \, U(\Lambda) } | p, \sigma \rangle  \nn\\
	& \hspace{6em}
		= U(\Lambda) \bb{ \Lambda^{\mu}{}_{\nu} \, \hat{P}^{\mu} } | p, \sigma \rangle
		= \bb{ \Lambda^{\mu}{}_{\nu} \, p^{\mu} } U(\Lambda) | p, \sigma \rangle,
\end{align*}
the state $U(\Lambda) | p, \sigma \rangle$ has a momentum $\Lambda p$, and hence can be expanded as
\beq[eq:lorentz-rep]
	U(\Lambda) | p, \sigma \rangle 
	= \sum_{\sigma'} | \Lambda p, \sigma' \rangle \, D_{\sigma' \sigma}(\Lambda, p).
\eeq 
This forms the unitary representation of the Lorentz group, which is infinitely dimensional,
\beq
	U(\Lambda) | p, \sigma \rangle
	= \bb{ \sum_{\sigma'} \int\frac{d^3\bm{p}'}{(2\pi)^3 2E_{p'}} } | p', \sigma' \rangle
		\bb{ (2\pi)^3 (2E_{p'}) \delta^{(3)}(\bm{p}' - \Lambda\bm{p} ) \, D_{\sigma' \sigma}(\Lambda, p) },
\eeq
where $\Lambda\bm{p}$ is the three-vector part of $\Lambda p$.
By choosing the normalization
\beq
	\langle p' , \sigma' | p, \sigma \rangle = (2\pi)^3 \, (2E_{p})\, \delta^{(3)}(\bm{p} - \bm{p}' ) \, \delta_{\sigma \sigma'},
\eeq
we can write the representation matrix as
\beq
	\langle p', \sigma' | U(\Lambda) | p, \sigma \rangle
	= (2\pi)^3 \, (2E_{p'}) \, \delta^{(3)}(\bm{p}' - \Lambda\bm{p} ) \, D_{\sigma' \sigma}(\Lambda, p)
\eeq
in the $(p, \sigma)$ space. The matrix $D_{\sigma' \sigma}(\Lambda, p)$ is also unitary,
\beq
	\sum_{\sigma'} D^{\dag}_{\sigma'' \sigma'}(\Lambda, p) D_{\sigma' \sigma}(\Lambda, p) = \delta_{\sigma'' \sigma},
\eeq	
and satisfies the multiplication rule,
\beq
	D_{\sigma'' \sigma} (\Lambda \Lambda', p) 
	= \sum_{\sigma'} D_{\sigma'' \sigma'}(\Lambda, \Lambda' p) D_{\sigma' \sigma}(\Lambda', p),
\eeq
with 
\beq
	D(1, p) = 1, \quad
	D^{-1}(\Lambda, p) = D(\Lambda^{-1}, \Lambda p).
\eeq

To find the representation matrix of the Lorentz group, it is necessary to first clearly define each particle state $|p, \sigma \rangle$, especially
the quantum number $\sigma$. 
For a particular particle, we define all its states by relating to the ``standard'' states with a canonical reference momentum $k$, 
of which the quantum number $\sigma$ is defined for all values. 
This can be chosen as
\beq[eq:ref-mom-m]
	k = (m, 0, 0, 0)
\eeq
for a massive particle with $m^2 = k^2 > 0$, or
\beq[eq:ref-mom-0]
	k = (E_0, 0, 0, E_0)
	\quad
	(\mbox{with $E_0 = 1~\GeV$ for example})
\eeq
for a massless particle.
Any other possible momentum $p$ is related to $k$ by a standard Lorentz transformation
$L(p) \equiv L(p; k)$,
\beq
	p^{\mu} = L^{\mu}{}_{\nu}(p) k^{\nu},
\eeq
where we suppress the dependence on $k$ since it is common for all the states of a particular particle.
$L(p)$ can be standardly defined by first boosting along the $+\hat{z}$ direction by $U(\Lambda_z(\beta))$ such that 
$p_1 = \Lambda_z(\beta) k$ has the same energy as $p$, and then rotating $p_1$ to the same direction as $p$ by
first rotating around the $y$ axis by $\theta$ and then around the $z$ axis by $\phi$,
\beq[eq:Lp]
	L(p) = R(\theta, \phi) \Lambda_z(\beta) = R_z(\phi) R_y(\theta) \Lambda_z(\beta) 
		= \Lambda_{\hat{p}}(\beta) R(\theta, \phi),
\eeq
where $\theta$ and $\phi$ are the polar and azimuthal angles of $p$, respectively.
The last step gives an alternative definition of $L(p)$ 
that first rotates $k$ to the direction of $p$ by $R(\theta, \phi)$ 
and then boosts along the direction of $p$ by 
$\Lambda_{\hat{p}}(\beta) = R(\theta, \phi) \Lambda_z(\beta) R^{-1}(\theta, \phi)$ 
to reach the same energy. 
The induced Lorentz transformation $U(L(p))$ in the Hilbert space is thus
\beq[eq:Lp-quantum]
	U(L(p)) = U(R(\theta, \phi)) U(\Lambda_z(\beta))
		= U(\Lambda_{\hat{p}}(\beta)) U(R(\theta, \phi)),
\eeq
with 
\beq
	U(R(\theta, \phi)) = e^{-i J_z \phi} e^{-i J_y \theta},
	\quad
	U(\Lambda_{\hat{n}}(\beta)) = e^{-i \bm{K} \cdot \hat{n} \beta}.
\eeq

The state $|p, \sigma\rangle$ is {\it defined} as the Lorentz transformation of $|k, \sigma\rangle$ under $U(L(p))$,
\beq[eq:1-particle-state-def]
	|p, \sigma\rangle \equiv U(L(p)) |k, \sigma\rangle.
\eeq
Under an arbitrary Lorentz transformation $U(\Lambda)$, the state $|p, \sigma\rangle$ becomes
\begin{align}\label{eq:Lorentz-rep-0}
	U(\Lambda) |p, \sigma\rangle
	&= U(\Lambda) U(L(p)) |k, \sigma\rangle = U(\Lambda L(p)) |k, \sigma\rangle  \nn\\
	&= U(L(\Lambda p)) \, U(L^{-1}(\Lambda p) \Lambda L(p)) |k, \sigma\rangle.
\end{align}
Note that although the transformation $\Lambda L(p)$ brings the momentum $k$ to $\Lambda p$, 
it is not necessarily equal to $L(\Lambda p)$. 
But it does imply that the transformation 
\beq[eq:little-group-W]
	W(\Lambda, p) \equiv L^{-1}(\Lambda p)\Lambda L(p)
\eeq
keeps $k$ invariant,
\beq
	W^{\mu}{}_{\nu}(\Lambda, p) k^{\nu} = k^{\mu}.
\eeq
Now for a specific momentum $k$, all the Lorentz transformations that 
leave it invariant form a subgroup of the Lorentz group,
which is called the {\it little group}. 
A little group transformation $W$ thus only mixes the quantum number $\sigma$, 
and its representation can be easily obtained,
\beq[eq:little-group-rep]
	U(W) |k, \sigma\rangle = \sum_{\sigma'} |k, \sigma' \rangle D_{\sigma' \sigma}(W),
\eeq
where $D(W)$ is a unitary matrix. 
Plugging \eq{eq:little-group-rep} back into \eq{eq:Lorentz-rep-0} then 
gives the Lorentz group representation,
\begin{align}\label{eq:Lorentz-rep}
	U(\Lambda) |p, \sigma\rangle
	&= U(L(\Lambda p)) U(W(\Lambda, p)) |k, \sigma\rangle
	= U(L(\Lambda p)) \sum_{\sigma'} |k, \sigma' \rangle D_{\sigma' \sigma}(W(\Lambda, p)) \nn\\
	&= \sum_{\sigma'} |\Lambda p, \sigma' \rangle D_{\sigma' \sigma}(W(\Lambda, p)).
\end{align}
Compared with \eq{eq:lorentz-rep}, we get
\beq
	D_{\sigma' \sigma}(\Lambda, p) = D_{\sigma' \sigma}(W(\Lambda, p)),
\eeq
which explains why the same symbol ``$D$'' is used.
In this way, the Lorentz group representation is induced from its little group representation. 
The task of obtaining the transformation behavior of $|p, \sigma\rangle$ under $U(\Lambda)$ is 
then reduced to finding the corresponding little group transformation $W(\Lambda, p)$.
For this purpose, we need to clearly define $k$ for each particle and the corresponding $L(p)$.
We will do that separately for massive and massless particles.

Before that, let us first work out the Lorentz transformation behavior of a scattering amplitude.
The helicity amplitude of a scattering 
$(p_1, \sigma_1; p_2, \sigma_2; \ldots) \to (q_1, \lambda_1; q_2, \lambda_2; \ldots)$ 
is obtained from the scattering $S$-operator by 
\beq
	\M_{\sigma_1, \sigma_2, \ldots; \; \lambda_1, \lambda_2, \ldots}(p_1, p_2, \ldots; q_1, q_2, \ldots)
	\sim 
	\langle q_1, \lambda_1; q_2, \lambda_2; \ldots | S | p_1, \sigma_1; p_2, \sigma_2; \ldots \rangle.
\eeq
If we transform the scattering system to another frame by $\Lambda$, the new helicity amplitude becomes
\beq[eq:lorentz-amplitude-2]
	\M_{\sigma_1, \sigma_2, \ldots; \, \lambda_1, \lambda_2, \ldots}
		(\Lambda p_1, \Lambda p_2, \ldots; \Lambda q_1, \Lambda q_2, \ldots)
	\sim 
	\langle \Lambda q_1, \lambda_1; \Lambda q_2, \lambda_2; \ldots | 
		S | \Lambda p_1, \sigma_1; \Lambda p_2, \sigma_2; \ldots \rangle.
\eeq
Their relations can be obtained by using $S = U(\Lambda) S U^{-1}(\Lambda) $ in \eq{eq:lorentz-amplitude-2},
\begin{align}\label{eq:lorentz-amplitude-21}
	\langle \Lambda q_1, \lambda_1; \ldots | S | \Lambda p_1, \sigma_1; \ldots \rangle
	= \big[ \langle \Lambda q_1, \lambda_1; \ldots | U(\Lambda) \big] S \bb{ U^{-1}(\Lambda) | \Lambda p_1, \sigma_1; \ldots \rangle }
\end{align}
Now \eq{eq:Lorentz-rep} gives
\beq[eq:U-1lambda]
	U^{-1}(\Lambda) | \Lambda p, \sigma \rangle 
	= \sum_{\sigma'} | p, \sigma' \rangle D_{\sigma' \sigma}(W(\Lambda^{-1}, \Lambda p))
	= \sum_{\sigma'} | p, \sigma' \rangle D^{-1}_{\sigma' \sigma}(W(\Lambda, p)),
\eeq
where we used $W(\Lambda^{-1}, \Lambda p) = W^{-1}(\Lambda, p)$ by the definition in \eq{eq:little-group-W}. 
So then \eq{eq:lorentz-amplitude-21} becomes
\begin{align}\label{eq:lorentz-amplitude-relation}
	&\langle \Lambda q_1, \lambda_1; \ldots | S | \Lambda p_1, \sigma_1; \ldots \rangle	\nn\\
	&\hspace{1em}
	= \sum_{\lambda_1', \ldots; \, \sigma_1', \ldots} 
		\big[ D_{\lambda_1 \lambda_1'}(W(\Lambda, q_1)) \ldots \big]
		\langle q_1, \lambda_1'; \ldots | S | p_1, \sigma_1'; \ldots \rangle
		\bb{ D^{\dag}_{\sigma_1' \sigma_1}(W(\Lambda, p_1)) \ldots }.
\end{align}
Therefore, helicity amplitudes in different frames are connected by a unitary transformation,
\begin{align}
	&\M_{\sigma_1, \ldots; \, \lambda_1, \ldots}(\Lambda p_1, \ldots; \Lambda q_1, \ldots)	\nn\\
	&\hspace{.2em}
	= \sum_{\lambda_1', \ldots; \, \sigma_1', \ldots} 
		\bb{ D_{\lambda_1 \lambda_1'}(W(\Lambda, q_1)) \ldots }
		\M_{\sigma_1', \ldots; \, \lambda_1', \ldots}(p_1, \ldots; q_1, \ldots)
		\bb{ D^{\dag}_{\sigma_1' \sigma_1}(W(\Lambda, p_1)) \ldots }.
\end{align}
Because of the unitarity of the representation matrices $D$'s, multiplying the helicity amplitude by its complex conjugate
and summing over all helicities leads to a Lorentz invariant unpolarized amplitude square.


\section{Massive case: $m > 0$}
\label{sec:little-group-massive}

The little group for a massive particle is the three-dimensional rotation group, SO(3).
For the reference momentum $k$, we define the quantum number $\sigma$ to be the angular momentum component along $\hat{z}$.
Particles can be further decomposed into different species according to different irreducible representations $D^j$ of SO(3). 
This introduces a total spin quantum number $j$, so we label the state as $|k, j, \sigma\rangle$,
\beq
	J_z |k, j, \sigma \rangle = \sigma |k, j, \sigma \rangle,
	\quad
	U(W) |k, j, \sigma \rangle = \sum_{\sigma' = -j}^j |k, j, \sigma' \rangle D^j_{\sigma' \sigma}(W),
\eeq
where $D^j(W)$ is the $(2j+1)$-dimensional irreducible representation matrix of the SO(3) group.
For an arbitrary momentum $p$, the quantum number $\sigma$ in the state $|p, j, \sigma \rangle$, 
which is related to $|k, j, \sigma\rangle$ by \eqs{eq:Lp}{eq:1-particle-state-def}, is defined as the helicity,
\begin{align}
	(\bm{J} \cdot \hat{p}) |p, j, \sigma \rangle
	& = U(R(\theta, \phi)) J_z \, U^{-1}(R(\theta, \phi)) |p, j, \sigma \rangle
	= U(R(\theta, \phi)) J_z \, U(\Lambda_z(\beta)) |k, j, \sigma \rangle \nn\\
	& = U(R(\theta, \phi)) U(\Lambda_z(\beta)) J_z  |k, j, \sigma \rangle
	= \sigma |p, j, \sigma \rangle,
\end{align}
where we have used \eq{eq:Lp-quantum} with $\beta = |\bm{p}| / \sqrt{\bm{p}^2 + m^2}$, and that $J_z$ commutes with $U(\Lambda_z(\beta))$. 
By \eq{eq:Lorentz-rep}, we can get the
representation of a general Lorentz transformation $U(\Lambda)$, which mixes different helicity states of a given particle,
\beq
	U(\Lambda) |p, j, \sigma \rangle = \sum_{\sigma' = -j}^j |\Lambda p, j, \sigma' \rangle D^j_{\sigma' \sigma}(W(\Lambda, p)).
\eeq

The little group transformation $W(\Lambda, p)$ for a general $\Lambda$ and $p$ is not easily worked out. 
Here we only consider two special cases. 

\subsection{Pure Rotation: $\Lambda = \hat{R}$}
The first is for a pure rotation $\Lambda = \hat{R}$. It only changes the direction $\hat{p}$ to $\hat{R} \hat{p}$, but does not change its energy. 
Following our notation of $R(\hat{n})$ as the standard rotation that takes $\hat{z}$ to $\hat{n}$, we have 
\beq[eq:rot-little-group-0]
	W(\hat{R}, p) = \Lambda_z^{-1} (\beta) \bb{ R^{-1}(\hat{R} \hat{p}) \hat{R} R(\hat{p}) } \Lambda_z(\beta).
\eeq
The rotation matrices in the square bracket first take $\hat{z}$ to $\hat{p}$, then to $\hat{R} \hat{p}$, and then back to $\hat{z}$, so it is at most a
rotation around $\hat{z}$,
\beq
	R^{-1}(\hat{R} \hat{p}) \hat{R} R(\hat{p}) = R_z(\delta(\hat{R}, \hat{p})).
\eeq
Inserting this back to \eq{eq:rot-little-group-0} gives
\beq[eq:little-group-rotation]
	W(\hat{R}, p) = R_z(\delta(\hat{R}, \hat{p})).
\eeq
So the little group for a rotation $\hat{R}$ is merely a rotation around $z$. 
The corresponding Lorentz representation is thus a pure phase under,
\beq[eq:lorentz-rep-massive-R]
	U(\hat{R}) |p, j, \sigma \rangle = e^{-i \sigma \delta(\hat{R}, \hat{p}) } |p, j, \sigma \rangle,
\eeq
which keeps $\sigma$ invariant. We consider three special examples of $\hat{R}$: 
\begin{enumerate}
\item [(1)] $\hat{R} = R(\theta, \phi) R_z(\gamma) R^{-1}(\theta, \phi)$ is a rotation around $\hat{p}$ by $\gamma$, which gives
	$\delta(\hat{R}, \hat{p}) = \gamma$;
\item [(2)] $\hat{R} = R_z(\gamma)$ is a rotation around $\hat{z}$ by $\gamma$, which gives
	$\delta(\hat{R}, \hat{p}) = 0$;
\item [(3)] $\hat{R} = R_z(\phi) R_y(\gamma) R^{-1}_z(\phi)$ is a rotation of the $\hat{z}$-$\hat{p}$ plane 
	(usually defined as the inclusive scattering plane) 
	by $\gamma$, which gives $\delta(\hat{R}, \hat{p}) = 0$.
\end{enumerate}

\subsection{Boost along $\hat{z}$: $\Lambda = \Lambda_z(\hat{\beta})$}
The second case is for a pure Lorentz boost along the $z$ direction. This is useful in two circumstances. 
First, the spin state of a particle produced from a hard scattering can be usually calculated easily in the c.m.~frame. 
But at a hadron collider such as the LHC, each hard scattering event in the lab frame differs from the c.m.~frame event by a longitudinal 
boost along $\hat{z}$. Second, if the particle is produced from a heavy particle decay, its spin state is easily worked out in the rest frame
of the mother particle. But the latter is likely boosted in the lab frame. 
The connection between the two frames requires a boost 
along the momentum of the mother particle which we can define as $\hat{z}$.

Denote $v$, $\theta$, and $\phi$ as the speed, polar angle, and azimuthal angle of $p$. 
The boost $\Lambda_z(\hat{\beta})$ transforms it to $p' = \Lambda_z(\hat{\beta})p$, with speed $v'$, polar angle $\theta'$,
and azimuthal angle $\phi'$. They are related by
\begin{align}
	\tan\theta' = \frac{v \sin\theta \sqrt{ 1 - \hat{\beta}^2 } }{\hat{\beta} + v \cos\theta}, \quad
	\phi' = \phi, \quad
	v' = \sqrt{ 1 - \frac{(1 - \hat{\beta}^2) ( 1 - v^2) }{ (1 + \hat{\beta} v\cos\theta)^2} }.
\end{align}
By \eq{eq:little-group-W}, the little group transformation is
\begin{align}\label{eq:little-group-boost-0}
	W(\Lambda, p) 
		&= \Lambda_z^{-1}(v') R_y^{-1}(\theta') R_z^{-1}(\phi) \Lambda_z(\hat{\beta}) R_z(\phi) R_y(\theta) \Lambda_z(v) \nn\\
		&= \Lambda_z^{-1}(v') R_y^{-1}(\theta') \Lambda_z(\hat{\beta}) R_y(\theta) \Lambda_z(v),
\end{align}
where the $\phi$ dependence cancels since $R_z$ commutes with $\Lambda_z$.
Note that \eq{eq:little-group-boost-0} only involves boosts along $\hat{z}$ and rotation around $\hat{y}$, which all keep the vector
$y^{\mu} = (0, 0, 1, 0)$ unchanged. So the resulting little group must be a rotation $R_y(\chi)$ around $\hat{y}$. 
This is verified by an explicit calculation, which gives
\begin{align}\label{eq:little-group-chi}
	W(\Lambda, p) = R_y(\chi), \quad
	\cos\chi &= \frac{v + \hat{\beta} \cos\theta}{\sqrt{(1 + \hat{\beta} v\cos\theta)^2 - (1 - \hat{\beta}^2) (1 - v^2)}}, 
	\quad
	\chi \in [0, \pi].
\end{align}
Such a nontrivial little group transformation causes the boost $\Lambda_z(\hat{\beta})$ to mix the helicity states,
\beq[eq:boost-transformation-massive]
	U(\Lambda_z(\hat{\beta})) \, |p, j, \sigma \rangle 
		= \sum_{\sigma' = -j}^j |\Lambda_z(\hat{\beta}) p, j, \sigma' \rangle \, d^j_{\sigma'\sigma}(\chi(\hat{\beta}, p)),
\eeq
where $d^j$ is the Wigner-$d$ function, being the representation matrix of $U(R_y(\chi))$.

\section{Massless case: $m = 0$}
\label{sec:little-group-massless}

The little group that keeps invariant the standard reference momentum vector $k = (1, 0, 0, 1)$ (suppressing the irrelevant $E_0$ factor in this section)
is isomorphic to ISO(2), the two-dimensional translation and rotation group. 
We will follow \citep{Weinberg:1995mt} for the derivation.

First introduce an auxiliary vector $t^{\mu} = (1, 0, 0, 0)$. The little group transformation $W$ has the properties
\beq[eq:little-group-W-massless]
	W^{\mu}{}_{\nu} k^{\nu} = k^{\mu}, \quad
	(Wt)^{\mu} k_{\mu} = (Wt) \cdot (Wk) = t \cdot k = 1, \quad
	(Wt)^{\mu} (Wt)_{\mu} = t^2 = 1.
\eeq
The second property of those implies
\beq
	W^{\mu}{}_{\nu} \, t^{\nu} = (1 + \zeta, \alpha, \beta, \zeta)
\eeq
and the third one constrains
\beq
	\zeta = \frac{\alpha^2 + \beta^2}{2}.
\eeq
This determines the first column of the $W$ matrix, $W^{\mu}{}_0$. The first condition in \eq{eq:little-group-W-massless} further constrains the last column,
$W^{\mu}{}_3$. The remaining two columns can be determined by Lorentz group properties up to some degrees of freedom. One solution for $W$ is
\beq[eq:little-group-S]
	S = \pp{ S^{\mu}{}_{\nu}(\alpha, \beta) }
	= 	\begin{pmatrix}
			1 + \zeta 	& \alpha 	& \beta 	& 	-\zeta \\
			\alpha 		& 1			&	0		&	-\alpha \\
			\beta			& 0			&	1		&	-\beta	\\
			\zeta			& \alpha 	& \beta	&	1 - \zeta
		\end{pmatrix}.
\eeq
To find the most general form of $W$, we notice that by $Wt = St$, the transformation $S^{-1} W$ leaves $t$ invariant, so $S^{-1} W \in {\rm SO(3)}$.
On the other hand, $S^{-1} W$ also leaves $k$ invariant, so it can only be a rotation $R_z(\theta)$ around $\hat{z}$. Hence, we have the general expression
for the little group element,
\beq[eq:little-group-W=SR]
	W = W(\alpha, \beta, \theta) = S(\alpha, \beta) R_z(\theta),
\eeq
which has three parameters $\alpha, \beta$, and $\theta$.

The little group multiplication properties can be worked out straightforward, and we get:
\begin{enumerate}
\item the subgroup formed by $S$ is Abelian and has a simple addition rule for the parameters $(\alpha, \beta)$: 
	$S(\alpha, \beta) S(\alpha', \beta') = S(\alpha + \alpha', \beta + \beta')$;
\item the subgroup formed by $R_z$ has the same property: 
	$R_z(\theta) R_z(\theta') = R_z(\theta + \theta')$; 
	and
\item the parameters $(\alpha, \beta)$ have a simple rotation property under the action of $R_z(\theta)$:
	$R_z(\theta) S(\alpha, \beta) R_z^{-1}(\theta) = S(\alpha \cos\theta - \beta \sin\theta, \alpha \sin\theta + \beta \cos\theta)$.
\end{enumerate}
The first and third properties together mean that the elements $S$ form an invariant Abelian subgroup, so that the little group is not semi-simple.
If we denote the $\bm{v} = (\alpha, \beta)$, the multiplication rules will become more transparent,
\beq[eq:little-group-multiplication]
	S(\bm{v} ) S(\bm{v}') = S(\bm{v} + \bm{v}'), \quad
	R_z(\theta) S(\bm{v}) R_z^{-1}(\theta) = S(R_z(\theta) \bm{v}),
\eeq
where in the expression $R_z(\theta) \bm{v}$, $R_z(\theta)$ is the rotation matrix adapted to the $x$-$y$ plane in an obvious way.
This clearly shows its isomorphism to ISO(2) on the $x$-$y$ plane, which transforms a point $(x, y)$ to $R_z(\theta) (x, y) + \bm{v}$.
with $\bm{v}$ corresponding to the two-dimensional translation vector, and $\theta$ the rotation angle around $\hat{z}$.

In the neighborhood of the identity element, the little group element $W(\alpha, \beta, \theta)$ can be expanded around $\alpha = \beta = \theta = 0$, 
which gives
\begin{align}\label{eq:little-group-small-expansion}
	W & \simeq 1 + 
	\begin{pmatrix}
		0 & \alpha & \beta & 0 \\
		\alpha & 0 & -\theta & - \alpha \\
		\beta & \theta & 0 & -\beta \\
		0 & \alpha & \beta & 0
	\end{pmatrix} \nn\\
	&= 1 - i ( K_1 \alpha + K_2 \beta + J_1 \beta - J_2 \alpha + J_3 \theta ) \nn\\
	&\equiv 1 - i (A \, \alpha + B \, \beta + J_3 \, \theta),
\end{align}
such that the little group is spanned by three generators,
\beq[eq:ISO2-generators]
	A = K_1 - J_2, \quad
	B = K_2 + J_1, \quad
	J_3.
\eeq
Here $K_i$ and $J_i$ are the representation matrices of the Lorentz group in the vector space.
The corresponding Lie algebra is
\beq[eq:ISO2-Lie-algebra]
	[J_3, A] = i B, \quad
	[J_3, B] = -i A, \quad
	[A, B] = 0.
\eeq
A finite little group element $W(\alpha, \beta, \theta)$ can then be generated from the exponential
\beq
	W(\alpha, \beta, \theta) = e^{-i(A\alpha + B \beta)} e^{-i J_3 \theta}.
\eeq
So far we have been working on the 4-dimensional Lorentz group representation in the Minkowski space. 
This also induces a unitary representation on the Hilbert space, 
\beq
	U(W(\alpha, \beta, \theta)) = e^{-i(\hat{A} \alpha + \hat{B} \beta)} e^{-i \hat{J}_3 \theta},
\eeq
where $\hat{A}$, $\hat{B}$, and $\hat{J}_3$ are Hermitian operators and 
have the same properties as Eqs.~\eqref{eq:ISO2-generators} and \eqref{eq:ISO2-Lie-algebra}.

The little group ISO(2) contains all Lorentz group elements that leave $k$ invariant. The transformation property of the $\sigma$ index
in the state $|k, \sigma \rangle$ under the little group gives a physical definition of $\sigma$.
Because only two of the Hermitian generators, $A$ and $B$, commute, we may orient the reference state to be a simultaneous eigenstate,
$|k, a, b \rangle$, of $\hat{P}^{\mu}$, $A$, and $B$,
\beq[eq:A-B-eigen]
	A |k, a, b \rangle = a |k, a, b \rangle, \quad
	B |k, a, b \rangle = b |k, a, b \rangle,
\eeq
with $(a, b)$ the quantum numbers charactering the state, together with the momentum $k$. 
Now we define
\beq
	(A_{\theta}, B_{\theta}) = e^{-i J_3 \theta} (A, B) e^{i J_3 \theta}.
\eeq
Applying a derivative with respect to $\theta$ using \eq{eq:ISO2-Lie-algebra} gives
\beq
	\frac{d}{d\theta}(A_{\theta}, B_{\theta}) 
	= (A_{\theta}, B_{\theta}) 
		\begin{pmatrix}
		0 & -1 \\
		1 & 0
		\end{pmatrix}
	= -i (A_{\theta}, B_{\theta}) \sigma_2.
\eeq
The solution is obtained by an exponentiation,
\beq
	(A_{\theta}, B_{\theta})  = (A, B) \, e^{ -i \sigma_2 \theta} = (A, B) R_z(\theta)
		= (A \cos\theta + B \sin\theta, -A \sin\theta + B \cos\theta).
\eeq
Then from \eq{eq:A-B-eigen}, we have
\begin{align} 
	A e^{-iJ_3 \theta} |k, a, b \rangle 
		&= e^{-iJ_3 \theta} A_{-\theta} |k, a, b \rangle 
		= (a \cos\theta - b \sin\theta) e^{-iJ_3 \theta} |k, a, b \rangle, \nn\\
	B e^{-iJ_3 \theta} |k, a, b \rangle 
		&= e^{-iJ_3 \theta} B_{-\theta} |k, a, b \rangle 
		= (a \sin\theta + b \cos\theta) e^{-iJ_3 \theta} |k, a, b \rangle,
\end{align}
and so a rotation $R_z(\theta)$ mixes the two quantum numbers $a$ and $b$,
\beq
	e^{-iJ_3 \theta} |k, a, b \rangle = |k, a \cos\theta - b \sin\theta, a \sin\theta + b \cos\theta \rangle.
\eeq
Such a continuous spectrum is not observed in nature, and hence we must have $a = b = 0$.

While $J_3$ does not commute with $A$ or $B$ and thus they generally cannot have simultaneous eigenstates, 
now $A$ and $B$ have zero eigenvalues, so the state can also be a simultaneous eigenstate of $J_3$,
\beq
	|k, \sigma \rangle \equiv |k, (a, b) = (0, 0), \sigma \rangle,
\eeq
with
\beq
	A |k, \sigma \rangle = 0, \quad
	B |k, \sigma \rangle = 0, \quad
	J_3 |k, \sigma \rangle = \sigma |k, \sigma \rangle,
\eeq
without violating \eq{eq:ISO2-Lie-algebra}. The quantum numbers associated with $A$ and $B$ thus become redundant and
$\sigma$ has the physical meaning of helicity. The little group transformation is
\beq
	U(W(\alpha, \beta, \theta)) |k, \sigma \rangle = e^{-i \sigma \theta} |k, \sigma \rangle.
\eeq
This then induces the Lorentz group representation,
\beq[eq:Lorentz-rep-massless]
	U(\Lambda) |k, \sigma \rangle = e^{-i \sigma \theta(\Lambda, p)} |k, \sigma \rangle,
\eeq
with $\theta(\Lambda, p)$ determined by
\beq[eq:little-group-massless]
	W(\Lambda, p) = L^{-1}(\Lambda p) \Lambda L(p) = S(\alpha, \beta) R_z(\theta),
\eeq
where $S(\alpha, \beta)$ is the little group element {\it defined} to have the form in \eq{eq:little-group-S}.

Similar to \sec{sec:little-group-massive}, now we give two special cases where the little group can be explicitly evaluated.

\subsection{Pure Rotation: $\Lambda = \hat{R}$}
A pure rotation $\Lambda = \hat{R}$ on massless states has the same effects as it applies on massive states, since it only 
involves the momentum directions. The little group transformation is the same as \eq{eq:little-group-rotation}. 
Compared with \eq{eq:little-group-massless}, we have
\beq
	\alpha = \beta = 0, \quad
	\theta = \delta(\hat{R}, \hat{p}),
\eeq
which gives the same Lorentz representation as \eq{eq:lorentz-rep-massive-R},
\beq
	U(\hat{R}) |p, \sigma \rangle = e^{-i \sigma \delta(\hat{R}, \hat{p}) } |p, \sigma \rangle.
\eeq

\subsection{Boost along $\hat{z}$: $\Lambda = \Lambda_z(\hat{\beta})$}
A pure Lorentz boost along $\hat{z}$ results in a similar expression like \eq{eq:little-group-boost-0}, just with 
different values for $\theta'$, $v$, and $v'$. Before evaluating it, we make the observation that none of the transformations
in \eq{eq:little-group-boost-0} changes $y^{\mu} = (0, 0, 1, 0)$. 
On the other hand, the little group element in \eq{eq:little-group-massless} 
takes it to
\beq
	y^{\prime\mu} = (\beta \cos\theta - \alpha \sin\theta, -\sin\theta, \cos\theta, \beta \cos\theta - \alpha \sin\theta).
\eeq
Therefore, we must have
\beq
	\beta = \theta = 0.
\eeq
This is easily verified by an explicit evaluation of \eq{eq:little-group-boost-0}, which gives
\beq
	\alpha = - \frac{E_0}{E} \frac{\hat{\beta} \sin\theta_p}{1 + \hat{\beta} \cos\theta_p},
\eeq
where $E_0$ and $E$ are the energies of $k$ and $p$, respectively, and $\theta_p$ is the polar angle of $p$.
As a result, even if the corresponding little group transformation is not identity, the Lorentz boost along $\hat{z}$ does leave the helicity invariant,
\beq[eq:boost-transformation-massless]
	U(\Lambda_z) |p, \sigma \rangle = | \Lambda_z p, \sigma \rangle,
\eeq
which is in contrast to the massive case in \eq{eq:boost-transformation-massive}.


\chapter{Polarization of fermions at high-energy colliders}
\label{ch:pol-fermion}


\section{Fermion spin density matrix}
\label{sec:spin-dm-f}

At high-energy colliders, fermion spins are usually described in the helicity basis $\cc{ |p, \pm \rangle }$. 
A general fermion spin state is described by the density matrix, defined as 
\beq
	\rho^{1/2}_{\alpha \alpha'}(p) =  \langle p, \alpha | \hat{\rho}^{1/2} | p, \alpha' \rangle,
	\quad
	\alpha, \alpha' = \pm 1/2,
\eeq
with $\hat{\rho}^{1/2}$ being the spin density operator.
It is a $2 \times 2$ Hermitian matrix with a unity trace, so can be decomposed in terms of the Pauli matrices 
$\bm{\sigma} = (\sigma_1, \sigma_2, \sigma_3)$,
\beq[eq:f-spin-den-mtx]
	\rho^{1/2}_{\alpha \alpha'}(p) 
	= \frac{1}{2} \pp{ 1 + \bm{s}(p) \cdot \bm{\sigma} }_{\alpha \alpha'} 
	= \frac{1}{2} 
		\begin{pmatrix}
			1 + \lambda(p) & b_1(p) - i b_2(p) \\
			 b_1(p) + i b_2(p) & 1 - \lambda(p)
		\end{pmatrix}_{\alpha \alpha'} ,
\eeq
which defines the spin vector $\bm{s}(p) = (b_1(p), b_2(p), \lambda(p))$ for the fermion. 
From now on, we will suppress the momentum dependence of the density matrix and spin vector, unless necessary.
The positivity condition requires 
\beq[eq:dm-positivity-f]
	\det [ \rho^{1/2} ]  = 1 - \bm{s}^2 \geq 0, 
\eeq
	which means 
\beq
	\bm{s}^2 = b_1^2 + b_2^2 + \lambda^2 \leq 1,
\eeq
where $\bm{s}^2 = 1$ refers to a pure state and $\bm{s}^2 < 1$ to a mixed state. 

Under a general Lorentz transformation $\Lambda$, the density matrix becomes
\begin{align}\label{eq:denmtx-lorentz}
	\rho_{\alpha \alpha'}(p) \;\to\; \langle \Lambda p, \alpha | \bb{ U(\Lambda) \, \hat{\rho} \, U^{-1}(\Lambda) } | \Lambda p, \alpha' \rangle,
\end{align}
where we have temporarily suppressed the superscript ``$1/2$'' because it applies to all cases. 
Then using \eq{eq:U-1lambda}, we have
\beq[eq:denmtx-lorentz-f]
	\rho_{\alpha \alpha'}(p) 
	\;\to\; 
	\sum_{\bar{\alpha}, \bar{\alpha}'} 
		D_{\alpha \bar{\alpha}}(W(\Lambda, p)) 
		\rho_{\bar{\alpha} \bar{\alpha}'}(p) 
		D^{\dag}_{\bar{\alpha}' \alpha'}(W(\Lambda, p)),
\eeq
which transforms in a similar way to the helicitiy amplitude [\eq{eq:lorentz-amplitude-relation}].
As a result, a general Lorentz transformation mixes different components of the density matrix.

The physical meaning of $(b_1, b_2, \lambda)$ can be examined through their properties under a rotation $\hat{R}(\phi)$ around the 
momentum direction. That gives the little group $W(\hat{R}(\phi), p) = R_z(\phi)$, by \eq{eq:little-group-rotation}, and thus
\beq
	\rho^{1/2}_{\alpha \alpha'} (\bm{s}') 
	= e^{-i \, \alpha \, \phi} \, \rho^{1/2}_{\alpha \alpha'} (\bm{s})\, e^{+i \, \alpha' \, \phi},
\eeq
which gives 
\beq
	\lambda' = \lambda, \quad
	b_1' = b_1 \cos\phi - b_2 \sin\phi, \quad
	b_2' = b_1 \sin\phi + b_2 \cos\phi.
\eeq
Hence $\lambda = \rho^{1/2}_{++} - \rho^{1/2}_{--}$ is the ``net" helicity of the fermion, which is unchanged under the rotation $\hat{R}(\phi)$,
and $\bm{b}_T \equiv (b_1, b_2)$ is the {\it transverse spin} of the fermion, which rotates as a two-dimensional vector.

Let us choose the particle momentum direction as the $z$ direction, and the two perpendicular directions as $x$ and $y$ directions.
Since each Pauli matrix $\sigma^i$ can be decomposed into the spin eigenstates along the $i$-th direction,
\beq
	\sigma^i = | i \rangle \langle i | -  | -i \rangle \langle -i |,
	\quad
	(i = x, y, z)
\eeq
where the bra and ket notations are abused to refer to two-component spinors, then \eq{eq:f-spin-den-mtx} implies
\begin{align}
	\rho^{1/2} 
	&= \frac{1}{2} \bb{ 1 + b_1 \pp{ |x\rangle \langle x | - |-x\rangle \langle -x | } \right.\nn\\
	& \left. \hspace{3em}
			+ \, b_2 \pp{ |y\rangle \langle y | - |-y\rangle \langle -y | } 
			+ \lambda \pp{ |z\rangle \langle z | - |-z\rangle \langle -z | } },
\end{align}
where the $z$ direction is along the particle momentum.
This gives a clear physical meaning for each component of the spin vector, 
\beq[eq:spin-params-f]
	b_1 = \rho^{1/2}_{x} - \rho^{1/2}_{-x}, \quad
	b_2 = \rho^{1/2}_{y} - \rho^{1/2}_{-y}, \quad
	\lambda = \rho^{1/2}_{z} - \rho^{1/2}_{-z},
\eeq
where $\rho^{1/2}_i \equiv \langle i | \rho^{1/2} | i \rangle$.
That is, $(1 \pm s_i) / 2$ is the probability for the particle spin to be along the $i$ or $-i$ direction.

\section{Singly polarized fermion production: general discussion}
\label{sec:fermion-spin}

In this section, we focus on the production of a singly polarized fermion, that is, we only observe the polarization of a certain fermion in 
the final state, and inclusively sum over all the other particles' spins. Intuitively, such an unpolarized scattering should not produce a singly
polarized particle. However, there is an interesting correlation between spins and momenta, which can yield singly polarized particle.

We consider a $2\to2$ scattering
\beq
	a(p_1, \alpha_1) + b(p_2, \alpha_2) \to c(p_3, \alpha_3) + f(p, \alpha),
\eeq
in which $f$ is the fermion whose spin $\alpha$ we observe. 
In the c.m.~frame, we choose $a$ to be along the $\hat{z}_\lab$ direction, which
together with the fermion momentum direction $\hat{p}(\theta_f, \phi_f)$ defines a scattering plane, whose normal is $\hat{z}_\lab \times \hat{p}$. 
The spin density matrix of $f$ can be obtained form the helicity amplitude 
$\M_{\alpha_1 \alpha_2 \alpha_3 \alpha}(p_1, p_2, p_3, p)$
by
\beq[eq:spin-dm-f-2to2]
	\rho_{\alpha\alpha'}(p) 
	= \frac{\sum_{\alpha_1, \alpha_2, \alpha_3} 
				\M_{\alpha_1 \alpha_2 \alpha_3 \alpha} \,
				\M^*_{\alpha_1 \alpha_2 \alpha_3 \alpha'}
			}{	\sum_{\alpha_1, \alpha_2, \alpha_3, \alpha_4} 
				\abs{ \M_{\alpha_1 \alpha_2 \alpha_3 \alpha_4} }^2
			},
\eeq
which in turn defines the spin vector $\bm{s} = (\bm{b}_T, \lambda)$ through \eq{eq:f-spin-den-mtx}.
A nonzero $\lambda$ implies the asymmetry between productions of a right-handed $f$ and a left-handed $f$.
The transverse spin $\bm{b}_T$ is provided by the off-diagonal elements of $\rho$, 
which is given by the interference of two amplitudes,
$\M_{\alpha_1 \alpha_2 \alpha_3 +}$ and $\M_{\alpha_1 \alpha_2 \alpha_3 -}$, 
which differ by only flipping the helicity of $f$.
Before going further, let us first clarify with respect to which axes the transverse spin is defined. 

Note that the state vector $|p, \alpha\rangle$ used in the calculation of the helicity amplitude is constructed from a reference state $|k, \alpha\rangle$
in the standard way as specified in \eq{eq:Lp}. 
Similarly, the transverse spin eigenstates $|p, \perp, \varphi \rangle$, defined as linear superpositions of the helicity eigenstates,
\beq
	|p, \perp, \varphi \rangle = \frac{1}{\sqrt{2}} \pp{ e^{-i \varphi / 2} |p, + \rangle + e^{i \varphi / 2} |p, - \rangle },
\eeq
are obtained from $|k, \perp, \varphi \rangle$ by the same set of transformations in \eq{eq:Lp}. 
Since the same definitions of $\bm{b}_T$ in \eq{eq:spin-params-f} hold for the density matrix in \eq{eq:spin-dm-f-2to2}
using the transverse spin basis $|p, \perp, \varphi \rangle$, 
the reference directions $\hat{x}$ and $\hat{y}$ with respect to which $\bm{b}_T$ is defined
are obtained from the lab frame $\hat{x}_\lab$ and $\hat{y}_\lab$ by first rotating around 
$\hat{y}_\lab$ by angle $\theta_f$ and then around $\hat{z}_\lab$ by angle $\phi_f$. 
And the $\hat{z}$ direction referred to by $\lambda$ in \eq{eq:spin-dm-f-2to2} is the direction $\hat{p}$. Therefore, we have
\beq[eq:fermion-xyz]
	\hat{z} = \hat{p}, \quad
	\hat{y} = \frac{\hat{z}_\lab \times \hat{z}}{| \hat{z}_\lab \times \hat{z} |}, \quad
	\hat{x} = \hat{y} \times \hat{z},
\eeq
such that $\hat{x}$ and $\hat{y}$ are perpendicular to the particle momentum, with $\hat{x}$ lying on the scattering plane, and $\hat{y}$ perpendicular.

\subsection{Constraints from parity conservation}
\label{ssec:parity-fermion}
Assuming parity conservation, the helicity amplitude has the property
\beq
	\M_{\alpha_1, \alpha_2, \alpha_3, \alpha}(p_1, p_2, p_3, p)
	= ({\rm phase}) \times \M_{-\alpha_1, -\alpha_2, -\alpha_3, -\alpha}(\bar{p}_1, \bar{p}_2, \bar{p}_3, \bar{p}),
\eeq
where the overall phase is independent of $\alpha$'s and $\bar{p}_i^{\mu} = p_{i, \mu}$.
The parity inversion not only flips all the helicities, but also flips all the momenta. 
To relate back to the original scattering, we need to perform a further rotation on the scattering plane by $\pi$. 
This rotation will restore all the momenta but retain the flipped helicities.
So overall we are examining the symmetry transformation
\beq[eq:UP]
	U_P = U(R_3(\phi_f)) U(R_2(\pi)) U^{-1}(R_3(\phi_f)) P,
\eeq
where $P$ is the parity operator. 
The rotation operation in \eq{eq:UP} is similar to the third rotation case below \eq{eq:lorentz-rep-massive-R}, which gives an identity little group transformation.
However, one key difference is that the rotation $R_2(\pi)$ will change the 
polar angle $\theta_i$ of a particle to $\theta_i + \pi$, 
which will cross the boundary of the $\theta$ domain: $\theta \in [0, \pi]$. 
The discontinuity of the SO(3) topology will thus play a nontrivial role. 
It introduces an extra phase that depends on the helicity,
\beq[eq:UP-relation-amp]
	\M_{\alpha_1, \alpha_2, \alpha_3, \alpha}(p_1, p_2, p_3, p)
	= ({\rm phase}) \cdot e^{i \delta_{123}} \, (-1)^{\alpha - 1/2} \M_{-\alpha_1, -\alpha_2, -\alpha_3, -\alpha}(p_1, p_2, p_3, p),
\eeq
where $\delta_{123}$ is the phase associated with the particles $a, b, c$, which may depend on $\alpha_{1,2,3}$, but will eventually cancel when we multiply 
$\M$ by its complex conjugate in \eq{eq:spin-dm-f-2to2}.
The phase $(-1)^{\alpha - 1/2}$ for the particle $f$ gives an extra minus sign when $\alpha = -1/2$. 
Such phase will also cancel in the diagonal elements of $\rho$, but not in the off-diagonal elements, and therefore will set a special constraint on the transverse spin.

Using \eq{eq:UP-relation-amp}, we can get the parity relation for the density matrix,
\begin{align}\label{eq:UP-relation-dm-f}
	\rho_{\alpha\alpha'}(p) 
	& = \frac{\sum_{\alpha_1, \alpha_2, \alpha_3} (-1)^{\alpha + \alpha' - 1}
				\M_{-\alpha_1, -\alpha_2, -\alpha_3, -\alpha} \,
				\M^*_{-\alpha_1, -\alpha_2, -\alpha_3, -\alpha'}
			}{	\sum_{\alpha_1, \alpha_2, \alpha_3, \alpha_4} 
				\abs{ \M_{\alpha_1 \alpha_2 \alpha_3 \alpha_4} }^2
			} \nn\\
	& = (-1)^{\alpha + \alpha' - 1} \rho_{-\alpha, -\alpha'}(p),
\end{align}
which means
\beq
	\rho_{++} = \rho_{--}, \quad
	\rho_{+-} = - \rho_{-+},
\eeq
or equivalently, 
\beq
	\lambda = b_1 = 0.
\eeq
Therefore, $b_2$ is the only allowed spin degree of freedom if parity is conserved. 
In the case with parity violation, all three components are not forbidden.

\subsection{Constraints from the amplitude structure}
\label{ssec:interference-fermion}
In the general case, the scattering amplitude has both real and imaginary parts,
\beq
	\M_{\alpha_1 \alpha_2 \alpha_3 \alpha}
	= \Re \M_{\alpha_1 \alpha_2 \alpha_3 \alpha} + i \Im \M_{\alpha_1 \alpha_2 \alpha_3 \alpha}.
\eeq
Introducing the shorthand notation
\beq[eq:prod-notation-f]
	A_\alpha * B_{\alpha'} 
	\equiv \sum_{\alpha_1 \alpha_2 \alpha_3 } 
	A_{\alpha_1 \alpha_2 \alpha_3 \alpha} B_{\alpha_1 \alpha_2 \alpha_3 \alpha'} , \quad
	|A|^2 = \sum_{\alpha_1 \alpha_2 \alpha_3 \alpha} |A_{\alpha_1 \alpha_2 \alpha_3 \alpha}|^2,
\eeq
we have
\beq
	\rho_{\alpha\alpha'} = \frac{(\Re \M_{\alpha} + i \Im \M_{\alpha}) * (\Re \M_{\alpha'} - i \Im \M_{\alpha'})}{
		|\Re \M |^2 + |\Im \M |^2},
\eeq
which gives the spin vector
\begin{align}
	\lambda & = 
		\frac{M_{+} * M^*_{+} - M_{-} * M^*_{-}}{M_{+} * M^*_{+} + M_{-} * M^*_{-}}, \nn\\
	b_1 & = 
		2 \frac{\Re \M_+ * \Re \M_- + \Im \M_+ * \Im \M_-}{|\Re \M |^2 + |\Im \M |^2}, \nn\\
	b_2 & = 
		2 \frac{\Re \M_+ * \Im \M_- - \Im \M_+ * \Re \M_-}{|\Re \M |^2 + |\Im \M |^2}.
\end{align}
Therefore, $b_2$ can only exist if the amplitude has an imaginary part. 
In a parity-conserving perturbation theory, such a phase can only occur through loops, which necessarily suppresses $b_2$ 
by the coupling constant. In contrast, there is no such constraint for $\lambda$ and $b_1$, which can be produced at tree level
as long as there is parity violation.

\subsection{Constraints from chiral symmetry}
\label{ssec:chiral-fermion}
It is crucial that transverse spin arises from the interference between two amplitudes which only differ in the observed fermion helicity.
In terms of the cut diagram notation, a fermion line always forms a closed loop. A nonzero interference thus requires that the fermion 
helicity be flipped at some point of the fermion loop. This can only happen if 
(1) the fermion is massive, or
(2) there is a Yukawa or tensor interaction vertex.
In the SM, there is no tensor interaction, and the only Yukawa interaction is the source for the fermion mass, 
so the necessary condition to generate a single transverse spin is to have the fermion being massive.
This in turn means that the magnitude of the transverse spin will be proportional to the fermion mass $m_f$, which is compensated by
the scattering energy $\sqrt{s}$. That means, 
\beq
	b_T \propto \frac{m_f}{\sqrt{s}},
\eeq
which will be highly suppressed at high energies.

On the other hand, in a new physics or effective field theory scenario with tensor interactions, one may have a 
single transverse spin that does not suffer from such suppression. One recent work~\citep{Wen:2023xxc} employed this effect 
to explore the impacts of single transverse spin asymmetry at transversely polarized lepton colliders on dimension-six electron dipole operators.
Similar study can be done for final-state single fermion spin by measuring the heavy fermion spins through their decays.

\subsection{Short summary}
\label{ssec:ssa-fermion}
Summarizing this section, we note that the transverse spin $b_2$ is the only spin degree of freedom allowed by parity, but it must require 
an imaginary part of the amplitude, which only occurs beyond tree level. 
In contrast, the other two spin degrees of freedom, $\lambda$ and $b_1$,
can occur at tree level as long as parity is violated.
Furthermore, a single transverse spin can only happen to massive fermions, not to massive ones, and the magnitude is suppressed by the fermion mass.

We note that the derivation for the parity relation in \eq{eq:UP-relation-dm-f} only applies to $2\to2$ scattering and inclusive one-particle production. 
It cannot be trivially extended to more complicated final states. But it is generally true that a single transverse spin is allowed by symmetries.

\section{Example: $s$-channel single top production}
\label{sec:s-top}

In this section, we illustrate the single spin production with the example of $s$-channel single top quark production at the LHC.
This is well suited for illustrating all the points discussed in \sec{sec:fermion-spin} because
(1) the top quark is a massive fermion whose mass $m_t$ is not negligible at the LHC energy, 
	which makes the production of a transverse spin possible;
(2) the top quark is produced via an $s$-channel $W$ boson, whose interaction violates parity, 
	such that a nonzero $\lambda$ and $b_1$ can be produced at tree level;
(3) beyond tree level, the one-loop QCD correction can trigger a threshold effect to 
	generate an imaginary part in the amplitude, so that a $b_2$ can be produced.

At LO, we consider the partonic process $u(p_1, \alpha_1) + \bar{d}(p_2, \alpha_2) \to t(q_1, \sigma_1) + \bar{b}(q_2, \sigma_2)$, 
which happens through an $s$-channel $W^+$ boson, in its c.m.~frame, with $u$ along $\hat{z}$, and $t$ along $\hat{n}(\theta, \phi)$. 
The kinematics can be easily worked out by momentum conservation,
\beq
	p_{1,2} = \frac{\sqrt{s}}{2}(1, 0, 0, \pm 1), \quad
	q_{1,2} = \frac{\sqrt{s}}{2} \pp{ \frac{s \pm m_t^2}{s}, \pm \frac{s - m_t^2}{s} \hat{n} },
\eeq
with $s = (p_1 + p_2)^2$ being the partonic c.m.~energy squared.
The scattering amplitude $i\M$ is
\begin{align}
	i \M_{\alpha_1 \alpha_2 \sigma_1 \sigma_2}
	& = \bar{u}(q_1, \sigma_1) \pp{ \frac{-i g}{\sqrt{2}} \gamma^{\mu} P_L } v(q_2, \sigma_2)
			\frac{-i g^{\mu\nu}}{s - m_w^2}
			\bar{v}(p_2, \alpha_2) \pp{ \frac{-i g}{\sqrt{2}} \gamma^{\nu} P_L } u(p_1, \alpha_1) \nn\\
	& = \frac{i g^2}{2(s - m_w^2)} 
			\bb{ \bar{u}(q_1, \sigma_1) \gamma^{\mu} P_L v(q_2, \sigma_2) }
			\bb{ \bar{v}(p_2, \alpha_2) \gamma_{\mu} P_L u(p_1, \alpha_1) },
\end{align}
where $g$ is the SU(2) gauge coupling.
The helicity structure is greatly simplified by the left-handed vector current interaction and that $u$, $\bar{d}$, and $\bar{b}$ are taken massless.
This constrains $\alpha_1 = -1/2$ and $\alpha_2 = \sigma_2 = +1/2$. 
So we are only reduced to two helicity amplitudes, with $\sigma_1 = \pm 1/2$, wherein only the $-1/2$ helicity can exist if we take $m_t \to 0$,
and thus the amplitude with $\sigma_1 = +1/2$ is proportional to $m_t$. 
By explicit calculation, we obtain
\beq
	\M_{-+-+} = N \, (1 + \cos\theta), \quad
	\M_{-+++} = N \, \frac{m_t}{\sqrt{s}} \sin\theta, \quad
	N = - \frac{g^2}{2} \frac{s \, e^{-i \phi}}{s - m_w^2} \sqrt{ 1 - \frac{m_t^2}{s} }.
\eeq
A few remarks are in order.
\begin{itemize}
\item The initial $(u \bar{d})$ state has a nonzero spin, $\alpha_1 - \alpha_2 = -1$, along $\hat{z}$,
	which gives a phase factor $e^{i (\alpha_1 - \alpha_2) \phi} = e^{-i \phi}$. This phase applies to both amplitudes, $\M_{-+-+}$ and $\M_{-+++}$,
	and will cancel when we take a product between an amplitude and a complex conjugate amplitude.
\item The factor $N$ contains a threshold factor $\sqrt{ 1 - m_t^2 / s }$ to suppress the amplitude as $s \gtrsim m_t^2$. As $s \gg m_t^2$, $N$ approaches
	a constant $- (g^2 / 2) e^{-i \phi}$.
\item $\M_{-+-+}$ is the only amplitude that survives as $s \gg m_t^2$, and it favors production of the top quark in the forward region, controlled by the angular
	function $d^1_{-1, -1}(\theta) \propto (1 + \cos\theta)$, as a result of the left-handed coupling. In contrast, the amplitude $\M_{-+++}$ flips the top quark helicity 
	by a mass insertion, and so is only significant when $s$ is not much greater than $m_t^2$. The angular distribution is controlled by
	$d^1_{1, 0}(\theta) \propto \sin\theta$, which is symmetric between forward and backward regions.
\end{itemize}

The density matrix of the top quark can be easily calculated by \eq{eq:spin-dm-f-2to2},
\begin{align}
	\rho^t_{\alpha \alpha'} = \frac{\M_{-, +, \alpha, +} \M^*_{-, +, \alpha', +} }{|\M_{-+-+}|^2 + | \M_{-+++} |^2},
\end{align}
which gives
\beq[eq:s1t-dm]
	\rho^t = 
	\pp{ \frac{m_t^2}{s} \sin^2\theta + (1 + \cos\theta)^2 }^{-1}
	\begin{pmatrix}
		\frac{m_t^2}{s} \sin^2\theta & \frac{m_t}{\sqrt{s}} \sin\theta (1 + \cos\theta) \\
		\frac{m_t}{\sqrt{s}} \sin\theta (1 + \cos\theta) & (1 + \cos\theta)^2
	\end{pmatrix},
\eeq
where the factor in front plays the role of normalizing $\rho^t$. 
Clearly, due to the fact that there are only two non-zero helicity amplitudes, the density matrix has the structure
\beq
	\rho^t \sim 
	\begin{pmatrix}
		a^2 & ab \\
		ab & b^2
	\end{pmatrix},
\eeq
up to a normalization. This immediately leads to $\det\rho^t = 0$ such that the polarization vector $|\bm{s}_t| = 1$, recalling \eq{eq:dm-positivity-f}.
As a result, the top quark must be at a pure spin state at LO; in other words, it is 100\% polarized. 
From \eq{eq:s1t-dm}, one can obtain the polarization vector,
\begin{align}\label{eq:s1t-pols}
	\lambda = - \frac{s - m_t^2 + ( s + m_t^2 ) \cos\theta}{s + m_t^2 + ( s - m_t^2 ) \cos\theta}, \quad
	b_1  = \frac{m_t}{\sqrt{s}} \, \frac{2s \cdot \sin\theta}{s + m_t^2 + ( s - m_t^2 ) \cos\theta}, \quad
	b_2  = 0,
\end{align}
from which we can easily verify $\lambda^2 + b_1^2 = 1$.

We note that the polarization vector expression in \eq{eq:s1t-pols} holds only in the partonic c.m.~frame. When the whole system is boosted 
along $\hat{z}$ by $\Lambda_z(\beta)$, the density matrix transforms according to \eq{eq:denmtx-lorentz-f}. The little group corresponding to
such boost has been obtained in \eq{eq:little-group-chi}, which is a rotation around $\hat{y}$ by $\chi$, with
\beq
	\cos\chi = \frac{v + \beta \cos\theta}{\sqrt{(1 + \beta v\cos\theta)^2 - (1 - \beta^2) (1 - v^2)}},
	\quad
	v = \frac{s - m_t^2}{s + m_t^2}.
\eeq 
Therefore, \eq{eq:denmtx-lorentz-f}
becomes
\beq
	\rho^t(\bm{s}_t) \to d^{1/2}(\chi) \cdot \rho^t(\bm{s}_t) \cdot \bb{ d^{1/2}(\chi) }^{\dag},
\eeq
which keeps $b_2$ invariant but mixes $b_1$ with $\lambda$,
\beq[eq:s1t-mixing]
	\lambda \to \lambda \cos\chi - b_1 \sin\chi,
	\quad
	b_1 \to b_1 \cos\chi + \lambda \sin\chi.
\eeq
Because of this mixing, it is necessary to analyze the top polarization in the partonic c.m. frame event by event.

The mixing [\eq{eq:s1t-mixing}] does not alter the fact $\lambda^2 + b_1^2 = 1$, 
and also provides a physical understanding for the full polarization.
If we take an infinite boost with $\beta = 1$, which gives
\beq
	\cos\chi = \frac{v + \cos\theta}{1 + v \cos\theta}, \quad
	\sin\chi = \frac{\sqrt{1 - v^2} \sin\theta}{1 + v \cos\theta},
\eeq
then \eq{eq:s1t-mixing} gives
\beq
	\lambda \to -1, \quad
	b_1 \to 0,
\eeq
so that the top quark becomes completely left-handed. This agrees with the physical picture that the infinite boost takes $t$ and $\bar{b}$ to be collinear along
$\hat{z}$. Their spins sum up to $-1$ along $\hat{z}$, equal to the spin of the initial state. Such ``infinite momentum frame" explains why the top quark is 100\%
polarized, and the nonzero $b_1$ in the ``finite momentum frame" is a result of polarization mixing when going from the ``infinite momentum frame" to the 
``finite momentum frame".

\section{Observing the fermion spin}
\label{sec:f-spin-observe}

Spins are usually not directly measured at high-energy colliders, so that information is lost in merely constructing the production rates.
However, if the polarized particle decays, the kinematic distributions of the decay products are likely to retain the spin information of the mother particle.
This is the case for the heavy fermions in the SM, especially for the top quark.

This is best illustrated in the rest frame of top quark, constructed by 
boosting along $-\bm{p}_t$ with the same set of coordinate system $\hat{x}$-$\hat{y}$-$\hat{z}$ as \eq{eq:fermion-xyz},
with $\hat{p} = \bm{p}_t / |\bm{p}_t|$ the direction of the top quark momentum $\bm{p}_t$.
The helicity amplitude of the decay $t(\alpha_t) \to W^+(\alpha_w) b(\alpha_b)$ can be written generally as~\citep{Tung:1985na}
\beq[eq:tbw-amplitude-A]
	\M_{\alpha_t \alpha_w \alpha_b}(\theta^{\star}, \phi^{\star}) 
	= A_{\alpha_w, \alpha_b} D^{1/2*}_{\alpha_t, \alpha_w - \alpha_b}(\phi^{\star}, \theta^{\star}, 0)
	= A_{\alpha_w, \alpha_b} \, e^{i \alpha_t \phi^{\star}} \, d^{1/2}_{\alpha_t, \alpha_w - \alpha_b}(\theta^{\star}).
\eeq
where $\alpha$'s denote the helicities, with $\alpha_t$ with respect to $\hat{z}$, and $d^{1/2}$ is the Wigner $d$-function.
The angles $\theta^{\star}$ and $\phi^{\star}$ characterize the $W$ boson direction in the top rest frame.
The coefficient $A_{\alpha_w, \alpha_b}$ does not depend on top helicity or the angles.
The angular distribution of the $W$ is given by
\begin{align}\label{eq:tbw-decay-angular-0}
	\frac{d\Gamma_t}{d\cos\theta^{\star} d\phi^{\star}}
	\propto
		\rho^t_{\alpha_t \alpha_t'}(\bm{s}_t) 
		\M_{\alpha_t \alpha_w \alpha_b}(\theta^{\star}, \phi^{\star}) 
		\M^*_{\alpha_t' \alpha_w \alpha_b}(\theta^{\star}, \phi^{\star}),
\end{align}
where the summation over repeated indices is implied, and $\bm{s}_t = (b_1, b_2, \lambda)$ is the top spin vector.
Substituting \eq{eq:tbw-amplitude-A} for the amplitudes in \eq{eq:tbw-decay-angular-0} gives
\begin{align}\label{eq:tbw-decay-angular}
	\frac{1}{\Gamma_t} \frac{d\Gamma_t}{d\cos\theta^{\star} d\phi^{\star}}
	&= \frac{1}{4\pi} \bb{ 1 + \kappa_w \, \bm{s}_t \cdot \bm{\Omega}^{\star} } ,
\end{align}
where $\bm{s}_t \cdot \bm{\Omega} = b_1 \sin\theta^{\star}\cos\phi^{\star} + b_2 \sin\theta^{\star}\sin\phi^{\star} + \lambda \cos\theta^{\star}$ and
\beq[eq:kappa-W]
	\kappa_w 
		= \frac{|A_{1, 1/2}|^2 - |A_{0, 1/2}|^2 + |A_{0, -1/2}|^2 - |A_{-1, -1/2}|^2
			}{ |A_{1, 1/2}|^2 + |A_{0, 1/2}|^2 + |A_{-1, -1/2}|^2 + |A_{0, -1/2}|^2 }
\eeq
is the spin analyzing power for the $W$. 
As a result, the nonzero polarization $\bm{s}_t$ of the mother particle leads to 
an asymmetric decay distribution with respect to the polarization direction. 
In the chosen $\hat{x}$-$\hat{y}$-$\hat{z}$ frame, the longitudinal polarization $\lambda$ leads to 
a forward-backward asymmetry while the transverse spin $\bm{b}_T$ introduces an azimuthal asymmetry.

It is worth noting that using such single-particle distribution as a spin observable belongs to 
the single-spin phenomenon. It relies on parity violation in the $tbW$ interaction. 
As can be obviously noticed from \eq{eq:kappa-W}, if the top decay process preserves parity, 
one would have $\kappa_w = 0$, which offers no ability to probe the top polarization.
In the SM with a purely left-handed $tbW$ current interaction and neglecting the $b$ mass, 
the LO coefficients $A_{\lambda_w, \lambda_b}$ are 
\beq
	A_{-1, -1/2} = - g \sqrt{m_t^2 - m_w^2}, \quad
	A_{0, -1/2} = - g \sqrt{m_t^2 - m_w^2} \frac{m_t}{\sqrt{2} m_w},
\eeq
with all the others being $0$. This gives the spin-analyzing power for $W$ as 
\beq
	\kappa_w = \frac{m_t^2 - 2 m_w^2}{m_t^2 + 2 m_w^2} \simeq 0.4.
\eeq
It is positive, so the $W$ prefers to be along the spin direction of $t$, as a result of being dominantly produced with a longitudinal polarization.

In the real situation, the $W$ boson from top decay rapidly decays to another fermion-anti-fermion pair $f\bar{f}'$, so that 
the top quark decays into three particles, $t \to b f \bar{f}'$. 
The single-particle distribution in \eq{eq:tbw-decay-angular} simply generalizes to the three-body decay,
\begin{align}\label{eq:tbw-decay-angular}
	\frac{1}{\Gamma_t} \frac{d\Gamma_t}{d\cos\theta_i^{\star} d\phi_i^{\star}}
	&= \frac{1}{4\pi} \bb{ 1 + \kappa_i \, \bm{s}_t \cdot \bm{\Omega}^{\star}_i },
\end{align}
where $i$ can be $b$, $f$, or $\bar{f}'$, 
$\bm{\Omega}^{\star}_i = (\sin\theta^{\star}_i \cos\phi^{\star}_i, \sin\theta^{\star}_i \sin\phi^{\star}_i, \cos\theta^{\star}_i)$ is its direction in the top rest frame,
and $\kappa_i$ is the corresponding spin-analyzing power. 
This general form [\eq{eq:tbw-decay-angular}] holds because of rotational invariance and the fact that spin vectors $\bm{s}_t$
appears at most in a linear form.

\eq{eq:tbw-decay-angular} can be marginalized to give a $\cos\theta^{\star}$ distribution that exclusively probes the helicity $\lambda$,
\beq[eq:t-f-polar-dist]
	\frac{1}{\Gamma_t} \frac{d\Gamma_t}{d\cos\theta_i^{\star}}
	= \frac{1}{2} (1 + \kappa_i \, \lambda \cos\theta_i^{\star} ),
\eeq
or a $\phi^{\star}$ distribution that only probes the transverse spin $\bm{b}_T$,
\beq[eq:t-f-azimuthal-dist]
	\frac{1}{\Gamma_t} \frac{d\Gamma_t}{d\phi_i^{\star}}
	= \frac{1}{2\pi} \bb{ 1 + \frac{\pi}{4} \kappa_i \,  (b_1 \sin\theta^{\star}_i \cos\phi^{\star}_i + b_2 \sin\theta^{\star}_i \sin\phi^{\star}_i ) }.
\eeq
While the $\theta_i^{\star}$ and $\phi_i^{\star}$ distributions play similar roles in the rest frame, 
they do not when the top quark is boosted.
In the boosted frame, the $\theta_i$ distribution becomes highly distorted such that 
all the decay products prefer to be collinear with $t$ regardless of the value of $\lambda$.
Thus \eq{eq:t-f-polar-dist} loses its power as a spin polarimeter in the boosted frame.
The azimuthal angle $\phi_i$, on the other hand, remains unchanged by the boost, 
so \eq{eq:t-f-azimuthal-dist} still gives a good method for measuring the transverse spin. 
Since at the LHC, the top quark can be produced with a large boost, 
the transverse spin stands out against the longitudinal polarization,
which is one of the reasons why we study the transverse spin in this thesis. 
While one can convert the polar angle distribution [\eq{eq:t-f-polar-dist}] into the 
energy fraction distribution of the daughter particles and produces a new
polarimeter for $\lambda$, that is not the focus of our discussion in this thesis. 
We will instead give another method for measuring $\lambda$ for a highly boosted top quark, 
based on the azimuthal correlation among the three daughter particles of the top decay, 
in \sec{sec:linear-pol-W}.

\section{Compared to fermion spin correlation}
\label{sec:f-spin-correlation}

We have seen that the single transverse spin of a fermion is produced via a single helicity flip in the amplitude.  
This comes with a mass penalty in the SM. So as we go to higher energies, a single transverse spin is usually suppressed.
However, if we measure the transverse spins of two fermions at the same time, the effects of double helicity flips 
in the product of the amplitude and conjugate amplitude cancel each other and do not necessary require mass insertions. 
We call this {\it transverse spin correlation}. It is not necessarily suppressed as one goes to high energies.
In this section, we lay down the general formalism for describing the spin correlations between two fermions. 

Suppose a pair of fermions, $f_1$ and $f_2$, are produced in a certain process, 
\beq
	a(p_1) + b(p_2) \to f_1(k_1, \alpha_1) + f_2(k_2, \alpha_2) + X,		\quad
	(\alpha_1, \alpha_2 = \pm	1 / 2)
\eeq
where we suppress the helicities of all other particles, which will be traced over. 
The helicity amplitude is denoted as
\beq
	\M_{\alpha_1 \alpha_2}(p_1, p_2; k_1, k_2).
\eeq
Similar to \eq{eq:spin-dm-f-2to2}, but now with both helicity indices open, we can get the multi-dimensional spin density matrix
for the two fermions,
\beq[eq:spin-den-mtx-ff]
	\rho_{\alpha_1, \alpha_2; \alpha_1', \alpha_2'}
	= \frac{ \M_{\alpha_1 \alpha_2} \M^*_{\alpha_1' \alpha_2'}}{
		\sum_{\bar{\alpha}_1 \bar{\alpha}_2} | \M_{\bar{\alpha}_1 \bar{\alpha}_2} |^2 },
\eeq
where the momentum dependence has been suppressed. This matrix can be thought of the direct product of two single-fermion
density matrices, so can be decomposed as~\citep{Bernreuther:2015yna}
\beq[eq:spin-den-mtx-ff-decomp]
	\rho_{\alpha_1, \alpha_2; \alpha_1', \alpha_2'}
	= \frac{1}{4} \bb{ \delta_{\alpha_1 \alpha_1'} \delta_{\alpha_2 \alpha_2'} 
		+ \bm{s}_1 \cdot \bm{\sigma}_{\alpha_1 \alpha_1'} \delta_{\alpha_2 \alpha_2'} 
		+ \delta_{\alpha_1 \alpha_1'}  \bm{s}_2 \cdot \bm{\sigma}_{\alpha_2 \alpha_2'}
		+ \sum_{i, j = 1}^3 C_{ij} \sigma^i_{\alpha_1 \alpha_1'} \sigma^j_{\alpha_2 \alpha_2'}
		},
\eeq
or symbolically,
\beq
	\rho = \frac{1}{4} \bb{ \mathbf{1} \otimes \mathbf{1} 
		+ (\bm{s}_1 \cdot \bm{\sigma}) \otimes \mathbf{1}
		+ \mathbf{1} \otimes (\bm{s}_2 \cdot \bm{\sigma})
		+ C_{ij} \, \sigma^i \otimes \sigma^j
		}.
\eeq
This defines two single fermion spin vectors, $\bm{s}_1$ and $\bm{s}_2$, and a spin-spin correlation matrix $C_{ij}$.
They are all real-valued by the Hermiticity of $\rho$.

Note that by a chosen convention of the fermion helicity spinors, 
the Pauli matrix decomposition in \eq{eq:spin-den-mtx-ff-decomp}
automatically determines the $x$-$y$-$z$ systems for the spin parameters 
$\bm{s}_1 = (s_1^x, s_1^y, s_1^z)$,
$\bm{s}_2 = (s_2^x, s_2^y, s_2^z)$, and 
$C_{ij} = (C_{xx}, C_{xy}, \cdots, C_{zz})$.
The $x$-$y$-$z$ system for either fermion is determined according to \eq{eq:fermion-xyz}. 
In the c.m.~frame of the exact $2\to2$ kinematics, the two systems have the same $x$ direction and opposite $y$ and $z$ directions,
with $x$ on the scattering plane and $y$ perpendicular.

Evidently, the single spin information in $\bm{s}_1$ and $\bm{s}_2$ is independent from that in the spin correlation $C_{ij}$.
Tracing over the helicity of $f_1$ or $f_2$ in \eq{eq:spin-den-mtx-ff} or \eqref{eq:spin-den-mtx-ff-decomp} reduces to
the single fermion spin density matrix in \eq{eq:spin-dm-f-2to2} or \eqref{eq:f-spin-den-mtx}.

Similar to the parity constraint for a single fermion density matrix in \eq{eq:UP-relation-dm-f}, in an exact $2\to2$ kinematics,
parity symmetry (if it holds) constrains \eq{eq:spin-den-mtx-ff-decomp} by
\beq[eq:parity-ff-den-mtx]
	\rho_{\alpha_1, \alpha_2; \alpha_1', \alpha_2'}
	= (-1)^{\alpha_1 + \alpha_1' + \alpha_2 + \alpha_2' - 2} 
		\, \rho_{-\alpha_1, -\alpha_2; -\alpha_1', -\alpha_2'}.
\eeq
In terms of the Pauli matrix decompositions in \eq{eq:spin-den-mtx-ff-decomp}, the right-hand side of \eq{eq:parity-ff-den-mtx}
differs from the left-hand side by adding a minus sign to all occurrences of $\sigma^1$ and $\sigma^3$, so this constrains 
the spin parameters by 
\beq
	s_1^x = s_1^z = s_2^x = s_2^z = 0, \quad
	C_{xy} = C_{yx} = C_{yz} = C_{zy} = 0.
\eeq
Different from the single fermion spins, for which parity only allows the component ($s^y$) perpendicular to the scattering plane,
the correlation of two helicity polarizations ($C_{zz}$) or helicity and transverse spin ($C_{xz}$ and $C_{zx}$) is allowed by parity.
Also, the transverse spin correlations $C_{xx}$ and $C_{yy}$ along each transverse direction are allowed.

The spin parameters can be obtained from \eq{eq:spin-den-mtx-ff-decomp} by tracing with corresponding Pauli matrices,
\begin{align}
	s_1^i = \sigma^i_{\alpha_1' \alpha_1} \rho_{\alpha_1, \bar{\alpha}_2; \alpha_1', \bar{\alpha}_2}, 
	\quad
	s_2^i = \sigma^i_{\alpha_2' \alpha_2} \rho_{\bar{\alpha}_1, \alpha_2; \bar{\alpha}_1, \alpha_2'},
	\quad
	C_{ij} = \sigma^i_{\alpha_1' \alpha_1} \sigma^j_{\alpha_2' \alpha_2} \rho_{\alpha_1, \alpha_2; \alpha'_1, \alpha_2'},
\end{align}
where repeated indices are summed over and $i, j$ run from $1$ to $3$.
This gives, e.g.,
\begin{align}
	s_1^y & = i \pp{ \rho_{+, \bar{\alpha}_2; -, \bar{\alpha}_2} - \rho_{-, \bar{\alpha}_2; +, \bar{\alpha}_2} }
		= -2 \Im\bb{  \M_{+ \bar{\alpha}_2} \M^*_{- \bar{\alpha}_2} } 
			\big/ \,  \overline{ | \M |^2}, 	\nn\\
	C_{zz} & = \pp{ |\M_{++}|^2 + |\M_{--}|^2 - |\M_{+-}|^2 - |\M_{-+}|^2 } 
			\big/ \, \overline{ | \M |^2},	\nn\\
	C_{xx} & = 2 \Re\bb{ \M_{++} \M_{--}^* + \M_{+-} \M_{-+}^* } 
			\big/ \, \overline{ | \M |^2},	\nn\\
	C_{xy} & = -2 \Im\bb{ \M_{++} \M_{--}^* - \M_{+-} \M_{-+}^* } 
			\big/ \, \overline{ | \M |^2}.
\end{align}
Thus,
a single transverse spin corresponds to the interference of different helicity states for a single fermion,
whereas a transverse spin correlation is due to the interference of different helicity states for {\it a fermion pair}. 
Of course, in another language, this is the entanglement of two fermions in the helicity space.
While a single transverse spin requires a single fermion helicity flip so is suppressed at high energy limit by the fermion mass 
(at least for the current interactions in the SM),
the transverse spin correlations flip the fermion helicity twice so are not necessarily suppressed.


\chapter{Linear polarization of vector bosons at high-energy colliders}
\label{ch:pol-vboson}


The transverse spin of fermions is an interesting phenomenon that is readily overlooked by only examining the total production rate.
It encodes the quantum interference information at high-energy scattering experiments, and reveals itself as an azimuthal distribution,
$\cos\phi$ or $\sin\phi$, which can be easily measured at high-energy colliders. The natural question, then, is whether a similar
phenomenon holds for vector bosons. 

Unlike fermions, the spin of a vector boson cannot be pictured as an arrow pointing to a certain direction in the space, 
but we have the same understanding that 
(1) a transverse spin is an interference of different helicity states, and
(2) it sets a special direction in the transverse plane, which breaks the rotational invariance around the particle momentum.
For a massless vector boson, superposition of the two helicity states $|\pm\rangle$ can make up the familiar {\it linear polarization} states,
\beq[eq:linear-pol-state]
	|x\rangle = \frac{-1}{\sqrt{2}} \pp{ |+ \rangle - | - \rangle }, \quad
	|y\rangle = \frac{i}{\sqrt{2}} \pp{ |+ \rangle + | - \rangle },
\eeq
which transform as a transverse vector under a rotation $R_z(\phi)$ around the momentum direction,
\beq
	|x\rangle \to \cos\phi \, |x\rangle + \sin\phi \, |y\rangle, \quad
	|y\rangle \to \cos\phi \, |y\rangle - \sin\phi \, |x\rangle.
\eeq
They, therefore, play the counterpart roles of the fermion transverse spins for the vector bosons.
A massive vector boson, on the other hand, has one additional helicity state $| 0 \rangle$, 
which can make up extra ``transverse spin" states by superposing with $|x\rangle$ and $|y\rangle$.
	
The mass suppression of single fermion transverse spin arises from the chiral symmetry inherent in a massless fermion. 
In contrast, for a vector boson, no chiral symmetry protects the helicity from flipping. 
Consequently, a vector boson's linear polarization is more easily produced. 
Given that the helicity can now flip by up to two units, we can observe 
$\cos2\phi$ and $\sin2\phi$ 
azimuthal patterns in the decay products of the vector boson. 
This makes the linear polarization phenomenon more intricate than the fermion transverse spin. 
In this section, we will introduce the formalism for describing linear polarization phenomena 
and present two physical examples where it can manifest and yield intriguing observational signals.


\section{Vector boson spin density matrix}
\label{sec:spin-dm-v}

\begin{figure}[htbp]
	\centering
		\includegraphics[scale=0.85]{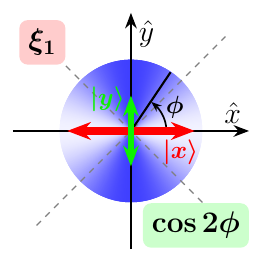} 
		\hspace{4em}
		\includegraphics[scale=0.85]{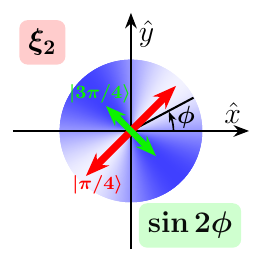}
	\caption{Interpretations of the polarization $\xi_1$ and $\xi_2$ in the linear polarization basis, 
	and the associated azimuthal angular distributions.
	}
	\label{fig:g-linear-pol}
\end{figure}

Massless vector bosons, such as gluons and photons, have only two helicity states, so their spin density matrix is also a $2 \times 2$ Hermitian 
matrix with a unity trace, like the fermion case. 
We can also decompose this into Pauli matrices, thereby defining three polarization parameters
$\bm{\xi} = (\xi_1, \xi_2, \xi_3)$ in the helicity basis,
\beq[eq:denmtx-vector-g]
	\pp{ \rho^1_{\lambda\lambda'}(p) }
	= \frac{1}{2} ( 1 + \bm{\xi}(p) \cdot \bm{\sigma} )
	= \frac{1}{2} \begin{pmatrix}
				1 + \xi_3(p)  &  \xi_1(p) - i \xi_2(p) \\
				\xi_1(p) + i \xi_2(p)  &  1 - \xi_3(p)
			\end{pmatrix}.
\eeq
As for the fermion case, $\xi_3 = \rho^1_{++} - \rho^1_{--}$ is the net helicity, while the off-diagonal elements $(\xi_1, \xi_2)$, being interference
of different helicity states, are the linear polarization. In terms of the linear polarization state 
\beq
	| \phi \rangle = \frac{-1}{\sqrt{2}} \bb{ e^{-i \phi} | + \rangle - e^{i \phi} | - \rangle }
\eeq
along the $\phi$ direction (in the transverse plane), $(\xi_1, \xi_2)$ can be represented as
\bse\begin{align}
	\xi_1 & = \rho^1_{+-} + \rho^1_{-+} = \langle \pi / 2 | \hat{\rho}^1 | \pi / 2 \rangle - \langle 0 | \hat{\rho}^1 | 0 \rangle = \rho^1_{yy} - \rho^1_{xx}, \\
	\xi_2 & = i( \rho^1_{+-} - \rho^1_{-+}) = \langle 3\pi / 4 | \hat{\rho}^1 | 3\pi / 4 \rangle - \langle \pi/4 | \hat{\rho}^1 | \pi/4 \rangle.
\end{align}\ese
Thus they are differences of the linear polarization degrees along two orthogonal directions, as shown in \fig{fig:g-linear-pol},
with respect to the same $\hat{x}$-$\hat{y}$-$\hat{z}$ coordinate system as defined in \eq{eq:fermion-xyz}.
Under a rotation around the gluon momentum direction by $\phi$, the density matrix changes as 
\beq[eq:denmtx-rotation-z-g]
	\rho^1_{\lambda \lambda'}(\bm{\xi}) \to \rho^1_{\lambda \lambda'}(\bm{\xi}') = e^{-i (\lambda - \lambda')\phi} \rho^1_{\lambda \lambda'}(\bm{\xi}),
\eeq
so that $\bm{\xi}$ transforms as
\beq
	\xi_3' = \xi_3, \quad
	\xi_1' = \cos2\phi \, \xi_1 - \sin2\phi\, \xi_2, \quad
	\xi_2' = \cos2\phi \, \xi_2 + \sin2\phi \, \xi_1.
\eeq
This shows the difference of the linear polarization $\bm{\xi}_{\perp} = (\xi_1, \xi_2)$ from the fermion transverse spin $\bm{b}_T$:
the former transforms like a spin-2 tensor, whereas the latter like a spin-1 vector.

Massive vector bosons, on the other hand, have one extra longitudinal polarization $|0 \rangle$, which extends the density matrix to $3\times3$.
We parametrize it as
\beq[eq:denmtx-massive-1]
	\pp{ \rho^M_{\lambda\lambda'} } = 
		\begin{pmatrix}
			\frac{1}{3} + \frac{\delta_L}{6} + \frac{J_3}{2} & \frac{J_1 + Q_{xz} - i(J_2 + Q_{yz})}{2\sqrt{2}} & \frac{\xi - i Q_{xy} }{2}		\\
			\frac{J_1 + Q_{xz} + i(J_2 + Q_{yz})}{2\sqrt{2}} & \frac{1 - \delta_L}{3} & \frac{J_1 - Q_{xz} - i(J_2 - Q_{yz})}{2\sqrt{2}}  \\
			\frac{\xi + i Q_{xy} }{2} & \frac{J_1 - Q_{xz} + i(J_2 - Q_{yz})}{2\sqrt{2}} & \frac{1}{3} + \frac{\delta_L}{6} - \frac{J_3}{2}
		\end{pmatrix},
\eeq
on the helicity basis, in terms of the eight real polarization parameters ($J_1$, $J_2$, $J_3$, $Q_{xy}$, $Q_{yz}$, $Q_{xz}$, $\delta_L$, $\xi$). 
We have suppressed their dependence on the vector boson momentum $p$. 
Under the rotation by $\phi$ around $\hat{p}$, the same transformation in \eq{eq:denmtx-rotation-z-g} holds for $\rho^M$, 
which gives the transformation behaviors of the polarization parameters,
\beq[eq:massive-params-rotation-z]
	\begin{pmatrix}
		J_1' & Q_{xz}'		\\
		J_2' & Q_{yz}'
	\end{pmatrix}
	= \begin{pmatrix}
		\cos\phi & -\sin\phi \\
		\sin\phi & \cos\phi
		\end{pmatrix}
		\begin{pmatrix}
			J_1 & Q_{xz}		\\
			J_2 & Q_{yz}
		\end{pmatrix},
	\quad
	\begin{pmatrix}
		\xi' \\
		Q'_{xy}
	\end{pmatrix}
	= \begin{pmatrix}
		\cos2\phi & -\sin2\phi \\
		\sin2\phi & \cos2\phi
		\end{pmatrix}
		\begin{pmatrix}
			\xi \\
			Q_{xy}
		\end{pmatrix},
\eeq
with $J_3$ and $\delta_L$ unchanged. 
In this way, all the parameters in off-diagonal elements behave like transverse spins.
As interference between $|\pm\rangle$ and $|0\rangle$, the parameters
($J_1$, $J_2$, $Q_{xz}$, $Q_{yz}$) behave like transverse vectors with spin 1. 
Similarly, as interference between $|+\rangle$ and $|-\rangle$,
($\xi$, $Q_{xy}$) are like transverse vectors with spin 2. 
The density matrix for a massive vector boson reduces to the massless case by taking 
($J_1$, $J_2$, $Q_{xz}$, $Q_{yz}$) $ \to 0$, and equating $(\xi, Q_{xy}, J_3)$ with
$(\xi_1, \xi_2, \xi_3)$. 

The physical meaning of $(\xi_1, \xi_2, \xi_3)$ carry through to
$(\xi, Q_{xy}, J_3)$ for the massive case, as the linear polarization states and helicity.
The parameter $\delta_L$ characterizes the longitudinal polarization state.
In the rest frame of the massive vector boson, $J_1$, $J_2$, and $J_3$ are the angular momentum (spin)
components along the $x$, $y$, and $z$ directions, which can be obtained by tracing $\rho^M$ with the 
spin operators,
\beq
	\hat{J}_1 = \frac{1}{\sqrt{2}} 
		\begin{pmatrix}
			0 & 1 & 0 \\
			1 & 0 & 1 \\
			0 & 1 & 0
		\end{pmatrix},
	\quad
	\hat{J}_2 = \frac{1}{\sqrt{2}} 
		\begin{pmatrix}
			0 & -i & 0 \\
			i & 0 & -i \\
			0 & i & 0
		\end{pmatrix},
	\quad
	\hat{J}_3 =  
		\begin{pmatrix}
			1 & 0 & 0 \\
			0 & 0 & 0 \\
			0 & 0 & -1
		\end{pmatrix}.
\eeq
The other five parameters can be made analogous to the electric quadrupole moments. By transforming \eq{eq:denmtx-massive-1}
into the linear polarization basis, constituted by \eq{eq:linear-pol-state} and $|z \rangle = |0 \rangle$, we have 
\beq[eq:denmtx-massive-1-linear-basis]
	\pp{ \rho^M_{ij} }= 
		\frac{1}{2} \begin{pmatrix}
			\frac{2}{3} + \frac{\delta_L}{3} - \xi  &  - Q_{xy} - i J_3 &  - Q_{xz} + i J_2		\\
			-Q_{xy} + i J_3   &  \frac{2}{3} + \frac{\delta_L}{3} + \xi &  - Q_{yz} - i J_1 \\
			-Q_{xz} - i J_2  &  -Q_{yz} + i J_1 & \frac{2}{3} (1 - \delta_L)
		\end{pmatrix}.
\eeq
This immediately gives the quadrupole moments,
\beq[eq:quadrupole-Q]
	Q_{xy} = -(\rho^M_{xy} + \rho^M_{yx}), \quad
	Q_{yz} = -(\rho^M_{yz} + \rho^M_{zy}), \quad
	Q_{xz} = -(\rho^M_{xz} + \rho^M_{zx}),
\eeq
for the off-diagonal elements, and
\beq[eq:quadrupole-xi]
	\xi = \rho^M_{yy} - \rho^M_{xx}, \quad
	\delta_L = \rho^M_{xx} + \rho^M_{yy} - 2 \rho^M_{zz},
\eeq
for the diagonal elements. This representation also gives a physical picture for the transformation in \eq{eq:massive-params-rotation-z}.
	
Under a general Lorentz transformation $\Lambda$, the spin-1 density matrices also transform as \eq{eq:denmtx-lorentz-f}, 
with the transformation matrix $D$ determined by the little group $W(\Lambda, p)$ in the same way.
For massless vector bosons, this matrix $D$ is only a phase, 
$D_{\lambda\lambda'} = e^{-i \lambda \theta(\Lambda, p)} \delta_{\lambda\lambda'}$ [\eq{eq:Lorentz-rep-massless}], 
which only mixes between $\xi_1$ and $\xi_2$, but does not change $\xi_3$. 
In particular, for the special Lorentz boost along $\hat{z}$, 
we have $\theta(\Lambda, p) = 0$ and $D$ is an identity matrix [\eq{eq:boost-transformation-massless}], 
which does not change $\bm{\xi}$ at all.
For massive vector bosons, in contrast, $D$ can be an arbitrary rotation that mixes among various components of the polarization, 
especially it can mix the linear polarization $(\xi, Q_{xy})$ with other components. 
As a result, even if the linear polarization is 0 in one frame, it is likely to be nonzero in some other frame.
We will make further use of this fact in \sec{sec:linear-pol-W}.

\section{Parity constraint on the vector boson polarization}
\label{sec:parity-vector-spin}

We observed in \sec{sec:fermion-spin} that the single transverse spin of a fermion can only be produced if it is massive. 
Parity-conserving cases further constrain $b_2$ to be the only possible polarization degree of freedom, 
which can only appear through threshold effects at a loop level so is destined to be small.
The linear polarization of a vector boson, on the other hand, does not suffer from these constraints, 
because there is no counterpart of chiral symmetry to protect the helicity of a vector boson from being flipped. 
So, in general, we should expect a nonzero linear polarization to be produced for a vector boson.

Similar to \eq{eq:UP-relation-dm-f}, if the vector boson is produced in a $2\to 2$ process via a parity-conserving interaction, 
its polarization density matrix would satisfy
\beq[eq:parity-vector-denmtx]
	\rho^1_{-\lambda, -\lambda'}(p) = (-1)^{\lambda + \lambda'} \rho^1_{\lambda \lambda'}(p),
\eeq
which is obtained by performing the same $U_P$ transformation in \eq{eq:UP} and
using the transformation behavior of a vector boson state,
\beq[eq:parity-vector-state]
	U_P | p, \lambda \rangle = (-1)^{\lambda} | p, -\lambda \rangle.
\eeq
\eq{eq:parity-vector-denmtx} applies to both massless and massive vector bosons. 
For a massless vector boson, it implies $\rho^1_{++} = \rho^1_{--}$ and $\rho^1_{+-} = \rho^1_{-+}$, such that 
only the linear polarization $\xi_1$ is allowed to be nonzero, while $\xi_2$ and the helicity $\xi_3$ are forbidden. 
This reduces \eq{eq:denmtx-vector-g} to 
\beq[eq:denmtx-vector-g-P-conserve]
	\pp{ \rho^1_{\lambda\lambda'} } 
	= \frac{1}{2} \begin{pmatrix}
				1  &  \xi_1 \\
				\xi_1 &  1
			\end{pmatrix},
	\quad
	\mbox{(if parity conserves.)}
\eeq
Since $\xi_1 = 2\Re(\rho^1_{+-})$, it does not require an imaginary part from the amplitude, so it can appear at tree level.
The same conclusion also holds for a massive vector boson, for which \eq{eq:parity-vector-denmtx} means
$J_1 = J_3 = Q_{yz} = Q_{xy} = 0$, and we are only allowed to have nonzero $\delta_L$, $Q_{xz}$, $J_2$, or $\xi$.
Then \eq{eq:denmtx-massive-1} is reduced to 
\beq[eq:denmtx-massive-1-P-conserve]
	\pp{ \rho^M_{\lambda\lambda'} } = 
		\frac{1}{2}
		\begin{pmatrix}
			\frac{2 + \delta_L}{3} & \frac{Q_{xz} - i J_2}{\sqrt{2}} & \xi	\\
			\frac{Q_{xz} + i J_2}{\sqrt{2}} & \frac{2(1 - \delta_L)}{3} & \frac{- Q_{xz} - i J_2}{\sqrt{2}}  \\
			\xi & \frac{- Q_{xz} + i J_2}{\sqrt{2}} & \frac{2 + \delta_L}{3}
		\end{pmatrix},
	\quad
	\mbox{(if parity conserves.)}
\eeq

There are no other general symmetries to constrain the density matrix. For a particular situation, one only needs to examine
whether a single helicity flip is allowed for the vector boson under study.

The parity-conserving cases include pure QED and/or QCD production, but not the processes involving EW or other 
parity-violating new physics interactions.
In the latter case, all the polarization parameters are in principle not forbidden, and they are mixed under transformations between
different frames. 
Then the parity-violating polarization parameters would be sensitively dependent on the parity-violating interactions, 
so can serve as useful probes. 
This will be illustrated in \sec{sec:linear-pol-gluon} for the gluon polarization.

\section{Linearly polarized gluon and $CP$ violation}
\label{sec:linear-pol-gluon}

As we see in \eq{eq:denmtx-vector-g-P-conserve}, linear gluon polarization is generally allowed for the $\xi_1$ degree of freedom,
in the parity-conserving case. Since gluons are charge neutral, the parity property is equivalent to the $CP$ property, so that
$\xi_1$ is $CP$ even whereas $\xi_2$ and $\xi_3$ are $CP$-odd polarization degrees of freedom.
Measuring nonzero $\xi_2$ and/or $\xi_3$ can therefore serve as new probes of $CP$-violating interactions. 
In this section, we illustrate this by considering the linear gluon polarization in the associative Higgs and gluon jet production,
which happens in the SM from a gluon fusion through a top quark loop. 

Even though the LHC is an unpolarized proton-proton collider (so the gluon partons are also unpolarized), 
the hard scattering $gg \to hg$ can serve as a ``polarizer'' to produce a gluon with substantial polarization $\bm{\xi}$.
As this gluon further fragments into a jet, its polarization will modulate the kinematic distribution of the jet constituents;
in particular, the linear polarization breaks the rotational invariance around the jet direction to leave a nontrivial 
azimuthal distribution, which can be projected out by weighting each event with some azimuth-sensitive observable. 
The azimuthally weighted cross section $\sigma_w$ of the inclusive $h+g$ production at a $pp$ collider
can be factorized into a hard scattering coefficient multiplied by a polarized gluon jet function, 
in much the same way as the factorization for an unpolarized jet function~\citep{Berger:2003iw, Almeida:2008yp, Almeida:2008tp} 
or fragmentation function~\citep{Nayak:2005rt, Collins:2011zzd},
\begin{align}\label{eq:hg-factorization}
	\frac{ d\sigma_{w} }{dy_g \, dp_T^2 \, d m_J^2 \, d \phi}
			&= \frac{d\hat{\sigma}}{dy_g \, dp_T^2} \, \frac{d J(\bm{\xi}(p_T, y_g), m_J^2, \phi)}{d \phi}
				+ \order{\frac{m_J}{p_T}, R} \,,
\end{align}
in terms of the rapidity $y_g$ and transverse momentum $p_T$ of the gluon jet in the c.m.~frame of the $hg$ system 
and the jet mass $m_J$ and azimuthal substructure $\phi$, which is to be defined shortly.

\subsection{Production of polarized gluons in the hard scattering}
\label{ssec:hard-hg}

In \eq{eq:hg-factorization}, the hard coefficient
\beq[eq:hg-hard]
	\frac{d\hat{\sigma}}{ dy_g \, dp_T^2} = \lum(s, \hat{s}) \, \frac{\overline{\abs{\M}^2}}{ 16\pi E_h \sqrt{\hat{s}} }
\eeq
is the differential cross section for the on-shell gluon production, 
with respect to the gluon rapidity $y_g$ and transverse momentum $p_T$ in the partonic c.m.~frame.
The $s$ and $\hat{s}$ are the c.m.~energies squared for the $pp$ and $hg$ systems, respectively,
and $E_h$ is the Higgs energy in the partonic c.m.~frame; they are sufficiently determined by $p_T$ and $y_g$,
\beq
	E_h = (m_H^2 + p_T^2 \cosh^2 y_g)^{1/2}, \quad
	\sqrt{\hat{s}} = p_T \cosh y_g + E_h,
\eeq
where $m_H$ is the Higgs mass.
In \eq{eq:hg-hard},
\beq
	\lum(s, \hat{s}) = \frac{1}{s} \int_{\hat{s}/s}^1 \frac{dx}{x} \, f_{g/{\rm p}}(x, \, \mu_F) f_{g/{\rm p}}\pp{ \frac{\hat{s}}{x s}, \, \mu_F }
\eeq 
is the gluon-gluon parton luminosity, 
with the factorization scale chosen at $\mu_F = p_T$ 
in the parton distribution function (PDF) $f_{g/{\rm p}}(x, \, \mu_F)$ of the proton.
We have used the LO kinematics to integrate out the Higgs phase space.

Since the linear gluon polarization degrees are produced from the hard polarizer and depend sensitively on the interaction structure of the hard scattering, 
the azimuthal substructure of the gluon jet can help probe the $CP$ property of the Higgs-top interaction, which we parametrize by 
an effective operator, 
\beq[eq:L-hg]
	\mathcal{L} \supset - \frac{y_t}{\sqrt{2}} \, h \, \bar{t} \, (\k + i \,\kt\, \gamma_5) \, t 
		= - \frac{y_t}{\sqrt{2}} \, \k_t \, h \, \bar{t} \, (\cos\alpha + i \sin\alpha\, \gamma_5) \, t \, ,
\eeq
where $y_t = \sqrt{2} m_t / v$ is the Yukuwa coupling of Higgs and top quark in the SM, and $(\k, \kt)$ parametrize 
the $CP$-even and $CP$-odd $\htt$ interactions, respectively, 
which are usually reparametrized as $(\k, \kt) = \k_t (\cos\alpha, \sin\alpha)$, with $\alpha$ being the $CP$ phase. 
Pinning down the $CP$ nature of this interaction is an important program being pursued at the LHC
~\citep{CMS:2021nnc, CMS:2020cga, ATLAS:2020ior, CMS:2022dbt, ATLAS:2020ior, ATLAS:2023cbt}. 
Any deviation from a Standard-Model-like $\htt$ coupling, i.e., $(\k, \kt) = (1, 0)$ or $(\k_t, \alpha) = (1, 0)$, 
could indicate new physics as well as provide a potential source for the $CP$ violation as required by the baryogenesis~\citep{Sakharov:1967dj}.
Unlike $CP$-violating Higgs interactions with vector bosons, which arise from dimension-six operators, 
$CP$-violating effects in \eq{eq:L-hg} occur via a dimension-four operator and can be potentially larger.

Numerous approaches have been proposed for determining the $CP$ phase, 
either directly via associated Higgs and top production~\citep{Ellis:2013yxa, Boudjema:2015nda, Buckley:2015vsa, Gritsan:2016hjl, Mileo:2016mxg, AmorDosSantos:2017ayi, Azevedo:2017qiz, Li:2017dyz, Goncalves:2018agy, Faroughy:2019ird, Bortolato:2020zcg, Cao:2020hhb, Goncalves:2021dcu, Patrick:2019nhv},
or indirectly via Higgs or top induced loop effects~\citep{Brod:2013cka, Dolan:2014upa, Englert:2012xt, Bernlochner:2018opw, Englert:2019xhk, Gritsan:2020pib, Bahl:2020wee, Martini:2021uey}. 
The sensitivity to $\alpha$ can be enhanced by using observables that are odd 
under $CP$ transformation~\citep{Mileo:2016mxg, Goncalves:2018agy}. 
Machine learning techniques have also been considered~\citep{Patrick:2019nhv, Ren:2019xhp, Bortolato:2020zcg, Bahl:2021dnc, Barman:2021yfh} 
aiming to optimize the sensitivity.
The current experimental bounds from direct measurements are 
$\abs{\alpha} \leq 35^\circ$~\citep{CMS:2020cga}, 
$\abs{\alpha} \leq 48^\circ$~\citep{CMS:2022dbt}, 
and $\abs{\alpha} \leq 63^\circ$~\citep{ATLAS:2023cbt} at $68\%$ C.L.,
and 
$\abs{\alpha} > 43^\circ$ has been excluded at $95\%$ C.L.~\citep{ATLAS:2020ior},
by various Higgs detection channels.
This makes it necessary to have more complementary observables to further constrain $\alpha$, at the upcoming 
High-Luminosity LHC (HL-LHC)~\citep{Apollinari:2120673} and a possible future $pp$ collider at $100~\TeV$ (FCC-hh)~\citep{Mangano:2270978}.

In the following, we examine how $\xi_1$ and $\xi_2$ can serve as useful probes of the $CP$ phase;
especially, we propose $\xi_2$ as a new $CP$-odd observable.
This $\xi_2$ is a genuine $CP$-odd observable that is constructed purely out of the kinematic information in the gluon jet, 
and not via a neutral state of charged particles and antiparticles~\citep{Han:2009ra}. 
Such $CP$ sensitivity would not be possible in the $hg$ production process without the gluon jet substructure, 
which has not been considered previously.
Furthermore, we note that associated Higgs-top production and indirect measurements via $hV$ or $VV$ production 
also depend on the $hVV$ couplings and require assumptions on the latter, whereas $hg$ production only depends on the $\htt$ coupling.

\begin{figure}[htbp]
	\centering
		\includegraphics[trim={0.45cm -0.32cm 0.3cm 0.4cm}, clip, scale=0.8]{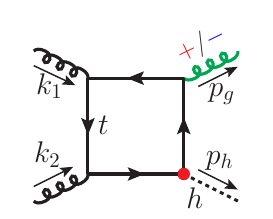}
		\includegraphics[trim={0cm -0.45cm 0cm 0cm}, clip, scale=0.8]{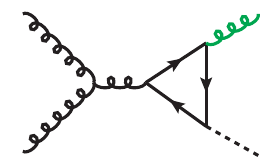}
		\includegraphics[trim={0cm 0cm 0cm 0cm}, clip, scale=0.8]{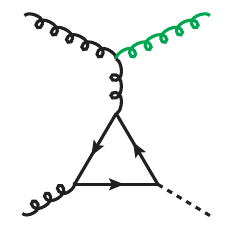}
		\includegraphics[scale=0.55]{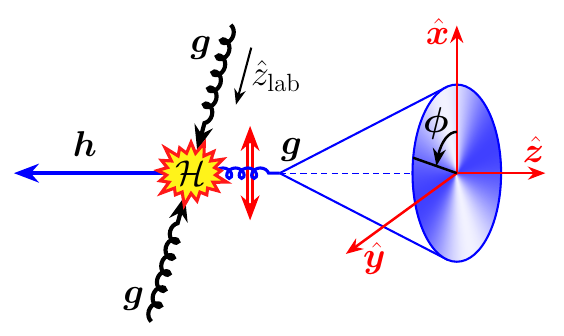}
	\vspace{-3mm}
	\caption{Left three: Representative diagrams for $gg\to hg$ via a top loop. 
		Rightmost one: the gluon $\hat{x}$-$\hat{y}$-$\hat{z}$ frame defined in the same way as \eq{eq:fermion-xyz}.}
	\label{fig:hg}
\end{figure}

With the Lagrangian in \eq{eq:L-hg}, we can calculate the polarization degree of the gluon.
At LO, both $gg$ fusion and $q\bar{q}$ annihilation contribute via a top loop, 
as exemplified in \fig{fig:hg} (left) for the $gg$ channel. 
Even though the $q\bar{q}$ channel can also produce a substantially polarized gluon, 
its contribution to the total cross section is much smaller and will be neglected.
Parametrizing the helicity amplitudes $g(\lambda_1)\, g(\lambda_2)\to h\, g(\lambda_3)$ in the partonic c.m.~frame 
in terms of the gluon's transverse momentum $p_T$, rapidity $y_g$, and azimuthal angle $\phi_g$, we have 
\begin{align}\label{eq:hg-amp}
	&\M_{\lambda_1\lambda_2\lambda_3}(p_T, y_g, \phi_g) = f^{abc} \,e^{i(\lambda_1 - \lambda_2)\phi_g}
			\bb{ \k \, \A_{\lambda_1 \lambda_2 \lambda_3}(p_T, y_g) + i \, \kt \, \At_{\lambda_1\lambda_2\lambda_3}(p_T, y_g) },
\end{align}
with $f^{abc}$ the color factor, and $\lambda_i$ the gluon helicities.
In \eq{eq:hg-amp}, $\A$ and $\At$ are the $CP$-even and $CP$-odd helicity amplitudes, respectively, constrained by their $CP$ properties as
\beq\label{eq:CP amp}
	(\A, \, \At)_{-\lambda_1,-\lambda_2,-\lambda_3}(p_T, y_g) 
		= (-\A, +\At)_{\lambda_1\lambda_2\lambda_3}(p_T, y_g),
\eeq
given by \eq{eq:parity-vector-state}.
The gluon density matrix is then determined through
\beq[eq:hg-amp-rho]
	\frac{1}{4 \, N_{c,g}^2} \sum\nolimits_{\lambda_1, \lambda_2} 
		\M_{\lambda_1 \lambda_2 \lambda} \M^{*}_{\lambda_1 \lambda_2 \lambda'}
		\equiv \rho_{\lambda\lambda'}(\bm{\xi}) \, \overline{\abs{\M}^2},
\eeq
where the convention of summing over repeated indices is taken, and $\overline{\abs{\M}^2}$ is the unpolarized squared amplitude, averaged/summed over the spins and colors, with $N_{c, g} = 8$.
Due to their $CP$ properties in \eq{eq:CP amp}, $\A$ and $\At$ individually only contribute to $\xi_1$, 
while it is their interference that contributes to $\xi_2$.
In terms of the $CP$ phase $\alpha$, $\bm{\xi}$ can be expressed as
\beq[eq:xi12]
	\xi_1 = \frac{ \omega + \beta_1 \cos2\alpha }{1 + \Delta \cos2\alpha },
	\quad
	\xi_2 = \frac{ \beta_2 \sin2\alpha }{1 + \Delta \cos2\alpha }	,
	\quad
	\xi_3 = \frac{\beta_3 \sin2\alpha}{1 + \Delta \cos2\alpha},
\eeq
where we have defined the polarization parameters
\begin{gather}
	\Delta = \frac{ |\A|^2 - |\At|^2 }{ |\A|^2 + |\At|^2 }	,	\quad
	\omega = \frac{2( \A_+ * \A^*_{-} + \At_+ * \At^*_{-} )}{ |\A|^2 + |\At|^2 }	,	
	\nn\\
	\beta_1 = \frac{2( \A_+ * \A^*_{-} - \At_+ * \At^*_{-} )}{ |\A|^2 + |\At|^2 }	,	\quad
	\beta_2 = \frac{ 4 \Re ( \A_+ * \At^*_{-} )}{  |\A|^2 + |\At|^2 }	, \quad
	\beta_3 = \frac{4 \Im(\A_+ * \At^*_+)}{ |\A|^2 + |\At|^2 },
\label{eq:hg-pol-params}
\end{gather}
with the notations defined similarly to \eq{eq:prod-notation-f},
\beq[eq:prod-notation-g]
	A_{\lambda} * B_{\lambda'} \equiv \sum\nolimits_{\lambda_1, \lambda_2} 
		A_{\lambda_1\lambda_2\lambda} B_{\lambda_1\lambda_2\lambda'},\quad
	\abs{A}^2 \equiv \sum\nolimits_{\lambda_1, \lambda_2, \lambda_3} 
		A_{\lambda_1\lambda_2\lambda_3} \, A^*_{\lambda_1\lambda_2\lambda_3}.
\eeq
Parametrizing $\bm{\xi}$ as in \eq{eq:xi12} clearly shows that the polarization only depends on the $CP$ phase $\alpha$, 
but not on the coupling strength $\k_t$, 
which only controls the event rate.
The helicity polarization $\xi_3$ requires an imaginary part from the amplitudes so is nonzero only at $\sqrt{\hat{s}} > 2m_t$.
Its value is generally small compared to $\xi_1$ and $\xi_2$, and will not be discussed in the following.

\subsection{Polarized gluon jet function}
\label{ssec:gluon-jet-f}
With the polarization $\bm{\xi}$ produced from the hard scattering in \eq{eq:hg-amp-rho}, 
the gluon then fragments into a polarized jet.
In the partonic c.m.~frame, the gluon jet momentum $k$ defines the jet mass $m_J^2 = k^2$ 
and direction $\hat{z}$ as in \eq{eq:fermion-xyz} and \fig{fig:hg} (right).
By defining two lightlike vectors $n^{\mu} = (1, -\hat{z})/\sqrt{2}$ and $\bar{n}^{\mu} = (1, \hat{z})/\sqrt{2}$, we can 
approximate the gluon momentum in the hard part to be on shell by only retaining the large component, $p_g^{\mu} = (k\cdot n) \bar{n}^{\mu}$,
which then defines the rapidity $y_g$ and $p_T = k\cdot n / (\sqrt{2} \cosh y_g)$.
To the leading power of $m_J/p_T$, the polarized gluon fragmentation is described by the polarized jet function in \eq{eq:hg-factorization},
\begin{align}\label{eq:pol-g-jet}
	\frac{d J(\bm{\xi}, m_J^2, \phi)}{d \phi}
		&= \frac{1}{2\pi N_{c,g}(k \cdot n)^2} \sum_{X} \int d^4 x \, e^{i k\cdot x}  \bb{ \rho_{\lambda\lambda'}(\bm{\xi})  O(\phi, X) } 
				\varepsilon_{\lambda \mu}(p_g) \, \varepsilon_{\lambda' \nu}^*(p_g) 
				\nn\\
			&\hspace{2em} \times 
			 \langle 0 | W_{ac}(\infty, x; n) \, n_{\sigma} G_c^{\sigma \nu}(x) | X \rangle
			 \, \langle X | W_{ab}(\infty, 0; n) \, n_{\rho} G_b^{\rho \mu}(0) | 0 \rangle 	\,,	
\end{align}
where $X$ denotes the state of the particles within the jet, in accordance with the jet algorithm~\citep{Almeida:2008tp, Ellis:2010rwa}, whose momenta are dominantly along $\bar{n}$. 
$G_c^{\mu\nu}$ is the gluon field strength tensor, and 
\beq
	W_{ab}(\infty, x; n)
	= \P \exp\cc{-ig \int_0^{\infty} d\lambda \, n \cdot A^c(x + \lambda n) ( T^{{\rm adj}, c}_{ab} ) }
\eeq
is the Wilson line in the adjoint representation from $x$ to $\infty$ along $n$, 
with $\P$ denoting the path ordering and $T^{{\rm adj}, c}$ the SU(3) generator in the adjoint representation.
The repeated color indices $a, b$, and $c$ in \eq{eq:pol-g-jet} are summed over.
In \eq{eq:pol-g-jet}, the gluon polarization states are projected using the on-shell polarization vectors 
$\varepsilon_{\lambda}^{\mu}(p_g)$ with helicity $\lambda = \pm 1$,
which are then averaged with the density matrix $\rho_{\lambda\lambda'}(\bm{\xi})$.
The resultant azimuthal distribution is extracted by inserting the observable
\beq[eq:O]
	O(\phi, X) = \frac{1}{\sum_{i\in X} p_{i,T}} \sum_{i\in X} p_{i,T} \delta(\phi - \phi_i),
\eeq
where $p_{i,T}$ and $\phi_i$ are, respectively, the transverse momentum and azimuthal angle of the jet constituent $i$ 
with respect to the $\hat{x}$-$\hat{y}$ plane defined in \eq{eq:fermion-xyz} and shown in \fig{fig:hg} (right).
Such a $\phi$ distribution is a new jet substructure observable introduced by the linear polarization. 
The dependence on $\xi_3$ would vanish due to parity invariance of $O(\phi, X)$.

As a result of the $p_{i,T}$ weight, the observable $O(\phi, X)$ is IR safe, and hence
the polarized gluon jet function is insensitive to hadronization effects and becomes perturbatively calculable, with a predictable $\phi$ dependence. 
Nevertheless, it was noted long before~\citep{DeGrand:1980yp, Hara:1988uj} that the gluon polarization information will be 
greatly washed out by the cancellation between the $g\to gg$ and $g\to q\bar{q}$ channels, 
which was also found recently in a similar situation~\citep{Chen:2020adz, Chen:2021gdk, Larkoski:2022lmv}. 
It is possible to mitigate these effects by using jet flavor tagging techniques%
~\citep{Gallicchio:2011xq, Gallicchio:2012ez, FerreiradeLima:2016gcz, Frye:2017yrw, Banfi:2006hf,
Gras:2017jty, Metodiev:2018ftz, Larkoski:2014pca, Bhattacherjee:2015psa, Kasieczka:2018lwf, Larkoski:2019nwj, Bright-Thonney:2022xkx}.
For example, one may recluster the identified gluon jet into two subjets, and only keep those gluon jets with their two subjets tagged as quarks.
At $\order{\alpha_s}$, requiring a tagged quark in the gluon jet leaves $g \to q\bar{q}$ as the only diagram, 
giving the polarized gluon jet function,
\begin{align}\label{eq:g-jet-qq}
	\frac{d J^{(q)}}{d\phi} 
		= \frac{\alpha_s T_F}{6 \pi^2 m_J^2}
			\bb{ 1 + \frac{1}{2} \pp{ \xi_1 \cos2\phi + \xi_2 \sin2\phi  } },
\end{align}
where the jet algorithm dependence does not come in at this order to the leading power of $m_J$.
\eq{eq:g-jet-qq} needs to be multiplied by the tagging efficiency when used in \eq{eq:hg-factorization}.
Although flavor tagging reduces the statistics significantly, 
it enhances the gluon spin analyzing power from $\order{1\%}$ to about 
$50\%$~\citep{Hara:1988uj} and will improve the statistical precision.

It is worth noting that even though we choose to express the event kinematics in the partonic c.m.~frame for simplicity, 
a boost along the beam direction does not change the polarization of the gluon jet to the leading power of $m_J$, 
due to the special transformation property of massless particle state in \eq{eq:boost-transformation-massless}. 
Therefore, we may equally describe each event in the lab frame, and the azimuthal jet anisotropy structure in \eq{eq:g-jet-qq} 
stays the same.

Before closing this section, we note the difference of the gluon polarization from a quark. 
While a transversely polarized light (massless) quark can also be produced from hard scattering processes, 
as discussed in \ch{ch:pol-fermion}, its transverse spin cannot be conveyed via the {\it perturbative} quark jet function 
due to the chiral symmetry of a massless quark. 
It is hence related to chiral symmetry breaking and must require the presence of some 
non-perturbative functions~\citep{Collins:1993kq, Collins:2011zzd, Kang:2020xyq}.

\subsection{Numerical results for the polarization degrees}
\label{ssec:hg-numerics}

\begin{figure*}[htbp]
	\centering
		\includegraphics[scale=0.6]{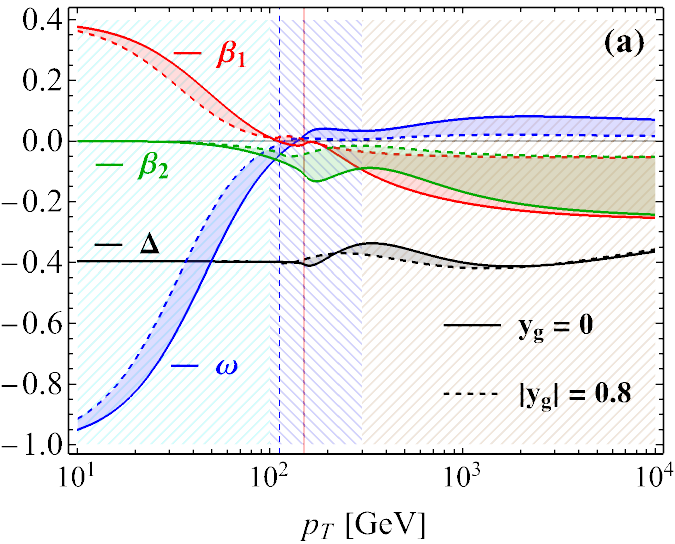} \quad
		\includegraphics[scale=0.6]{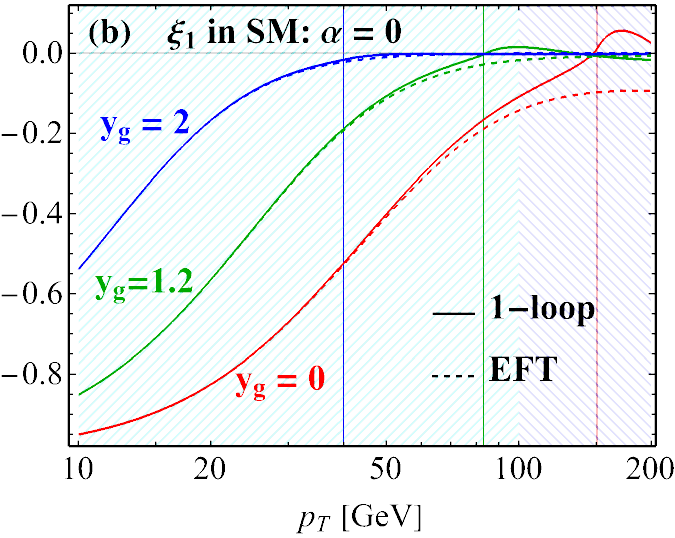} \\
		\vspace{1em}
		\includegraphics[scale=0.6]{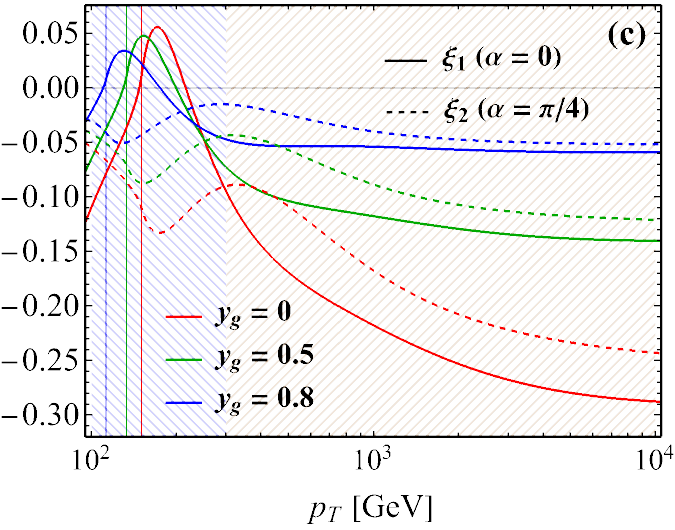} 
	\caption{
	(a) Polarization parameters $\Delta$, $\omega$, $\beta_1$, and $\beta_2$, as functions of the gluon $p_T$ in the partonic c.m.~frame. Each parameter is shown as a shaded region constrained by $\abs{y_g} \leq 0.8$, bounded by a solid curve and a dashed curve, corresponding to $y_g = 0$ and $\abs{y_g} = 0.8$, respectively. The two vertical lines stand for the $\sqrt{\hat{s}} = 2m_t$ threshold for $y_g = 0$ (red, solid) and $\abs{y_g} = 0.8$ (blue, dashed), respectively. The three hatching-shaded regions are the low-$p_T$ region (cyan) for $p_T < 100~\GeV$, transition region (blue) for $p_T \in (100, 300)~\GeV$, and high-$p_T$ region (brown) for $p_T > 300~\GeV$. 
	(b) $\xi_1$ in the low-$p_T$ region with the SM Lagrangian ($\alpha = 0$) for three values of $y_g$, where the full one-loop calculation (solid) is compared with the EFT result (dashed). The three vertical lines are the $\sqrt{\hat{s}} = 2m_t$ threshold for $y_g = 0$ (red), $y_g = 1.2$ (green) and $y_g = 2$ (blue). 
	(c) $\xi_1$ and $\xi_2$ in the transition and high-$p_T$ regions, for $CP$ phase $\alpha = 0$ and $\pi/4$, respectively, at which $\xi_1$ and $\xi_2$ peak respectively.
	}
	\label{fig:pols}
\end{figure*}

The parameters $(\Delta, \omega, \beta_1, \beta_2)$ in \eq{eq:hg-pol-params} are functions of $p_T$ and $y_g$, 
as shown in \fig{fig:pols}(a) for some benchmark phase-space points.
While the parameter $\Delta$, which describes the relative difference between the $CP$-even and $CP$-odd amplitudes squared, stays relatively flat around $-0.4$ in the range $p_T < 10~\TeV$, the parameters $\omega$, $\beta_1$, and $\beta_2$, which control the sizes of the polarizations $\xi_1$ and $\xi_2$, vary sizably with $p_T$.
Based on their $p_T$ dependence, we divide the phase space into three $p_T$ regions and discuss them in turn.

\emph{1. Low-$p_T$ region, with $p_T \lesssim 100~\GeV$.}
In this region, both $\abs{\omega}$ and $\beta_1$ have large values, whereas $\beta_2 \simeq 0$. The linear polarization is thus dominated by $\xi_1$, with $\xi_2 \simeq 0$. The dominance of $\omega$ over $\beta_1$ further implies that $\xi_1$ does not depend sensitively on $\alpha$. 
Being well below the $\sqrt{\hat{s}} = 2m_t$ threshold, this region 
can be well approximated by the infinite-top-mass effective field theory (EFT)~\citep{Dawson:1990zj, Djouadi:1991tka},
\beq
	\mathcal{L}_{\rm EFT} \supset
		- \frac{h}{4v} \pp{ \lambda_{hgg} \,  G^{a\mu\nu} G^{a}_{\mu\nu}
			+ \widetilde{\lambda}_{hgg} \, G^{a\mu\nu} \widetilde{G}^{a}_{\mu\nu} },
\eeq
which is matched onto \eq{eq:L-hg} by
\beq
	\lambda_{hgg} = - \frac{2\alpha_s}{3\pi} \, T_F \, \k,
    \quad
    \widetilde{\lambda}_{hgg} = - \frac{\alpha_s}{\pi} \, T_F \, \kt \,.
\eeq
In \fig{fig:pols}(b), the SM predictions for $\xi_1$ are shown for both the full one-loop calculation 
and the EFT approximation, 
\bse\begin{align}
	\xi^{\rm EFT}_1(p_T, y_g) &= - \frac{1}{U(p_T, y_g)}
    		\bb{ \pp{\frac{m_H}{p_T}}^2 + \pp{\frac{p_T}{m_H}}^2 
    		\frac{\lambda_{hgg}^2 - \wt{\lambda}_{hgg}^2}{\lambda_{hgg}^2 + \wt{\lambda}_{hgg}^2} },    
	\label{eq:xi1 pt eta}\\
	\xi^{\rm EFT}_2(p_T, y_g) &= - \frac{1}{U(p_T, y_g)}
    		\bb{  \pp{\frac{p_T}{m_H}}^2 
			\frac{2 \lambda_{hgg} \wt{\lambda}_{hgg} }{\lambda_{hgg}^2 + \wt{\lambda}_{hgg}^2} },
\end{align}\ese
with 
\beq
	U(p_T, y_g) = 2 + \bb{ \frac{p_T}{m_H} \pp{1+2\cosh2 y_g} + \frac{m_H}{p_T} }^2.
\eeq
One can see that $\xi_1$ generally has a large negative value, which means that the produced gluon is dominantly polarized along the $\hat{x}$ direction in the production plane, cf.~\fig{fig:hg} (right).
Furthermore, it is not dramatically dependent on the gluon rapidity $y_g$.

Since the low-$p_T$ region contains most of the $hg$ events, it is suitable for testing the gluon linear polarization phenomenon. 
Here we expect a significant $\cos2\phi$ jet anisotropy due to the dominant $\xi_1$, as shown in \fig{fig:g-linear-pol}(a).
Its insensitivity to $\alpha$ also means that this region can serve as a calibration region for experimentally measuring the linear polarization, which is important to ensure its viability and to understand the systematic uncertainties of the measurement since such phenomenon has not been observed before.

\emph{2. Transition region, with $100~\GeV \lesssim p_T \lesssim 300~\GeV$.}
In this region, $\beta_1$ and $\omega$ rapidly go to 0 and flip their signs, while $\abs{\beta_2}$ starts growing to an appreciable value. 
Hence, the linear polarization is dominated by $\xi_2$ if $\alpha$ is not too small,
as illustrated in \fig{fig:pols}(c) for $\xi_1$ at $\alpha = 0$, and $\xi_2$ at $\alpha = \pi/4$, which corresponds to a maximal $CP$ mixing. 
A nonzero $\alpha$ would then lead to a linearly polarized gluon jet that features a $\sin2\phi$ anisotropy,
whose measurement provides a good opportunity for constraining the $CP$-odd coupling. 
Furthermore, this region covers the $\sqrt{\hat{s}} = 2m_t$ threshold, so the EFT is no longer a good approximation, as indicated in the right half of \fig{fig:pols}(b). In this region, both $\xi_1$ and $\xi_2$ are sensitive to $y_g$, and their magnitudes are larger for gluon jets at more central rapidity region.

\emph{3. High-$p_T$ region, with $p_T \gtrsim 300~\GeV$.}
Here, both $\beta_1$ and $\beta_2$ have appreciable negative magnitude. Their values grow and approach each other as $p_T$ increases. 
Moreover, $\omega$, being smaller than $\abs{\beta_1}$, becomes less important in $\xi_1$. Qualitatively, we can interpret this region by taking $\omega$, $\Delta \to 0$ and $\beta_1$, $\beta_2 \to \beta$, which gives
$(\xi_1, \xi_2) \sim \beta (\cos2\alpha, \sin2\alpha)$.
Then the jet anisotropy in \eq{eq:g-jet-qq} can be recast as
\beq[eq:oscillation]
	1 + \frac{1}{2} \, 
	\beta \cos2(\phi - \alpha) \,,
\eeq
so that the main axis direction of the jet image gives a direct measure of the $CP$ phase. It can be shown that as $\hat{s} \to \infty$, this qualitative simplification becomes exact in the one-loop calculation.
The quantitative behavior of $\xi_1$ and $\xi_2$ in the high-$p_T$ region is shown in the right half of \fig{fig:pols}(c), where we see that they drop rapidly to 0 as $\abs{y_g}$ increases, and a simple kinematic cut $\abs{y_g} < 0.8$ yields the polarization $\abs{\beta_{1,2}} \gtrsim 0.05$.

\subsection{Phenomenology}
\label{ssec:hg-pheno}
The gluon jet azimuthal anisotropy in \eq{eq:g-jet-qq} can be experimentally measured by simply constructing the 
asymmetry observables,
\begin{align}
	A_{i} &= \frac{\int_0^{2\pi} d\phi  \, (d\sigma_w / d\phi) \cdot \sgn{ F_i(\phi) } 
				}{ \int_0^{2\pi} d\phi \, (d\sigma_w / d\phi)} = \frac{\xi_{i}}{\pi},
\label{eq:asym}	
\end{align}
where $i \in \{1,2\}$, $F_1(\phi) = \cos2\phi$ and $F_2(\phi) = \sin 2\phi$.
The uncertainties of the asymmetries $A_{1,2}$ are dominated by statistical ones, given by 
$1/\sqrt{N}$ with 
$N$ being the number of the observed events. 
Now we provide a simple demonstration of the constraining power of the gluon linear polarization on the $CP$ phase, by 
confining ourselves to the transition region 
for both the HL-LHC at $14~\TeV$ and FCC-hh at $100~\TeV$, with integrated luminosities $3~\ab^{-1}$ and $20~\ab^{-1}$, respectively.

The $hg$ cross section in the transition region is estimated 
for the Lagrangian [\eq{eq:L-hg}] 
using CT18NNLO PDFs~\citep{Hou:2019efy}
with {\tt MG5\_aMC@NLO 2.6.7}~\citep{Alwall:2014hca} by first generating the $hg$ events with $p_T \in [100, 300]~\GeV$ and $\abs{\eta_g} \le 2.5$ in the lab frame, and then boosting to the partonic c.m.~frame 
with a further cut $\abs{y_g} \le 0.8$, which gives 
$\k_t^2 (0.57 \cos^2\alpha + 1.3 \sin^2\alpha)~\pb$ for the HL-LHC and 
$\k_t^2 (13.7  \cos^2\alpha + 30.7 \sin^2\alpha)~\pb$ for the FCC-hh.
While both $\k_t$ and $\alpha$ affect the total production rate and can be constrained by the measurement of the latter, only $\alpha$ determines the polarization. In the following, we take $\k_t = 1$ and consider the constraint on $\alpha$ from the polarization data.

We are interested in final states where the (fat) gluon jet is composed of a pair of quark subjets. 
While it is possible to also discriminate light quark subjets from gluon subjets,
here we only provide a conservative estimate by 
restricting to the bottom ($b$) and charm ($c$) quark tagging as
used in experiments~\citep{CMS-PAS-BTV-15-002, ATLAS-CONF-2016-002, CMS:2017wtu, ATLAS:2018sgt, ATLAS:2018mgv, ATLAS:2019bwq, ATLAS:2019lwq, CMS:2021scf, ATLAS:2021cxe, ATL-PHYS-PUB-2022-027, ATL-PHYS-PUB-2022-010}.
We estimate the branching fraction $f_{g_{b\bar b}}$ ($f_{g_{c\bar c}}$) of $g\to b\bar{b}$ ($g\to c\bar{c}$)
through parton shower simulation using \texttt{Pythia 8.307}~\citep{Sjostrand:2014zea}, 
which gives $f_{g_{b\bar b}} = 0.013$ and $f_{g_{c\bar c}} = 0.019$ in the selected kinematic region. 
Following Refs.~\citep{ATLAS:2019bwq, ATLAS:2018mgv}, 
we take $b$-tagging efficiency $\epsilon_b = 0.7$ and $c$-tagging efficiency $\epsilon_c = 0.3$.
We consider the diphoton decay channel of the SM Higgs boson and assume a Higgs tagging efficiency $\epsilon_h = 0.002$.
This then gives about  
$(51\cos^2\alpha + 115 \sin^2\alpha)$ reconstructed events at the HL-LHC and 
$(8100\cos^2\alpha + 18200\sin^2\alpha)$ events at the FCC-hh.

\begin{figure}[htbp]
	\centering
		\includegraphics[trim={0 0.1cm -0.5cm 0}, clip, scale=0.65]{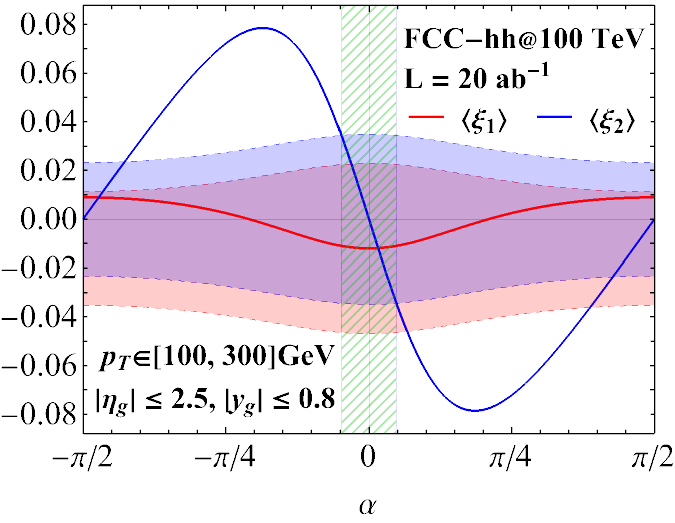} 
	\caption{Constraining power of the FCC-hh gluon polarization data, in the transition region, 
	on the $CP$ phase $\alpha$. $\langle \xi_{1,2} \rangle$ are the average values of $\xi_{1,2}$ in the specified kinematic region.
	Their statistical uncertainties are indicated by the red and blue bands, respectively, around the SM prediction (with $\alpha = 0$).
	The green-hatched region is the $\alpha$ range allowed by the $\xi_2$ measurement.
	}
	\label{fig:fcc}
\end{figure}

In \fig{fig:fcc}, we display the predicted average values of $\xi_{1,2}$ in the transition region at the FCC-hh as functions of the $CP$ phase $\alpha$, 
together with their uncertainty bands around the SM central values.
As expected, it is $\xi_2$ that constrains small values of $\alpha$, whereas $\xi_1$ is too small to have an impact in this region.
Assuming the SM scenario with $\xi_2 = 0$, we can project the constraint $\abs{\alpha} \le 8.6^\circ$.
In this estimate, we have only used the gluon polarization information with Higgs decaying to diphotons. 
In order to make a significant impact with data from the HL-LHC, 
one will have to include other Higgs decay channels and 
light quark flavor tagging in the gluon jets, as well as data from the low-$p_T$ and high-$p_T$ regions, 
which will significantly improve the constraints. 
A more careful phenomenological study is left for future.

\subsection{Summary}
A precise understanding of the $CP$ property of the Higgs boson is important both to test the SM and to probe new physics. 
In this section, we proposed a novel way of probing the $CP$ structure of the Higgs-top interaction, 
by measuring the azimuthal anisotropy substructure of the gluon jet produced in association with a Higgs boson, 
which originates from the linear polarization of the final-state gluon. 
We have introduced a factorization formalism and defined a perturbative polarized gluon jet function with 
insertion of an IR-safe azimuthal observable. 
Experimental measurement of the linearly polarized gluon jet will be an important test of the SM 
and can also serve as a new tool to search for new physics.

\section{Linearly polarized $W$ boson and boosted top quark jet substructure}
\label{sec:linear-pol-W}

As we have seen in \sec{sec:linear-pol-gluon}, linearly polarized gluons can be produced from hard partonic scattering at the unpolarized LHC
and induce azimuthally anisotropic jet images which can serve as useful probes for the hard interaction structures, especially the $CP$ property.
Obviously, similar phenomena exist for heavy gauge bosons like the $W$ and $Z$. The inherent parity-violating interactions can give rise to
richer phenomenology. Instead of studying their prompt production from the hard scattering, however, in this section, we examine the linear
polarization of the $W$ boson in the boosted top quark decay. As can be easily inferred from \sec{sec:linear-pol-gluon}, in the absence of 
$CP$-violating interactions, only the linear polarization $\xi_1$ can be produced, characterizing the difference between the linear polarization degrees
along and perpendicular to the top quark decay plane. Nevertheless, as we will see in the following,
due to the large parity violation in the $tbW$ interaction, $\xi_1$ will be sensitively dependent on the longitudinal polarization of the top quark, 
so its measurement will serve as a useful boosted top quark polarimeter.

\subsection{Polarization of the $W$ boson in the top rest frame}
\label{ssec:w-pol-R}
For the sake of a generic discussion, we work with the Lagrangian 
\beq[eq:L-tbw]
	\mathcal{L} \supset 
	-\frac{g}{\sqrt{2}} \, \bar{b} \gamma^{\mu} 
		\pp{ g_{L} P_{L} + g_{R} P_{R} } t\, W_{\mu} + {\rm h.c.}
\eeq
to describe the $t \to bW$ decay.
When a boosted top quark is produced in the lab frame, we construct the coordinate system 
$\hat{x}$-$\hat{y}$-$\hat{z}$ as in \eq{eq:fermion-xyz},
with $\hat{z}$ along the top quark momentum direction, 
and $\hat{x}$-$\hat{y}$ quantifying the transverse plane.
The decay process is most conveniently analyzed in the top rest frame, where
the amplitude $\M^R_{\lambda_t\lambda_w\lambda_b}$ takes a simple form,
\beq[eq:tbw-MR]
	\M^R_{\lambda_t\lambda_w\lambda_b}(\theta_w, \phi_w) 
		= A_{\lambda_w, \lambda_b} D^{1/2 *}_{\lambda_t, \,\lambda_w-\lambda_b}(\phi_w, \theta_w, 0)
		= A_{\lambda_w, \lambda_b} e^{i \lambda_t \phi_w} d^{1/2}_{\lambda_t, \,\lambda_w-\lambda_b}(\theta_w),
\eeq
where $\theta_w$ and $\phi_w$ describe the angles of the $W$
with respect to the $\hat{x}$-$\hat{y}$-$\hat{z}$ coordinate system.
The constant coefficients $A_{\lambda_w, \lambda_b}$ can be explicitly calculated from \eq{eq:L-tbw},
with four independent components,
\begin{align}
	A_{1,1/2} &= i N(g_R -  f_L / r), \quad&&
	A_{-1,-1/2} = i N(g_L - f_R / r),	\nn\\
	A_{0, 1/2} &= \frac{iN}{\sqrt{2}} \pp{ g_R / r - f_L }, \quad&&
	A_{0, -1/2} = \frac{iN}{\sqrt{2}} \pp{ g_L / r -  f_R },
\end{align}
with $N = g \sqrt{m_t^2 - m_w^2}$ and $r = m_w / m_t \simeq 0.46$.
We neglect the $b$ mass throughout the whole discussion. 
Importantly, because of the spin-half nature of the top quark, $W$ cannot simultaneously have both left- and right-handed
polarization states for a given $b$ state, regardless of whether $b$ is massless or massive. 
There is thus no interference of those states, implying a vanishing linear polarization.

For a general top spin state described by a density matrix $\rho^t = (1 + \bm{s}_t \cdot \bm{\sigma}) / 2$ with 
$\bm{s}_t = (b_1, b_2, \lambda_t)$ being the top spin vector, 
the unnormalized $W$ density matrix can be obtained explicitly from \eq{eq:tbw-MR} by
\begin{align}
	&W^R_{\lambda_w\lambda_w'}(\theta_w, \phi_w) 
		= \sum_{\lambda_t,\lambda'_t, \lambda_b} \rho^t_{\lambda_t\lambda'_t} 
			\mathcal{M}^R_{\lambda_t\lambda_w\lambda_b}
			\mathcal{M}^{R*}_{\lambda'_t\lambda'_w\lambda_b}	\nn\\
		& \hspace{2em}
		= \,\frac{N^2}{2}
			\begin{pmatrix}
				g_R^2 \pp{ 1 + \bm{n}_w \cdot \bm{s}_t } &
				\frac{g_R^2}{\sqrt{2} r} \, L(\theta_w, \phi_w) &
				0 	\\
				\frac{g_R^2}{\sqrt{2} r} \, L^*(\theta_w, \phi_w) &
				\frac{1}{2\,r^2} \bb{ g_{+}^2  + g_{-}^2 \bm{n}_w \cdot \bm{s}_t } &
				\frac{g_L^2}{\sqrt{2} r}  \, L(\theta_w, \phi_w)	\\
				0&
				\frac{g_L^2}{\sqrt{2} r} \, L^*(\theta_w, \phi_w) &
				g_L^2 \pp{ 1 - \bm{n}_w \cdot \bm{s}_t }
			\end{pmatrix},
\label{eq:tbw-w-denmtx-R}
\end{align}
where we defined $g_{\pm}^2 \equiv g_L^2 \pm g_R^2$ and $\bm{n}_w = (\sin\theta_w \cos\phi_w, \sin\theta_w \sin\phi_w, \cos\theta_w)$,
and 
\beq
	L(\theta_w, \phi_w) \equiv b_1 \pp{\cos\theta_w \cos\phi_w + i\sin\phi_w } 
		+ b_2 \pp{\cos\theta_w \sin\phi_w - i \cos\phi_w } - \lambda_t \sin\theta_w.
\eeq
Compared with \eq{eq:denmtx-massive-1}, we can extract the unnormalized polarization parameters,
\bse\label{eq:W-pol-R}\begin{align}
	{\rm tr}W^R &= W^R_{++} + W^R_{00} + W^R_{--}
		&&= \pp{ \frac{1}{2\,r^2} + 1 } g_{+}^2 + \pp{ \frac{1}{2\,r^2} - 1 } g_{-}^2 \,\bm{n}_w \cdot \bm{s}_t,\\
	J_1 &= \sqrt{2} \,{\rm Re}\pp{W^R_{+0} + W^R_{0-} }
		&&= \frac{ g_{+}^2 }{ r} \cdot {\rm Re}\bb{ L(\theta_w, \phi_w) },	\\
	J_2 &= - \sqrt{2} \,{\rm Im}\pp{W^R_{+0} + W^R_{0-} }
		&&= - \frac{ g_{+}^2 }{ r } \cdot {\rm Im}\bb{ L(\theta_w, \phi_w) },	\\
	J_3 &= W^R_{++} - W^R_{--}
		&&= - g_{-}^2 + g_{+}^2 \,\bm{n}_w \cdot \bm{s}_t, 	\\
	Q_{yz} &= - \sqrt{2} \,{\rm Im}\pp{W^R_{+0} - W^R_{0-} }
		&&= \frac{g_{-}^2}{r}  \cdot {\rm Im}\bb{ L(\theta_w, \phi_w) },	\\
	Q_{xz} &= \sqrt{2} \,{\rm Re}\pp{W^R_{+0} - W^R_{0-} }
		&&= - \frac{g_{-}^2}{r} \cdot {\rm Re}\bb{ L(\theta_w, \phi_w) },	\\
	\delta_L &= W^R_{++} + W^R_{--}  - 2W^R_{00} 
		&&= - \pp{ \frac{1}{r^2} - 1 }  g_{+}^2 -  \pp{ \frac{1}{r^2} + 1 }  g_{-}^2 \,\bm{n}_w \cdot \bm{s}_t,	\\
	\xi &= 2 {\rm Re} \pp{W^R_{+-} } 
		&&= 0,	\\
	Q_{xy} &= -2 {\rm Im} \pp{W^R_{+-} } 
		&&= 0,
\end{align}\ese
where a common factor $N^2 / 2$ has been omitted.
Note that when the top quark is unpolarized ($\bm{s}_t = \bm{0}$), only the diagonal elements in \eq{eq:tbw-w-denmtx-R} survive,
so only $J_3$ and $\delta_L$ are nonzero; if parity further conserves ($g_L = g_R$), only $\delta_L$ is allowed.

\subsection{Polarization of the $W$ boson in the boosted top frame}
While one can (in principle) always keep the full dynamic information by analyzing the top decay events in their rest frame 
and constructing the full $W$ decay distributions, it is more desirable to analyze boosted top quarks within the boosted frame.
Boosted top quarks are likely to originate from the decay of heavy particles beyond the SM, so both their production rate
and polarization information can serve as useful probes of new physics~\citep{Schatzel:2013wsr}. 
In this kinematic regime, a top quark decays into collimated particles, exhibiting a cone signature that 
can greatly enhance the selection efficiency of boosted top quark events with respect to the 
intrinsic $W+j$ background, which compensates for the small production rate~\citep{Abdesselam:2010pt}. 
On the other hand, the semileptonic decay mode no longer retains special advantages over the hadronic mode, 
and one ought to take the latter into account to enhance the statistics.
Identifying hadronically decayed boosted top quarks first requires distinguishing them from light QCD jet background.

Following the tagging procedures, one may identify the top decay products as different subjets within the fat jet.
Because the finite granular size of the detector leads to large uncertainties of the angular separations (especially in polar angles) 
among the subjets inside the top jet, going back to the top rest frame is not a valid choice, and it is more appropriate to analyze the
top quark jet substructure with directly measured kinematic observables. 
This makes the azimuthal correlation like \eq{eq:g-jet-qq} a good candidate for such analysis, which is very suitable for hadronic
decay mode since $\cos2\phi$ and $\sin2\phi$ do not require identifying particle types but can be measured from the azimuthal
energy deposition anisotropy. In contrast, $\cos\phi$ and $\sin\phi$ signatures do not have such advantages.

As we have seen in \sec{sec:linear-pol-gluon}, $\cos2\phi$ and $\sin2\phi$ arise from linear polarization of a vector boson.
In the top rest frame, the linear $W$ polarization $\xi = Q_{xy} = 0$ independent of the interaction details.
Now we boost the top decay system along the $\hat{z}$ direction by $\Lambda_t = \Lambda(\beta_t)$
to recover the boosted top momentum in the lab frame, 
\beq
	p^{\mu}_t = \Lambda_t (m_t, 0,0,0)^T = (E_t, 0, 0,p_t),
\eeq
where $\beta_t = p_t / E_t$ is determined by top momentum $p_t$ and energy $E_t$. 
This does not change the top helicity state but changes the $b$ and $W$ states, 
which transform according to their little groups, such that in the boosted top frame, 
the $t\to bW$ amplitude $\mathcal{M}^B_{\lambda_t\lambda_w\lambda_b}$ is related to 
$\mathcal{M}^R_{\lambda_t\lambda_w\lambda_b}$ in \eq{eq:tbw-MR} by
\begin{align}
	\mathcal{M}^B_{\lambda_t\lambda_w\lambda_b}(\theta_w, \phi_w) 
		= e^{-i \lambda_b \Theta(\Lambda_t, p_b) }
			\sum_{\lambda_w'}
				D^1_{\lambda_w\lambda_w'}\pp{ W(\Lambda_t, p_w) }
				\mathcal{M}^R_{\lambda_t\lambda_w' \lambda_b}(\theta_w, \phi_w) ,
\label{eq:tbw-MB}
\end{align}
where we describe each amplitude in terms of the $W$ angles in the top rest frame, 
but the helicity of each particle is with respect to the specified frame.
The explicit forms of the little group elements $W(\Lambda_t, p_w)$ and $\Theta(\Lambda_t, p_b)$ have been worked out
in Eqs.~\eqref{eq:little-group-chi} and \eqref{eq:boost-transformation-massless}, respectively, with $\Theta(\Lambda_t, p_b) = 0$
and $W(\Lambda_t, p_w)$ being a rotation $R_y(\chi)$ around the $\hat{y}$ direction by an angle $\chi$, 
as determined by the top quark speed $\beta_t$ and $W$ kinematics in the top rest frame,
\beq
	\cos\chi(\beta_t, \theta_w) = 
	\frac{v_w + \beta_t \cos\theta_w}{
		\sqrt{ (1+\beta_t \, v_w \cos\theta_w)^2 - (1-\beta_t^2)(1-v_w^2) }
	},
\eeq
where $v_w = (1 - r^2) / (1 + r^2) \simeq 0.64$ is the $W$ speed in the top rest frame. 
Then \eq{eq:tbw-MB} becomes
\beq[eq:tbw-MB-d]
	\mathcal{M}^B_{\lambda_t\lambda_w\lambda_b}(\theta_w, \phi_w) 
	= \sum_{\lambda_w'} d^1_{\lambda_w\lambda_w'} \big( \chi(\beta_t, \theta_w) \big)
		\mathcal{M}^R_{\lambda_t\lambda_w' \lambda_b}(\theta_w, \phi_w) .
\eeq

The $W$ density matrix $W^B$ in the boosted top frame can be obtained from \eq{eq:tbw-MB-d} in the same way as \eq{eq:tbw-w-denmtx-R}.
It is related to $W^R$ by a rotation like for a rank-2 tensor,
\begin{align}
	W^B_{\lambda_w \lambda_w'}(\theta_w, \phi_w) = 
		\sum_{\bar{\lambda}_w, \, \bar{\lambda}_w'}
		d^1_{\lambda_w \bar{\lambda}_w} \big( \chi(\beta_t, \theta_w) \big)
		d^1_{\lambda_w' \bar{\lambda}'_w} \big( \chi(\beta_t, \theta_w) \big)
		W^R_{\bar{\lambda}_w \bar{\lambda}_w'}(\theta_w, \phi_w).
\label{eq:tbw-w-denmtx-B}
\end{align}
This translates into the transformations of the $W$ polarization parameters,
\bse\label{eq:tbw-spin-mixing}\begin{align}
	J_1' & = J_1 \cos\chi + J_3 \sin\chi, 	\\
	J_3' & = J_3 \cos\chi - J_1 \sin\chi, 	\\
	J_2' & = J_2,	\\
	Q_{yz}' & = Q_{yz} \cos\chi - Q_{xy} \sin\chi, 	\\
	Q_{xy}' & = Q_{xy} \cos\chi + Q_{yz} \sin\chi,	 \\
	Q_{xz}' & = Q_{xz} \cos2\chi - \frac{\xi - \delta_L}{2} \sin2\chi, \\
	\xi' & = \frac{3\xi + \delta_L}{4} + 
		\frac{1}{2} \pp{ Q_{xz} \sin2\chi + \frac{\xi - \delta_L}{2} \cos2\chi }, \\
	\delta_L' & = \frac{3\xi + \delta_L}{4} 
		- \frac{3}{2}\pp{ Q_{xz} \sin2\chi + \frac{\xi - \delta_L}{2} \cos2\chi }.
\end{align}
\ese
where the primed labels refer to the ones in the boosted top frame.
The last three equations can be rewritten as
\bse
\begin{align}
Q_{xz}' & = Q_{xz} \cos2\chi - \frac{\xi - \delta_L}{2} \sin2\chi,
\\
\frac{\xi' - \delta_L'}{2} &= Q_{xz} \sin2\chi + \frac{\xi - \delta_L}{2} \cos2\chi,
\\
3\xi' + \delta_L' &= 3\xi + \delta_L.
\end{align}
\ese
In this way, the polarization parameters mix under the little group rotation. 
$(J_1, J_3)$ and $(Q_{xy}, Q_{yz})$ mix among themselves, so that $J_1^2 + J_3^2$ and $Q_{xy}^2 + Q_{yz}^2$ are two rotational invariants. 
$Q_{xz}$ and $(\xi - \delta_L)/2$ mix with each other, so $Q_{xz}^2 + (\xi - \delta_L)^2/4$ is rotationally invariant. 
Another two invariants are $J_2$ and $3\xi + \delta_L$.
These transformations are reminiscent of the quadrupole interpretations of the polarization parameters in 
Eqs.~\eqref{eq:quadrupole-Q} and \eqref{eq:quadrupole-xi}.

The mixing of those parameters does not alter the physics information of the $tbW$ interaction encoded in the $W$ polarization, 
but does change the angular distribution of the $W$ decay. Especially, we see that the linear polarizations $\xi$ and $Q_{xy}$ can generate
nonzero values in the boosted top frame via mixing with other parameters. The resultant $\cos2\phi$ and $\sin2\phi$ distributions can be more 
easily measured to reveal hidden interaction information.
This mixing is the source of generating {\it new} kinds of polarization observables in the boosted top system, 
which are absent in the rest frame of top quark. It is due to the massiveness (or off-shellness) of the $W$. 
If we are looking at an on-shell photon or gluon, its helicity will be conserved under the boost, so that if $\xi$ or $Q_{xy}$ 
(or $\xi_1$ and $\xi_2$ in the notations for massless spin-1 particle in \eq{eq:denmtx-vector-g}) is 0 in a certain frame, 
it keeps being 0 in any other frame, cf. \eq{eq:boost-transformation-massless}. 

Using Eqs.~\eqref{eq:tbw-spin-mixing} and \eqref{eq:W-pol-R}, we can get the explicit forms of the polarization parameters in the boosted top frame,
\bse\label{eq:W-pol-B}\begin{align}
	J_1' &= g_{+}^2 \bb{ 
			\frac{\Re\bb{L(\theta_w, \phi_w)}}{r} \cos\chi + (\bm{n}_w\cdot \bm{s}_t) \sin\chi }
			- g_{-}^2 \sin\chi,	\\
	J_2' &= - g_{+}^2 \, \frac{ \Im\bb{ L(\theta_w, \phi_w) } }{ r } ,	\\
	J_3' &= - g_{+}^2 \bb{ 
			 \frac{{\rm Re}\bb{ L(\theta_w, \phi_w) }}{r} \sin\chi
			- (\bm{n}_w\cdot \bm{s}_t) \cos\chi }
			- g_{-}^2 \cos\chi,	\\
	Q_{yz}' &= g_{-}^2 \, \frac{ \Im\bb{ L(\theta_w, \phi_w) } }{ r } \cos\chi,	\\
	Q_{xy}' &= g_{-}^2 \, \frac{ \Im\bb{ L(\theta_w, \phi_w) } }{ r } \sin\chi,	\\
	Q_{xz}' &= - g_{-}^2 \bb{ \frac{\Re\bb{L(\theta_w, \phi_w)}}{r} \cos2\chi
			+ \frac{1+r^2}{2r^2} (\bm{n}_w\cdot \bm{s}_t) \sin2\chi }
			- g_{+}^2 \frac{1-r^2}{2r^2}  \sin2\chi,	\\
	\xi' &= -g_{-}^2 \bb{ \frac{\Re\bb{L(\theta_w, \phi_w)}}{2r} \sin2\chi
			+ \frac{1 + r^2}{2 r^2} (\bm{n}_w\cdot \bm{s}_t) \sin^2 \chi }
			- g_{+}^2 \frac{1-r^2}{2r^2} \sin^2 \chi,	\\
	\delta_L' &= g_{-}^2 \bb{ \frac{3 \Re\bb{L(\theta_w, \phi_w)} }{2r} \sin2\chi
			- \frac{1+r^2}{2r^2} (\bm{n}_w\cdot \bm{s}_t) \frac{1+3\cos2\chi}{2} }	
			- g_{+}^2 \frac{1 - r^2}{2r^2}\frac{1+3\cos2\chi}{2},
\end{align}\ese
which are again expressed in terms of the $W$ angles in the top rest frame, and where we have
omitted an overall factor $N^2 / 2$.

\subsection{Decay distribution of the top quark}
The top quark decays into a $b$ and $W$, which further decays into a fermion-anti-fermion pair $f\bar{f}'$. 
In the boosted top frame, this amplitude can be written as 
\beq[eq:t-decay-MB]
	\mathcal{M}^B_{\lambda_t \lambda_b \lambda_f \lambda_{f'}}
	= \frac{i}{(p_w^2 - m_w^2)^2 + i m_w \Gamma_w} 
		\sum_{\lambda_w} \mathcal{M}^B_{\lambda_t \lambda_w \lambda_b}(\theta_w, \phi_w) 
		\mathcal{M}^B_{\lambda_w \lambda_f \lambda_{f'}}(\theta_f^{\star}, \phi_f^{\star}),
\eeq
where all helicities are defined in the boosted top frame, but the event is described by the $W$ angles $(\theta_w, \phi_w)$ in the top rest frame
and the fermion angles $(\theta_f^{\star}, \phi_f^{\star})$ in the $W$ rest frame. The latter is obtained by boosting along $-\bm{p}_w$ and has the  
coordinate system $\hat{x}_w$-$\hat{y}_w$-$\hat{z}_w$,
\beq[eq:w-frame]
	\hat{z}_w = \frac{\bm{p_w}}{ |\bm{p_w}|}, \quad
	\hat{y}_w = \frac{\bm{p_t} \times \bm{p_w}}{|\bm{p_t} \times \bm{p_w}|}, \quad
	\hat{x}_w = \hat{y}_w \times \hat{z}_w,
\eeq
where all momenta are in the boosted top frame. 
The kinematics in the boosted top frame is shown in \fig{fig:top-frame}(a), where it is
clear that the angle $\phi_f$ in the boosted frame is equal to $\phi_f^{\star}$ in the $W$ rest frame 
and characterizes the relative angle between the two successive decay planes of $t\to bW$ and $W \to f \bar{f}'$, respectively.
In \eq{eq:t-decay-MB}, we have separated the propagator of the intermediate $W$ boson and converted its numerator to a polarization sum,
\beq
	- g^{\mu\nu} + \frac{p_w^{\mu} p_w^{\nu}}{m_w^2}
	= \sum_{\lambda_w = \pm, 0} \epsilon^{\mu *}_{\lambda_w}(p_w) \epsilon^{\nu}_{\lambda_w}(p_w) ,
\eeq
with $p_w = p_f + p_{f'}$ being the $W$'s 4-momentum and $\Gamma_w$ its decay width.
Averaging the square of the amplitude in \eq{eq:t-decay-MB} with the top quark spin density matrix gives
\begin{align}
	&\sum_{\lambda_t,\lambda'_t,\lambda_b, \lambda_f, \lambda_{f'}}
		\rho^t_{\lambda_t\lambda'_t} 
		\M^B_{\lambda_t\lambda_b\lambda_f\lambda_{f'}}
		\M^{B *}_{\lambda'_t\lambda_b\lambda_f\lambda_{f'}}	\nn\\
	&=
		\frac{1}{\pp{ p_w^2 -m_w^2 }^2 + m_w^2 \Gamma_w^2}	\,
		\sum_{\lambda_w, \lambda_w'}
		\bb{
			\sum_{\lambda_t, \lambda'_t, \lambda_b} \rho^t_{\lambda_t\lambda'_t} 
			\M^B_{\lambda_t\lambda_w\lambda_b}
			\M^{B *}_{\lambda'_t\lambda'_w\lambda_b}
		}
		\bb{
			\sum_{\lambda_f, \lambda_{f'}}
			\M^B_{\lambda_w\lambda_f\lambda_{f'}}
			\M^{B *}_{\lambda_w'\lambda_f\lambda_{f'}}
		}	\nn\\
	& \simeq 
		\frac{\pi}{m_w \Gamma_w} \delta\pp{ p_w^2 -m_w^2 } \,
		\sum_{\lambda_w, \lambda_w'}
			W^B_{\lambda_w, \lambda_w'}(\theta_w, \phi_w)
			\cdot
				\sum_{\lambda_f, \lambda_{f'}}
				\M^B_{\lambda_w\lambda_f\lambda_{f'}}
				\M^{B *}_{\lambda_w'\lambda_f\lambda_{f'}}
			(\theta_f^{\star}, \phi_f^{\star}),
\label{eq:tbw-ff-M2}
\end{align}
where we used narrow width approximation for the $W$ and the definition of the polarization matrix $W^B$ in the last step.
Using \eq{eq:tbw-ff-M2} and integrating over the top decay phase space, we can express its width as
\begin{align}
	\Gamma_t =& \frac{1}{2E_t} \frac{1-r^2}{2m_w \Gamma_w}
		\int\frac{\dd\cos\theta_w\dd\phi_w}{32\pi^2}
		\cdot
		\int\frac{\dd\cos\theta_f^{\star}\dd\phi_f^{\star}}{32\pi^2}
		\nn\\
		&\hspace{4em}\times
		\sum_{\lambda_w, \lambda_w'}
		W^B_{\lambda_w\lambda_w'}(\theta_w, \phi_w)
			\cdot
				\sum_{\lambda_f, \lambda_{f'}}
				\M^B_{\lambda_w\lambda_f\lambda_{f'}}
				\M^{B *}_{\lambda_w'\lambda_f\lambda_{f'}}
			(\theta_f^{\star}, \phi_f^{\star}),
\label{eq:top-width-angles}
\end{align}
where we have set the $W$ on shell using the $\delta$-function in \eq{eq:tbw-ff-M2}.

Boosting the $W$ rest frame for the $W\to f\bar{f}'$ system changes the helicity states of $f$ and $\bar{f}'$ according to their 
little group rotations, but does not change the $W$ helicity state. Due to the sum over $\lambda_f$ and $\lambda_{f'}$, such
rotation matrices reduce to identity by their unitarity, so the $W$ decay amplitude square can be equally described in the $W$ 
rest frame, 
\beq
	\sum_{\lambda_f, \lambda_{f'}}
		\M^B_{\lambda_w\lambda_f\lambda_{f'}}
		\M^{B *}_{\lambda_w'\lambda_f\lambda_{f'}}(\theta_f^{\star}, \phi_f^{\star})
	= \sum_{\lambda_f, \lambda_{f'}}
		\M^R_{\lambda_w\lambda_f\lambda_{f'}}
		\M^{R *}_{\lambda_w'\lambda_f\lambda_{f'}}(\theta_f^{\star}, \phi_f^{\star}),
\eeq
with $R$ denoting the $W$ rest frame.
Then we can use the simple result
\begin{align}
	\M^R_{\lambda_w\lambda_f\lambda_{f'}}(\theta_f^{\star}, \phi_f^{\star})
	= C_{\lambda_f, \lambda_{f'}} D^{1*}_{\lambda_w, \,\lambda_f-\lambda_{f'}}(\phi_f^{\star},\theta_f^{\star}, 0)
	= C_{\lambda_f, \lambda_{f'}} e^{i \lambda_w \phi_f^{\star}} d^{1*}_{\lambda_w, \,\lambda_f-\lambda_{f'}}(\theta_f^{\star}),
\label{eq:W helicity decay}
\end{align}
where the constant coefficient $C_{\lambda_f, \lambda_{f'}}$ only has two nonzero components 
$C_{-1/2,\, 1/2} \equiv C_-$ and $C_{1/2,\, -1/2} \equiv C_+$ when neglecting the fermion masses. 
In the SM only the former exists, but we keep both here for a general discussion. 
Then the matrix element in \eq{eq:top-width-angles} (second line) can be directly obtained from 
\eq{eq:denmtx-massive-1} with the boosted polarization parameters in \eq{eq:W-pol-B}, 
\begin{align}
& \sum_{\lambda_w, \lambda_w'}
		W^B_{\lambda_w\lambda_w'}(\theta_w, \phi_w)
			\cdot
				\sum_{\lambda_f, \lambda_{f'}}
				\M^R_{\lambda_w\lambda_f\lambda_{f'}}
				\M^{R *}_{\lambda_w'\lambda_f\lambda_{f'}}
			(\theta_f^{\star}, \phi_f^{\star}) 
\nn\\
&\hspace{0.2em} = 
	\frac{|C_+|^2 + |C_-|^2}{2} \frac{N^2}{2} \bigg\{
		\frac{2}{3} \bb{ g_{+}^2 \pp{ \frac{1}{2\,r^2} + 1 }  + g_{-}^2 \pp{ \frac{1}{2\,r^2} - 1 } \,\bm{n}_w \cdot \bm{s}_t }
		\nn\\
			& \hspace{11em}
			- \frac{\delta_L'\pp{\theta_w, \, \phi_w}}{6} \pp{ 1- 3\cos^2\theta_f^{\star} }
			+ \frac{1}{2} \xi'\pp{\theta_w, \, \phi_w} \sin^2\theta_f^{\star} \cos2\phi_f^{\star} 
			\nn\\
			& \hspace{11em}
			+ \frac{1}{2} Q_{yz}'\pp{\theta_w, \, \phi_w} \sin2\theta_f^{\star} \sin\phi_f^{\star}
			+ \frac{1}{2} Q_{xz}'\pp{\theta_w, \, \phi_w} \sin2\theta_f^{\star} \cos\phi_f^{\star}	\nn\\
			& \hspace{11em}
			+ \frac{1}{2} Q_{xy}'\pp{\theta_w, \, \phi_w} \sin^2\theta_f^{\star} \sin2\phi_f^{\star} \bigg\}	\nn\\
			& \hspace{1.5em} +
		\frac{|C_+|^2 - |C_-|^2}{2} \frac{N^2}{2} \bigg[
			J_1'\pp{\theta_w, \, \phi_w} \sin\theta_f^{\star}\cos\phi_f^{\star} 
			+ J_2'\pp{\theta_w, \, \phi_w} \sin\theta_f^{\star}\sin\phi_f^{\star} 	\nn\\
			& \hspace{11em}
			+ J_3'\pp{\theta_w, \, \phi_w} \cos\theta_f^{\star}
		\bigg].
\label{eq:matrix element}
\end{align}
Inserting this into \eq{eq:top-width-angles} gives the full angular distribution of the top quark decay.

Focusing on the azimuthal distribution of the $f$, which is the same in the $W$ rest frame and boosted top frame, we integrate out 
$\theta_f^{\star}$ and $(\theta_w, \phi_w)$ in \eq{eq:top-width-angles}. The former kills the angular components for $\delta_L'$, $Q_{yz}'$, $Q_{xz}'$,
and $J_3'$, and the latter renders $Q_{xy}'$ and $J_2'$ to vanish. In the end, we can express the azimuthal distribution as
\begin{align}
	\frac{1}{\Gamma_t}\frac{\dd{\Gamma_t}}{\dd\phi_f} 
		= \frac{1}{2\pi} \bb{ 1 + \frac{1}{2} \vv{\xi'} \cos2\phi_f + \frac{3\pi}{8} f_w \vv{J_1'} \cos\phi_f },
\label{eq:top-phi}
\end{align}
where $f_w \equiv \pp{ |C_+|^2 - |C_-|^2} / \pp{|C_+|^2 + |C_-|^2} $ describes the parity violation degree in $W$ decay, which is $-1$ in the SM,
and 
\beq
	\vv{O'}  \equiv \frac{2r^2}{g_+^2 (1 + 2r^2)}
		\int \frac{\dd\cos\theta_w\dd\phi_w}{4\pi} \bb{ O'(\theta_w,\phi_w) },
\eeq 
is the averaged $W$ polarization. 
In \eq{eq:top-phi}, we have changed $\phi_f^{\star}$ to the azimuthal angle $\phi_f$ in the boosted top frame since they are the same. 
Using Eq.~\eqref{eq:W-pol-B}, we have
\begin{align}
	\vv{\xi'}
		&= \frac{1}{1+ 2 r^2} \int_{-1}^1 \frac{\dd\cos\theta_w}{2} 
		\cc{ \lambda_t f_t \bb{ r \sin\theta_w \sin2\chi - \pp{ 1 + r^2 } \cos\theta_w \sin^2 \chi }
			- \pp{ 1 - r^2 } \sin^2 \chi
		},
		\nn\\
	\vv{J_1'}
		&= \frac{2 r}{1 + 2r^2} \int_{-1}^1 \frac{\dd\cos\theta_w}{2} 
		\bb{ \lambda_t \pp{ - \sin\theta_w \cos\chi + r \cos\theta_w \sin\chi }
			- f_t \, r \sin\chi
		},
\label{eq:xi-J1}
\end{align}
where $f_t\equiv g_{-}^2/g_{+}^2$ is the parity violation degree in $t\to bW$ decay, which is equal to 1 in the SM.
We see that the angle between the two successive decay planes in the top quark events exhibits $\cos2\phi$ and $\cos\phi$ distributions. 
The $\cos2\phi$ component is due to the linear polarization of $W$ that is parallel or perpendicular to the $t\to bW$ plane, 
which is generated due to the boost effect. 
The $\cos\phi$ component is due to the angular momentum $J_1$ of $W$, which is the interference between $W_L$ and $W_T$, 
and is only present because the $W$ decay violates parity. 

In the case for antitop quark, we have the same angular correlation as in \eq{eq:top-phi}, 
but the coefficient $\vv{ \bar{\xi}' }$ and $\vv{ \bar{J}_1' }$ differ from \eq{eq:xi-J1} by $\lambda_t \to -\lambda_t$ due to {\it CP} invariance.

\subsection{Phenomenology of the azimuthal correlation}
\begin{figure}[htbp]
	\centering
		\includegraphics[scale=0.68]{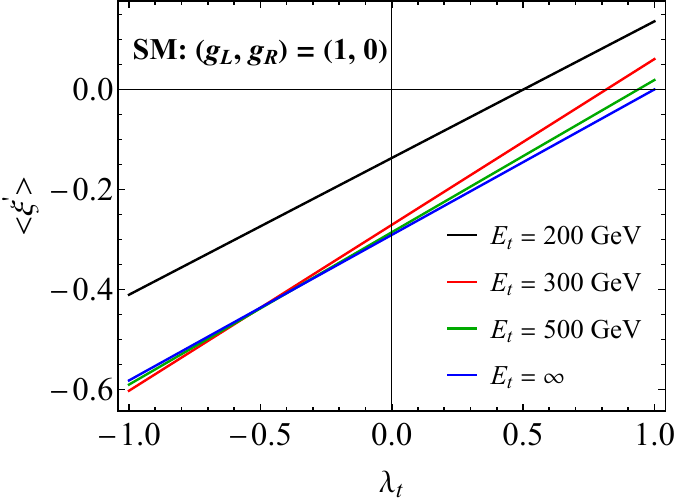}\quad
		\includegraphics[scale=0.68]{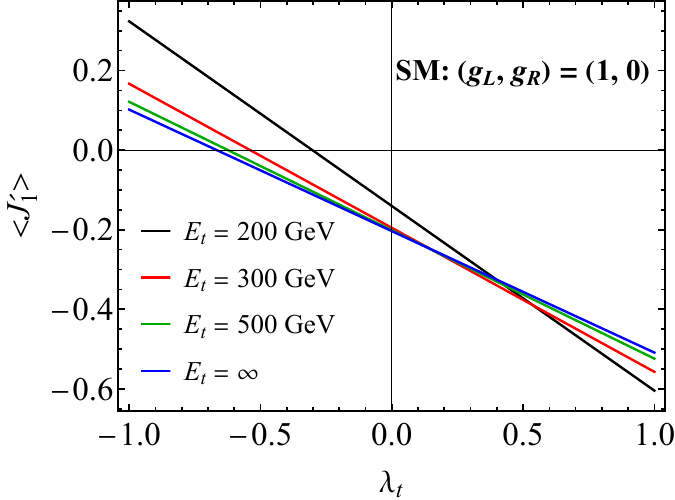}
	\caption{$\vv{\xi'}$ (left) and $\vv{J_1'}$ (right) as functions of $\lambda_t$ at different top quark energies for the SM couplings $f_t = 1$.}
\label{fig:top-xi-J1}
\end{figure}

Since the quantities $\vv{\xi'}$ and $\vv{J_1'}$ in \eq{eq:xi-J1} depend on the top quark energy through its velocity, this dependence
saturates very quickly to infinitely boosted limit, as shown in \fig{fig:top-xi-J1}.
As a good approximation, we define top quarks with $E_t \gtrsim 500~\GeV$ as being boosted, and approximate their $\vv{\xi'}$ and $\vv{J_1'}$
by the infinitely boosted limit ($\beta_t = 1$), at which \eq{eq:xi-J1} gives simple analytic expressions,
\begin{align}
	\vv{\xi'} & = \frac{8r^2}{(1-r^2)(1+2r^2)} \bb{ \frac{1+r^2}{1-r^2}\ln\frac{1}{r} - 1 } \pp{\lambda_t f_t - 1}
		&&\simeq 0.291 \pp{ \lambda_t \, f_t  - 1 },	 \nn\\
	\vv{J_1'} & = - \frac{\pi \, r}{2 (1+r)^2 (1+2r^2)} \bb{ 4 f_t r^2 + \lambda_t (1-r)(1+3r) } 
		&&\simeq -0.203 f_t - 0.305 \lambda_t,
\label{eq:top-xi-J1-N}
\end{align}
where the second equalities are the numerical expressions with $r \simeq 0.46$. In the SM, we have $f_t = 1$.

\subsection{Azimuthal correlation as a boosted top polarimeter}
\begin{figure}[htbp]
	\centering 
	\begin{tabular}{cc}
		\includegraphics[trim={0.4cm -0.6cm 0 0}, clip, scale=0.6]{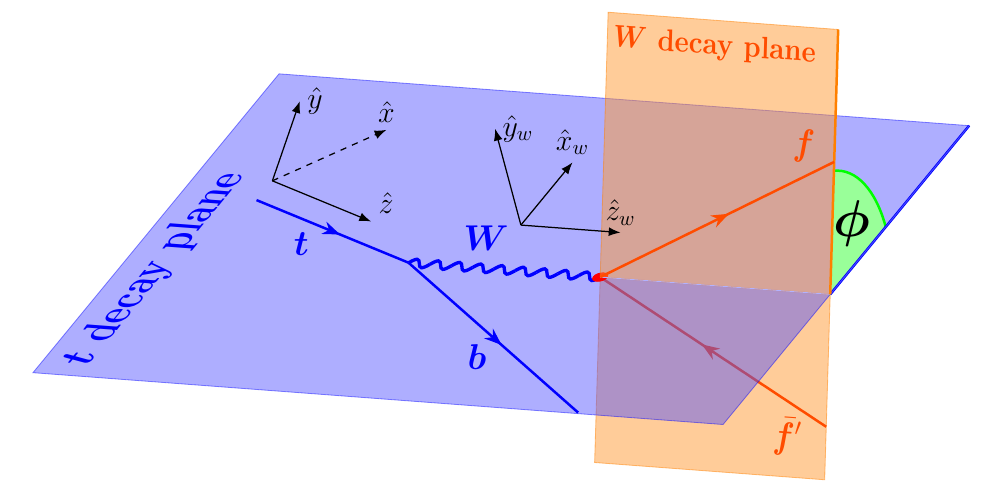} &
		\includegraphics[trim={0 -1.5cm 0 0}, clip, scale=0.65]{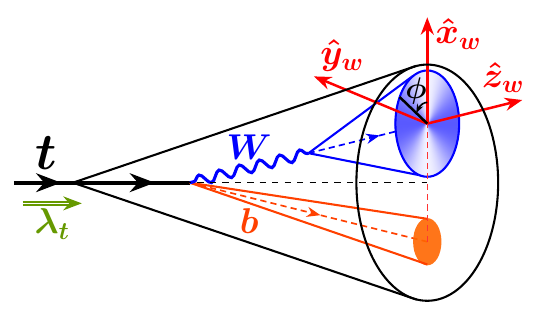} \\
		(a) & (b)
	\end{tabular}
	\caption{(a) The two successive decay planes in $t\to bW (\to f\bar{f}')$ decay process. 
	The coordinate systems of top frame and $W$ frame are shown separately. 
	The $\hat{x}_w$ axis of $W$ frame lies on the $t$ decay plane, while the $\hat{x}$ axis of the top frame may not.
	(b) The azimuthal correlation in the boosted top quark jet is reflected as energy deposition anisotropy, favoring a rounder top jet image. 
	}
	\label{fig:top-frame}
\end{figure}

If the top quark events have been identified and distinguished from the background, then it serves as a definite prediction 
for the azimuthal correlation, in which the linear dependence on $\lambda$ can be used to measure the longitudinal polarization of boosted top quarks
within the boosted frame. 
This can be done for both semileptonic and fully hadronic decay modes of top quark. 
For the semileptonic mode, one needs to first reconstruct the missing neutrino three-momentum by imposing kinematic constraints 
of the event~\citep{ATLAS:2013nki, CMS:2012jea}. Then \eq{eq:top-phi} is exactly the azimuthal correlation for the neutrino (the fermion, not antifermion). 
For the fully hadronic decay mode, however, one cannot distinguish the up-type and down-type quarks in $W$ decay products, 
so can only use $\phi$ in the range of $[0,\pi)$ and sum over the events with $\phi \to \phi+\pi$. 
This kills the $\cos\phi$ component and gives
\beq
	P^j_t(\phi) \equiv \frac{\pi}{\Gamma_t} \frac{\dd\Gamma}{\dd\phi} 
		= 1 + \frac{1}{2} \vv{\xi'} \cos2\phi, \quad
	\phi\in [0,\pi),
\label{eq:top-phi-j}
\eeq
where we have thrown the subscript ``$f$'' in $\phi_f$ since this angle is now reinterpreted as the 
angle between the two successive decay planes, as shown in \fig{fig:top-frame}(a).
This distribution is shown in \fig{fig:top-frame}(b) for a few different values of $\lambda_t$. 
We note particularly that for the unpolarized case ($\lambda_t = 0$), the fluctuation is about $10\%$, which is significant enough for experimental measurement.
By fitting the top event ensemble to the corresponding distribution, one can determine the value of $\lambda_t$.

The measurement of top quark polarization is important for testing the SM and exploring new physics models~\citep{Kane:1991bg,Berger:2011hn}, 
which is commonly done in the top rest frame for the semileptonic decay mode~\citep{ATLAS-CONF-2021-027, ATLAS:2013gil, Jezabek:1994qs, Brandenburg:2002xr, CMS:2019nrx, Mahlon:2010gw, Schwienhorst:2010je, Aguilar-Saavedra:2017wpl}. 
For such measurements in the boosted regime, some methods have been designed~\citep{Shelton:2008nq, Krohn:2009wm, Kitadono:2015nxf, Godbole:2019erb} 
by making use of the energy or polar angular distribution of the decay products.
It is the first time in \citep{Yu:2021zmw} to employ the $\cos2\phi$ correlation as a boosted top polarimeter in the hadronic mode. 

Even though we only performed a LO calculation in the analysis, the $\cos2\phi$ correlation arises from the $W$ boson polarization, 
which is robust against perturbative QCD correction~\citep{Do:2002ky} and parton showering. 
In reality, we need to take the latter into account by defining an infrared safe observable. 
Note that the energies of $W$ decay products are not correlated with the azimuthal angle $\phi$, 
and therefore \eq{eq:top-phi-j} can directly translate into energy distribution in the transverse plane of the $W$ frame,
\begin{align}
	\frac{\dd E}{\dd \phi} = 
		\frac{E_{\rm tot}}{2\pi} \pp{ 1 + \frac{1}{2} \langle \xi' \rangle \cos2\phi } ,
		\quad
		\phi \in [0, 2\pi),
	\label{eq:energy phi distribution}
\end{align}
where $E$ can also be taken as the transverse momentum in the $W$ frame, which is equally infrared safe, and we have extended $\phi$ to $[0, 2\pi)$.
From \eq{eq:top-xi-J1-N}, we see that $\vv{\xi'}$ is always negative, so that more energy is deposited perpendicular to the $t\to bW$ plane, making 
the top quark jet image tend to be round, as shown in \fig{fig:top-frame}(b).

\begin{figure}[htbp]
	\centering
	\includegraphics[scale=0.4]{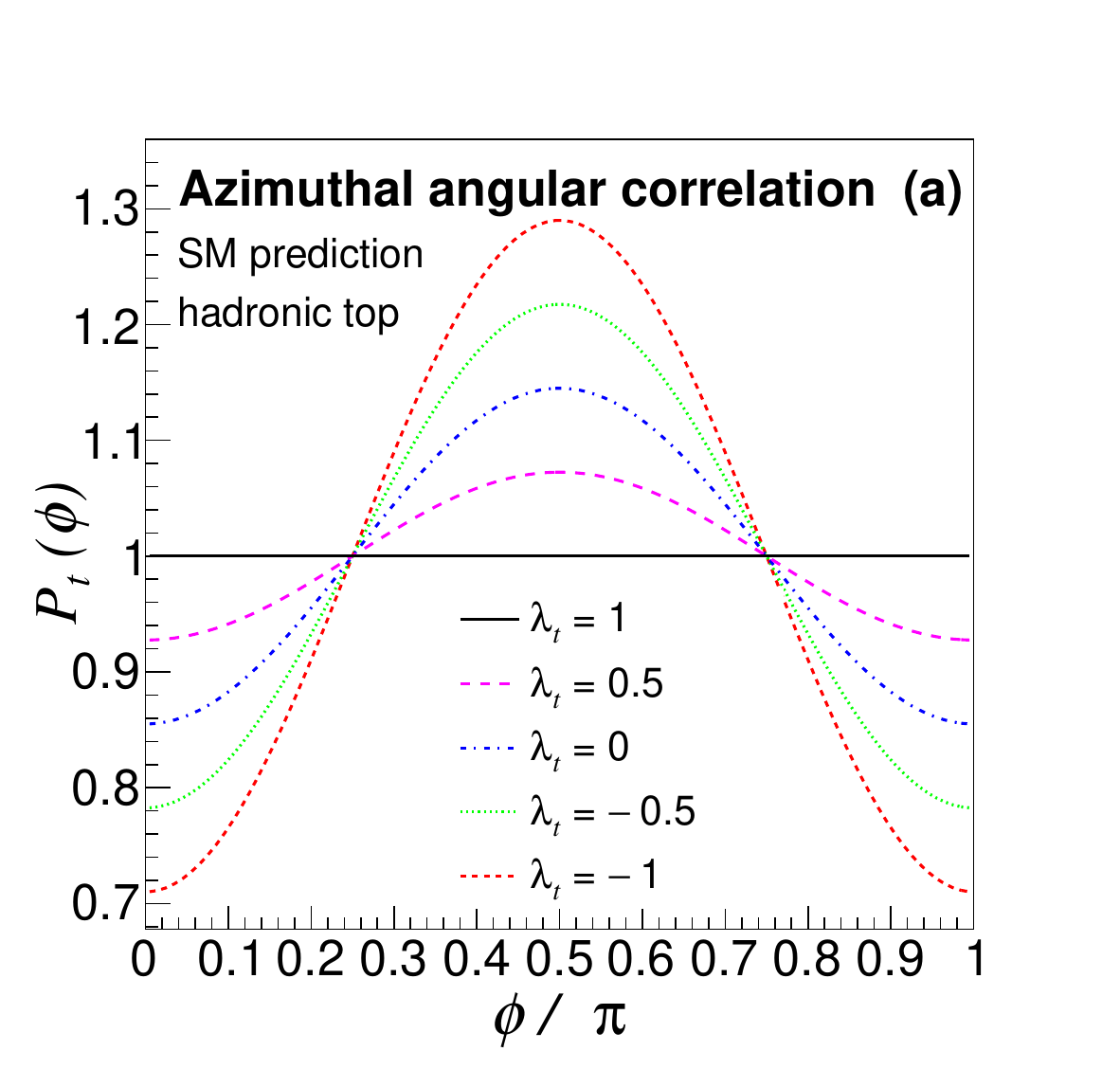}
	\includegraphics[scale=0.4]{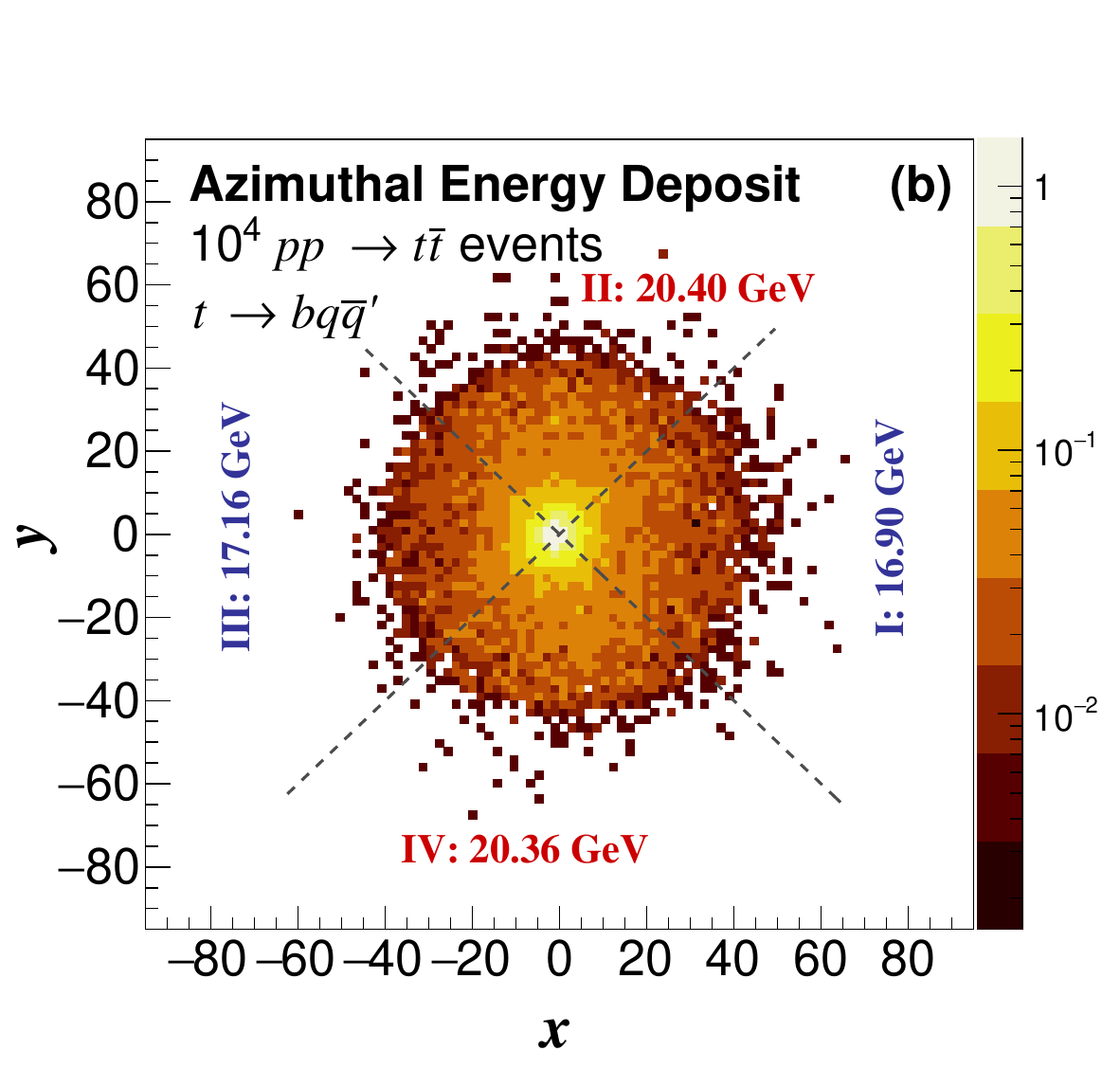}
	\caption{(a) Azimuthal angular correlation in the decay of boosted top quark for different values of top longitudinal polarization $\lambda_t$. 
		(b) The transverse momentum distribution of $W$ decay products in the azimuthal plane of $W$ frame, 
		viewed from the $\hat{z}_w$ direction in \fig{fig:top-frame}. 
		The accumulated transverse momentum, averaged over $10^4$ events, has been indicated in each quadrant.}
\label{fig:top-phi}
\end{figure}

The $\cos2\phi$ distribution leads to an asymmetry of azimuthal energy deposition between the regions 
with $\cos2\phi > 0$ and $\cos2\phi < 0$, which divides the transverse plane into four quadrants, 
as shown by the two dashed diagonal lines in \fig{fig:top-phi}(b). 
This consideration motivates the following method to extract the coefficient $\langle \xi' \rangle$ that is suitable in experimental analysis:
\begin{enumerate}
\item[(1)] construct the top jet and its four-momentum $p_t^{\mu}$;
\item[(2)] use jet substructure technique with $b$ tagging to reconstruct the $b$ subjet with its four-momentum $p_b^{\mu}$;
\item[(3)] determine the $W$'s four-momentum $p_w^{\mu} = p_t^{\mu} - p_b^{\mu}$;
\item[(4)] construct the $W$ frame coordinate system ($\hat{x}_w$-$\hat{y}_w$-$\hat{z}_w$) 
	as in \eq{eq:w-frame} and \fig{fig:top-frame}; and
\item[(5)] remove the particles in the $b$ subjet and determine the energy distribution 
	of the rest of top quark jet in the transverse plane ($\hat{x}_w$-$\hat{y}_w$).
\end{enumerate}
This method does not require identifying the quarks or subjets from $W$ decay. 
As a demonstration, in \fig{fig:top-phi}(b) we show the transverse energy deposit distributed in the azimuthal plane of $W$ frame, 
which is the average of $10^4$ hadronically decayed top quarks with $p_T \in (500, 600)~{\rm GeV}$ from the $t\bar{t}$ pair production 
in proton-proton collision at $\sqrt{s} = 13~{\rm TeV}$. 
The decayed events are generated with {\tt MG5\_aMC@NLO 2.6.7}~\citep{Alwall:2014hca} 
at leading order and passed to {\tt Pythia 8.307}~\citep{Bierlich:2022pfr} for parton showering, 
with full initial and final state radiations. 
Since hadronization is not correlated with the azimuthal distribution, it will not change the infrared-safely defined azimuthal asymmetry. 
A similar argument also holds for the effect of underlying events that cancel in the asymmetry observable.
The anti-$k_T$ algorithm~\citep{Cacciari:2008gp} implemented in {\tt FastJet 3.4.0}~\citep{Cacciari:2011ma, Cacciari:2005hq} 
is used for the jet analysis, with a radius parameter $R = 1.0$ for finding the top jets and $R= 0.2$ for 
reclustering the top jets and identifying the $b$-tagged subjets. 
The energy deposits in the four quadrants are denoted as $E_1, \cdots, E_4$, sequentially, 
which have been indicated in \fig{fig:top-phi}(b). 
Evidently, there are more energy deposits in the $\hat{y}_w$ direction, perpendicular to the $tbW$ plane, 
than the $\hat{x}_w$ direction, which is parallel to the $tbW$ plane. Then we have
\begin{align}
	\langle \xi' \rangle 
		= \pi \cdot \frac{(E_1 + E_3) - (E_2 + E_4) }{(E_1 + E_3) + (E_2 + E_4) }.
\label{eq:top-xi-exp-def}
\end{align}
This gives $\langle \xi' \rangle = -0.282 \pm 0.032$ in the simulated $t\bar{t}$ events, which agrees well with analytic calculation 
in \eq{eq:top-xi-J1-N} for top helicity $\lambda_t = 0$. 
The quoted uncertainty is only of statistical origin, which is the dominant uncertainty in 
asymmetry observables~\citep{ATLAS:2019fgb, CMS-PAS-SMP-21-002}.
When using the same event selection criteria as in \citep{ATLAS:2022mlu}, which yields $17\, 261$ boosted $t\bar{t}$ events 
at the LHC Run-2 with $139~\ivfb$ integrated luminosity, we obtain an uncertainty $\delta{\langle \xi' \rangle} = 0.024$. 
Hence, the azimuthal correlation can already be observed with the Run-2 data. 
Since $\delta{\langle \xi' \rangle} \propto 1/\sqrt{N_{\rm events}}$, we can project an uncertainty of 0.016 for 
$300~\ivfb$ at the LHC Run-3 and 0.0052 for $3000~\ivfb$ at the High-Luminosity LHC~\citep{Apollinari:2120673}. 
It is evident that the LHC data allow the precision measurement of such azimuthal correlation,
making the latter a good polarimeter for boosted top quarks.

\subsection{Azimuthal correlation as a top quark jet tagger}

The derivation of Eqs.~\eqref{eq:tbw-spin-mixing} and \eqref{eq:top-phi} makes it clear that the $\cos2\phi$ azimuthal correlation 
is not only relevant to boosted top quarks, but also to any boosted $1\to 3$ decay systems as long as 
they are mediated by virtual vector bosons, such as boosted QCD jets with a virtual gluon, 
boosted $b\to s l^+ l^-$ decay through a virtual photon or $Z$ boson, or $b\to c \bar{\nu}_l l^- $ decay via a virtual $W$. 
In more general cases with {\it CP} violation, there will also be an additional $\sin2\phi$ correlation.

A particular example is the three-pronged QCD jets, for which the azimuthal angular correlation 
$P_j(\phi) = 1 + (\langle \xi_j \rangle /2) \cos2\phi$ has been pointed out for the three-point energy correlator~\citep{Chen:2020adz}. 
This is relevant to boosted top quarks because QCD jets can be a source of background of the latter and 
needs to be distinguished when studying the hadronically decayed boosted top quarks.
However, there are more diagrams contributing to the three-point energy correlator of QCD jet that are not mediated by a virtual gluon.
Furthermore, for the diagrams that {\it are} mediated by a virtual gluon, the splittings of $g^*\to gg$ and $g^*\to q\bar{q}$ 
are not distinguishable if no flavor tagging criterion is imposed, and their contributions to the $\cos2\phi$ correlation have opposite signs to each other~\citep{Chen:2020adz, Hara:1988uj}, as we have seen in the polarized gluon jet in \sec{ssec:gluon-jet-f}.
As a result, the $\langle \xi_j \rangle$ is rather small. The analytic formula in the collinear limit is given by Eq.~(3) of~\citep{Chen:2020adz}.
For an active fermion number $n_f = 5$, $\langle \xi_j \rangle$ is $-0.02$ for quark jets and $-0.012$ for gluon jets.

In fact, since the $W$ only decays only to a fermion pair, \eq{eq:energy phi distribution} exactly resembles the energy deposition anisotropy for 
quark-subjet-tagged gluon jets in \eq{eq:g-jet-qq}. Without such (sub)jet flavor tagging, the gluon jet fragmentation is dominated by the $g\to gg$
splitting, which greatly washes out the azimuthal correlation. 

This observation motivates the use of the azimuthal correlation as a new top quark jet tagger.
While there have been many tagging algorithms proposed and applied to discriminate boosted top quark events 
from QCD jets and they have reached rather high efficiencies~\citep{CMS-PAS-JME-13-007, Plehn:2010st, ATLAS:2018wis},
the current top quark taggers mainly make use of the top and $W$ mass conditions and the three-subjet structure,
and are highly dependent on machine learning methods ~\citep{10.21468/SciPostPhys.7.1.014, Bhattacharya:2020aid, ATLAS:2022mlu} 
trained on Pythia simulated top and QCD jet events.
Since Pythia fails to incorporate the spin correlations in the parton showering, it is very likely that the azimuthal correlation feature is absent
in both the top and QCD jet events. It is then not clear whether machine learning captured the right physics, and there is a mismatch when it is
applied to real data. Therefore, it is worthwhile to feed the additional azimuthal correlation information to proper machine learning algorithms,
either by improving the parton showering simulator, or by directly examining on real data. We leave this study to future.

Here, instead of constructing an event-by-event top tagger against QCD jets, 
we propose a simpler ``tagger" that acts on the whole ensemble of boosted top candidates to determine the fraction of top quark events.
In this ensemble, one can 
first measure the azimuthal asymmetry coefficient $\xi_0$ following the same strategy discussed above. This $\xi_0$ is not the same as the one for pure top quark events, as given in Eqs.~\eqref{eq:energy phi distribution} and \eqref{eq:top-xi-J1-N}, but is for a mixture of top and QCD jet events.
Then, if the top quark events account for a fraction $\delta_t$ of the whole ensemble, we should have
\beq
	\xi_0 =  \delta_t \, \langle \xi' \rangle + (1 - \delta_t) \, \langle \xi_j \rangle,
\eeq
from which we can determine 
\beq
	\delta_t = \frac{ \xi_0 - \langle \xi_j \rangle }{\langle \xi' \rangle - \langle \xi_j \rangle },
\eeq
where $\langle \xi_j \rangle$ is obtained by averaging over the light quark and gluon jet contributions and only depends on their relative fraction 
in the boosted QCD jet events. As an example, for single top quarks produced via $s$-channel SM-like heavy resonance $W^\prime$ with a mass $> 1~{\rm TeV}$, $\langle \xi' \rangle \sim -0.58$,
while the magnitude of $\langle \xi_j \rangle \lesssim 0.01$. As long as the top quark yield is not more than an order of magnitude smaller than the QCD jet background rate, $\delta_t$ can be precisely determined from the measurement of $\xi_0$ to constrain the parameter space of this new physics model, such as the $W^\prime$-$t$-$b$ coupling strength. 

\subsection{Conclusion}
In this section, we examined the linear $W$ polarization in a boosted top quark decay and showed that
it leads to a nontrivial $\cos2\phi$ azimuthal correlation between the $t\to bW$ and $W\to f\bar{f}'$ decay planes, 
which induces a new substructure observable in the boosted top quark jet.
In the hadronic decay mode, this correlation translates into an energy deposition asymmetry in the azimuthal plane. 
Interestingly, such linear polarization is not present in the top rest frame but only emerges under the boost 
as a result of mixing with other polarization parameters.
This property is due to the massiveness of the $W$ boson, differing from the linear polarization of gluons in \sec{sec:linear-pol-gluon}.
As phenomenological applications, we demonstrated that such correlation can be used to 
either measure the longitudinal polarization of a boosted top quark for testing the SM and probing new physics 
or distinguish a boosted top quark from the QCD jet background.


\chapter{Summary and Outlook}
\label{ch:spin-summary}

Spin property is a long-studied subject throughout the history of particle physics but 
remains relatively poorly explored in the context of high-energy unpolarized colliders such as the LHC. 
Following the early works, we re-emphasized the importance of transverse polarization phenomena
at the LHC, which correspond to the quantum interference effects between different helicity states,
entail information about the hard scattering that is not probed by the unpolarized production rate, and
can bring out a wealth of new physical observables. 
In particular, in the boosted regime, a heavy unstable particle produced with a transverse polarization
can lead to a jet of decay products with a new substructure characterized by certain azimuthal
correlations.

We have discussed two kinds of single transverse polarization productions. 
The first kind is to have the polarized particle directly produced from the hard scattering 
with a hard transverse momentum. This applies to 
both spin-half quarks and spin-one massless gluons and massive $W$ and $Z$ bosons. 
For the quark case, chiral symmetry requires a nonzero quark mass, which strongly suppresses the degree of 
the transverse spin, except for heavy quarks like the top quark.
No such suppression exists for the linear polarization of vector bosons, and one generally expects a large degree
of linear polarization. 
At the hadron colliders, however, the lab frame generally differs from the c.m.~frame of the hard scattering by a 
longitudinal boost, under which transverse polarizations of massive particles mix with other polarization components
but those of massless particles remain invariant. 
As a result, it is better to measure the polarization in the partonic c.m.~frame.
The second kind is for linearly polarized vector bosons that are not directly produced from the hard scattering,
but appear from the decay of a boosted heavy object. One example is the linear gluon polarization in a parton 
showering. The other example, as we discussed in detail, is for the boosted top quark that decays into a collimated
pair of bottom quark and $W$ boson. The linearly polarized $W$ leads to a nontrivial azimuthal correlation, which
is in turn reflected as a new boosted top quark jet substructure.

The main idea of linear polarization is the resultant azimuthal correlation caused by helicity interference effects.
Therefore, the subject of this thesis can be readily extended to much broader physical contexts. 
A direct application is to use azimuthal correlations to determine the spin of the mother particle.
Further, in the context with new physics extensions of the SM, possible fermionic tensor interaction can lead to 
different fermion helicity structures from the SM interactions, which can interfere and generate nonzero fermion
transverse spin. Similarly, new physics operators can also generate linearly polarized vector bosons. 
In particular, with possible $CP$-violating new physics interactions, new correlation functions can appear due to
$CP$-violating transverse polarization components, such as the $\sin2\phi$ components in the polarized gluon
jet. The measurements of transverse polarizations thus provide new opportunities to probe possible new physics.

\end{doublespace}

\bibliographystyle{plainnat}
\bibliography{reference}

\end{document}